%% file: TP_SHiP.tex
\documentclass[12pt,a4paper,openany]{book}
\usepackage{amssymb}
\usepackage{amsmath}
\usepackage{xspace}
\usepackage{pbox}
\usepackage{tabularx,colortbl}
\usepackage[figuresright]{rotating}
\usepackage{dcolumn}
\usepackage{url}
\usepackage{hyperref}
\usepackage{multirow}
\usepackage{booktabs}
\usepackage{fancyhdr}
\usepackage{citesort}
\usepackage{enumerate}
\usepackage[titletoc]{appendix}
\usepackage{mciteplus}

\usepackage[]{units}
\hypersetup{ pdfborder=0 0 0, urlbordercolor=1 0 0 }
\usepackage{cite}

\usepackage{pbox}
\usepackage{tabularx,colortbl}
\usepackage{graphicx}% Include figure files
\usepackage{graphics}
\usepackage{color}
\usepackage{titlesec}
\usepackage[nottoc]{tocbibind}
\usepackage{dcolumn}% Align table columns on decimal point
\usepackage{bm}% bold math
\usepackage{afterpage}
\usepackage{citesort}
\usepackage{subfigure}

\usepackage{afterpage}

\newcommand{\gev}{\ \rm GeV}
\newcommand{\mev}{\ \rm MeV}

\newcommand{\gevc}{\ensuremath{\textrm{GeV}\!/\textrm{c}}}
\newcommand{\gevctwo}{\ensuremath{\textrm{GeV}\!/\textrm{c}^2}}

\newcommand{\mevctwo}{\ensuremath{\textrm{MeV}/\textrm{c}^2}}

\titleformat{\paragraph}
{\normalfont\normalsize\bfseries}{\theparagraph}{1em}{}
\titlespacing*{\paragraph}
{0pt}{3.25ex plus 1ex minus .2ex}{1.5ex plus .2ex}

\begin{document}
                        %==================================================================================
%  Title page

\begin{titlepage}

\belowpdfbookmark{Title page}{title}

\pagenumbering{roman}
\vspace*{-2.5cm}
\centerline{\large EUROPEAN ORGANIZATION FOR NUCLEAR RESEARCH (CERN)}
\vspace*{0.2cm}
\hspace*{-5mm}\begin{tabular*}{17cm}{lc@{\extracolsep{\fill}}r}
%\vspace*{-12mm}\mbox{\!\!\!\epsfig{figure=SHiP-logo,width=.15\textwidth}}& & \\
\vspace*{-10mm}\mbox{\!\!\!\includegraphics[angle=-90,width=0.15\textwidth,clip=]{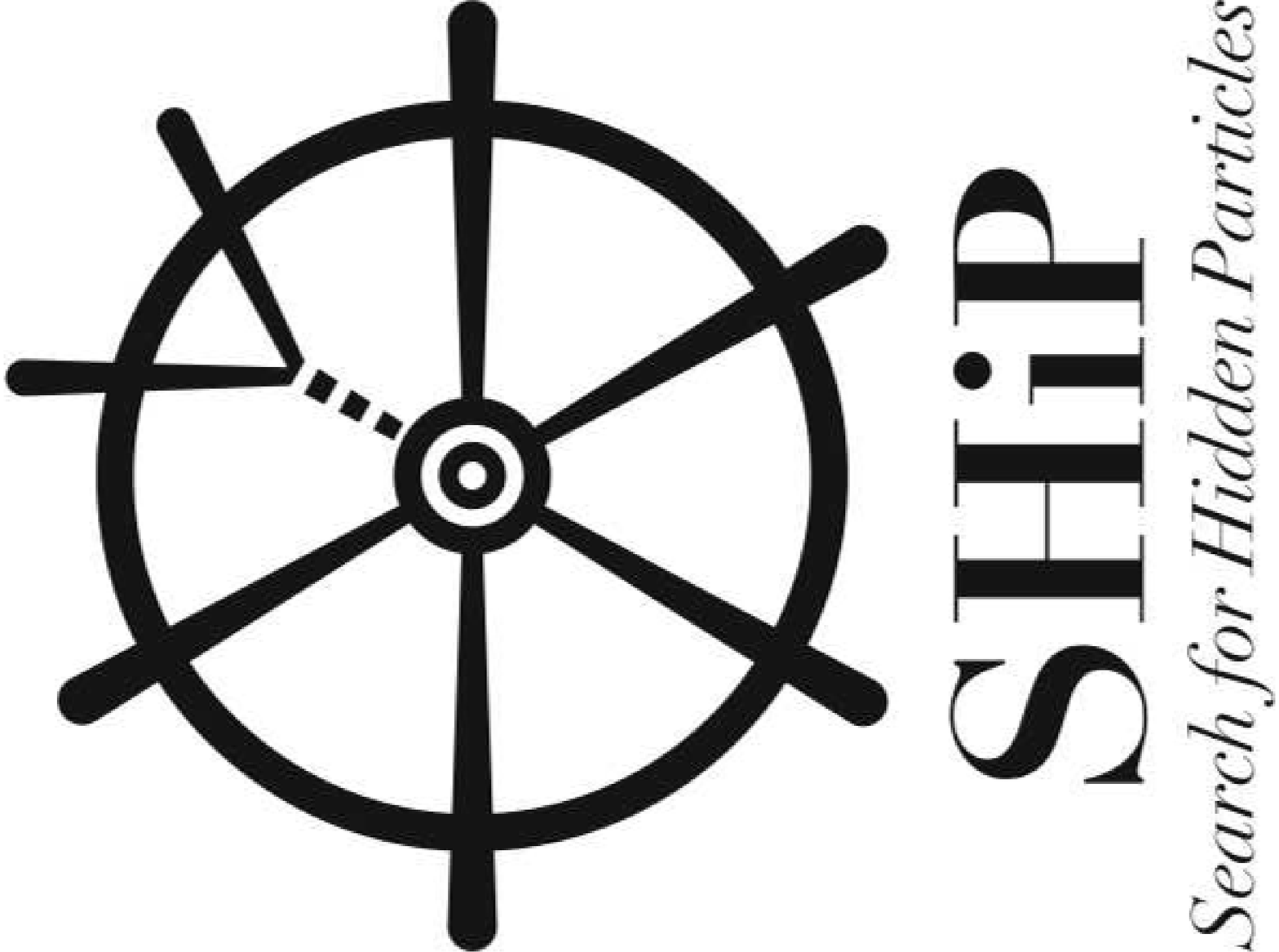}}& & \\
&& CERN-SPSC-2015-016\\
&& SPSC-P-350\\
&& 8 April 2015\\
\end{tabular*}
\vspace*{1cm}
\begin{center}
{\bf\huge\boldmath Technical Proposal \\} 
\vspace*{9mm} 
{\bf\huge\boldmath A Facility to Search for Hidden Particles (SHiP) at the CERN SPS \\}
\vspace*{0.9cm}
\normalsize {
The SHiP Collaboration\footnote{Authors are listed on the following pages.}
%========================================================================%
}
\end{center}
\vspace{\fill}
\centerline{\bf Abstract}
\vspace*{3mm}\noindent  {\small 
A new general purpose fixed target facility is proposed at the CERN SPS accelerator which
is aimed at exploring the domain of hidden particles and make measurements with tau neutrinos.
Hidden particles are predicted by a large number of models beyond the Standard Model.
The high intensity of the SPS 400~GeV beam allows probing a wide variety of models containing 
light long-lived exotic particles with masses below ${\cal O}$(10)~GeV/c$^2$, including very
weakly interacting low-energy SUSY states. The experimental programme of the proposed facility 
is capable of being extended in the future, e.g. to include direct searches for Dark Matter 
and Lepton Flavour Violation.}
\vspace{\fill}

\end{titlepage}

%\clearpage
%\mbox{\ } 
%\clearpage
%\setcounter{page}{2}
%\belowpdfbookmark{SHiP author list}{authors}
\input{SHiP-authorlist.tex}
\input{SHiP-acknowledgments.tex}

\tableofcontents
\clearpage
\setcounter{page}{1}
\pagenumbering{arabic}
\newpage
\input{summary/executive_summary}
\newpage
\chapter{Introduction and Physics Motivation}
\input{intro/Intro}

\input{requirements/experimental_requirements}

\input{facility/Facility}

\input{detector/Detector}

\input{performance/Performance}

\input{cost_schedule/Cost_Schedule}

\input{annexes/Annexes}

\bibliography{TP_SHiP,biblio/bib-intro,biblio/bib-req,biblio/bib-bgtagger,biblio/bib-nutaudet,biblio/bib-tracker,biblio/bib-calo,biblio/bib-muon,biblio/bib-sensitivity}     
\bibliographystyle{SHiP}    % SHiP: Bibliography: Author-Date system

\end{document}

%% file: SHiP-authorlist.tex
%
\centerline{\Large\bf The SHiP Collaboration}
\vspace*{1mm}
\begin{flushleft}
%-- 
%-- SHiP Authorlist as of 26 March 2014
%-- 

M. Anelli$^{14}$,
S.~Aoki$^{17}$,
G.~Arduini$^{33(BE)}$,
J.J.~Back$^{40}$,
A.~Bagulya$^{24}$, 
W.~Baldini$^{11}$, 
A.~Baranov$^{30}$, 
G.J.~Barker$^{40}$,
S.~Barsuk$^{4}$,
M.~Battistin$^{33(EN)}$
% ...
\\

\clearpage

{\footnotesize \it

% ...
$ ^{4}$LAL, Universit\'{e} Paris-Sud, CNRS/IN2P3, Orsay, France\\
% ...
$ ^{11}$Sezione INFN di Ferrara, Ferrara, Italy\\
% ...
$ ^{14}$Laboratori Nazionali dell'INFN di Frascati, Frascati, Italy\\
% ...
$ ^{17}$Kobe University, Kobe, Japan\\
% ...
$ ^{24}$P.N.~Lebedev Physical Institute (LPI), Moscow, Russia\\
% ...
$ ^{30}$Yandex School of Data Analysis, Moscow, Russia\\
% ...
$ ^{33}$European Organization for Nuclear Research (CERN), Geneva, Switzerland\\
% ...
$ ^{40}$Department of Physics, University of Warwick, Coventry, United Kingdom\\
% ...
}
\end{flushleft}

%% file: SHiP-acknowledgments.tex
\section*{Acknowledgments}
The SHiP Collaboration is greatly indebted to all the technical and
administrative staff for their important contributions to the design, testing
and prototype activities. We are grateful for their dedicated work and are aware
that the successful construction and commissioning of SHiP will also in future
depend on their skills and commitment.

We would like to thank E.~Delagnes, E.~Ferrer-Ribas, F.~Jeanneau, 
P.~Schune and M.~Titov for their work on the
$\nu_{\tau}$-detector tracking system and the SAMPIC electronics.

We would like to thank I.~Bezshyiko for her work on the
optimisation of the active muon shield.

We express our gratitude to I.~Bereziuk  for performing 
the initial GARFIELD simulation studies presented in section~\ref{sec:tracker}.
We would like to thank R.~Veenhof for his GARFIELD support, 
H.~Danielsson  for providing useful information about the NA62 straw tracker, Y.~Ermoline and E.~Usenko 
for participating in the discussions on straw tracker electronics. 
In addition, we are grateful to I.~Guz for his help on preparing the R\&D straw tracker test setup.

We would like to thank A.~Chumakov for his help with the integration of GENIE into FairShip. We 
acknowledge the contribution of A.~Chukanov for his help in the generation of neutrino interactions 
in the tau neutrino target. 

We thank very much M.~Al-Turany and F.~Uhlig for their indispensable support for \textsc{FairRoot} and for providing training. 
S.~Neubert for integrating Genfit into \textsc{FairShip} and helping with the track fitting including a tutorial. 
A.~Gheata for helping out to solve issues with the ROOT TGeo package.

We are extremely grateful to
R. Saban and L. Gatignon for the key contribution to the design of the SHiP facility, and also to
R. Losito, R. Folch and A. Ferrari for providing invaluable support for the design of the target complex.

We acknowledge the contributions from
P. deNiverville for providing the plots in Figure~\ref{fig:sig_DM} and G. Cibinetto for Figure~\ref{residuals}.

%% file: summary/executive_summary.tex
\chapter*{Executive Summary}
\addcontentsline{toc}{chapter}{Executive Summary}
\label{sec:ExSummary}

A new general purpose fixed target facility is proposed at the CERN SPS accelerator which
is aimed at exploring the domain of hidden particles and make measurements with tau neutrinos.
Hidden particles are predicted by a large number of models beyond the Standard Model.
The high intensity of the SPS 400~GeV beam allows probing a wide variety of models containing 
light long-lived exotic particles with masses below ${\cal O}$(10)~GeV/c$^2$, including very
weakly interacting low-energy SUSY states. The experimental programme of the proposed facility 
is capable of being extended in the future, e.g. to include direct searches for Dark Matter 
and Lepton Flavour Violation.
% which are capable of providing solutions to the observed shortcomings of the Standard Model and its theoretical problems.

The facility will be serviced by a new dedicated beam line branched off the splitter section 
on the North Area. It is followed by a new target station and a magnetic shield to suppress 
beam induced background. The proposed orientation of the beam line and the underground complex 
allows reserving more than 100~m of space beyond the experimental hall to host
future extensions of the experimental program.
In the first phase, the facility will host a detector to search for hidden particles and a 
compact tau neutrino detector. The hidden particle detector consists of a long, evacuated 
decay volume with a magnetic spectrometer, calorimeters and muon detectors at the far end. 
This allows full reconstruction and particle identification of hidden particle decays. The 
robustness against various types of background is guaranteed by background taggers and a 
dedicated timing detector. The tau neutrino detector consists of an emulsion target equipped 
with tracking in a magnetic field followed by a muon spectrometer.

Under nominal conditions the current SPS is capable of providing an integrated total of 
2$\cdot$10$^{20}$ protons on target in five years of operation. This allows access to a 
significant fraction of the unexplored parameter space for the Hidden Sector, at nearly 
zero background level, with sensitivities which are several orders of magnitude better than 
previous experiments. The associated tau neutrino detector will allow performing a number 
of unique measurements with tau neutrinos, including a first direct experimental observation 
of the anti-tau neutrino interactions.

The SHiP facility will provide a unique experimental platform for physics at the Intensity 
Frontier which is complementary to the searches for New Physics at the Energy Frontier.
A discovery of a very weakly interacting Hidden Sector would lead to a dramatic breakthrough 
in our understanding of particle physics and of the Universe. In addition, it would open up a 
new field of experimental particle physics, which would benefit from accelerators capable of 
exceeding the already outstanding performance of the SPS.

%% file: intro/Intro.tex
\section{Introduction}
\label{sec:intro}

The completion of the particle content of the Standard Model (SM) with the discovery of the Higgs boson, and advances in cosmology highlight the necessity for a new level of understanding
of physics Beyond the Standard Model (BSM). 
At the same time, neither experiment nor theory provide clear hints of the nature or the scale of this new physics.

Over the next decades the Fermi-mass scale, and even beyond, 
will be comprehensively explored either directly by ATLAS and CMS at the LHC (and possibly at a 
hadron Future Circular Collider (FCC) facility), or indirectly, assuming 
generic couplings, at experiments like LHCb, Belle2 and 
NA62\footnote{Experiments at Linear Collider (LC) and at an electron FCC can contribute to both
direct and indirect searches}.
Hidden particles, which interact very weakly with the SM particles, are predicted in 
many theoretical models capable of explaining the shortcomings of the SM. A large part of 
their accessible parameter space remains unexplored.

In this proposal a new general purpose fixed target facility at the SPS, SHiP, is described.
In the initial phase, SHiP aims to explore the domain of hidden particles, such as Heavy Neutral Leptons (HNL), 
dark photons, ligth scalars, supersymmetric particles, axions etc., with masses 
below ${\cal O}$(10)~GeV/c$^2$ and couplings which are orders of magnitude smaller than previously constrained.
In addition it will make measurements with tau neutrinos with three orders of magnitude more 
statistics than available in previous experiments combined.
The SHiP facility provides a unique combination
of intensity and energy capable of producing the large yields required. The experimental conditions 
will be optimized for hidden particles in this mass range.
For example, the charm yield
with the 400 GeV beam on a fixed target will provide by more than an order of magnitude more charm than what can be obtained
at the high luminosity LHC over the same period of time.

The SHiP facility will maximize the physics use of the proton intensity available at the SPS while respecting the demands of the LHC and
the other fixed target experiments. 

This Technical Proposal (TP) focuses on the experimental aspects of the SHiP facility. The complete physics programme is described in the 
Physics Proposal~\cite{PP}. The Physics Proposal also discusses other possible experiments requiring high intensity beams, such as the
search for Charged Lepton Flavour Violating (CLFV) $\tau \rightarrow 3\mu$
decays and the direct search for Dark Matter (DM). The CLFV search requires
a dedicated thin proton target operated in parallel with the SHiP target, while the direct DM search 
can be considered an extension of the SHiP experiment.

Following the SHiP Expression of Interest (EOI)~\cite{Bonivento:2013jag}, the design of the facility and the experiment
has been optimized (see Section~\ref{sec:eoi_diff}). 
This SHiP TP gives a brief physics motivation and a comparison with previous experiments, 
followed by a description of the SHiP experimental objectives and the general requirements.
The conceptual design of the facility and the detectors for the hidden particle 
search and the tau neutrino measurements are described in detail.
The chapter on the physics performance presents a detailed background 
estimation and the sensitivity reach for representative signal channels for both the hidden sector and for the 
tau neutrino physics. The TP is concluded with the schedule and the cost of the fixed target facility and the 
SHiP detectors, followed by details on the organization of the Collaboration, and the interest 
from the collaborating institutions for the construction of the different detector components.

The remaining choices of the baseline technologies 
will be made prior to the submission of the Technical Design Reports in 2018. The  
detector construction and five years of data taking to accumulate
2$\cdot$10$^{20}$ protons on target can be achieved in a period
of 10 years.

%A discovery of a very weakly interacting hidden sector
%would lead to a dramatic breakthrough in our understanding of particle physics and of the Universe. 
%In addition, it would open up a new field of experimental particle physics, which would benefit from accelerators
%capable of exceeding the already unique performance of the SPS.

\section{Physics motivation}
\label{sec:physics}

With the discovery of the Higgs boson \cite{Aad:2012tfa,Chatrchyan:2012ufa} all the predicted constituents of the SM have now been observed.
% and with the crucial observation of neutrino mass present
% experimental results show significant deviation from the predictions of the basic SM.
It seems that the SM provides a phenomenological description of the dynamics of
the electroweak and strong interactions at the Fermi scale, including the Higgs sector of electroweak symmetry 
breaking. Moreover, the mass of the Higgs boson is such that the lifetime of the SM vacuum greatly exceeds the 
age of the Universe~\cite{Altarelli:1994rb,Casas:1994qy,Casas:1996aq,Ellis:2009tp,EliasMiro:2011aa,Bezrukov:2012sa,Degrassi:2012ry,Buttazzo:2013uya}. In addition, it suggests that the Landau pole in the Higgs self-interaction potential, 
which indicates when the theory would become strongly coupled, is well above the Planck scale~\cite{Maiani:1977cg,Cabibbo:1979ay,Lindner:1985uk,Hambye:1996wb,Ellis:2009tp}.

\begin{figure}[tb]
\begin{center}
\includegraphics[width=0.6\linewidth]{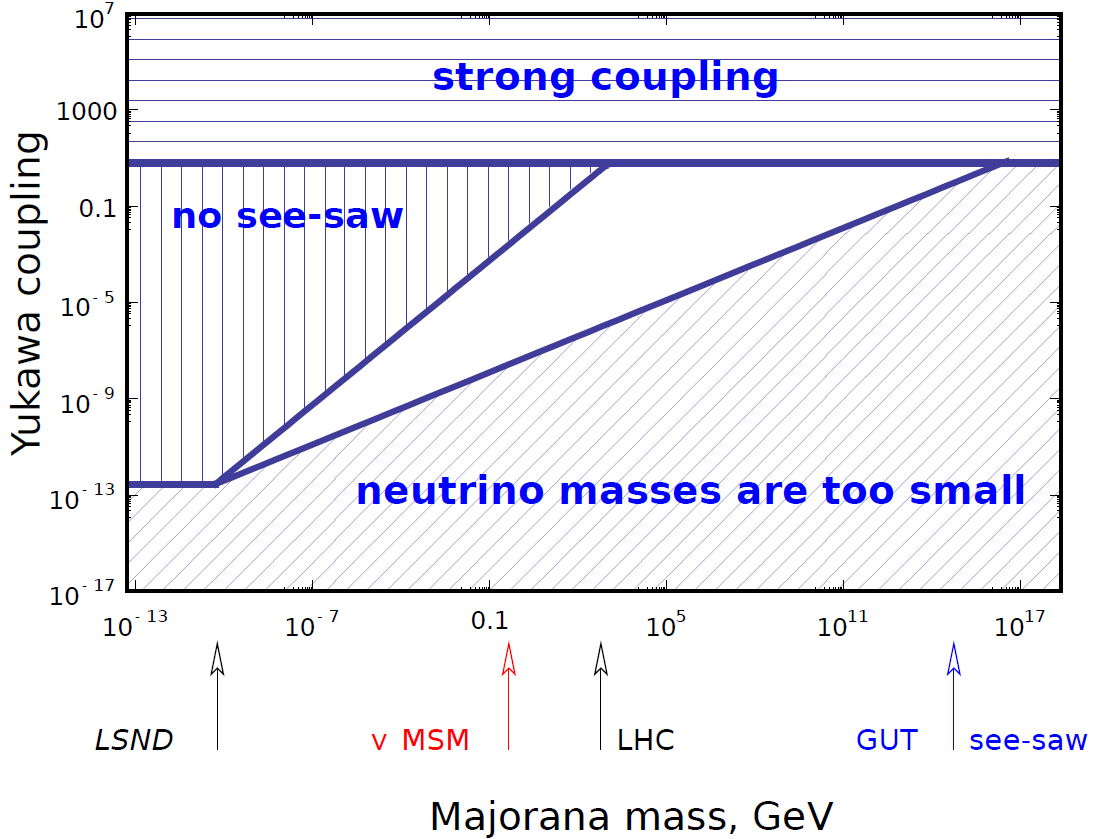}
\caption{Possible values of the Yukawa couplings and Majorana masses of HNLs in seesaw models~\cite{Abazajian:2012ys}.}
\label{fig:intro_seesaw}
\end{center}
\end{figure}

% where the SM must break-down in order to accommodate gravitational interactions.
It is therefore possible that the SM could be an effective, weakly-coupled field theory all the way up to the Planck 
scale. However, the non-zero neutrino mass, the existence of dark matter (DM), and the baryon asymmetry of the 
universe (BAU) show that the SM is an incomplete description of Nature. 
There is also no consensus view on a mechanism to account for 
cosmic inflation. Thus there are compelling experimental and theoretical 
reasons\footnote{Such as a lack of explanation for the gap between the Fermi and the Planck scales, 
dark energy, the connection to gravity, the resolution of the strong CP problem, the naturalness 
problem of the Higgs mass and the pattern of masses and mixings in the quark and lepton sectors (for recent discussuion, 
see e.g.~\cite{Altarelli:2014xxa}).} to extend the theory.

There is therefore an extensive list of shortcomings but no clear guidance on the scale of any new physics, or on the 
coupling strength of any new particles to the SM particles. For example,
neutrino oscillations as a result of non-zero neutrino mass hint at the existence of new degrees of freedom such as 
right-handed Majorana neutrinos which can provide an elegant way to incorporate non-zero neutrino mass through the 
seesaw mechanism~\cite{Minkowski:1977sc,Yanagida:1979as,Glashow:1979nm,GellMann:1980vs,Mohapatra:1979ia,Mohapatra:1980yp}. The smallness of the neutrino masses could be driven by either the existence of super-heavy 
Majorana neutrinos, which have ${\cal O}$(1) Yukawa couplings, or by the existence of Majorana neutrinos with masses at the 
Fermi scale but with Yukawa couplings smaller than that of the electron  (see Figure~\ref{fig:intro_seesaw}). This lack 
of theoretical guidance necessitates experimental searches for new physics at both the energy and the intensity 
frontiers.

From the point of view of this proposal, there are two kinds of BSM theories of interest:
\begin{enumerate}
\item{BSM theories with no new physics between the Fermi and the Planck scales,}
\item{BSM theories with a new energy scale which may also incorporate light particles.}
\end{enumerate}

Models with no new physics between the Fermi and the Planck scales try to extend the SM using the smallest 
possible set of fields and renormalizable interactions. For example this $"Minimality\ principle"$ 
motivates the $\nu$MSM \cite{Asaka:2005an,Asaka:2005pn}  which attempts to explain the pattern of neutrino masses, DM and the 
observed BAU by introducing three HNLs. The lightest of these, $N_1$, 
provides the DM candidate, while $N_{2,3}$ are responsible for the baryon asymmetry. Through the seesaw  mechanism these HNLs also allow the pattern of neutrino masses and oscillations to be explained. 

Supersymmetry (SUSY) is an example of a theory which has some new energy scale but could still have light new particles. SUSY is a broken symmetry but the energy scale at which the symmetry is broken is unknown. 
% The symmetry could be broken at some very high energy scale, far beyond the energies that are directly accessible at accelerators. In certain models the breaking of the symmetry is accompanied by the appearance of s-goldstinos.
If the masses of SUSY particles are determined by the conventional naturalness argument 
(for reviews see~\cite{Giudice:2008bi,Giudice:2013yca}), then SUSY partners with 
masses comparable to the Higgs mass are needed to protect against quadratic radiative corrections without significant fine-tuning.
In certain models (see, e.g. \cite{Ellis:1984kd} and for a review \cite{Giudice:1998bp}) the breaking of the symmetry is  
accompanied by the appearance of light sgoldstinos~\cite{Perazzi:2000id},
which are the superpartners of the Nambu-Goldstone fermion, goldstino, emerging in the spontaneous breaking of SUSY.
The couplings of these sgoldstinos are inversely proportional to the square of the scale of the SUSY breaking and hence the couplings 
could be significantly suppressed. 
The resulting very weak couplings mean that light sgoldstinos may have evaded detection at previous experiments.
The new SUSY scale may therefore have light particles with masses at the Fermi scale, not observed by previous experiments, 
in addition to the conventional heavy SUSY partners. 

% Such models were pioneered by Bjorken~\cite{Bjorken:1988as} 
%and Okun [ref]
%and often involve new particles which are singlets with respect to the SM gauge group. 

A further class of theoretical models with a new energy scale are models with a dark sector~\cite{Kobzarev:1966qya,Foot:1991bp,Foot:1991kb,Patt:2006fw}. The only interactions of such dark sector particles with the SM particles are through so-called messenger or portal particles which can either mix with their SM counterparts, or, alternatively, are charged under both the SM and the dark sector fields~\cite{Batell:2009di}. These messenger or portal particles then enable very weak interactions between the SM and the dark sector particles. These very weak interactions and a potentially isolated sector of new particles means that the dark sector naturally provides DM candidates. 
The dark sector may have a rich structure of light hidden messengers described by various operators with vector, Higgs, neutrino and axion forms. Again, there is no theoretical input on the dark sector mass-scale and the relevant energies may well be accessible at future experiments, provided that their sensitivity  is sufficient. 

The high energy frontier will be comprehensively investigated in the next few decades, while
the searches for alternative low mass BSM physics at the intensity frontier have been neglected in recent years. 
Light new particles may have remained undetected by previous experiments because of the very small couplings involved.  
New data at the intensity frontier will therefore be particularly useful in exploring portal models with light new physics, and in searching for Majorana neutrinos. A new intensity frontier experiment, SHiP, is consequently very timely for direct searches for very 
weakly interacting new physics.

\section{SHiP experimental prospects}

\begin{figure}[htb]
\begin{center}
\includegraphics[width=0.46\linewidth]{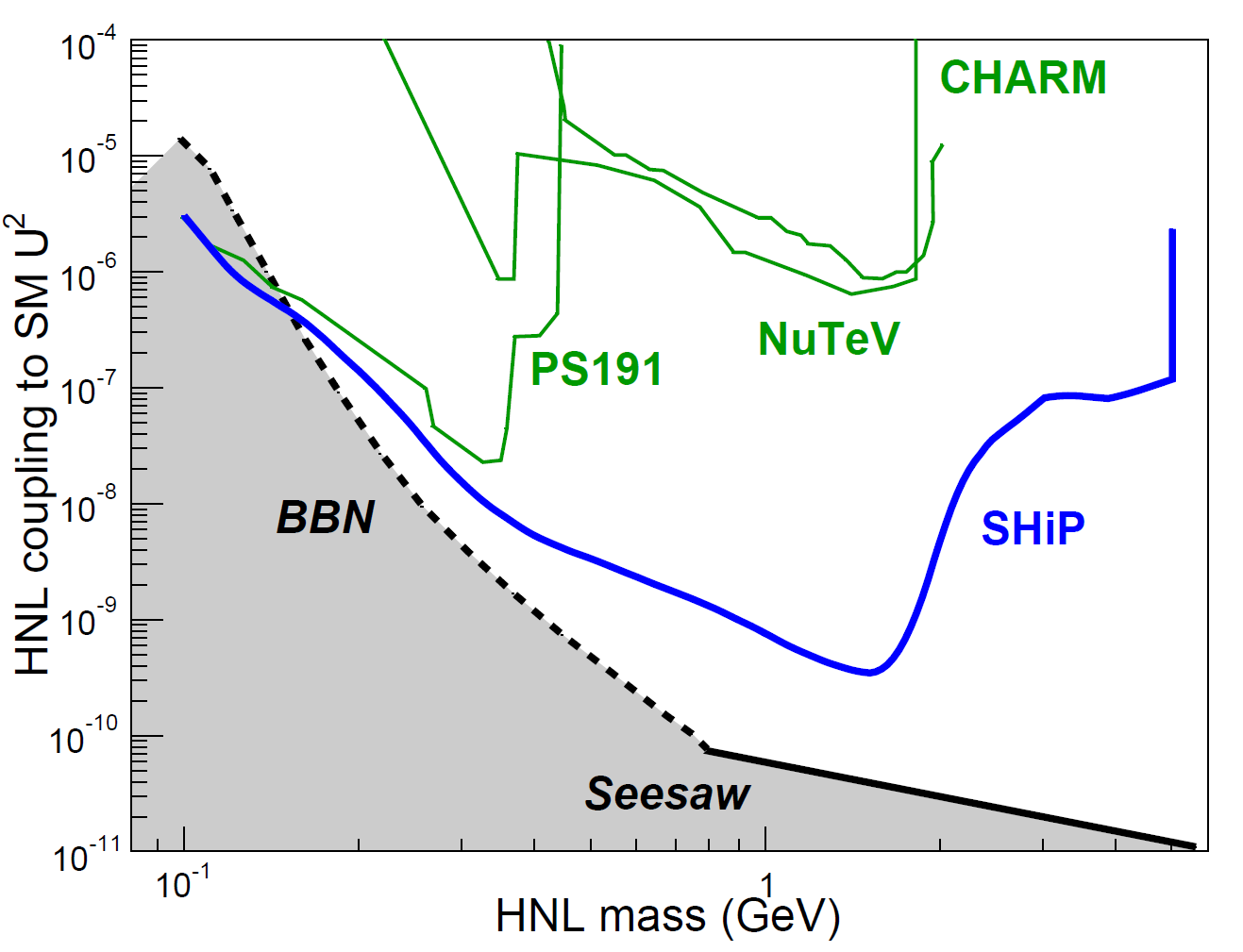}
\includegraphics[width=0.51\linewidth]{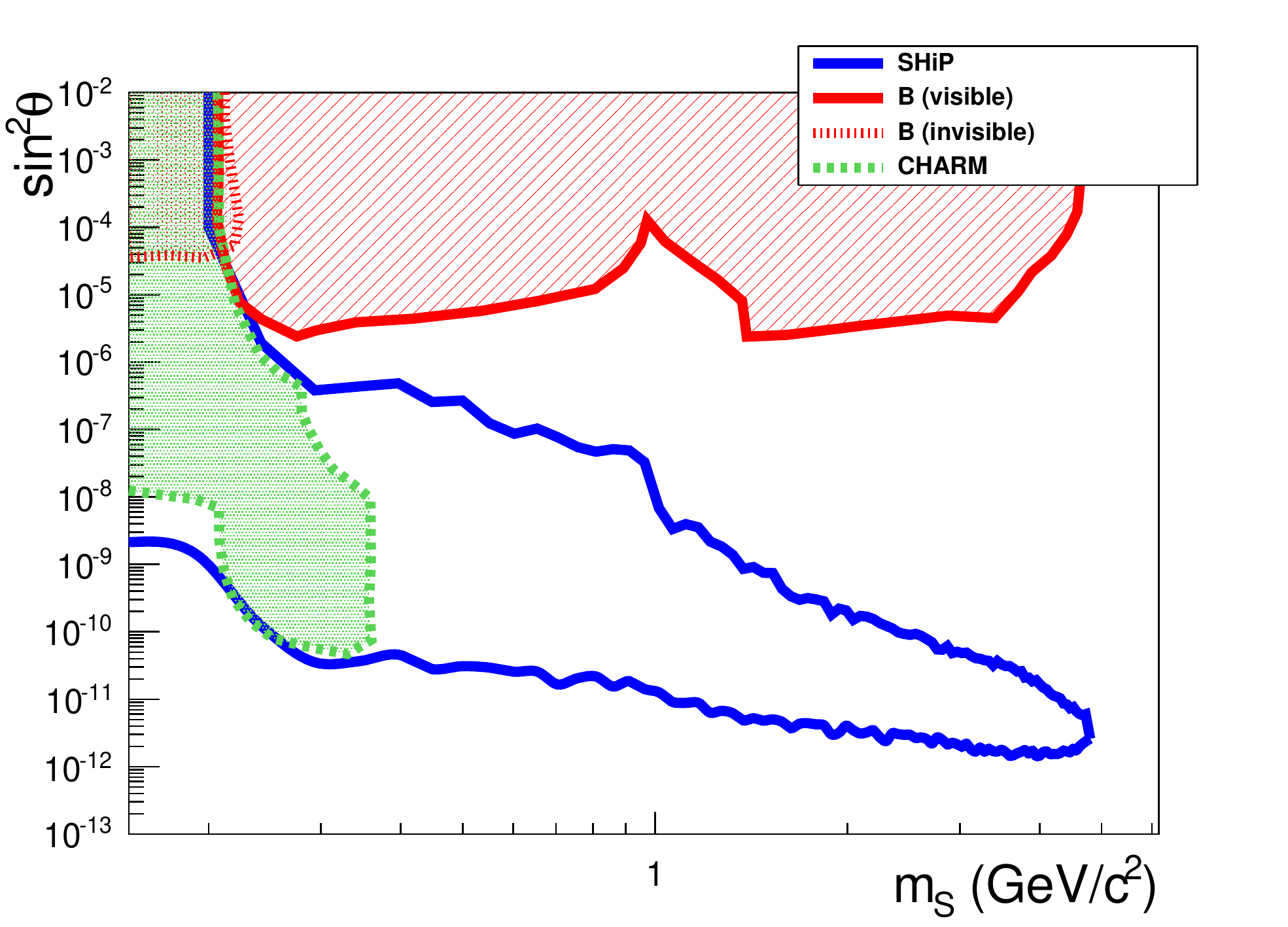}
\caption{(Left) Sensitivity contours for the HNL coupling to active neutrino, $U^2 = U_e^2 + U_{\mu}^2 + U_{\tau}^2$ as function of the HNL mass assuming
$U_e^2 : U_{\mu}^2 : U_{\tau}^2$ = $1 : 16 : 3.8$. (Right) Sensitivity contours for a light hidden scalar particle of mass $m_S$ coupling to the 
Higgs with $sin^2 \theta$ mixing parameter and decaying in $e^+e^-,\mu^+\mu^-, \pi^+\pi^-, K^+K^-$ final states (solid blue line). Red dashed area is the 
excluded region from $B$-factories in the visible modes, red dotted area is the excluded region from $B$-factories in the invisible modes and green shaded 
area in the exclusion region from the CHARM experiment.}
\label{fig:intro_limits}
\end{center}
\end{figure}

Previous experiments have made important contributions to constraining the parameter space for 
axions, dark photons and HNLs. The most significant limits below the charm mass have been obtained 
in the fixed target experiments PS191~\cite{Bernardi:1985ny,Bernardi:1987ek,Vannucci:2008zz}, CHARM~\cite{Bergsma:1986is} and NuTeV~\cite{Vaitaitis:1999wq}, shown in Figure~\ref{fig:intro_limits}(left) 
for 
HNLs\footnote {Searches in $B$ and $Z^0$ decays, and in electron beam dump experiments, are in general sensitive 
but cover a different region of the parameter space.}
and in Figure~\ref{fig:intro_limits}(right) for hidden scalar particles. 
Table~\ref{tab:intro_pastexp} lists the relevant parameters of these three experiments, 
in comparison with those planned for SHiP.

\begin{table*}[!htp]
\begin{center}
\caption{Comparison of the experimental conditions for search experiments dedicated to HNLs, with SHiP.}
\label{tab:intro_pastexp}
\begin{tabular}{l  r r r r }\\\hline
Experiment & PS191 &NuTeV & CHARM & SHiP\\
\hline
Proton energy (GeV)&19.2&800&400&400\\
Protons on target ($\cdot 10^{19}$)& 0.86 & 0.25 &0.24 & 20\\
Decay volume (m$^3$)& 360 &1100 & 315& 1780\\  
Decay volume pressure (bar)& 1 (He) &1 (He) &1 (air) &$10^{-6}$ (air)\\
Distance to target (m)&128 &1400&480&80-90\\
Off beam axis (mrad)&40&0&10&0\\
%$\#$ background events& $<1$& $0.57\pm0.15$& ? &$<10^{-2}$\\
\hline
\end{tabular}
\end{center}
\end{table*}

The pioneering accelerator experiment, PS191, was specifically designed to search for HNLs at the CERN PS.
No signal candidates survived rigorous visual tests, after the final stage of the 
event selection, leading to the limits on $U^2$, as shown in 
Figure~\ref{fig:intro_limits}(left). Due to 
the relatively low energy of the PS beam, the results of PS191 are limited to masses below 
450~MeV/c$^2$.
The high energies available at the CERN SPS and the FNAL accelerators allowed extending the searches for 
HNLs to higher masses using the HNL production via mixing to active neutrinos from charmed decays.
The CHARM experiment searched for HNL decays in the $e^+e^-\nu$, $\mu^+\mu^-\nu$ 
and $e\mu\nu$ final states. No signal candidate was found resulting in the limit on $U^2$
shown in Figure~\ref{fig:intro_limits}(left).  
The search for light hidden scalar particles in the $\mu^+ \mu^-$ final states by the CHARM experiment resulted in 
the limit shown in Figure~\ref{fig:intro_limits}(right).
NuTeV searched for events with a muon and a charged track originating from 
a common decay vertex in the helium volume. No signal events passed the selection criteria. This 
is consistent with the expected background of 0.57$\pm$0.15 events originating mainly from active 
neutrino interactions within the decay volume and in its vicinity. The corresponding NuTeV limit on 
$U^2$ is shown in Figure~\ref{fig:intro_limits}(left).

SHiP will greatly improve the sensitivity of the previous experiments using the production of 
heavy hadrons at the SPS. In particular, a data sample of 8$\cdot$10$^{17}$ $D$~mesons is expected in
about 5 years of nominal SPS operation, as well as a data sample of 3$\cdot$10$^{15}$ $\tau$ leptons.
Despite being suppressed in production
by four orders of magnitude with respect to the charmed hadrons, 
the beauty hadrons will also contribute to the physics sensitivity between the beauty hadron masses and 
the charm hadron masses. The beauty hadrons will also be the dominant source of light scalars mixing 
with the Higgs boson. Below the beauty mass, the SHiP experiment will be able to exceed the sensitivity 
of previous experiments for the neutrino portal
by several orders of magnitude. This will allow SHiP to explore a range of phenomena with unprecedented 
reach - for example,  HNL couplings could be probed close to the ultimate seesaw limit, as shown in 
Figure~\ref{fig:intro_seesaw}. The sensitivities of other existing or planned experiments to the
exploration of hidden sector particles are more than an order of magnitude lower than the SHiP sensitivity, even under the experimentally challenging
assumption of zero background (see Appendix~\ref{app:others}). 

% A large gain in sensitivity can be achieved using in particular a 
% data sample of 7$\cdot$10$^{17}$ $D$~mesons (and 6$\cdot$ 10$^{15}$ $\tau$ leptons), which can be 
% accumulated in $\sim$5 years of SPS operation. Below the charm mass, the SHiP experiment will be 
% able to exceed the sensitivity of previous experiments, described in section~\ref{sec:pastexp},
%  by several orders of magnitude. 

The SHiP detector will have the unique potential to explore the physics of $\nu_{\tau}$ and $\bar{\nu}_{\tau}$,
the only missing piece in the SM. In fact the tau anti-neutrino is the only SM particle that has never been directly observed. Whereas the DONUT experiment at Fermilab observed nine tau neutrino candidates~\cite{Kodama:2007aa}, and the OPERA experiment at LNGS observed four candidates from oscillations~\cite{Agafonova:2014ptn}, the SHiP experiment will collect about 10$^4$ tau neutrino interactions during its operation outlined above. 
Both the $\nu_{\tau}$ and $\bar{\nu}_{\tau}$  cross sections will be measured independently for the first time. This allows extracting the $F_4$ and $F_5$ structure functions~\cite{Albright:1974ts}, which are negligible in electron and muon neutrino interactions.
SHiP is unique in its capability to accumulate large statistics of all three active neutrino and anti-neutrino events. 
The $\nu_e$ cross section will be studied at high energies and, since its production is dominated by charmed hadron decays, this measurement will also provide the normalization for the search for hidden particles.
In addition the strange-quark content of the nucleon will be measured with unprecedented accuracy by means of the charmed hadron production in anti-neutrino interactions and the collection of a few million muon neutrinos will contribute to the studies of structure 
functions and to the measurement of the Weinberg angle as a further test of the Standard Model. 
%
%\section{Other facilities}
%\label{sec:pastexp}

\section{Evolution since the Expression of Interest}
\label{sec:eoi_diff}

Since the Expression of Interest~\cite{Bonivento:2013jag}, the basic concept of the SHiP facility has remained unchanged. 
However, the general layout underwent a few significant modifications.
Firstly, the passive muon shield has been replaced with an active shield based on magnetic deflection. 
A compact neutrino detector has been added in between the muon shield and the Hidden Sector detector, thus extending the physics programme of the facility.
The two HS detector elements have been replaced with a single decay volume of a larger cross section.
The muon shield deflection is horizontal and produces two muon plumes left and right of the detector. 
For this reason, the vertical dimension of the decay volume was enlarged while keeping the transverse dimension.
As a consequence, the HS spectrometer dipole has been turned by 90$^{\circ}$. 
The cross sections of the HS detectors have been increased accordingly.
The robustness of the HS detector against various backgrounds has been improved by adding a dedicated timing 
detector and event taggers. Finally, a hadronic calorimeter has been added to the calorimeter system.

%%%%%%%%%%%%%%%%%%%%%%%%%%%%%%%%%%%%%%%%

% \bibitem{PP}
% S. Alekhin et al.
% ``A facility to search for hidden particles (SHiP) at the SPS: the physics case''.  An accompanying paper.
% \bibitem{Bjorken}
% " Search for neutral metastable ...", J.D. Bjorken et al, Phys. Rev. D 38 V11, p. 3375, 1988
% \bibitem{ps191}
% EOI[50]
% \bibitem{charm}
% EOI[49]
% \bibitem{nutev}
% EOI[51]
% \bibitem{donut}
% K. Kodama et al., Final tau-neutrino results from the DONuT experiment, Phys. Rev. D 78 (2008) 052002.  
% \bibitem{opera}
% N. Agafonova et al., Observation of tau neutrino appearance in the CNGS beam with the OPERA experiment, Prog. Theor. Exp. Phys. 10 (2014) 101C01. 
% \bibitem{f4f5}
% ask Giovanni

%% file: requirements/experimental_requirements.tex
\chapter{Overview of the Experiment}
\label{sec:requirements}

\section{Experimental objectives}

The primary physics goal consists of exploring hidden portals and extensions of the Standard 
Model which incorporate long-lived and very weakly interacting particles through the direct 
detection of their decays to SM particles. A general purpose fixed target facility at the 
SPS accelerator which uses high intensity and slowly extracted proton spills at an energy of 
about 400~GeV is ideally suited to carry out this physics programme.

% as outlined in the physics motivation (Section~\ref{sec:physics}

Since many of the hidden particles are expected to be accessible through the decays of heavy
hadrons, the facility must maximize their production and the detector acceptance in the 
cleanest possible environment. Hadrons, electrons, and photons must be stopped, the 
production of neutrinos from light hadron decays must be minimized, and the detector must 
be shielded from the residual muon flux.

% Consequently the neutrinos from charm hadrons share the kinematical properties with, for 
% instance, the HNLs of the neutrino portal. 

\begin{figure}[htb]
\begin{center}
\includegraphics[width=0.95\linewidth]{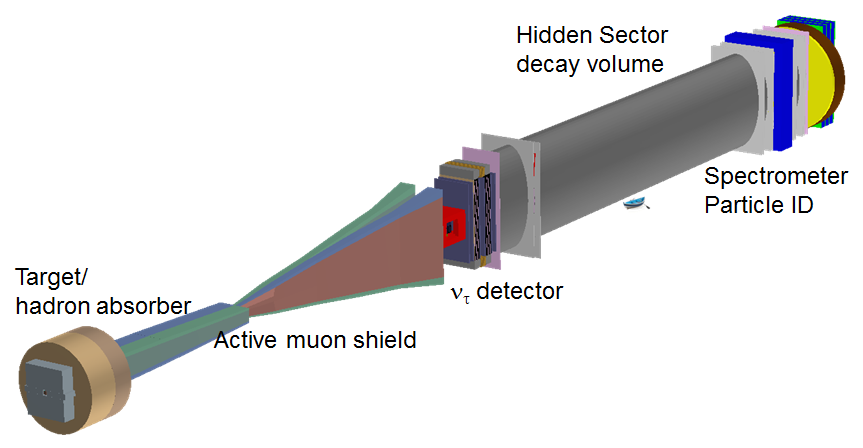}
\caption{Overview of the SHiP facility.}
\label{fig:SHiP_facility_overview}
\end{center}
\end{figure}

In this energy range, the $D_s$ meson decays are the principal source of tau neutrinos. The 
charm hadrons decays are also a source of electron and muon neutrinos. Thus, while the 
optimization of the facility as driven by the requirements 
for the hidden particle search makes it unsuitable for neutrino oscillation physics, the 
experimental setup is ideally suited for studying interactions of tau neutrinos and neutrino 
induced charm production by neutrinos and anti-neutrinos of all flavours. The facility will 
therefore host a detector to search for hidden particles in combination with a tau neutrino 
detector located upstream of the decay volume of the hidden particles detector. 
Figure~\ref{fig:SHiP_facility_overview} shows a schematic overview of the SHiP facility from 
the proton target to the end of the Hidden Sector (HS) detector as proposed in this document. 

The design of the facility assumes the SPS accelerator in its current state with a 400 GeV 
beam and a nominal spill intensity of 4$\cdot$10$^{13}$ protons on target. Over five years
of data taking this results in 2$\cdot$10$^{20}$ protons on target (Section~\ref{sec:op_scenario}).

While it is not part of the current Technical Proposal, the design of the facility 
takes into account future extensions which are addressed in the SHiP Physics 
Proposal~\cite{PP}.

%XXXMutual benefit: upstream NC/CC neutrino tagger, electron neutrino counting to get charmed 
%hadron production for normalization of hidden particle search, ++?, XXX\\

\section{Hidden Sector detector} 
\label{sec:hs_reqs}

% The goal of the Hidden Sector (HS) detector is the direct detection of the hidden particles 
% through their decays to SM particles. 

The phenomenologies of the Hidden Sector models share a number of unique and common physics 
features which dictate the specifications for the design of the beamline and the detector. 
 
As described in detail in the Physics Proposal~\cite{PP}, at the energy accessible at the 
SPS, the hidden particles are predominantly produced in decays of hadrons, in particular
in decays of charmed and beauty hadrons above the kaon mass, and in proton bremsstrahlung. 
In comparison with the couplings between the particles of the SM, the hidden sector couplings with
SM particles are very suppressed leading to expected production rates of $O(10^{-10})$ or less. 

% The aim of directly detecting 
% the decays means that the acceptance factor in the physics sensitivity also involves
% a suppression by the coupling. 
 
% The aim of directly detecting the decays leads to another suppression factor  the 
% due to the acceptance 

% \begin{figure}[htb]
% \begin{center}
% \includegraphics[width=0.6\linewidth]{requirements/req_lifetime.png}
% \caption{HNL lifetime as a function of the HNL mass derived from $U^2$.  PLOT WITH ONLY SEE-SAW LIMIT}
% \label{fig:req_lifetime}
% \end{center}
% \end{figure}

% Typically, the models predict lifetimes in the order of fractions of a microsecond to hundreds of 
% microseconds in the mass range of interest. 
% The decay signature is thus characterized by a $V^0$. 

\begin{figure}[htb]
\begin{center}
\includegraphics[width=0.6\linewidth]{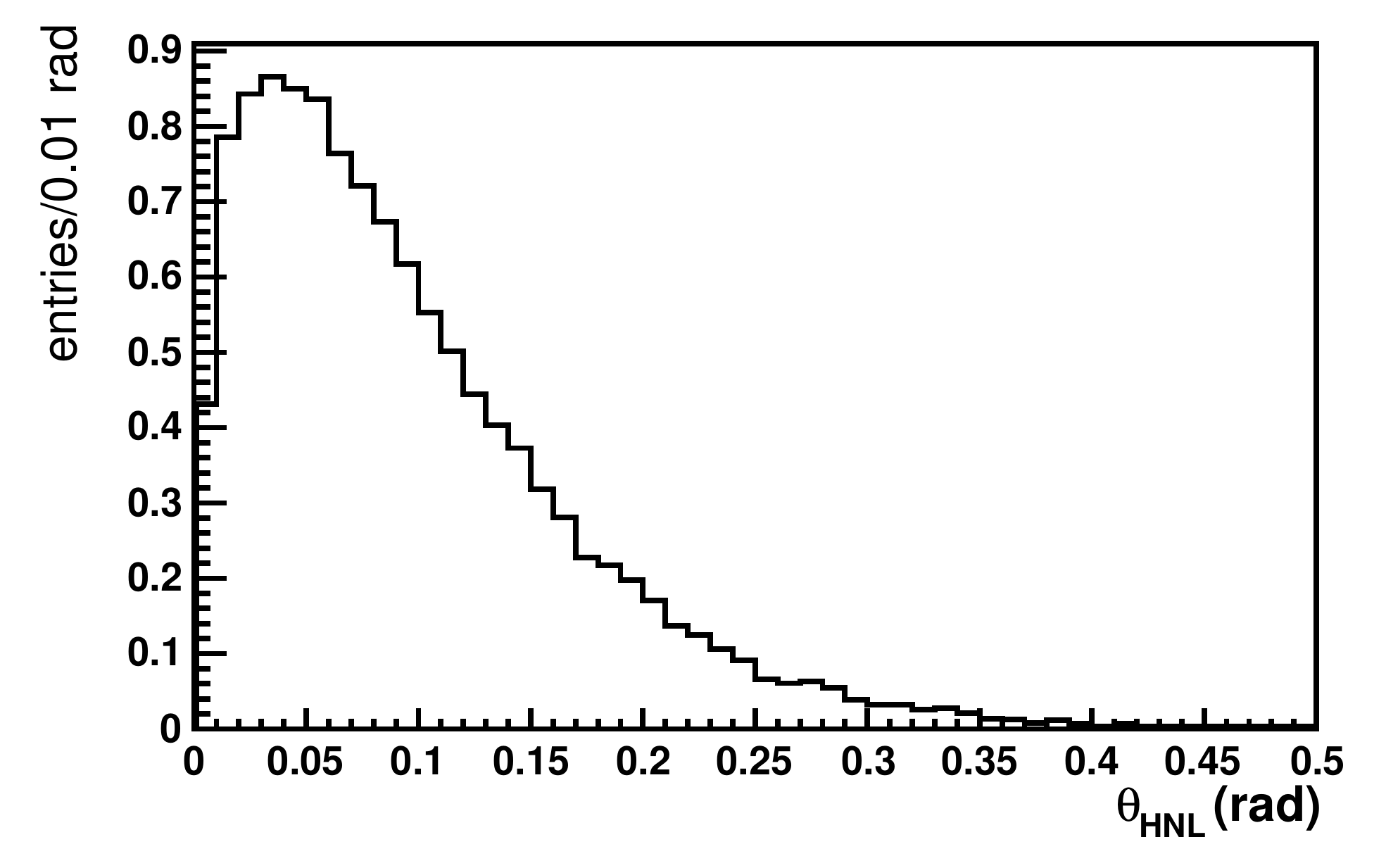}
\caption{Polar angle distribution for HNLs with a mass of 1 GeV/c$^2$ from decays of charmed hadrons.}
\label{fig:hnl_kinematics}
\end{center}
\end{figure}

The smallness of the couplings implies that the objects are very long-lived compared to 
the bulk of the unstable Standard Model particles. While the long lifetimes allude to an experimental 
design with a distant detector, the small boost of the heavy hadrons implies relatively 
large polar angles for the hidden particles. Figure~\ref{fig:hnl_kinematics} shows the polar 
angle distribution for HNLs from decays of charmed hadrons. Hence, in order to maximise the 
acceptance, the decay volume should be placed as close as possible to the target. The distance 
is only constrained by the requirement of reducing the beam induced background to a manageable level. 

The detector for the direct detection of the hidden particles is designed to fully reconstruct 
their exclusive decays. Table~\ref{table:req_decaymodes} summarizes the main decay modes of the 
hidden particles in the various models considered. In order to ensure model independence, 
and allow models to be distinguished, the detector must be sensitive to as many 
decay modes as possible. Other decay modes involving charm, strangeness and $\tau$ will be 
addressed at a later stage. 

\begin{table}[htb]
\begin{center}
\caption{Summary of the main decay modes of hidden particles in various models ($\ell = e, \mu$).}
\label{table:req_decaymodes}
\vspace{2mm}
\begin{tabular}{ll}
\hline
Models                         	& Final states              \\
\hline
Neutrino portal, SUSY neutralino                 & $\ell^{\pm}\pi^{\mp}, \ell^{\pm} K^{\mp}, \ell^{\pm}\rho^{\mp},\,\,\,\,\, \rho^{\pm}\rightarrow \pi^{\pm}\pi^0$ \\
Vector, scalar, axion portals, SUSY sgoldstino   & $\ell^+\ell^-$ \\
Vector, scalar, axion portals, SUSY sgoldstino   & $\pi^+\pi^-, K^+K^-$ \\
Neutrino portal ,SUSY neutralino, axino          & $\ell^+\ell^-\nu$ \\
Axion portal, SUSY sgoldstino                    & $\gamma\gamma$ \\
% Axino ?                                          & $\gamma$... \\ 
SUSY sgoldstino                                  & $\pi^0\pi^0$ \\
\hline
\end{tabular}
\end{center}
\end{table}

The principal background to the hidden particle decay signal originates from the inelastic scattering
of neutrinos and muons in the vicinity of the detector producing long-lived $V^0$ particles. 
Table~\ref{table:req_backgrounds} gives a list of possible decay modes of $V^0$ particles and the 
associated decay modes that present the most difficult residual backgrounds in view of the signal decay 
modes considered. The accompanying fragments, $X$, help tagging these interactions.
Another source of background comes from random combinations of tracks in the fiducial 
volume from the residual muon flux, or other charged particles from interactions in the proximity, 
which enter the decay volume and together mimick signal events. The contribution from cosmics to both
types of background is expected to be small.

\begin{table}[htb]
\begin{center}
\caption{List of background sources with $V^0$ particles.}
\label{table:req_backgrounds}
\vspace{2mm}
\begin{tabular}{ll}
\hline
Background source                         	 & Decay modes              \\
\hline
$\nu$ or $\mu\,\, + $ nucleon $\rightarrow X + K_L$          & $K_L \rightarrow \pi e\nu, \pi\mu\nu, \pi^+\pi^-, \pi^+\pi^-\pi^0$ \\
$\nu$ or $\mu\,\, + $ nucleon $\rightarrow X + K_S$          & $K_S \rightarrow \pi^0\pi^0, \pi^+\pi^-$ \\
% $\nu$ or $\mu\, + $ nucleon $\rightarrow X + n$            & $n \rightarrow pe^-\bar{\nu}_e$ \\
$\nu$ or $\mu\,\, + $ nucleon $\rightarrow X + \Lambda$      & $\Lambda \rightarrow p\pi^-$ \\ 
$n$ or $p\,\, + $ nucleon $\rightarrow X + K_L$, etc\,\,\,            & as above \\
\hline
\end{tabular}
\end{center}
\end{table}

It is essential that the beam line is designed to 
minimize the background sources. The proton interaction in the target gives 
rise to a copious direct production of short-lived resonances, and pions and kaons. While 
a hadron stopper of a few metres of iron is sufficient to absorb the hadrons and the 
electromagnetic radiation emerging from the target, the decays of pions, kaons and 
short-lived resonances result in a large flux of muons and neutrinos. In order to reduce the 
flux of neutrinos, in particular the flux of muon neutrinos and the associated muons,
the pions and kaons should be stopped as efficiently as possible before they decay. 
The target must therefore be made of a material with the shortest possible interaction length 
and be sufficiently long to contain the hadronic showers with minimum leakage. Since 
the production angle of the hidden particles is relatively large, there is no requirement to 
minimize the beam spot. 

% This allows the adequate dilution of the deposited beam energy by 
% sweeping the beam on the target.

\begin{figure}[htb]
\begin{center}
\includegraphics[width=0.8\linewidth]{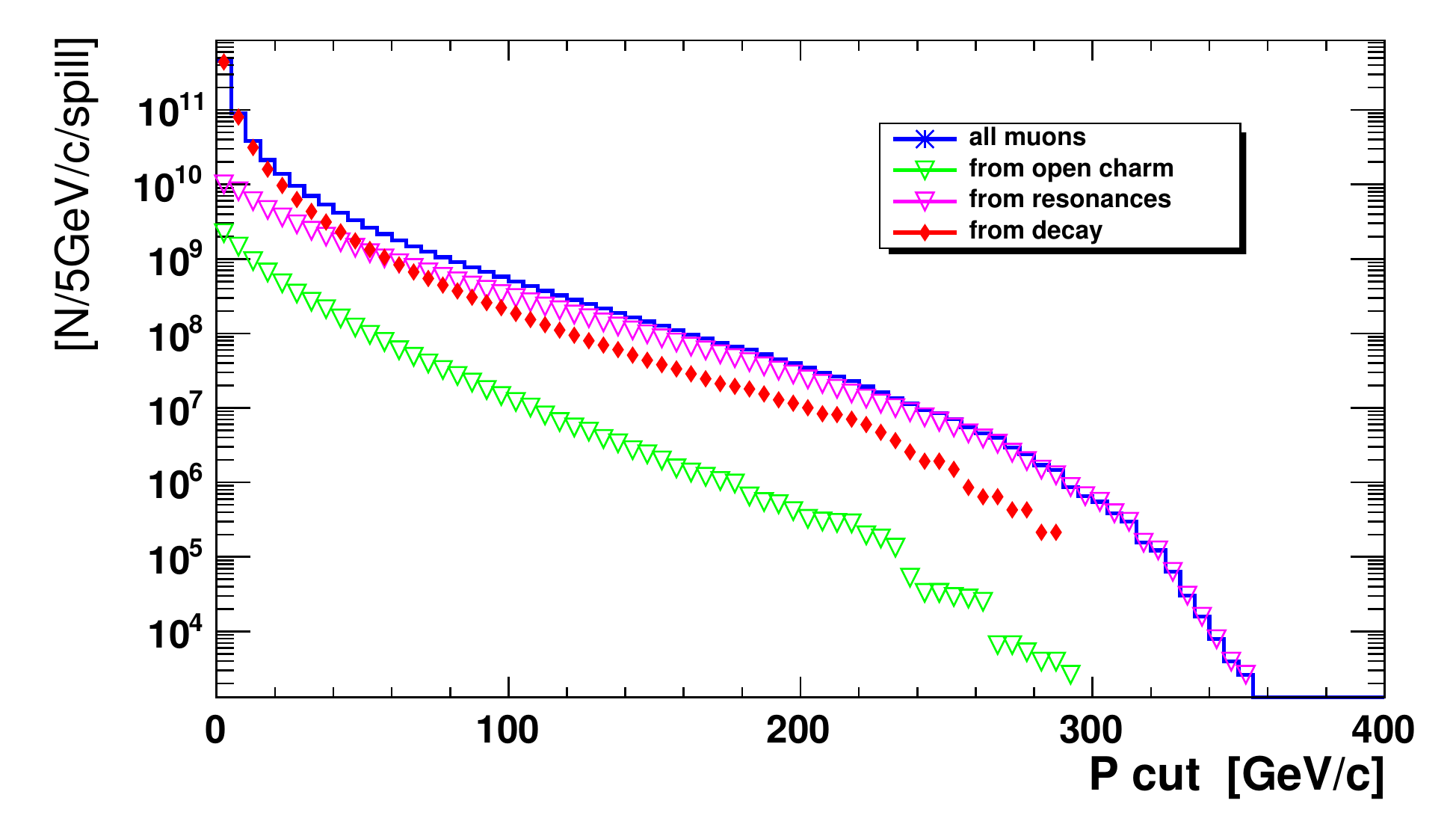}
\caption{Integrated muon flux per spill (here  5$\cdot$10$^{13}$ protons) after the hadron 
absorber above a momentum cut as a function of the momentum cut with a beam of 400 GeV protons 
on a fixed target consisting of molybdenum and tungsten (Section~\ref{sec:target})}
\label{fig:req_muon_flux}
\end{center}
\end{figure}

The short-lived resonances and the residual flux of decaying pions and kaons still give rise to a large 
flux of muons. Figure~\ref{fig:req_muon_flux} shows the integrated muon flux above a momentum cut 
as a function of the muon momentum cut. Below about 40~GeV/c, muons from pion and kaon decays dominate.
The muon flux at higher momenta comes predominatly from prompt decays. The muons from charm decays 
represent a negligible fraction of the total muon flux. 
This flux must be efficiently cleared from the detector fiducial volume by either a passive shield 
or through an active shield based on magnetic deflection. 
The aim is to reduce the flux to a level at which the muon induced $V^0$ background is less than 
the neutrino induced background. The residual flux should also be low enough so not to compromise the 
occupancy limit in the tau neutrino detector. As illustrated in Figure~\ref{fig:SHiP_facility_overview},
in the baseline design a 5~m horizontally wide region respecting these requirements has been achieved 
with a 48~m long active muon shield based on magnetic deflection of the muons in the horizontal 
plane (see detailed discussion in Section~\ref{sec:muon_shield}). In order to reduce the probability 
of combinatorial background from the residual muons entering the decay volume and muons deflecting 
off the cavern walls, the proton spills must be prepared with a slow beam extraction of about one second, 
and the extractions should be as uniform as possible. 

\begin{figure}[htb]
\begin{center}
\includegraphics[width=0.6\linewidth]{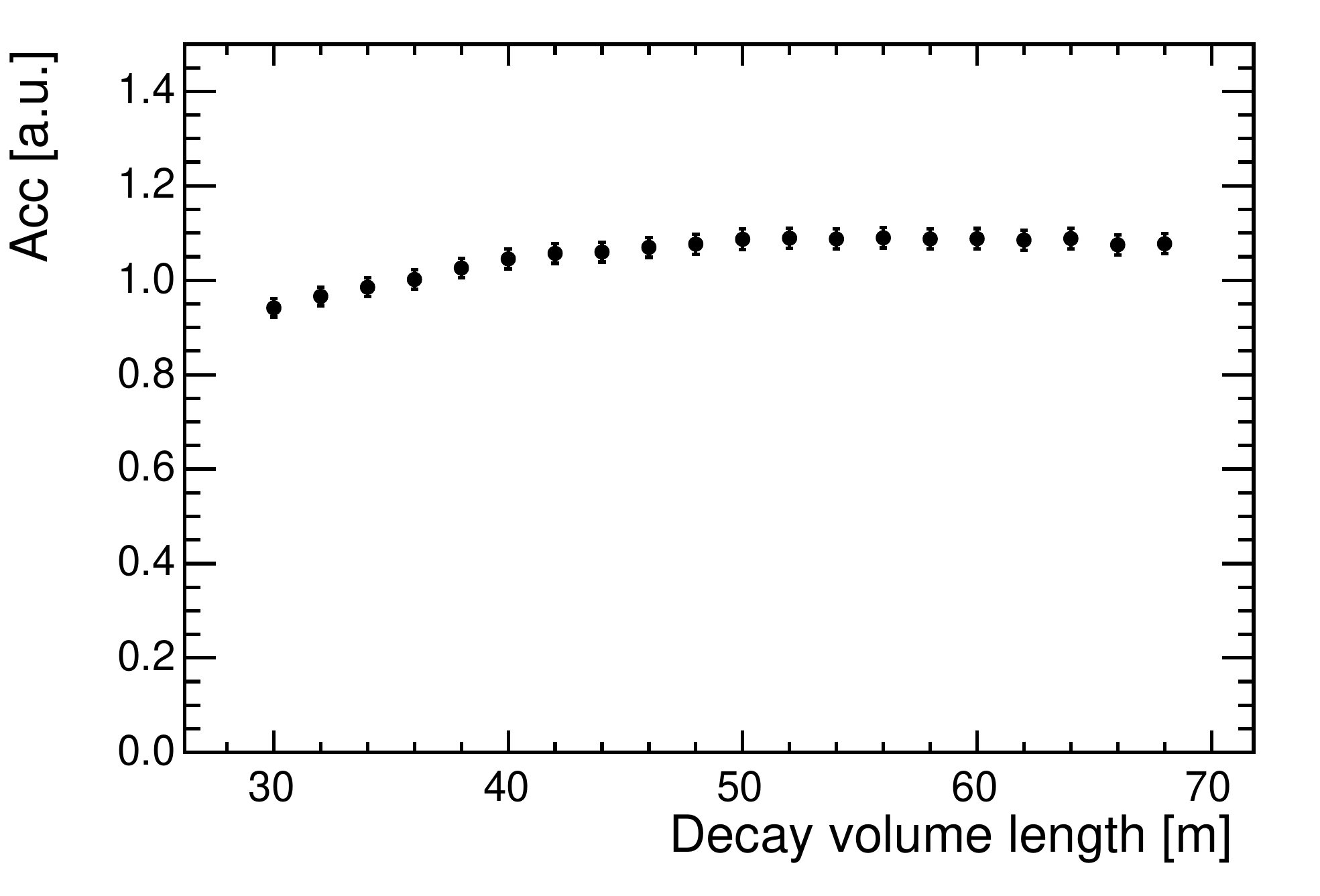}
\caption{Acceptance for the hidden particle decays as a function of the length of the decay volume
for a vessel with a transversal size of 5~m~$\times$~10~m.}
\label{fig:req_acceptance}
\end{center}
\end{figure}

The muon shield is followed by the 10~m long tau neutrino detector, which puts the start of the 
HS decay volume at about 64~m. The dimensions of the tau neutrino detector must be adapted
to the region cleared by the muon shield to avoid intercepting the muon flux and induce background
in the HS detector, and to avoid large occupancy in the neutrino detector. In order to minimize the 
background induced by the residual flux of neutrinos and muons inside the HS detector and in its 
proximity, the decay volume must be under vacuum and must be located at a distance away from cavern 
walls and heavy structures. The width of the decay volume is defined by the region that can be 
shielded from the muon flux. The vertical dimension of the decay volume is mainly driven by the 
cost of the vacuum vessel. The current optimization of muon shield and cost, results in a decay 
volume with an elliptical shape of 5~m width and 10~m height. The length of the decay volume is 
obtained by maximizing the acceptance to the hidden particle decay products given the transversal 
size. The acceptance includes the probability of decay on that length. Figure~\ref{fig:req_acceptance} 
shows the acceptance as a function of the length of the decay volume for the 5~m~$\times$~10~m vessel. 
As a result a length of 50~m has been defined for the decay volume.

\begin{figure}[htb]
\begin{center}
\includegraphics[width=0.6\linewidth]{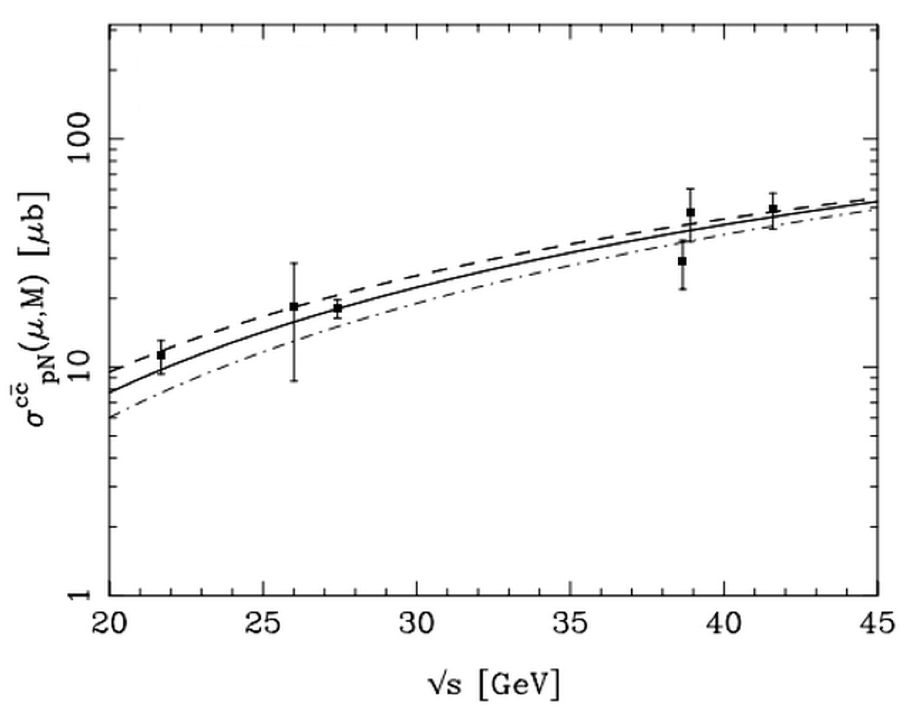}
\caption{Total $c\overline{c}$ production cross sections at fixed target energies~\cite{Lourenco:2006vw,f_ds,pdg}.}
\label{fig:req_cross_section}
\end{center}
\end{figure}

Despite the advantages of a shorter muon shield that a lower proton beam energy would allow, 
the larger detector geometric acceptance does not compensate the rapid drop of the heavy hadron production
cross sections below $E_{cms}\sim$ 25~GeV (see Figure~\ref{fig:req_cross_section}). A beam energy 
of around 400~GeV, equivalent to $E_{cms}\sim$27.4~GeV, results in a 
compromise between the signal production rates and the geometric acceptance of the detector 
at an affordable price.

In order to maximize the sensitivity over a large range of phase-space, great attention has to be paid to 
achieve near zero level of background after the analysis of the full data set from 
2$\cdot$10$^{20}$ protons on target.

% Consequently, the ideal experimental setup consists of a fixed target experiment
% with a large decay volume. The experiment requires critically as large a number of protons on 
% target as possible to maximize the yield for the HS and the $\nu_{\tau}$ physics signals. 

% In order to claim a 3$\sigma$ signal evidence in case of single event observation in a data sample 
% of 2$\cdot$10$^{20}$ protons on target, the hidden particle experiment must be designed to achieve 
% a virtually zero level of background and should hence aim for the best possible performance 
% at an affordable price (TO BE WRITTEN ACCORDING TO THE FINAL STRATEGY). A higher level of background 
% will translate into an effective loss of sensitivity which is equivalent to a reduction of the 
% effective number of protons on target.
% This goal requires having <0.2 expected background events.

Counters surrounding the vacuum vessel are necessary to tag muons which are sent back to the detector
by scattering, and to tag hadronic showers from inelastic interactions in the vessel walls by neutrinos and by muons.

% Muons deflecting off the cavern walls still present a potentially dangerous source of both
% combinatorial background and production of long-lived $V^0$ particles. 

% A spill intensity monitor with high bandwidth needs to be included on the transfer beamline 
% to detect spikes in the spills. 

% which may have produced long-lived $V^0$ particles of the types given 
% in Table~\ref{table:req_backgrounds}. The surrounding background tagger should also allow 
% detecting neutrons down to low energies. 

CC and NC $\nu$-interactions in the material upstream of the HS fiducial volume may produce 
$V^0$ particles. In order to provide efficient protection against these, a combination of 
light taggers should be located upstream and at the beginning of the HS fiducial volume. 

% These upstream detectors must also be capable of providing nanosecond 
% timing to allow correlation between charged particles at the beginning of the decay 
% volume with the particles reconstructed in the tracking system for background rejection.

\begin{figure}[htb]
\begin{center}
\includegraphics[width=0.49\linewidth]{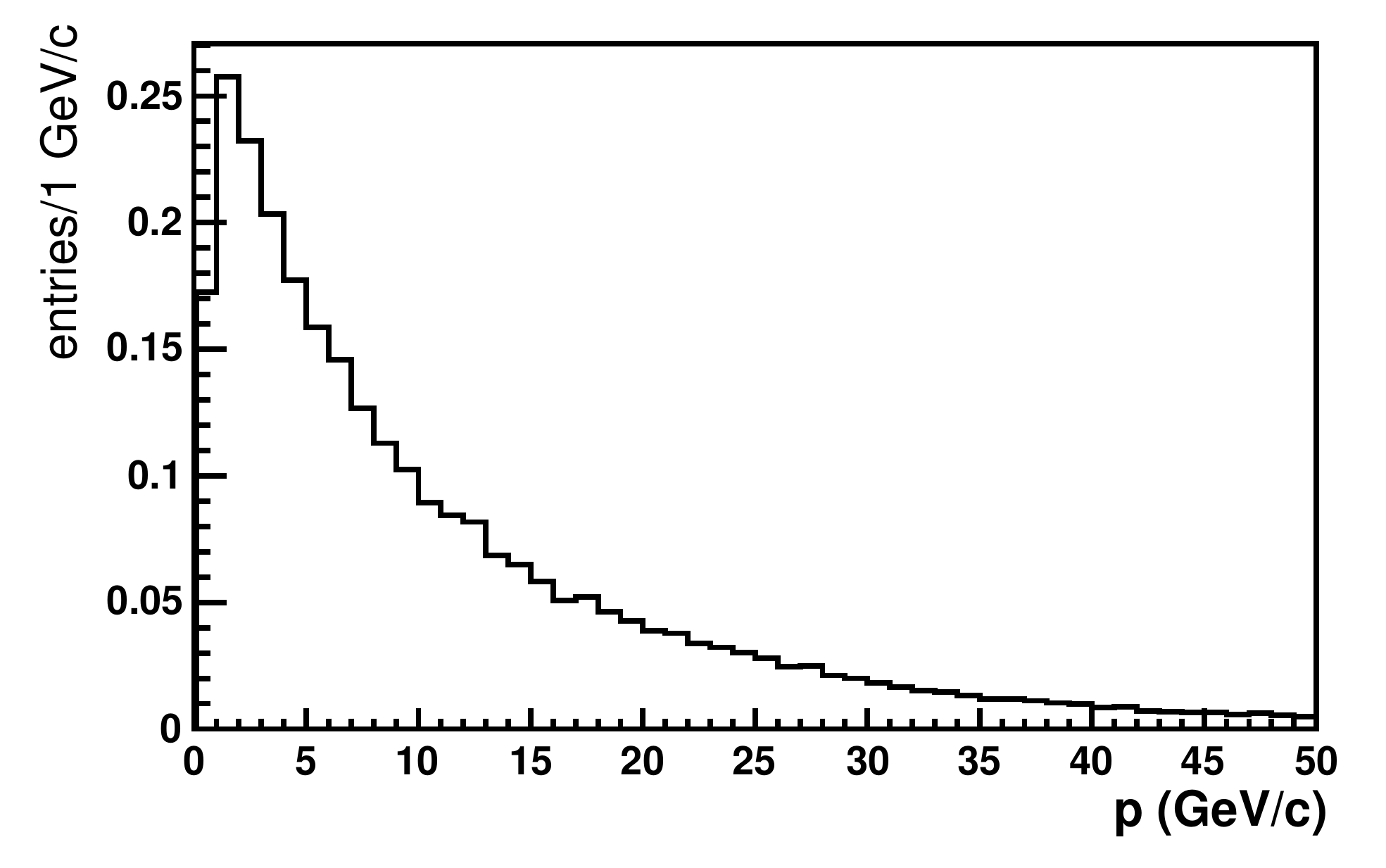}
\includegraphics[width=0.49\linewidth]{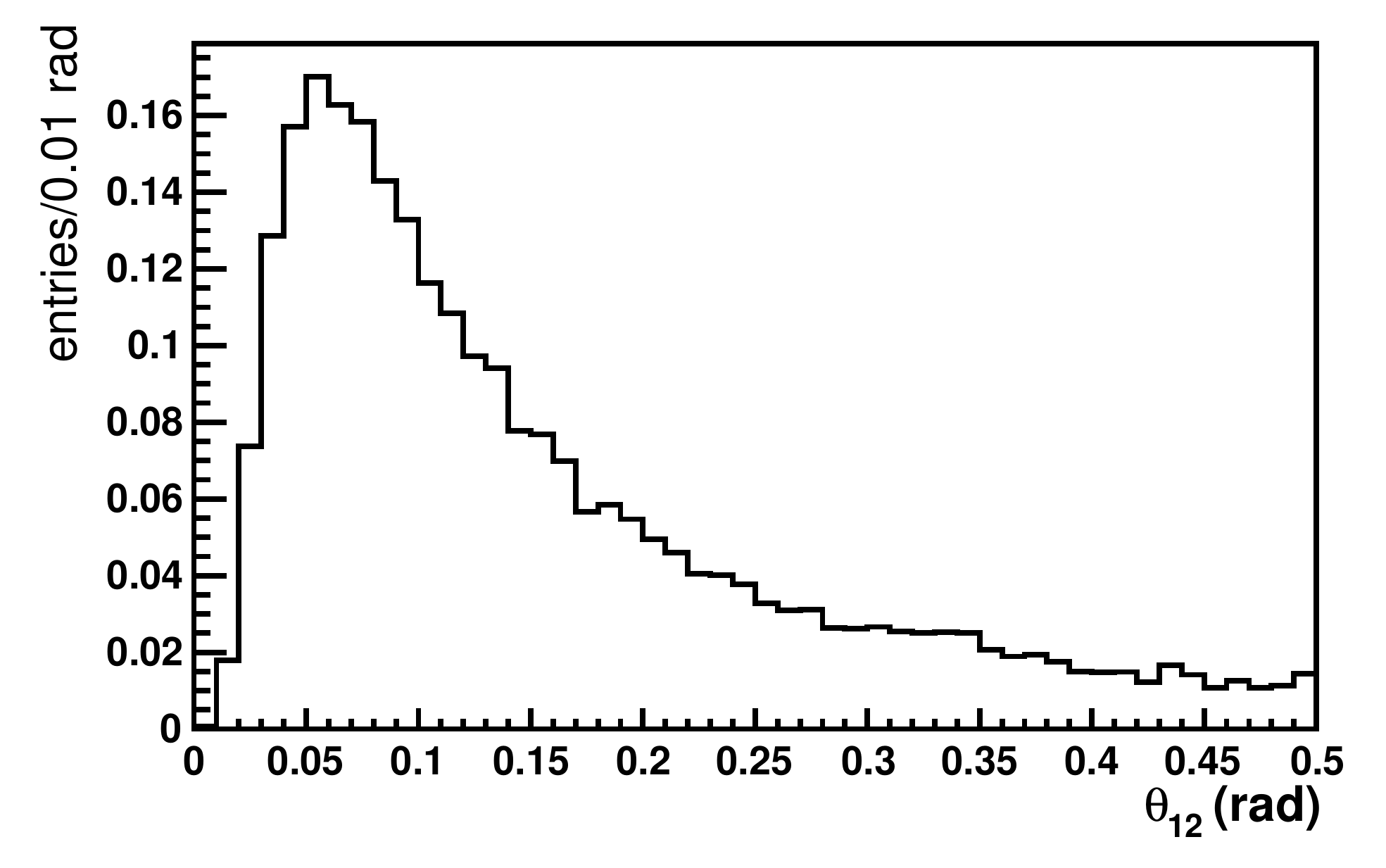}
\caption{Momentum distribution (left) and opening angle distribution (right) of the
decay products in $\pi\mu$ decays of HNLs with mass 1 GeV/c$^2$ from charmed hadrons.}
\label{fig:daughter_kinematics}
\end{center}
\end{figure}

The full reconstruction of the hidden particle decays requires a magnetic spectrometer 
and a system for particle identification at the end of the decay volume. 
Figure~\ref{fig:daughter_kinematics} shows the momentum distribution (left) and the opening angle
distribution (right) of the decay products in $\pi\mu$ decays of HNLs from charmed hadrons.
In order to provide discrimination against background, the tracking system of the 
spectrometer must provide good tracking and mass resolution, and precise determination of the 
position of the decay vertex and the impact parameter at the target for the particle
producing the decay vertex. A timing detector with sub-nanosecond time resolution as part of 
the hidden particle reconstruction system is further required to exclude fake decay vertices 
from combinations of random tracks. 

%  as well as for obtaining constraints on missing momentum from hypothetical neutrinos for decay modes with neutrinos.

The particle identification system requires an electromagnetic calorimeter for $e/\gamma$ 
identification with sufficient granularity and energy resolution in order to reconstruct 
$\pi^0$'s, and a hadron calorimeter in combination with a muon detector for $\pi/\mu$ 
separation.

\section{Tau neutrino detector}
\label{sec:nu_reqs}

% A large neutrino flux is produced in a beam dump facility like the one we are considering. 
% Muon and electron neutrinos are produced by the decay of both prompt charmed hadrons and 
% pions and kaon mesons induced as secondary particles in proton interactions. Tau neutrinos 
% are at the level of one permille and come practically only from the leptonic decay of $D_s$ 
% mesons into $\tau$ and  the subsequent $\tau$ decay. \\

The main purpose of the tau neutrino detector is to perform the first direct observation of the 
$\bar{\nu}_\tau$, and to study the properties and the cross section of $\nu_\tau$ and $\bar{\nu}_{\tau}$.

% The number of $\nu_\tau$ and $\bar{\nu}_\tau$  emerging from the molybdenum target can be estimated as follows:
% \begin{equation}
% N_{\nu_{\tau} + \bar{\nu}_{\tau}} = 4 N_p \dfrac{\sigma_{c\bar{c}}}{\sigma_{pN}} f_{D_s} Br(D_s \rightarrow \tau) = 2.85 \times 10^{-5} N_p
% \end{equation}
%  where 
% \begin{itemize}
% \item $N_p$ is the number of interacting protons (all incoming ones)
% \item$\sigma_{c\bar{c}} = 18.1 \pm1.7$ $\mu$barn~\cite{doublecharm} is the associated charm production  per nucleon. This cross-section does not depend on $A$; 
% \item $\sigma_{pN} = 10.7$ mbarn is the hadronic cross-section per nucleon in a Mo target\footnote{The hadronic cross-section per nucleon on a target with A nucleons can be expressed in terms of the nuclear interaction lengths ($\lambda_{int}$), the target density $\rho$ (g/cm$^{3}$) and the Avogadro number $N_A$ as $\sigma_{pN} = \sigma_{pA}/A = 1/\lambda_{int} \rho N_A$.} 
% \item $f_{D_s} = (7.7 \pm 0.6 ^{+0.5}_{-0.4})\%$~\cite{f_ds} is the fraction of $D_s$ mesons produced
% \item $Br(D_s \rightarrow \tau) = (5.54 \pm 0.24)$\%~\cite{pdg} is the $D_s$ branching ratio into $\tau$ 
% \item The factor of 4 accounts for the charm pair production and the two $\nu_\tau$ produced in each $D_s$ decay.
% \end{itemize}

\begin{figure}
\begin{center}
% \begin{minipage}{14pc}
\includegraphics[width=0.49\linewidth]{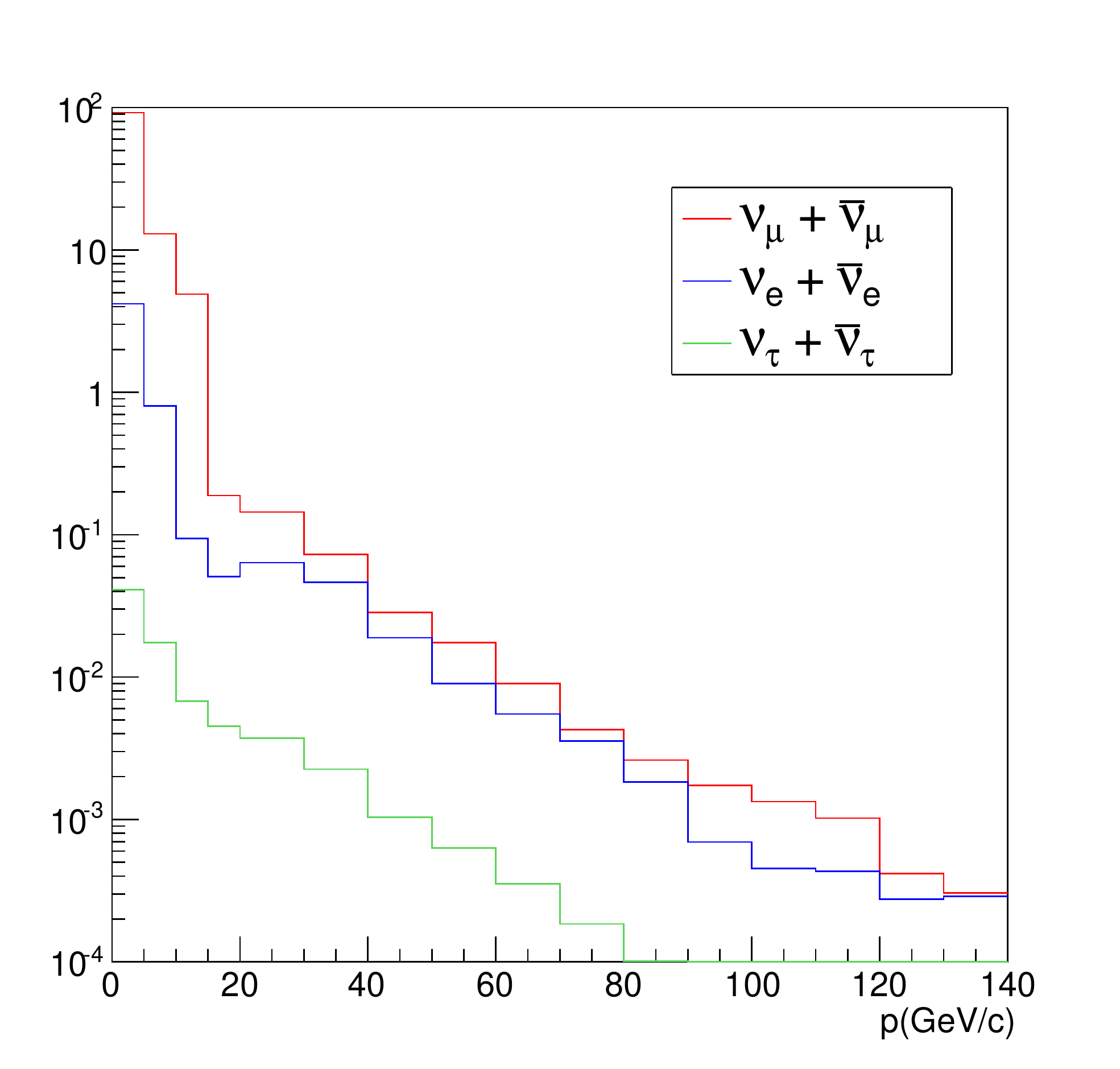}
%\end{minipage}
% \begin{minipage}{14pc}
\includegraphics[width=0.49\linewidth]{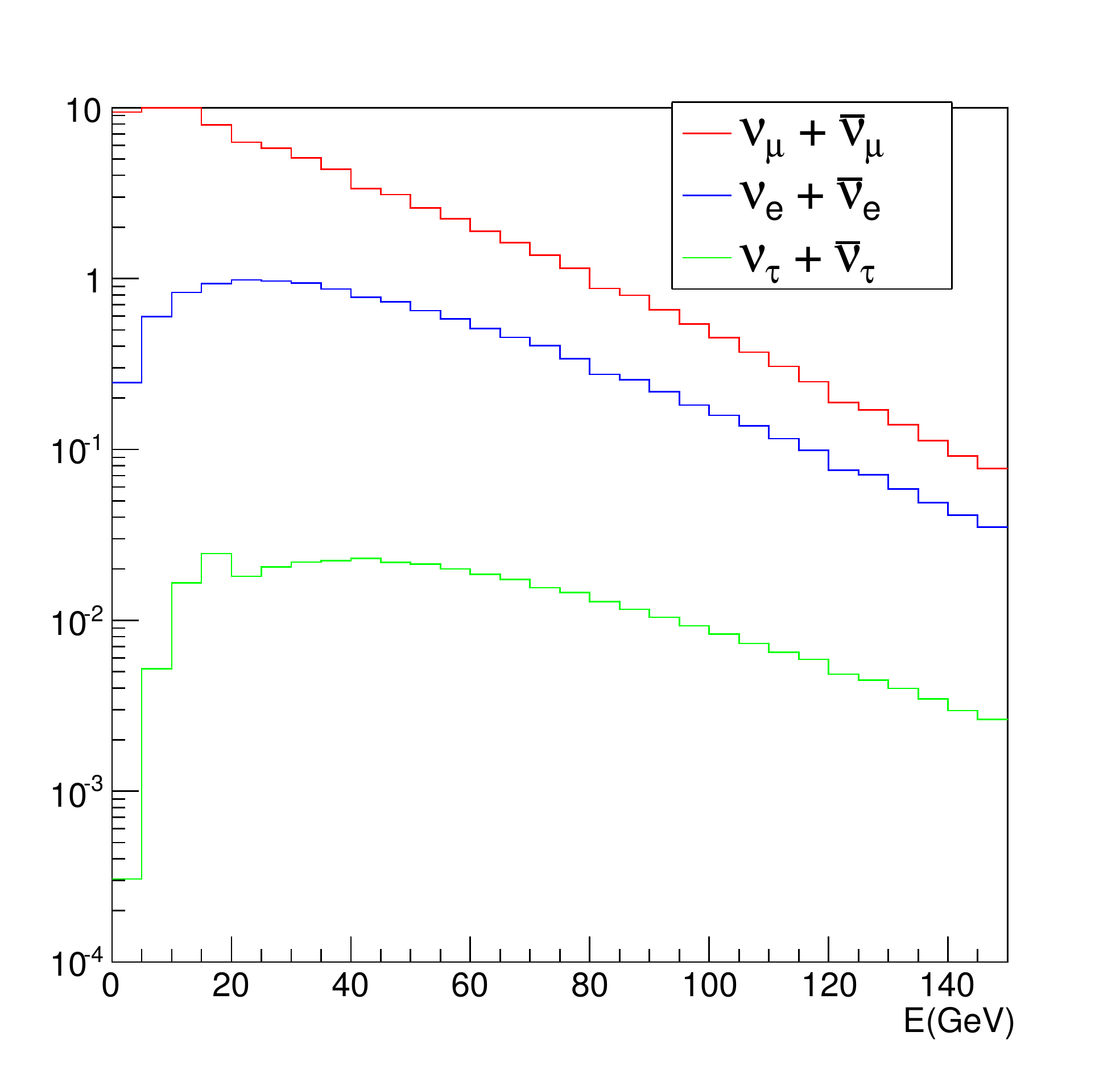}
% \end{minipage}
\end{center}
\caption{Energy spectra of the different neutrino flavours produced in the 
beam dump (left) and interacting in the neutrino detector (right). A 0.5~GeV/c cut
is applied for $\nu_e$ and $\nu_{\mu}$. The total number of neutrinos is normalized to one hundred. } \label{fig:spectrum}
\end{figure}

Five years of nominal operation with 2$\cdot$10$^{20}$ protons on target is expected to produce a total of 
$5.7 \cdot 10^{15}$  $\nu_\tau$ and $\bar{\nu}_\tau$ in about equal proportions. The expected number of 
muon and electron neutrinos above 0.5~GeV is 
$7.2 \cdot 10^{18}$ and $3.7 \cdot 10^{17}$, respectively. Figure~\ref{fig:spectrum} shows the momentum spectrum 
of the neutrinos produced in the primary proton target (left) and the energy spectrum of the neutrinos interacting in the 
neutrino detector (right) for all three neutrino flavours.

The tau neutrino detector should have a neutrino target optimized to induce a maximum number of 
$\nu_{\tau}$ interactions. At the same time it must be compact to respect the constraints from
the design of the muon shield so not to intercept the deflected muon flux. A micron position resolution is required 
to disentangle the $\tau$ lepton production and decay vertices. Since all flavours of neutrinos are 
produced, the detector design is aimed at being able to identify all of them. The detector is also 
designed to have the unique capability of distinguishing neutrinos from anti-neutrinos. 

% The expected number of interacting neutrinos with 2$\cdot$10$^{20}$ protons on target for a $\sim 10$ tonne detector located 
% immediately downstream of the muon shield with cross section of about 1.3~$m^2$, is reported in 
% Table~\ref{tab:neutrino_interact} together with their average energies. The numbers reported in the table refer only to 
% deep-inelastic charged-current (CC) neutrino interactions. The energy spectra are shown 
% in the right plot of Figure~\ref{fig:spectrum}. In the spectra of interacting neutrinos, only deep-inelastic CC interactions are considered. 
%
\begin{table}[htdp]
\begin{center}
\caption{Expected number of interactions in the target and their average energies for the different neutrino flavours for 2$\cdot$10$^{20}$ protons on target. Only CC deep-inelastic interactions are listed. }
\label{tab:neutrino_interact}
\begin{tabular}{c c c}
\hline
& $<$E$>$ (GeV) &  Number of $\nu$ \\
 \hline
 $N_{\nu_e}$ & 46 & $2.5 \cdot 10^{5}$ \\
 $N_{\nu_\mu}$ & 29 & $1.7 \cdot 10^{6}$ \\
 $N_{\nu_\tau}$ & 59 & $6.7 \cdot 10^{3}$ \\
 $N_{\overline{\nu}_e}$ & 46 &$9.0 \cdot 10^{4}$ \\
 $N_{\overline{\nu}_\mu}$ & 28 & $6.7 \cdot 10^{5}$ \\
 $N_{\overline{\nu}_\tau}$ & 58 & $3.4 \cdot 10^{3}$ \\
 \hline
\end{tabular}
\end{center}
\end{table}
%

% The $\nu_\tau$ signature is both topological and kinematical: a $\tau$ neutrino interaction is characterized 
% by the presence of two vertices, one from the neutrino charged current interaction and the other one from 
% the $\tau$ lepton decay. The short-lived  (0.3~ps) $\tau$  produces a secondary vertex a few hundred 
% microns downstream of the neutrino interaction vertex, hence defining the position accuracy 
% needed to clearly distinguish the two vertices. The $\tau$ lepton mass being 1.78~GeV/c$^2$, the measurement
% of the transverse momentum of its daughter particles is used to suppress residual background.

% Being the tau lepton mass almost 2 GeV, the transverse 
% momentum of its decay is large as well as the momentum of its daughter particles. Thus, besides the 
% topological identification, the measurement of kinematical variables is needed to exploit the kinematical 
% characteristics of the tau neutrino interactions and further suppress residual background.

% The Emulsion Cloud Chamber technology, alternating high density material plates and nuclear emulsion films, 
% is the ideal solution. The technology has already been successfully used in the OPERA experimetn~\cite{bopera}. 

For the tau neutrino search, the main background comes from $\nu_\mu$ and $\nu_e$ charged current interactions 
with the production of charged hadrons and where the primary lepton is not identified. Charmed hadrons, having masses and 
lifetimes similar to those of the $\tau$ lepton, can mimic its decay. If the primary lepton is not detected, 
the charged charmed hadron can be misidentified as the $\tau$ lepton. Several approaches can be used to reduce 
the background: maximizing the identification efficiency for the primary lepton, measuring the charge of the 
decaying particle, and exploiting the different kinematics of the two processes. 

Muons are produced in the fully leptonic $\tau$ decay and in muon neutrino charged current interactions. The 
identification of muons, and the measurement of their charge and momentum are therefore an important handle to 
discriminate between the signal and the background processes. This task has to be accomplished by a muon magnetic 
spectrometer located downstream of the emulsion target. 

Similarly, electrons are also produced in fully leptonic $\tau$ decays and in electron neutrino charged current 
interactions. The identification of electrons is therefore mandatory. The dense structure of the neutrino target 
means electromagnetic showers develop within the target itself. This makes  electrons identifiable through the 
detection of the shower by the use of a Emulsion Cloud Chamber~\cite{bopera}.  

The need to distinguish $\nu_\tau$ from $\bar{\nu}_\tau$ sets the requirement for a neutrino target in a magnetic 
field. In order to allow a compact design of the target, the field has to be about 1~T. If one exploits the micron 
accuracy of the emulsion, the detection of the charge of the $\tau$ daughter particles can be accomplished within 
an emulsion chamber of a few centimeters.   

Besides the muon magnetic spectrometer, real-time detectors are important also in the target region to provide the 
time stamp of the neutrino interaction. They are essential when no muon is produced in the final state as well as 
when the muon is so soft that it stops within the target without reaching the muon spectrometer. 

A high efficiency in the lepton identification is necessary to reduce the background induced by 
interactions with charmed hadron production. The muon identification can also  profit from the high 
target density by exploiting the correlation between range and momentum. 

The background rejection requires the momentum measurement of $\tau$ daughters. This measurement can profit from the 
need of a neutrino target in a magnetic field. 

The detector, being optimized for the $\tau$ lepton identification, is also well suited for the observation of 
the production and the decay of charmed hadrons. The study of  charm production cross section will considerably 
improve the measurement of the strange quark content of the nucleons.  

Electron neutrinos with energies above 40~GeV come predominantly from charmed hadron decays in the beam dump. 
The fraction of charm induced electron neutrinos among the incident flux at the neutrino target is about 
95\% above 40~GeV. Therefore, their detected interaction rate can be used to normalize the charm production in 
the target. 

Given the extremely high neutrino flux, a compact detector of a few cubic meter volume would 
detect a few millions of neutrino interactions including several thousands $\nu_\tau$ 
interactions. Being located between the muon shield and decay vessel, it can profit of the low 
muon rate without causing significant loss of acceptance for the hidden particle search. 
Nuclear emulsions record  all charged particle tracks integrated during their exposure. In order 
not to degrade the performances achievable in terms of vertex reconstruction and kinematics 
measurements, the integrated rate should not exceed $10^3$ particles/mm$^2$. 
This goal can only be achieved by replacing the films during the run period. With the current estimate 
of the muon flux immediately downstream of the muon shield, it is estimated the films would 
need to be replaced ten times during the five years run. The soft electron component accompanying the muon tracks is 
expected to be ten times larger with an energy spectrum softer than 20 MeV. A lead box 
with a thickness of 7 mm surrounding the neutrino target would make this component negligible.

%% file: facility/Facility.tex
\chapter{The SHiP Facility at the SPS}
\label{sec:facility}

\section{Introduction and operational scenario}
\label{sec:op_scenario}

\begin{figure}[htb]
\begin{center}
\includegraphics[width=0.8\linewidth]{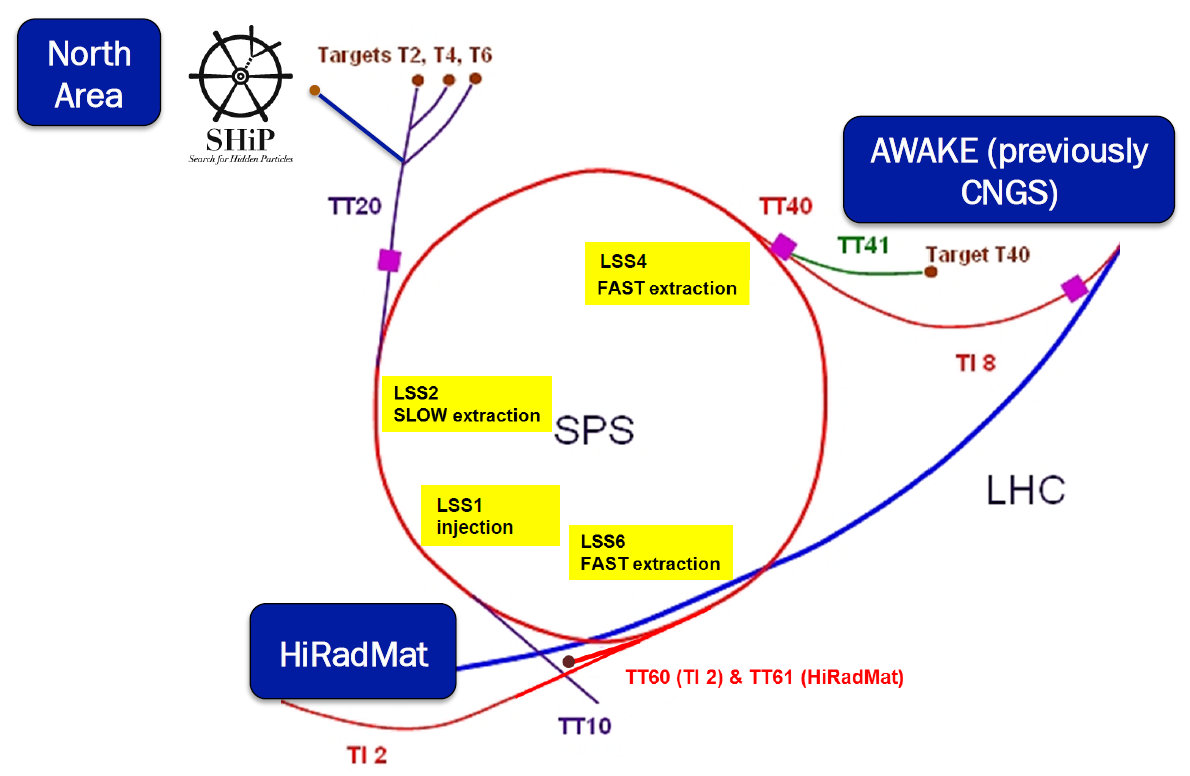}
\caption{Overview of the SPS accelerator complex. The SHiP facility is located
on the North Area and shares the TT20 transfer line with the fixed target programmes.}
\label{fig:facility_acc_complex}
\end{center}
\end{figure}

\begin{figure}[htb]
\begin{center}
\includegraphics[width=0.85\linewidth]{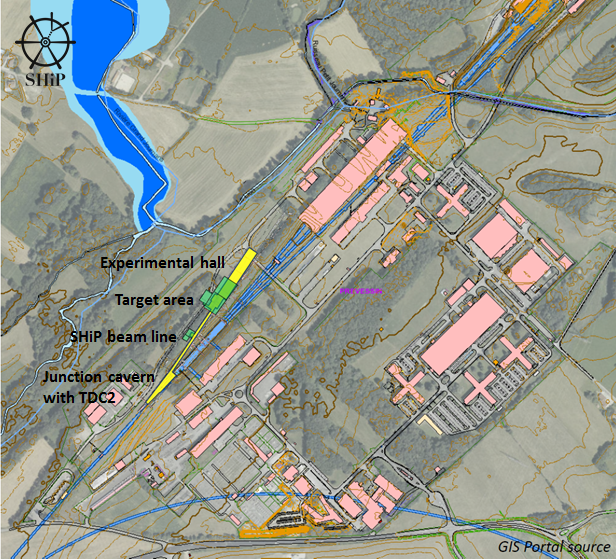}
\caption{Aerial view of the geographic location of the SHiP facility.}
\label{fig:facility_aerial_view}
\end{center}
\end{figure}

The proposed implementation of the SHiP operational requirements is based on minimal 
modification to the SPS complex and maximum use of the existing transfer lines. 
Figure~\ref{fig:facility_acc_complex} shows schematically the proposed location of the 
SHiP facility at the CERN Prevessin site. Figure~\ref{fig:facility_aerial_view} shows the
geographical location of the facility. The location is ideally suited since it allows a 
full integration on CERN land with almost no impact on the existing facilities. The SHiP 
facility shares the TT20 transfer line with the other North Area facilities, allowing 
both use of slow extraction for SHiP and switching on a cycle-by-cycle basis with 
other North Area beam destinations.

The SHiP operational scenario assumes the same fraction of beam time starting from 2025 
(Section~\ref{sec:schedule}) as the recently completed CERN Neutrinos to Gran Sasso 
(CNGS) programme. It also assumes the current SPS performance. CNGS made 
use of 400~GeV protons from the SPS. A nominal year of operation corresponded to about 
200 days of physics during which 60\% of the SPS super-cycle was dedicated to CNGS on 
average. Typical machine availability of about 80\% was achieved in these conditions.
This allowed producing an annual total of 4.5$\cdot$10$^{19}$ protons on target in parallel 
to the operation of the test beams and the physics programmes on the North Area, and the LHC.
In the same way, the operation of the SHiP facility is compatible with the operation of 
both the LHC and the North Area facilities for test beams and fixed target programmes as 
they are currently implemented.

In accordance with the general requirements described in Chapter~\ref{sec:requirements}, the
most favourable experimental conditions for SHiP is obtained with a beam energy of around 
400~GeV. Machine Protection considerations require defining different extraction momenta 
for the beams circulating in the SPS according to their destination to allow proper 
interlocking. For this reason the momentum of the proton beam extracted from the SPS for 
SHiP must differ from 400~GeV/c by at least 5~GeV/c.

The physics sensitivity of the experiment is based on acquiring a total of 2$\cdot$10$^{20}$ 
protons on target. The operational mode consists of continuous 24-hour data taking throughout 
the operational year with the exception of maintenance during technical stops of the SPS and 
limited accesses for faults. 
  
In order to respect the maximum beam induced instantaneous particle flux in the SHiP detectors 
and the limits on the power density deposition in the target, the SHiP proton spill is 
transferred to TT20 using a slow resonant extraction on the third integer resonance at a time scale of 
one second with a flat top of 1.2 seconds. Based on past experience, a beam intensity of 
4$\cdot$10$^{13}$ protons on target per spill is assumed as the baseline for the SHiP facility, 
and for the design of the critical components like the target, the detectors 
and the general layout of the experimental area.
 
\begin{figure}[htb]
\begin{center}
\includegraphics[width=0.7\linewidth]{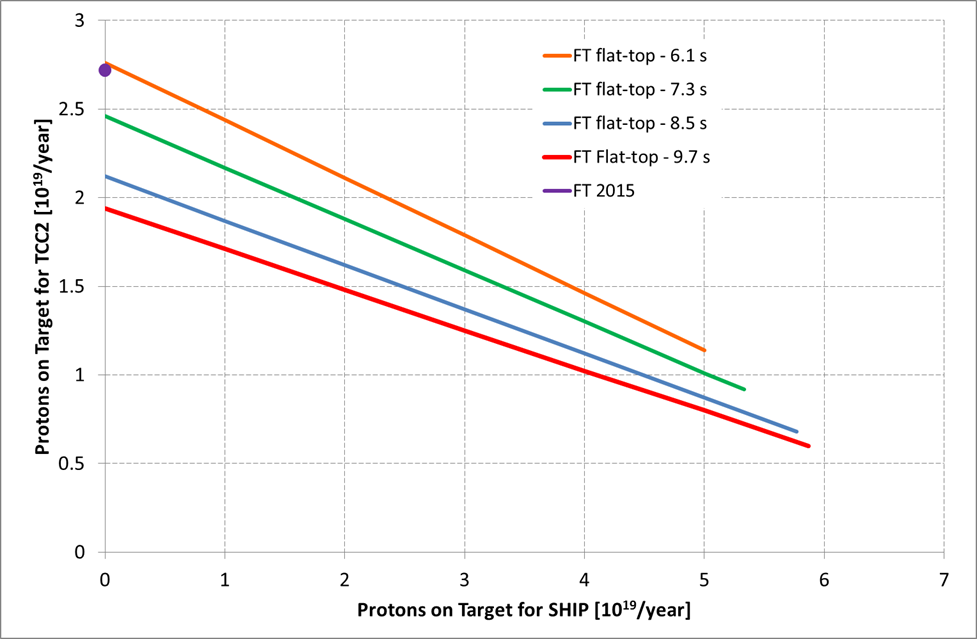}
\caption{The expected number of protons on the current North Area targets as a function of the 
number of protons on target for the SHiP facility. The plot shows the performance for different 
flat-top durations of the cycle for the current targets. The expected integrated number of 
protons for exclusive operation of the North Area targets with the super-cycle planned for 
2015 is indicated as a dot.}
\label{fig:facility_pot}
\end{center}
\end{figure}

The minimum cycle length which is compatible with the parameters above is 7.2~s 
(Appendix~\ref{sec:support_documents}:A2). The flat-top 
duration of the cycles dedicated to fixed target physics with the current North Area targets must 
be of the order of 9~s to profit from the maximum number of protons that can be accelerated per 
cycle in the SPS to 400 GeV, and be compatible with the maximum event rate acceptable at 
present by the North Area experiments and with the thermo-mechanical stress on the splitter 
magnets. Taking this into account, realistic configurations of the SPS super-cycle have been 
elaborated using the operational efficiency and the injector schedule from 2011 and 2012 for 
the operation of the North Area, CNGS and LHC, and Machine Developments (MD). It has been assumed that 
10\% of the SPS scheduled physics time is devoted to run LHC pilot cycles and another 
10\% to run LHC nominal cycles. Figure~\ref{fig:facility_pot} shows the number of protons on the 
current North Area targets as a function of the number of protons on the SHiP target for 217 
days of physics, corresponding to the situation for the 2011 run. 
The targeted working point delivers an annual total of 4$\cdot$10$^{19}$ protons on target for the 
SHiP facility and a total of 1$\cdot$10$^{19}$ protons on the current North Area targets with 
extractions of 0.43$\cdot$10$^{13}$ protons per second (0.7$\cdot$10$^{13}$ protons per second in 2015) 
during a plateau of 9.7~s with a cycle length of 15.6~s as in 2012. As an example, the SPS 
super-cycle with SHiP would consist of the 15.6~s fixed target cycle followed by five 7.2~s SHiP cycles, 
and a 4.8~s cycle (daytime only) dedicated to machine studies at low energy, leading to a total 
super-cycle length of 56.4~s. Hence, this corresponds to an 18\% spill duty cycle for the current 
North Area programmes and 11\% for SHiP\footnote{In terms of the full SPS cycles, the duty cycles 
for the current North Area programmes and SHiP are 28\% and 64\%, respectively.}. 

Consequently, the projected integrated flux of 2$\cdot$10$^{20}$ protons on target for SHiP can 
be provided over five years of fully efficient operation in conditions which may be considered 
nominal. In practice, including commissioning, upgrades and possible extensions of the programme, 
the facility has been designed for at least ten years of operation with the described equipment.

% Compared to entirely dedicated operation without SHiP, the operational scenario with the 
% delivery of $4\cdot 10^{19}$ protons per year on the SHiP target implies a 40\% reduction of the 
% beam availability for the North Area fixed-target programme from 2023

The sections which follow below describe briefly the entire facility and its main aspects 
and parameters. More details are to be found in the complementary documents listed in 
Appendix~\ref{sec:support_documents}.

\section{Primary beam line and beam transfer}
\label{sec:beam_line}

The slow extraction from the SPS LSS2 begins by first creating a large momentum spread 
and de-bunching. The de-bunching is essential for SHiP to produce a spill which is as
uniform as possible. The continuous beam extraction is then accomplished with a 
set of suitably located extraction sextupoles which are used to create a stable area in 
the horizontal phase space of the beam. This initial phase space area is larger than the 
area occupied by the beam. A dedicated servo-quadrupole consisting of four short QMS 
quadrupoles moves the tune towards QH = 26.666 to shrink the stable phase space area. 
Protons with coordinates outside the stable area move away from the beam core along 
the outward going separatrices, and eventually cross the wires of the electrostatic 
septa (ZS), into its high field region. The ZS deflects the particles into the magnetic 
elements of the extraction channel consisting of thin MST septum magnets and thick 
MSE septum magnets, which move the beam into TT20 transfer line.

Unavoidably the extraction involves beam losses of about 0.5 - 1\% on the wires as the 
beam is driven across the ZS septum. This leads to radioactivation of the LSS2 
extraction region, and the inherent risk of sparking, damage to the wires in the ZS due 
to beam heating, vacuum pressure rise, radiation damage to sensitive equipment and cables, 
and finally also increased cooldown times for interventions. While extractions of up to 
4$\cdot$10$^{13}$ have been used routinely, the performance of the ZS septa and the induced 
radioactivity in the LSS2 are key factors in the overall SHiP performance. The experience 
from the SPS operation for WANF, during which approximately half of the total integrated 
number of protons for SHiP was extracted with fast-slow extraction during a 5-years period, and 
during 2007 when over 2$\cdot$10$^{19}$ protons were slow extracted to the North Area 
through LSS2 has been studied to allow extrapolating to SHiP operation. The studies
indicate that improvements can be expected from improved stability of the extraction
and the machine, and improved beam quality. Additional instrumentation will be added 
along the extraction region to allow fast monitoring and optimization of the quality of 
the extraction to minimize the overall losses. MD studies and simulation efforts are 
required for this. Studies of improvements in the extraction technique will also
be pursued. Mitigations to reduce the activation and reduce 
exposures to personnel come from adaptation of the materials used in the equipment 
in the extraction region and improved equipment reliability and redundancy, as well 
as alternative intervention techniques. A serious effort will be needed to 
investigate these. Clearly, improvements in these areas are all of general interest
to the CERN fixed target programmes.

\begin{figure}[htb]
\begin{center}
\includegraphics[width=0.8\linewidth]{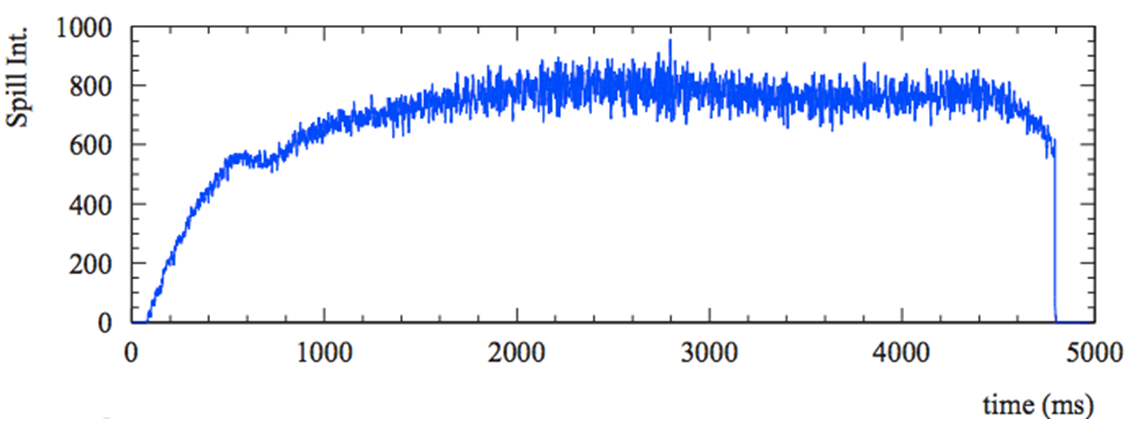}
\caption{Example of a properly adjusted quasi-trapezoidal spill with good uniformity.}
\label{fig:facility_spill_uniformity}
\end{center}
\end{figure}

As shown in Figure~\ref{fig:facility_spill_uniformity}, the extracted spill will not be 
perfectly trapezoidal, especially at the beginning of the spill, where an 
initial spike in the proton rate can be expected. This may have some impact on the 
energy deposition in the target and increase the probability of combinatorial background 
from the instantaneous increase in the beam induced particle flux. The spike could be 
mitigated by modifying the start of the sweep but the design of critical components 
should take into account possible spikes in the proton rate of up to a factor of two to 
three. Also, a high bandwidth intensity monitor installed on the SHiP transfer line is 
aimed at detecting spikes and provide the information to the SHiP detector readout.

Machine development (MD) sessions for SHiP are envisaged at the end of the 2015 proton 
run to develop a 7.2~s cycle with 1~s slow-extraction flat-top and to study beam losses,
optimize spill uniformity, and check the switch of powering of TT20 between the SHiP
cycles and the general fixed target cycle.

\begin{figure}[htb]
\begin{center}
\includegraphics[width=0.9\linewidth]{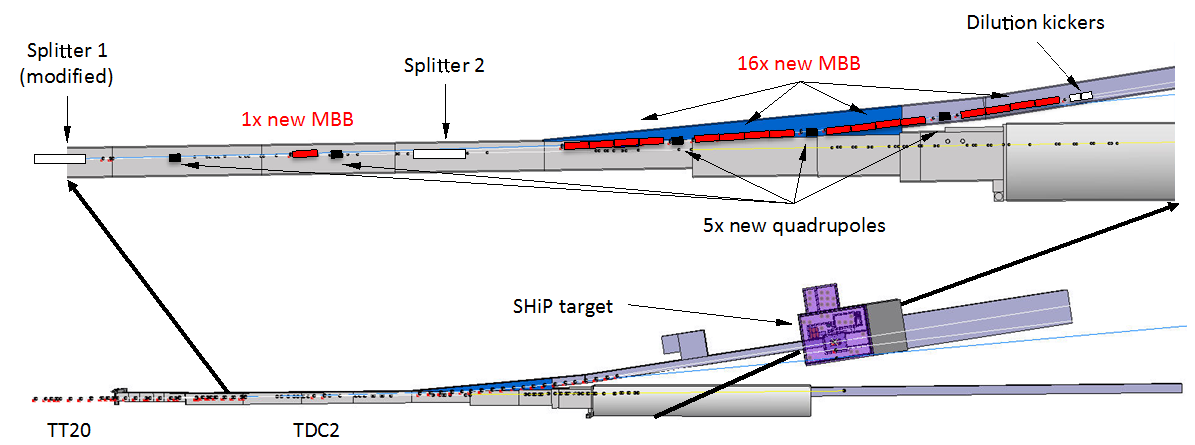}
\caption{Layout of the switch region from TT20 to the new beam line towards the
SHiP target.}
\label{fig:facility_ship_switch}
\end{center}
\end{figure}

The location of the SHiP facility allows reuse of about 600~m of the present TT20 transfer 
line, which has sufficient aperture for the slow-extracted beam at 400~GeV. As illustrated in
Figure~\ref{fig:facility_ship_switch} the new 
dedicated SHiP beam line branches off at the top of the existing TT20, in the 
TDC2 cavern with the help of a set of newly proposed bi-polar magnets which 
replace the current M2 splitter magnets. The current M2 splitter, which consists 
of three MSSB2117 magnets, deflects part of the beam to the Eastern side of the EHN1 
North Area hall towards the T6 production target for the EHN2 experimental building 
currently hosting the COMPASS experiment. The rest of the beam continues towards the second 
splitter which distributes the beam to EHN1 and to EHN3 for the NA62 experiment. The new 
splitter will ensure on a cycle-to-cycle basis, both the current beam splitting function 
and the beam switching function required to transfer the entire beam to the dedicated beam 
line for the SHiP facility. 

% The principal of the bi-polar three-way magnet is shown 
% in Figure~\ref{fig:facility_splitter}. 

The present MSSB design is an in-vacuum Lambertson septum, built with radiation robust 
materials and low-maintenance assembly, with a yoke machined from solid iron. The coil 
technology is special, using compacted MgO powder around a central copper current carrying 
a water-cooled tube, mechanically supported by an external grounded copper sheath. The 
new magnet has to have a larger horizontal aperture. It has
to be built with a laminated yoke in order to perform the polarity switch between SPS 
cycles, which implies ramping reliably in about 2~s. Although existing technology (e.g. 
1.5 mm punched laminations, blue-steamed for insulation) is proposed, early R\&D and 
prototyping is needed to make sure that the design can achieve the very tight mechanical 
tolerances needed in the septum region, which are critical for beam losses. The magnet
has also to be built from radiation robust materials. The feasibility of this MSSB-S magnet 
has been investigated in a pre-study (Appendix~\ref{sec:support_documents}:A3).

% \begin{figure}[htb]
% \begin{center}
% \includegraphics[width=0.7\linewidth]{facility/facility_splitter.png}
% \caption{Schematic view of principle of the new bi-polar splitter/switch.}
% \label{fig:facility_splitter}
% \end{center}
% \end{figure}

As shown in Figure~\ref{fig:facility_ship_switch} the splitter/switch magnet is followed 
by a set of 17 main dipole magnets which produces a 
136~mrad deflection of the beam out of TDC2 towards the Western side
via an 86m long junction cavern into a new 150~m extraction 
channel towards the target complex for the SHiP secondary beam line. These magnets will 
consist of SPS main bends of either type MBBs or MBNs. The beam deflection has 
been optimized in order for the SHiP facility to stay clear of the existing installations 
with safe margins, while maintaining the SHiP transfer line as short as possible. The 
current clearance allows civil engineering to take place along the entire SHiP facility 
starting from the access point in the middle of the SHiP transfer line even during North 
Area beam operation. It also avoids excavations in radioactivated soil along this 
entire stretch.

The SHiP beam line is entirely in the horizontal plane. Bending magnets and a
set of quadrupoles are used to minimize dispersion and to suppress motion induced by momentum
variations during extraction. The transfer line is mainly composed of 120~m of drift 
space. In order to produce a sufficient dilution of the energy deposited in the target,
the drift space is used to produce a beam spot of at least 6mm RMS together with 
the critically important beam sweep on the target face. The beam sweep is implemented with 
a pair of orthogonal deflection magnets similar to the existing SPS MPLH and MPLV bumper 
magnets, with a fast Lissajous powering function to produce an Archimedian spiral sweep 
with constant arm separation and uniform speed. With a drift of about 120~m and a bending 
power of 0.25~mrad per plane, it is possible to achieve a sweep radius of up to 30~mm and 
a total sweep path of $>$~600~mm. In order to protect the target and the downstream equipment 
and detectors, the SPS beam extraction is interlocked on the beam dilution system.

As the SHiP extraction tunnel contains active elements, it is ventilated and has a material 
and personnel access point located in the middle. Beam loss monitors will detect 
accidental losses of the beam near this access shaft. An associated surface building 
houses the infrastructure systems for the transfer beam line.

\subsection{R\&D related to beam transfer}
\label{sec:splitter_RandD}

There are two main areas requiring R\&D related to the beam transfer, the first of which 
concerns the newly proposed magnet to perform both the switching function to the current
North Area targets, and the second which concerns the beam sweep on the SHiP target.

In particular, the MSSB-S splitter/switch requires an extended R\&D in order to investigate 
the challenges related to both the laminated yoke and the coils. This coil technology also 
needs extra lead-time to identify and qualify suppliers. The overall schedule for the
R\&D, design, production and testing of these magnets has been estimated at 48 months. 

In addition, there are two main Machine Developments involved in the setup for 
SHiP, the first of which concerns the performance of the SPS extraction and beam losses,
and the second which concerns the spill quality and uniformity. In addition, MDs are
needed to setup the SHiP SPS cycle and the TT20 optics. All of these MDs can be done without
the SHiP beam line.

\section{Target and target complex}

\subsection{Proton target}
\label{sec:target}

In order to maximize the production of heavy mesons and minimize the production of 
neutrinos and muons, the target must be designed with a material with the shortest 
possible nuclear interaction length and with dimensions that contain the proton 
shower. In addition, to further minimize the flight of pions and kaons, the target should be
designed with minimum space for cooling channels seen by the proton shower. Consequently,
the SHiP production target is one of the most challenging aspects of the facility
due to the very high average beam power (up to 350~kW) deposited on the target.
The peak power during the spill amounts to 2.56~MW. For this reason, the design of the 
target relies on the energy dilution produced by the large beam spot and the beam sweep 
produced at the beginning of the SHiP beam line. 

\begin{figure}[htb]
\begin{center}
\includegraphics[width=0.49\linewidth]{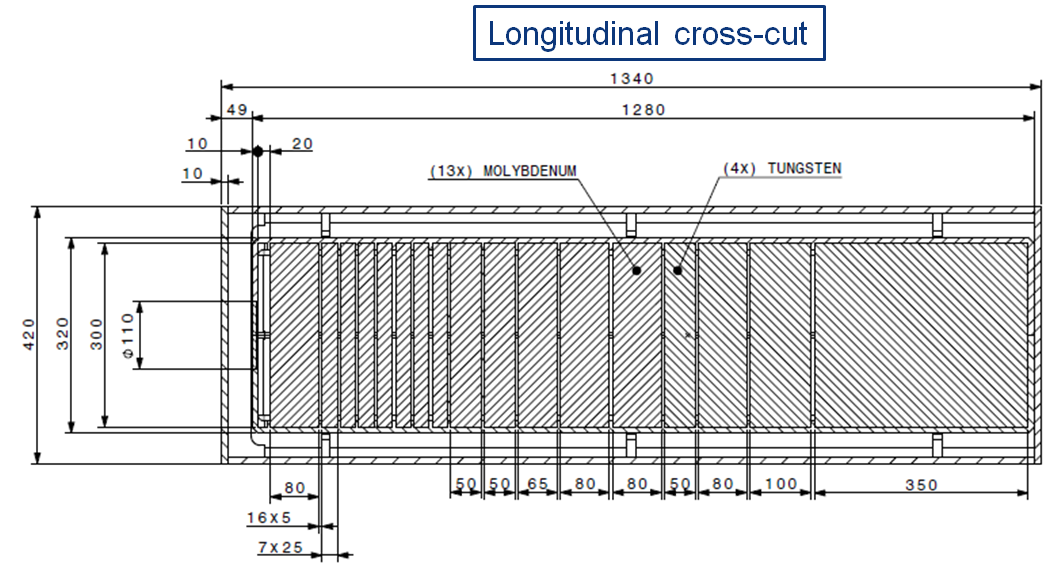}
\includegraphics[width=0.49\linewidth]{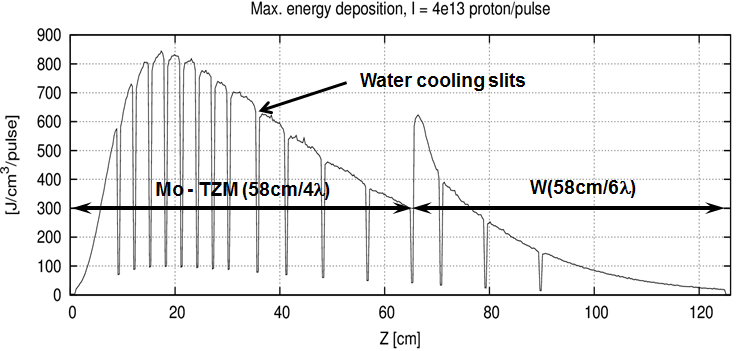}
\caption{(Left) Target configuration. (Right) Peak energy 
deposition during a spill.}
\label{fig:facility_target_config}
\end{center}
\end{figure}

These requirements have been the subject of a detailed conceptual study, which has
demonstrated that the target is feasible (Appendix~\ref{sec:support_documents}:A4). 
The required performance is achieved with a longitudinally segmented hybrid target 
consisting of blocks of four interaction lengths of titanium-zirconium doped molybdenum (TZM) alloy (58~cm) in 
the core of the shower followed by six interaction lengths of pure tungsten (58~cm). 
The blocks are interleaved with 16 slits of 5~mm thickness for water cooling. In order to respect the material limits on 
the thermo-mechanical stresses, the thickness of each slab together with the location of 
each cooling slit has been optimized to provide a uniform energy deposition and to 
guarantee sufficient energy extraction. This limits the peak power density in 
the target to around 850~J/cm$^3$/spill and compressive stresses below 400~MPa in 
the core of the shower. Figure~\ref{fig:facility_target_config}(left) shows the target 
design. The total dimensions of the target is 1.2~m in length with transverse dimensions 
of 30x30~cm$^2$ to maximize the shower containment. Figure~\ref{fig:facility_target_config}(right) 
shows the maximum energy density per spill of 4$\cdot$10$^{13}$ protons on target.
Specifically, the molybdenum blocks consists of a doped alloy (TZM) to obtain the required 
material properties, while the tungsten blocks are made of pure tungsten.

\begin{figure}[htb]
\begin{center}
\includegraphics[width=0.8\linewidth]{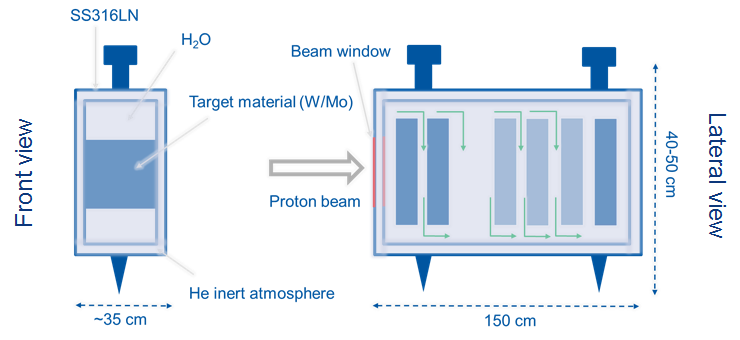}
\caption{Schematic view of the SHiP target assembly.}
\label{fig:facility_target_compartment}
\end{center}
\end{figure}

The target pile is assembled in a double-walled compartment as shown schematically in 
Figure~\ref{fig:facility_target_compartment}. The inner vessel enforces the high flow 
water circulation between the target slabs and ensures a pressurized water cooling of
15-20~bar in order to increase the water boiling point. A flow rate of 180~m$^3$/h is envisaged.
An alternative cooling based on 
helium gas cooling will be investigated in a TDR phase. The external vessel provides a 
primary circulation of helium around the inner container, in order to guarantee an 
inert atmosphere to reduce corrosion effect and to provide a protection layer capable 
of detecting potential water leakages from the internal containment. The assembly is 
fitted with connections and hooks which ensure completely remote handling by means of 
the target area crane. 

\subsection{Target complex}
\label{sec:target_complex}

\begin{figure}[htb]
\begin{center}
\includegraphics[width=0.7\linewidth]{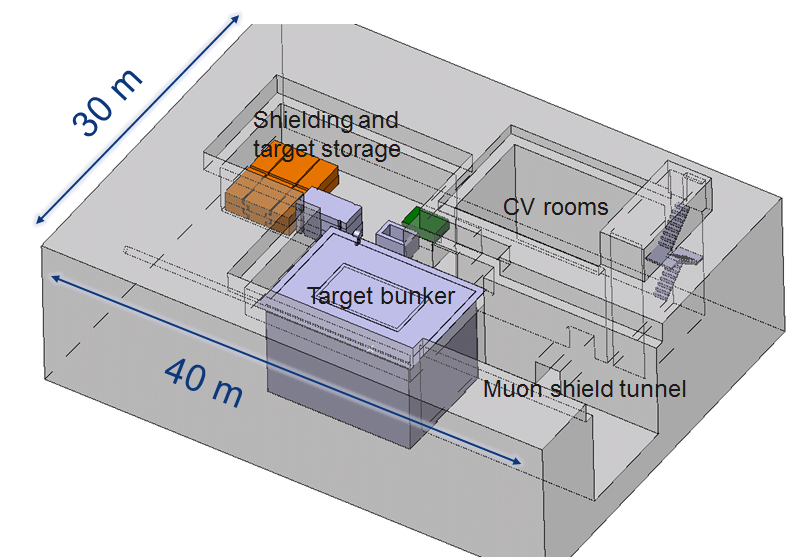}
\caption{View of the SHiP target complex. A 30$\times$40~m$^2$ surface building houses
the entire complex.}
\label{fig:facility_target_complex}
\end{center}
\end{figure}

The SHiP target complex is shown in Figure~\ref{fig:facility_target_complex} 
(Appendix~\ref{sec:support_documents}:A4). It has been designed with the aim of being 
multi-purpose, i.e. allowing exchange of target and shielding configuration for 
alternative uses in future experiments. 

The target is embedded in a massive cast iron bunker (440~m$^3$), with an inner core 
consisting of water-cooled cast iron blocks with embedded stainless steel water cooling 
pipes (about 20~m$^3$). All the iron blocks are specially designed 
for completely remote handling. Studies show that about 20~kW of cooling capacity is required. 
The downstream proximity shielding which has a thickness of 5~m also acts as a hadron 
stopper. The hadron stopper has the double objective of absorbing the secondary hadrons 
and the residual non-interacting protons emerging from the target, and to significantly reduce 
the exposure of the downstream experiment muon shield to radiation. In total, roughly 
450~m$^3$ of cast iron will be required. 

In order to minimize the radiation issues in the primary beam line, the upstream shielding has 
only a limited passage for the beam vacuum chamber of about a 20~cm diameter. An absorber 
for neutral particles protects the upstream beam line from neutrons and other 
neutral particle radiation. 

\begin{figure}[htb]
\begin{center}
\includegraphics[width=0.7\linewidth]{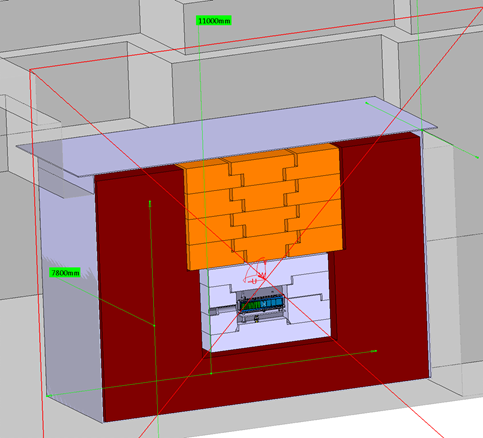}
\caption{View of the target bunker.}
\label{fig:facility_target_bunker}
\end{center}
\end{figure}

The target bunker consist of a compact multi-compartment complex with a floor size of 
7.8$\times$10~m$^2$ at a depth of 11~m to comply with the radiological requirements 
(Figure~\ref{fig:facility_target_bunker}). It is designed for access from the top only. The bunker 
containing the production target is embedded inside a  secondary helium atmosphere with a 
circulation system and an online purification to prevent production of high mass isotopes generated by the 
interaction of secondary neutrons with air and oxidation. A shielding and target storage is 
also foreseen at a depth of about 4~m with respect to the floor of the target hall to allow 
temporary storage of the bunker shielding blocks during interventions and to store a target 
for cooldown.

A service pit on the top of the helium vessel allows human interventions on the various 
feed-throughs (water, helium, controls) between the service areas and the internal part 
of the helium vessel and target bunker. An extra 240~cm of concrete slabs separate the 
service pit and the shielding area from the target hall. The slabs are removable 
to allow crane access to the target bunker and to the storage zone.

The cooling and ventilation equipment required for the operation of the target is 
located in an underground area connected to the target bunker via the service pit. 
A stair-case allows access from the target hall. A material access is also envisaged
from the target hall by means of the target area crane.

A 30$\times$40~m$^2$ target surface building covers the target bunker, the storage zones 
for the activated target and for the shielding blocks during interventions, the 
cooling and ventilation bunker, and the first 15~m section of the muon shield. An 
adjacent 25$\times$15~m$^2$ building with supervised access holds all other 
services related to the operation of the target. It houses the PAD/MAD to the target 
hall, RP monitoring control systems, intermediate heat exchangers for the target and 
shielding cooling circuit, target hall ventilation system as well as ancillary 
systems such as UPSs, control racks for access and safety systems, etc. The service 
building will also house all the electrical cabinets, control PLCs as well as all 
electronics sensitive to radiation.

To prevent any leakage into the environment, the design of the ventilation system of the target 
hall and all related underground areas is to be based on the ISO17874:2004 norms with a 
pressure cascade between the various compartments. The target hall is underpressurized 
by 60~Pa with respect to the external environment while the most exposed zones are at 
roughly -220~Pa. The air-tightness of the various areas is interlocked with beam operation.
The design will be subject to a dedicated approval by the corresponding safety commission.

A fully redundant 40~tonne crane services the target installation and the first section 
of the muon shield.

\subsection{R\&D related to the target and target complex}
\label{sec:target_RandD}

Following the concepts developed in the present design, several essential R\&D items have
been identified. They include

\begin{itemize}
\item{target material mechanical properties as function of irradiation,} 
\item{the assembly, configuration and manufacturing of the cladded refractory metal blocks for the target,}
\item{the water-cooled cast iron shielding blocks,} 
\item{the water cooling circuit of target assembly, including in particular the fully metal, high-flow 
rate compatible and high-pressure plugin systems for remote and fast connection/disconnection of water pipes,}
\item{the alternative of a helium cooled target.}
\end{itemize}

% \section{Muon shield}

\input{shield/muon_shield.tex}

\section{Experimental area}
\label{sec:exp_area}

By maintaining the entire beam line horizontal and at the same level as the switching in 
TDC2, the experimental hall ends up conveniently at a depth which avoids radiation 
problems, even with horizontal magnetic sweeping of the muon flux, while still allowing easy 
direct access from top without a shaft.

The experimental area consists of a 120~m long and 20~m wide underground hall.
It hosts the second section of the muon shield, the tau neutrino detector, and the 
decay volume together with the spectrometer and the rear detectors. A thin sealed 
wall ensures that the air volume around the first section of the muon shield located 
in the 8$\times$15~m$^2$ trench is isolated from the experimental hall. 

\begin{figure}[htb]
\begin{center}
\includegraphics[width=0.7\linewidth]{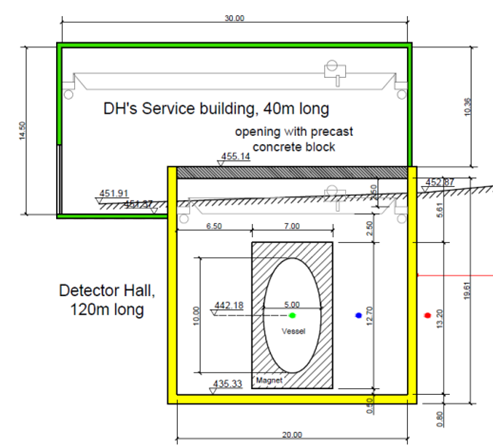}
\caption{Cross section of the experimental hall together with the detector surface 
building.}
\label{fig:facility_det_hall}
\end{center}
\end{figure}

The underground experimental hall is serviced along the entire length by an overhead crane with 
a 40~tonne capacity. In order to house the large vacuum vessel and the spectrometer 
magnet spanning a height of 12.7~m, and allowing sufficient room for the crane and 
hook, the experimental hall has a height of about 19~m. This puts the ceiling at 
the level of the ground (Section~\ref{sec:civil_engineering}). The hall is ventilated 
for temperature control and fresh air. The emulsion used in the tau neutrino detector 
requires that the temperature remains stable to a few degrees. Due to the prompt muon dose, 
access to the experimental area will be prohibited during the operation of the facility. 

A 30$\times$40~m$^2$ surface building adjacent to the target surface building covers the 
first section of the underground experimental hall. In order to allow  
top access to first 20$\times$20~m$^2$ of the experimental hall for installation and 
maintenance, the underground hall is covered with removable precast concrete blocks.
This section is serviced by a 40~tonne crane. For radiological reasons, the opening to 
the underground hall is closed during operation. Despite the concrete blocks, the 
residual muon flux on top of the underground hall gives rise to a prompt dose rate of a 
few $\mu$Sv/h during operation (see Figure~\ref{fig:facility_surface_dose}). For this 
reason, this section of the surface building is classified as Supervised Radiation Area~\cite{FWirth:2006}. 
A PAD/MAD and a large access door ensures the controlled access to this area. The other 
section of the surface building houses the general services and the experiment control 
room. A cross section of the experimental hall and the surface building is shown in 
Figure~\ref{fig:facility_det_hall}. A secondary personnel access shaft is located at 
the rear of the experimental hall.
 
% What is the limit of the supervised area for the surface building? turn around building?

A zone of restricted access will be ensured in the proximity of the spectrometer magnet in 
the experimental hall and, if necessary, on top of the ceiling of the hall 
according to the CERN regulations.

An open storage area in connection with the surface buildings is envisaged to allow 
temporary storage during the installation phase.

%\begin{itemize}
%\item{Electrical infrastructure}
%\item{Cooling water}
%\item{Gases}
%\item{IT networks}
%\item{All the alarms (fire detection, gas detection}
%\end{itemize}

\section{Civil engineering}
\label{sec:civil_engineering}

The SHiP facility consists of a new beam extraction tunnel on the North Area starting from a 
new junction cavern alongside the splitter section in TDC2, leading to a new target complex 
and a new underground experimental hall. It also includes several surface buildings for 
services and accesses. A preliminary study (see Appendix~\ref{sec:support_documents}:A6 for 
details) has been performed to 
investigate the feasibility and the construction techniques, with the aim of producing a work 
plan and a cost estimate. The location of the facility, the work strategy and the plan has 
been optimized to take into account the clearance of minimum 8~m soil separation between 
the civil engineering works and the existing facilities during North Area beam operation, 
as required by radiation protection.

\begin{figure}[htb]
\begin{center}
\includegraphics[width=0.75\linewidth]{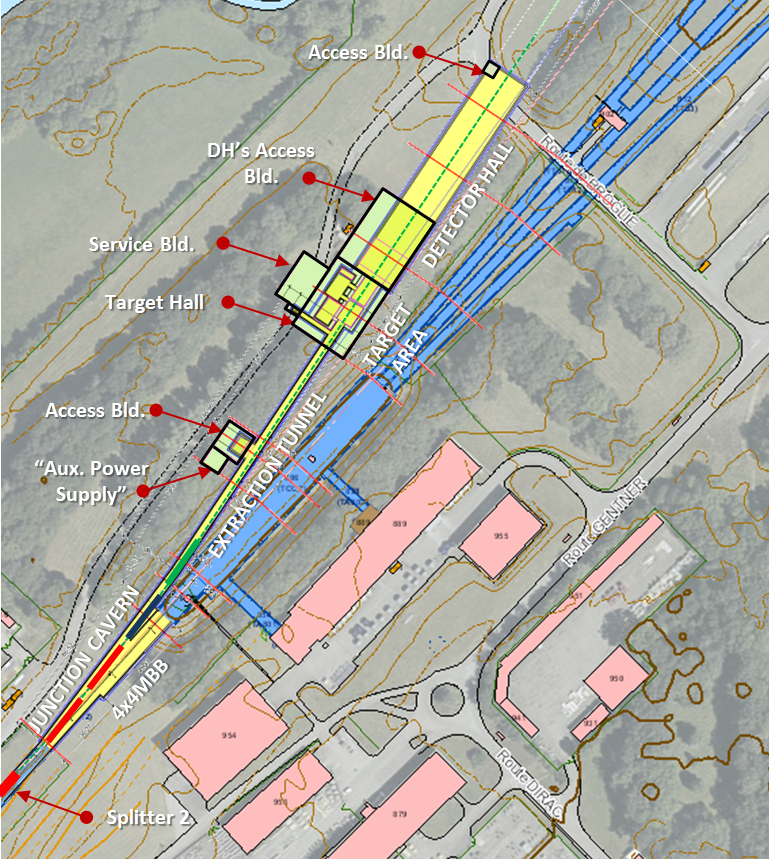}
\caption{Overview of the SHiP facility and civil engineering works.}
\label{fig:facility_ce_works}
\end{center}
\end{figure}

Figure~\ref{fig:facility_ce_works} shows an overview of the civil engineering for the 
SHiP facility. All civil engineering works are fully located within existing CERN 
land on the Prevessin campus. This location is extremely well suited to housing the SHiP 
facility, with the very stable and well understood ground conditions, and very limited 
interference with the current buildings, galleries and road structures. Detailed geological 
records exist and have been consulted to minimize the costs and risk to the project. 

\begin{figure}[htb]
\begin{center}
\includegraphics[width=0.95\linewidth]{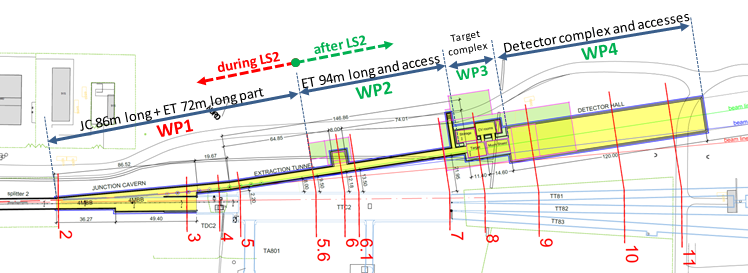}
\caption{The SHiP project work packages. The cross sections corresponding to the numbers
can be found in the Appendix~\ref{sec:support_documents}:A6.}
\label{fig:facility_wp}
\end{center}
\end{figure}

As shown in detail in Figure~\ref{fig:facility_wp}, the civil engineering works as well as 
the installations are naturally split in four work packages to respect the North Area beam 
operation and to allow parallelizing the design, the integration and the construction works:

\begin{enumerate}
\item{WP1: TDC2 junction cavern and the first section of the SHiP extraction beam line necessary 
to continue civil engineering end installation downstream in parallel to beam operation on the North Area}
\item{WP2: Beam extraction tunnel}
\item{WP3: Target area} 
\item{WP4: Experimental area}
\end{enumerate}

All underground works are excavated using the “cut-and-cover” method, rather than tunneling
methods. Due to the depth and the proximity to existing facilities, it is assumed that the 
construction will require excavation with retaining walls (diaphragm walls) along the entire
length of the facility. To avoid having to dispose of large quantities of spoil, it will be 
used to cover the experimental hall. This will also eliminate any problems related to the 
muon dose levels on top of the underground detector hall after the surface bulding.

The floor level of the extraction tunnel is set at a constant 441~masl (metre above sea level), 
which is same as the TDC2/TCC2 complex. Consequently the excavated depth to the invert of the beam 
line is approximately 10~m, which means that there is approximately 5.5~m of land fill over the 
top of the completed extraction tunnel.

The 8~m limit on the soil distance to the excavations during beam operation corresponds to 
the beginning of WP2 in Figure~\ref{fig:facility_wp}, which is 66~m downstream of the junction
cavern. A shielding will be installed in this tunnel section to allow continued construction and 
installation downstream during beam operation (see Section~\ref{sec:radioprotection}).  

\begin{figure}[htb]
\begin{center}
\includegraphics[width=0.9\linewidth]{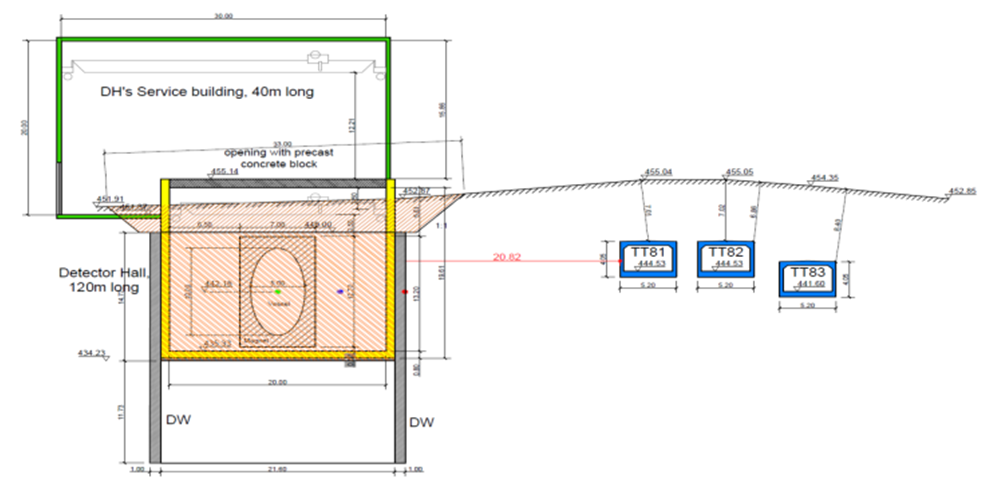}
\caption{Cross section of the experimental area and the TT81-83 transfer lines.}
\label{fig:cross_section_9}
\end{center}
\end{figure}

With the proposed beam bending out of TDC2, the target area and the experimental hall 
are at safe distance (Figure~\ref{fig:cross_section_9}) from the TT81 - 83 transfer lines 
both from the point of view of construction and radiation from North area fixed target 
operation and from SHiP operation.

Due to the vertical size of the SHiP vacuum vessel of about 11~m, including supports and the
centering on the beam axis, the floor level of the underground experimental hall is about 
4.5~m lower than the beam extraction tunnel floor level, that is at 436.5~masl. At the location 
of the SHiP spectrometer and the rear detectors, the floor level is another meter lower over 
a distance of about 15~m to provide space for the magnet, the calorimeters and the 
muon detector.

In order to take into account future extensions of the experimental programme, and re-use 
of the facility, the proposed orientation of the beam line and the underground complex 
allow reserving 100~m of space beyond the experimental hall
which is currently free from important structures.

\section{Radiation protection and safety}
\label{sec:radioprotection}

\begin{figure}[htb]
\begin{center}
\includegraphics[width=0.99\linewidth]{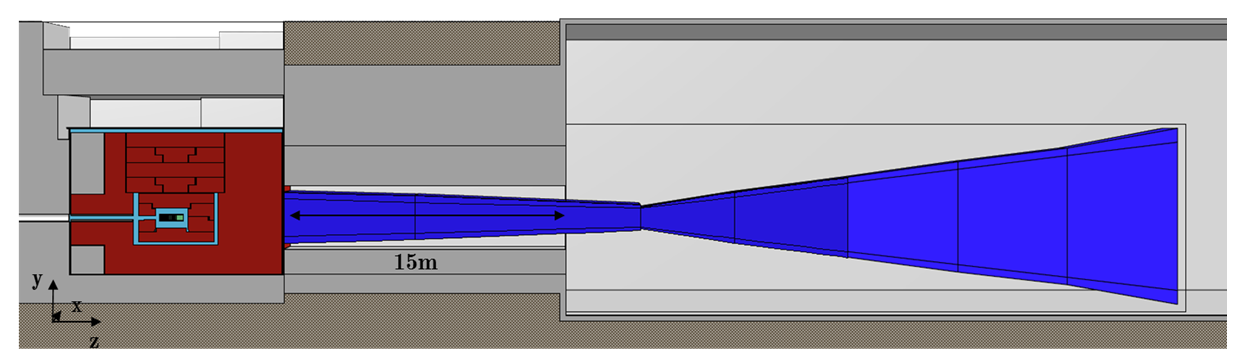}
\caption{Layout of the facility as implemented in FLUKA~\cite{Ferrari:2005zk, Battistoni:2007zzb}.}
\label{fig:facility_fluka_geometry}
\end{center}
\end{figure}

The large number of protons on target required by the SHiP facility during five nominal years 
of operation poses challenges to the radiation protection in several locations. In order to reduce 
the effect and mitigate the impact, the radiological aspects have been carefully addressed at the 
design stage of all components (Appendix~\ref{sec:support_documents}:A5). The studies include 
expected prompt and residual dose rates in the various accessible areas of the SHiP facility 
as well as the levels 
of activation and stray radiation in the surrounding experimental and public areas. The studies 
are based on past measurements and extensive simulations with the FLUKA Monte Carlo particle 
transport code~\cite{Ferrari:2005zk, Battistoni:2007zzb}. 
Figure~\ref{fig:facility_fluka_geometry} shows the layout of the facility as implemented in FLUKA.
All studies assume the SHiP operational baseline of 4$\cdot$10$^{13}$ protons on 
target per spill and an integrated total of 2$\cdot$10$^{20}$ protons on target per five nominal years.
%4$\cdot$10$^{19}$ protons per year. 
%As a saturation level is reached with this annual integrated flux, the number of years of operation is mainly relevant for the longevity of equipment (CHECK). 
%-> a saturation for a given radionuclide is roughly reached after an operation period of 6 half-lifes -> thus for long-lived radionuclides differences between 1 and 5 years of nominal operation are non-negligible. Since we have used the more precise (and more conservative) operation scenario of 5 years for our calculations, I think that we should not give the impression here that we have neglected contributions from long-term operation

The investigations identified the following critical areas which required detailed radiological assessment in view of the SHiP operation:
\begin{itemize}
\item{LSS2 extraction region}
\item{M2 splitter in TDC2, including accident scenarios} % included the paragraphs on CE, removal/installation of the new M2 splitter?
\item{Target complex}
\item{Experimental area}
\item{Surrounding experimental areas}
\item{Surrounding environment and public areas}
\item{First section of muon shield}
\end{itemize}

In addition, the studies included drawing up guide lines for the civil engineering in the 
proximity of the existing facilities during beam operation in the North Area and the 
procedure to respect during the construction of the junction cavern along-side TDC2.\\

%PARAGRAPH ON LSS2 EXTRACTION REGION TO COME.\\
The induced radioactivity in the LSS2 extraction region is expected to increase significantly with SHIP in addition to the North Area operation. 
Based on data from a radiation protection survey made 17 days after the end of the 2010 proton run, an increase by a factor of about 6 has been estimated.
This would imply doses of approximately 70~mSv/h at a distance of 40~cm to the ZS septum. 
To allow for a reduction of the doses, a decrease of the losses and failure rates should be pursued. 
Accidental losses may for example be decreased by more interlocking in general, such as intensity interlocking. 
Also the instrumentation may be improved, as the present system is more than 40 years old.
In addition, the reliability of specific components has been or can be improved (e.g. ZS feedthroughs). 
Intervention times and methods should further be optimised, for example by using robots and remote handling to lower the exposure of personnel to radiation.\\

% PARAGRAPH ON THE M2 SPLITTER AND ACCIDENT SCENARIOS TO COME.\\ % included in the paragraphs on CE, removal/installation of the new M2 splitter?

The current design of the target bunker and the target hall reduces the prompt dose rates 
in the target hall above the concrete shielding to a few $\mu$Sv/h, which allows designating 
the hall as Supervised Radiation Area~\cite{FWirth:2006}. At the bottom and the sides of the target bunker, the 
dose rates are reduced to about 1~mSv/h or below. Thus, it does not exceed the envisaged 
limit which has been defined to keep ground activation at an acceptable level~\cite{Strabel:2014}. \\

\begin{figure}[htb]
\begin{center}
\includegraphics[width=0.49\linewidth]{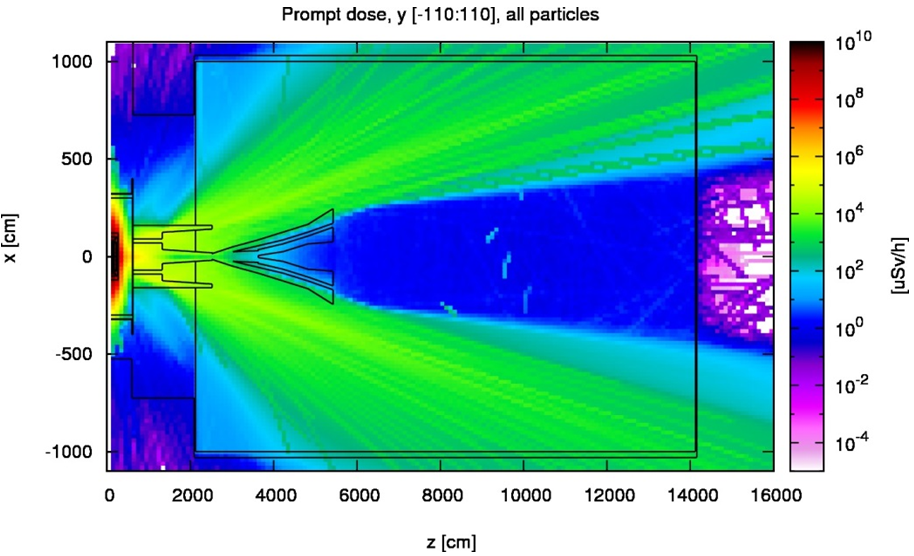}
\includegraphics[width=0.49\linewidth]{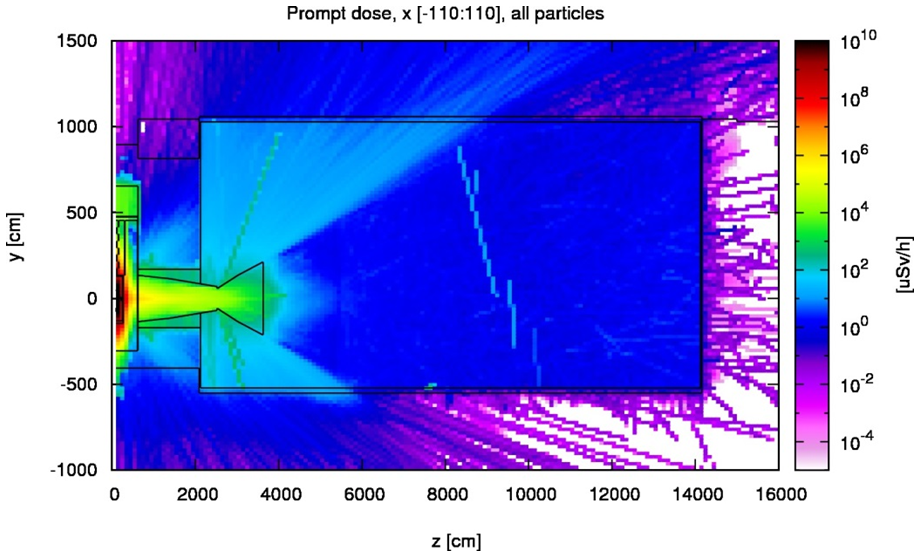}
\caption{Prompt dose rates in $\mu$Sv/h in the SHiP experimental hall for all particles. 
Left plot shows the top view and the right plot shows the side view cut through the beam line.}
\label{fig:facility_hall_dose}
\end{center}
\end{figure}

Figure~\ref{fig:facility_hall_dose} shows the prompt dose rates from all particles along 
the muon shield and in the experimental hall and surface building in the horizontal plane 
(left) and vertical plane (right) cut through the beam line. During operation the dose rates 
reach a few mSv/h along the walls of the experimental hall behind the muon shield and 
drop below 1 mSv/h in the surrounding soil such that levels of soil activation are 
considered acceptable. The side view of the experimental hall illustrates that the muons 
are also bent towards the top of the experimental hall. A few $\mu$Sv/h are reached in the 
surface building on top of the underground hall requiring this area to be classified as 
Supervised Radiation Area~\cite{FWirth:2006}. 

\begin{figure}[htb]
\begin{center}
\includegraphics[width=0.6\linewidth]{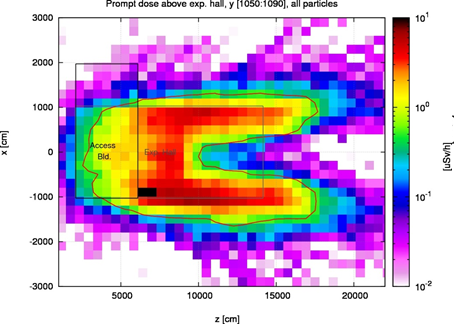}
\caption{Prompt dose rates in $\mu$Sv/h at the ground level (40~cm) above the experimental area with 
30~cm of concrete. The red line indicates the 0.5 $\mu$Sv/h contour. The soil from the excavations will be used to cover the ceiling of 
the underground hall to get below 0.5~$\mu$Sv/h.}
\label{fig:facility_surface_dose}
\end{center}
\end{figure}

Further simulations demonstrate that the existing beam lines TT81, TT82 and TT83 are not 
affected by the prompt dose rates originating from the SHiP facility. The SHiP operation 
neither influences the present area classification of the EHN1 experimental hall, which 
corresponds to a Supervised Radiation Area ($<$~3~$\mu$Sv/h). It should be noted
that these results on long-range effects are conservative due to the assumption of a moraine 
density which is 20\% lower than measured in ground samples. The operation of the SHiP 
facility does therefore not have any impact on the neighbouring experimental areas.

Figure~\ref{fig:facility_surface_dose} shows the prompt dose rates from all particles above the 
underground experimental hall with only 30~cm of concrete ceiling. It shows that a few $\mu$Sv/h
are expected on top of the ceiling of the underground experimental hall after the detector
surface building. As it is planned to cover the ceiling with the remaining soil from the
excavations, the level is brought down below 0.5~$\mu$Sv/h allowing for a Non-Designated Area~\cite{FWirth:2006}.

\begin{figure}[htb]
\begin{center}
\includegraphics[width=0.6\linewidth]{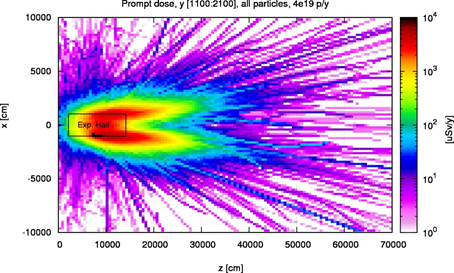}
\caption{Prompt dose rates in $\mu$Sv/year at the ground level (averaged over the first 10~m above ground) 
around the experimental area with 30~cm of concrete ceiling only.}
\label{fig:facility_environment_dose}
\end{center}
\end{figure}

Figure~\ref{fig:facility_environment_dose} shows the annual dose rates at the ground level 
around the experimental area with a 30~cm concrete ceiling of the underground hall. At CERN’s fence, 
which at the closest point is approximately 70~m from the SHiP beam line, the annual dose rates 
are expected to be less than 5~$\mu$Sv/year, which is an acceptable 
dose objective for the facility. Note again, that in case the soil from the excavations will 
be used to cover the experimental hall, the dose rates are expected to be reduced further. 
A standard stray radiation monitor for photons, muons and neutrons is envisaged at the fence 
closest to the most exposed area.\\

%PARAGRAPH ON THE FIRST SECTION OF THE MUON SHIELD AND AIR TO COME.\\
%for the moment the residual dose rates of the muon shield magnets and the air activation have not been estimated, but, as stated in the paragraph below, both should not be a critical issue.

Assuming five years of operation the residual dose rates in the SHiP target bunker are expected to be at the level of a few $\mu$Sv/h around the helium vessel after one week of cooling, thus allowing for access to this area. The residual dose rates originating from the removable cast iron shielding and the surface of the target reach up to a few 10$^{5}$~$\mu$Sv/h and 10$^{7}$~$\mu$Sv/h after 1 week of cooling, respectively. 
Due to these extremely high dose rates remote handling and designated storage areas are imperative for these elements (Section~\ref{sec:target_complex}).
Note that the residual dose rates at the beginning of the active muon shield are expected to be negligible ($<$~0.5~$\mu$Sv/h). 
The committed effective dose due to activated air and helium should be evaluated at a later stage of the project. 
When comparing the design and prompt dose rates of SHiP and CENF one can however conclude that significantly lower doses can be expected for SHIP~\cite{Strabel:2014, Vojtyla:2013}.

%PARAGRAPH ON THE PROCEDURES FOR CE WORKS FOR JUNCTION CAVERN. Levels of radioactivity in the soil, personal protection by workers during excavations and concrete cutting.\\
The civil engineering of the dedicated SHiP extraction tunnel is constrained such that a minimum soil thickness of 8~m must be kept around the tunnel walls of TDC2 and TCC2 during beam operation in the North Area. During shut-down periods the civil engineering is influenced by soil activation, which is expected to drop to below 0.1~$\mu$Sv/h (corresponding to approximately 1~LE of the new design exemption limits) at 3~m and 2~m from the tunnel walls of TDC2 and TCC2, respectively. To verify the amount of soil activation it is advised to take soil samples in this area. Any civil engineering (CE) work beyond the above-given limits requires CE workers classified as Radiation Workers~\cite{Cite:2014}.

Also the construction of the SHiP junction cavern, which involves partial demolishing of the activated concrete tunnel structure of TDC2, demands Radiation Workers. Additional exposure of the CE workers to radiation originating from activated beam line components in TDC2 should be minimised by installation of shielding or temporary removal of components.
A detailed RP survey has been 
performed in September - October 2013 after 10 months of cooling time in TDC2 and TCC2 to allow a thorough 
work and dose planning for the cable exchange activities in those areas in 2013/2014. The results 
of these surveys can be found in~\cite{Survey:2013}.  The SHiP removal and re-installation activities in and 
around the SHiP junction cavern, as well as the ones connected to the M2 splitter, will be planned according to this information in order to prepare 
an optimized work and dose plan. Ending the last year before LS2 by an ion run will allow
additional cool down before the works should start.

%PARAGRAPH ON THE REMOVAL/INSTALLATION OF THE NEW M2 SPLITTER. 
% added to the paragraph above

%PARAGRAPH ON THE CONSTRAINTS ON THE CIVIL ENGINEERING DURING OPERATION\\
Once the SHiP junction cavern and extraction tunnel are built, the civil engineering and installation of the SHiP facility requires shielding during beam operation in the North Area.
About 50~m of concrete or 20~m of iron are necessary to reduce the doses behind the extraction tunnel down to 0.5~$\mu$Sv/h allowing for a Non-Designated Area. A total shielding volume of 800~m$^{3}$ of concrete or 320~m$^{3}$ of iron in the extraction tunnel is therefore required. Here the iron shielding actually foreseen for the SHiP target facility could be utilised, which adds up to approximately 390~m$^{3}$ of shielding usable for this purpose. To avoid activation of the iron shielding and facilitate its installation later on in the target facility about 1~m of concrete should be placed upstream of it. A total of 16~m$^{3}$ of concrete is thus additionally required, which could as well be reused as shielding in the SHiP facility.

%% file: shield/muon_shield.tex
\section{Muon shield}
\label{sec:muon_shield}

\subsection{Introduction}

The protons that impinge on the target produce some $\sim$5$\cdot $10$^{9}$ muons/ spill. The 
momentum spectrum expected for these muons is shown in Figure~\ref{fig:req_muon_flux}. The 
simulation used to generate this spectrum is discussed in Section~\ref{sec:simulation}. 
The reduction of this flux needs to be achieved in as short a length along the beam line 
as is possible, so that the transverse size of detector can be smaller to cover a  given bite in solid angle. Both active and passive solutions for the muon shield have been investigated and the studies are summarised below. The design of the system depends critically on the muon spectrum predicted by the simulation. Tests to demonstrate the validity of this simulation are detailed in Section~\ref{sec:muon_simulation}. 

% Even neglecting 
% the occupancy resulting from such a flux of muons, the muons can produce neutrinos or $K_L$ 
% particles in the proximity of the detector which can give rise to backgrounds in the fiducial 
% volume (see section~\ref{sec:vessel}).
% It is therefore desirable to reduce the muon flux entering the decay volume to a level where the background from residual muons is comparable to that which will in any case be induced by neutrinos. This requires removing muons with momenta up to $\sim$350\,GeV and leaving a residual rate of $\sim10^5$  muons per SPS spill. 

% The majority of the muons must also be directed to avoid the emulsion of the tau neutrino detector which requires  $<$Xk muons/spill (=$10^4$ particles/mm$^2$)  (see section~\ref{}).

\subsection{Active shield}
\subsubsection{Conceptual design}

\begin{figure}[htb]
\begin{center}
\includegraphics[width=0.55\linewidth]{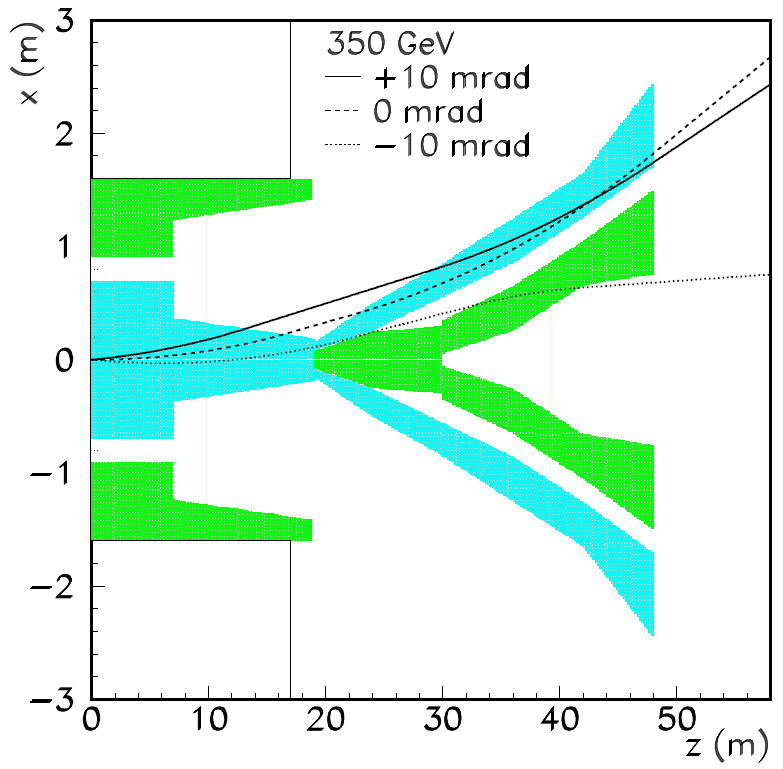}
\caption{The $x,z$ configuration of a possible active muon shield showing the trajectory of three 350\,GeV muons with a range of initial angles. The blue and dark green show the regions of field and return field respectively. While the muons with initial directions 0 and $10\,$mrad are swept out, the muon with -10\,mrad will traverse the detector. The field (return-field) is shown in light blue (green).}
\label{fig:Mitesh-350-1page}
\end{center}
\end{figure}

A total field of $B_y=40$\,Tm is required to bend out 350\,GeV muons beyond the 5\,m aperture of the vacuum vessel.
CERN beam lines already include 1.8\,T conventional magnets of 5.66\,m length and 23\,m of such magnets would be required to generate the required field.
However, the return fields of such a long sequence of magnets present a problem.
These return field tends to bend back towards the detector muons which 
have been bent out by the first part of the shield. 
The method proposed to prevent this is shown in Figure~\ref{fig:Mitesh-350-1page}.
A first section of field $0<z<19\,$m long is used to separate $\mu^+$ to one side and $\mu^-$ to the other side (regardless of their initial direction). The field orientations are shown by the two colours and the return field is placed at $x\sim 1.5\,$m (shown in green). This separation creates enough room to place the return field for subsequent magnets $19<z<48\,$m in the central region, where there are then no charged particles. The field orientation is therefore reversed in the second region, $19<z<48\,$m, with the return field (green) closer to the $z$ axis and the regular field (blue) at larger $x$.
%%%%%%%%%%%%%%%%%%%%%

\begin{figure}[htb]
\begin{center}
\includegraphics[width=0.55\linewidth]{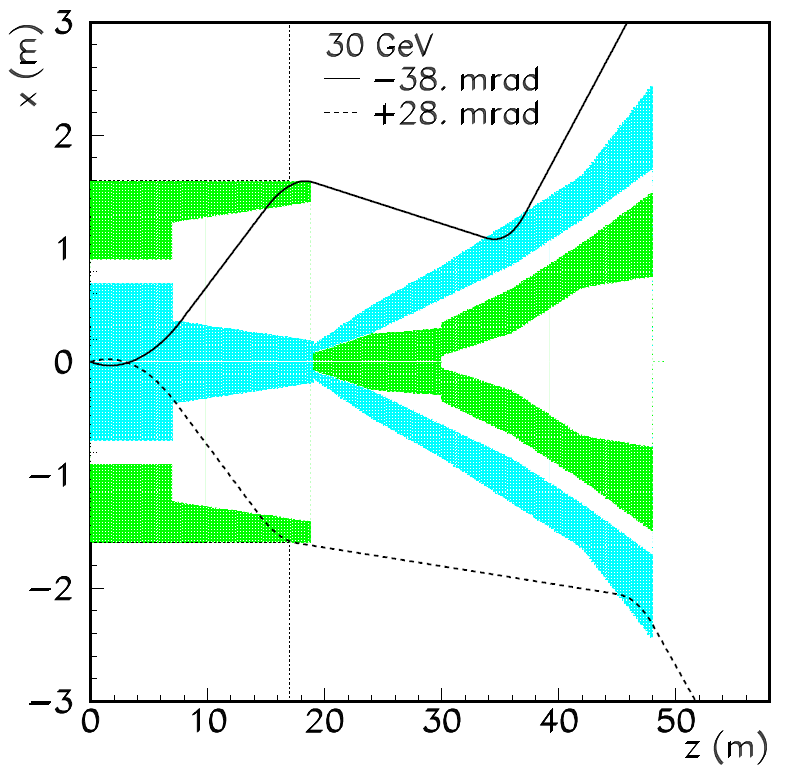}
\caption{The $x,z$ configuration of a possible active muon shield showing the trajectory of two 30\,GeV muons with a range of initial angles. The blue and dark green show the regions of field and return field respectively. The muons are bent out sufficiently by the first part of the shield such that they encounter the return field. This return field directs the muons back towards the detector. The field (return-field) is shown in light blue (green).}
\label{fig:Mitesh-30-1page}
\end{center}
\end{figure}

%%%%%%%%%%%%%%%%%%%%%

\begin{figure}[htb]
\begin{center}
\includegraphics[width=0.95\linewidth]{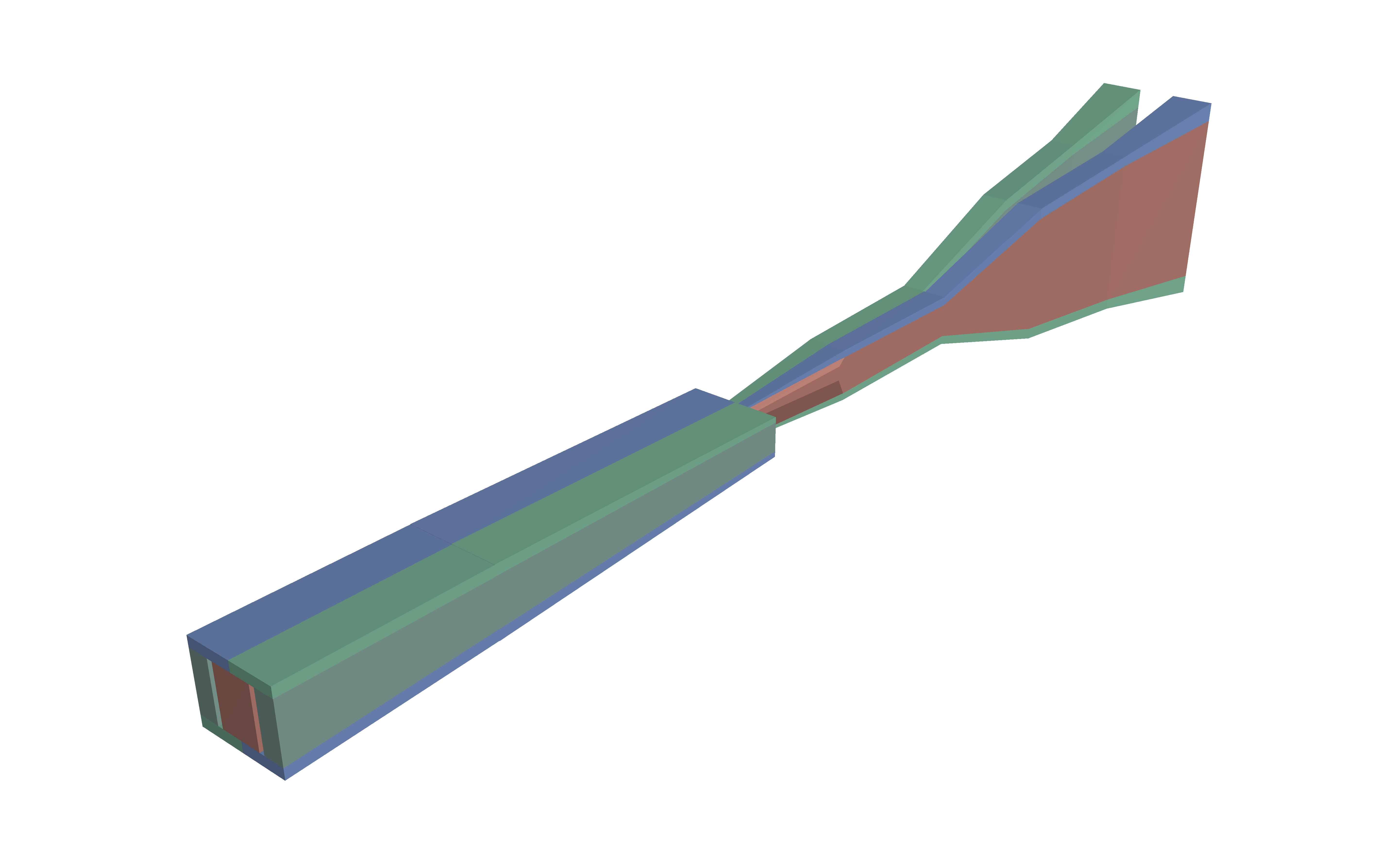}
\caption{Three dimensional view of the active shield. The different colours show the field orientation in the iron, for example, considering the horizontal plane, the red field can be seen around the $z$ axis for  $0<z<19\,$m and at large $x$ for $19<z<48\,$m.} 
\label{fig:humbug}
\end{center}
\end{figure}

Muons bent out by the first part of the shield are then bent 
further outward, rather than back towards detector, solving the problem of a single long sequence of magnets discussed above. 
The large winged shape section with $19<z<48\,$m is needed to sweep out particles that still traverse the return field of the first section of the shield (see Figure~\ref{fig:Mitesh-30-1page}). Multiple scattering effects muons traversing the shield and a full three-dimensional GEANT simulation is used to assess the performance. The simulation indicates that with the proposed configuration it is possible to reduce the number of muons to the required level. 
A three-dimensional view of the shield as simulated is shown in Figure~\ref{fig:humbug}. The different colours again indicate the different field directions in the iron. 

The shield is divided into a number of different magnets, each of which is approximately 6\,m in length. 
The implementation of two of these magnets, magnet~1, which is located at the start of the shield $0<z<6$\,m and magnet~4, which is located in the region $19<z<25$\,m, is discussed in the next section. 
After the first 19\,m in $z$, it is necessary to move walls of the experimental hall out to $x=\pm10\,$m in order to stop  muons from being scattered back towards the detectors.
% The residual combinatorial background is considered in Section~\ref{}.
% While the detector will still experience a relatively high occupancy from low momentum muons and delta rays, these do not produce long-lived $K_L$ particles that might generate significant backgrounds. 

% In order to check the robustness of the system, the flux of muons that would 
% enter the detector has also been evaluated for reduced field strengths. 
% At fields of X.X\,T the system is able to deliver a larger but acceptable flux of muons of $XX /times 10^5$ / spill.
% [point missing somewhere: impact on vacuum vessel design, first 5\,m have to 
% be 40\,cm smaller radius - ask hans's for the detail?? Also, total amount of iron required should be added].

\subsubsection{Engineering design}
In order to demonstrate the feasibility of constructing an active muon shield 
system as proposed above, finite element simulations of two of the key magnets, 1 and 4, have been made in the software OPERA-3d (http://operafea.com/).
These simulations demonstrate that such magnets could be fabricated 
and that they can give the required performance.

\begin{figure}[htb]
\begin{center}
\includegraphics[width=0.6\linewidth]{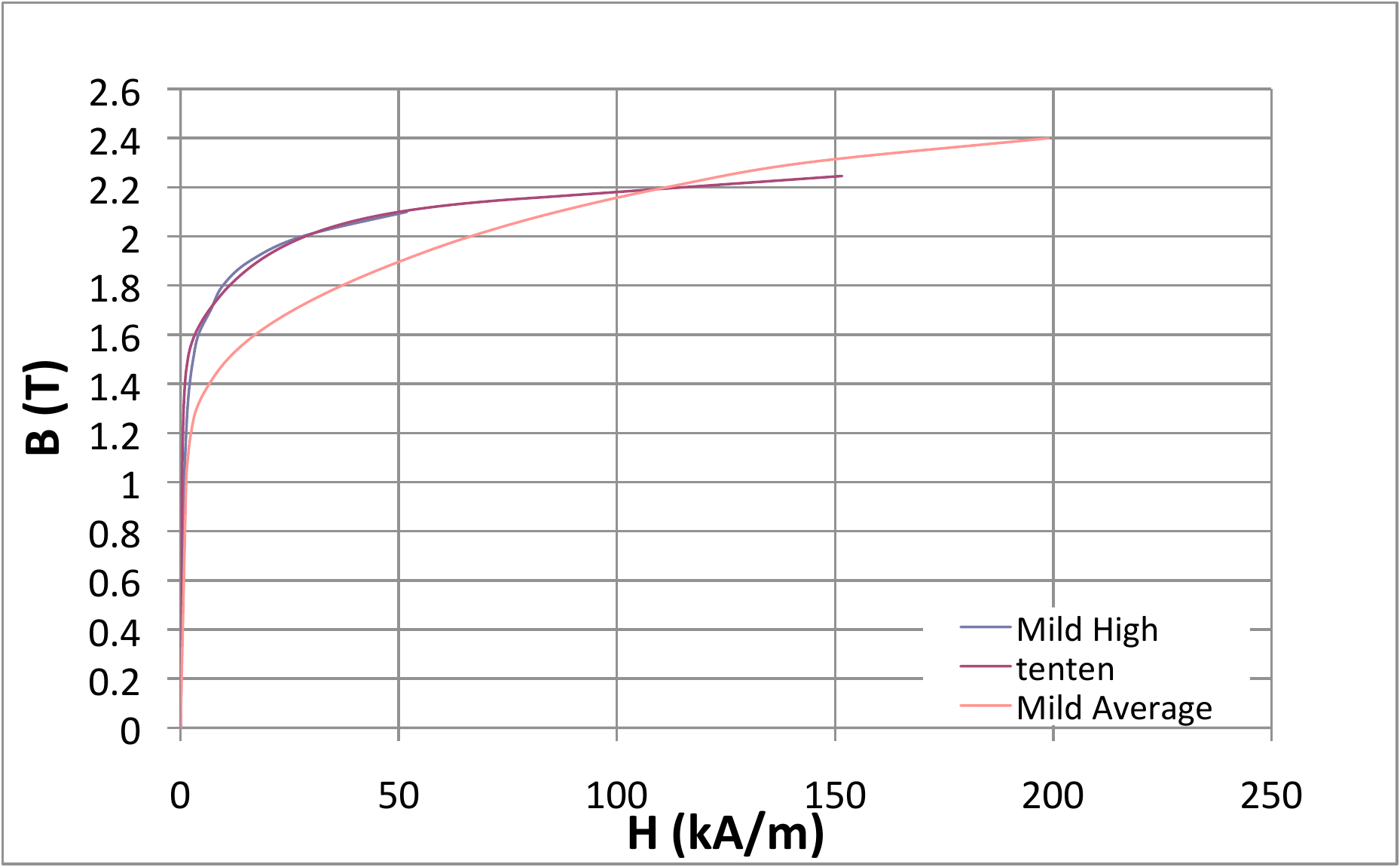}
\caption{B-H curve for various kinds of iron. The material shown as ``ten-ten'' is used for the present simulation.}
\label{fig:BH1}
\end{center}
\end{figure}

The B-H curve shown in Figure~\ref{fig:BH1} is for AISI 1010 iron which is readily available in  the quantities required and has suitably high permeability. This data was input into OPERA-3d. 
The field saturates at 2.2\,T but starts to require much larger currents above 1.8\,Tm. The cost of the iron will need to be optimised against the 
field that can be achieved for the eventual choice of material.
Magnetic modelling has thus far used current densities in the range $1<J($A/mm$^2)<10$, depending on the coil geometry; the cooling requirements are therefore relatively low and air cooling could be considered in the lower part of this range of current densities. In addition, the forces between the coils are manageable. For example, for the second of the magnets detailed below the force between the top and bottom coils is 0.7 tonnes over the 6\,m length. 

%%%%%%%%%%%%%%%%%%%%%

%%%%%%%%%%%%%%%%%%%%%

\paragraph{Magnet 1}

\begin{figure}[htb]
\begin{center}
\includegraphics[width=0.6\linewidth]{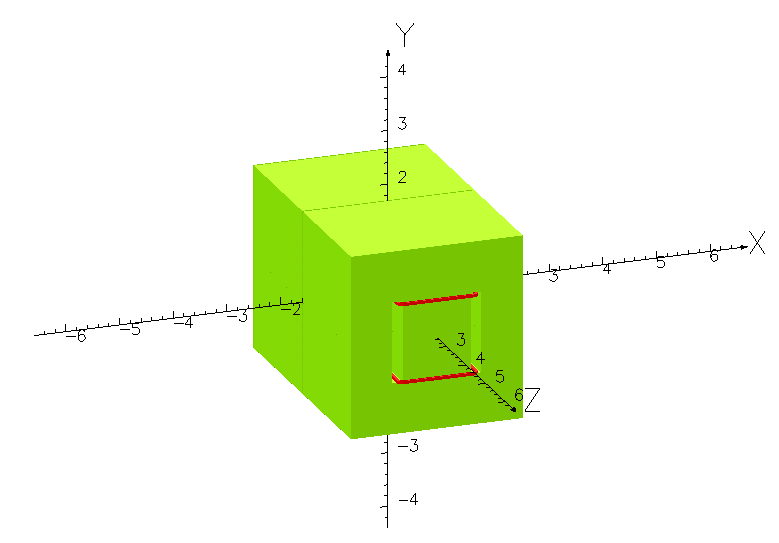}
\caption{3D view of the Opera model for magnet 1. The coils are shown in red and an air gap can be seen on the front face in light green.}
\label{fig:Magnet1_5}
\end{center}
\end{figure}

\begin{figure}[htb]
\begin{center}
%\begin{floatrow}
\begin{minipage}[c]{.27\linewidth}
\includegraphics[width=\linewidth]{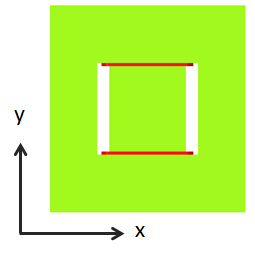}
\end{minipage}
\begin{minipage}[c]{.45\linewidth}
\includegraphics[width=\linewidth]{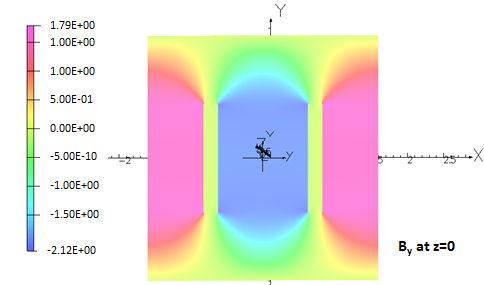}
\vspace{2em}
\end{minipage}
\caption{Schematic of magnet 1 in the $x,y$-plane (left) and map of the field strength in the y-direction in the same plane and at the $z$ coordinate corresponding to the centre of the magnet (right).}
\label{fig:mag1_xy}
% \end{floatrow}
\end{center}
\end{figure}

\begin{figure}[htb]
\begin{center}
\includegraphics[width=0.42\linewidth]{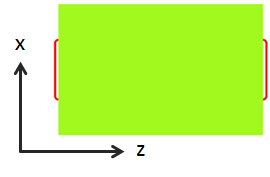}
\includegraphics[width=0.42\linewidth]{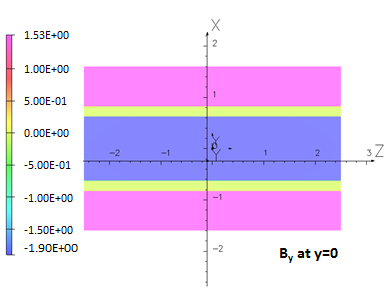}
\caption{Schematic of magnet 1 in the $x,z$-plane (left) and map of the field strength in the y-direction at $y=0$ (right).}
\label{fig:mag1_xz}
\end{center}
\end{figure}

\begin{figure}[htb]
\begin{center}
\includegraphics[width=0.6\linewidth]{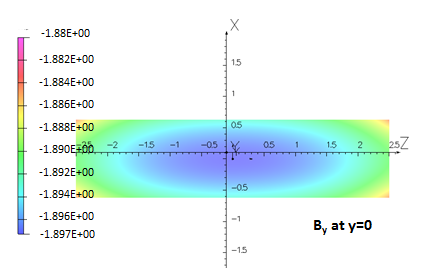}
\caption{Map of the field strength in the y-direction at $y=0$ with a modified scales to show the uniformity. The selection of scale means the return field is not visible.}
\label{fig:mag1_xz_zoom}
\end{center}
\end{figure}

Magnet 1 is a dipole magnet with a 1.8\,T field. The configuration of the yoke and coils is shown in Figure~\ref{fig:Magnet1_5}. The coils shown at the top and bottom are 1.4\,m apart and each have $\sim$20~kA-turns.  The field strength given by the finite element simulation is shown in Figures~\ref{fig:mag1_xy},~\ref{fig:mag1_xz}~and~\ref{fig:mag1_xz_zoom}. The air gap used to separate field and return-field can be seen in  Figures~\ref{fig:Magnet1_5}~and~\ref{fig:mag1_xy}.

\paragraph{Magnet 4}

\begin{figure}[htb]
\begin{center}
\includegraphics[width=0.7\linewidth]{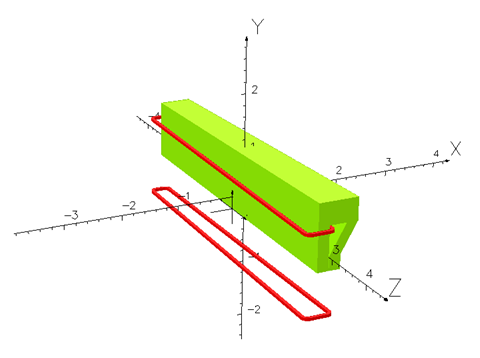}
\caption{3D view of the Opera model for magnet 4. The coils are shown in red and an air gap can be seen on the front face in light green.}
\label{fig:Magnet4_1}
\end{center}
\end{figure}

\begin{figure}[htb]
\begin{center}
\includegraphics[width=0.49\linewidth]{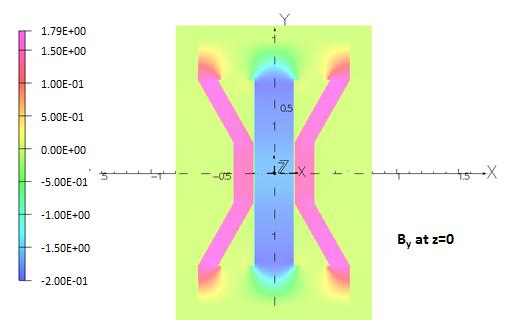}
\includegraphics[width=0.49\linewidth]{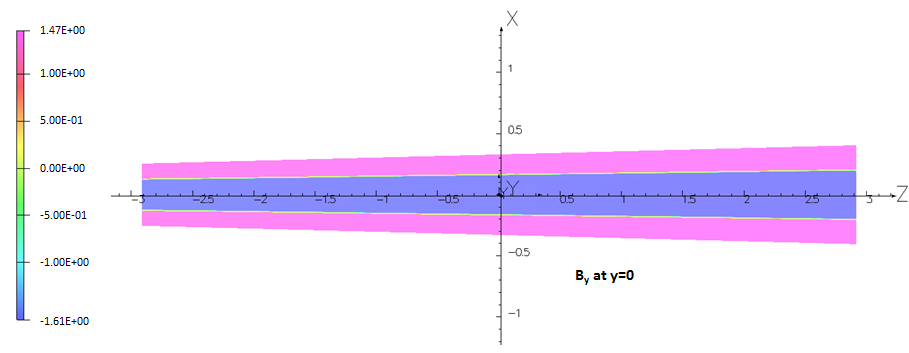}
\caption{Map of the field strength in the y-direction in the $x,y$  plane and at the $z$ coordinate corresponding to the centre of the magnet and field strength in the ($x,z$) plane at (y=0). }
\label{fig:Magnet4_field}
\end{center}
\end{figure}

Magnet 4 has a similar coil to magnet 1 but a more complex iron yoke, which is designed to maximise the flux linking the coils while bringing the inner and return fields as close as possible in the centre region. The magnet has an aperture of 1.4\,m height and the configuration of the yoke and coils is shown in Figure~\ref{fig:Magnet4_1}. As for magnet 1, each of the coils shown at the top and bottom has $\sim$20~kA-turns. The field given by the finite element simulation is shown in Figure~\ref{fig:Magnet4_field}. 

\subsection{Future R\&D}

Several issues will require further investigation for a future TDR:
\begin{itemize} 
\item{The conceptual design described above will be further refined. The concept presented has been optimised mainly in a 2D fast-simulation, with  the muon shielding performance evaluated in the full 3D simulation. A full 3D optimisation will be made to see if the required amount of iron can be reduced;}
\item{The magnets described above have only been simulated in isolation. A full magnet model encompassing all magnetic elements that are in close proximity will be required to consider the interactions between the different magnets and between other elements such as the iron hadron absorber;}
\item{The system will require a full engineering design which includes mechanical and thermal stress/strain calculations and which considers alignment issues and robustness/reduncancy in the operating fields with respect to the muon shielding performance;}
\item{The optimal choice of pole piece material will be considered further;}
\item{Recent tests of LHC dipole magnets have failed to quench the magnets despite impinging some $10^{10}$ particles. It may therefore be possible to consider a superconducting magnet to achieve higher field strengths than those proposed above.}
\end{itemize}

\subsection{Passive shield}

\begin{figure}[htb]
\begin{center}
\includegraphics[width=0.6\linewidth]{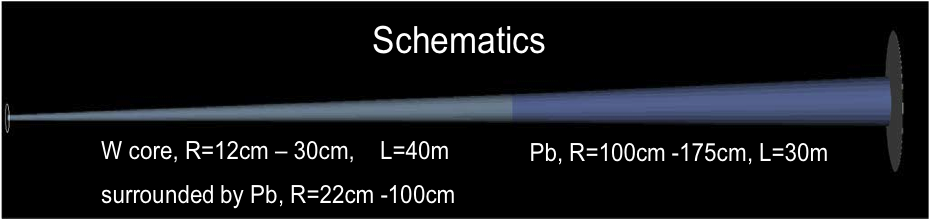}
\caption{Schematic of a possible material arrangement for a passive shield.}
\label{fig:passive}
\end{center}
\end{figure}

\begin{figure}[htb]
\begin{center}
\includegraphics[width=0.42\linewidth]{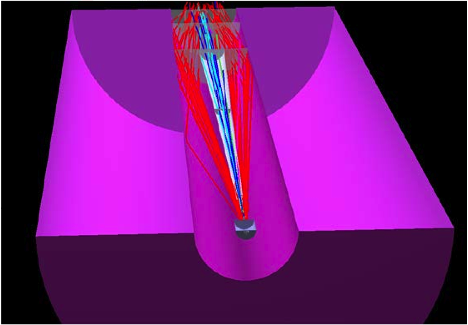}
\caption{Simulation showing the trajectory of muons leaving the passive shield. The shield itself is invisible under the trajectories of the large number of muons simulated.
The trajectory of high momentum muons are shown in red. A large number of these muons can be seen leaving the passive shield volume, backscattering from the concrete of the experimental hall (shown as the solid purple) and returning in the direction of spectrometer which is towards the top of the figure.}
\label{fig:backscatter}
\end{center}
\end{figure}

\begin{figure}[htb]
\begin{center}
\includegraphics[width=0.42\linewidth]{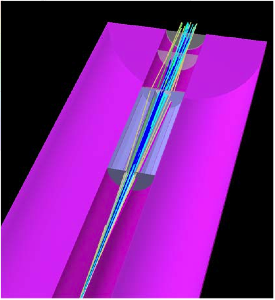}
\caption{Simulation showing the trajectory of muons leaving the passive shield. The end of the shield region is filled with a 17000 tonne iron plug which is shown in grey. Even with this plug in place, muons continue to be scattered towards the spectrometer, the first stations of which are visible as grey disks.}
\label{fig:plug}
\end{center}
\end{figure}

The use of a volume of dense material to slow the muons down has also been investigated. Tungsten is the most readily available of the materials which give high energy loss. %  $\sim$X\,GeV/m. 
A passive muon filter might then consist of $\sim$40\,m of tungsten arranged in a conical shape, as shown in Figure~\ref{fig:passive}. Such a volume contains 110\,tonnes of tungsten
%  and is therefore already prohibitively expensive ($\sim$10~MCHF). It would also be
but this is insufficient to stop 350\,GeV muons. A volume of 2500~tonnes of lead is therefore added both after and around the tungsten volume, giving a total length of 70\,m for the passive shield.
Simulation studies show that  a large number of muons leave the tungsten volume before they are completely stopped and can be backscattered from walls into the spectrometer (see Figure~\ref{fig:backscatter}). As in the case of the active shield, the walls of the experimental hall can be moved out to 10\,m radius to reduce the effect but it is still necessary to shield the entire detector face to stop a very large flux of muons entering the decay volume. To combat this, a further large iron plug is required between the shield and the walls of the hall just before the detector (see Figure~\ref{fig:plug}). The plug requires $\sim$17,000 tonnes of iron. Simulation studies indicate that even in this case, there is still an unacceptably large flux of hard muons 
which leave the filter material, backscatter from the walls of the experimental hall and then enter the detector. 

%% file: detector/Detector.tex
\chapter{Experiment Conceptual Design}
\label{sec:detector}

\input{detector/General_layout.tex}

\input{nutaudet/tp_neutrino.tex}

\input{vessel/Vessel.tex}

\input{taggers/Taggers.tex}

\input{taggers/timing/Upstream_VETOtiming.tex}

\input{tracking/Tracking.tex}

\input{magnet/SpectrometerMagnet.tex}

\input{taggers/timing/SHiP_TP_TimingDetectors.tex}

\input{calorimeter/Calorimeters.tex}

\input{muon/Muon.tex}

\input{online/Online.tex}

\input{computing/Computing.tex}

%% file: detector/General_layout.tex
\section{General experimental layout}
\label{sec:GeneralLayout}

\begin{figure}[htb]
\begin{center}
\includegraphics[width=0.95\linewidth]{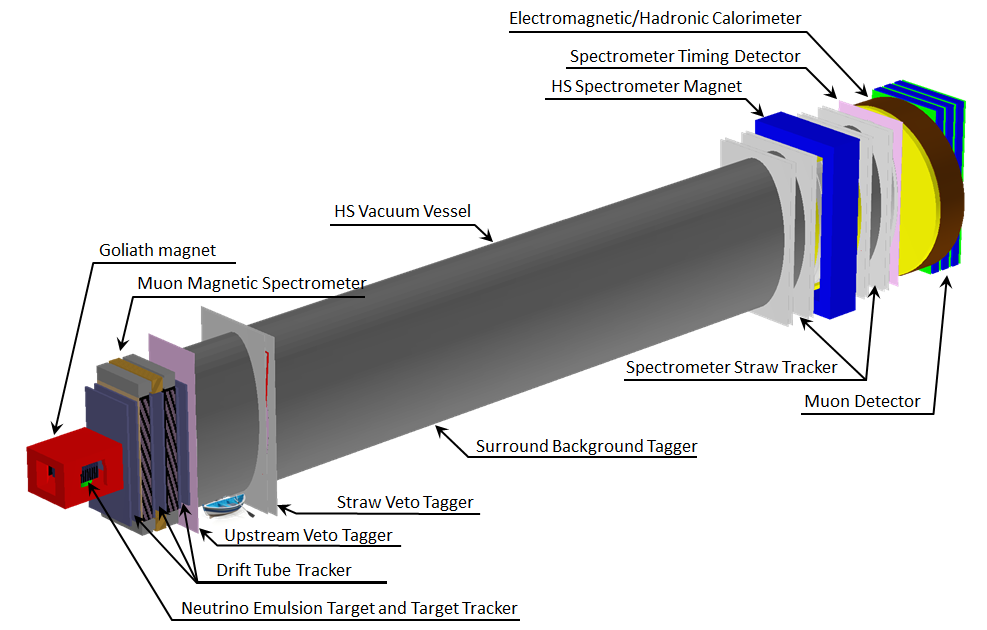}
\caption{The SHiP detector layout.}
\label{fig:detector_layout}
\end{center}
\end{figure}

Following the experimental objective and requirements outlined in Chapter~\ref{sec:requirements}
this chapter describes the conceptual design of the experiment and the details of each sub-system
in the order that they are arranged. The overall SHiP detector layout is shown in Figure~\ref{fig:detector_layout}.

\begin{figure}[h]
\centering
\includegraphics[width = 0.8\textwidth]{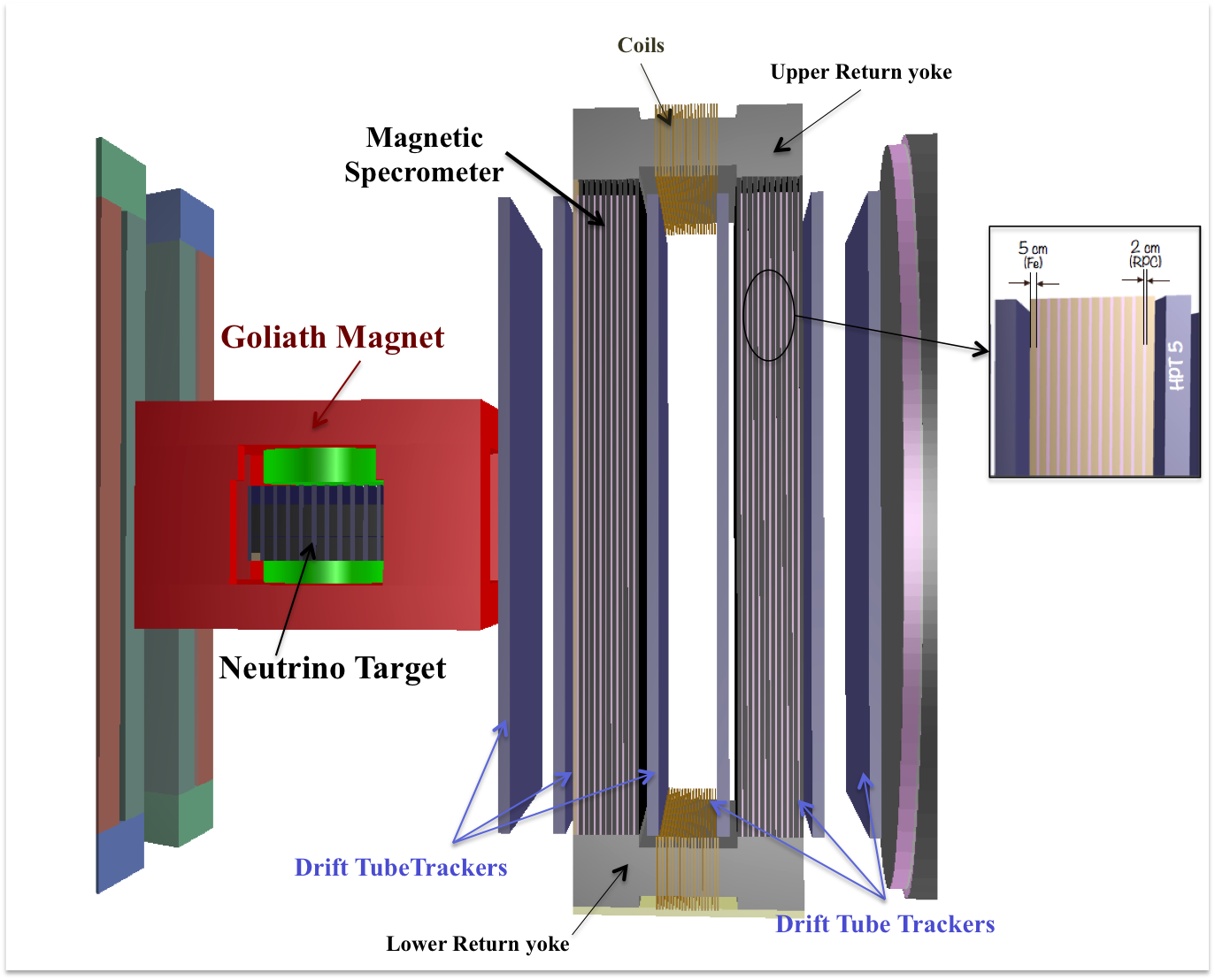}
\caption{Zoomed view of the tau neutrino detector.}
\label{fig:nutaudet}
\end{figure}

The tau neutrino detector shown in Figure~\ref{fig:nutaudet} is located immediately downstream 
of the active muon shield. It consists of the Neutrino Emulsion Target (NET) in a magnetic 
field followed by a Muon Magnetic Spectrometer (MMS). The Neutrino Emulsion Target is made of Opera-type modules 
which employ the Emulsion Cloud Chamber (ECC) technology. The charges of the hadrons produced in the 
neutrino interactions are identified using the Target Tracker (TT) consisting of active planes which 
interleave the ECC modules. The magnetic field is provided by the Goliath magnet which is expected 
to be free after 2018.

The Muon Magnetic Spectrometer of the tau neutrino detector consists of a warm iron dipole magnet 
instrumented with active layers based on the Opera Resisitive Plate Chambers (RPC) and the 
Drift Tube Tracker (DTT)~\cite{bopera}. 

The Hidden Sector (HS) detector searches for decay vertices of hidden particles in a single large 
decay volume under vacuum. The HS Vacuum Vessel has an elliptical cross-section of 5\,m width 
by 10\,m height, and a total length of $\sim$60~m. 

The decay volume is covered by the Surrounding Background Tagger (SBT) which is based on liquid 
scintillator cells read out by PMTs using the wave-length-shifting (WLS) technique.

To provide redundancy against background initiated by interactions of neutrinos and residual muons 
in the tau neutrino detector, a dedicated Upstream Veto Tagger (UVT) is located in 
front of the vacuum vessel. Background induced by interactions in the entrance window of the 
vacuum vessel is further tagged by the Straw Veto Tagger (SVT) located in vacuum at $\sim$5~m downstream 
of the entrance window. This second tagger has very low material budget, at the expense of a slightly 
lower tagging efficiency.

% The design of the HS detector tracking system is driven primarily by the requirement for robust 
% identification of the hidden particle decay vertex and good invariant mass resolution in the 
% regime dominated by multiple scattering in low material environment.% Good momentum resolution 
% is also essential for pointing which helps to discriminate the hidden particle signal against 
% backgrounds produced far away from the beam target. 

The HS spectrometer for the reconstruction of the hidden particle decays 
consists of a large aperture dipole magnet with a moderate magnetic field and the Spectrometer Straw
Tracker (SST) based on two planes of straw tubes in vacuum located upstream and downstream of the magnet. 

% The straw tubes developed for the NA62 experiment are
% ideal tracking detector for this appliaction in the SHiP HS detector. Having very low material 
% budget they are fast and can give very good spatial resolution.

A dedicated Spectrometer Timing Detector (STD) is placed directly downstream the vacuum vessel 
behind the last tracking station to eliminate combinatorial background.

% by measuring the difference 
%in time arrivals of the tracks originating from hypothetical decay vertices in the vacuum.

Finally the particle identification system comprises an electromagnetic (ECAL) and a hadron (HCAL)
calorimeter followed by the muon system (MUON).

% The following sections describe in detail each of the subdetectors and their performance.

%% file: nutaudet/tp_neutrino.tex
\section{Neutrino target}

\label{sec:nutaudet}

\begin{figure}
\centering
\includegraphics[width=0.7\linewidth]{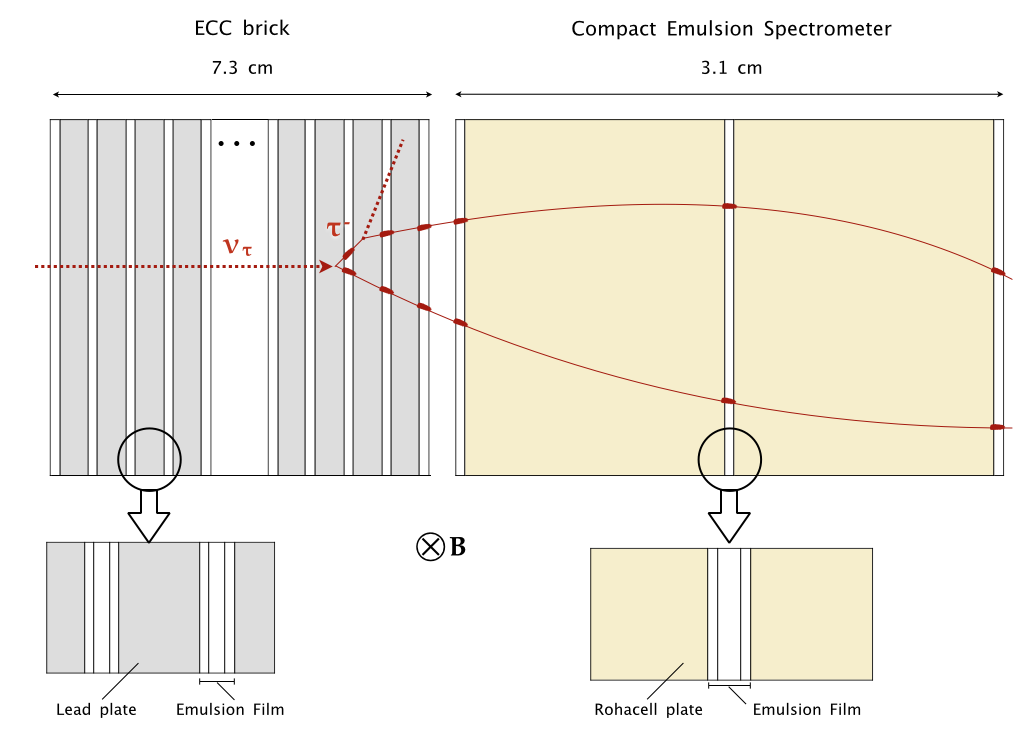}
\caption{Schematic representation of the neutrino detector unitary cell.}
\label{fig:general_schema}
\end{figure}

The neutrino detector is based on the Emulsion Cloud Chamber (ECC) technique, a structure made of a sequence of passive material plates interleaved with emulsion films. The target is modular and its units are made of two parts as shown in Figure~\ref{fig:general_schema}: the \emph{brick} and the \emph{Compact Emulsion Spectrometer} (CES). 
The brick, using lead as passive material, combines the micrometric tracking accuracy of nuclear emulsions and the high lead density as required to maximise the number of neutrino interactions in a compact detector.  
The CES is made of a sandwich of light material plates (e.g.~Rohacell) and emulsion films. It is designed to distinguish $\nu_\tau$ and $\bar{\nu}_{\tau}$ by performing the electric charge measurement of the $\tau$ decay products.

By assembling hundreds of target units, it is possible to achieve the tonne scale for a very high resolution device optimized for the $\nu_\tau$ detection.
The target is complemented by planes of electronic detectors to provide the time stamp of the event and to identify the target unit where the neutrino interaction occurred. 

So far only DONUT~\cite{DONUT_Tau} and OPERA~\cite{bopera} experiments succeeded in detecting a few $\nu_\tau$ interactions. The OPERA experiment observed for the first time the $\nu_\mu \to \nu_\tau$ oscillation in appearance mode, by detecting the decay of the $\tau$ lepton produced in $\nu_\tau$ charged current interactions. The four $\nu_\tau$ candidates~\cite{OPERA_tau1,OPERA_tau2,OPERA_tau3,OPERA_tau4} were identified in ECC detectors analogous to the bricks planned for SHiP.

The ECC structure is well suited for the measurement of charged particles momenta~\cite{MCS} and for electron identification~\cite{operanue}, as largely exploited with the OPERA experiment. The momentum is measured by the multiple coulomb scattering in the lead plates; electrons can be identified by detecting the electromagnetic showers.

The $\nu_\tau$ and anti-$\nu_\tau$ interactions are identified through the detection of the $\tau$ lepton production and decay.  The $\tau$ decay channels investigated are the
electron, muon and hadron channels, as reported in Table~\ref{tab:br} with the relative branching ratios.

\begin{table}[hbtp]
\begin{center}
\caption{\label{tab:br}Branching ratio for different $\tau$ decay channels.}
\vspace{2mm}
\begin{tabular}{lc}
\hline
channel & branching ratio \\
\hline
$\tau^- \rightarrow e^- \nu_\tau \overline{\nu}_e$ & 17.8 \% \\
$\tau^- \rightarrow \mu^- \nu_\tau \overline{\nu}_\mu$ & 17.7 \% \\
$\tau^- \rightarrow h^- \nu_\tau (n\pi^0)$ & 49.5 \% \\
$\tau^- \rightarrow h^- h^- h^- \nu_\tau (n\pi^0)$ & 15.0 \% \\
\hline
\end{tabular}
\end{center}
\end{table}

\begin{figure}
\begin{center}
% \begin{minipage}{14pc}
\includegraphics[width=0.45\linewidth]{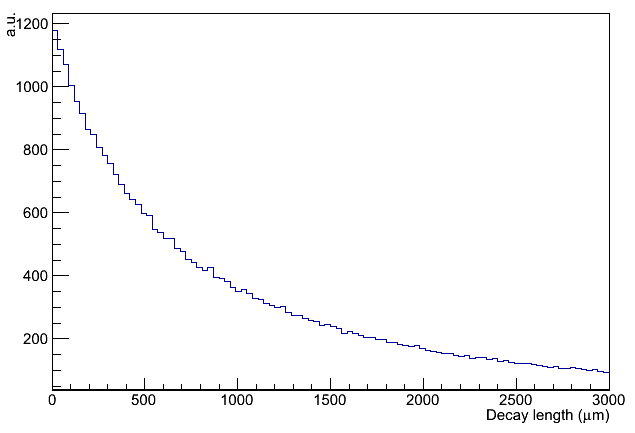}
% \end{minipage}
% \begin{minipage}{14pc}
\includegraphics[width=0.45\linewidth]{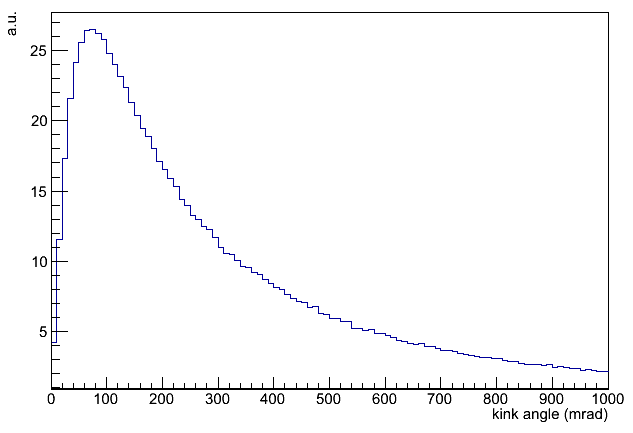}
% \end{minipage}
\end{center}
\caption{(Left) $\tau$ decay length distribution. (Right) $\tau$ kink angle distribution for the $\tau \to h$ decay channel.}
\label{fig:decay}
\end{figure}

At the SHiP energies, the expected $\tau$ decay length distribution is shown in Figure~\ref{fig:decay} (left), where about 50\% of the distribution is below 1~mm. 
$\tau$ decays are classified as \emph{long} and
\emph{short}. Short decays correspond to  $\tau$ lepton decays in the same lead plate where the neutrino interaction occurred. These $\tau$ candidates are selected on the basis of the impact parameter of the $\tau$ daughter track with respect to the interaction vertex. Long $\tau$ decays occur further downstream. $\tau$ candidates are selected in this case on the basis of the measured 
kink angle between the $\tau$ and its daughter. Figure~\ref{fig:decay} (right) shows the kink angle distribution for the hadronic decay channel. 
\\

	\subsection{Emulsion target}

		\subsubsection{Nuclear emulsion films}  

Nuclear emulsion films are the thinnest and lightest three dimensional tracking detectors with sub-micrometric position and milli-radian angular resolution. 
The use of such detector made it possible to observe $\nu_\tau$ charged current interactions in DONUT~\cite{DONUT_Tau} and OPERA~\cite{bopera} experiments.

Nuclear emulsions consist of AgBr crystals scattered in a gelatin binder. The AgBr crystals, with a diameter of $\rm 0.2 \, \mu m$, are sensitive to minimum ionizing particles (MIP). 
The trajectory of a MIP is recorded by a series of sensitised AgBr crystals along its path acting as \emph{latent image}  centers. A chemical process, known as development, enhances latent images inducing the growth of silver clusters (grains) with a diameter of $\rm 0.6 \, \mu$m, visible by an optical microscope. A MIP typically leaves about $\rm 36 \,  grains / 100 \, \mu$m (grain density), defined as the sensitivity of the emulsion.

A nuclear emulsion film has two sensitive layers on both sides of a transparent plastic base.  By connecting the two hits left by a charged particle on both sides of the base, the slope of the track can be measured with milli-radian accuracy. The OPERA experiment, the largest nuclear emulsion experiment  ever built, used the so-called OPERA films~\cite{OPERA_Film}, made of two $\rm 44 \, \mu m$-thick emulsion layers and a plastic base of  $\rm 205 \, \mu m$. The total thickness is $\rm 293 \, \pm \, 5 \, \mu m$. The transverse size is $\rm 124.6 \, \times \, 99.0 \, mm^2$. 
The overall emulsion area is about  $\rm 110,000 \, m^2$ and the films were industrially produced by the Fuji Film company\footnote{Fuji Photo Film Co., Ltd, Minamiashigara, Kanagawa 250-0193, Japan.}. The largest handmade emulsion film production was performed more than 20 years ago for the CHORUS experiment~\cite{Eskut:1997ar,CHORUS_Film}, with a total area of about $\rm 300 \, m^2$.

Nuclear emulsions integrate all tracks including cosmic rays and natural radiations from their production to development. The CHORUS emulsion film production was carried out on-site to minimise the  accumulation of background tracks.  Moreover, the fading effect, namely a partial cancellation of the latent images caused by oxidation, was controlled along the two-year exposure time by keeping the emulsion target at the stable temperature of $\rm 5 \, ^\circ C$.
OPERA was run at the room temperature of the underground Gran Sasso cavern, around $15 \, ^\circ $C: indeed the fading was not an issue since the emulsion analysis was performed soon after the occurrence of the neutrino interaction. In SHiP, the target will be replaced twice a year and therefore the fading effect will be negligible. Nevertheless, the temperature is required to be around $20 \, ^\circ $C or lower, and stable within one degree, to prevent thermal deformations. 

 With the help of experienced engineers of the Fuji Film company, the University of Nagoya has  recently developed a project for large-scale production of films with improved sensitivity. At present, films with a sensitivity of $\rm 41 \, grains /100\, \mu$m have been successfully produced. 
Assuming to replace the target ten times during the data-taking, the total emulsion area required by SHiP is of about $\rm 8700 \, m^2$, corresponding to 1155 bricks replaced ten times. 
The foreseen developments of the emulsion production facility at the Nagoya University would cope with the need for an overall amount of 8700~m$^2$ emulsion films. 

		\subsubsection{Bricks}  

Based on the successful experience of the OPERA experiment, the $\nu_\tau$ target will be segmented into units, called \emph{bricks}, each one made of 57 thin emulsion films interleaved with 56 lead plates of $\rm 1 \, mm$ thickness. This structure, combining sub-micron resolution trackers with layers of high density passive  material, makes use of the so-called \emph{Emulsion Cloud Chamber} (ECC) technique %~\cite{ECC} 
and allows for a compact modular target with each single brick acting as a stand-alone detector. The transverse sizes  of each unit are $\rm 128 \, \times \, 102 \, mm^2$, the mass is  $\rm 8.3 \, kg$ and the longitudinal dimension is $\rm 79 \, mm$, corresponding to about 10 radiation lengths. Given the short lifetime of the $\tau$ lepton ($3 \cdot 10^{-13}$~s), both the neutrino interaction and the $\tau$ lepton decay vertices are in the same brick and a full event reconstruction can be thus achieved exploiting the characteristic features of the brick structure. The excellent spatial resolution of nuclear emulsions makes it possible to obtain a precise tri-dimensional reconstruction of the event. 
The momenta of all charged particles produced in the interaction can be measured through the deviations they undergo due to multiple Coulomb scattering along their trajectories~\cite{MCS,mcstrk}.  Electrons may be separated from pions through the detection of the electromagnetic shower produced  while traversing the brick~\cite{ELE-PI}, given the overall thickness of 10~$X_0$. 

SHiP $\nu_\tau$ target will consist of 11 \emph{walls} of $\rm 15 \, \times \, 7$ bricks, perpendicular to the beam direction,  with a total mass of about $\rm 9.6$ tonnes. In between walls, planes of electronic trackers will be installed with the task of providing the time stamp to the interactions  and associating the muon tracks measured in the bricks with those reconstructed in the magnetic spectrometer. 

Bricks will be hosted on very light stainless-steel frames mounted on rails for brick insertion and extraction. Emulsion films and lead plates will be piled up by means of dedicated assembly machines with an accuracy of $\rm 0.1 \, mm$. Each unit will be packed in special aluminum-laminated paper to ensure light-tightness. Low radioactivity lead (PbCa $\rm 0.04 \, \%$ alloy), with adequate mechanical properties and chemical compatibility with nuclear emulsions~\cite{LEAD}, might become available from the OPERA detector decommissioning. 

		\subsubsection{Compact emulsion spectrometer} 
The capability of measuring the electric charge of the $\tau$ daughters is a key point in the $\nu_\tau$ and anti-$\nu_\tau$ identification. 
$\tau$ hadronic decay modes have the largest branching ratio but, given that a brick corresponds to about one third of interaction length, their charge has to be measured inside the target. 
A compact and light detector capable of measuring the hadron track curvature and a magnetized target are therefore needed.

For this purpose, a Compact Emulsion Spectrometer (CES) is attached immediately downstream of each ECC brick. It is made of three emulsion films interleaved with two, 15-mm thick, Rohacell layers. The passive material, with a density of $57$ mg/cm$^3$, allows the necessary spacing between two consecutive emulsion films minimizing the effect of multiple coulomb scattering on the measurement of the magnetic deflection.

\begin{figure}[htb]
\centering
\includegraphics[scale=0.25]{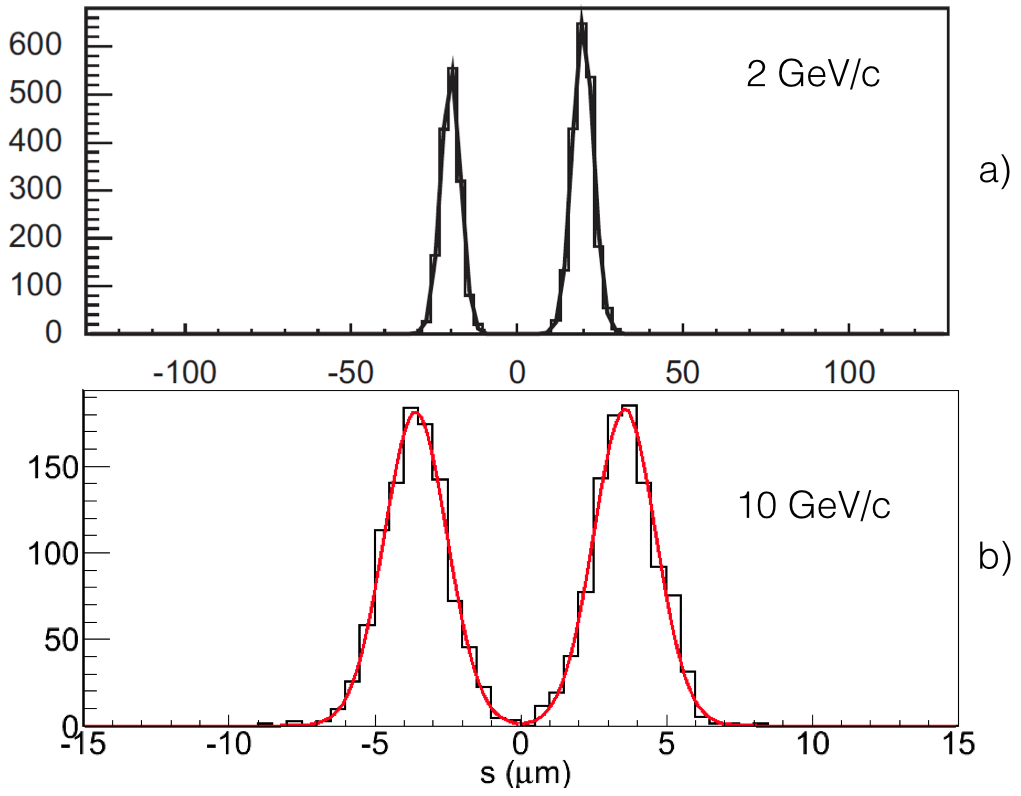}
\caption{Measured sagitta distributions of positive and negative 2.0~GeV/c pions~\cite{ces} (a). Simulations show that the electric charge can be determined with better than three standard deviation level up to 10~GeV/c (b). }
\label{fig:sagitta}
\end{figure}

The layout of the detector is optimized in order to have good performances in the charge measurement for hadrons produced in the $\tau$ decay. Efficiencies as high as $90 \%$ for hadronic $\tau$ daughters reaching the end of ECC brick can be achieved with a magnetic filed of 1~T.
The CES detector can also improve the charge measurement efficiency of about 20\% for muon tracks, when it is not possible with the magnetic spectrometer.\\
The usage of the Emulsion Cloud Chamber technique in a magnetic field was tested in 2005 with a pion test beam performed at the KEK facility in Japan~\cite{ces}. 
A spectrometer prototype made of three emulsion films interleaved with two spacers, each producing an air gap of 15 mm, was located in a dipole magnet with a magnetic field of about 1~T. The momentum was determined for charged pions of 0.5, 1.0 and 2.0 GeV/c with a resolution of 13\%. Results show that in this momentum range, the electric charge can be determined with a significance level better than the three standard deviations (see figure \ref{fig:sagitta}(a)).

A Geant4 based simulation was performed, assuming a 1~$\mu$ position accuracy on the alignment of the emulsion films.  This simulation reproduces well the sagitta measurements, both for its mean  and RMS values. 

%, as shown in figure \ref{fig:sagitta}  for 2.0 GeV pions.

\begin{figure}[htb]
\centering
\includegraphics[scale=0.25]{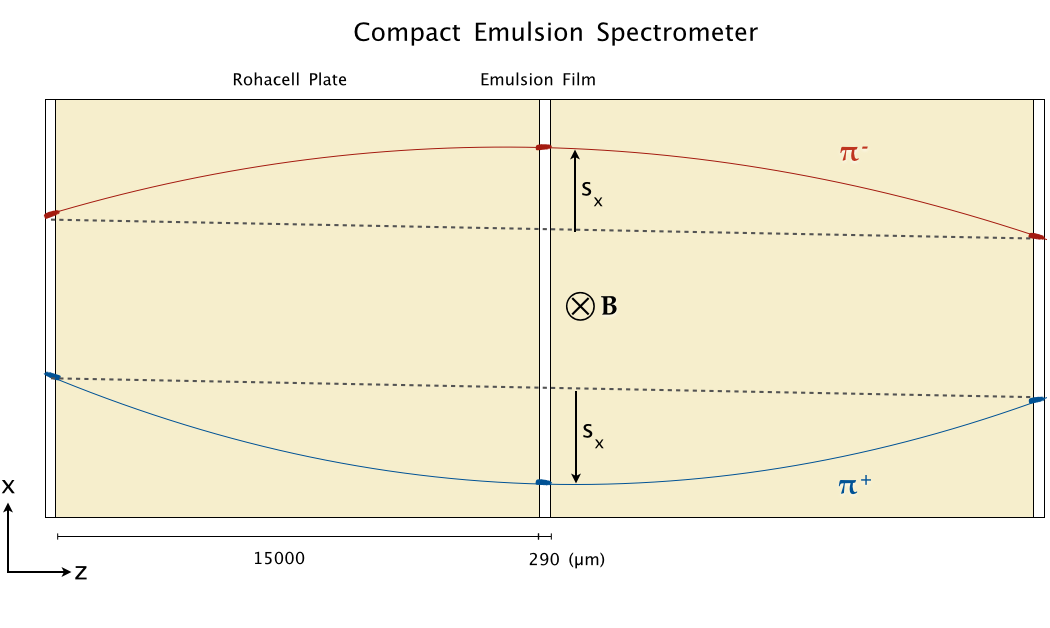}
\caption{Schematic representation of the sagitta measurement using the Compact Emulsion Spectrometer.}
\label{fig:schema}
\end{figure}

An analogous  simulation was performed to evaluated the performances of the CES. $\pi^+$ and $\pi^-$ in the 1-15 GeV/c  momentum range were used.
The particle beam is directed along the $z$-axis, orthogonal to the emulsion film surface.
The simulated CES is immersed in a 1~T magnetic field oriented along the $y$ axis. 

\begin{table}[htb]
\begin{center}
\caption{The mean value of the sagitta and its RMS in the CES for simulated positive pions of different momenta.}
\label{tab:sagitta}
\begin{tabular}{ccc}
\hline 
\multicolumn{3}{c}{$\pi^+$} \\ 
\hline 
P (GeV/c) & $s$ ($\mu$m) & RMS ($\mu$m) \\ 
\hline 
%0.5 & 69.3 & 10.6  \\ 
%1 & 34.7 & 5.4  \\ 
2 & 17.3 & 2.8 \\ 
4 & 8.6 & 1.6  \\ 
%6 & 5.7 & 1.6 \\ 
8 & 4.3 & 1.2  \\ 
10 & 3.4 & 1.0 \\ 
\hline
\end{tabular} 
\end{center}
\end{table}

The sagitta method is used to discriminate between positive and negative charge. The two-dimensional sagitta $s=(s_x,s_y)$ is defined as the distance between the track position in the middle sheet and the intercept in this sheet of the straight line joining the track positions in the two external sheets (see figure \ref{fig:schema}).
The sagitta $s_x$ is very nearly equal to $s$ and reflects the curvature due to magnetic field, while the $s_y$ is centered at zero and its width reflects the deviations due to multiple scattering.  Arithmetic mean
(average) and RMS values are summarized in Table~\ref{tab:sagitta}. %, where RMS is the root mean square deviation of sagitta values from their average. 
The results show that with a magnetic field of 1 T it is possible to measure the electric charge of particles with momenta up to 10 GeV/c, as shown in figure \ref{fig:sagitta}(b).

Using the linearity of the relation between the average sagitta
values in the bending plane and the electric charge, the momentum estimation can also be performed with a 20\% resolution.

\subsubsection{Target magnetisation} 	
\begin{figure}[h]
\centering
\includegraphics[width = 0.45\textwidth]{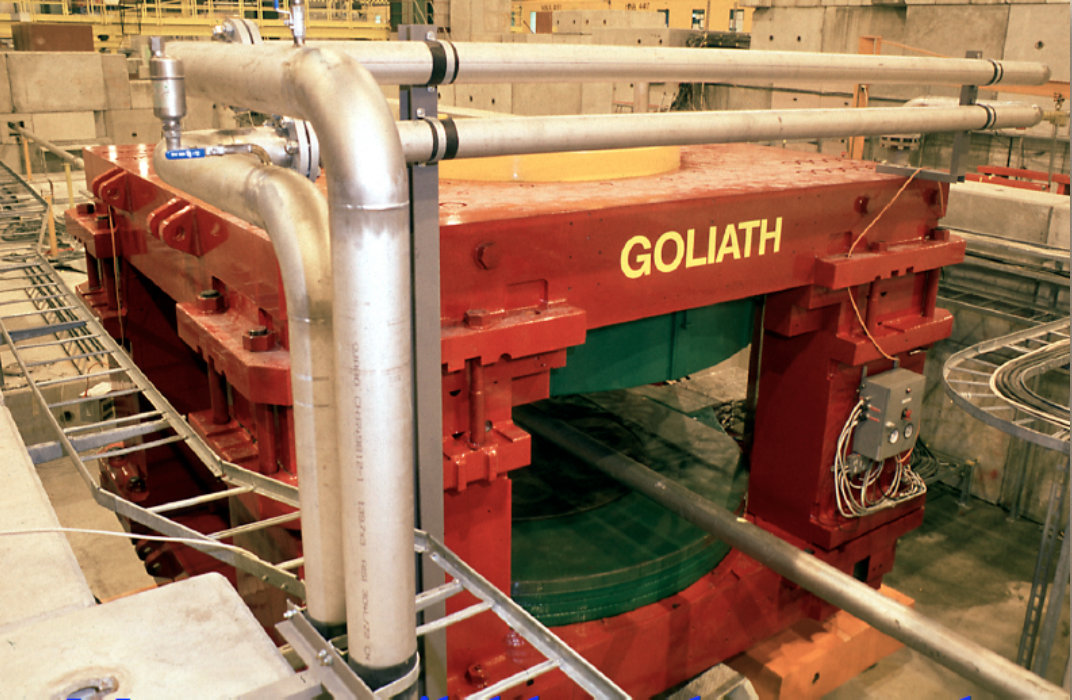}
\caption{Photograph of the Goliath magnet in the H4 experimental area at CERN.}
\label{fig:Goliath}
\end{figure}

In order to perform charge measurements for the hadrons produced in a neutrino interaction, the neutrino target has to be magnetized. Given the target volume of a few m$^{3}$ and the need for a Tesla field along the vertical axis, the Goliath magnet (on the CERN H4 beam line inside PPE134 zone, see Figure~\ref{fig:Goliath}), has been identified.

Produced in 1976, its overall dimensions are 4.5$\times$3.6$\times$2.79~m$^3$. It exhibits an open window of 2.4~m on two of the four sides, thus allowing to replace target bricks during the run. The magnet is composed of two coils (with a diameter of 2~m) mounted in serie. The superior coil is made of copper and it is 45~cm high; its total resistence is $R= 0.043\,\Omega$ when considering also the DC cables. The inferior coil is made of aluminium, it is 30~cm high and its total resistance is of $R= 0.035\,\Omega$. The distance between the two coils is 1.05~m extendable up to 1.5~m.

Starting from a map obtained by the NA57 experiment~\cite{Antinori:2001rt}, we have derived the behaviour of the magnetic field. It shows a cylindrical symmetry, with a rather fast drop of its intensity when approaching the radius of the coils, while it is approximately constant and equal to its maximum value (1.5~T) in the central region.

%\begin{figure}[h]
%\centering
%\includegraphics[width = 0.4\textwidth]{nutaudet/PICTURES/CampoGoliath.png}
%\caption{Map of the magnetic field inside the Goliath magnet by NA57 %Collaboration }
%\label{fig:mapNA57}
%\end{figure}

\begin{figure}[h]
\centering
\includegraphics[width = 0.45\textwidth]{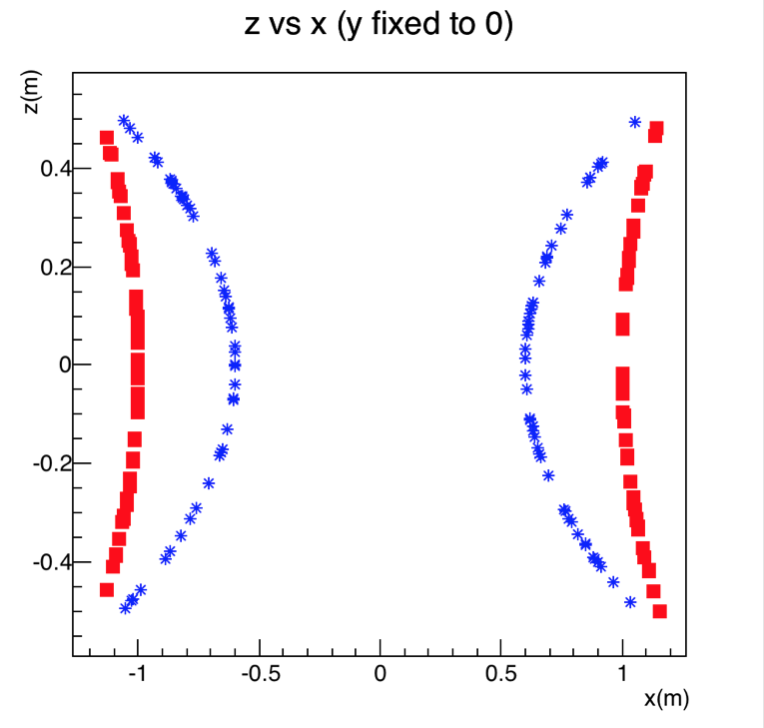}
\caption{Magnetic field behaviour in the target region. Within the blue lines the field is approximately 1.5~T. Within the red lines the field is always above 1~T.}
\label{fig:fregions}
\end{figure}

We have parametrized its behaviour at different heights (z coordinate) between the two coils and we have identified a region where the field is approximately constant. The results of this study are shown in Figure~\ref{fig:fregions}. The plot shows the magnetic field behaviour in the target region. The y coordinate is fixed to 0 so that $r = \vert x \vert$ and two regions are identified in terms of z and x ranges. The region within the  blue lines defines a volume where the field is approximately constant and equal to 1.5~T. The wider region identified by the red lines defines a volume were the magnetic field is above 1~T.

The need for a magnetic field of at least 1~T defines a cylinder of $r = 1$~m and a height of 1~m.

		\subsubsection{Emulsion scanning and analysis}  

The acquisition and processing of the information recorded in the emulsion films will be accomplished by means of automated optical microscopes that have undergone tremendous progress in past decades starting from the first conceptual design in 1974 in Japan~\cite{NIWA}. 

The automatic nuclear emulsion scanning system, Super Ultra Track Selector (S-UTS)~\cite{S-UTS}, developed at the University of Nagoya,  is the latest evolution of the Track Selector~\cite{TS}, the first computer-controlled system able to acquire sequences of tomographic images of the emulsion and apply on-the-fly image processing techniques to recognise in real time the tracks left by charged particles. Differently from the previous versions of the system, the S-UTS grabs the tomograpich images while the microscope stage is in motion. The synchronisation between the image acquisition along the optical axis and the movement of the stage in the plane of the emulsion is obtained by means of piezo actuators that allow limiting the blurring 
of the images to within the sensor pixel size. The architecture of the S-UTS thus overcomes the speed limit inherent in the old \emph{stop-and-go} mechanism that requires a dead time for dumping the mechanical vibrations araising when the microscope stage is stopped before image acquisition. The S-UTS can take data at a speed as high as $\rm 72 \, cm^2/h/layer$, with a tracking efficiency of about 95\% and a positioning precision of about $\rm 0.4 \, \mu m$. The system is currently being used for the location and study of the neutrino interactions in OPERA, together with the ESS~\cite{ESS_hard,ESS_soft,ESS_track,ESS_prec}. 

The European Scanning System (ESS), based on the original project of the SySal system~\cite{SySal}, has been designed with the contribution of several European institutions. It is a software-based system that makes use of commercial hardware components. The track finding efficiency is above 90\% for particles with $\rm tg \theta \, \leq \, 0.7$, $\theta$ being the angle with respect to the perpendicular to the film. The track positioning error, mostly limited by the absolute stage positioning accuracy over large distances, is about $\rm 0.3 \, \mu m$. The tracking is performed by a cluster of computing machines. A high-level modular infrastruture~\cite{ESS-INFR}, running on a local network and including ESS microscopes, Plate Changers~\cite{PlCh}, tracking and data processing servers, a database server and a Process Manager machine, makes the scanning of emulsion films a fully automated operation, as required by nowadays large-scale experiments. 

Both the S-UTS and the ESS are being upgraded to improve the  scanning speed by (at least) one order of magnitude with respect to present systems and prototypes ~\cite{lasso} 
already exist.  

About 300 $\nu$ interactions are expected in each single  brick in a 
six-month exposure. Given the relatively high number of events, the 
fastest and most efficient way of collecting the full statistics is the 
\emph{general scanning} of all films, i.e. the measurement of the whole 
emulsion surface to track all particles traversing the films within a 
given angular range. With the above-mentioned upgrades of the existing microscopes in the Collaboration, the scanning time is such that the analysis of all the films will be performed well within the six months envisaged as the replacing period.

Sub-micron accuracy can be achieved  thanks to 
beam muons that cross the bricks, thus allowing relative misalignments 
between consecutive emulsion films to be precisely estimated. Tracks are 
then associated to form vertices. In order to tag neutrino-induced 
interactions, all tracks originating inside the bricks are matched to 
those reconstructed in the target tracker and in the muon spectrometer 
that provide the event time stamp.
The matching requires a two-step procedure.
General scanning is applied to each of the three emulsion films of the 
CES. High energy beam muons ($E  \geq 15$~GeV), practically 
undeflected by the magnetic field, can be used for brick - CES alignment 
and for fine CES film  intercalibration at $ 1 \, \mu$m level. 
Tracks produced in reconstructed neutrino interactions and reaching the 
most downstream film of the brick are then connected to the CES, i.e. 
they are searched for in the most upstream CES film  within given 
tolerances depending on the achieved alignment accuracy between the 
brick and the CES itself. Each connected track is iteratively searched 
for in the second CES film. Given the orientation of the magnetic field 
and being the sign of the particle charge unknown, two possible 
deflections in the horizontal plane have to be accounted for. The same 
procedure is repeated for the last CES film. Reconstructed tracks can be 
finally matched to the electronic tracker. Only those interactions 
having at least one matched track can be regarded as beam related 
$\nu$-induced interactions and are thus fully analysed to detect the 
sub-sample of $\nu_\tau$-induced charged current interactions with its 
characteristic topology.

	\subsection{Target tracker}

\label{sec:target_tracker}

\begin{figure}
\centering
\includegraphics[scale=0.5]{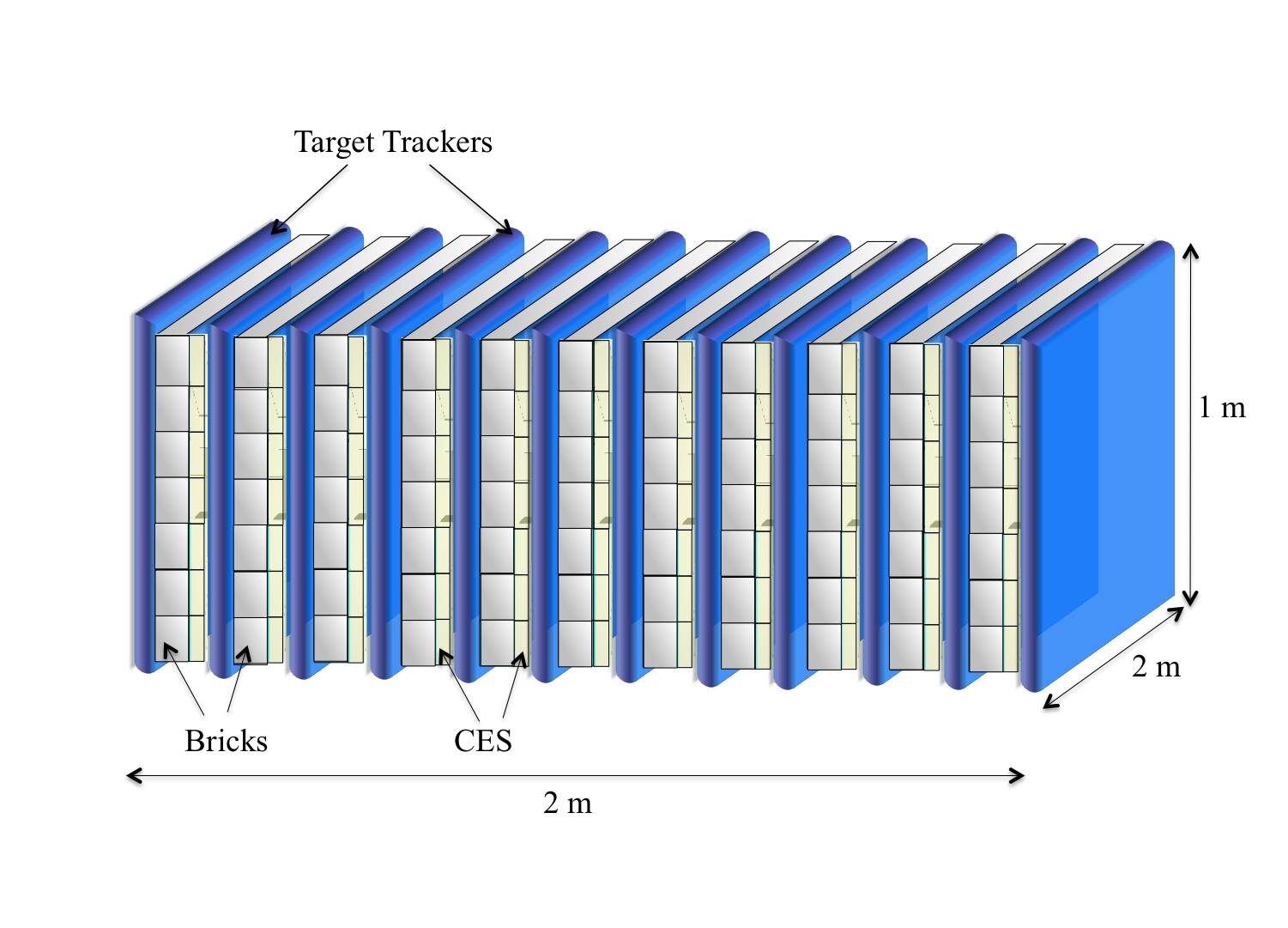}
\caption{Schematic view of the target with the 12 target tracker layers.}
\label{fig:ttneutrino}
\end{figure}

The target tracker system provides the time stamp to the events reconstructed in the emulsion bricks and predicts the target unit where the neutrino interaction occurred. 
The neutrino emulsion target is made of 11 walls, each interleaved with a Target Tracker (TT) plane. Considering also a TT plane upstream acting as a veto, 12 target planes are foreseen. A schematic view of the target tracker planes with the neutrino target is shown in Figure~\ref{fig:ttneutrino}. 

In order to fit the brick wall size, each plane must have a transverse size of about $\mathrm{2\times1~m^2}$, with the longest side being horizontal. 

The distance between the target tracker plane and the emulsion bricks must be a few mm and kept uniform over the whole surface, to assure constant tracking capabilities. Possible dead spaces, at most 1~mm wide, are allowed only if aligned with the dead zones of the emulsion bricks, in such a way to not further decrease the overall efficiency.
The thickness of a TT plane must be contained within about 5~cm.

The physics performances, to be obtained in a magnetic field with a strength  between 1.0 and 1.5~T, are: 100~$\mu$m position resolution on both coordinates; high efficiency ($>$99\%) for angles up to 1~rad. The required accuracy comes from several arguments. The first one is the need to disentangle tracks from the same interaction that would possibly come from two separate vertices: this is the case for instance for a $D^0$ produced in neutrino interactions. In addition, the high level of integrated muon flux makes the tracking more challenging in the compact emulsion spectrometer and it would benefit from a high accuracy in the target tracker to veto penetrating muons. 
The measurement of the track slope in each plane should be also considered as a possible benefit.

We consider in the following two basic options for the target tracker: scintillating fibres and micro-pattern gas detectors. 

		\subsubsection{Scintillating fibre tracker} 

A highest spatial resolution option, proposed for the tau neutrino detector tracker, 
    is the sub-detector based on scintillating fibres.
    This technology is chosen as the baseline option for upgrade of the large downstream
    tracker in LHCb experiment. A lot of R$\&$D studies have been performed for the LHCb
    SciFi project at CERN, TU of Dortmund, RWTH of Aachen and NRC Kurchatov Institute.
    The SHiP experiment could benefit from these studies to a very large extent, in particular,
    using many elements of the LHCb SciFi tracker design and already developed production
    infrastructure. A detailed description of the LHCb SciFi tracker design is given in \cite{nutaudet:SciFiTrackerTDR}. 
    
		%\vfill
    \begin{figure}[tp]
    \centering
    \includegraphics[height=10cm]{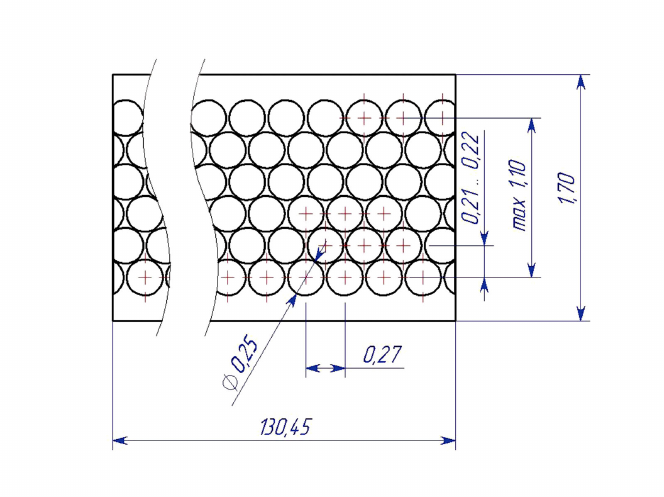}
    \caption{Scintillating fibre ribbon cross section. Units are mm.}\label{nutaudet:FibreMatXsection}
    \end{figure}%
    %
   
    %\vfill
    \begin{figure}[tp]
    \centering
    \includegraphics[height=8cm]{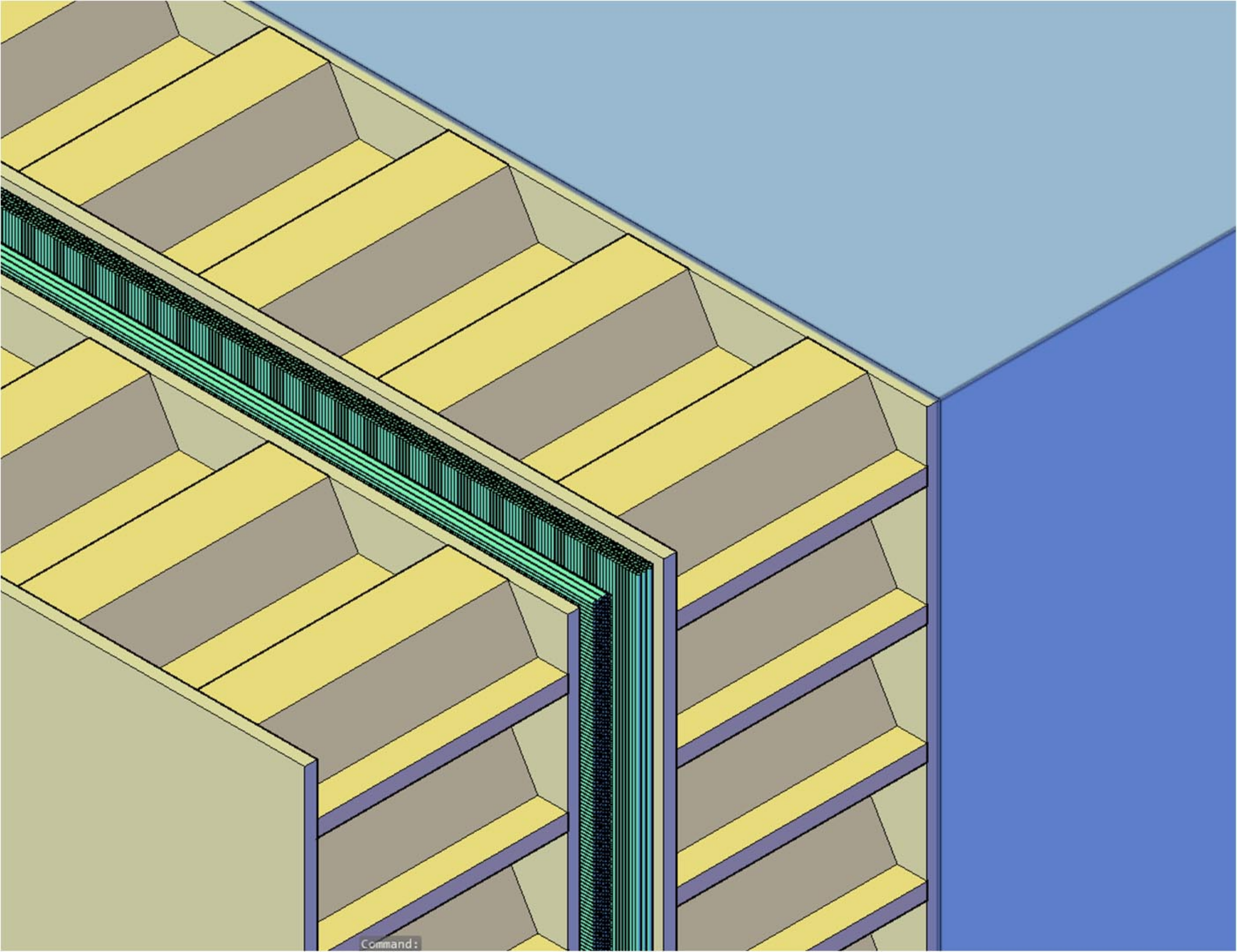}
    \caption{SciFi module structure: 3D cut view.}\label{nutaudet:SciFiT3DCut}
    \end{figure}%

    The main active element of the tracker is a fibre ribbon. Its cross section is shown in Figure~\ref{nutaudet:FibreMatXsection}.
    The ribbon consists of  250 $\mu$m scintillating fibres arranged in 5-6 layers and bound together
    by the epoxy bonding, impregnated with TiO$_2$, in a 132 mm wide and 1980 mm long fibre mat (512 fibres-wide)
    for the horizontal plane, and 924 mm long for the vertical plane. The module consists of 7 ribbons in the horizontal direction,
    and 15 ribbons in the vertical one, placed orthogonally to achieve the best track spatial resolution in both $X$ and $Y$ planes.
     The resolution is estimated \cite{nutaudet:SciFiTracker} to be better than 70 $\mu$m 
    for ribbons with 6 layers, worsening for a smaller number of layers. 
    The perpendicular layers of ribbons are glued to a composite carrier (holder) panels, made of a closed-cell
    foam (i.e. Rohacell) or a honeycomb (i.e. Nomex), and thin carbon fiber sheets (Figure~\ref{nutaudet:SciFiT3DCut})
    forming a single SciFi tracker module. Each module, therefore, has a continuous active area of 1980 x 924 mm$^2$ without
    any dead region or any other structural non-uniformity. 
  
    %\vfill
    \begin{figure}[tp]
    \centering
    \includegraphics[height=8cm]{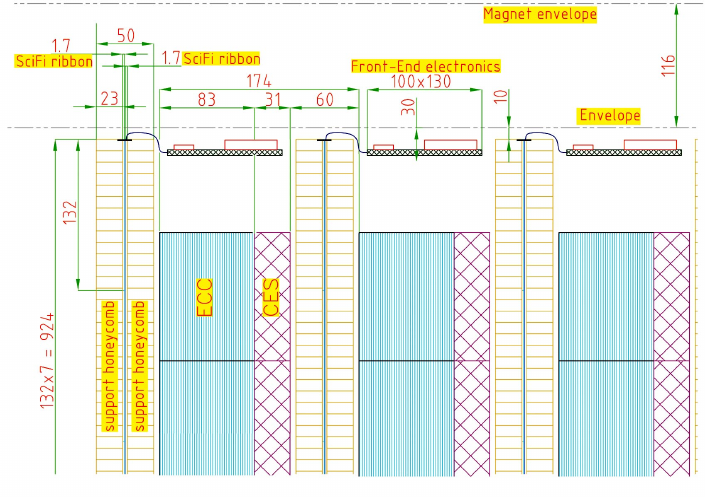}
    \caption{SciFi Tracker longitudinal cut view. Units are mm.}\label{nutaudet:SciFiT2DStruct}
    \end{figure}%

    The scintillating fibres from Kuraray, type SCSF-78MJ with double cladding used in the LHCb SciFi, are
    a viable option for the SHiP tracker. An alternative option could be the BCF-12 fibres from Saint-Gobain
    company. The light yield value for these types of fibres is typically around 8000 photons/MeV. A nominal emission 
    spectrum for the SCSF-78MJ fibre extends from about 400 to 600 nm and peaks at 450 nm, with a bulk optical absorption
    length greater than 3 m. These properties are more than sufficient to achieve an excellent tracking efficiency with
    the proposed module design. 
    
    In the current design 12 SciFi tracker modules of 50 mm thickness are installed in the Goliath magnet, alternating
    the ECC brick walls of the $\tau$-neutrino detector. The SciFi tracker longitudinal cut view with its envelop inside
    the Goliath magnet is shown in Figure~\ref{nutaudet:SciFiT2DStruct}. 
      
    %\vfill
    \begin{figure}[tp]
    \centering
    \includegraphics[height=8cm]{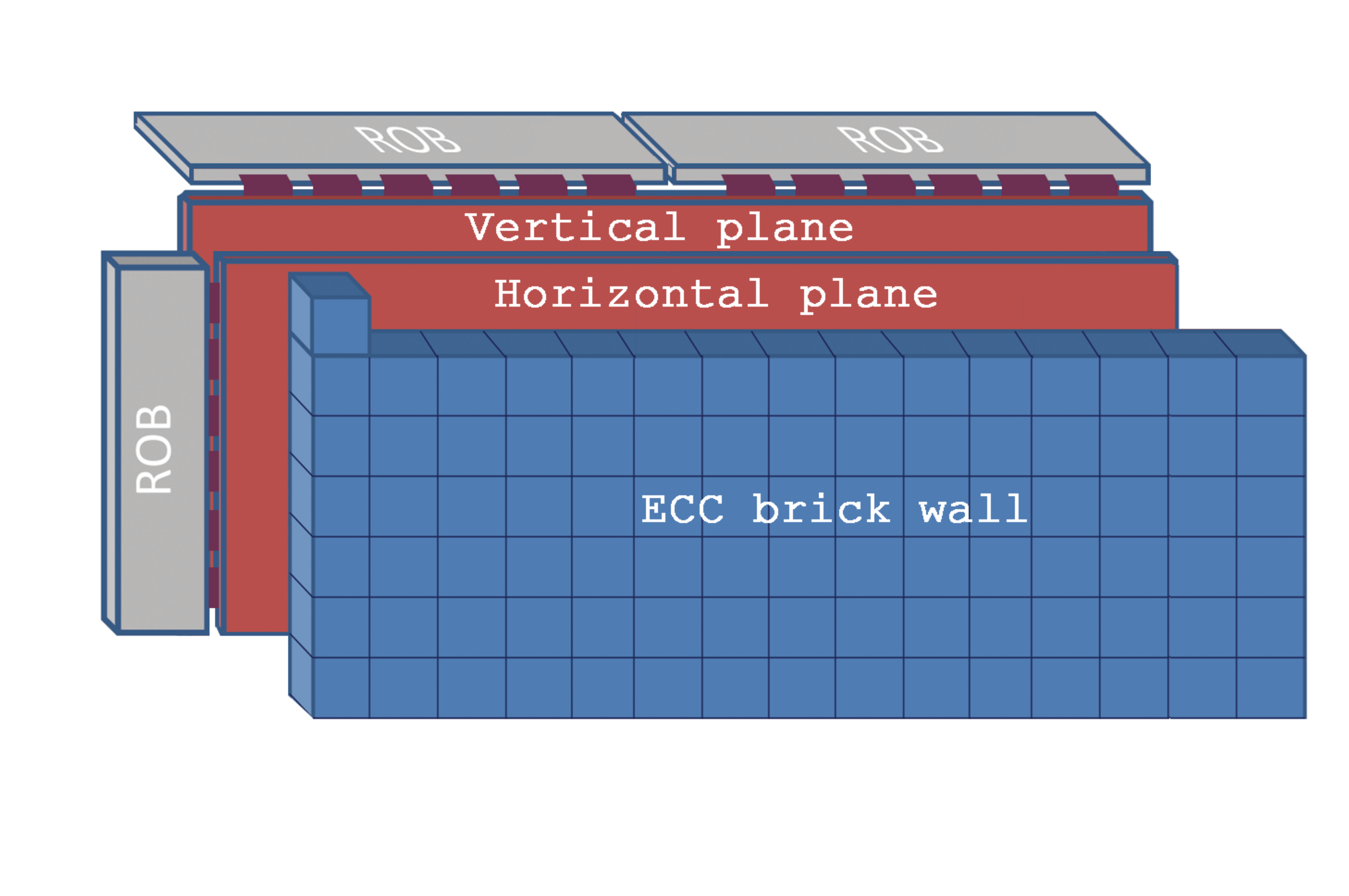}
    \caption{Exploded view of the Scintillating Fibre Tracker module.}\label{nutaudet:SciFiNuTrackerModule}
    \end{figure}%

    Light detection is carried out by SiPM chips. Two SiPM manufacturers, Hamamatsu 
    and KETEK, have developed dedicated devices for the SciFi Tracker. These custom devices
    provide high photon detection efficiency (PDE) in a large wavelength range, high reliability
    and low cross-currents. The readout end-face of the module is polished, while mirror
    is put on the opposite face. The SiPM photo sensors are connected directly to the 
    readout side.
    Each ribbon is readout by four 128 channel SiPM chips. The design of the readout
    boxes should account for a lack of space in vertical direction and, hence, allow
    a flexible connection to the module as shown in Figure~\ref{nutaudet:SciFiNuTrackerModule}. 
    The SiPM cooling is not so crucial for the SHiP $\tau$-neutrino detector as it is for
    the LHCb SciFi tracker, where SiPM will operate at -40$^{\circ}$C with the thermal uniformity
    of about 1$^{\circ}$C. The SiPM irradiation is negligible in our case. However cooling would be beneficial to reduce the SiPM dark noise. 
    The actual temperature regime optimal for the tracker operation has to be determined.

		\subsubsection{GEM tracker} 
One of the options for a gas detector is a triple-GEM detector. GEM detectors are nowadays a well established technology with the broader family of Micro-pattern gas detectors, instrumenting several experimental high-energy systems.

\begin{figure}[ht!]
\centering
\includegraphics[width=90mm]{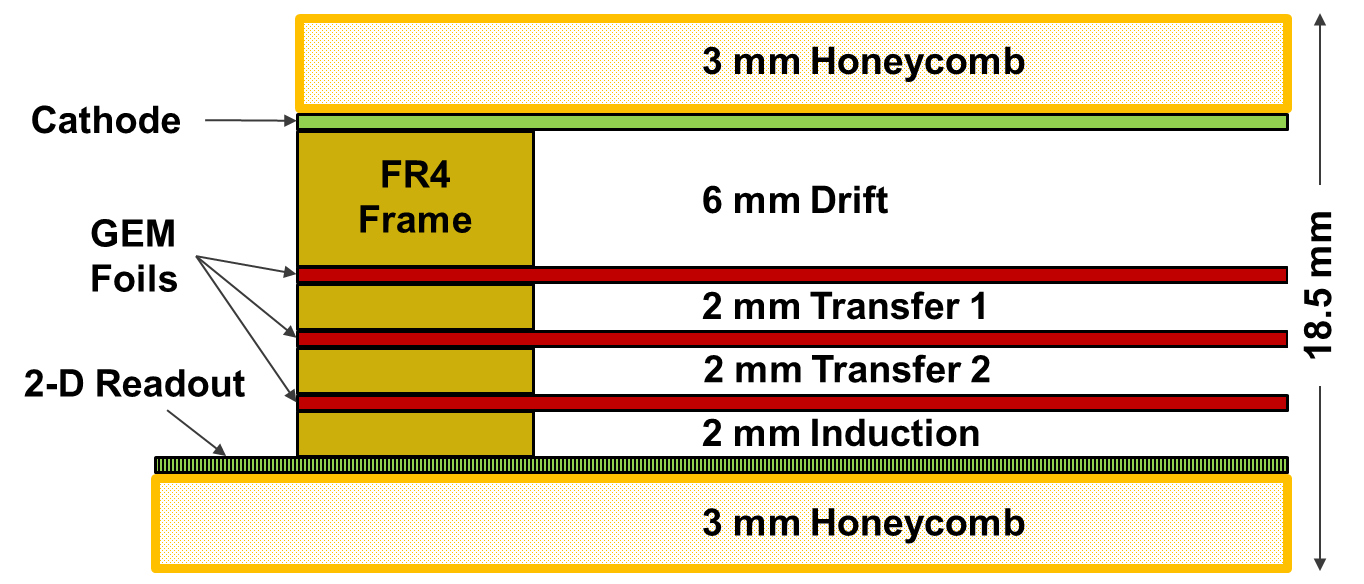}
\caption{Cross section of a triple-GEM detector. }
\label{fig:gemSection}
\end{figure}

A GEM (Gas Electron Multiplier)\cite{sauli} is made of a $\mathrm{50\,\mu m}$ thick polyimide foil, copper clad on each side and perforated with a high density of holes ($\mathrm{70\,\mu m}$ diameter, $\mathrm{140\,\mu m}$ pitch). When a voltage difference of typically 400~V is applied between the two copper faces, a field as high as 100 kV/cm is created within the holes, acting as multiplication channels for the electrons produced by the gas ionization. A triple stages multiplication structure allow to reach a gain as high as $\mathrm{10^4}$ while minimizing the discharge probability. Figure~\ref{fig:gemSection} shows a cross-section of a triple-GEM detector. The large drift gas of 6~mm provides a high efficiency and allows to reconstruct the track angle, working as a micro-TPC. One plane corresponds to $X/X_0 \simeq 0.05$\% and it shows an efficiency of about 98\% for non-inclined particles. Future tests will study the efficiency for inclined tracks. 

Recently, within the RD51 Collaboration, a consistent effort has been made to overcome the past size limitation, in order to fulfill the increasing requests of large-area detectors. Presently the largest GEM foil ever produced is 200~cm long and 60~cm wide. In particular the limitation in the width comes directly from an intrinsic raw material manufacturing limit that cannot be exceeded.

\begin{figure}[ht!]
\centering
\includegraphics[width=90mm]{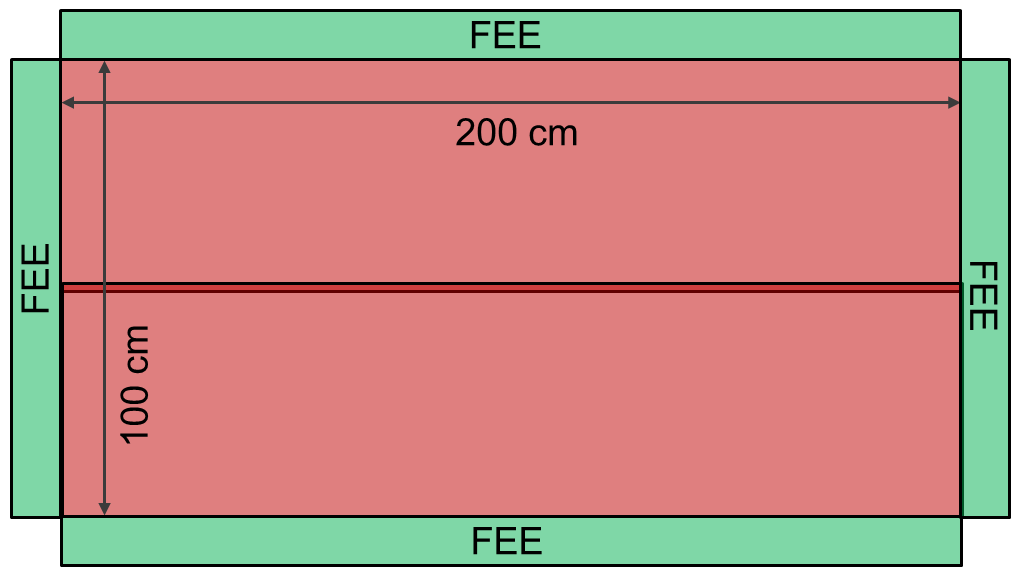}
\caption{Layout of a Target Tracker station. }
\label{gemLayout}
\end{figure}

The most compelling challenge for the exploitation of GEM detectors as Target Trackers is represented by the unprecedented size required. The foil needed to cover the TT area can be obtained by splicing together two smaller foils, according to a technique developed and tested in the construction of the KLOE-2 Inner Tracker\cite{kloe2}. The single foils will be as large as $\mathrm{200\times50~cm^2}$, and glued along the long side with a narrow overlap of about 3~mm (see Figure~\ref{gemLayout}). The same procedure would be used to produce the cathodes and the readout electrodes, the other parts composing a triple-GEM detector.

\begin{figure}[ht!]
\centering
\includegraphics[width=40mm]{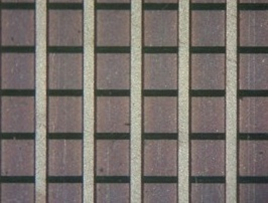}
\caption{Detail of the readout circuit. The X and Y strips have the same pitch but different widths, to equally share the signal being at different distance from the last GEM foil.}
\label{readout}
\end{figure}

The readout plane is a multi-layer circuit on a polyimide substrate, patterned with a XY structure of copper strips engraved at two different levels (Figure~\ref{readout}).

\begin{figure}[ht!]
\centering
\includegraphics[width=80mm]{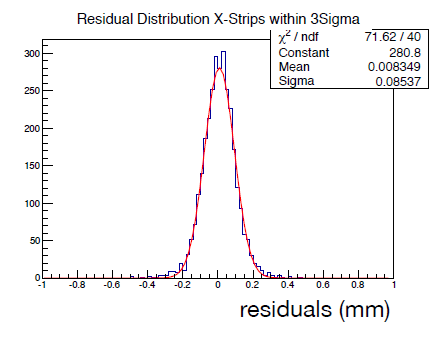}
\caption{Distribution of the residuals of a GEM prototype with $\mathrm{650~\mu m}$ pitch strips readout by analog front-end amplifier (APV25).}
\label{residuals}
\end{figure}

This solution has been already used by other experiments. For instance, Figure~\ref{residuals} shows the space resolution of a GEM prototype for the BESIII Inner Tracker with $\mathrm{650~\mu m}$ of pitch between the strips and analog signal amplifier. The residual value of 85~$\mu$m microns in Figure~\ref{residuals} corresponds to the resolution of 70~$\mu$m after subtracting the contribution from external trackers. 

In order to ensure a $\mathrm{100~\mu m}$ resolution, and taking into account the natural spread of the charge induced by the magnetic field, a pitch of $\mathrm{500~\mu m}$ should be chosen. 

		\subsubsection{Micromegas tracker} 
\begin{figure}[h!]
\centering
			\includegraphics[scale=0.9]{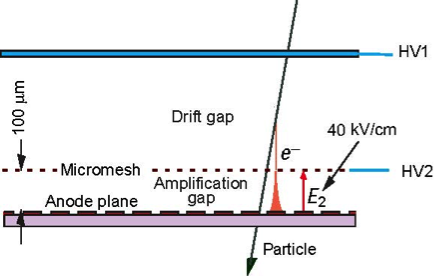}
			\includegraphics[scale=1.2]{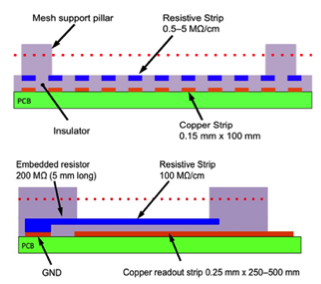}
			\caption{(Left) Basic principle of a Micromegas (MM) detector. A metallic micromesh separates a few mm of low-field ($\sim 1$~kV/cm) drift region from the high-field ($\sim 50$~kV/cm) amplification region. (Right) Sketch of the resistive Micromegas principle, illustrating the resistive protection. }
			\label{fig:micromegas1}
		\end{figure}

 The concept of the Micromegas was developed in the ‘90s~\cite{MM1} in the context of R\&D on Micro Pattern Gaseous Detectors (MPGD)~\cite{titov}. A sketch of the MM operating principle is shown in the left plot of Figure~\ref{fig:micromegas1}. It consists of a few millimeters of drift region and a narrow multiplication gap (25$\div$150~$\mu$m) between a thin metal grid (micromesh) and the readout electrode (strips or pads of conductor printed on an insulator board). Regularly spaced supports (insulating pillars) guarantee the uniformity of the gap between the anode plane and the micromesh. The electric field is homogeneous both in the drift (electric field $\sim 1$~kV/cm) and amplification ($\sim$40$\div 70$~kV/cm) gaps. By proper choice of the applied voltages, most of the positive ions are collected by the micromesh; this prevents space-charge accumulation and induces very fast signals. The small amplification gap produces a narrow avalanche, giving rise to excellent spatial resolution: $\sim 30$~$\mu$m accuracy, limited by the micromesh pitch, has been achieved for minimum ionizing particles.

The industrial assembly process of Micromegas allows regular production of large, robust, and inexpensive detector modules. Since a few years, the spark-protected Micromegas has been built by adding above the anode readout strips a layer of resistive ones, separated by an insulating layer.  The readout strips are individually connected to the ground through the large resistor. The principle of the resistive spark protection is schematically shown in the right plot of Figure~\ref{fig:micromegas1}. A rather advanced proponent of resistive Micromegas is the MAMMA collaboration working on the New Small Wheel (NSW) upgrade for the ATLAS Muon Spectrometer. The largest detectors built so far and operating smoothly have dimensions of $\sim 2.2 \times 1.0$~m$^2$ with the potential to go even to the larger module sizes.

\begin{figure}
\centering
\includegraphics[scale=1.0]{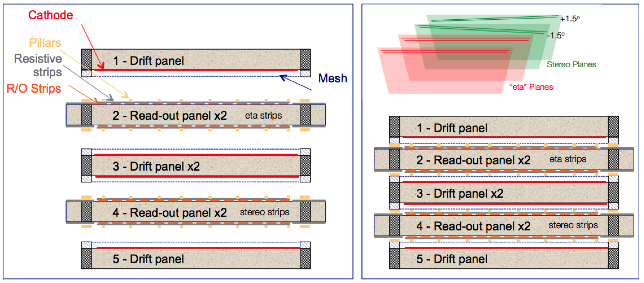}
\caption{Sketch of the assembly of a quadruplet. (Left) Five panels before assembly. (Right) The assembled quadruplet Micromegas chamber. A sketch of the orientation of the strips is shown~\cite{Iodice}.}
\label{fig:micromegas2}
\end{figure}

The design and the construction procedures for the NSW Micromegas are the result of detailed studies of panels construction with extensive prototyping and tests, and of a campaign of measurements of deformations in comparison with mechanical simulations models. The sketch of the main components of the Micromegas quadruplet chamber is shown in Figure~\ref{fig:micromegas2}~\cite{Iodice}. The readout (r/o) planes are disposed in a “back-to-back” configuration onto two stiffening panels with an aluminum honeycomb internal structure (r/o panels). Four different strips orientations are used in a single module: one read-out panel with orthogonal XY strips while the other r/o planes of the second panel are tilted by 1.5$^{\circ}$ (stereo strips or UV planes). The pitch (width) of the strips is approximately 450 (300)~$\mu$m respectively. The total thickness of the Micromegas chamber (containing 4 drift gaps/strip orientations) is about 7 cm.

\begin{figure}
\centering
\includegraphics[width=0.45\linewidth]{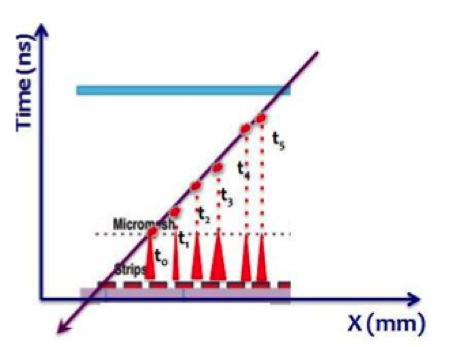}
\includegraphics[width=0.45\linewidth]{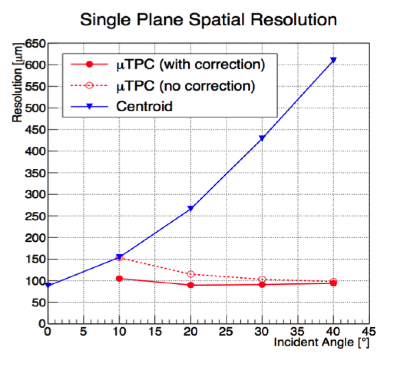}
\caption{(Left) Concept of $\mu$TPC mode. Given the drift speed, the measurement of the strip signal arrival times provides the position of ionization clusters. (Right) Spatial resolution as a function of the particle incidence angle. Resolutions obtained using the charge centroid and the $\mu$TPC mode are compared, together with the one obtained by combining the two methods~\cite{Bini}.}
\label{fig:micromegas3}
\end{figure}

The spatial resolution can be determined in two ways: either making the charge centroid of the fired strips, optimal for almost perpendicular tracks, or using the so-called micro-TPC ($\mu$TPC) method~\cite{Alexopoulos} for tracks emitted at large angles. The micro-TPC concept is described in the left plot of Figure~\ref{fig:micromegas3}. Measuring the arrival time of the ionized electrons with a time resolution of a few nanoseconds allows reconstructing positions of the ionization clusters and the segment of the track inside the drift gap. The $\mu$TPC method has been successfully applied in many test-beam data. The right plot of Figure~\ref{fig:micromegas3} shows the spatial resolution achieved with both algorithms~\cite{Bini}. The cluster centroid behaves better at small angles (small cluster size) while the μTPC method reaches best performance for larger angles. Combination of the cluster centroid and $\mu$TPC algorithms is optimal to achieve a spatial resolution well below 100~$\mu$m in the full angular range.

\subsubsection{Front-end electronics for micro-pattern gas detectors}	
 Both digital and analog readout methods can be used to achieve the goal of a spacial resolution better than 100~$\mu$m.
The digital readout identifies clusters by detecting adjacent strips with a collected charge above a fixed threshold. The reconstructed position of the track is the geometrical center of the cluster and the resolution is defined as the pitch size divided by $\sqrt{12}$. 

The analog readout allows to set a threshold on both the single strips and the total charge, in such a way to improve the ghost hit rejection. Moreover, the encoding of the charge collected on a strip provides the reconstruction of the charge centroid, thus improving the resolution by more than the pitch/$\sqrt{12}$ as obtained with the binary readout method. The analog readout method achieves the required spatial resolution using a lower number of readout channels, then lowering the cost/channel and the power dissipation of the  system; moreover it is safer as it improves the noise rejection. While the digital readout method is easy to implement as it manages a single hit information per channel, the analog readout method requires encoding the collected charge, i.e.~each readout channel must include a digitization section.

Each view (X and Y) of the proposed detector will be instrumented with 500~$\mu m$ pitch readout strips, thus obtaining 8000 channels per layer (Y view will be spit in the middle to limit the parasitic capacitance) as shown in Figure~\ref{gemLayout}.

Due to the 112000 readout channels involved, a highly integrated and low power-consuming system must be used to instrument the detector. Moreover,  due to the experiment trigger-less data acquisition, readout devices capable of event time-stamp must be used. 

A very promising development in this direction is the front-end asic for the ATLAS muon upgrade, the 64 channels VMM2 chip envisaged to instrument both TGC (small Strip Thin Chamber) and Micromegas.

\begin{figure}[h!]
			\begin{center}
			\includegraphics[angle=0, scale = 0.8]{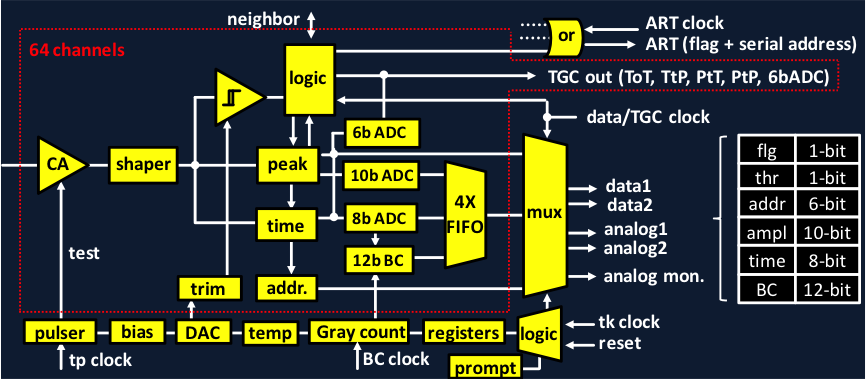}
			\caption{\small{VMM2 block diagram}}
			\label{fig:vmm2_block_diagram}
			\end{center}
		\end{figure}

The chip single channel block diagram is shown in Figure~\ref{fig:vmm2_block_diagram}. Each channel is made of a low noise charge amplifier with adaptive feedback, test capacitor and adjustable polarity; it can manage up to 200 pF input capacitance and the peaking time is 25 ns. After the preamplifier there is a shaper with adjustable peaking time (25, 50, 100 and 200 ns) \cite{degeronimo_1} together with a stabilized band-gap reference baseline; the shaper gain is adjustable as  well.

Along the chain there are the discriminator with adjustable threshold (also for pulses smaller than the discriminator hysteresis), the peak detector and the time detector. The VMM2 chip foresees three different operation modes: direct output, two-phases or analog and continuous (digital). 
When operating in continuous (digital) mode, the peak and time detector outputs are, respectively, routed to a 10 bits ADC for peak amplitude measurement and to a 8 bits ADC to measure, by means of a TAC circuit, the time difference from the time of the peak to a stop signal. The stop signal is generated each clock cycle of a (gray-code) counter; an external clock signal (BC) is used to increment the common counter. When the stop signal occurs, the value of the counter (12 bits) is latched thus providing a 20-bit timestamp word (12 bits from counter + 8 bits from TAC value conversion).

The  continuous (digital) single channel output frame is made of 38 bits as shown in Table~\ref{tab:VMM2_datastream}: the threshold crossing indicator allows discrimination between above-threshold and neighbor events.

\begin{table} [h]
	\caption {VMM2 output data stream in {\it continuous mode} operation} 
	\begin{center}
	\begin{tabular}{|l|l|}
	\hline
		bit 1 & readout flag \\ \hline
		bit 2 & threshold crossing indicator \\  \hline
		bit 3-8 & 6 bits - channel address\\ \hline
		bit 9-18 & 10 bits - peak amplitude\\ \hline
		bit 19-38 & 20 bits - time measurement\\ \hline
	\end{tabular}
	\end{center}
	\label{tab:VMM2_datastream}
\end{table}
A 4-events deep derandomizing FIFO is used to store output events; FIFOs containing valid events are readout in sequence and data are multiplexed to the digital outputs {\it data0} and {\it data1}; {\it data0} is used as a readout flag from off-detector electronics. A dedicated digital output (Address in Real Time - ART)  is generated every bunch crossing, providing the address of the channel with the earliest signal above threshold; the output can be used as a FAST-OR as well. All digital I/O are LVDS (600 mV baseline and $\pm$ 150 mV swing) and the power dissipation per channel ranges from about 4 to 8~mW according to the selected operating mode.

\subsubsection{An R\&D option: the $\mu$-RWELL }

\begin{figure}
  \begin{minipage}[t]{.46\textwidth}
    \centering
   \includegraphics[scale=0.28]{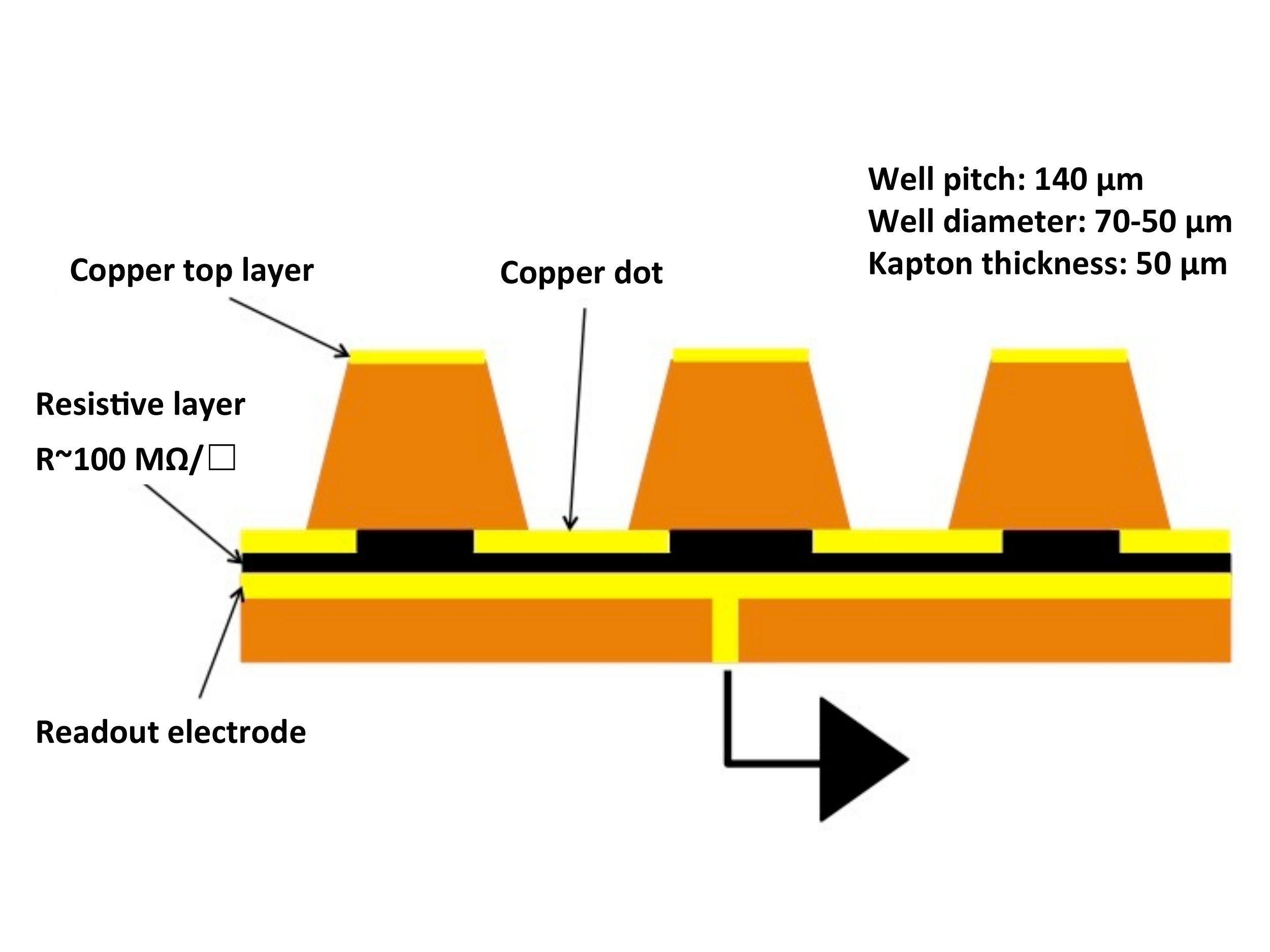}
    \caption{Schematic drawing of the $\mu$-RWELL PCB.}
    \label{fig:blind-gem-substrate}
  \end{minipage}
    \hspace{5mm}
  \begin{minipage}[t]{.46\textwidth}
   % \centering
    \hspace{-7mm}
         \includegraphics[scale=0.35]{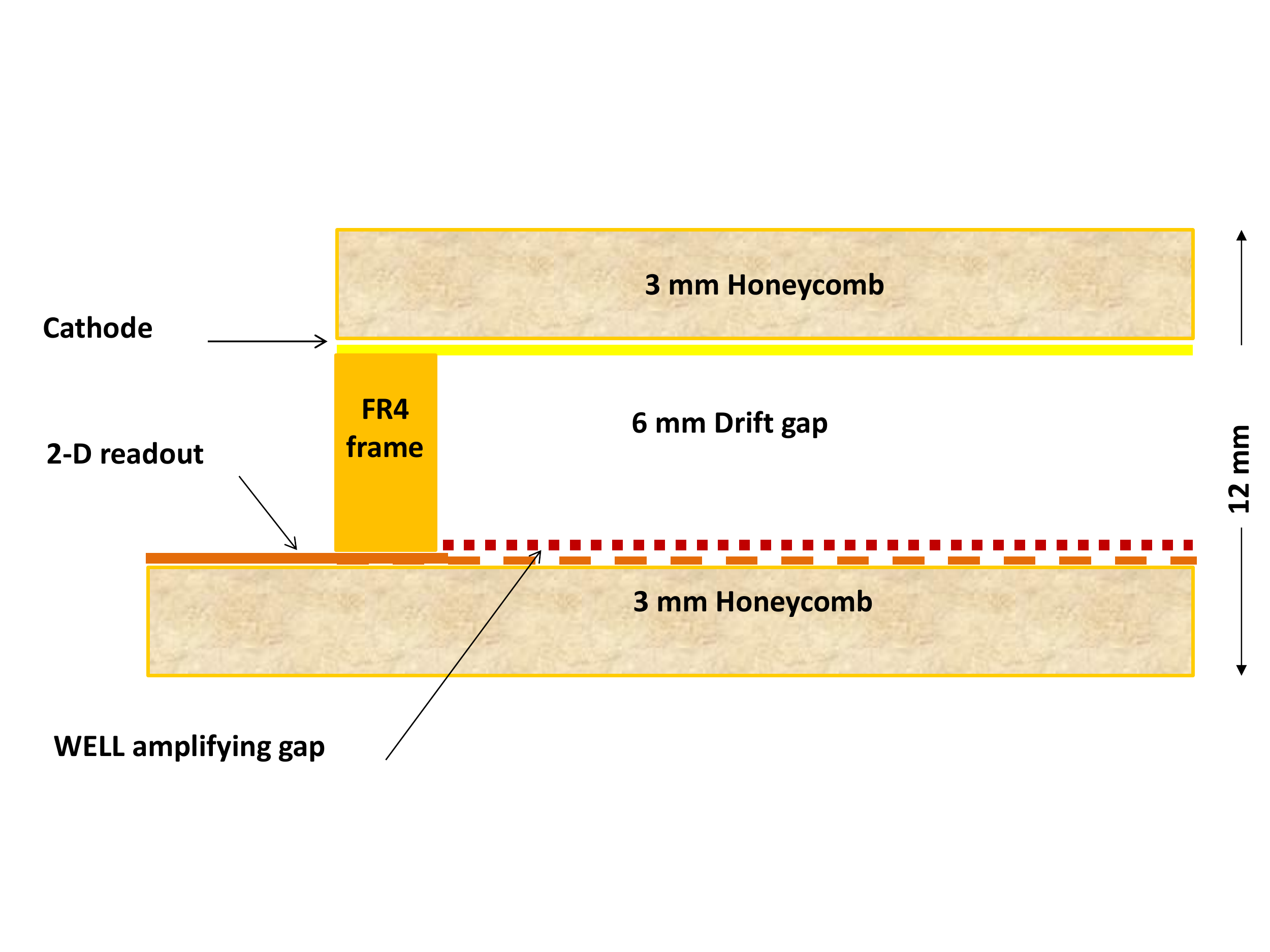}
         \caption{Schematic drawing of the $\mu$-RWELL detector.}
    \label{fig:blind-gem-picture}
  \end{minipage}
\end{figure}

Among the very recent developments in the MPGDs, the micro-Resistive WELL ($\mu$-RWELL) combines in a unique approach the solutions and improvements achieved in the last years in the field of gas detectors.
The $\mu$-RWELL is a very compact detector structure, very robust against discharges and exhibiting large gains (up to 10$^4$), easy to build, cost effective and suitable for mass production. 
%\cite{micro-RWELL}
%
The novel detector has been designed at the Laboratori Nazionali di Frascati and it was firstly built in  2009 by TE-MPE-EM Workshop at CERN in parallel with the CERN-GDD group~\cite{alfonsi2009performance,croci}. 
A similar device based on THGEM technology has been  recently proposed by other groups~\cite{breskin2}.
The $\mu$-RWELL, as sketched in Figure~\ref{fig:blind-gem-substrate}, is realized by merging a suitable etched GEM foil with the readout printed circuit board (PCB) plane coated with a resistive deposition. The copper on the bottom side of the foil has been patterned in order to create small metallic dots in correspondence of each WELL structure. The resistive coating has been performed by screen printing technique.
The WELL matrix geometry is realized on a 50$~\mu$m thick polyimide foil, with conical channels 70$~\mu$m (50$~\mu$m) top (bottom) diameter and  140$~\mu$m pitch (of course different geometries can be considered in order to optimize the detector performance, especially in terms of gain amplitude).
A cathode electrode, defining the gas conversion/drift gap, completes the detector mechanics (Figure~\ref{fig:blind-gem-picture}).

The $\mu$-RWELL has features in common either with GEMs~\cite{GEM1} or Micromegas~\cite{MM1}:
\begin{itemize}
\item[-]  from GEM it takes the amplifying scheme with the peculiarity of a {\it "well defined amplifying gap"}, thus ensuring very high gain uniformity.

\item[-]  from Micromegas it takes the resistive readout scheme that allows a strong suppression of the amplitude of the discharges.
The principle is the same of the resistive electrode used in Resistive Plate Counters (RPCs) \cite{Pestov, santo, bencivenni}.
\end{itemize}
Even though the amplifying element of the $\mu$-RWELL is practically the same as the one of GEM, its signal formation mechanism is completely different.
Unlike GEM detectors where the signal is mainly due to the electron motion, in a $\mu$-RWELL the signal is produced also by the ionic component. 
In this sense the signal of a resistive-WELL is more similar to the one of a micromegas.

The assembly aspect of the resistive-WELL technology is obviously a strong point in favor of this architecture.
The $\mu$-RWELL is composed by only two components: the readout-PCB, with the amplifying part embedded on it, and the cathode.
Its assembly does not require gluing or stretching of foils or meshes: a very critical and time-consuming construction step of both GEM and MM technologies. The stretching of a GEM foil as well as a metallic mesh for a 2 m long device requires up to 200 kg mechanical tension, that clearly must be supported by suitable rigid  mechanical structures.

Significant advantages introduced by the $\mu$-RWELL is the simplified construction for large area planes and the stability in terms of discharges. It  is extremely compact, it does not require very stiff (and large) support structures, allowing large area covering based on PCB splicing (with a dead space within 0.2$\div$0.3mm) while keeping a {\it "well defined amplifying gap"}.

For the Neutrino Detector at SHiP we foresee planes of about 2$\times$1~m$^2$ TT layer, with 2D-view and a 6 mm large conversion/ionization gap, could have a total thickness within 24~mm.

\begin{figure}
%  \begin{minipage}[t]{.65\textwidth}
    \centering
   \includegraphics[width=0.8\textwidth]{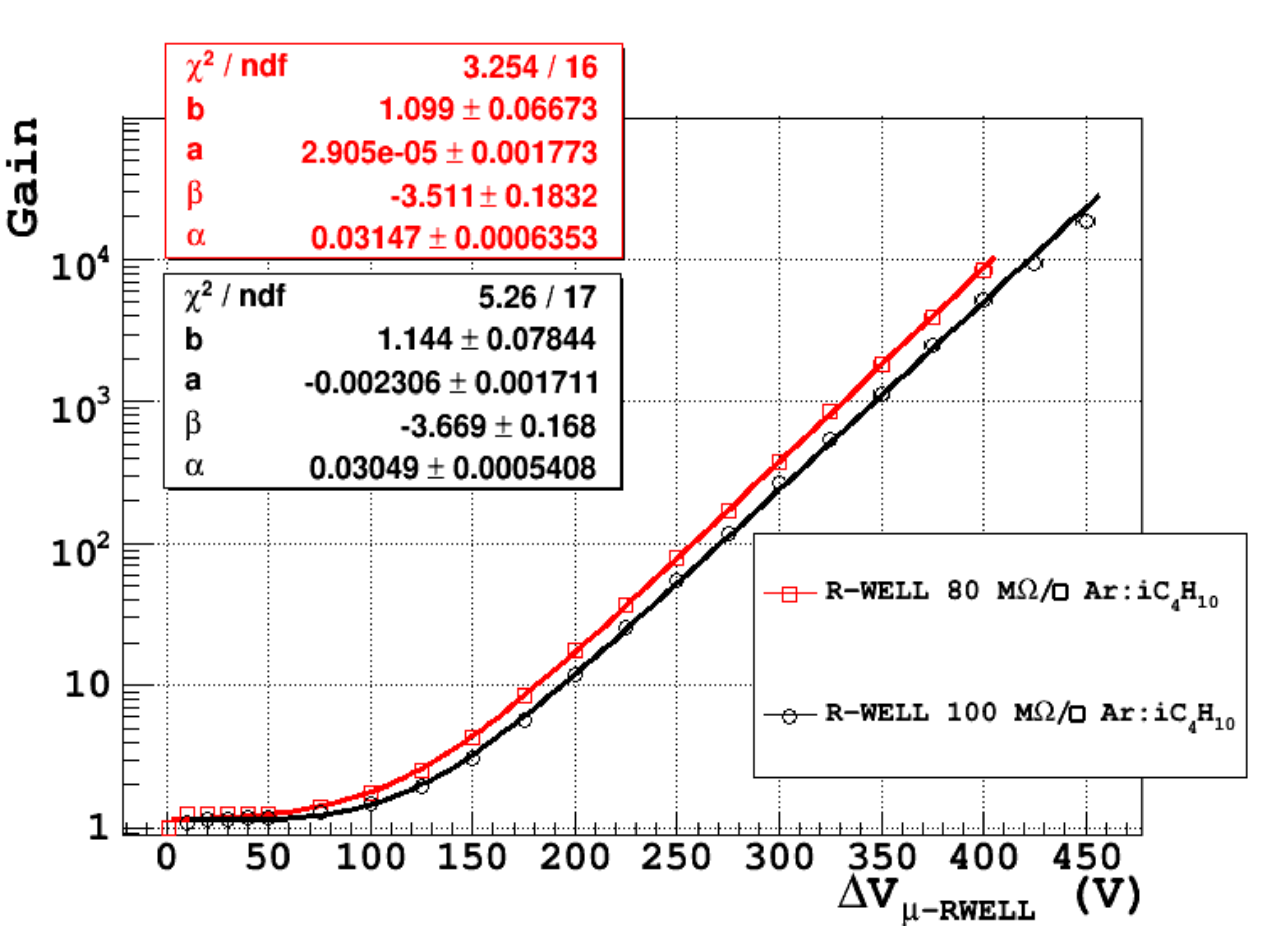}
   \caption{Gas gain in Ar:iso-buthane (90:10), for two different $\mu$-RWELL geometry: 80 M$\Omega$/$\square$ (red points) and 100 M$\Omega$/$\square$ (black points).}
   \label{fig:gain-ARISO}
%  \end{minipage}
\end{figure}

In the following we report some results about recent tests performed on the $\mu$-RWELL. 
The detector gain has been measured with X-rays, in current mode, as a function of the potential applied across the amplification stage and the resistive layer.
As shown in Figure~\ref{fig:gain-ARISO}, with a voltage difference of 400~V, the operation with an iso-buthane based gas mixture allows to achieve a gas gain larger than 10$^4$, comparable with the gain at which standard triple-GEM and MM are normally operated.
The use of thicker kapton foil (125~$\mu$m thick polyimide foil is commercially available, from KANEKA ltd in Japan) for the realization of the amplifying component of the detector should allow to achieve gas gain sensibly larger than those obtained with the 50$\mu$m thick amplification gap.

The typical  discharge amplitude for the $\mu$-RWELL is of the order of few tens of nA, while for a GEM detector discharges with amplitude of the order of $\mu$A are observed at high gas gain.

%(Figure~\ref{fig:discharge_gem} and Figure~\ref{fig:discharge_R-WELL}).

%
Very preliminary results from a test at the H4-SPS beam line with two different $\mu$-RWELL (with 80 M$\Omega$/$\square$ and 1 G$\Omega$/$\square$), with a 400~$\mu$m strip pitch, immersed in a dipole magnetic field (GOLIATH), show that:
\begin{itemize}
\item [-] a space resolution of $\sim$100$\mu$m  is achievable for magnetic field up to 1 Tesla;
\item [-] a detection efficiency of the order of $\sim$98\%  (with 4 mm gas gap).
\end{itemize}
Measurements at large particle incident angle require  further studies.

The $\mu$-RWELL is a very promising technology showing important advantages with respect to classical GEMs and MMs: the detector is {\it "thin - large - simple"} and fulfills the main requirements for the TT of the Neutrino Detector at SHiP.

Some R\&D is required, in particular to study the position resolution as a function of the different resistive layer configurations and as a function of the incident particle angle, together with the construction of a large size prototype with a technique similar to the one used for micromegas.

\section{Muon magnetic spectrometer}
The SHiP muon magnetic spectrometer has to accomplish the following tasks:

\begin{itemize}

\item identify with high efficiency the muons produced in neutrino interactions and in $\tau$ decays;

\item complement the muon momentum measurement performed in the magnetised target, extending the accessible kinematic  range.

\end{itemize}

The muon spectrometer consists of active detectors (RPC and Drift-Tubes) and a warm dipolar magnet.
Given the similar muon energy spectrum, the design of the magnet is based on the iron dipoles built and successfully operated in the OPERA experiment at LNGS~\cite{bopera}.

	\subsection{Magnet} 
\begin{figure}[!h]
\centering
\includegraphics[width=12cm]{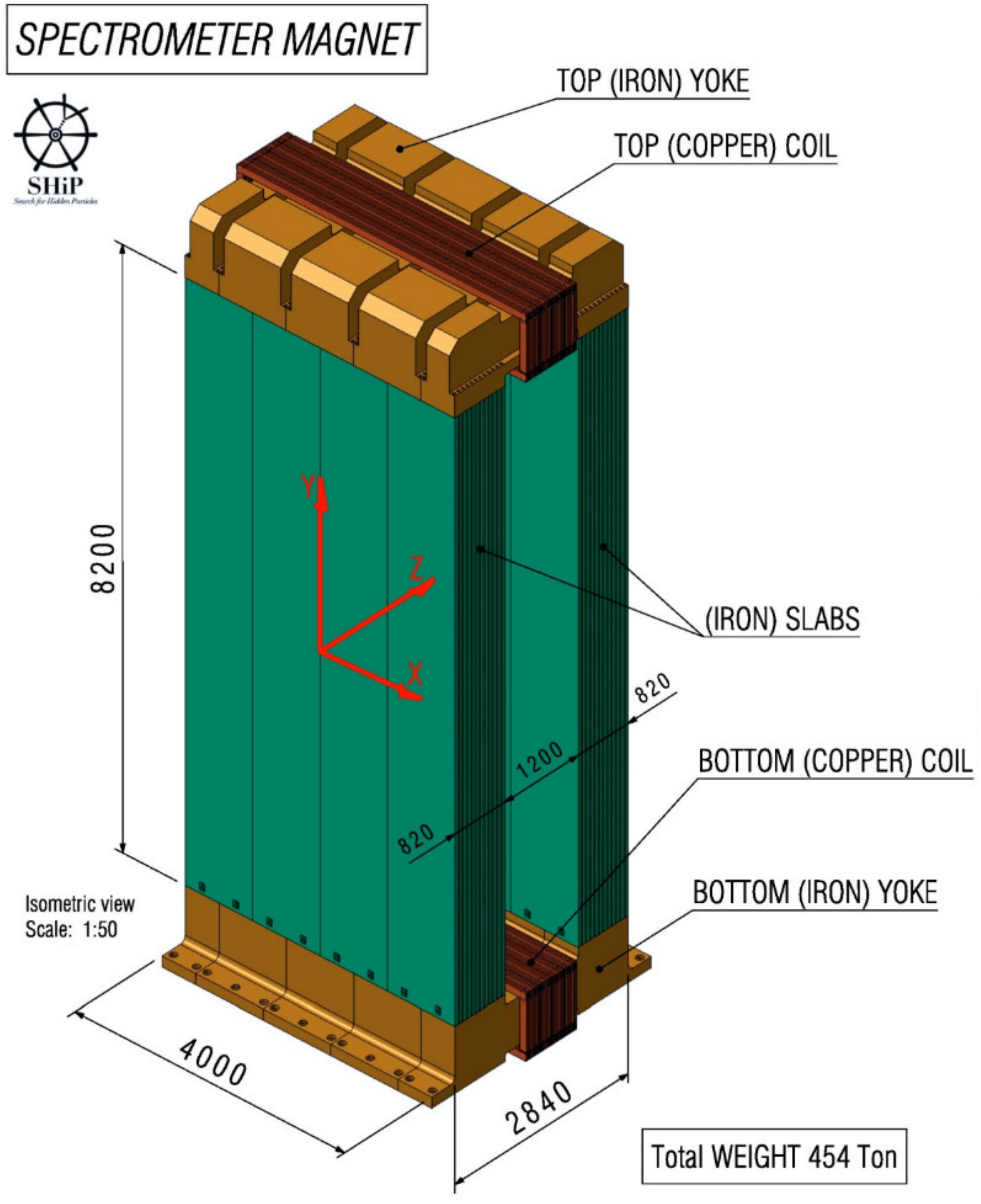}
\caption{Neutrino detector magnet sketch.}
\label{fig:magnete_3d}
\end{figure}

The magnet shown in Figure~\ref{fig:magnete_3d} is made up of two vertical walls of rectangular cross section and of top and bottom flux return paths. 
The width of the magnet is 4 m; its height, including the top and bottom yokes, is about 10 m. The transverse size of the magnet is driven by the requirement to measure muons emitted with angles up to $\pi/4$ with respect to the incoming neutrino direction. In the horizontal direction, the size is limited to 4~m to avoid the external high muon rate.  
 
Each wall is built lining twelve iron layers (5~cm thick) interleaved with 2~cm of air, allocated for RPC housing.  
These layers are bolted together to increase the compactness and the mechanical stability.
The bolts are spaced by 625~mm horizontally and 1150~mm vertically. The nuts holding the bolts also serve as spacers and fix the 20~mm air gap where the RPCs are installed. The layers are bolted at the top and bottom edges to the return flux paths.  
To ensure mechanical stability, the basements need to be fixed to the floor by means of bolts cemented to the floor concrete.
Lateral support may also be needed. 
The 24 iron layers are divided into slabs; the top and bottom return yokes are segmented as well, in order to ease their handling with cranes during installation.

The total iron thickness of the two walls is 1.2~m. This is an optimal tradeoff between the need to reduce the multiple scattering and to amplify the bending for the accuracy in the momentum measurement. RPC are intended for a course tracking of muons inside the iron, while the measurement of the bending is performed with the more precise drift-tubes.

The overall weight of the SHiP neutrino detector magnet is 454~t.  

The dipole is magnetized by means of two coils, 20 turns each, installed in the top and bottom flux return paths.  The nominal current flowing in the coils is 1600~A, corresponding to an overall magneto-motive force of 64000 A$\cdot$turns. The expected average field along the walls is 1.57~T. The field non-uniformity along the magnet height is within 5\%.

Detailed studies about the field map inside OPERA magnets have been performed, based both on measurements on prototypes and on the TOSCA \cite{ref:TOSCA} code.

\begin{figure}
\centering
\includegraphics[width=10cm]{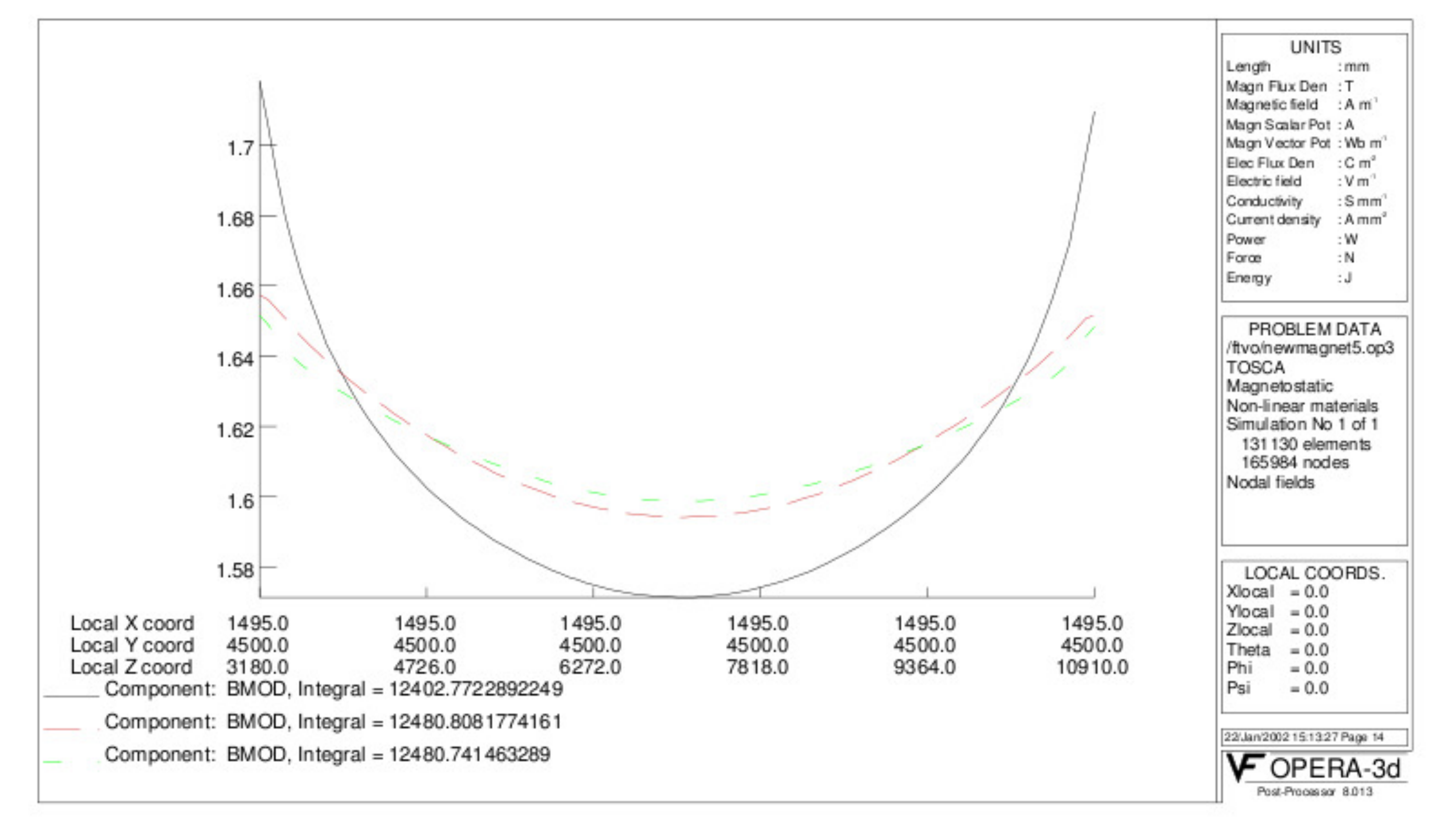}
\caption{Magnetic field along the height.}
\label{fig:heigh_1} \end{figure}

\begin{figure}
\centering
\includegraphics[width=10cm]{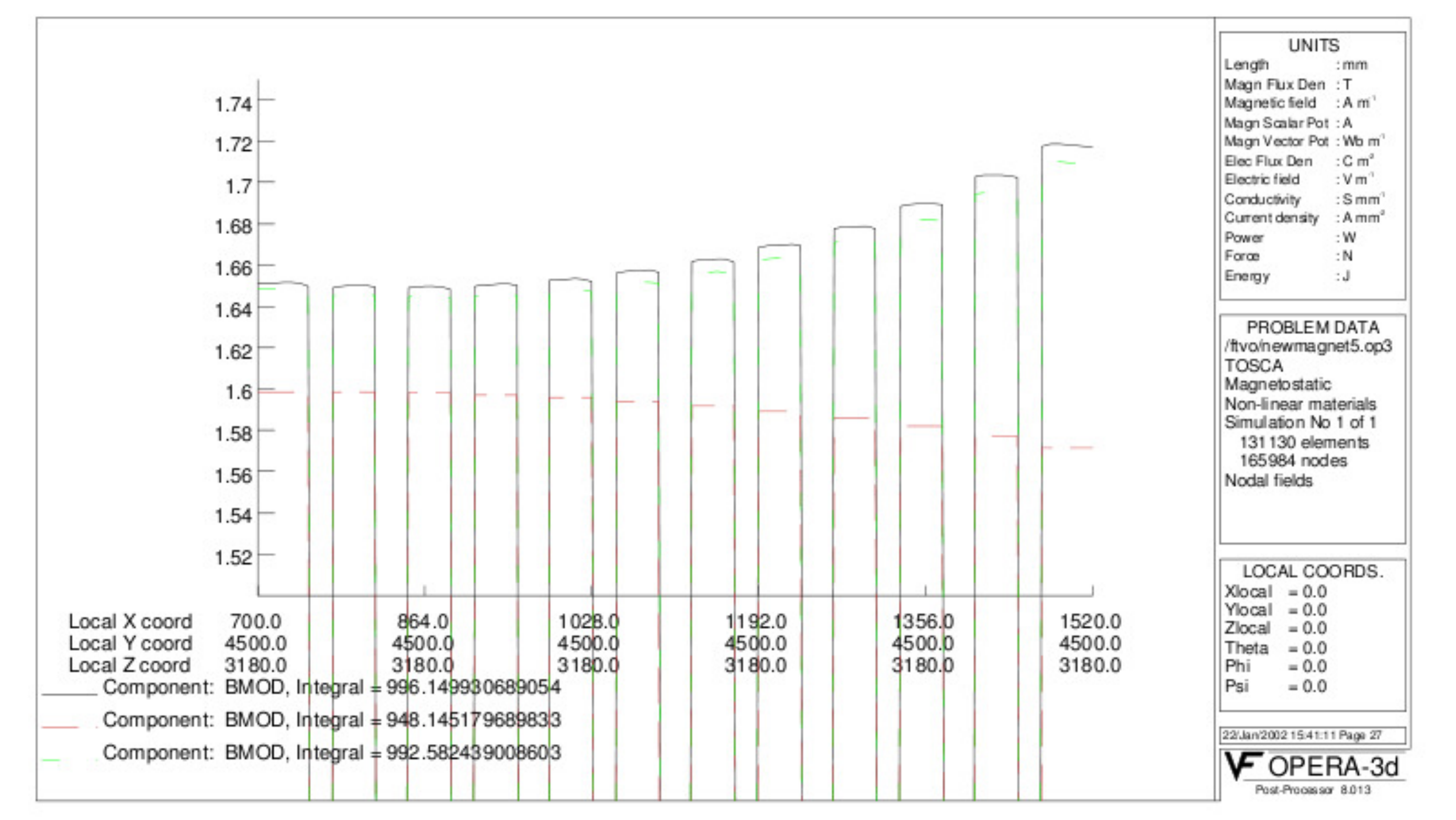}
\caption{Magnetic field along the neutrino direction from the outermost to the innermost layer.}
\label{fig:horiz_1} \end{figure}

The simulated strength of the $\vec{B}$ field along the height of the spectrometer is shown in Figure~\ref{fig:heigh_1} for the nominal current $I=1600$~A. 
The three lines correspond to the most internal (black), central (red) and most external (green) layer. The field is computed at the center of each layer.  The inner layers suffer from a lower uniformity. The non-uniformity of the field along the height, averaged over all the slabs is better than 5\%.  The uniformity among the various layer at fixed height can be inferred from Figure~\ref{fig:horiz_1} and it is smaller than $4\%$. Here, the three lines represent the field along the direction of incoming neutrinos through the twelve layers of the  spectrometer wall. The field is taken at three different heights: 1.2~m from the floor (black), 5~m (red) and 8.8~m (green). 

During OPERA data taking the magnetic flux was measured {\it in situ} by a dedicated field monitoring system with a precision of a few {\it per cent}.
For a current of 1600 A, the minimum field value measured at half height was $1.46 \pm 0.01$ T, with non-uniformities along the height not exceeding 3\%~\cite{ref:Bmeas}.
The measured field, averaged along the height, was $\sim$ 1.53 T, with a slight flux deficit observed with respect to the simulation (3-5\%).

	\subsection{RPC tracking system} 

\label{sec:rpc_tracker}

Profiting from the experience gained in OPERA as well as of the intensive R\&D activity carried out for the upgrade of LHC detectors, we propose to adopt the Resistive Plate Chamber  technology to perform the coarse tracking between the magnetized iron slabs.  The role of this detector is to track particles within the magnetised iron. A resolution at the level of 1~cm is enough, since the measurement of the momentum through the bending angle is done with the more accurate drift-tube stations. Sampling with a plane every 5~cm of iron sets less stringent requirements on the efficiency of each layer, $90\div 95$\% being enough. 

The Resistive Plate Chamber (RPC), originally developed by the group led by Santonico~\cite{bsant1,bsant2}, is employed in LHC experiments as well as in neutrino and cosmic ray physics experiments.

\begin{figure}[!h]
\centering
\includegraphics[width=10cm]{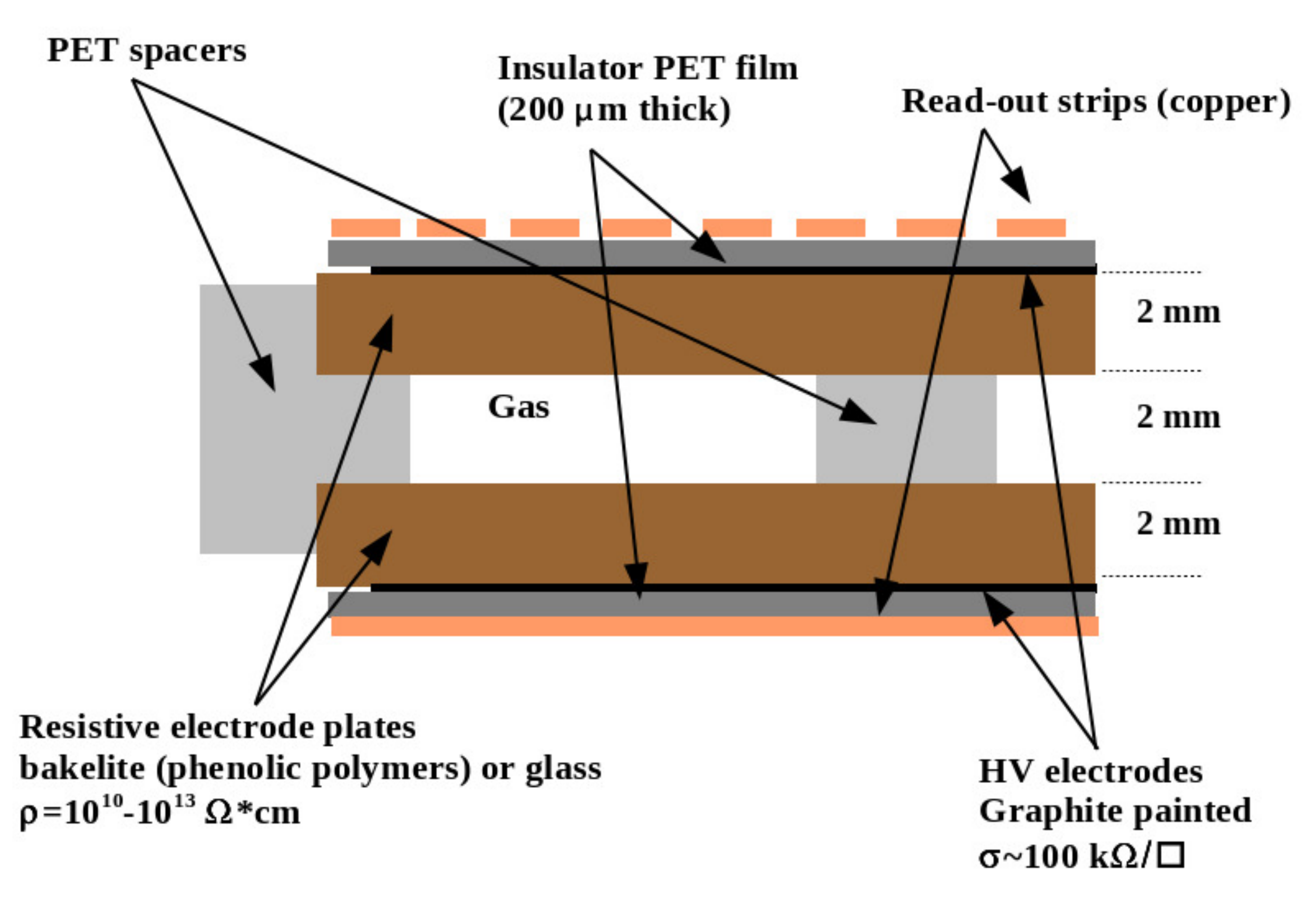}
\caption{Cross section of an RPC with the read-out electrodes.}
\label{fig:RPCsketch}
\end{figure}

A cross section of the detector is shown in Figure~\ref{fig:RPCsketch}, with its associated crossed read-out strips.
The electrodes are made of $\rm 2 \, mm$ thick High Pressure Plastic Laminate\footnote{The High Pressure Plastic Laminate is made by an overlay of kraft paper (with phenolic resin) and  layers of melaminic resins pressed at high temperature.} (HPL), commonly known as bakelite, with a volume resistivity in the range $\rm 5 \cdot 10^{10} \div 10^{13} \, \Omega \, cm$, painted on the external surface with graphite of high surface resistivity ($\rm \sigma \approx 100 \mbox{k} \Omega/\square$).
The graphite coating is insulated by $\rm 190 \, \mu$m PET layers. The internal surfaces of the gap are coated with linseed oil, in order to improve the smoothness on a microscopic scale.
The uniformity of the gas gap in ensured by a lattice of polycarbonate (LEXAN) spacers with $\rm 1 \, cm^2$ area and $\rm 10 \, cm$ pace.
The passage of a charged particle inside the gas gap produces a quenched (by the resistivity of the electrodes) spark, which induces signals on the read-out strips, typically $\rm 3 \, cm$  wide and made of copper. 
The gas gap of the RPCs is typically $\rm 2 \, mm$ thick. Gaps and electrodes of smaller thickness are recently being studied to increase the detector rate capability \cite{newRPC}.

SHiP RPCs will operate in a muon flux around $\rm 4 \, kHz/m^2$,  about 4 orders of magnitude lower than the flux sustained by LHC RPC systems, but one order of magnitude higher than 
typical counting rates at cosmic ray fluxes ($\rm 300 \, Hz/m^2$). 
At SHiP rates, streamer operated RPCs with low resistivity electrodes as well as avalanche operated RPCs can be used. 
The first option is preferable because in the latter case part of the electronics (preamaplifiers and discriminators embedded in the strip panels) would be located between the iron slabs of the magnet, thus constituting
a source of heat, and would not be accessible in case of failures during the experiment data taking.
For streamer operated RPCs, instead, given the high signal amplitude ($\rm \sim \, 100 \,mV$), the Front-End electronics can be located outside of the magnet, connected through flat cables to the read-out strips.

%=========================================================
% 
%Given the fact that OPERA RPC chambers are $\rm 2.9 \, m$ 
%large, in order to fit the $\rm 4 \, m$ lateral dimension of %the SHiP magnet a new production is anyway needed. 
%OPERA RPCs could be partially recuperated, especially those 
%with lower resistivity electrodes 
%($\rho$ $\sim$ $5 \times 10^{11}$ $\Omega$ cm).
%
%=========================================================

OPERA RPCs~\cite{OPERAcs} could be reused in SHiP. They have been operated in the CNGS beam from 2006 to 2012, flushed with a gas mixture $\rm Ar/C_2H_2F_4/i-C_4H_{10}/SF_6$ in the volume ratios $75.4/20.0/4.0/0.6$~\cite{OPERAgas}, suited for streamer working regime.
The operating voltage, around $\rm 5.7 \, kV$, was rescaled during the run according to the atmospheric pressure.

Typical measured currents for one RPC row (three chambers for a surface of about $\rm 9 \, m^2$) were $\rm \sim \, 900 \, nA$, corresponding to counting rates lower than $\rm 20 \, Hz/m^2$. Along the years, few RPC rows showed an increase of current (up to few $\rm \mu \,A$) and noise, without a significant degradation of the detector performance.

\begin{figure}[htb]
\begin{center}
\includegraphics[width=0.9\linewidth]{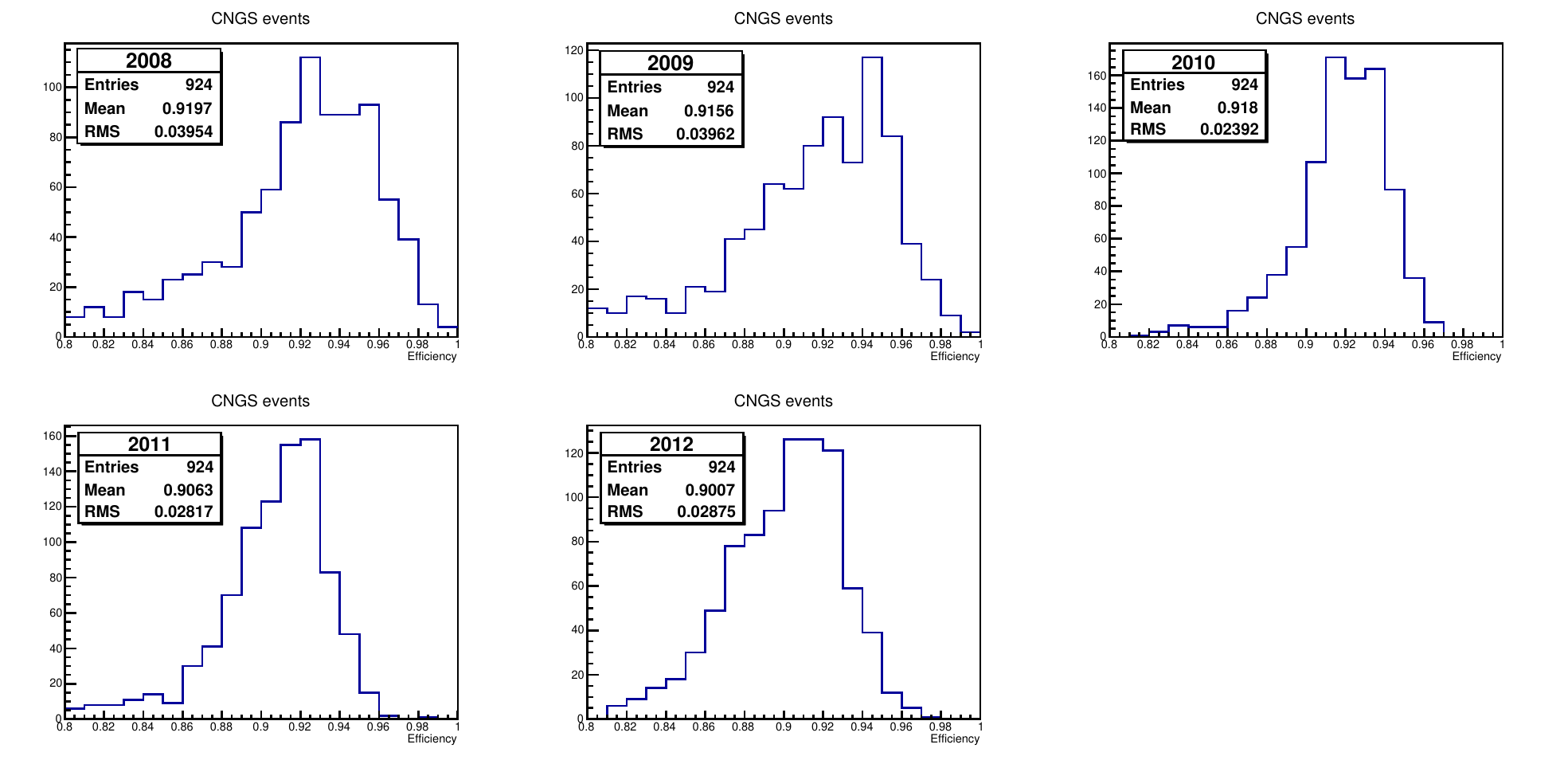}
\caption{Efficiency values of OPERA RPC chambers measured during the  CNGS run from 2008 to 2012.}
\label{fig:RPCeff}
\end{center}
\end{figure}

Efficiency values measured with muons produced by the interaction of CNGS neutrinos with the OPERA target are shown in Figure~\ref{fig:RPCeff} for each of the five CNGS runs from 2008 to 2012. The distributions show a peak above $90\%$, slightly decreasing along the years.
The tail at lower values is mostly due to RPC rows with high voltage cables in short-circuit, fixed at the end of the each year run.  Typical efficiency values measured with cosmic rays are about $4\%$ higher, as a consequence of the larger  inclination  and of the higher energy. 
The decrease of the average efficency with the years is due to the flushing with a dry mixture, an effect known to be reversible by flushing the detectors with a gas mixture added with water vapour~\cite{rpc_recover}. After the end of the CNGS run, the flushing of the chambers with pure Nitrogen at $\rm 30 \, \%$ relative humidity started aiming at recovering  the original resistivity value. Aging tests of RPC chambers dismounted from OPERA are foreseen. In addition, tests on RPC prototypes with electrodes of the same resistivity as OPERA ($> 5 \cdot 10^{11}$~$\Omega$cm), exposed to SHiP rates, will be performed in order to test their functionality. In case they would turn to be not adequate to the muon rates, new chambers with lower resistivity will be developed. 

Typical time resolution values around $\rm 3.5 \, ns$ have 
been achieved during CNGS runs.

%=========================================================
% 
%A typical \cite{OPERAQC} chain of quality control tests is 
%composed by:

%\begin{itemize}
%\item mechanical tests (verification of gas leakage rate and  %spacer gluing);
%\item measurement of electrical parameters in Argon (ohmic 
%current at low voltage and average electrode resistivity);
%\item efficiency and counting rate tests with the operating 
%gas mixture. 
%\end{itemize}

%All the RPCs installed in OPERA have been selected according %to the results of these tests. In Figure \ref{fig:RPCrho} the %average electrode resistivities are shown for the whole OPERA %RPC production. 

%\begin{figure}
%\centering
%\includegraphics[width=8cm]{nutaudet/PICTURES/rpc_rho.pdf}
%\caption{Average detector resisitivity measured on the whole %RPC production for OPERA.}
%\label{fig:RPCrho}
%\end{figure}
%
%=========================================================

The layout of the magnet inner trackers for the SHiP neutrino detector will be the similar to the one adopted in OPERA.
Planes of RPCs will be placed in the 22 gaps between the iron slabs, each 2 cm wide, with 8 m height and 4 m width.
The RPC system total area will be thus $\rm 704 \, m^2$.

\begin{figure}
\centering
\includegraphics[width=10cm]{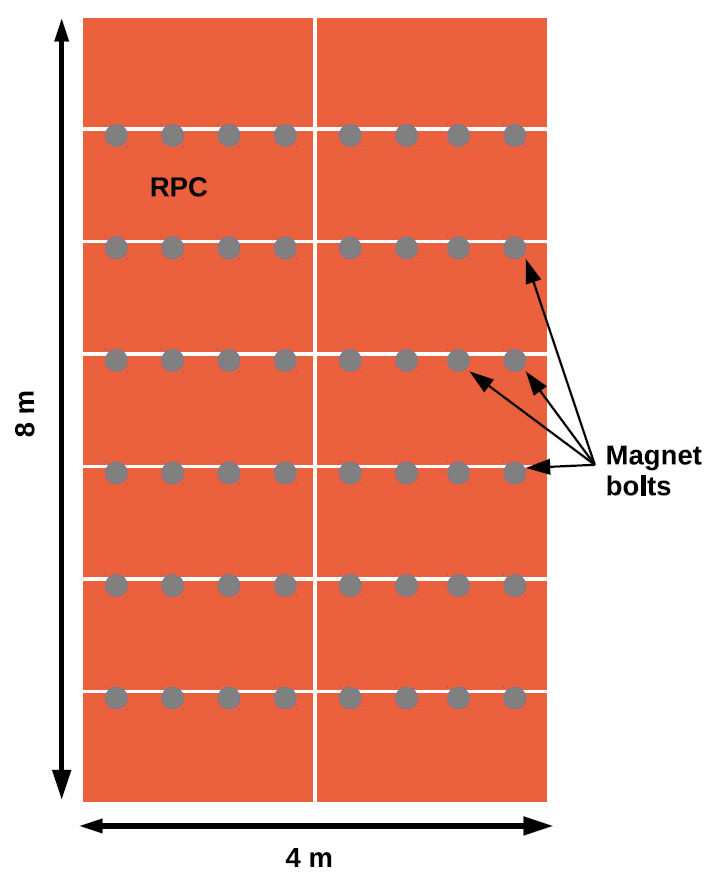}
\caption{Sketch of one RPC layer inside the spectrometer magnet.}
\label{fig:layout_sketch}
\end{figure}

According to its design, the stability of the magnet is ensured with 48 bolts per layer. Therefore, in order to maximize the sensitive instrumented area, 14 RPC chambers, each of area ($\rm 2 \times 1.1$) m$^2$,  will be arranged as in Figure~\ref{fig:layout_sketch}. The lowest RPC row will be placed on the magnet basement, while the bolt rows will act as the support for the other RPC detectors. A total of 308 chambers will be needed.

The RPC chambers will be read out by means of orthogonal strip panels. The horizontal strips will be 4 m long. Each RPC row is served by 32 strips with 3.5 cm pitch.
The vertical strips will be 8 m long, especially designed to turn around the bolts. For these strips, a lower pitch is advisable in order to reconstruct the bending coordinate. 
Given the length, the read-out strip panels will be made of a 3 mm thick plastic base, with copper strips glued on the side facing the RPCs and a copper layer on the other side.
The Front-End electronics, located outside of the iron walls, will be conneted to the strips through flat cables.
The residual space, about 7 mm inside the 2 cm gaps, will be filled by polyester fiber, widely used in the building industry for acoustic and thermal insulation.
Polyester fibers are CL1 according to CSE RF 2/75/A~\cite{buni1} and CSE RF 3/77~\cite{buni2}, showing first class reaction to fire (inflammable) and producing atoxic smoke while burning. Moreover they are scentless, atoxic, recyclable and not pulverulent.

%&%&%&%&

\subsubsection{Gas system and high voltage}
\label{rpc_gsystem}
The RPC chambers we intend to use have a total gas volume of $1.4 \mbox{ m}^3$, corresponding to a flux of about 300 l/h, assuming 5 refills/day. Typical mixtures for streamer operated RPCs are presently composed by Argon ($\rm Ar$), Tetrafluoroethane ($\rm C_2H_2F_4$), isobutane ($\rm i-C_4H_{10}$) and sulphur hexafluoride ($\rm SF_6$). The last three are in liquid phase under pressure inside the bottles, while the first one is a noble gas at normal temperatures, but it can be kept in liquid phase inside one cryostat to reduce the volume.
Isobutane is a flamable gas, limiting its volume percentage in the gas mixture. To preserve the bakelite resistivity,  RPCs will be flushed with a humified gas mixture.

The gas mixture will be chosen in compliance with  CERN safety rules. Should $\rm C_2H_2F_4$ and $\rm SF_6$ be banned in CERN experiments beacuse of their ozone depleting potential, an R\&D effort will be dedicated to the search for a new suitable gas mixture. We could profit of tests aready performed for LHC experiments~\cite{rpc_newgas}.

Pressure regulators will be used for lowering the pressure of the gas lines at the input of the mixing unit to about $\rm 1 - 1.5$ bar. The gas flows will be metered and regulated by means of mass flow controllers, with an absolute precision of the order of $1 \%$. The measured flows will be monitored by a computer. 
The gas consumption can be controlled registering the weight and output pressure of the bottles.
The quality of the gas mixture can be ensured during the operation by monitoring the cluster size \cite{OPERAcs}. 

The RPC system is composed by 308 chambers, supplied with gas 
in groups of two serially connected, resulting therefore in 
154 independent channels.
Each channel will be equipped with a ball valve, for closing the distribution line in case of gas leakages, a needle, in order to make sure of the impedance uniformity among different channels, and a safety bubbler for protecting the chambers against overpressure.
Another bubbler will be present at the output of the two chambers in order to check the gas flowing in the channel.
Profiting of the experience of ATLAS and CMS, a closed loop can be used to flush the chambers.

One RPC layer is made of 14 RPCs, arranged in 2 columns and 7 rows. OPERA RPCs were operated in streamer mode at about 6 kV with a gas mixture composed of Argon, Tetrafluoroethane, isobutane and sulphur hexafluoride. As in OPERA, it is convenient to split the voltage across the gas gap between
the negative charged cathode and the positive charged anode.
In this way, the voltage on each power line will not exceed 5~kV, depending on the final choice of the gas mixture, with a consequent simplification in the choice of  cables and connectors.

\begin{figure}
\centering
\includegraphics[width=10cm]{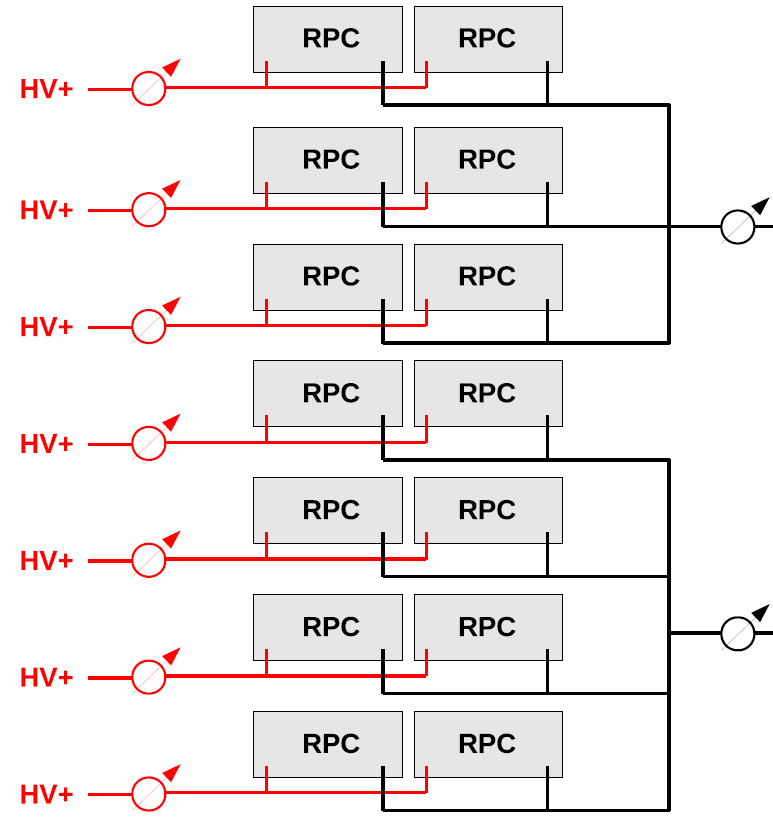}
\caption{High voltage distribution scheme for one RPC layer.}
\label{fig:HVsketch}
\end{figure}

A sketch of the high voltage distribution for one RPC layer is shown in Figure~\ref{fig:HVsketch}.
The basic unit of the system, as for the gas distribution system, is a row made of two RPCs.
It is not necessary to have a similar granularity in the two polarities, and one layer is served by two negative and 7 positive channels, with embedded ampere-meters.
The operating currents are indeed important parameters to monitor the performance and the aging of the RPC system.
Assuming the charge relased in the gas at the passage of one muon is of the order of 1 nC, a current lower than $\rm 20 \, \mu A$ is expected in each row for a muon rate of $\rm 4 \,  kHz/m^2$.  

\subsubsection{Front-End electronics and test facility}
\label{rpc_tdaq}  
The basic element of the read-out chain is the Front-End Board.
Each FEB has 64 input channels and manages both positive and negative signals. 

The input RPC signals are discriminated by means of 4 16-channel LVDS receivers acting as comparators with Ethernet programmable thresholds. Once discriminated, the signals are 
handled by a Field-Programmable Gate Array (FPGA) that stores the arrival times with a $\rm 5 \, \div \, 10 \, ns$ resolution time stamp. Data are stored in a 4096-sample circular buffer with a clock frequency of 100 MHz. Each FEB can generate two output fast-OR signals of 100 ns length for triggering purposes. A mask for the exclusion of noisy channels can be set. 
The firmware of the FPGAs can be updated remotely via Ethernet. Data are read by the DAQ system through an Ethernet interface implementing the UDP protocol. 

One RPC layer is served by 6 FEBs.
Up to 19 FEBs are hosted inside a single 6U Eurocard crate, together with a Controller Board (CB). 
The design of the CB is also based on an FPGA.
The CB has several tasks such as power supply management, slow control, control signal distribution, masking, FAST-OR collection and management. 

A test facility, housing up to 12 RPCs, is operating at Frascati INFN laboratories. This facility is intended for testing both the RPC chambers possibly available from the OPERA detector decommissioning and the new chambers that will be produced. The chambers are powered by CAEN SY127 units.
Up to 4 RPCs have the negatively charged electrode connected to the same channel, while the positive electrodes are served by one dedicated HV channel for each RPC, the measurement of the current being performed at 50 nA precision level.  
The read-out of the signals induced by the passage of cosmic rays is performed by means of 3 cm wide read-out copper strips, connected through 6 m long twisted flat cables to the FEBs described in Section~\ref{rpc_tdaq}. 

A batch of spare OPERA RPC chambers has been already tested, flushed both with pure Argon (to measure the resistivity from the slope of the current versus voltage characteristic above 2 kV) and with the OPERA gas mixture. 

\begin{figure}
\centering
\includegraphics[width=1.0\linewidth]{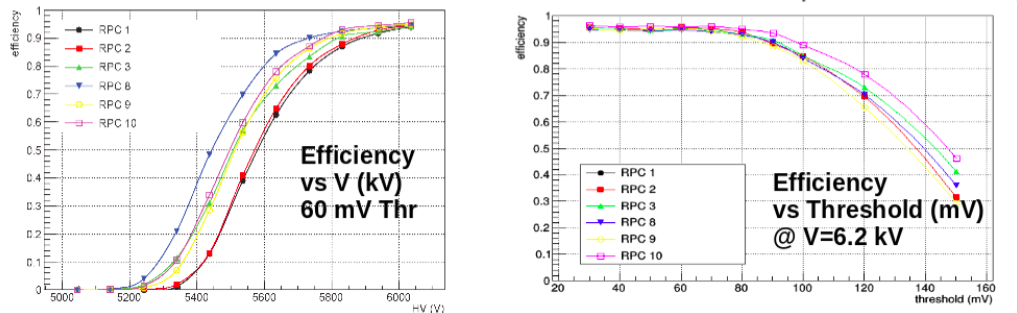}
\caption{RPC efficiency as a function of the operating voltage (left)
and of the discrimination threshold (right) at 6.2 kV operating voltage.}
\label{fig:RPC_test_efficiency}
\end{figure}

The efficiency is measured triggering cosmic rays with the coincidence of two chambers, whose efficiency is measured in a separate run using two other RPCs as trigger.
In the left plot of Figure~\ref{fig:RPC_test_efficiency}, the efficiencies measured as a function of the operating voltage are shown for an applied threshold of 60 mV. The right plot shows the 
efficiencies as a function of the threshold at 6.2 kV fixed operating voltage.  

%High statistics (of the order of $10^6$ tracks) efficiency maps have also been acquired at 6.2 kV operating voltage and 60 mV discrimination  threshold. 
%An example is shown in Figure~\ref{fig:RPC_efficiency_map}: the 10 cm pace spacer lattice is clearly visible, as well as the curved profile of the upper edge of the chamber.

%\begin{figure}
%\centering
%\includegraphics[width=7cm]{nutaudet/PICTURES/RPC_efficiency_map.pdf}
%%discrimination threshold of 60 mV. 
%The map has been obtained with a statistics of about $10^6$ reconstructed tracks.}
%\label{fig:RPC_efficiency_map}
%\end{figure}

	\subsection{Drift-tube tracking system} 

\label{sec:drift_tubes}

\begin{figure}
  \begin{center}
    \includegraphics[width=.45\textwidth]{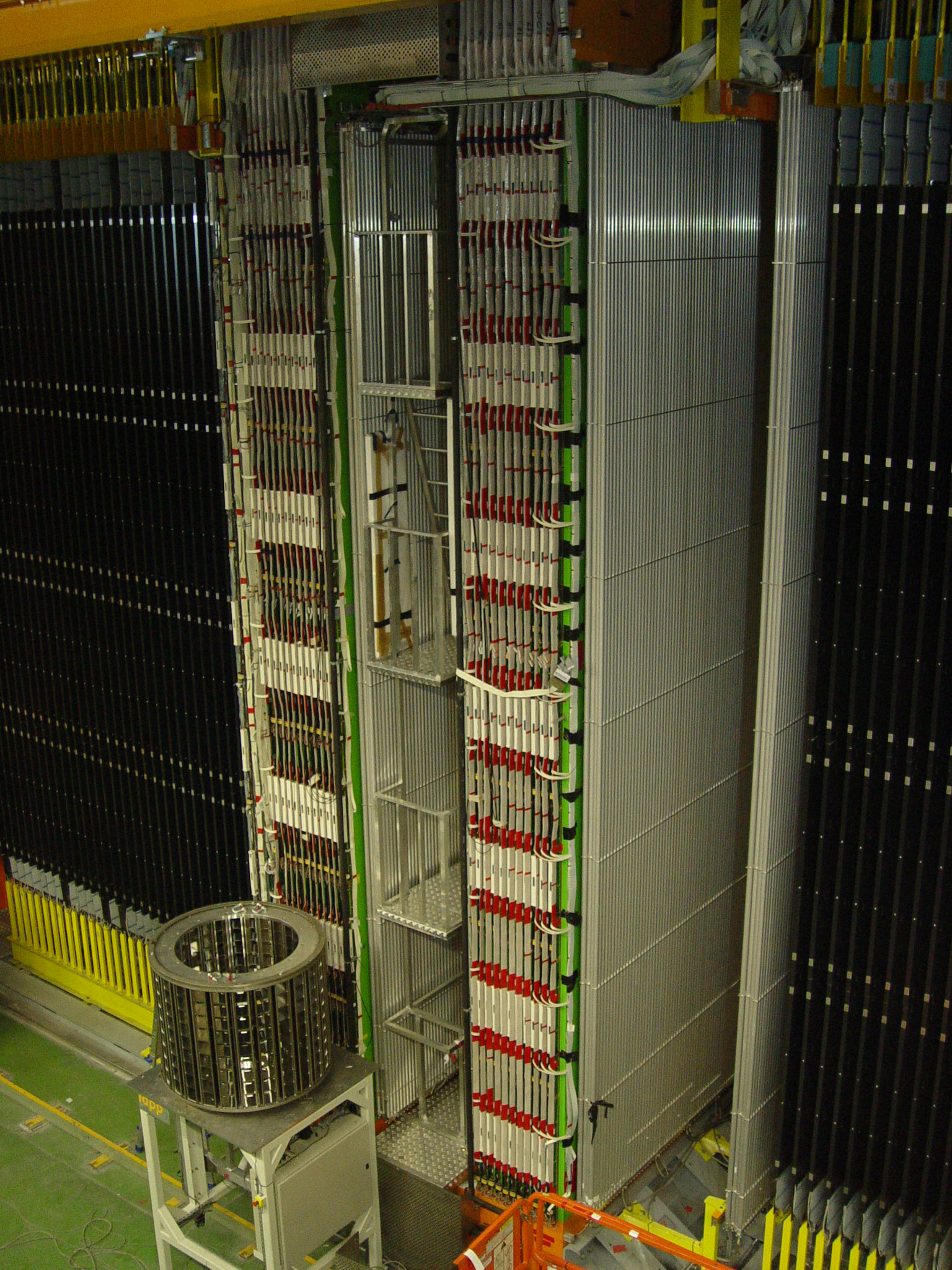}%
    \hfill%
    \includegraphics[width=.45\textwidth]{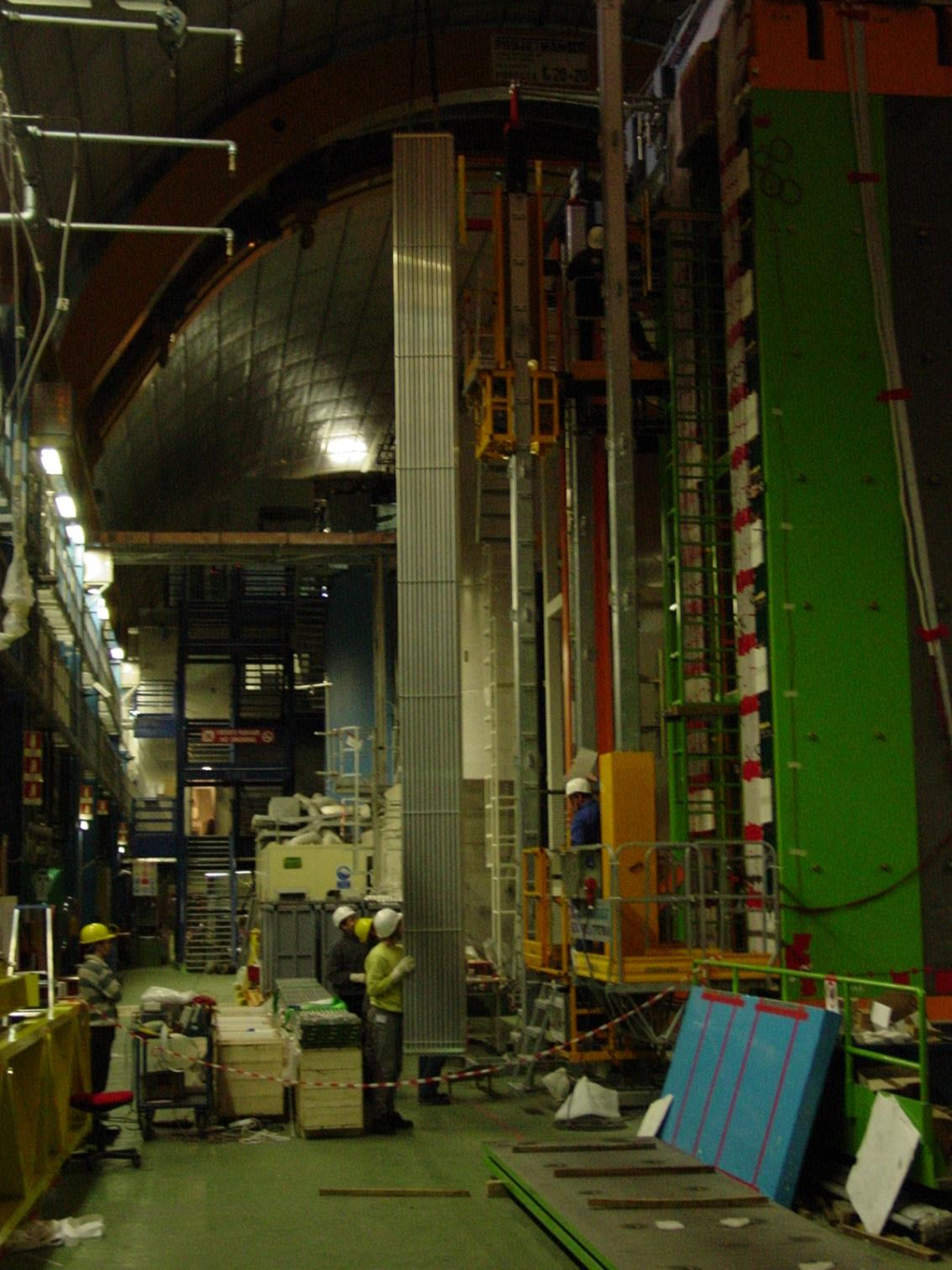}
  \end{center}  
  \caption{(Left) Photograph of the current setup of the drift-tubes in the spectrometer of OPERA. (Right) Installation of a drift-tube module in OPERA}
  \label{fig:dtopera}
\end{figure}

\begin{figure}
  \begin{center}
      \includegraphics[width=0.45\textwidth]{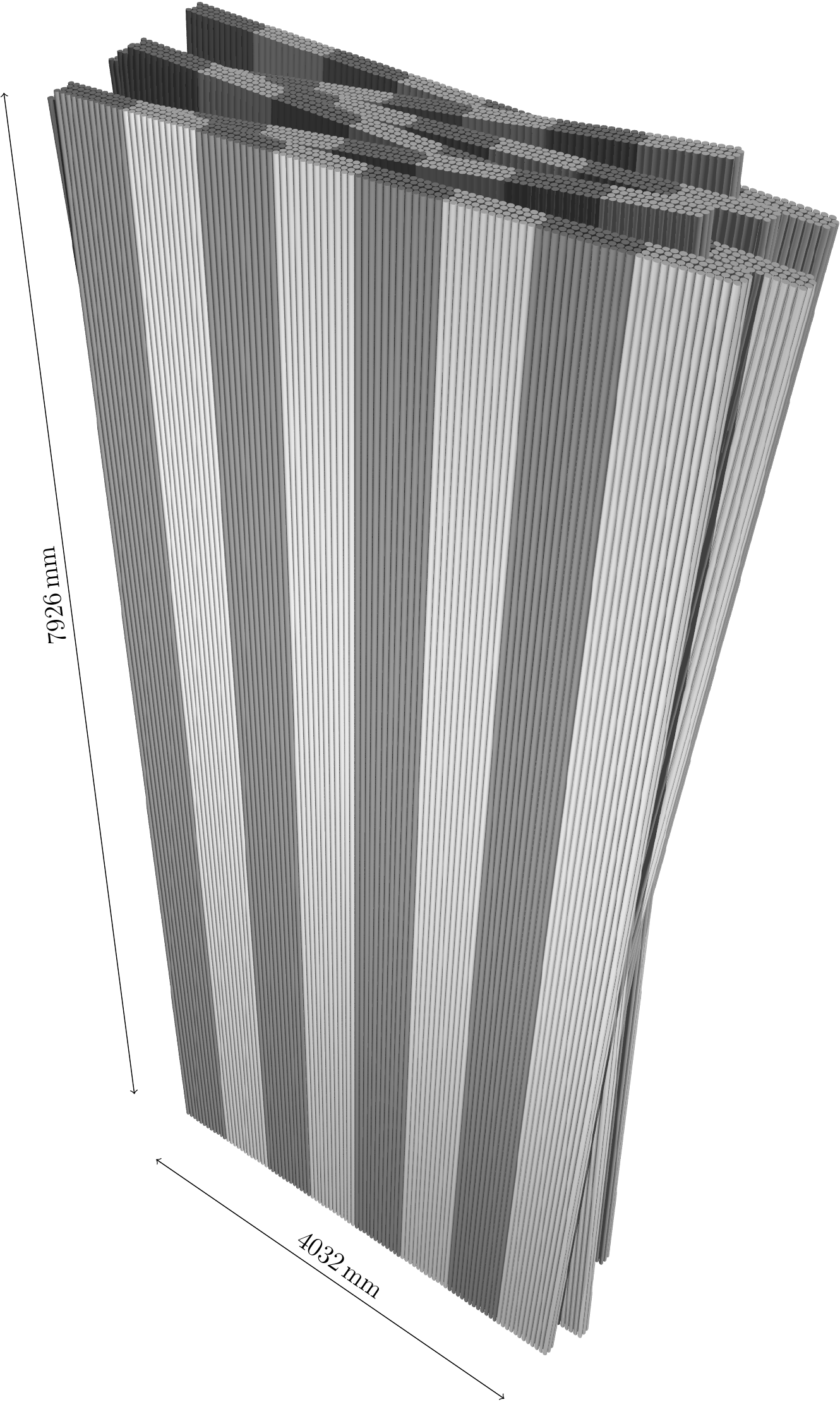}      
      \includegraphics[width=0.45\textwidth]{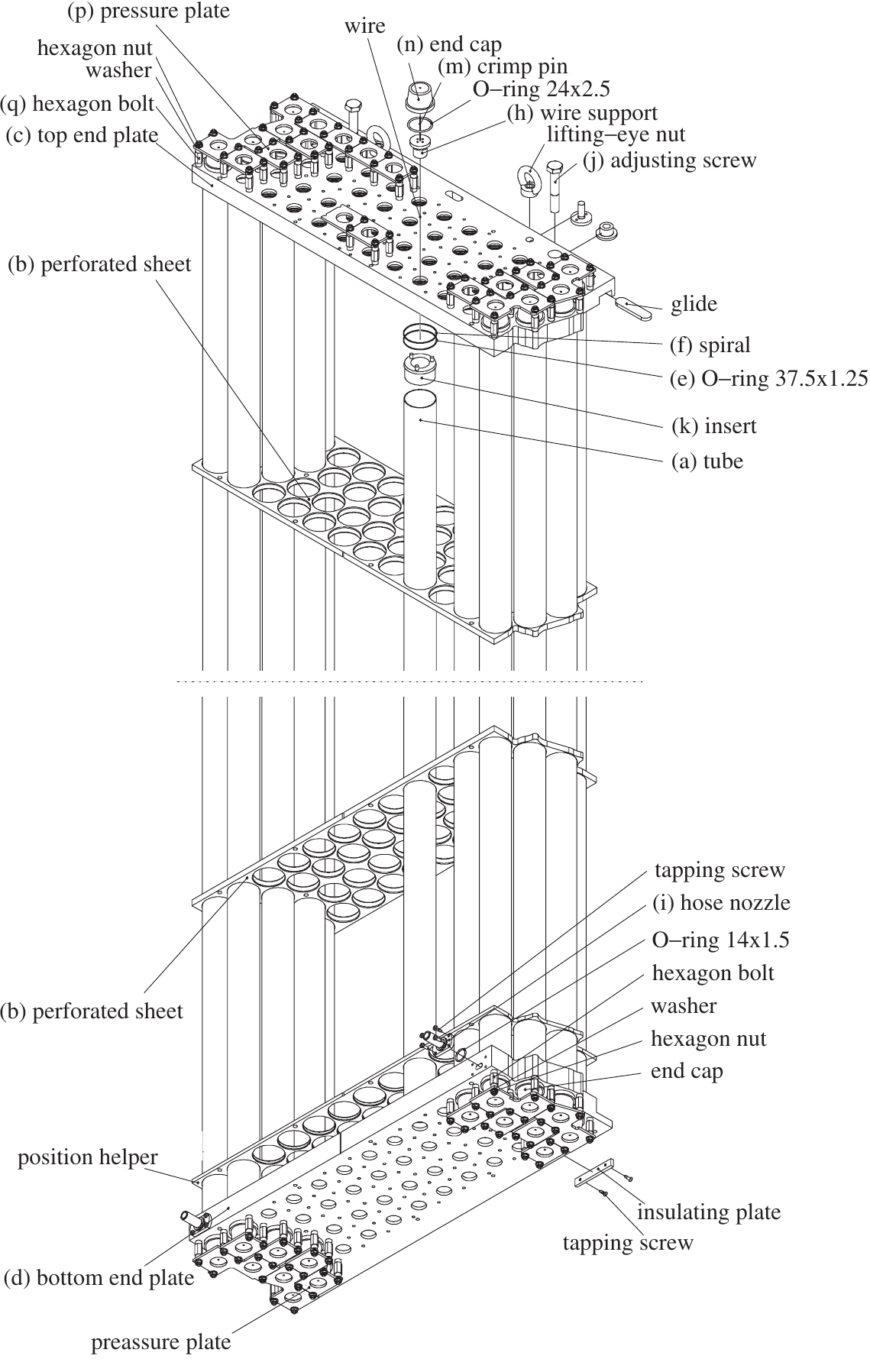}
  \end{center}
  \caption{(Left) Schematic draft of the drift-tube planes in front of the first magnet arm. Each of the six planes is made of eight modules (depicted in dark and light gray). The six planes are arranged in three projections, of which the second and third are rotated by a stereo angle of $\pm3.6^\circ$. (Right) Exploded view of a drift-tube module}
\label{fig:hptstereo}
\end{figure}

\begin{figure}
  \begin{center}
    \includegraphics[width=.8\textwidth]{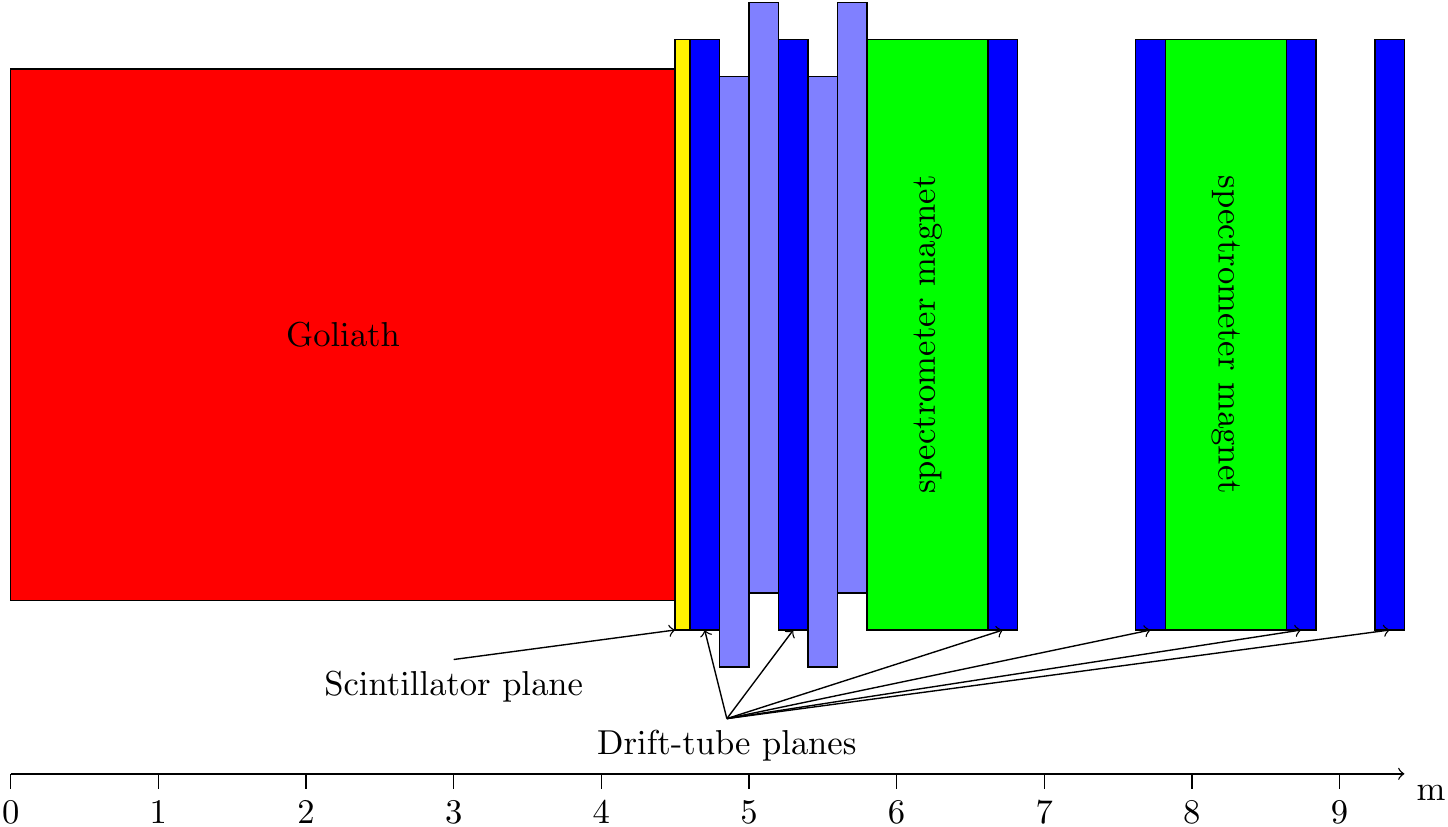}
  \end{center}
  \caption{Top view of the drift-tube setup within the spectrometer. The drift-tube planes shown in light blue will be rotated by a stereo angle of $\pm3.6^\circ$}
  \label{fig:dttopview}
\end{figure}

For a precise tracking of muons, the drift-tubes formerly used in the OPERA experiment and shown in Figure~\ref{fig:dtopera} will be used as part of the muon spectrometer. The main purpose of the drift-tubes is a precise measurement of the charge and momentum of muons emerging from neutrino interactions in the emulsion target. %
% Furthermore, muons from neutrino interactions that cross the pillars of the Goliath magnet should be fully reconstructed. These muons often have low energies when they reach the spectrometer %and thus do not completely cross the magnetic spectrometer. However, 
%and are thus absorbed in the magnet arm of the spectrometer. In order to still be able to determine the charge and momentum of these muons, the pillars of Goliath will be used as an additional part of the spectrometer. %
In order to maximise the muon identification efficiency, the iron yoke of the Goliath magnet is also used as a magnetized muon filter. Passing through muons are detected by the drift-tube setup between the Goliath magnet and the muon spectrometer magnet. %
In order to connect these muon tracks with the target, a three dimensional reconstruction will be done. This will also help to deal with a possible high occupancy in the drift-tubes from high multiplicity events.
%Furthermore, 16\% of the muons created in $\nu_\mu$-CC-charm interactions 
To achieve the requirement of a three dimensional reconstruction of the muon tracks, stereo planes of drift-tubes will be added between the Goliath magnet and the first magnet arm of the spectrometer, rotated around the beam axis by a stereo angle of $\pm3.6\,^\circ$. This angle is optimized for the reuse of the existing OPERA modules taking into account a high acceptance and low occupancy of the drift-tubes, without exceeding the lateral width where a significantly higher rate of background muons from the beam target is expected. %the width of the detector setup. %
A schematic drawing of the drift-tube planes in front of the first magnet arm is depicted in Figure~\ref{fig:hptstereo}. For the charge and momentum measurement by the spectrometer, two additional planes will be installed each inside and behind the magnet, so that there will be a total of ten planes. A schematic top view of the setup is shown in Figure~\ref{fig:dttopview}.

% . Moreover one should give an answer to the questions:, efficiency,    
%misidentification

For the use in SHiP, no mechanical modifications of the OPERA modules are needed. Each module consists of 48 aluminum tubes staggered in four layers. Each drift-tube is 7926\,mm long and has an outer diameter of 38~mm and a 0.85~mm wall thickness. A 45~$\mu$m gold-plated tungsten wire in the center serves as sense wire. Each module has a system width of 504~mm. The end-plates of a module have a depth of 190~mm at the top and 186~mm at the bottom. An additional space of approximately 20~mm will be needed between the planes for suspension. A detailed description can be found in Refs.~\cite{Zimmermann:2006xr} and~\cite{Ferber:2008zz}. 
%A schematic view of a drift-tube module can be seen in Figure~\ref{fig:dtmodule}. 
The muon spectrometer has ten stations consisting of eight modules each, leading to a total number of 3840 drift-tubes.

%\begin{figure}
%  \begin{center}
%    \includegraphics[width=.5\textwidth]{dt_explosion}
%  \end{center}
%  \caption{Schematic view of a drift-tube module}
%  \label{fig:dtmodule}
%\end{figure}

The drift-tubes were operated with a gas-mixture of Ar/CO$_2$ at a ratio of 80/20. This mixture yields maximal drift-times of approximately 1300\,ns. For the use in SHiP, an Ar/CO$_2$/N$_2$ mixture is foreseen, which will decrease the drift-times, allowing for faster collection times.

The signal attenuation  is  well described by an exponential function with an attenuation length of 32.1\,m over the entire length of a drift-tube. Thus single sided readout is sufficient. 

All electronics will be redesigned for the use in SHiP. The major difference will be a continuous triggerless read-out of the drift-tubes. For each hit, the drift-time and the time over threshold will be measured and sent to the central DAQ. The measurement of the time over threshold has proven to be a useful method for rejecting electronic noise and hits from X-rays. For a high precision track reconstruction, the drift times have to be known with a precision of 2\,ns rms. An event timestamp will be provided by the target tracker. To obtain a reliable redundant event timestamp, in addition a plane of scintillator tiles should be placed immediately after the Goliath magnet in front of the first drift-tube plane. 

\begin{figure}
  \begin{center}
    \includegraphics[width=.5\textwidth]{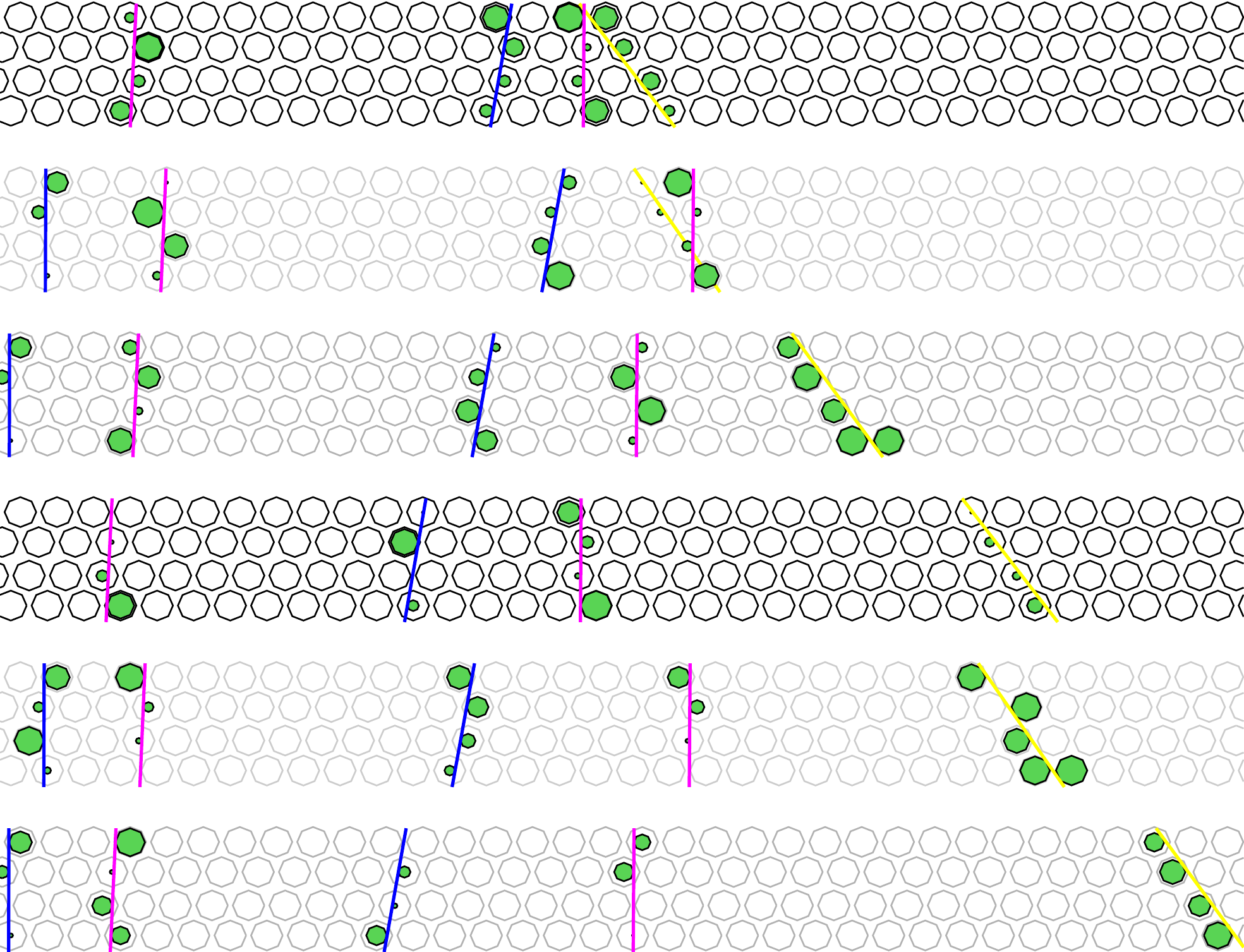}
  \end{center}
  \caption{Display of a simulated $\nu_\mu^\mathrm{CC}$-charm event in the first six drift-tube planes. The different shades of gray denote the three projections of the rotated planes. Pink tracks depict muons, yellow tracks electrons and blue tracks hadrons.}
  \label{fig:dtevent}
\end{figure}

A spatial resolution of 255~$\mu$m was achieved in OPERA after alignment.  
The momentum resolution of the muon spectrometer up to 20~GeV/c was thus entirely dominated by multiple scattering and it was measured to be $\Delta p/p =20$\%.
In order to calculate the reconstruction efficiency of the drift-tube setup after the Goliath magnet, neutrino events have been simulated in the emulsion target using Geant4. For a first test, all particle tracks from this simulation were used for a purely geometrical simulation of the drift-tube planes. Figure~\ref{fig:dtevent} shows the event display of tracks in the drift-tubes from a neutrino interaction simulated in the neutrino target.
The proposed setup yields an overall acceptance of 99.7\% for muon tracks emerging form the neutrino target. This value is reduced to 97.3\% for muons produced in charged-current neutrino interaction with the production of charmed hadrons ($\nu_\mu^\mathrm{CC}$-charm), due to the high occupancy in some cases. While 71\% of  muons from $\nu_\mu^\mathrm{CC}$-charm interactions reach the drift-tubes directly, 16\% cross the pillars of Goliath before they reach the first drift-tube plane. Simulations show, that a fully three dimensional reconstruction of these muons is possible in 99.0\% of the events.

\clearpage

%% file: vessel/Vessel.tex
\section{HS vacuum vessel}
\label{sec:vessel}

The discovery potential of the SHiP experiment depends critically on the background rejection
capability. To achieve the required level of background rejection, the decay volume has to be evacuated at
the level of $10^{-6}$ bar, to suppress the neutrino and muon interactions with the residual air,
and be instrumented with a highly efficient background event tagger. The background tagger is therefore designed
as an integral part of the vacuum vessel surrounding the decay volume. 
 
Based on its cost effectiveness, hermeticity, and high registration efficiency for the relativistic
charged particles and neutrons, a liquid scintillator is chosen as an active detection media of
the background event tagger. The vacuum vessel has a double-walled structure. The space between
the double walls of the vessel is filled with liquid scintillator, surrounding the decay volume 
with almost a full solid angle coverage. A detailed description of the liquid scintillator surround 
background tagger is given in Section~\ref{sec:taggers}.

\subsection{Design and construction}
\label{sec:vacuumvesseldesign}

%\vfill
\begin{figure}[tp]
\centering
\includegraphics[width=0.7\linewidth]{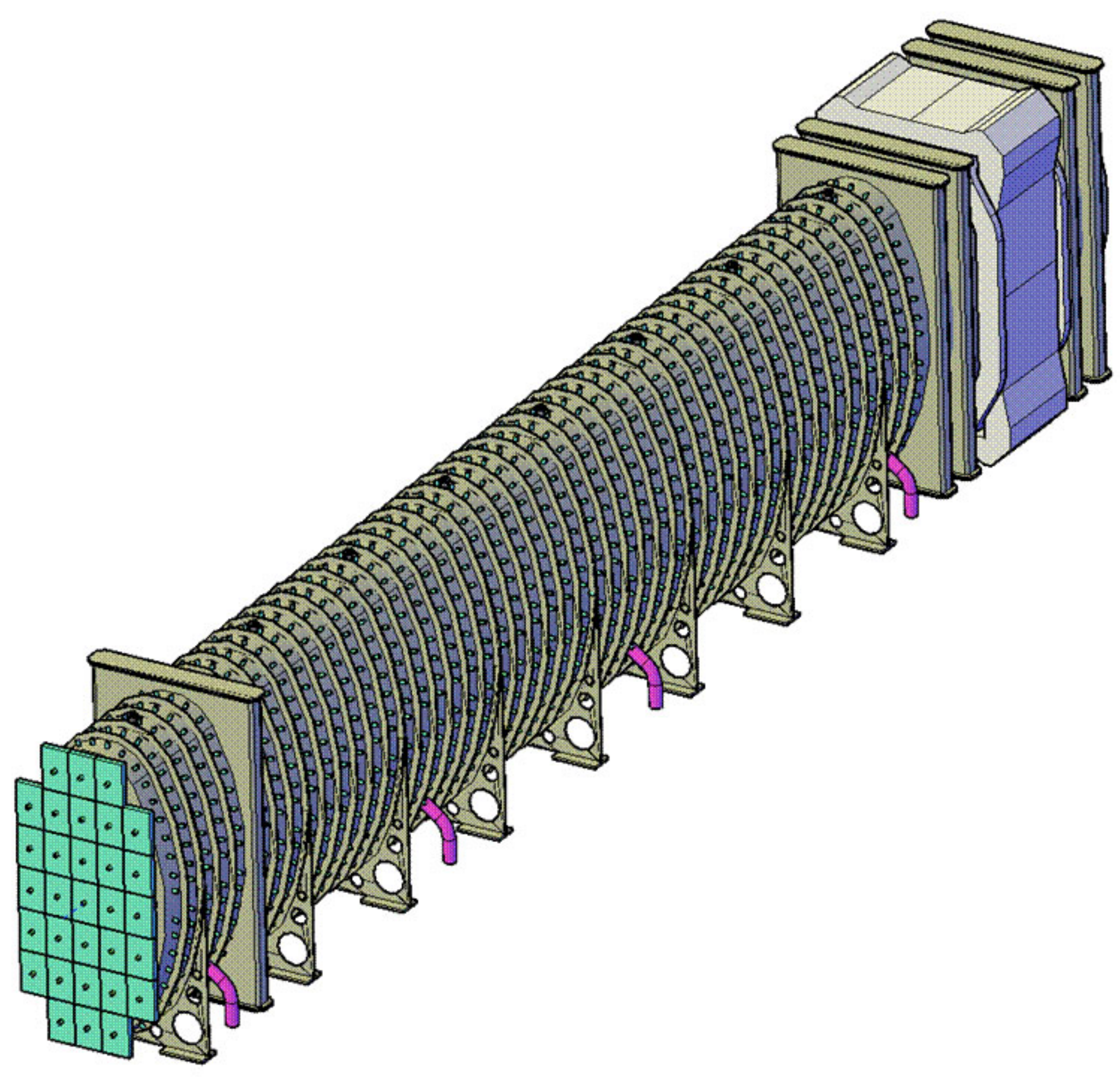}
\caption{A general layout of the baseline vacuum vessel design.}\label{vessel:VV3Dmodel}
\end{figure}%

A general layout of the baseline design of the vacuum vessel is shown in Figure~\ref{vessel:VV3Dmodel}. 
The vessel has an elliptical cross section with 10 m vertical and 5 m horizontal inner dimensions
(major and minor axis). The length of the vessel's elliptical tube is 62 m. It consists of a 50 m
long decay volume followed by a 12 m long magnetic spectrometer. The cross section of the elliptical vacuum
vessel, with the major axis oriented vertically, is chosen to avoid a large muon
flux deflected horizontally by the active muon shield to $>\pm$ 2.5 m from the beam axis at the
entrance of the vessel, yet maximizing the geometrical acceptance.

Since the vacuum vessel has to withstand an outside pressure of 1 bar, it has
a $\sim$3 cm thick strong wall with welded transverse and longitudinal reinforcing ribs. The ribs divide
the liquid scintillator volume between strong and external thin walls into individual cells. 
The pitch of the transverse ribs is 1 m. The longitudinal ribs are separated by approximately 1.5 m
distance. The exact thickness of the walls, the ribs size and spacing are being obtained by a finite
element analysis of the structure, taking into account the technological and budgetary considerations. 
Initial parameters were chosen using an approximate empirical criterion of the buckling stability of
vacuum vessel's structure. For the thickness of the vessel wall, the following equation was used:        

   \begin{equation}
  s = 0.47 \times  \frac{D}{100} \times \left\{ \frac{p}{10^{-6}\times E}\times \frac{L}{D}\right\}^{0.4} + c \hspace{0.5 cm} ;
    \label{stab_eqn}
   \end{equation}
   \vspace{0.3cm}
   
\noindent
where: $\it s$ - is the vessels wall thickness, in mm; $\it D$ - is the vessel's inner diameter, mm;
$\it p$ - external pressure, in MPa; $\it E$ - Young's modulus; $\it L$ - distance between wall edges, mm;
$\it c$ - corrosion margin, mm. 

The elliptical shape of the vessel's wall is approximated by four cylindrical surfaces: top 
and bottom of 1.6 m inner radius, and side of 8.5 m inner radius respectively. For the
elliptical shape chosen, the empirical structure buckling stability criteria (\ref{stab_eqn}) gives
the wall thickness $\it s$ = 19.5 mm, for a 8.5 m inner radius shell. With a safety margin factor
of 1.5, that $\it s$ value was used in the initial parameters set for the vessel's finite element 
analysis model. 

%\vfill
\begin{figure}[tp]
\centering
\includegraphics[height=10cm]{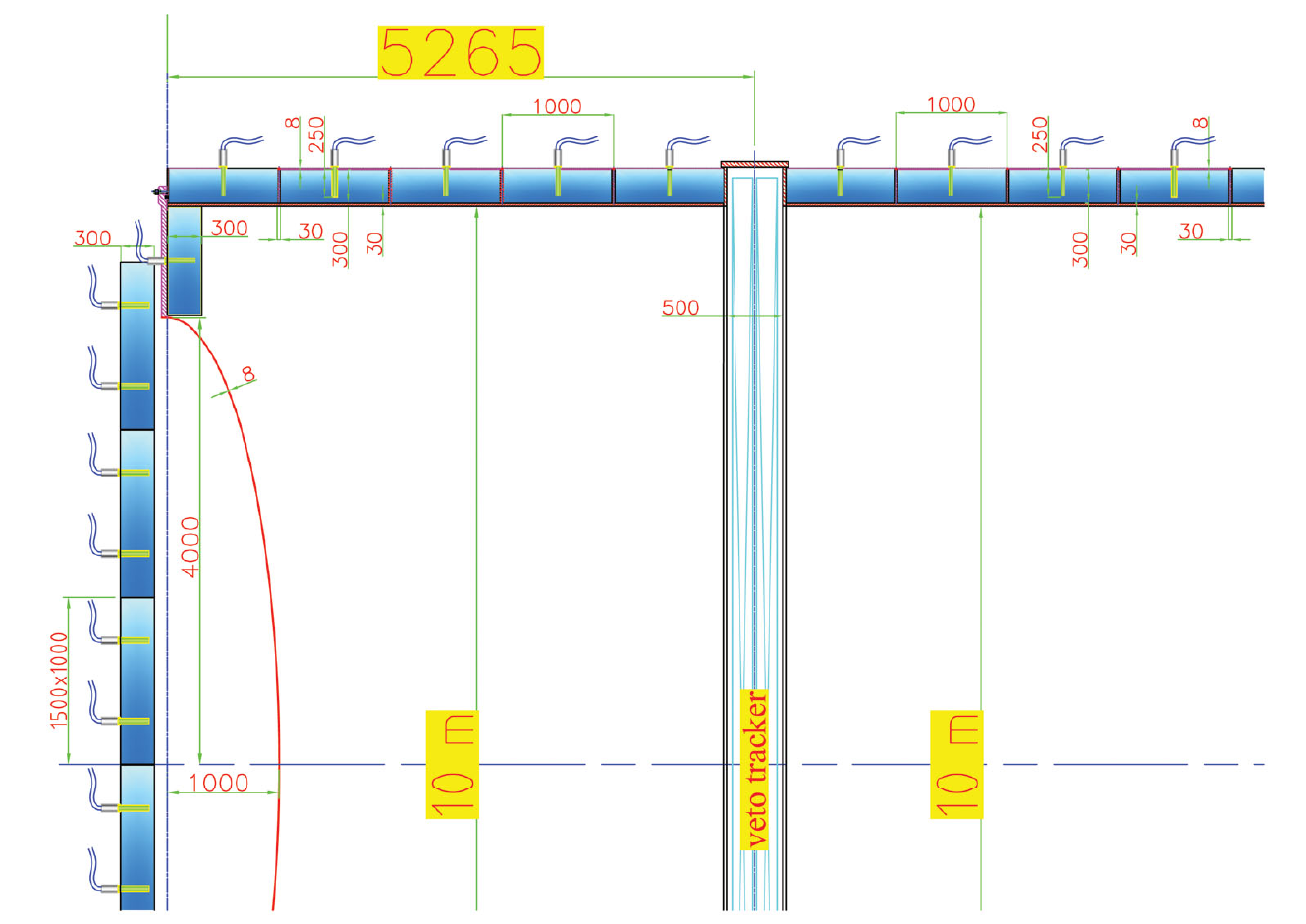}
\caption{A longitudinal cut view of the vacuum vessel forward part.}\label{vessel:VVForward}
\end{figure}%
%\

%\vfill
\begin{figure}[tp]
\centering
\includegraphics[height=10cm]{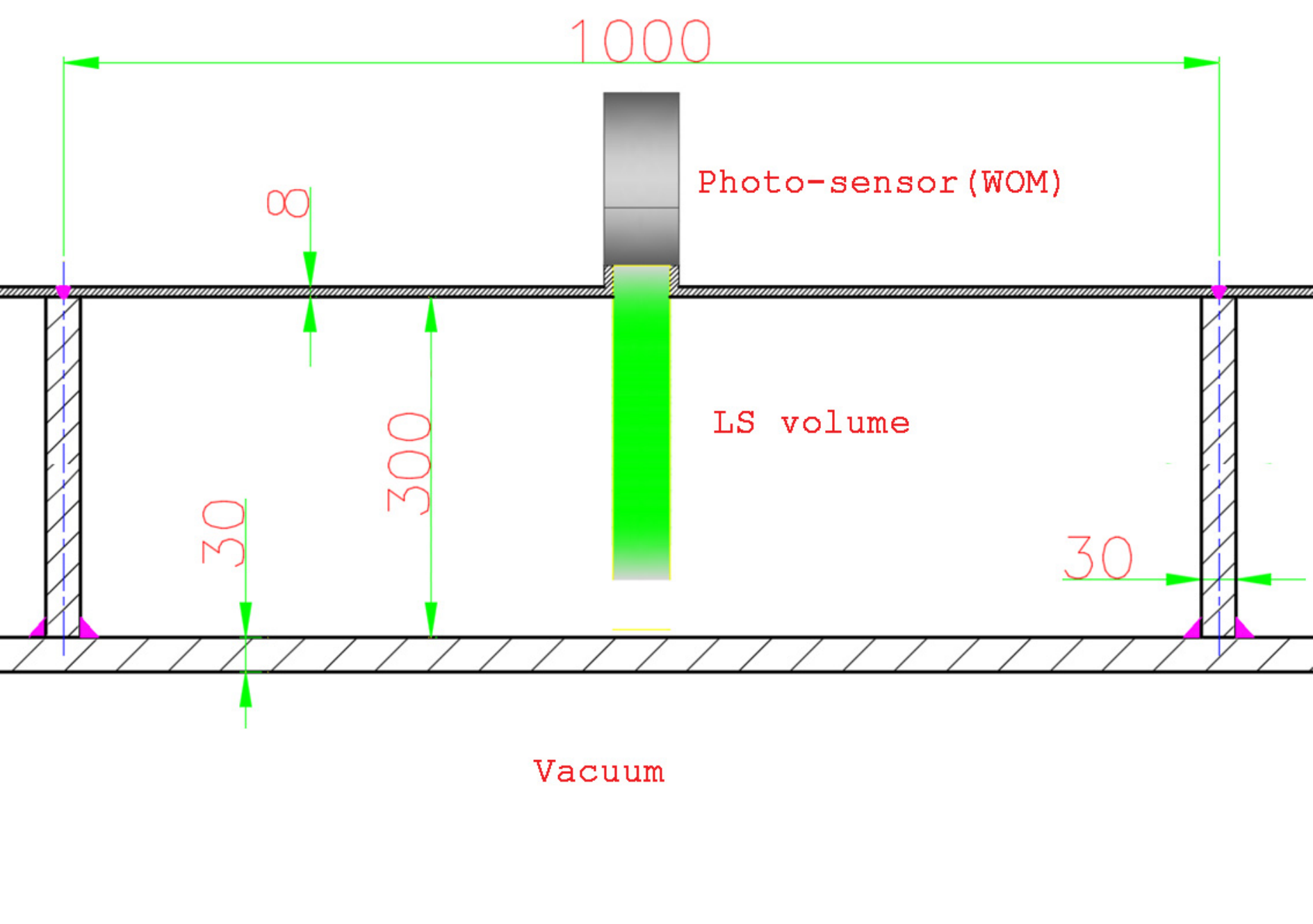}
\caption{ A longitudinal cut view of the vacuum vessel structure.}\label{vessel:VVSide_1}
\end{figure}%
%

%\vfill
\begin{figure}[tp]
\centering
\includegraphics[height=10cm]{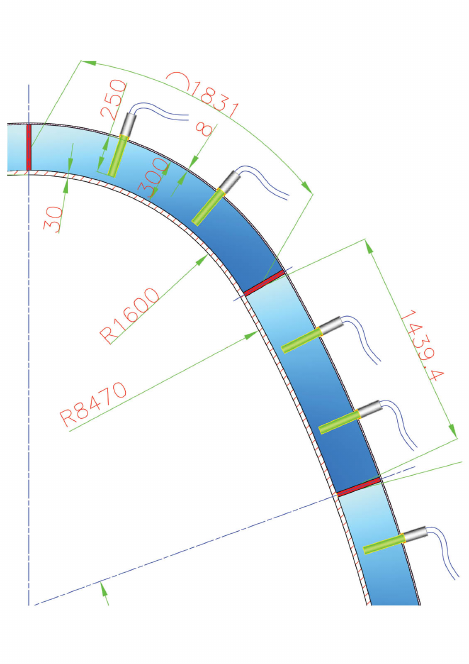}
\caption{ A transverse cut view of the vacuum vessel structure.}\label{vessel:VVSide_2}
\end{figure}%

The end lids of the vacuum vessel are closed by an ellipsoid shaped 8~mm thick fiber reinforced
aluminum membrane (Figure~\ref{vessel:VVForward}). The outer 8~mm aluminum shell seals the liquid 
scintillator volume outside the vessel's tube (Figure~\ref{vessel:VVSide_1}). This shell is made 
by welding large aluminum sheets together. The shell is attached to the transverse and longitudinal 
reinforcement ribs of the strong inner shell of the vessel (Figure~\ref{vessel:VVSide_2}). 
The strong flange, which holds the perimeter of the membrane is also welded to the strong shell. 
Optionally, the entrance window may be instrumented with liquid scintillator. In this case, the 
front background tagger consists of a standalone wall made of thin-walled aluminum cells filled
with 300 mm of liquid scintillator (shown upstream of the vacuum vessel in Figure~\ref{vessel:VV3Dmodel}), 
together with an additional 300 mm thick liquid scintillator volume right behind the strong front 
flange to tag interactions in the flange (Figure~\ref{vessel:VVForward}). 

The vacuum vessel is also crossed by five large prismatic volumes to house the tracking detectors. All
these volumes have 10 m $\times$ 5 m elliptical openings on both sides, so, that there is no metal 
obstructing the SHiP's decay vacuum volume. The tracking detectors are inserted and removed through
the sides of the prismatic volumes, covered with vacuum flanges. The first tracking station is located
at 5 m distance from the forward end of the vacuum vessel. It serves as a veto station to reject 
the residual background events from $K^0_s$ decays and any other residual charged background in the
forward region of the vessel. Other four tracking stations belong to the magnetic spectrometer after 
the decay volume.  

\subsection{Integration and infrastructure}
\label{sec:vacuumvesselintegration}

\begin{figure}[tp]
\centering
\includegraphics[height=10cm]{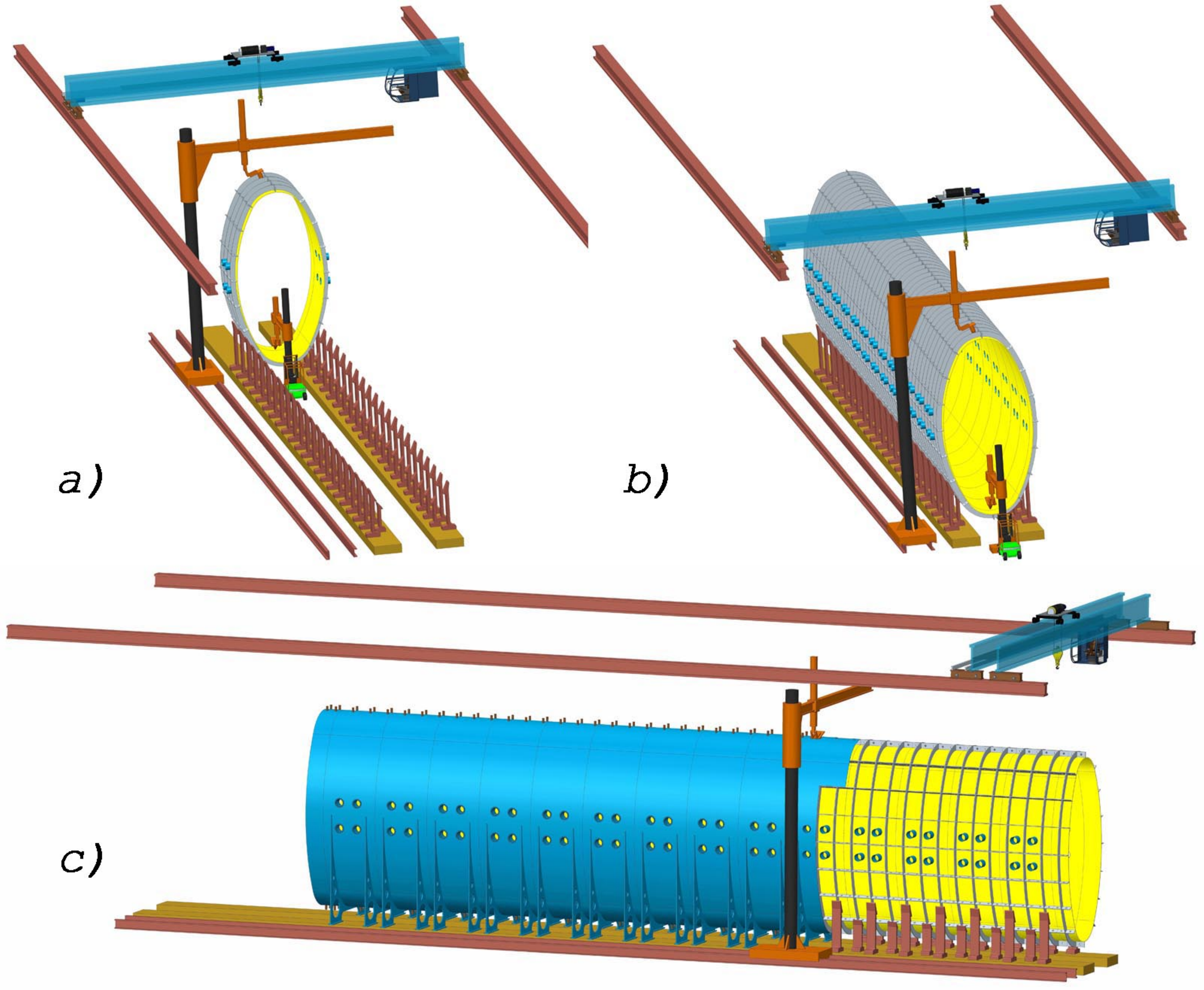}
\caption{A conceptual view of the vacuum vessel construction phases a), b) and c).}\label{vessel:VVAssembly}
\end{figure}

A large size of the vacuum vessel defines the manufacturing and integration strategy. The vessel is assembled directly
in the experimental hall (Figure~\ref{vessel:VVAssembly}). The building blocks (sections) are preassembled out of large 
stainless steel plates shaped as two different radii cylinder sectors.
The reinforcing ribs are welded to the outer surface of the sections.  The final integration of the preassembled
sections is performed on the assembly jig by the robotic welding technology.

%\vfill
%

%\vfill
\begin{figure}[tp]
\centering
\includegraphics[height=10cm]{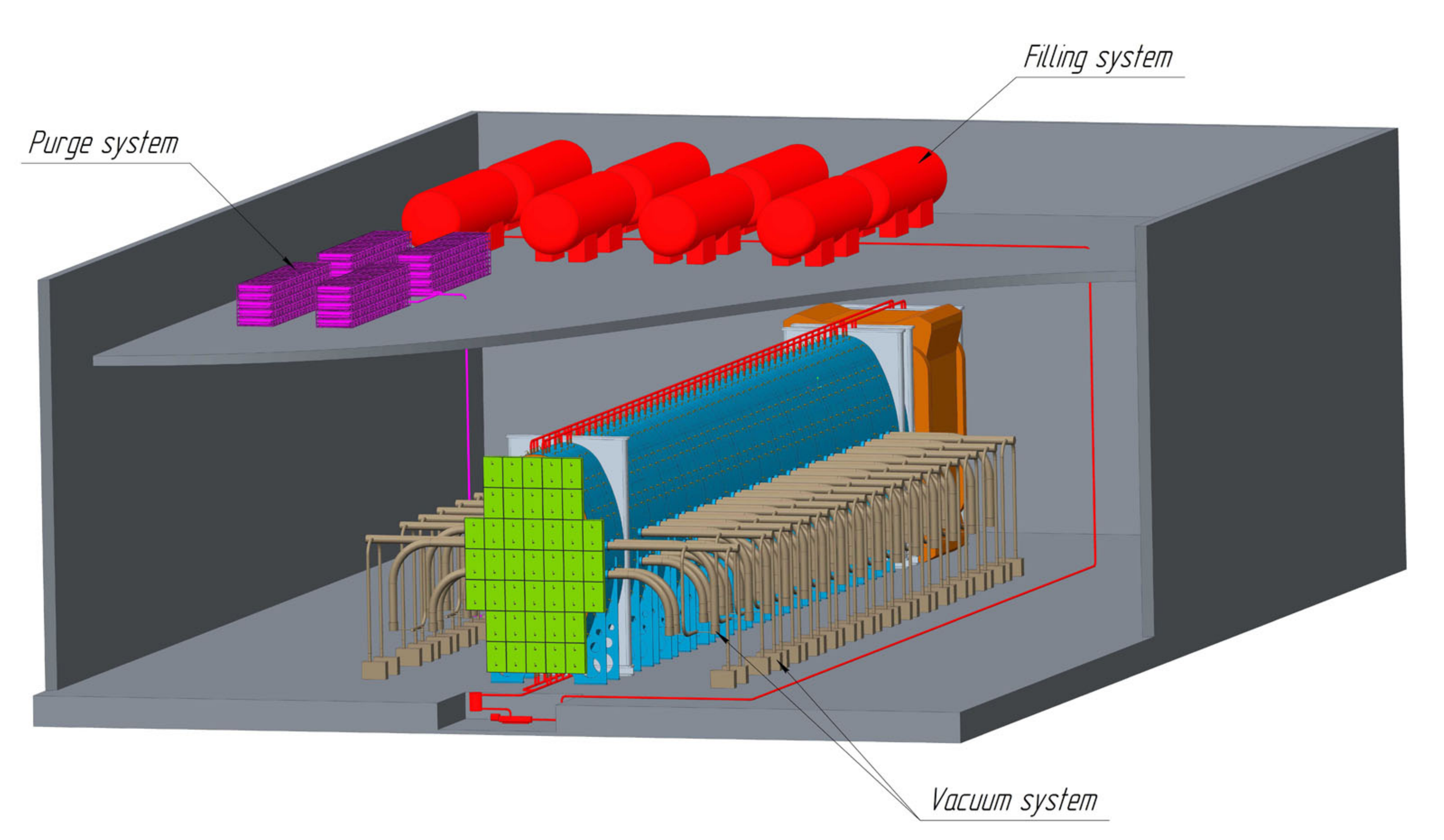}
\caption{A conceptual view of the vacuum vessel and background tagger infrastructure.}\label{vessel:VVInfrastructure}
\end{figure}%

The vessel's vacuum system and the background tagger's liquid scintillator infrastructure conceptual lay-out
is shown in Figure~\ref{vessel:VVInfrastructure}.
The vacuum system is based on pumps, ensuring an oil-free vacuum. Initial pumping down
is conducted by means of the stations, consisting of the dry screw pumps and Roots blowers.
High-vacuum is achieved by means of the turbomolecular pumps.
To ensure the required pressure limit of 10$^{-6}$ bar (0.1 Pa) turbomolecular pumps with a magnetic
levitation of the rotor are proposed.
The vacuum vessel pumping-out sequence is the following:
\begin{itemize}
\item Screw pump and Roots blower stations perform an initial pumping down to 1-10 Pa;
\item Then the inlets to the Roots blowers are switched to the turbomolecular pumps exhaust collectors;
\item Turbomolecular pumps together with the Roots blowers perform a high-vacuum vessel evacuation
and further maintain a nominal pressure of 10$^{-6}$ bar.
\end{itemize}

The liquid scintillator filling system consists of the following components: the stainless steel receivers
and distribution pipelines; the manifolds with the flow regulators; the collector manifold with individual drain
valves for each background tagger section, defined by the vessel's transverse ribs; the circulation pumps; and the liquid
scintillator regeneration system, with the dry nitrogen gas bottle racks and bubblers for purging out the residual
oxygen.

%\subsubsection{(Cost estimate)}
%\label{sec:vesselcostestimate}

%% file: taggers/Taggers.tex
\section{Surround background tagger}
\label{sec:taggers}

%Background neutrinos and muons originating from the target and cosmic muons can 
%enter the experimental hall. They can either directly enter the vacuum vessel or 
%produce secondaries in the proximity of the vacuum vessel that enter then into
%the vacuum vessel. For that reason the vacuum vessel will be surrounded by a 
%large-area detector, called background tagger, which will measure the energy 
%deposition from charged particles (see also Sec.~\ref{sec:backgrounds}). 
%Such particles could be: 
%\begin{enumerate}
%\item Charged particles like background muons and cosmic muons entering the 
%      vacuum vessel from outside
%\item Charged particles that are produced in neutrino or muon Deep Inelastic Scattering 
%      inside the inner and outer vacuum vessel wall or inside the supporting ribs 
%      of the vacuum vessel wall. These reactions are of particular
%      concern since they represent a potential source of $K_S$, $K_L$ or $\Lambda$
%      that could mimic a signal-like event in the tracking and calorimeter detector
%\item Recoil protons from neutron-proton scattering inside the liquid scintillator 
%      induced by neutrons pentrating the vacuum vessel walls.
%\end{enumerate}
The background tagger will potentially not only serve as a veto detector but offers
the possibility to decide offline whether activity in the tracking and calorimeter 
detectors is (partially) produced from particles entering from outside.

Since the outer surface of the vacuum vessel is very large, the background tagger
has to be very cost-effective but at the same time also highly efficient. 
For that reason we are proposing here a liquid scintillator based detector.

\subsection{Liquid scintillator}
\label{sec:taggerliquidscintillator}

The liquid scintillator technology has been demonstrated beneficial properties 
as targets of large masses detectors as well as an active material for the veto setup 
that enable real-time detection of charged particles and neutrons starting from a low energy 
threshold.

Liquid scintillators are mixtures of an organic solvent with the admixture of wavelength 
shifter or fluor as the solute. 
Charged particles which cross a solvent 
medium ionize and excite molecules on their track. In the molecular deexcitation process, 
the solvent transfers its excitation energy mainly nonradiatively to the solute emitting 
light in a wavelength region of 360-400 nm at which the solvent is transparent. 
The shape of the scintillation pulse has fast time components, usually of few ns decay time, 
and amplitudes that depend on the energy deposition per unit length. Therefore it provides 
a useful method for background rejection.

Photo-sensors are used to detect the scintillation photons. The efficiency of the photon 
detection is strongly dependent on the effective light yield (Y) of the scintillator, which 
drives the energy resolution and the energy threshold. Absorption and scattering processes 
decrease the total number of photons which arrive at the photo-sensors. For large-volume 
detectors, the impurities-free and transparent medium is required to reach long absorption 
lengths. Properties of the photo-sensors, and especially their spectral sensitivity (mainly 
in the region of 320-450 nm), play an important role for the performance of the detector.

\subsection{Choice of the liquid scintillator candidate}
\label{sec:taggerliquidscintillatorcandidate}

The most promising scintillator options consist of two scintillating organic compounds - 
a solvent and a solute powder. Such mixtures have to be compared taking into account 
the following properties:
\begin{itemize}
\item the light yield and scintillation pulse shape, which are important for the energy 
      and time response of the detector,
\item an emission spectrum guaranteeing large absorption and scattering lengths,
\item the compatibility with construction materials,
\item the flammability and other health (H), safety (S) and environmental (E) risks in handling the organic liquid scintillator~\footnote{Chemically, the solvent of an organic liquid scintillator is a hydrocarbon. Thus, flammability is the most important concern. The Hazardous Materials Identification System (HMIS) classifies the danger in handling the liquid (from 0 - safe to 4 - dangerous) concerning flammability, reactivity and health. The HMIS ratings for scintillators in flammability range from 1-3, reactivity (almost all 0) and health (0-1) are in general not a big concern. The flash point of the material is another characteristic parameter for its flammability.},
\item the price of components for synthesis of a large volume veto detector.
\end{itemize}
The mixtures under consideration providing high light yields are linearalkyl-benzene (LAB), 
phenyl-o-xylylethane (PXE) and pcedocumene (PC) with PPO or PMP as fluors. 
Tables~\ref{tab:LS_solvents_properties} and ~\ref{tab:LS_solutes_properties} summarize 
the properties of these solvents and fluors. 

\begin{table}[htb]
\caption{Summary table of properties of possible solvents for the liquid scintillator candidates.}
\label{tab:LS_solvents_properties}
{\footnotesize
\begin{center}
\begin{tabular}{ccccccccc}\hline
Solvent & Density                  & Relative     & Flash Point   & Emission   & Attenuation  &              & HMIS       &        \\
        & (${\rm g}{\rm cm}^{-3}$) & Light Yield  & ($^{\circ}$C) & Max. (nm)  & Length (m)   & Flammibility & Reactivity & Health \\
\hline
LAB     & $0.863$                  & $0.98$       &   140         &  283       & 20           &     1        &     0      &   1    \\
PXE     & $0.985$                  & $0.87$       &   145         &  290       & 12           &     1        &     0      &   1   \\
PC      & $0.889$                  & $1$          &    48         &            & 10           &     2        &     0      &   2   \\
\hline
\end{tabular}
\end{center}
}
\end{table}

\begin{table}[htb]
\caption{Summary table of properties of possible solutes for the liquid scintillator candidates.}
\label{tab:LS_solutes_properties}
\begin{center}
\begin{tabular}{ccc}\hline
Solute & Absorption Max.   & Emission Max.       \\
       & (nm)              & (nm)                \\
\hline
PPO    & $303$             & $370$         \\
PMP    & $294$             & $415$         \\
\hline
\end{tabular}
\end{center}
\end{table}

The first solvent, LAB, has been investigated in a number of the 
studies~\cite{LS-LAB-PPO,Kraus:2006qp,Ding:2008zzb}. 
The original LAB produced in distillation plants is usually rather pure. 
This choice is very appealing due to its high transparency, high light yield, and 
its low cost. Moreover, it is a non-hazardous liquid with relatively high flash point 
of $140^{\circ}$C.

The second solvent, PXE, has been tested in the Gran Sasso laboratory~\cite{Back:2008zz}. 
Results look very promising as it shows a high light yield and a high flash point of 
$145^{\circ}$C. However, the required large amount of solvent is rather expensive, 
and this fact makes PXE not competitive in relation to LAB.

PC is well known as a solvent used in the liquid scintillator of the Borexino 
detector~\cite{Alimonti:2000xc}. Unfortunately, this flammable liquid has a low flash 
point of $48^{\circ}$C and it leads to additional difficulties in operation.

In order to shift the wavelength to a region of about 400 nm, there are two fluor 
candidates - 2,5-diphenyl-oxazole (PPO) and 1-phenyl-3-mesityl-2-pyrazoline (PMP). 
PPO is used since a very long time and all operations with it worked out well. 
It has a Stokes shift of 60 nm and the absorption spectrum significantly overlaps with 
the emission spectra of the solvents. It is added in small quantities (1,5-3 g/l) to 
avoid self-absorption at large wavelengths. The fluor PMP is a solute with a large Stokes 
shift of about 120 nm, resulting in a marginal overlap of the absorption and emission 
spectra and thus a small self-absorption. Unfortunately, using PMP as fluor, makes the 
mixture much slower.

The final choice of a well-suited liquid scintillator mixture for the background-tagger 
was performed taking into account the above-mentioned properties, mainly the high light 
yield, the low auto-absorption, the optimal conformity between the emitted light spectrum 
and the photodetector quantum efficiency, as well as the relatively low cost.

As can be seen, the combination of LAB and PPO provides a good light yield of about 10000 
photons/MeV and its emission spectrum peaks at about 370 nm. 
%thus matching well the peak 
%efficiency of. 
With respect to the scintillation light yields, LAB and PXE show comparable results. 
However, in terms of both the optical properties and the cost, LAB is preferable to PXE. 
The LAB+PPO mixture shows a good time response: the first exponential, corresponding to 
the fastest decay time, accounts for most of the emitted light and has a decay time of 
approximately 5 ns resulting in a good time resolution of the background tagger.
LAB is a widely used liquid in chemical industry and it is produced on large scales by 
factories in many countries, particularly, in Russia. 
%The PO Kirishinefteorgsintez company (Kirishi, Leningrad region, Russia) provides a 
%high quality LAB sample with an initial attenuation length of more than 10 m at 400 nm 
%wavelength. This company provides LAB for 1 kEuro per 1 ton. 
A powder of commercially 
available PPO of suitable quality ($99~\%$) can be purchased from various companies
%such as
%%Sigma-Aldrich and 
%%Always Chem International Co (2,5-Diphenyloxazole, D4630 suitable for liquid scintillation 
%%spectrometry) for the price of 1075 Euro per 1 kg, from TCI America for 714 Euro per kg, 
%%and from ACROS Organics for 485 Euro per 1 kg. For larger quantities the price is
%%expected to drop significantly. Other experiments have obtained price offers from
%%a specific company for large-scale samples of very high quality PPO on the order of 
%%222 Euro per kg. Therefore the current estimate for the cost of 1 ton of the LS mixture 
%%is 1390 Euro where we have chosen a PPO admixture of 1.5 g/l of LAB.
%Sigma-Aldrich, Always Chem International Co (2,5-Diphenyloxazole, D4630 suitable for liquid scintillation                                  %                              spectrometry), TCI America, ACROS Organics, or PerkinElmer.               

\subsection{Liquid scintillator handling}
\label{sec:liquidscintillatorhandling}

The liquid scintillator handling comprises mainly the off-site scintillator fluid systems 
for production, purification and storage of the liquid scintillator, and the on-site 
systems situated on the experimental area, close to the background-tagger location.
The off-site scheme envisions the production and storage of the complete liquid scintillator. 
These systems will include ISO-containers for storage and subsequent transport to the 
experimental site, a purification column, a nitrogen purging unit, a mixing chamber, 
nitrogen blankets and auxiliary systems. 

The on-site systems will consist of an area close to the detector sites for the transport 
tanks which will be connected to the background tagger by a tubing system, which has to 
transfer the different liquids from their transport container into the detector volumes 
in a safe and clean way. The different detector volumes will be filled simultaneously 
and kept at equal hydrostatic pressures.

At all times, the liquid scintillator has to be kept away from oxygen contamination, 
which would induce a degradation of the liquid scintillator. Thus, ultra-clean nitrogen 
gas must be used to flush pipes and tanks, as well nitrogen purging is needed during 
the liquid filling. Any material that comes into contact with liquid scintillator has 
to be tested on its chemical compatibility. Possible material is passivated stainless steel. 
The cost of creation of the specified infrastructure has to be determined as a result of 
careful study of all conditions and terms of the experiment. It is necessary to pay attention 
that, as usual~\cite{Alimonti:2009zz}, similar plants for liquid handling will be developed 
within several years and include a contribution of the available laboratory and industrial 
equipments from participants of the collaboration.

\subsection{Design of the surround background tagger}
\label{sec:taggerdetectordesign}

As described in Section~\ref{sec:vessel} the volume between the inner 3 cm thick stainless 
steel and the outer 8 mm thick aluminium wall surrounding the vacuum vessel will be filled 
with a liquid scintillator (LS). For the chosen bending power and geometry of the active 
muon filter magnet the tail of the bent-out muons hitting the vessel walls on the left 
and right side should be as small as possible. For that reason the horizontal diameter 
of the vacuum vessel is reduced from 5 m to $3.724~{\rm m}$ for the first 5 m length of 
the vacuum vessel.
%
%The liquid scintillator has to provide good light yield, a long absorption length 
%for scintillation light, and given the large volume, be relatively cheap. Moreover,
%it should have low health (H), safety (S) and environmental (E) risks. 
%We consider as a good candidate a mixture of Linear Alkyl Benzene (LAB) with 
%an addition of 3 g/l of diphenyl oxazole (PPO) as used e. g. in the SNO+ 
%experiment~\cite{bgtagger:SNO-LS} for which we quote the properties in 
%Table~\ref{tab:LS_properties}.
%\begin{table}[htb]
%{\footnotesize
%\begin{center}
%\begin{tabular}{ll}\hline
%Quantity                              & Value                              \\
%\hline
%\quad Flashpoint                      & 140 $^{\circ}$C                    \\
%\quad Density                         & $0.86 {\rm g}{\rm cm}^3$           \\
%\quad Refraction Index                & 1.51 (at 400 nm)                   \\
%\quad Light Yield                     & 10000 photons/MeV                  \\
%\quad Absorption Length               & 5 m (at 380 nm), 12 m (at 420 nm)  \\
%\hline
%\end{tabular}
%\label{tab:LS_properties}
%\caption{Summary table of properties of the liquid scintillator.}
%\end{center}
%}
%\end{table}
%Despite its low HSE risks one has to foresee for the LS a retaining container 
%under the vessel in case of a LS leakage.

The liquid scintillator volume is planned to be separated into 30 cm thick sections 
of 100 cm length in $z$ direction and about 150 cm length in $x-y$ direction resulting 
in 863 sections. To guarantee good homogenity in scintillation-light collection, 
each section is viewed by two large-area photodetectors located at the same $z$ 
coordinate and separated from each other by a distance of 50 cm. Hence, the total 
amount of photodetectors is 1726.

One possible solution would be to choose spherical large-area photomulitpliers
viewing the LS volume in each section in order to collect the scintillation light.
A possible candidate would be a spherical 8-inch EMI9351KB PMT with a peak quantum 
efficiency of $31~\%$ at 370 nm. Another possibility which we are pursuing is a 
large-area photodetector consisting of a photomultiplier (PMT) of small-to-intermediate 
size viewing a large-area tube painted with a wavelength-shifting (WLS) material which 
absorbs the primary short-wavelength scintillation photons, converts them into 
longer-wavelength photons that are then transported by total internal reflection to 
the end of the tube where they are then detected by the PMT. 
This kind of photodetector has been proposed in~\cite{Schulte:2013dza} for a 
large-volume extension of the IceCube detector and is called a Wavelength-shifting
Optical Module (WOM). Such a WOM would be easier to mount and has the advantage
that the PMT can be significantly cheaper than in the large-area PMT solution.
The large detection area in this case comes from the large WLS-painted tube.
A proof-of-principle for this technique to work has been already 
shown~\cite{Schulte:2013dza, bgtagger:WOM-masterthesis}. A good choice for the WLS paint 
consists of the following mixture: 25 g Paraloid B72, 0.15 g Bis-MSB and 0.3 g p-Terphenyl 
in 100 ml Toluen on a PMMA tube~\cite{bgtagger:WOM-masterthesis}. 
With this choice the emission spectrum of the LS scintillator fits quite well the absorption 
spectrum of the WLS paint which itself has an emission spectrum with its maximum at 425 nm 
well adapted with typical PMT spectral sensitivities~\cite{bgtagger:WOM-masterthesis}.
The WLS-painted tube sits inside a slightly larger tube made out of quartz glass in order to 
separate the inner tube from the liquid scintillator and to let pass the scintillation light 
in the near-UV range. To enlarge the photon capture efficiency the tube end opposite to the 
PMT needs to be metallized so that wavelength-shifted photons emitted towards that tube end 
will be reflected and have eventually the chance to be detected as well by the PMT.

\begin{figure}
  \begin{center}
  \includegraphics[width=0.7\linewidth]{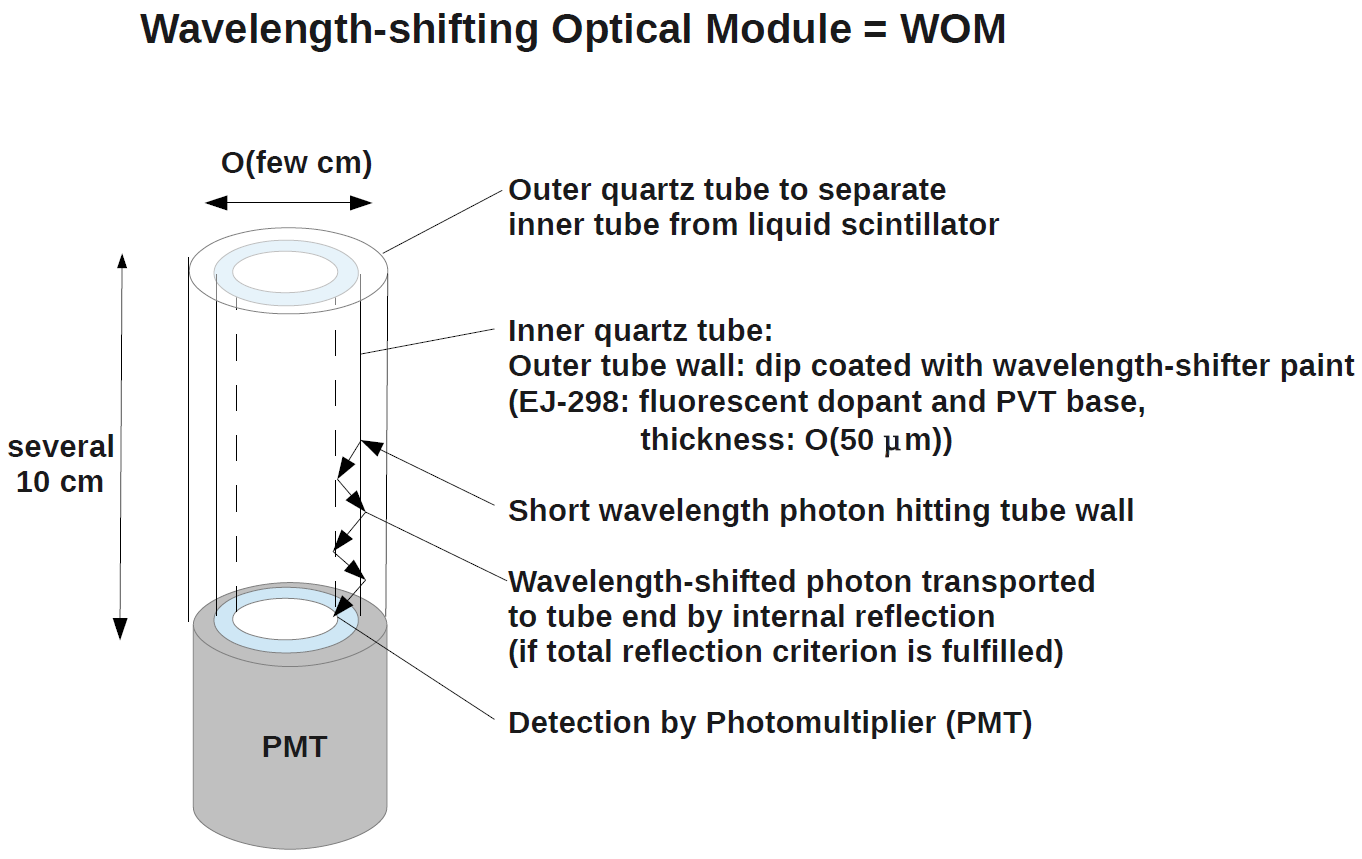}
  \end{center}
  \caption{Principle of the WOM technique for the detection of scintillation light.}
  \label{fig:bgtagger-wom}
\end{figure}

Figure~\ref{fig:bgtagger-wom} shows the principle layout of such a WOM and its photon
detection technique. To obtain good scintillation-light collection the length of the 
WLS-painted tube is as long as the distance between the inner and outer vessel wall 
of order 30 cm and the tube diameter is chosen to be 8 cm. The secondary light exiting 
the tube is then transported by an adiabatic PMMA light guide to the PMT of significantly 
smaller diameter. A possible choice would be e. g. the Hamamatsu R1924A PMT with a cathode 
diameter of 2.5 cm.

\subsection{Expected performance}
\label{sec:taggerperformance}

A minimum ionizing particle (MIP) traversing 1 cm of LS will loose 1.8 MeV of energy
resulting in 540000 scintillation photons in 30 cm thick liquid scintillator. 
With a photon capture-efficiency of O($35~\%$) for a WOM module as currently achieved
in laboratory measurements and a typical PMT photon detection-efficiency of O($25~\%$) 
one expects, from geometrical considerations, to detect several tens of scintillation 
photons by the WOM from direct light if the scintillation light is generated in the 
corners of the section. The theoretically achievable WOM photon capture-efficiency 
is about a factor of two larger than the currently achieved one and hence there is
quite some room for improvements.

\begin{figure}
  \begin{center}
  \includegraphics[width=0.7\linewidth]{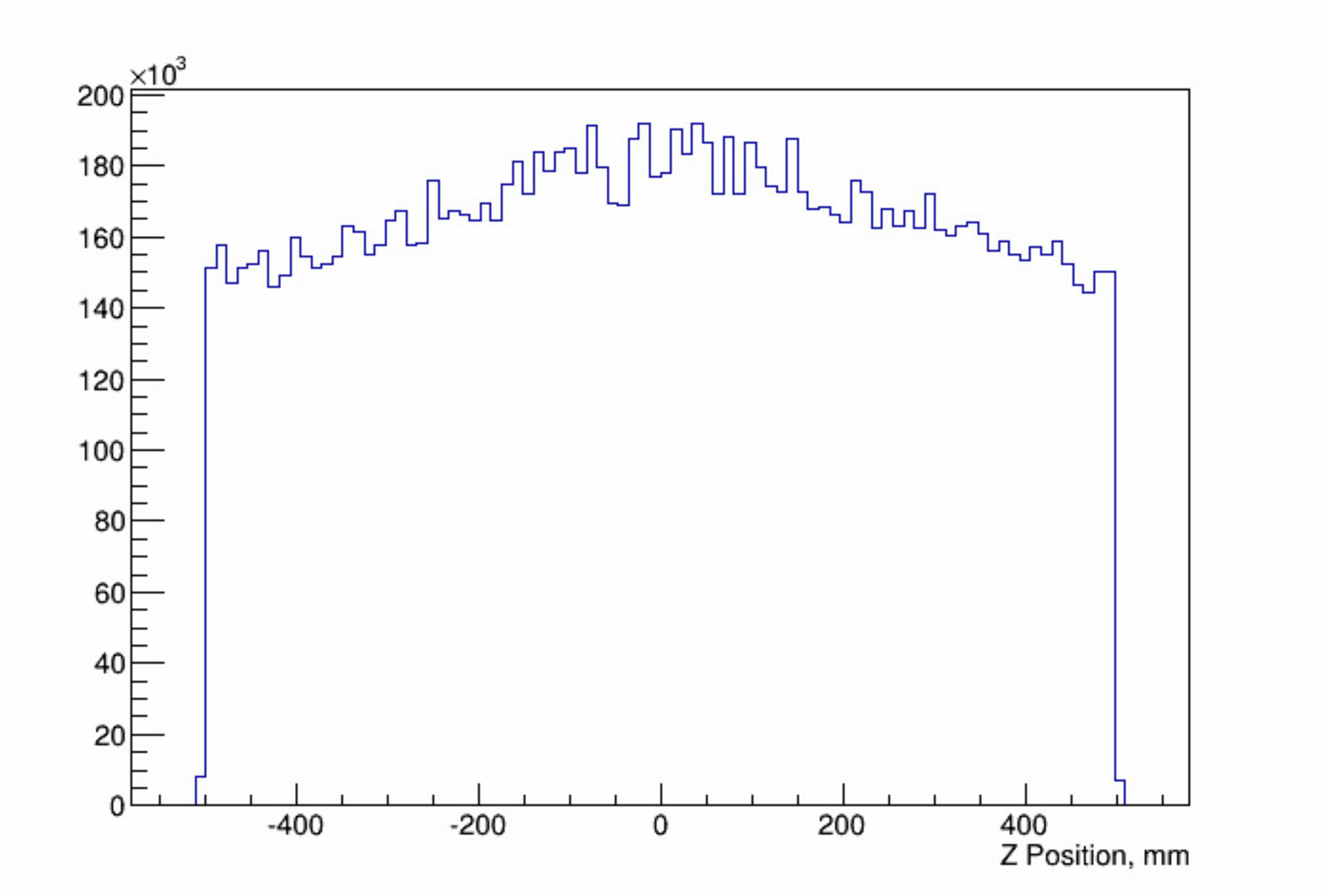}
  \end{center}
  \caption{Light collection uniformity in a section as a function of $z$
          for one photodetector.}
  \label{fig:bgtagger-uniformity}
\end{figure}

Detailed simulation studies with GEANT4 have been performed studying the detector
response for 500 MeV muons entering a LS section perpendicular to the vacuum vessel
symmetry axis (``perp'') and parallel to the vacuum vessel symmetry axis (``para'').
To enhance the reflectivity of the inner section walls it is assumed that they
are painted with a material of diffuse reflectivity like MgO or ${\rm TiO}_{2}$.
Since a full simulation of a WOM was not available at this stage, the light 
collection was simulated by two spherical 8-inch PMTs with the properties of the
EMI9351KB PMT. 
These simulation studies show that the average time to detect 200 photoelectrons 
with such a PMT (taking into account the wavelength-dependent PMT efficiency) 
is of order 4 ns (``perp''), respectively, 5 ns (``para''). The standard deviation 
on the average collection time is of order $\pm 1.2$ ns (``perp''), respectively, 
$\pm 2$ ns (``para''). 
For a detection threshold of 200 photoelectrons one expects to collect $99~\%$ of 
those after a time of about 20 ns. The corresponding light collection uniformity 
is shown in Figure~\ref{fig:bgtagger-uniformity}.
%
%The background tagger also allows to detect high-energy neutrons. For neutrons
%with kinetic energies between 100 MeV and 1 GeV the efficiency to collect 100 
%photoelectrons within a collection time of 20 ns is of order $65-70~\%$.
%The average collection time depends on the neutron energy. For neutron energies
%between 400 MeV and 1 GeV one finds in the simulation an average collection time
%of about 5 ns (``perp'') and between 4 and 6.5 ns (``para'') for the detection of
%200 photoelectrons. For these neutron energies, the standard deviation on the 
%average collection time is of similar size as for 500 MeV muons.\\

From laboratory measurements the WOM detection efficiency is expected to be similar 
to the pure large-area PMT solution thanks to the much larger WOM detection area. 
Compared to the pure large-area PMT solution the average collection time and its 
standard deviation for the WOM solution is expected to increase by a few nanoseconds 
as a result of a) the WLS-paint decay time of a few nanoseconds, and b) the varying 
photon paths inside the WOM.

%Background events are of concern if activity in the tracking detector and 
%calorimeter is observed but no detector response from the background tagger 
%is recorded. This can e. g. happen if muons that can either be cosmic rays 
%or debris from processes induced by collisions in the target produce 
%long-living neutral particles entering the vacuum vessel in such a way 
%that energy depositions from charged particles from such reactions don't 
%produce any or too small of a signal in the LS. 
%Particles like $K_S$, $\Lambda$ and $K_L$ produced in such reactions might 
%then decay inside the vacuum vessel mimicking a signal event. 
%Another background of concern comes from neutrino-induced reactions in the
%vessel walls and supporting ribs producing such long-lived particles.\\
%Using simulated cosmics, muon background and neutrino background samples 
%we have studied deep-inelastic scattering events taking place in the inner 
%and outer vessel wall and their supporting ribs, and also in the walls of 
%the cavern (see Sec.~\ref{sec:backgrounds}). 
From the background studies presented in Section~\ref{sec:backgrounds}) 
we conclude that the background tagger should have a detection efficiency 
of $99~\%$ per section which is well in reach with the proposed design.
The surround background tagger can be complemented by a liquid scintillator 
wall on the vessel front window as indicated in
Figure~\ref{vessel:VV3Dmodel},~\ref{vessel:VVForward},~and~\ref{vessel:VVInfrastructure},
if further background studies show that a higher redundancy in the
veto efficiency is required.

%% file: taggers/timing/Upstream_VETOtiming.tex
\section{Upstream veto tagger}
\label{sec:upstream_vetotiming}

An important background source comes from neutral kaons produced by
neutrino and muon interactions upstream of the vacuum
vessel, see Section~\ref{sec:backgrounds}. Mainly these interactions happen in the passive material of the tau neutrino detector, and can be in principle detected by the active layers forseen for the tau neutrino muon system. On the other hand this muon system is not designed and optimised to act as a veto system, and can therefore have a not optimum efficiency, noise rate and time resolution. Moreover, no active material is present in front of the yokes on top of this system ($|y|>400$ cm). Thus, the option of an upstream veto tagger station to be located between
the tau neutrino detector and the vacuum tank with a coverage area of $4\times 12$
m$^2$ is considered. This detector plane would be expected to reduce the backgrounds from neutral kaons produced upstream of the vessel down to negligible levels, and also to veto on muons entering the vessel from the front window. A redundance of vetoing systems is important at the current stage, before advanced studies are performed to quantify the level of importance of each particular system.
%Since SHiP wants to achieve a negligible background level, the redundance of vetoing systems is essential. On the other hand the presence of extra material is itself source of extra background. Therefore advance studies are planned to understand the level of importance of this detector plane. For the moment 

An array of plastic scintillator bars represents a technology that provides very high efficiency, good time resolution, low maintenance, and large area coverage for a relatively low cost.  
%A detector based on plastic scintillating bars is proposed. 
The design takes advantage of similarities with other large scintillator-bar detectors being proposed for SHiP, such as the muon detector (Section~\ref{subsec:muondet}) and the spectrometer timing detector (Section~\ref{sec:timing_detectors}), in order to profit from common R\&D and production. The main emphasis for the upstream-veto detector is on efficiency. The dead time introduced by a veto detector is proportional to the signal rate and the time resolution (for matching a veto signal to a signal in the spectrometer). The signal rate is expected to be at most $100$ kHz assuming a negligible noise rate. With a time resolution of 1 ns, this results in a dead time of $\sim 0.01\%$, low enough to allow for some margin. The aim of the upstream veto tagger station is to achieve an efficiency of 99.9\% or higher for detecting a muon or a charged hadron hitting the detector on the $4\times 12$ m$^2$ surface, a negligible noise rate, and a time resolution of 1 ns. 

%front_view_rot0_line_txt.png}

\begin{figure}[htb]
\begin{center}
\includegraphics[width=0.31\linewidth,angle=0]{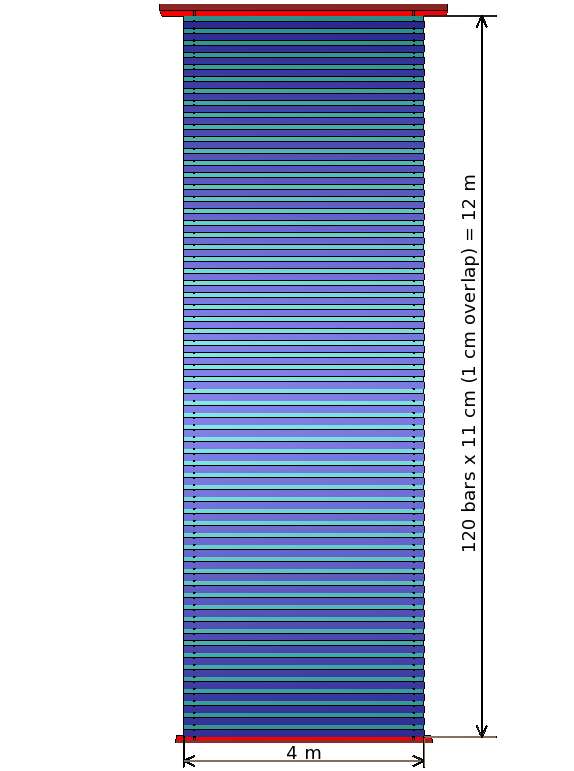}
\hspace{1cm}
\includegraphics[width=0.45\linewidth,angle=0]{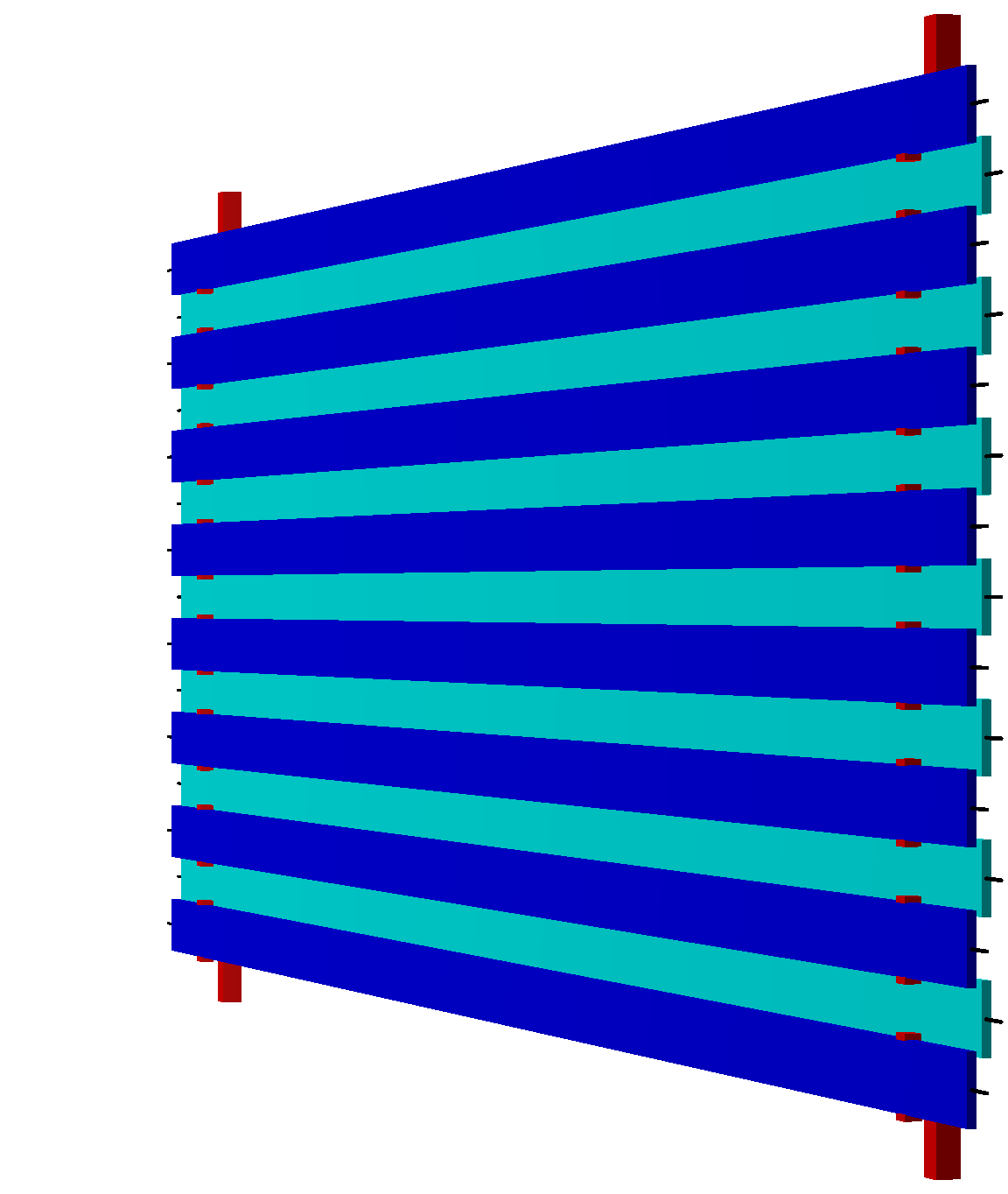}
\caption{
   (Left) Schematic front-view of the upstream-veto detector.
   The oval represent an acceptance of the vacuum vessel.
   (Right) Rotated 3D view of a section of the detector.}
\label{Upstream_Veto_scint_geometry}
\end{center}
\end{figure}

\subsection{Detector layout}

The time measurement is better for higher signal amplitude, which
itself is degraded by light attenuation inside the bar. In a test
using extruded scintillator slabs read out by a wavelength-shifting
(WLS) fibre running through the length of the bar and coupled to
photodiodes from both ends, acceptable signals could be measured along
bars of lengths up to 16 m, with light yield degraded by up to 80\%
for a distance of 8 m (middle of the bar)~\cite{BabyMIND2014}. For
bars of length 5 m or smaller, the decay time constant of the WLS
fibre (typically $\sim 10$ ns) becomes the dominant contributor to the
timing resolution, limiting it to $\sim 500$ ps for lenghts of $\sim 1$ m. Bars of 4 m length can thus achieve time 
resolutions below 1 ns, which perfectly suits the needs 
of the upstream veto tagger detector. The option of 4 m bars with WLS fibres read out on both ends by silicon photomultipliers is chosen here as a baseline. In case of interesting results obtained by the R\&D in the context of the timing detector (see Section~\ref{sec:timing_detectors}), a different solution may be considered in the future. 

% \begin{figure}[t]
% \centering
% \includegraphics[width=0.9\linewidth,angle=0]{taggers/timing/fig/SHiP_timing_detector.png}
% \caption{
%   Baseline design of the SHiP upstream-veto detector geometry. The light readout scheme depicted here uses wavelength-shifting (WLS) fibres coupled to SiPMs on both ends of each bar, a compact system with demonstrated time resolution of 1 ns for 4 m bars. \textit{Need to update the layout: only one column of 4 m bars, resize to 4 x 12 m2, and place the electronics on front} 
% }
% \label{timing_detector_scint_geometry}
% \end{figure}

%The baseline geometry of the proposed detector is depicted in Fig.~\ref{timing_detector_scint_geometry}. 
The baseline array comprises one 
column of horizontal bars 400 cm long, 12 cm wide and 1 cm thick. To cover the full $4\times 12$ m$^2$ 
area, a total of $120$ bars are needed. Each bar is read
out on both ends, thus requiring 240 electronic channels. The readout
electronics will be located directly upstream of the sensitive plane. 
The support structure will be designed such that no material extends laterally beyond $|x|=2$ m to avoid intercepting the flux of muons which could scatter back to the detector volume. Moreover, minimal amount of passive material should be present downstream of the detector.
%around the edges of the detector module. %%% Phil: no we cannot do that because we need to avoid muon back-scattering
Including the support structure, the whole detector would weigh about one tonne. 

\subsection{Front-end electronics}

The choice of electronics for the upstream veto tagger 
%is in large part informed by previous experience obtained in the NA61/SHINE experiment~\cite{NA612014b} and 
can be shared with the spectrometer timing detector, see Section~\ref{sec:timing_detectors}, in case the same technology is
chosen. Since the choice of the eletronic readout is more relevant for
the spectrometer timing detector more details about the readout electronics are
provided in Section~\ref{sec:timing_detectors}. 
Depending on costs, the upstream veto tagger planes, which requiring less
timing precision, could eventually be read out by potentially cheaper alternatives, such as
the CITIROC family of ASICs. Although the amplitude of the signal can only be extracted from the CITIROC at a limited rate, due to the multiplexing of the 32 analogue output channels, hit information can be extracted from the fast shapers on the chip at a high enough rate for this application.

%% file: tracking/Tracking.tex
\newcommand\mysection[2]{\section{#1 (#2)}}
\newcommand\mysubsection[2]{\subsection{#1 (#2)}}
\newcommand\mysubsubsection[1]{\subsubsection{#1}}
\newcommand\mysubsubsubsection[1]{{\bigskip\noindent\bf #1}\medskip}
\newcommand\mycomment[1]{\textcolor{red}{(\sl #1)}}
\def\dyfr{\displaystyle\frac}
\def\stereo{\theta_{\rm stereo}}
\def\ie{{\em i.e.}\xspace}
\def\sighit{\sigma_{\rm hit}}

\def\citIvan{\cite{Bereziuk:2005286}}
\def\citMakankin{\cite{Makankin2014649}}
\def\citBRR{\cite{BRR}}
\def\citShiPEOI{\cite{Bonivento:1606085}}
\def\citGARFIELD{\cite{GARFIELD}}
\def\citEricvH{\cite{VanHerwijnen:2005715}}
\def\citCARIOCA{\cite{Bonivento2002233}}
\def\citASDBLR{\cite{1239288}}
\def\citASDBLRnew{\cite{ASDBLRnew}}
\def\citPythiaEight{\cite{Sjöstrand2008852}}
\def\citGeantFour{\cite{1610988}}
\def\citTemurEnik{\cite{TemurEnik}}
\def\citLHCbUpgradeLOI{\cite{CERN-LHCC-2011-001}}
\def\citNAsixtytwoTD{\cite{Hahn:1404985}}
\def\citGenfit{\cite{Höppner2010518}}

%\mysection{Spectrometer straw tracker}{Massi Ferro-Luzzi}
\section{Spectrometer tracker}
\label{sec:tracker}

The purpose of the HS spectrometer is to reconstruct with high efficiency the tracks
of charged particles from the decay of hidden particles, while rejecting background events. 
Additionally, the spectrometer must provide an accurate determination of the track 
momentum and of the flight direction within the fiducial decay volume.
The precision of the extrapolated position of the tracks must be well matched with the 
segmentation of the timing detectors (see Section~\ref{sec:timing_detectors}) such that
the high accuracy of the associated track time can be used to remove combinatorial background.
The invariant mass, the vertex quality, the timing, the matching to background taggers
and the pointing to the production target are crucial
tools for rejecting background from spurious $V^0$ meson decays or from random combinations. 

\subsection{Detector layout}
\label{sec:tracker-layout}

\begin{figure}[tbp]
 \centering
 \includegraphics*[width=0.55\textwidth]{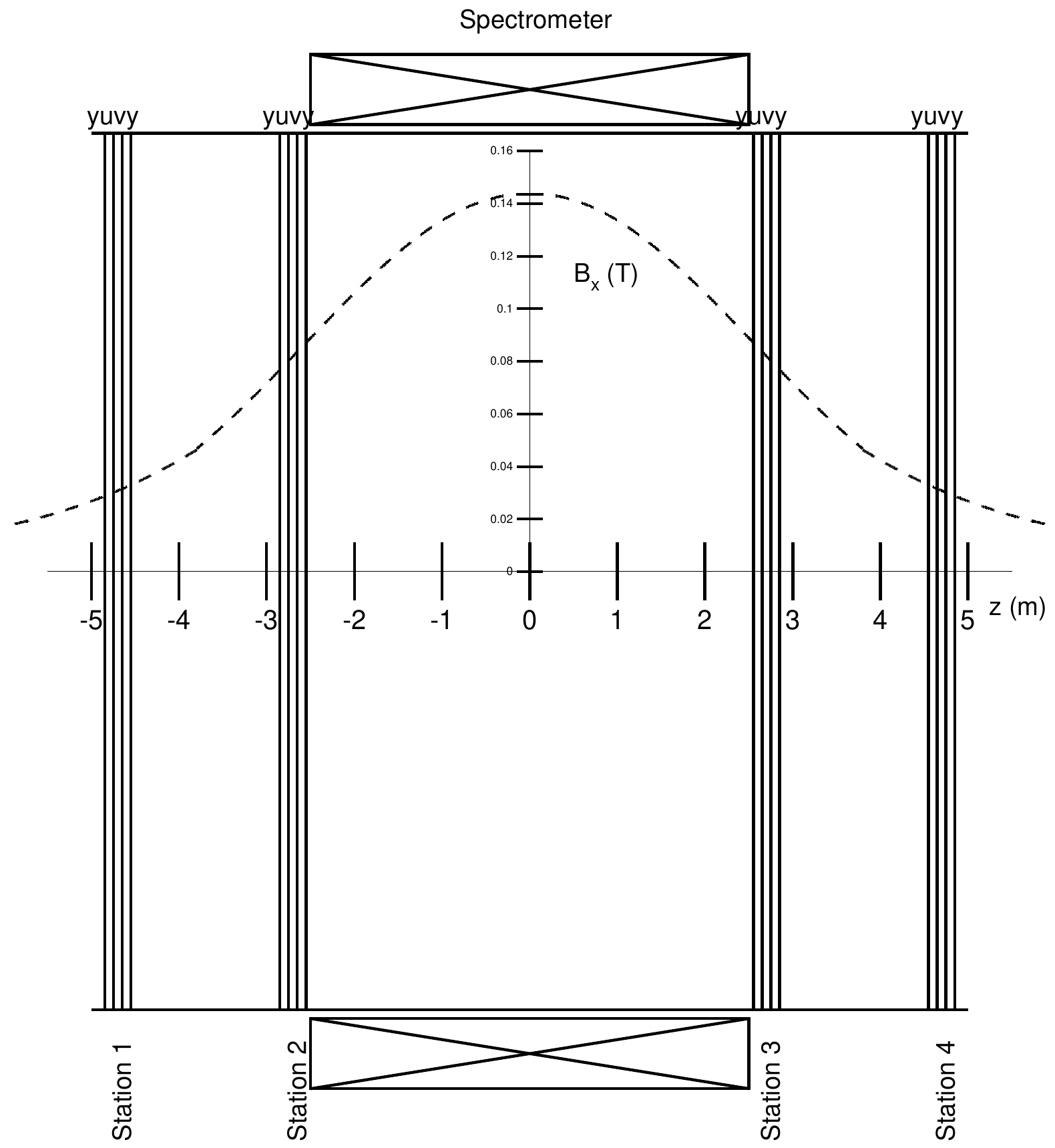}
 \includegraphics*[width=0.44\textwidth]{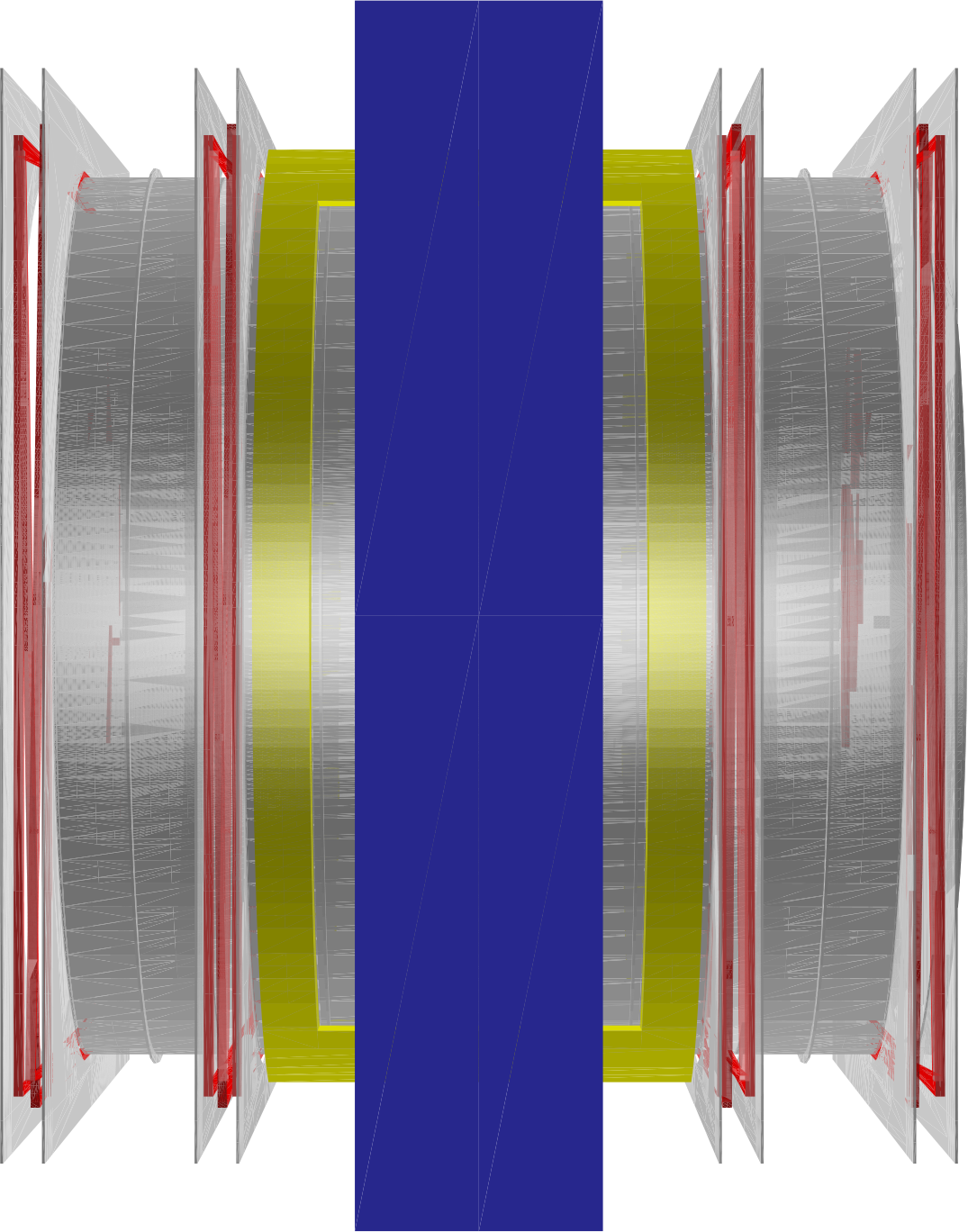}
 \caption{Spectrometer layout. (Left) Position of the tracking stations (each with four views)
  and dipole magnet,
  overlaid with magnetic field component $B_x$ as a function of $z$.
  (Right) 3D view of the spectrometer as implemented in 
  the FairShip simulation.}\label{fig:spectrometer-layout}
\end{figure}

The spectrometer consists of a large aperture dipole magnet (discussed in Section~\ref{sec:spectrometermagnet}) 
and two tracking telescopes on each side of the magnet.
A layout with four tracking stations symmetrically arranged around the dipole magnet,
as depicted in Figure~\ref{fig:spectrometer-layout},  is taken as a baseline.
The size and layout of the tracker stations is connected to the size of the magnet.
A dipole spectrometer magnet with a horizontal gap of 5~m, a height of 10~m 
and a length of 5~m provides good acceptance coverage and is considered
feasible at a reasonable cost. % (see section \ref{sec:spectrometermagnet}).
The $B$ field is about 0.14~T at its maximum and about 0.08~T at the location of the 
closest tracker stations, just outside the magnet.
On the longitudinal axis the field integral between the second and third station
is approximately 0.65~Tm.
 
%The size and layout of the tracker stations is connected to the size of the magnet.
Following the direction of the magnetic field, the measuring elements are oriented horizontally
to measure precisely the vertical (Y) coordinate. Two stereo views (U and V) are rotated by an
 angle $\pm\stereo$ for measuring the transverse coordinate X
   with an accuracy degraded by $\sim 1/\sin\stereo$.
The precision in X (i.e. the value of the stereo angle) is driven by the need of a good enough 
measurement of the decay vertex, opening angle of the daughter particles (which enters the invariant mass)
and impact parameter at the production target.
Each station contains 4 views (Y-U-V-Y). %, where Y is the vertical coordinate, U and V are stereo coordinates of Y).
The two stations on the same side of the magnet are separated by $\Delta=2$~m and a gap of 5~m 
is left between the second and third stations (i.e. each is 2.5~m away from the centre of the magnet).

A fifth `veto' station with only two views (Y-U) is located a few meters downstream of the 
vacuum vessel entrance lid, see Figure~\ref{fig:detector_layout}. 
It will use the same drift tube technology as the tracking stations.
However, its chief purpose is to tag tracks that are reconstructed in the spectrometer tracker,
but which originate from upstream of the veto station.

The tracking stations of the magnetic spectrometer must provide good spatial resolution and minimise
the contribution from multiple scattering.
In addition, the tracker must operate in vacuum (see Section~\ref{sec:hs_reqs}).
% and to minimize the tracker material
A straw tracker made of thin polyethylene terephthalate (PET) tubes is ideal to meet these goals. 
Gas tightness of these tubes has been demonstrated in long term tests and the mass production procedure
is also well established (see NA62 experiment \citNAsixtytwoTD).
%in the acceptance, the choice is made to use a straw tracker of a design inspired from the NA62 design
The main differences between the SHiP tracker and the NA62 tracker are the need for 5~m long straws
(vs 2.1~m in NA62), the much lower expected rate of particle hits and the less
stringent requirements on gas tightness.
Based on FairShip simulation studies, one expects about $10^{7}$~hits/station in 1~s 
(i.e. during the spill of $5\cdot10^{13}~p$ on target) which corresponds to 
2~kHz/straw,
%\footnote{%
%   T.Ruf 15dec2014 meeting, $10^{7}$hits/station in 1~s (i.e. $5\cdot10^{13}~p$) 
%   and (568~straws/layer) $\times$ (4~layers/view) $\times$ (4~views/station).},
while in NA62 the most exposed straws may see rates of up to 500~kHz.
As presented in Section~\ref{sec:GeneralLayout}, the vacuum pressure in the decay volume must 
be less than $10^{-3}~{\rm mbar}$ to suppress neutrino interactions.
This is at least two  orders of magnitude less demanding than the pressure used in the
NA62 vacuum tank ($<10^{-5}~{\rm mbar}$) and should therefore not 
represent a technological challenge for the straw tube tracker.
The requirement on straw efficiency is relatively modest for signal events, when using four views per station. 
%Assuming a particle traverses 8 straws (within its sensitive radius) per station, 
Assuming a particle traverses 8 straws (within its sensitive radius) per station\footnote{The MC 
simulation shows that
the average number of straw hits per particle per station is $\approx 8$~\citEricvH.}
with an average efficiency of 90\% per straw one obtains at least 5 hits in 99.5\%
of the cases.
% 8 hits in 43\% of the cases, 7 in 38\%, 6 in 15\%, 5 in 3\%, etc.
%There is room for optimization.
The efficiency requirement is more stringent for the straw station used for vetoing at the
start of the vacuum vessel.

\begin{figure}[tbp]
 \centering
 \includegraphics*[width=0.44\textwidth,angle=90]{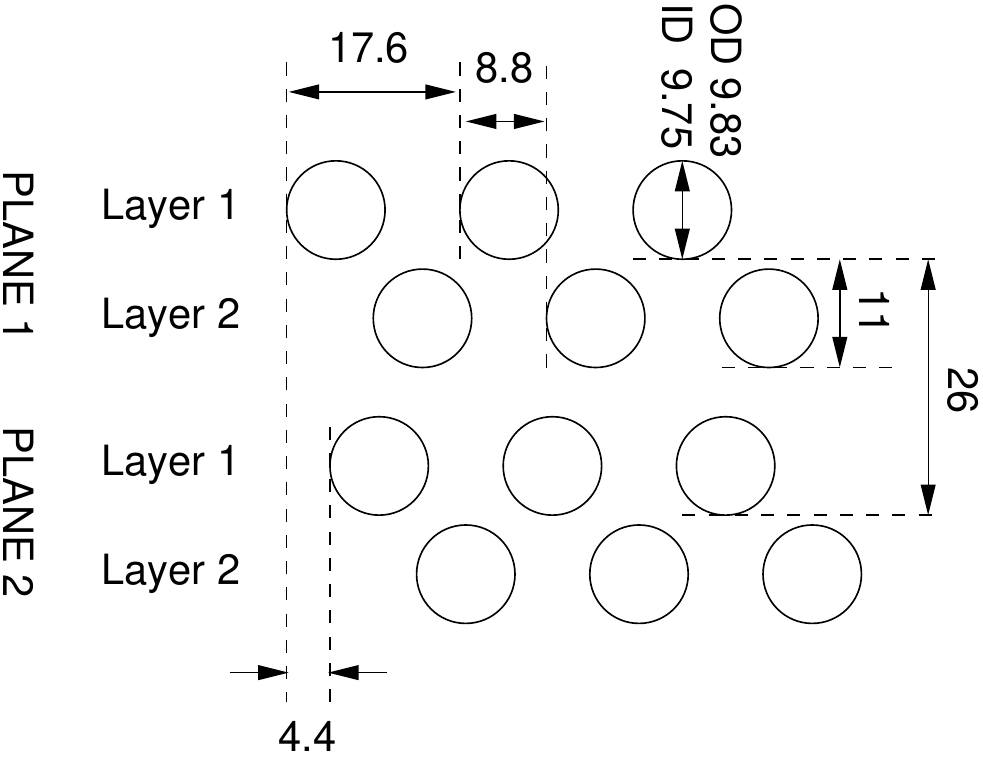}\hspace*{20mm}
 \includegraphics*[width=0.32\textwidth]{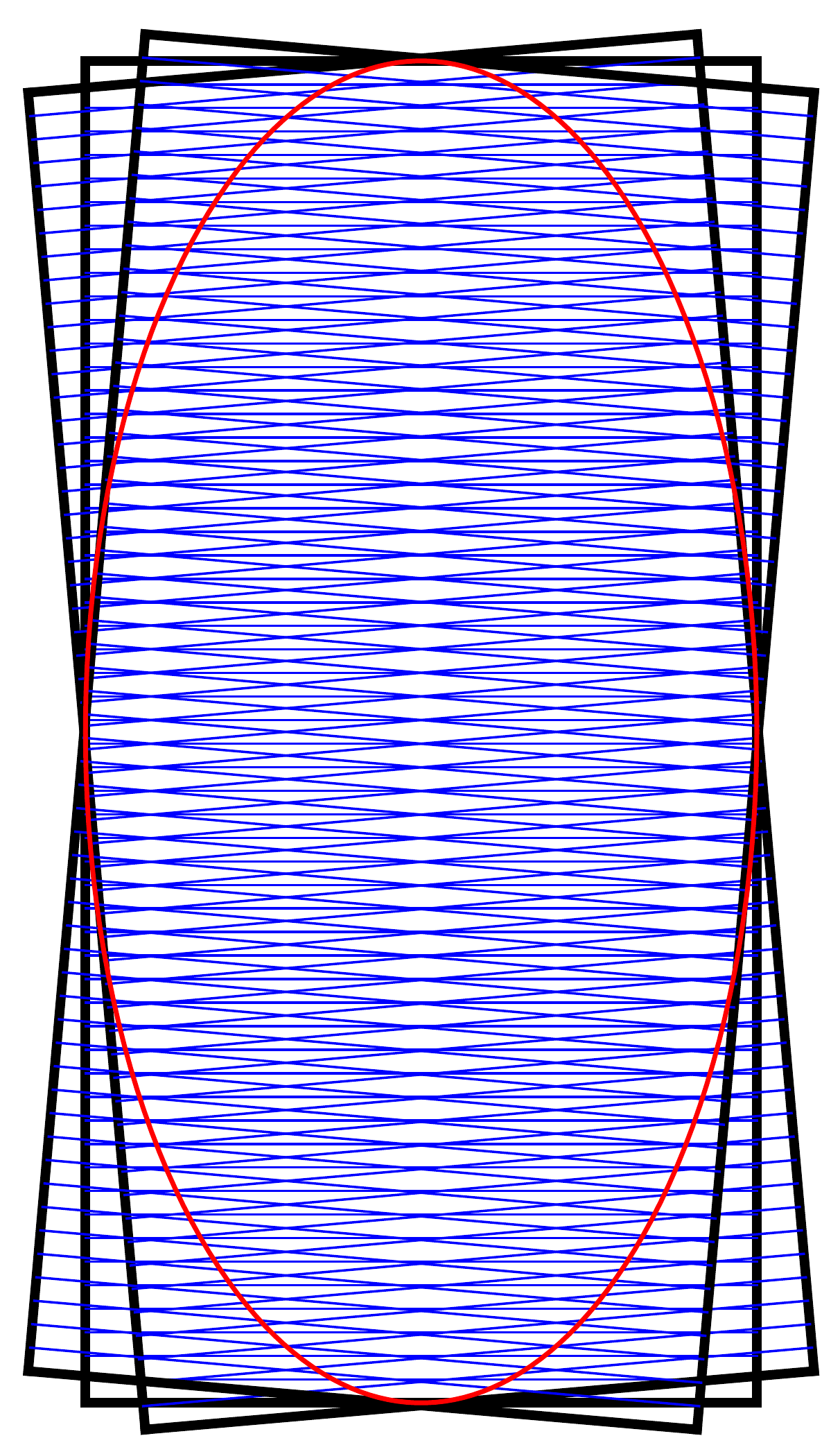}
 \caption{(Left) Straw layout for one tracker view, as currently used in the FairShip tracker simulation. 
%   Right: optional layout with less redundancy per view and less multiple scattering.
%   $\Delta y_{\rm straws}\approx 3\,r_{\rm eff} \approx 3\,\Delta y_{\rm planes}$.
  (Right) Four views in one station (not all straws shown, for the sake of clarity). 
  The nominal acceptance, defined by the vacuum vessel, is shown
   as a red ellipse.
  %\mycomment{figures to be improved. Add dimensions, angles ?}
   }\label{fig:straw-layout-NA62}
\end{figure}

Figure~\ref{fig:straw-layout-NA62} (left) shows the straw layout for one view, 
as currently implemented in the SHiP simulation (taken from the NA62 technical design report
without optimization for SHiP).  
Space around the straws is required for two reasons: 
(a) the straw diameter slightly increases when exposed to internal pressure and 
(b) some working space around the straw is needed when interfacing the straw to the flange.
Straws of 9.83 (9.75)~mm outer (inner) diameter are arranged in two planes of each two layers.
The vertical pitch between straws in one layer is 17.6~mm, approximately four times the
effective radius $r_{\rm eff}$ of the straw (the radius within which the straw efficiency is close to 100\%).
Within a plane, the second layer is staggered vertically with respect to the first layer 
by half a straw pitch (8.8~mm).
Plane 2 is a copy of Plane 1, shifted longitudinally by  26~mm and staggered vertically 
by a quarter straw pitch (4.4~mm).
In the current SHiP simulation, one tracking station contains four views (Y-U-V-Y)
with a longitudinal shift of 100~mm from view to view. 
The stereo angle is now set to $\stereo=+(-)5$ degrees for the U (V) views, but needs to be optimized.

\begin{table}[htb]
{\footnotesize
\begin{center}
\caption{Summary table of geometrical parameters for the straw tracker.}
%\mycomment{add somewhere gas and voltage parameters ?} }
\label{tab:straw-parameters}
\vspace{2mm}
\begin{tabular}{lll}\hline
Parameter                              & Value                &                Note            \\\hline
\multicolumn{3}{l}{Straw}                                                                      \\
\quad Length of a straw                      & 5~m            &                                 \\
\quad Outer straw diameter                   & 9.83~mm        &                                  \\
\multicolumn{3}{l}{\quad Straw wall (PET, Cu, Au)}                                               \\
\quad\quad   PET foil   thickness                & 36~$\mu$m  &\multirow{3}{*}{$\Bigg\}$~\parbox{20mm}{40~$\mu$m PET in simulation}}\\
\quad\quad   Cu coating thickness                & 50~nm      &                                 \\
\quad\quad   Au coating thickness                & 20~nm      &                                 \\
\multicolumn{3}{l}{\quad Wire (Au-plated Tungsten)}                                             \\
\quad\quad    diameter                            & 30~$\mu$m &                                \\
\multicolumn{3}{l}{Straw arrangement}                                                          \\
\quad Number of straws in one layer          & 568            &                                 \\
\quad Number of layers  per plane            & 2              &                                \\
\quad Straw pitch in one layer               & 17.6~mm        &                                \\
\quad Y extent  of one plane                 & $\sim 10$~m    &                                 \\
\quad Y offset between straws of layer 1\&2  & 8.8~mm         &                                 \\
\quad Z shift from layer 1 to 2              & 11~mm          &                                \\
\quad Number of planes  per view             & 2              &                                 \\
\quad Y offset between plane 1\&2            & 4.4~mm         &                                 \\
\quad Z shift from plane 1 to 2              & 26~mm          &                                \\
\quad Z shift from view to view              & 100~mm         &                                 \\
\multicolumn{3}{l}{Straw station}                                                               \\
\quad Number of views per station            & 4 (Y-U-V-Y)    &                                  \\
\quad Stereo angle of layers in a view Y,U,V & 0, 5, -5 degrees &                               \\
\quad Z envelope of one station              & $\sim 34~$cm     &                               \\
\quad Number of straws in one station        & 9088             &                               \\\hline
\multicolumn{3}{l}{Straw tracker}                                                               \\
\quad Number of stations                     &  4               & 2 before, 2 after the magnet   \\
\quad Z shift from station 1 to 2 (3 to 4)   & 2~m               &                                \\
\quad Z shift from station 2 to 3            & 5~m               &                               \\
\quad Number of straws in total              & 36352             &                               \\\hline
\end{tabular}
\end{center}
}
\end{table}

The key parameters of the currently considered straw tracker geometry
are summarized in Table~\ref{tab:straw-parameters}.
% As further discussed in Section~\ref{sec:tracker-performance-studies}, 
The main changes with respect to the Expression of Interest~\citShiPEOI
%\mycomment{EOI ref should be in main.bib}
follow from the changes applied to the spectrometer magnet.
The straw orientation has been turned from vertical to horizontal
and one transverse dimension has been increased from 5 to 10~m.
The longitudinal gaps have also been slightly modified. 
The straw tracker has been implemented in the SHiP Monte Carlo full simulation using
these parameters as a starting point~\citEricvH.

\subsection{Performance studies}  
\label{sec:tracker-performance-studies}  

The R\&D for the HS spectrometer tracker has focused thus far on three main activities.
For studies of the tracker performance, the FairShip Monte Carlo simulation package was
developed, see Section~\ref{sec:simulation}.
A pattern recognition algorithm and a generic track fitting routine % \mycomment{Genfit, ref ?}
are used to evaluate the performance, % of signal reconstruction,
% and background rejection, 
Section~\ref{sec:Tracker performance studies from simulation}.
Simulation studies of the basic detector element, the drift tube, have
been started using the GARFIELD package~\citGARFIELD.
The effect of wire sagging, gas mixture parameters and cross sectional 
dimensions of the drift tube will be investigated in details and compared  
with measurements on prototypes.
Preliminary results are discussed in Section~\ref{sec:simulation-straw-performance}.
The third key activity is the production and characterization of 
prototype drift tubes of 5~m length. 
Exploratory work has started and first results are presented 
in Section~\ref{sec:straw-prototypes}.

\subsubsection{Simulation of straw performance}
\label{sec:simulation-straw-performance}

\begin{figure}[tbp]
 \centering
 \includegraphics*[width=0.42\textwidth,angle=0]{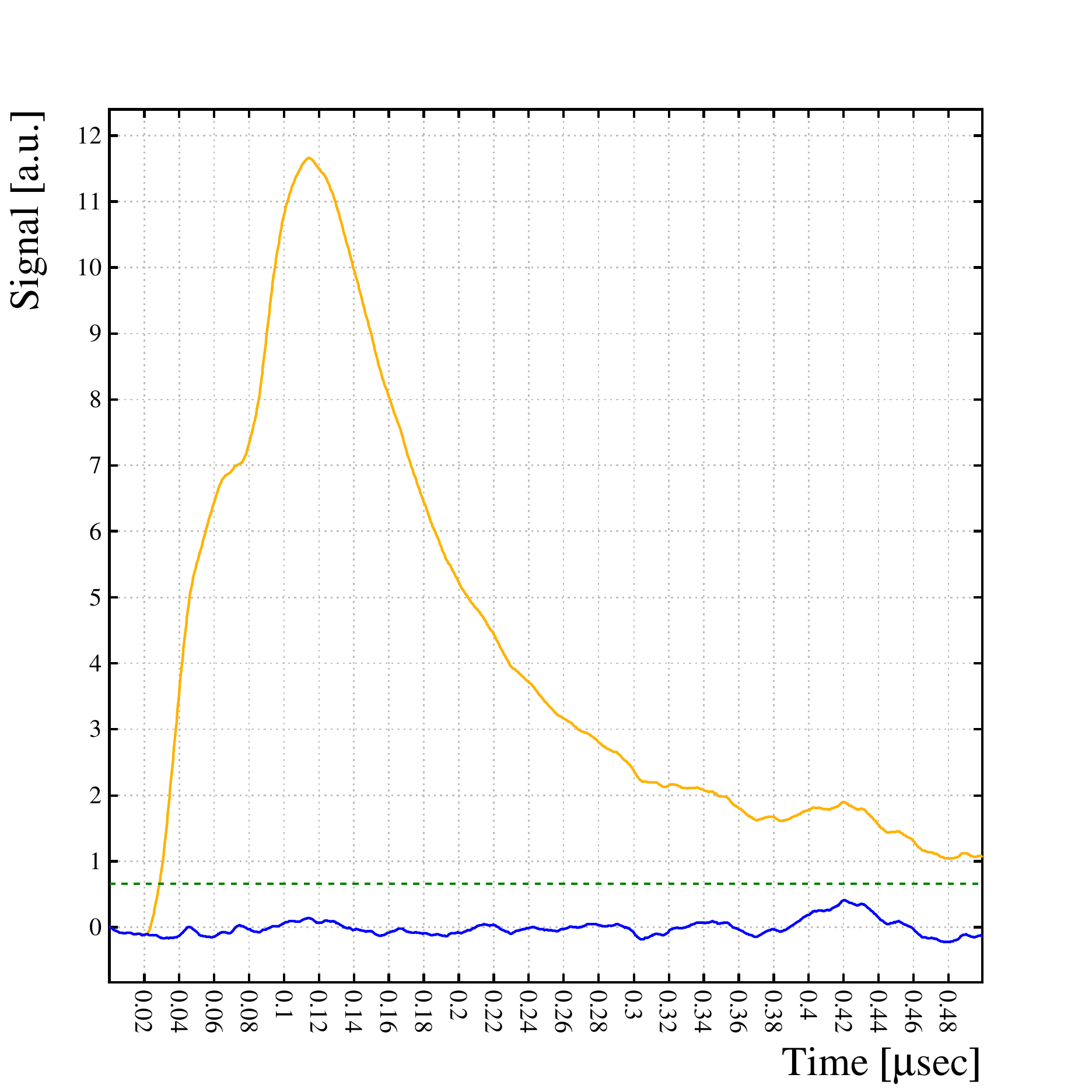}
 \includegraphics*[width=0.57\textwidth,angle=0]{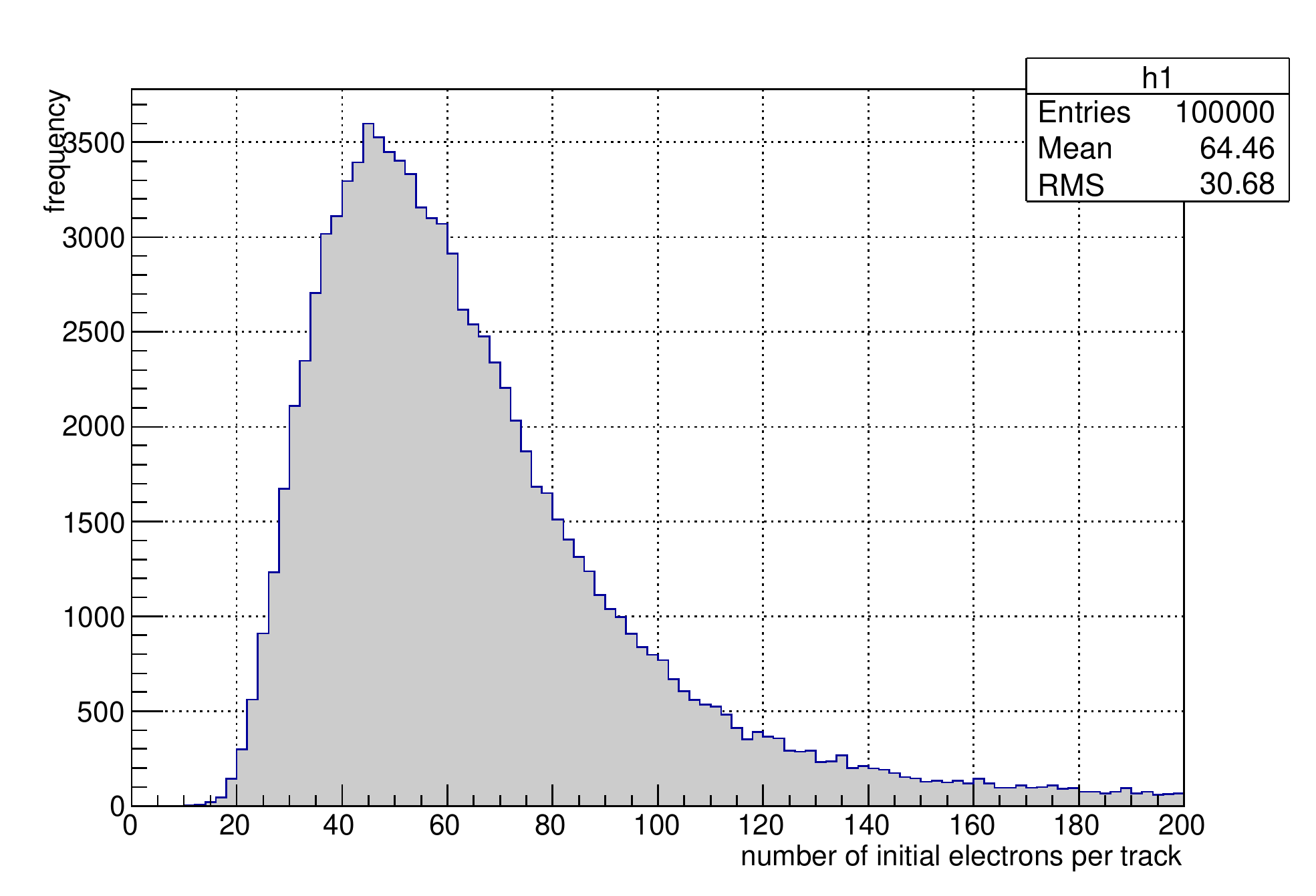}
 \caption{(Left) GARFIELD simulation results showing here the signal as a function
   of time for a 1~GeV muon traversing the drift tube.
   (Right) Distribution of the number of primary electrons produced in the straw
   when homogeneously irradiated.} % \mycomment{PLACE HOLDER}.}
  \label{fig:tracker-garfield-signal}
\end{figure}

%An important challenge of the chosen straw tracker technology is related to 
An important challenge for 5~m long horizontal straws is related to  
the fact that both the wire and straw will be subjected to forces (gravitational,
electrical, mechanical) which will cause sagging.
Because the straws are oriented horizontally (or almost, in case of stereo views),
sagging is expected to cause in most drift tubes a downward deflection, which 
might be exploited when applying a correction.
Sagging of a straw is caused by the overpressure which elongates the straw
while the extremities are fixed at the flange.
% which  because the straw is fixed at the extremities, elongates the straw. 
This effect will be studied
on prototypes and, if needed, countered by applying a pre-tension when mounting the straws
(in NA62 a pre-tension of about 2~kg was applied) and by using straw locators at discrete positions.
The gravitational sag $s$ for a horizontal wire of length $\ell$, density $\rho$,
cross sectional area $\sigma =\pi\,a^2$  ($a$ is the wire radius)
and strung under tension $T$  is given by
%\footnote{ The lowest resonance frequency of the wire is
%%\begin{equation}
%$ f_1 = \displaystyle\frac{1}{2\, \ell} \, \sqrt{\displaystyle\frac{T}{\rho\, \sigma}} = \sqrt{\displaystyle\frac{g}{32\, s}}$
%%\end{equation}
%must also be considered.}
%\begin{equation}
$  s = \rho\, g \, \sigma  \ell^2/(8T)$ \citBRR.
%\end{equation}
When applying high voltage, the sagitta increases.
For a perfect drift tube,  
%$E(b)$ the electric field at the tube wall (radius $r=b$) is
the electric field at the tube wall (radius $r=b$) is
%\begin{equation}
% E(b) =  \displaystyle\frac{U}{b\, \ln(b/a)}~$.
$ E_b =  U/(b\, \ln(b/a))~$.
%\end{equation}
%Defining $k^2=2\pi\varepsilon_0\,E^2(b)/T$ and $q=k\,\ell/2$, and assuming the wire
Defining $q=\sqrt{2\pi\varepsilon_0/T}\,E_b\,\ell/2$ and assuming the wire
is already deflected off-axis by gravity, the sagitta is then increased 
to 
% by the factor $s_k/s$
%\begin{equation}
  %\displaystyle\frac{s_k}{s} = \displaystyle\frac{2}{q^2} \, \Big( \displaystyle\frac{1}{\cos q} -1 \Big) 
  %s_k = \displaystyle\frac{2}{q^2} \, \Big( \displaystyle\frac{1}{\cos q} -1 \Big) \, s
$ s_q = 2q^{-2}\,[ (1/\cos q) -1 ] \, s$
%\end{equation}
when applying voltage. % The actual sagitta is $s_k$.
%If  $k\,\ell$ approaches $\pi$  (\ie $q\rightarrow \pi/2$), the wire becomes unstable.
If $q$ approaches  $\pi/2$, the wire becomes unstable.
Taking as example 
$U=1.75$~kV, $b=5$~mm, $a=0.015$~mm, 
% $\ln(b/a) = 5.8$,  $E(b) = 60$~kV/m,
%one expects
%Use maximum tension for W: $T/(\pi a^2) = T_c/\sigma =  410~{\rm kg/mm^2}\cdot g$, thus $T = 0.29~{\rm kg}\cdot g = 2.9$~N.
%This gives $s=144~\mu$m.
%
%Thus, $k^2=2\pi\varepsilon_0\,(60~{\rm kV/m})^2 /  2.9~{\rm N} = 0.07~{\rm m}^{-2}$ and, for $\ell=5~$m, $k\,\ell = 1.32$.
%
%This gives $q=0.659$ and $s_k/s = 1.22$, which is still reasonable.
%With a reduced and safer tension (i.e. using a $T_c$ below $180~{\rm kg/mm^2}$),
$T= % \pi a^2\, 150~{\rm kg/mm^2}\cdot g = 
1$~N, one expects 
%$s=418~\mu$m. Then follows $q = 1.12$ and  $s_k = 2.07\, s = 863~\mu$m which is 
$s\approx 0.4~$mm, $q = 1.12$ and  $s_q \approx 0.9~$mm which is considerable but
still acceptable.
The wire offset effects can be minimized by straw locators that allow one to adjust 
the amount of sagging of the straw tubes such as to be equal to the unavoidable sagging
of the wires.

\begin{figure}[tbp]
 \centering
 \includegraphics*[width=0.48\textwidth]{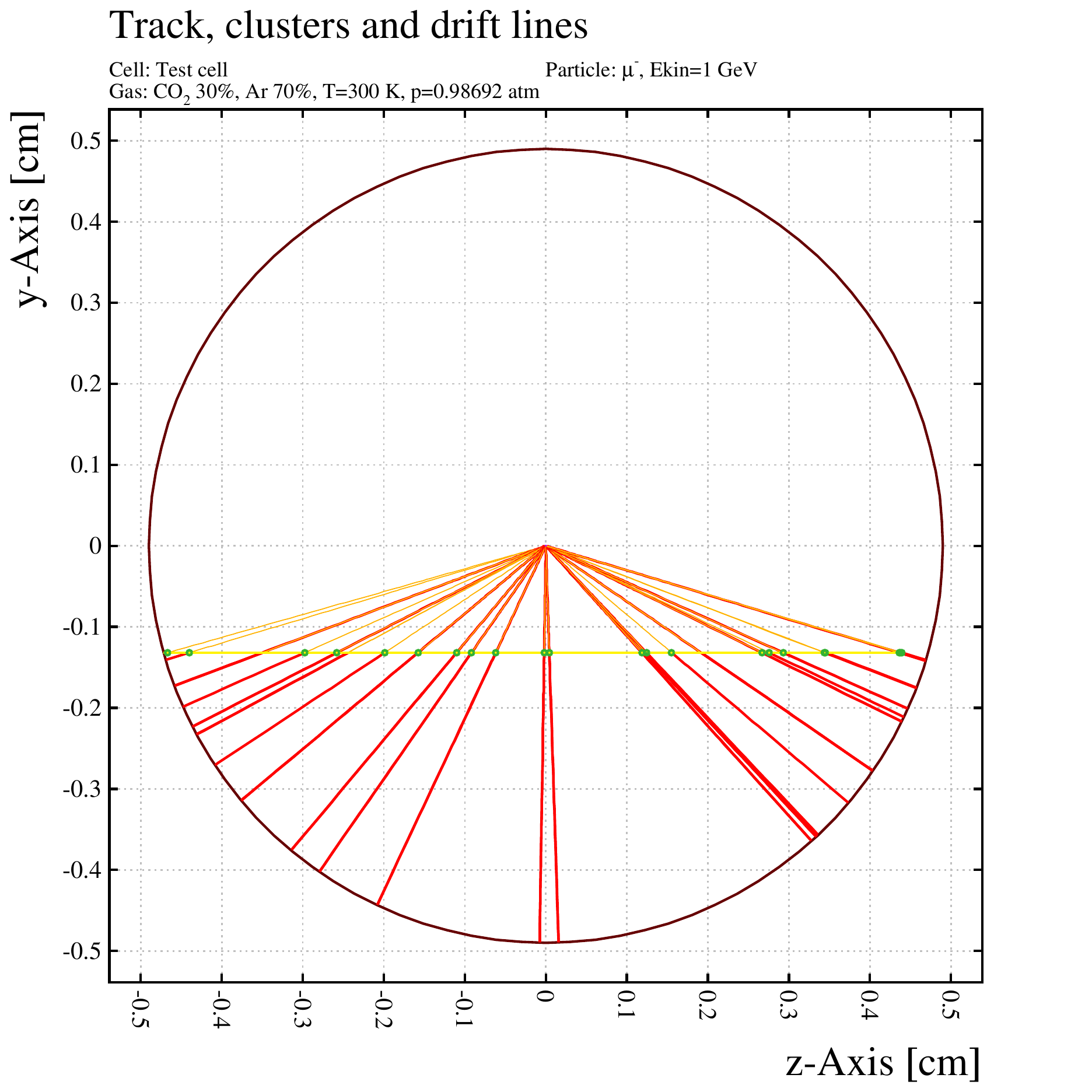}
 \includegraphics*[width=0.48\textwidth]{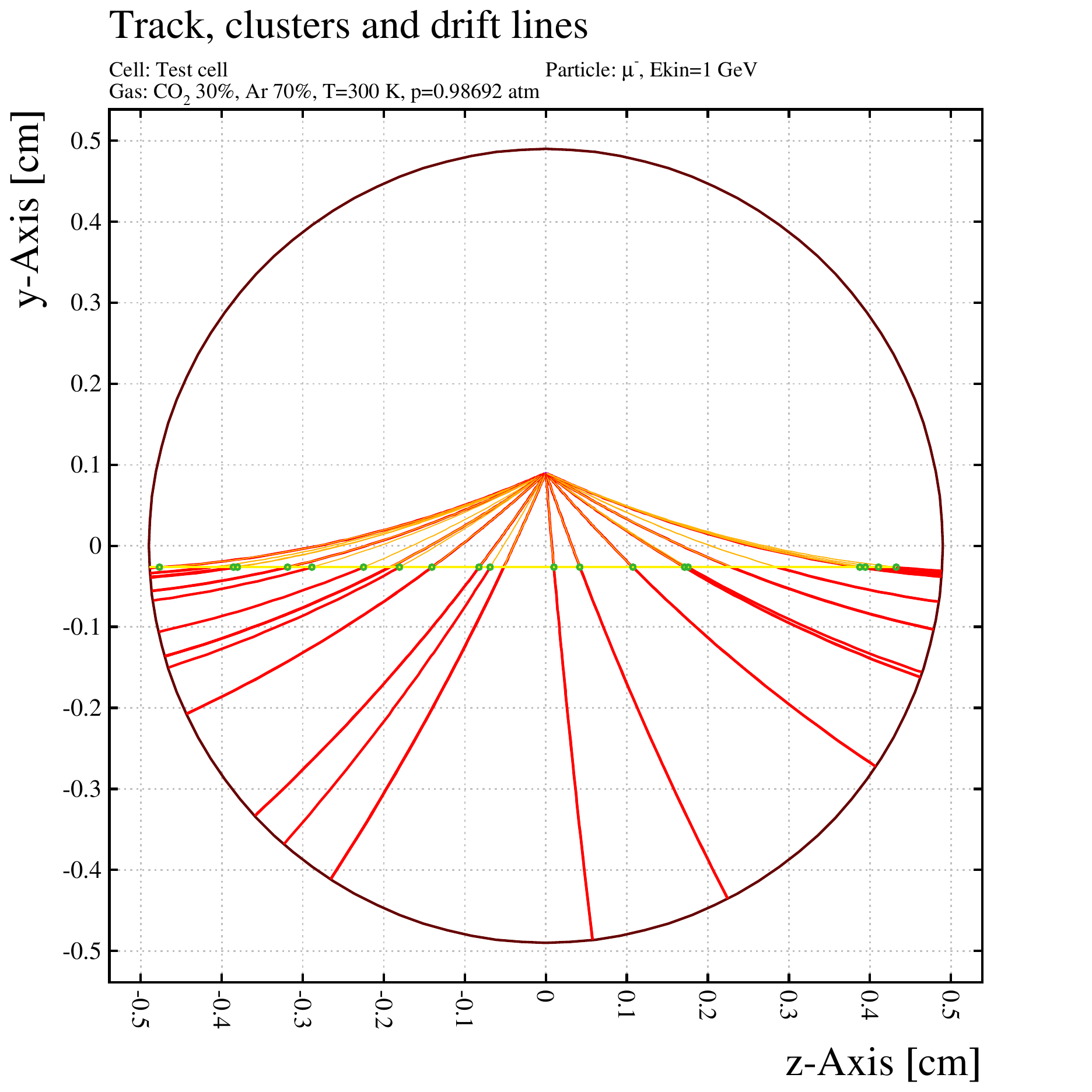}
 \includegraphics*[width=0.48\textwidth,height=50mm]{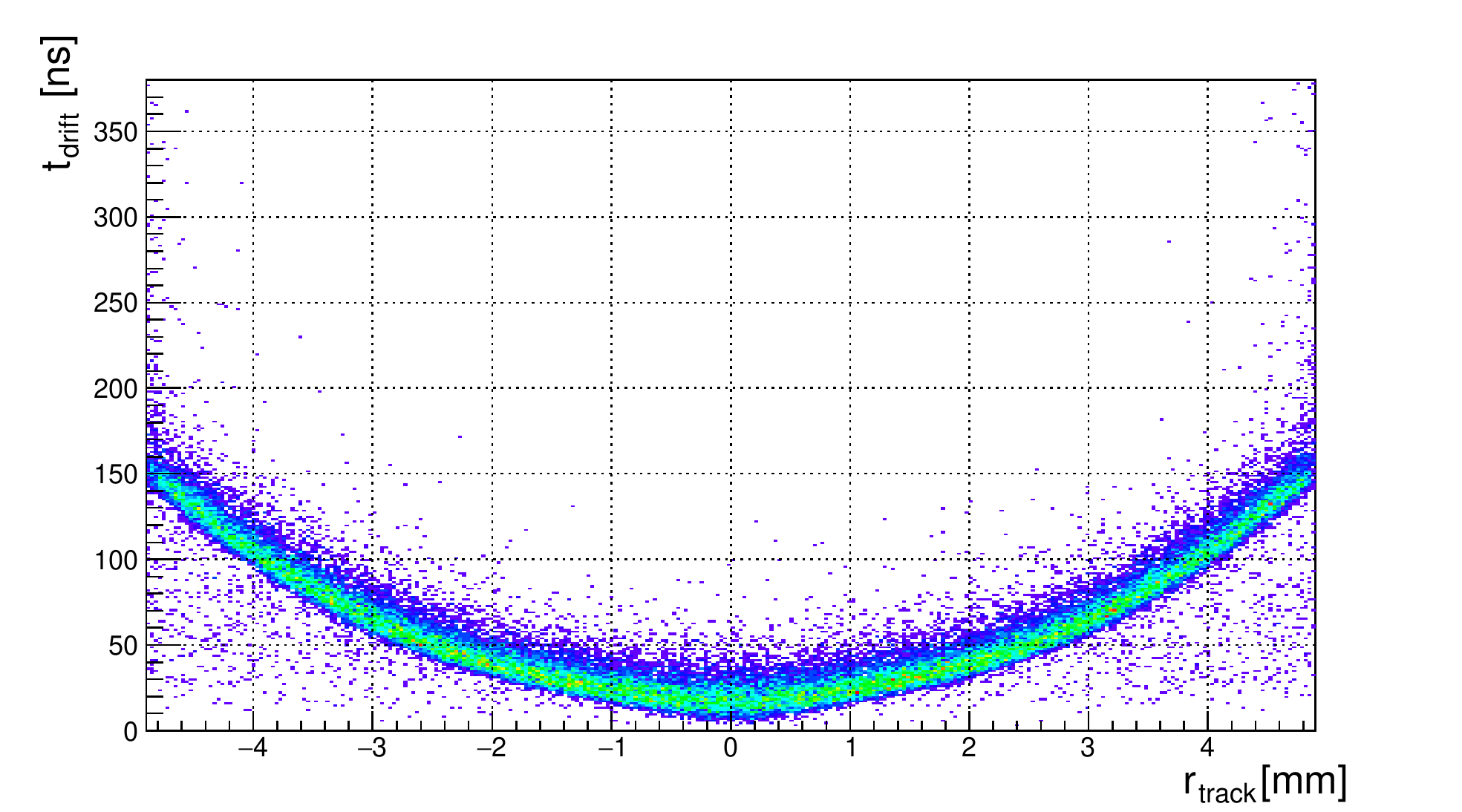}
 \includegraphics*[width=0.48\textwidth,height=50mm]{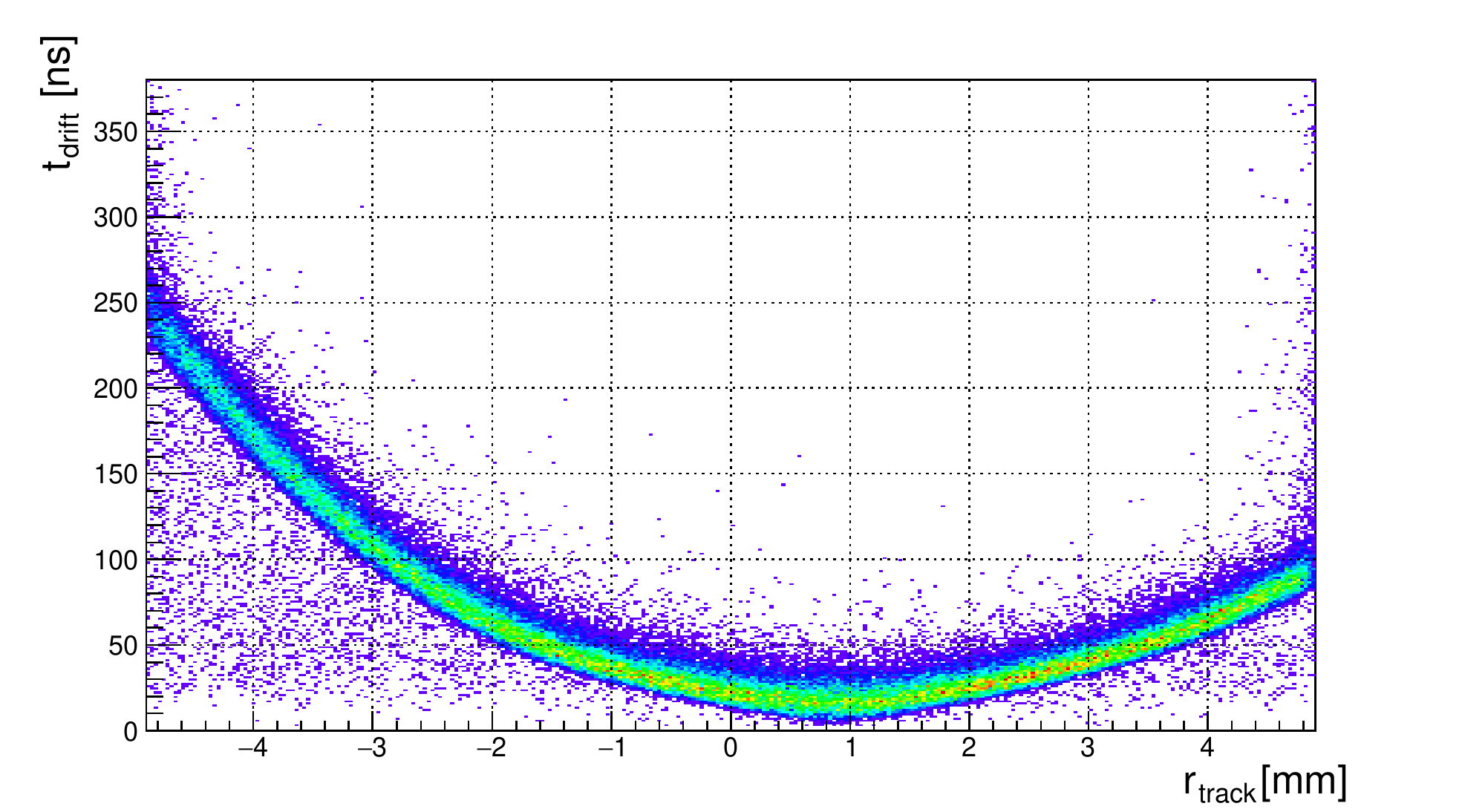}
 \caption{GARFIELD simulation results. The top graphs show the drift lines of
  electrons and ions for a centered (left) and 0.9~mm offset wire (right).
  The track comes from the left and green dots indicate the primary clusters.
  The bottom graphs show the drift time versus track offset for the same cases.}
  \label{fig:tracker-garfield}
\end{figure}

GARFIELD~\citGARFIELD simulations are performed to explore the effect of wire sagging~\citIvan.
In particular, the effects of sagging on gain and on the relation  between track position
and drift time are investigated. 
Studies are ongoing to define a method that allows one to recover the resolution assuming that 
one can calibrate locally the wire sag.
%, by some experimental measurement. % (and with some accuracy to be specified) ?
%For instance, by acquiring sufficient track hits and by
%projecting the measured drift time distribution, one may be able to distinguish an undeflected wire
%(no sag) from a deflected wire. %If so, with what precision ?
% But with  ? What about a 0.1mm-deflected ?
%Can one model that and fit out the deflection ?
%If yes, with what accuracy ? How many tracks does one need ? etc.
However, the effect of sagging is entangled with the effect of misalignment. 
Detailed studies and test beam measurements will be required to establish
the precision of this method.
%For instance, if the straw tube sags
%exactly like the wire sags, then the drift time distribution is exactly like
%for a straight tube with a straight wire. And yet the sagging is there.

%PS: the left-right ambiguity can be ignored for the time being.
%Just assume that you can determine (from a track fit of several
%measured hits) whether the hit is on the left or right of the wire.
%Later, we can study and refine that.

The parameters used for the first investigations are 
a voltage of 1750~V, a straw (wire) diameter of 10~mm (30~$\mu$m),
a gas mixture of Ar(70\%)/CO$_2$(30\%) at 1~bar, a temperature 
of 23~$^\circ$C and a Penning transfer probability set at 55\%.
Magnetic field effects will be checked in the future, but are expected to be negligible 
given the maximum $B$ field of 0.08~T in the stations.
Muons traversing the drift tube with an energy of 1~GeV are simulated.

\begin{figure}[tbp]
 \centering
 \includegraphics*[width=0.69\textwidth]{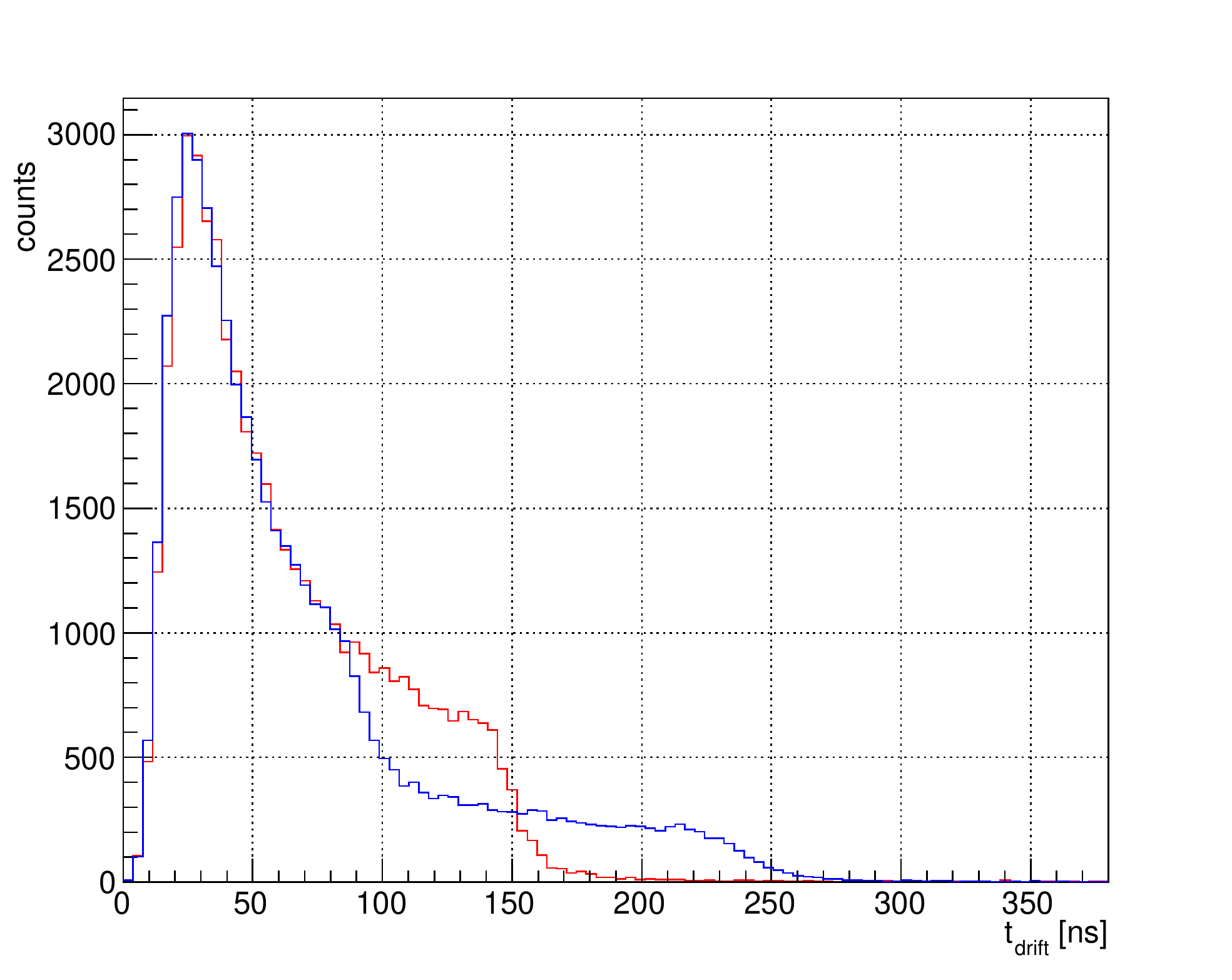}
 \caption{GARFIELD simulation results.  %For 50k events.  
      Drift time distribution for a homogeneous irradiation with
      a centered wire (red) and for a wire offset of 0.9~mm (blue).
     }\label{fig:tracker-garfield-tdrift}
\end{figure}

\def\signoise{\ensuremath{\sigma_{\rm noise}\xspace}}
Figure~\ref{fig:tracker-garfield-signal} (left) shows the signal as a function of time obtained
from a GARFIELD simulation which includes a random Gaussian noise (RMS $\signoise$).
A threshold is applied on the rising edge at $4\signoise$ to obtain the drift time.
Such signal responses are generated for a large number of tracks illuminating the straw homogeneously
and for different wire offsets perpendicular to the track direction.
Drift time distributions and gain are investigated.
The right plot shows the distribution of the number of primary electrons produced in the straw
with homogeneous irradiation.
The mean number of primary electrons is about 60 and the mean number of clusters is about 30.
In Figure~\ref{fig:tracker-garfield} the top graphs show %first results of such simulations.  The two top plots show 
the drifts lines from the clusters (green dots) generated
along the ionizing track, for 0~mm wire offset (left) and 0.9~mm wire offset (right).
The bottom plots show the drift time versus track offset, again for 0 and 0.9~mm wire offsets.
One observes that the large wire offset does not introduce much deformation
of the curve, apart from the obvious shift. 
In Figure~\ref{fig:tracker-garfield-tdrift} the projected drift time distribution 
is shown for the case of zero offset (red) and 0.9~mm offset (blue).
A clear difference between the two distributions is observed.
This property, if verified in the test measurements, could be used to measure the sag along the
straw from the drift time data.
%(here the offset is to be understood as the offset relative to the straw centre, which itself can have a sag).

\subsubsection{Tracker performance studies from simulation}
\label{sec:Tracker performance studies from simulation}
%\mysubsection{Requirements from simulation}{Ekaterina Kuznetsova}

The tracker performance in terms of resolution is driven by two effects, multiple scattering 
and detector hit resolution. In the case of momentum resolution, 
the error caused by both multiple scattering and detector resolution inversely scales 
with the magnetic field integral.
\def\numhits{N_{\rm hits}} % number of hits per view
\def\numview{N_{\rm views}} % number of views per station
Assuming the NA62 material budget, i.e. 0.5\% of a radiation length per station,
% $x/x_0\approx  0.12\%$ per view % 0.1136\%$ per view
%\mycomment{does this correspond to traversing 2 straws per view ??? seems too much},
a position resolution $\sighit=120~\mu{\rm m}$ per straw hit, 
8 hits per station on average, and the above station layout, one expects 
a momentum resolution 
\begin{equation}
\begin{array}{rclcc}
  \Big(\dyfr{\sigma_p}{p} \Big)^2 %&=& A_{\rm MS}^2     &+& B_{\rm Det}^2 \, p^2\\[4mm]
                              & \approx &  [ 0.49\% ]^2 &+& [0.022\%/({\rm GeV}/c)]^2 \cdot p^2 
\end{array}
\end{equation}
where the first term is due to multiple scattering and the second one to the detector resolution.
The momentum resolution of the spectrometer in this configuration is dominated by multiple scattering 
below about $22~{\rm GeV}/c$.

\begin{figure}[tbp]
 \centering
 \includegraphics*[width=0.56\textwidth]{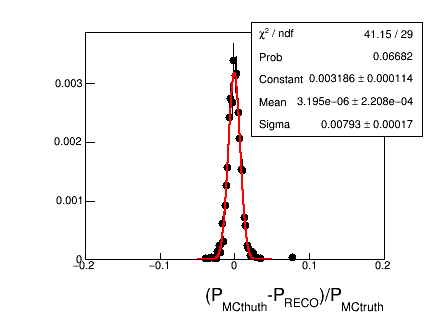}
 \includegraphics*[width=0.42\textwidth]{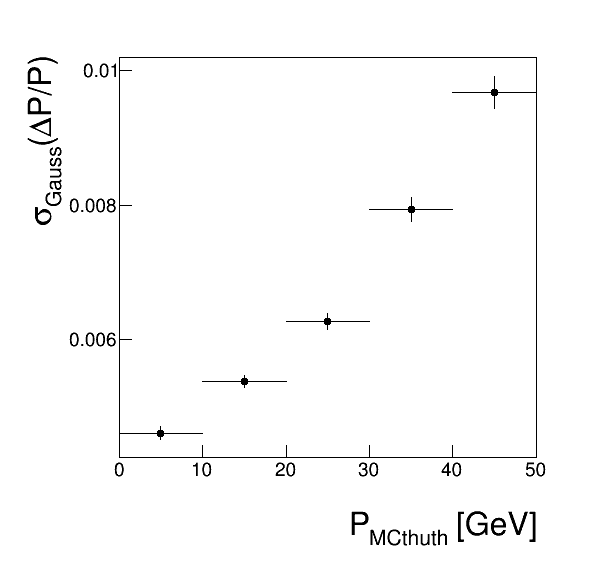}
 \caption{Tracker resolution for reconstructed HNL decays $N\rightarrow \ell^\pm\pi^\mp$. %with a mass $M_N = 1~{\rm GeV}/c^2$.
         (Left) Distribution of the difference between reconstructed and generated track momentum 
             for an example momentum bin (30 to $40~{\rm GeV}/c$).
         (Right) Relative error of the measured track momentum as a function of the generated track momentum.}
 \label{fig:tracker-perf-momresol}
\end{figure}

\def\sigms{\sigma_{\rm MS}}
\def\sigdet{\sigma_{\rm Det}}
The precision of the extrapolated track position at the decay vertex is also driven by 
the effects of multiple scattering and detector resolution.
Assuming the same detector parameters,  
% taking into account only the first and second station (at distances $z_1$ and $z_2$ from the decay vertex)
% and considering that the tracks have a small polar angle $\theta$, 
% then the impact parameter errors are approximately given by % $\sigma_w$ ($w=x,y$)
% $\sigma_w = ( \sigms^2 + \sigdet^2 )^\frac{1}{2}$ ($w=x,y$) 
% with
%  \begin{equation}
%    \sigms \approx   z_1\cdot      \dyfr{1}{1300 \, p~\mbox{[GeV]}/c}
%   \quad\quad\mbox{and}\quad\quad
%    \sigdet \approx  \sqrt{\dyfr{z_2^2+z_1^2}{2}} \cdot
%     \left\{
%      \begin{array}{l l}
%          \dyfr{1}{33000} & \quad\mbox{for }y \\[4mm]      % (0.12/(sqrt(2+2*(cos(5degrees))^2)*2000))^-1
%          \dyfr{1}{2000} & \quad\mbox{for }x               % (0.12/(sqrt(2*(sin(5degrees))^2)*2000))^-1
%   %
%   %\sigdet \approx  \sqrt{z_2^2+z_1^2} \cdot              % the same but with sqtr(2) absorbed...
%   % \left\{
%   %  \begin{array}{l l}
%   %      \dyfr{1}{47000} & \quad\mbox{for }y \\[4mm]      % (0.12/(sqrt(4+4*(cos(5degrees))^2)*2000))^-1
%   %      \dyfr{1}{2900} & \quad\mbox{for }x               % (0.12/(sqrt(4*(sin(5degrees))^2)*2000))^-1
%      \end{array}\right.
%  \label{eq:sigvtx_example}
%  \end{equation}
% indicating that for $p$ up to about 25~$(1.5)~{\rm GeV}/c$ multiple scattering effects dominate
% over detector resolution for the $y$ ($x$) coordinate.
one expects a two-track vertex position resolution of the order of one millimetre for
the $y$ vertex coordinate and about a factor of two to three more for the $x$ coordinate,
depending on the particle momentum and on the vertex $z$ distance from the first tracking station.
For the $z$ vertex coordinate the resolution is more than an order of magnitude 
worse due to a factor $1/\tan\theta$ ($\theta$ being the track polar angle in front of the tracker).

Since for an HNL decay $N\rightarrow \mu^+\pi^-$ with a mass $M_N = 1~{\rm GeV}/c^2$, 75\% of the decay
particles have both tracks with momentum $p < 20~{\rm GeV}/c$ (see Figure~\ref{fig:daughter_kinematics}),
there may be room for further optimization of the spectrometer, in particular
in the balance of $x$ versus $y$ and $p$ resolution.

%KUZN
\def\rirec{\ensuremath{r_{{\rm rec},i}}}
\def\rigen{\ensuremath{r_{{\rm gen},i}}}
\def\Pgen{\ensuremath{p_{\rm gen}}}
\def\Prec{\ensuremath{p_{\rm rec}}}
Simulation studies of the SHiP tracker performance are done for the currently assumed tracker geometry
%(NOT YET, will be updated with version from 03.02 with magnetic field etc) 
using two-prong decays ($N\rightarrow \ell^\pm\pi^\mp$).
Signal events are produced with the Pythia8~\citPythiaEight\, generator and the detector response is 
simulated using Geant4~\citGeantFour. 
A first version of the pattern recognition that uses only the wire positions
is described in~\citEricvH. 
This pattern recognition algorithm currently has an efficiency of $94.1\%$.
To take into account a realistic spatial resolution of an individual straw,
a Gaussian smearing of
120~$\mu$m is applied to each straw MC hit position. %\footnote{NB: resolution in code 100um (not 120um).}.
A `cheated' pattern recognition is applied using these hits.
The tracks are reconstructed from the straw hits using the Deterministic Annealing Filter 
of GENFIT~\citGenfit\, and matched to the generated particles. 
The reconstructed tracks that are matched to a signal decay particle are used to study the spatial 
and momentum resolution of the tracking system.

Figure~\ref{fig:tracker-perf-momresol} 
shows the track momentum resolution obtained for the reconstructed lepton and pion tracks
originating from HNL decays. The momentum resolution is obtained in bins of the generated
particle momentum $\Pgen$ from the sigma of a Gaussian fit to the $(\Pgen - \Prec)/\Pgen$
distribution, where $\Prec$ is the measured momentum of the reconstructed track.
The results are in agreement with the estimated value.
The resolution is better than 1\% for all tracks with a generated  momentum below $50~{\rm GeV}/c$.
\begin{figure}[tbp]
 \centering
 \includegraphics*[width=0.32\textwidth]{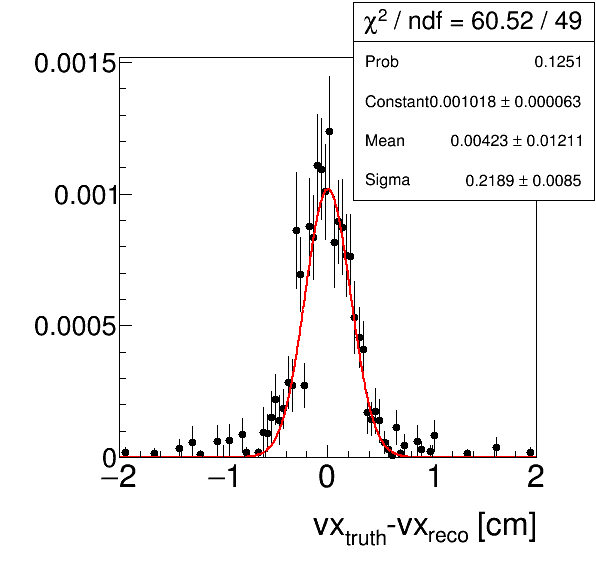}
 \includegraphics*[width=0.32\textwidth]{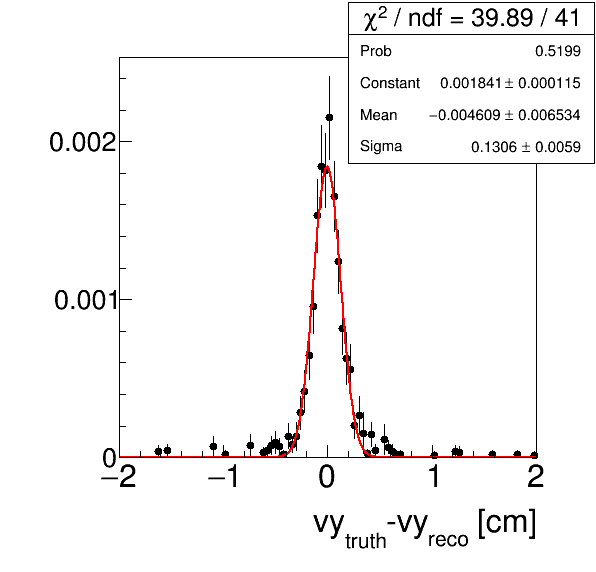}
 \includegraphics*[width=0.32\textwidth]{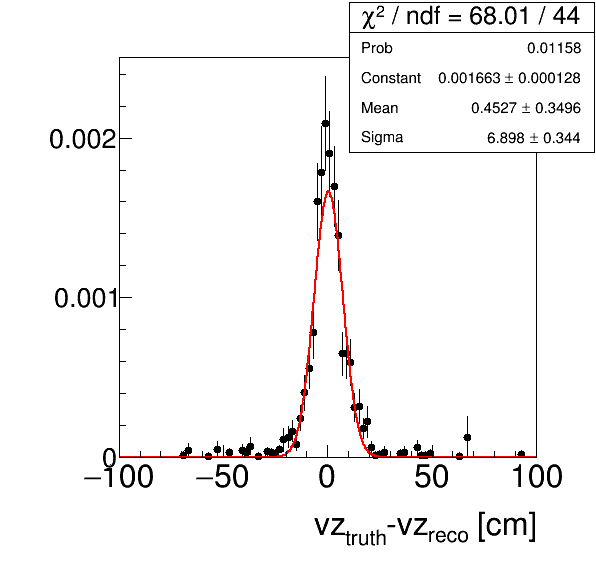}
% \caption{Distribution of the difference between reconstructed and generated vertex position 
%         for HNL two-prong decays,
%         for the $x$ (left), $y$ (middle) and $z$ (right) coordinates, obtained at 
%         a $z$ distance of 20--25~m in front of the first tracking station.
%         The corresponding Gauss curve fits are superimposed (red curve).}
% \label{fig:tracker-perf-vertexresol-zbin}
%\end{figure}
%\begin{figure}[tbp]
% \centering
 \includegraphics*[width=0.32\textwidth]{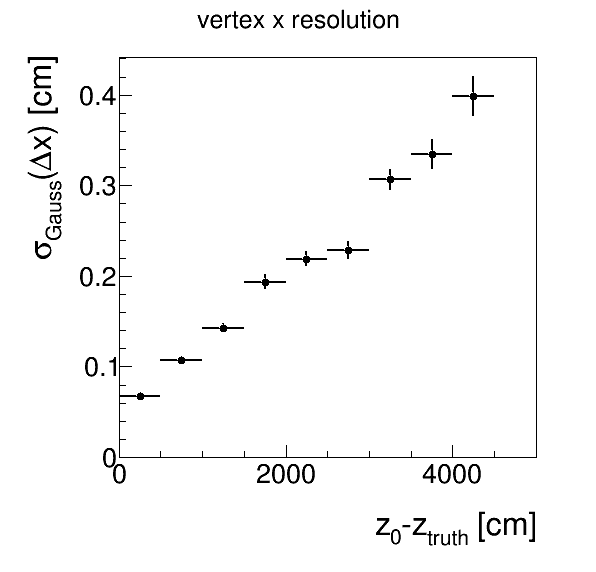}
 \includegraphics*[width=0.32\textwidth]{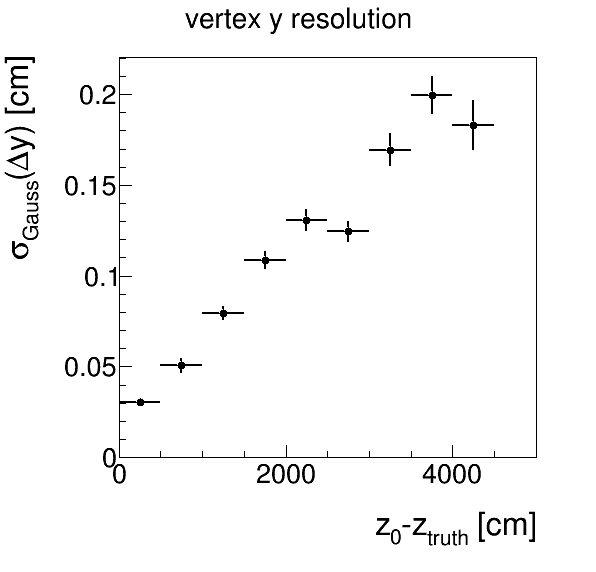}
 \includegraphics*[width=0.32\textwidth]{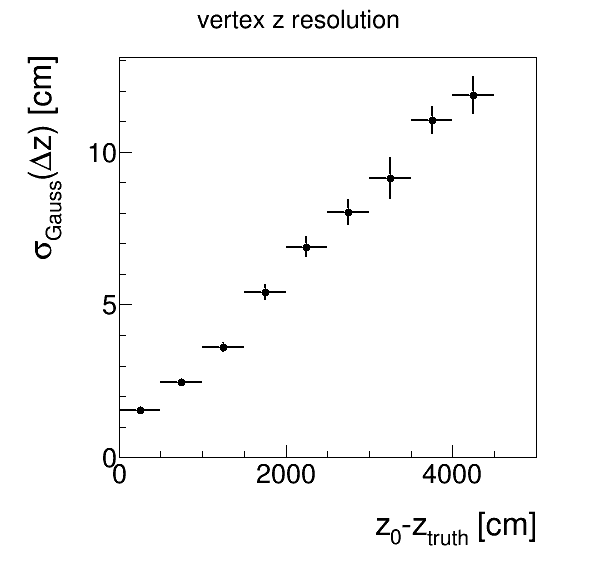}
% \caption{Resolution of vertex position as a function of $z$ distance.}  
%         %\mycomment{plots must be updated. A plot showing the $z$ dependence would be nice.}}
  \caption{Top: Distribution of the difference between reconstructed and generated vertex position 
          for HNL two-prong decays,
          for the $x$ (left), $y$ (middle) and $z$ (right) coordinates, obtained at 
          a $z$ distance of 20--25~m in front of the first tracking station.
          The corresponding Gauss curve fits are superimposed (red curve).
          Bottom: Standard deviation of  Gauss curve fits as a function of $z$ distance.}  
 \label{fig:tracker-perf-vertexresol}
\end{figure}

HNL decay vertices are assembled from the reconstructed tracks. The spatial resolution of a 
reconstructed vertex was obtained by fitting with a Gauss curve the distribution of the 
difference between the reconstructed and generated coordinates of the vertex $\rirec-\rigen$,
%see Figure~\ref{fig:tracker-perf-vertexresol-zbin}.
see top graphs of Figure~\ref{fig:tracker-perf-vertexresol}.
The transverse ($x$, $y$) resolution is found be of the order of a few mm, as approximately predicted,
and depends on the $z$ distance of the vertex from the first tracking station
%as is apparent in Figure~\ref{fig:tracker-perf-vertexresol}.
as is apparent on the bottom graphs of Figure~\ref{fig:tracker-perf-vertexresol}.
For the longitudinal position ($z$) the resolution is worse, as expected. 

The tracker performance in terms of signal reconstruction 
efficiency and background rejection is discussed in Section~\ref{sec:sensitivity}.

%\mysubsection{Straw prototypes}{Sergei Movchan / Temur Enik}
\subsubsection{Straw prototype developments}
\label{sec:straw-prototypes}

The current SHiP straw tube design largely builds on the experience of the NA62 experiment~\citNAsixtytwoTD.
Many of the NA62 ultra-light straws were produced at JINR, in Dubna.
The same tooling and base materials are now used to fabricate prototype straws of 5.3~m length
and 9.8~mm diameter for SHiP R\&D.
%The production of 5~m long prototype tubes is attempted with the same tooling and base materials.
The tubes are manufactured from $36~\mu$m thin PET foil, %polyethylene terephthalate (PET) foil,
coated on one side with two thin metal layers ($0.05~\mu$m of Cu and $0.02~\mu$m of Au)
to provide electrical conductivity on the cathode and to improve the straw tube gas 
impermeability.
NA62 has demonstrated that these straws can be operated in vacuum. 
A leak rate of about $7~{\rm mbar\,\ell/min}$ for the full detector (7168 straws) 
was measured~\citTemurEnik. 
A residual pressure inside the vacuum vessel of $10^{-5}~$mbar was achieved, 
which is  two orders of magnitude better than the pressure required for SHiP.
%Many of the NA62 ultra-light straws were produced at JINR, in Dubna.
%The same tooling was used to fabricate 50 straws of  5.3~m  length and 9.8~mm diameter  for SHiP R\&D,

\begin{figure}[tbp]
 \centering
 \includegraphics*[width=0.44\textwidth            ]{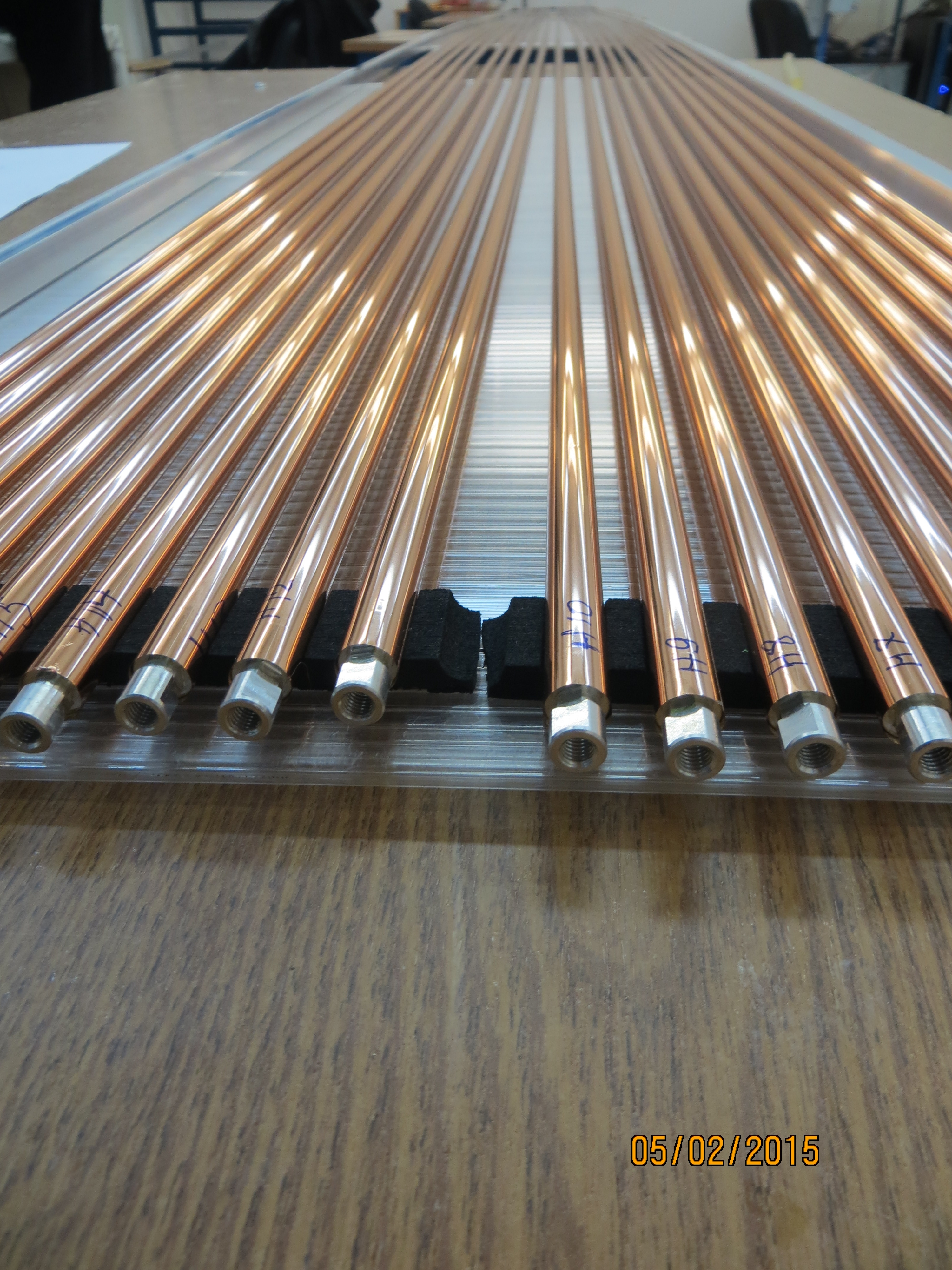}
 \includegraphics*[width=0.55\textwidth            ]{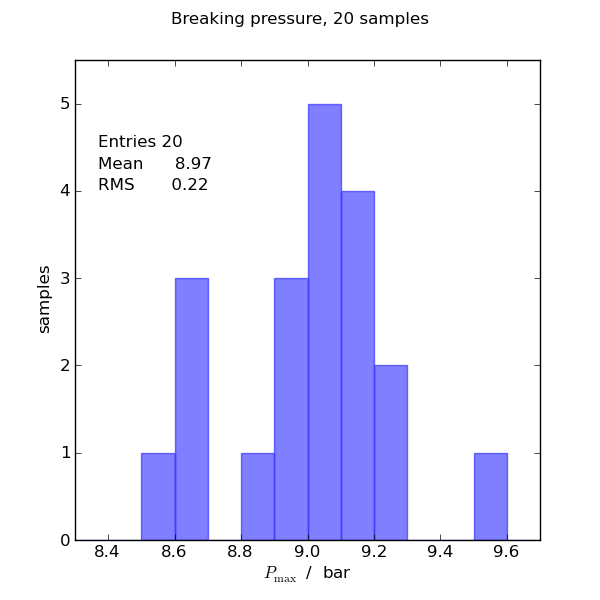}
 \caption{%Top left: picture of 5.1~m long prototype straws. 
          %Top middle: picture of the glued end-plugs on the straws.
          %Middle centre: Straws with glued temporary end plugs are ready for overpressure test.
          %Middle right: Straw machine common view (left - straw manufacturing in progress, right - set of  manufactured straws).
          %Bottom right: Straw overpressure test at $p=9~$bar of Ar. Straw shape before destruction.
          %\mycomment{Some pictures will be replaced by ``hot-off-the-press'' test results...}
          (Left) Picture of 5~m long prototype straws. 
          (Right) Breaking pressure $P_{\rm max}$ of 20 samples cut from a 5~m long straw.
}\label{fig:photo-straws}
\end{figure}

The first prototype straws for SHiP are shown in Figure~\ref{fig:photo-straws} (left).
A few straws are used for dedicated mechanical tests. They are cut in 20 segments of about 25~cm length
and tested under overpressure until the breaking point.
The other straws are cut to 5.3~m and the cut ends are stored for further analysis.
Figure~\ref{fig:photo-straws} (right) shows the distribution of the breaking pressure
measured for the 20 pieces of a test straw. The breaking pressure is 9~bar on average
and no sample broke under 8.5~bar.
%The straws undergo a long-term overpressure test with temporary end-plugs glued into
%both ends of each straw. %, as shown in Figure~\ref{fig:photo-straws} (right).

The quality control procedure will be the same as for NA62 straws.
During the ultrasonic welding process the seam quality is verified by a digital microscope 
(recorded to file for each straw). The seam quality is checked in real-time by the operator.
Of the 50 tubes produced so far, all had a good seam.
Post-fabrication, several measurements and tests are performed.
The seam width and straw inner diameter are measured by an optical method.
The cathode electrical DC resistance is measured.
The elongation and breaking force are both measured on the test samples (cut straw ends).
The straws undergo a long-term overpressure test with temporary end-plugs glued into
both ends of each straw. %, as shown in Figure~\ref{fig:photo-straws} (right).
An overpressure test to $\Delta p \approx 2~$bar is performed for a period of about 1~h.
Subsequently, the straw is subjected to a long term overpressure test at $\Delta p \approx 1~$bar 
for a period of at least 30 days. 
Gas leak estimation is obtained by measuring the loss of pressure over time.
The local straw deformation is measured under an applied weight of 300~g,
and the pressure is derived from the calibrated relation between loss of pressure and deformation.
%These trays are used for straw transportation to CERN and to the assembling area at JINR.
%The prototype straws will be transported to the assembly site under pressure
%in a long solid box with suitable packaging.

The straws are being prepared for transportation to CERN, where the 
straw array prototype will be assembled and tested with ionizing particles.

%\mycomment{Some first test results will be added: 
%  weld quality (camera), tensile test results, straw leak rate (pressure vs time),...}

\subsection{Readout electronics}
%\mysubsection{Readout electronics}{Yuri Ermoline/Iouri Guz} 
\label{sec:straw-readout-electronics}

\begin{figure}[tbp]
 \centering
 \includegraphics*[width=0.49\textwidth]{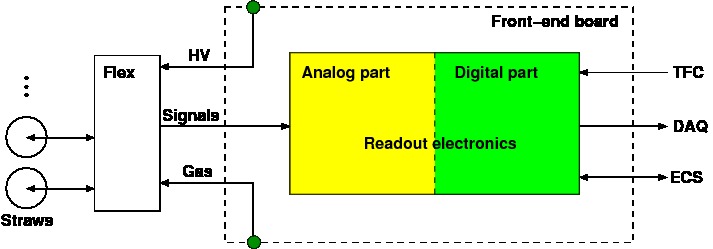}
 \includegraphics*[width=0.49\textwidth]{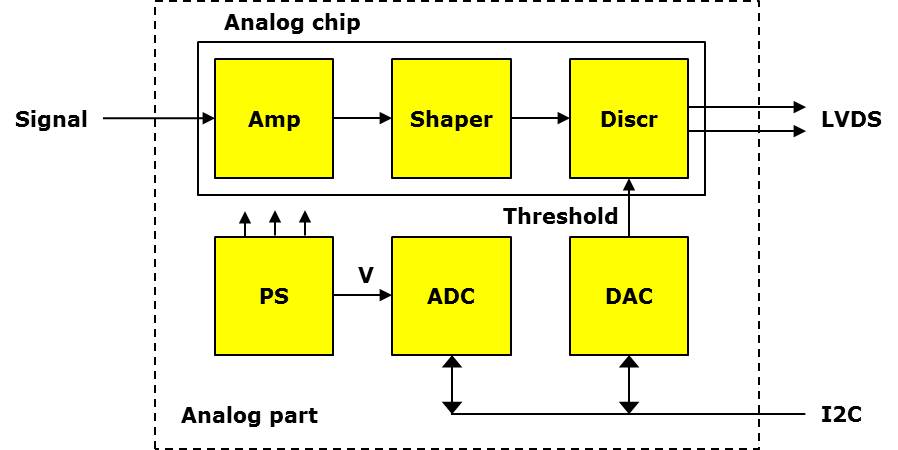}
 \caption{
  (Left) The general diagram of the tracker readout electronics.
  (Right) Functional diagram of the analog part of the readout electronics.
  }\label{fig:straw-electronics}
\end{figure}

The general diagram of the tracker readout electronics is shown on the left
 side of Figure~\ref{fig:straw-electronics}.
%Following the NA62 design, the tracker readout electronics 
They are split into
a front-end (FE) board which hosts the FE electronics and a flexible circuit
to interface with the straws. 
Since the radiation levels in SHiP are negligible, the circuits are located on 
the detector and no special radiation hard components are needed.
High voltage is fed to the straws via the FE board and flexible circuit.
The FE electronics are composed of an analog and a digital part. 
After analog processing of the drift tube signal, the arrival time of the
track signal is digitized and sent to the data acquisition system in the 
required format.
The fast signals (including the central clock) arrive from the
Timing \& Fast Control system (TFC) to the FE board.
An Experimental Control System (ECS) interface, e.g. using I2C, 
 allows one to configure and monitor the FE electronics.

The tracker detector is %a gaseous detector 
organized in four stations with a   number of readout electronics channels 
equal to the number of straws,
or double that amount if double-sided readout is used or if the 
5~m long straws are split into $2\times2.5~$m straws (in case it is found necessary).
The readout electronics modularity must conform with the straw mechanical fixation and gas 
distribution modularity. 
If using a modularity of 16 straws, 2272 readout boards are needed for the 36352 channels.
%The readout board comprises the readout electronics, high voltage and gas connections and 
%is connected to the straws via a flexible circuit.

As depicted in Figure~\ref{fig:straw-electronics} (right)
the analog part of the readout electronics 
contains an amplifier, shaper and discriminator.
%The second stage extracts timing information (a discriminator or digitizer). 
%Some parameters will need to be configurable (like thresholds), 
%other need only to be monitored (like power supply voltages). 
The requirements for this part are defined by the electrical properties
of the basic detector element, i.e. the drift tube, and by the position resolution 
to be achieved ($\sim 120~\mu$m). 

Assuming the parameters presented in Table~\ref{tab:straw-parameters},
the electrical properties of the %basic detector element, i.e. the
 drift tube are expected to be the following:
%\begin{itemize}
       capacitance $C=9.6~$pF/m, inductance $L=1.16~\mu$H/m;
       resistance $R=R_{\rm wire} + R_{\rm tube}$, 
       $R_{\rm wire}\approx80~\Omega/$m, $R_{\rm tube}\approx24~\Omega/$m; %, $R\approx115~\Omega/$m;
       %$R_{\rm wire}\approx80~\Omega/$m, $R_{\rm tube}\approx35~\Omega/$m; %, $R\approx115~\Omega/$m;
       characteristic impedance $Z_0=\sqrt{L/C}=350~\Omega$;
       characteristic frequency $f_0=R/(2\pi L) = 14~$MHz;
       %characteristic frequency $f_0=R/(2\pi L) = 16~$MHz;
       signal attenuation length at high frequencies $\lambda=2\,Z_0/R \approx 6.5~$m.
       %signal attenuation length at high frequencies $\lambda=2\,Z_0/R \approx 6~$m.
%\end{itemize}
The drift tube gain is expected to be in the order of a few  $10^4$.
Assuming an average of about 60 primary electrons per particle hit, the 
ionization charge is expected to be of the order of 100~fC.
The value of the input impedance of the preamplifier must be chosen such as
to provide good impedance matching at tolerable cross-talk.
%The value of the input impedance of the preamplifier is a compromise between a good straw impedance matching 
%and small cross-talk. A lower value is advantageous, as it guarantees lower cross-talk 
%and signal enhancement due to current increase during signal reflection on the low termination impedance.
%The reflected signal, which returns after travelling to the far end of the straw,
%is strongly attenuated and does not contribute significantly. % to the output. 

%Requirements for the analog part come from the required track position resolution,
%currently assumed to be 120~$\mu$m. 
The hit distance from the wire is extracted from the arrival time of the leading edge of the 
signal. %, which depends on the particle track distance from the anode wire. 
With a maximum drift distance of about 5~mm, i.e. an electron total drift time of the order of 150~ns, 
the minimum required drift time resolution would be 3-4~ns.
The shaping time should be short in order to get a response from the first primary cluster,
and thus a better time resolution, but should be sufficiently long to integrate further
clusters such that there is only a single output pulse per particle hit.
Assuming 0.3~mm average spacing between primary clusters, the shaping time should be 
much longer than about 15~(6)~ns for a slow (fast) gas.
%\mycomment{This may change in the future}.

The digital part of the readout electronics must provide time stamps for the leading
edges of the straw signals, processing, zero suppression, formatting and transmission
of digital data to the DAQ system via Ethernet. 
It will also include interfaces to the TFC system for timing
and synchronization via optical link, and to the ECS for control and monitoring
via Ethernet. 
The whole digital part of the readout electronics can be implemented in a single 
readout FPGA with high speed links. The optical transceivers and the Ethernet
physical layer chip will provide physical connections to the Online system.

The analog front-end chip could either be one produced for another experiment 
or a newly designed chip. 
%Design specifications will be drawn based on the output from the detector
%and signal studies. 
%An investigation of the usability and availability of existing chips
%will be carried out. A few examples are discussed in the following.
%
For the NA62 straw tracker the CARIOCA chip was used~\citCARIOCA.  
This is an 8-channel amplifier shaper discriminator developed for the LHCb muon chambers. 
The ASIC might be suitable for SHiP, but  is no longer available in sufficient quantities
and, because it was implemented using the IBM 0.25~$\mu$m CMOS process,
%Two time constants are used in the base line restoration, tailored for its specific application.
%It may be used with negative or positive input signal polarity.
%for the SHiP straw tracker 
cannot be re-manufactured without re-designing. %, due to the old technology.
Currently, there is no follow-up chip being developed.
Another candidate is the ASDBLR chip~\citASDBLR, designed in 2002 for the ATLAS TRT straw
detector and used also in the LHCb Outer Tracker (OT).
%ATLAS Upgrade Week 
%https://indico.cern.ch/event/108365/
%Development of a new front-end chip for the MDT chambers 15'
%Speaker: 	Robert Richter 
The ASDBLR acronym stands for Amplifier Shaper Discriminator with Base Line Restoration.
It  has two thresholds and two discriminator structures.
The 8-channel ASIC is implemented in bipolar technology,  in the phased-out DMILL process.
Due to the decommissioning of the LHCb OT during LHC Long Shutdown 2, several thousand chips 
may become available in 2018, probably enough to read out more than 40'000 electronics channels.
Although, the current OT digital part is not compatible with the current SHiP proposal,
new FPGA-based digitizer boards were developed at NIKHEF when the straw tube technology
was still being considered for the LHCb upgrade~\citLHCbUpgradeLOI.
Their design is compatible with the SHiP electronics architecture and first prototypes
were already produced.
Other candidate ASICs are being investigated.
%(an upgrade of the ATLAS ASD2 chip in 130~nm CMOS process
%is under development for the monitored drift tube chambers.

%The timing information is extracted from the signal shape, e.g. from the arrival time of the leading edge 
%of the straw signal with respect to a master clock. The required drift time resolution is 
%3-4 ns and a single-ended straw readout will be sufficient to achieve this resolution. 
%The extraction of timing information can be achieved in two different ways:
%(a) by discrimination of the shaped straw signal,  followed by edge detection and time stamping
%of the discriminator output pulse with sub-nanosecond resolution by TDC,
%or (b) by fast sampling and digitization of the shaped straw signal,
%followed by digital signal processing of the acquired samples to extract the 
%timing information.

%Currently, the first method (with TDC) seems the most realistic. %suitable.
% The TDC can be directly implemented in the digital part.
%Multiple channel sub-nanosecond TDC was recently implemented in low-cost FPGA.
%The second method will be tested in the laboratory and conclusions will be made 
%about its applicability.
%

Detector and signal studies will be performed in order to define the specifications 
for the input impedance of the preamplifier, the shaping
time and the ion tail cancellation.
The parameters and functionalities of all functional blocks of the FE electronics 
will be specified in details
on the basis of detector and signal simulation and test measurements.

%\mysubsection{R\&D Activities}{Massi Ferro-Luzzi}
\subsection{R\&D Activities}
\label{sec:straw-tracker-RandD}

An R\&D programme is underway
to establish the feasability of the currently considered SHiP straw tracker
geometry.  
The construction and operation of more than 36'000 drift tubes of 5~m length in vacuum 
remains a challenge and must be demonstrated on a prototype with several straws. 
The NA62 experiment has shown that close to 8'000 such elements with a length
of 2.1~m  can  be produced and operated successfully. 
For SHiP this means that the use of 2$\times$2.5~m straws, instead of 5~m long straws,
could be a  fall-back solution,  although the effect of the additional
material in the mid-plane of the experiment would have to be carefully evaluated. 

In view of preparing for a technical design, the R\&D for the 
SHiP tracker will concentrate on the following areas.

\subsubsection{Straw geometry optimization}
%\mysubsubsection{Straw geometry optimization}

   Critical design parameters of the tracker are the straw and wire dimensions, as those cannot be
   changed once built,  contrary to the gas mixture, gas pressure and anode high voltage 
   which can be adjusted within some range.
   Given the much reduced rate expected for the SHiP tracker when compared to the NA62 straw tracker,
   the use of a slower gas and/or larger straw diameter is  being considered,
   as well as a different operating voltage.
   Sagging effects may be reduced by a judicious choice of these parameters.
   The GARFIELD package will be used to simulate the performance of a drift tube,
   in terms of resolution, gain, signal shape, including the effect of sagging.
   The performance will be studied as a function of a few parameters such as straw and wire
   diameters, voltage, gas pressure and composition.
   The results will be compared to measurements obtained from straw prototypes using
   radioactive sources and high energy test beams.

   GARFIELD and full tracker simulation studies will be continued 
   to find optimal parameters and to specify how much sagging can be tolerated when 
   correcting the effect of wire offsets by 
   a calibration method that uses e.g. the local drift distribution.

\subsubsection{Tracker geometry optimization}
%\mysubsubsection{Tracker geometry optimization}

   Using the FairShip full simulation, the global performance of the spectrometer tracker
   will be investigated.
   We expect pattern recognition to be relatively easy at the expected occupancy. 
   The studies will focus on optimization of resolution performance such as
   to achieve best possible background rejection.
   For example, an increase of the stereo angle
   will improve the X resolution at a small cost for the Y resolution
   and may give a better balance between spatial resolution and momentum resolution.
   A reduction of the number of layers
   could also result in a better global performance if combined with a better straw resolution 
   and/or a larger lever arm, at a small cost in track efficiency.
   Design parameters that will be optimized are the stereo angle,
   the distance between planes,
   the intrinsic straw resolution
   and the number of layers (a way to reduce the amount of 
   straw tracker material in the acceptance).
   %The current design studies indicate that the performance (for the 1~GeV HNL study case)
   %dominated by multiple scattering. 
   %Optimize for $>90\%$ HNL reconstruction efficiency.
   %Try different straw layouts ? (larger diameter, less layers but better resolution, etc.)

   A more realistic description of the drift tube signal and digitization will be implemented  
   in the FairShip framework, based on GARFIELD results and test measurements.
   A strategy for a track-based alignment will be developed, taking into account
   the effect of sagging.

   These simulation studies will be used for guidance toward a final design.

%  How to reduce $A_{\rm MS}$ ? Reduce number of layers/view or number of views ?
%  Yes, but this also increases $B_{\rm Det}$ with the balanced $\sqrt{\numview}$ dependence...
%  So, one should simultaneously increase $\Delta$ or reduce $\sighit$  by some means...
% (slower gas) ???

\subsubsection{Straw prototypes}
%\mysubsubsection{Straw prototypes}

   Enough 5~m long straw prototypes have already been fabricated at JINR to construct
   a first tracker prototype with several layers of a single view (e.g. an array of
   $4\times 8$ straws).
   Quality control tests are ongoing and first results look promising.
   Several key quantities, such as straw leak rate, tensile strength and breakdown pressure
   are being or will be measured.

   Straw and wire sagging measurements will be performed in realistic conditions
   (horizontal, with high voltage) using optical systems with approximately 50~$\mu$m 
   precision, as was performed for the NA62 experiment.
   The straw prototypes will be exposed to high energy particles to check the 
   drift time distribution and investigate methods to correct for sagging effects 
   under controlled conditions.

   The total cost of electronics scales with the number of electronics channels, and therefore 
   with the straw diameter. 
   In addition, the risk of wire instabilities becomes smaller for larger straw diameters
   and the amount of material is also slightly reduced. % less wires, less straw boundary overlaps.
   Therefore, substantial effort will be devoted to the possibility of making
   drift tubes with different cross sectional dimensions (wall thickness, straw diameter, wire diameter),
   guided also by GARFIELD and FairShip simulation studies.

   An increase of the straw diameter also means a lower breakdown pressure.
   Preliminary results on the 9.8~mm diameter straws indicate that there may be
   enough margin to increase the diameter by a factor two.
   %Nevertheless, %Although challenging,  
   Improvements of the straw weld quality, for example by masking
   the PET foil over the seam surface when applying the metal coating, 
   will also be explored.

   Straw signals under varying conditions, such as high voltage, gas pressure and gas mixture,
   will be investigated, including attenuation and double-sided readout\footnote{A
   Double-sided readout could also give a measurement of the X coordinate by 
   time difference, see Ref.~\citMakankin. This will also be investigated.}. 
   For this purpose, FE PCBs with discrete components and a commercial readout system are being developed.
   Depending on the value of the measured signal attenuation, one may consider to alternate the readout 
   from one end to the other end of the straw for subsequent straws. This would guarantee
   an equalization of the number of hits with low and high attenuation for any given track
   and, if there is sufficient redundancy in the number of hits per track, could be sufficient
   to maintain a high track efficiency.

   A mechanical design with finite element analysis will be produced to validate that a 5~m~$\times$~10~m
   straw view can be constructed with acceptable straw and wire sagging.
   Different straw or wire tensions and suspension systems will be investigated.

%% file: magnet/SpectrometerMagnet.tex
\section{Spectrometer magnet}
\label{sec:spectrometermagnet}

As shown in Figure~\ref{vessel:VV3Dmodel} the spectrometer magnet surrounds the slightly smaller rear 
section of the vacuum vessel after the 50~m decay volume. The rear section consist of a single-wall 
non-magnetic steel chamber without surrounding liquid scintillator modules. For this reason, 
the rear section of the vessel has a transversal size of 5~m width and 10~m height. In order to cover the wide 
momentum range and properly match the spatial resolution and multiple scattering in the straw 
tracker, the dipole provides a bending power of 0.65~Tm with an option to go up to 1.0~Tm with 
an upgrade.

\subsection{Magnet design and performance}

The magnet has been designed as a low-power normal conducting dipole with a free aperture of 
about 5.10$\times$10.35~m$^2$ and a horizontal field in order to provide a vertical bending plane for the 
horizontally oriented straw tracker (Section~\ref{sec:tracker}). Currently there is no foreseen application 
for switching the polarity of the magnet. % (true? straw alignment?).

\begin{figure}[tb]
\begin{center}
\includegraphics[width=0.35\linewidth]{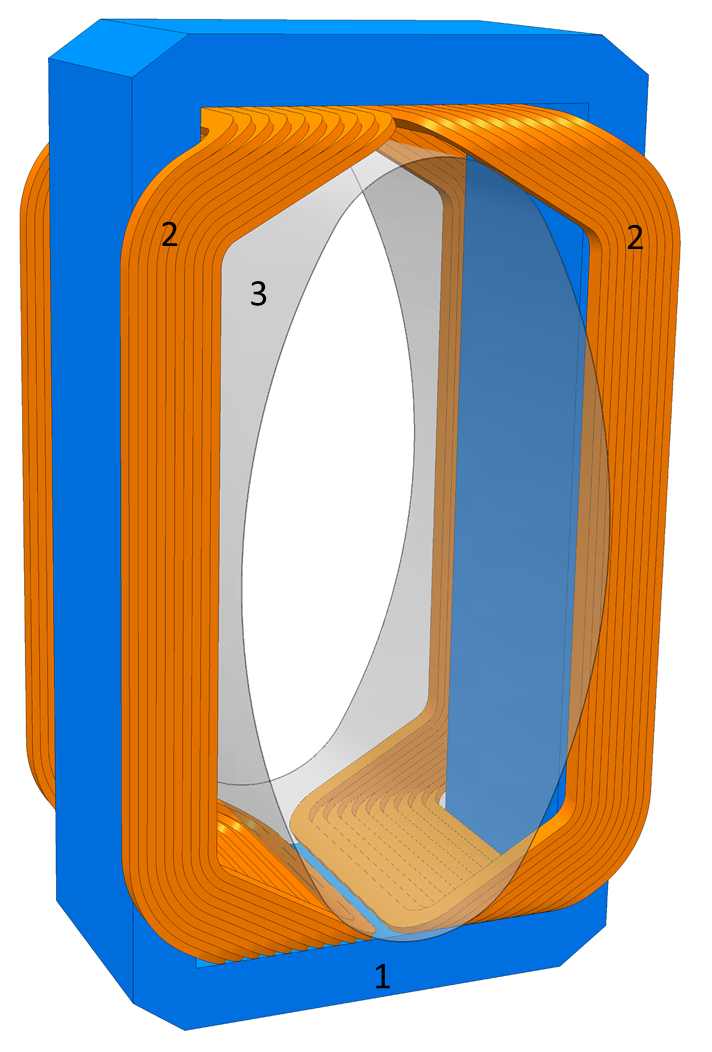}
\includegraphics[width=0.5\linewidth]{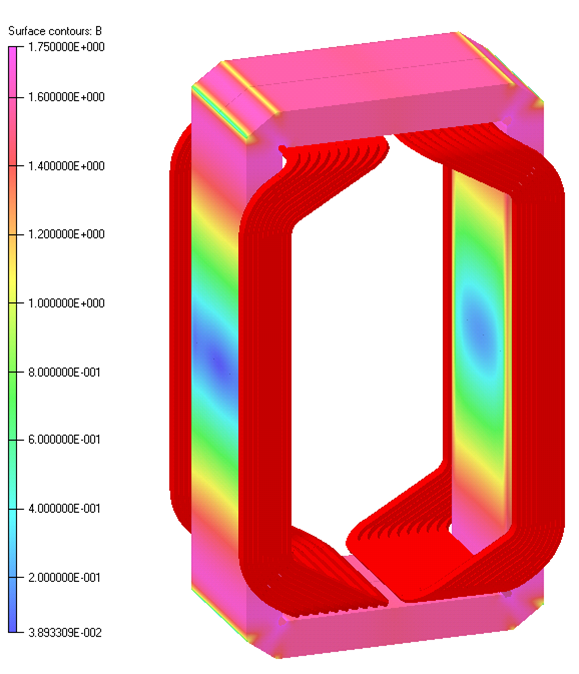}
\caption{(Left) 3D schematic view of the magnet with vacuum vessel. (Right) The field amplitude map inside the iron yoke.}
\label{fig:magnet_layout}
\end{center}
\end{figure}

As shown in Figure~\ref{fig:magnet_layout} (left) the magnet consists of two coils surrounded 
by a window-frame yoke structure. The coil shape has been designed to provide a uniform magnetic 
field inside the magnet aperture and at the same time minimize the coil length and thus the 
power consumption. The first objective 
has been achieved by placing the conductors inside the yoke frame and along the yoke backleg. 
The coil length has been minimized by driving the coil ends around the vacuum vessel contour. 
Each coil is made of ten pancakes, each of which is composed of two layers of six turns each, 
totalling 120 turns per coil. The conductor is similar to the one used for LHCb. It is made 
of Al-99.7 with a 50$\times$50~mm$^2$ cross section and a central bore hole of 25~mm diameter for 
water cooling. The conductor of each pancake has an average 
length of about 800~m. Thanks to the generous coil section, the magnet power at the nominal 
bending strength is about 1~MW. The 0.65~Tm overall magnet size is about 7~m in width, 
13~m in height and 5~m in length. 

\begin{figure}[tb]
\begin{center}
\includegraphics[width=0.38\linewidth]{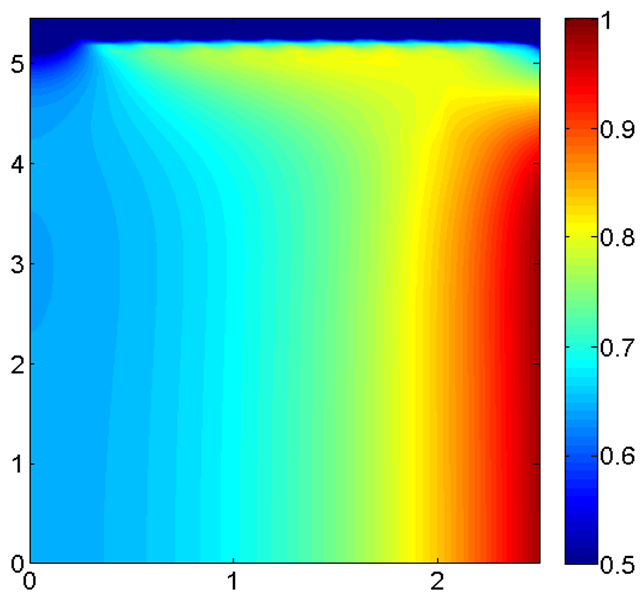}
\includegraphics[width=0.48\linewidth]{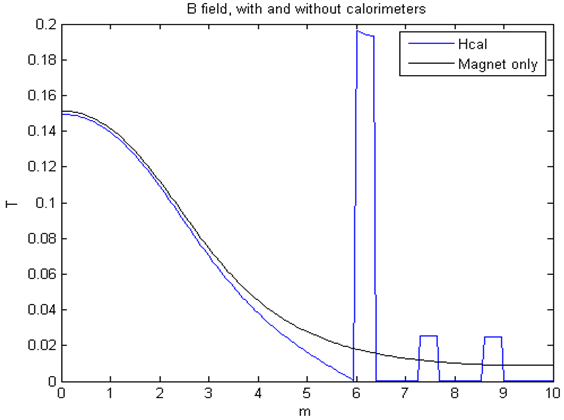}
\caption{(Left) The map of the field integrals in Tm over straight lines coming from 
         -2.5~m$<$z$<$+2.5~m from the magnet center. One fourth of the magnet aperture 
         (top-right quadrant) is pictured. (Right) The horizontal field component in Tesla 
         as a function of longitudinal distance from magnet center.}   
\label{fig:magnet_field}
\end{center}
\end{figure}

As the magnet is relatively short compared to its aperture, all magnetic field computations 
have been performed using OPERA-3d. The magnet yoke has been designed for a maximum field 
induction of 1.75~T at the nominal bending strength of 0.65~Tm (Figure~\ref{fig:magnet_layout} (right)). 
Within the magnet aperture, over straight lines going from -2.5~m to +2.5~m with respect to 
the magnet center, the integrated magnetic field varies from a minimum of 0.5~Tm to a maximum 
of 1~Tm close to the iron poles (Figure~\ref{fig:magnet_field} (left)). A magnetic field map
survey system will be necessary to precisely caracterize the magnet. Considering the large
magnet physical aperture compared to the distance to the iron of the hadron calorimeter, 
a 3D computation of the effect of the presence of the hadron calorimeter on the longitudinal 
profile of the magnetic field has been performed. The hadron calorimeter (Section~\ref{sec:calorimeters})
consists of about a metre of ferromagnetic steel interleaved with scintillators over a length of 
three metres. In order to ease the modelling of the calorimeter, it has been simulated as three 
ferromagnetic blocks 40~cm thick each. Figure~\ref{fig:magnet_field} (right) shows 
that the horizontal field amplitude between -2.5~m$<$z$<$+2.5~m from the magnet center is only 
marginally affected by the presence of the iron. Furthermore, it is also shown that the 
calorimeter completely shields the magnetic field. 

%If necessary, a detailed analysis of the 
%magnetic field distribution inside the calorimeter can be performed in the future by using a 
%refined magnetic model of the calorimeter.

The main magnet parameters are summarized in Table~\ref{tab:magnet_parameters}.

\begin{table}[htb]
\begin{center}
\label{tab:magnet_parameters}
\caption{Summary table of the main parameters of the baseline spectrometer dipole magnet.}
\vspace{0.2mm}
\begin{tabular}{ll}
\hline
Parameter                         	& Value              \\
\hline
Free aperture                     	& 53 m$^2$ \\
Current density                   	& 1.5 A/mm$^2$       \\
Conductor (Al-99.7)                   	& 50$\times$50~mm$^2$       \\
Central field                     	& 0.15 T             \\ 
Bending power $(0,0,\pm 2.5m)$    	& 0.65 Tm            \\
Operating current                 	& 3000 A             \\
Estimated power consumption       	&    1 MW             \\
Yoke mass                         	&  820 tonnes          \\
Coil mass			  	& 2$\times$32 tonnes          \\
Number of coils                   	&    2         \\
Number of pancakes per coil       	&   10             \\
Number of turns per pancake       	&   12             \\
Total water flow @ $\Delta$T=14$^{\circ}$C &   65 m$^3$/h             \\
Diameter of cooling hole          	&   25 mm             \\
Pressure drop		          	&   11 bar             \\
\hline
\end{tabular}
\end{center}
\end{table}

% \begin{itemize}
% \item{powering} 
% \item{heat dissipation in experimental hall}
% \item{assembly}
% \end{itemize}

For flexibility, the coil and the cooling circuit have been designed to cope with a 
possible magnet upgrade up to a 1.0~Tm bending strength with no modification of the coil or of 
the water flow. Keeping the same water flow, the operation at 1.0~Tm would increase the coil 
temperature from the reference 14$^{\circ}$C at 0.65~Tm to 35$^{\circ}$C at 1.0~Tm. In case the 
1.0~Tm upgrade option is pursued, the coil and the relevant support structure have to be 
designed to take into account the increased thermal deformations already at the initial stage. 
An upgrade to achieve a bending strength of up to 1.0~Tm would also require increasing the 
yoke size by fitting additional ferromagnetic plates on the outer sides of the yoke to accommodate 
the increased magnetic flux.  

The cooling parameters are set to cope with a possible 1.0~Tm upgrade. When operating at 
0.65~Tm it may be considered to decrease the cooling flow from the nominal 65~m$^3$/h to just 
30~m$^3$/h accepting to run the coils with a temperature increase of 30$^{\circ}$C instead of the 
nominal 14$^{\circ}$C. In order to keep the pressure drop in the range of 10~bar the pancakes have 
to be cooled in parallel. 

\subsection{Future R\&D}

While the baseline proposed here is a normal conducting magnet, a superconducting version is also 
being explored in parallel. It will be pursued in the TDR phase. The 
superconducting version may be based on 5~kA MgBo2 superconducting cables operating at 20~K.

% \subsection{(Cost)}
% The magnet cost is largely dominated by material costs and by the cost of the manufacturing of the coils.
% The cost of handling and assembly tools (in situ), ancillaries, field mapping and supports should be 
% contained within about 0.6~MCHF. The power converter will be around 0.5~MCHF plus the relevant 
% infrastructure and the cooling system (magnet + power converter) probably not far for a similar amount.
% The main unknown in the estimate is certainly the cost of the manufacturig of the coil, which will greatly 
% depend on the specific interest of the few potential bidders of doing the work at that specific time.
% Tentatively, we consider about 2~MCHF for the coil material and manufacturing, and about 1.7~MCHF for the 
% steel. This totals to: 0.6+0.5+0.5+2+1.7=5.3~MCHF.

%% file: taggers/timing/SHiP_TP_TimingDetectors.tex
\section{Spectrometer timing detector}
\label{sec:timing_detectors}

%As discussed in Section~\ref{backgrounds}, timing detectors are needed in addition to the surround background taggers and the tracking stations near the entrance of the vacuum vessel to reduce the backgrounds to the level of $0-1$ events (preferably $\ll1$) for the whole SHiP physics run duration. 

%One serious source of background is the random crossing of muons escaping the surround taggers, see Section~\ref{backgrounds}. Three major sources of muons entering the vessel from the sides are identified: (1) cosmic muons, (2) muons deflected by the magnetic muon shield which scatter back in the surrounding walls, and (3) secondary muons (and other charged particles) produced when the muons of type (1) and (2) undergo inelastic collisions inside the concrete walls. The rate of muons from these sources taken together which would leave tracks in the SHiP spectrometer is estimated to be of the order of 50 kHz. In addition, the magnetic muon shield is not perfect and there are also muons entering from the front window, at a rate of the order of 10 kHz -- these are less important thanks to the high efficiency of the upstream veto tagger (see Section~\ref{sec:upstream_vetotiming}). 
Background muons entering the vessel can potentially cross within the vertex reconstruction resolution and produce fake signals. One  efficient way to distinguish random crossings from genuine physics events is to require the measured particle signals in the SHiP spectrometer to be coincident in time. 
%With a timing resolution of 100 ps, the total number of coincident random crossing background events (for the whole SHiP physics run duration) is $5\times 10^5$. Taking surround tagger efficiency ($\sim 10^{-3}$ probability for vetoing at least one of the muons), tracking resolution ($\sim 10^{-3}$ probability for the tracks to form a vertex), and reconstructed signal pointing resolution (another factor $\sim 0.2$) into account, the background can be reduced to the desired level of $\sim 0.1$ event. Thus, a timing resolution of 100 ps (preferably better to allow for some margin) is necessary. Since the best timing resolution which can be achieved by other SHiP subdetectors is of the order of 1 ns, this requires the use of a dedicated timing detector placed in front of the ECAL. 
In order to reduce combinatorial di-muon background to an acceptable level, see Section~\ref{sec:backgrounds}, a timing resolution of 100 ps (preferably better to allow for some margin) is necessary. Since the best timing resolution which can be achieved by other SHiP subdetectors is of the order of 1 ns, this requires the use of a dedicated timing detector placed in front of the ECAL. 

Two alternative technologies are described below: plastic scintillators and multigap resistive plate chambers (MRPCs). Two options for the scintillator optical readout are also discussed: the first one uses existing technologies, while the second one has the potential to provide surpassing performances after a $1-2$ years R\&D programme. The use of the MRPC and scintillator technologies can both be based on existing and well-studied designs and reach the desired time resolution of 100 ps. The decision of which technology is chosen for the baseline solution will be taken after more detailed studies on the actual background rates are performed, and after the respective performances, the calibration strategy,  and costs are compared. 

%Another important background source comes from neutral kaons produced by neutrino and muon deep inelastic scattering upstream of the vacuum vessel, see Section~\ref{backgrounds}. While tracking chambers belonging to the muon system of the tau-neutrino detector (also in combination with the liquid scintillator and tracking chamber near the vacuum tank entrance) can tag a significant fraction of such reactions, they lack an efficient triggering station with large area coverage and good timing resolution. Thus, a veto station needs to be located between the tau-neutrino detector and the vacuum tank with a full $5\times 10$ m$^2$ area coverage. A low fake rate of background event tagging is achieved by using time matching between signals in the veto detector and the timing detector. Owing to similar requirements as the Veto-Timing detector (good timing resolution, same dimensions), the design of the veto-Veto-Timing detector is almost identical except that more emphasis is placed on efficiency and less on timing resolution. Below we describe options for the baseline design of the Veto-Timing detector and mention the slight modifications needed for the veto-Veto-Timing detector whenever relevant.

\input{./taggers/timing/timing_scintBar.tex}

\input{./taggers/timing/timing_MRPC.tex}

%% file: taggers/timing/timing_scintBar.tex
\subsection{Detector layout $-$ scintillator option}

Arrays of plastic scintillator bars of length ranging from 1 m to 6 m
can efficiently cover large areas for a relatively low cost, and have
the additional advantages of robust construction, low maintenance, and
reliability. Recent examples are the NA61/SHINE ToF detector~\cite{NA612014},
trigger hodoscopes in COMPASS \cite{COMPASS}
and the calorimeter of the MINOS neutrino
detector~\cite{MINOS2008}. Depending on bar length, scintillator type, and light readout
scheme, such detectors can achieve time resolutions in the range 50 ps $-$ 1 ns.  

The baseline option being considered here relies on light guides and photomultiplier tubes (PMTs) for the optical readout. The design is analogous to, and largely inspired from, the forward time-of-flight detector of the NA61/SHINE experiment~\cite{NA612014}, which was shown to achieve a time resolution of 110 ps for bars of dimensions $120\times 10\times 2.5$ cm$^3$. The plastic used in this case (Bicron BC-408) has an attenuation length of 210 cm, a rise time of 0.9 ns and a decay time of 2.1 ns. 
Even better performance was achieved in the TOF system of the MICE experiment  
(50-60 ps time resolution) \cite{MICE2010} where the ToF detector is based 
on plastic scintillator BC-404 with an attenuation length of 140 cm, a rise time of 0.7 ns 
and a decay time of 1.8 ns. 
Slabs are coupled to magnetic field-tolerant (few gauss) Hamamatsu fast PMTs R4998.
Using more transparent material, such as the EJ200 plastic (400 cm attenuation length), test-beam measurements with even longer bars (of dimensions $230\times 6\times 5$ cm$^3$) and PMT optical readout also yielded excellent results. In this case, a 100 ps time resolution was demonstrated with ADC+TDC electronics~\cite{Wu2005}, and an even better resolution -- down to 50 ps -- was achieved with waveform digitizer electronics~\cite{Wang2015}. 
Based on this experience, there is little doubt that a scintillator-bar system with 100 ps timing resolution or better can be built for the SHiP timing detector using well-known technologies.
 
\begin{figure}[t]
\centering
%\includegraphics[height=0.52\linewidth,angle=0]{taggers/timing/fig/fron_view_rot0_txt_TD.png}
%\hspace{2cm}
%\includegraphics[height=0.50\linewidth,angle=0]{taggers/timing/fig/section_PMT_2pic_TD.png}
\includegraphics[height=0.6\linewidth,angle=0]{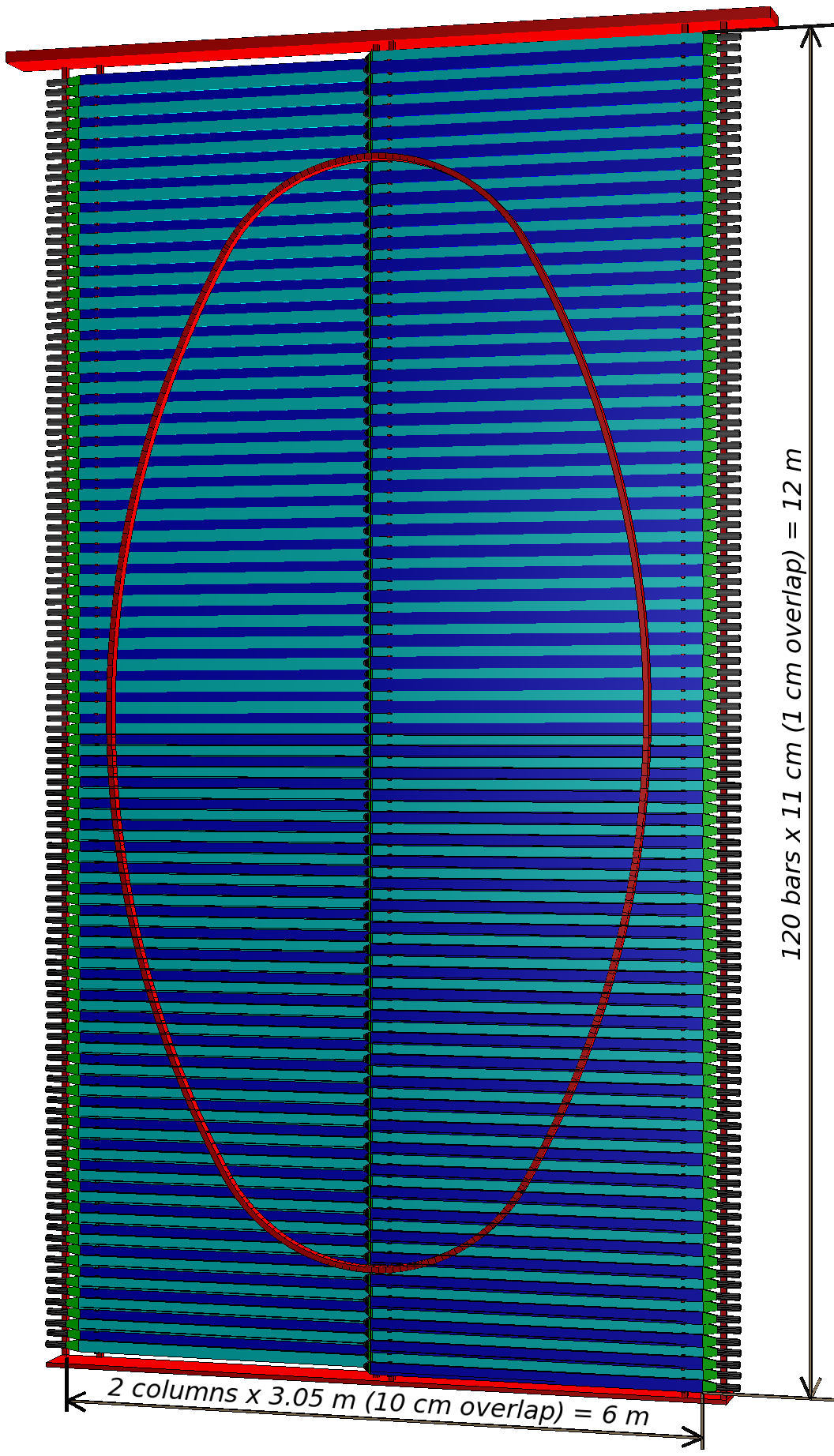}
\hspace{2cm}
\includegraphics[height=0.6\linewidth,angle=0]{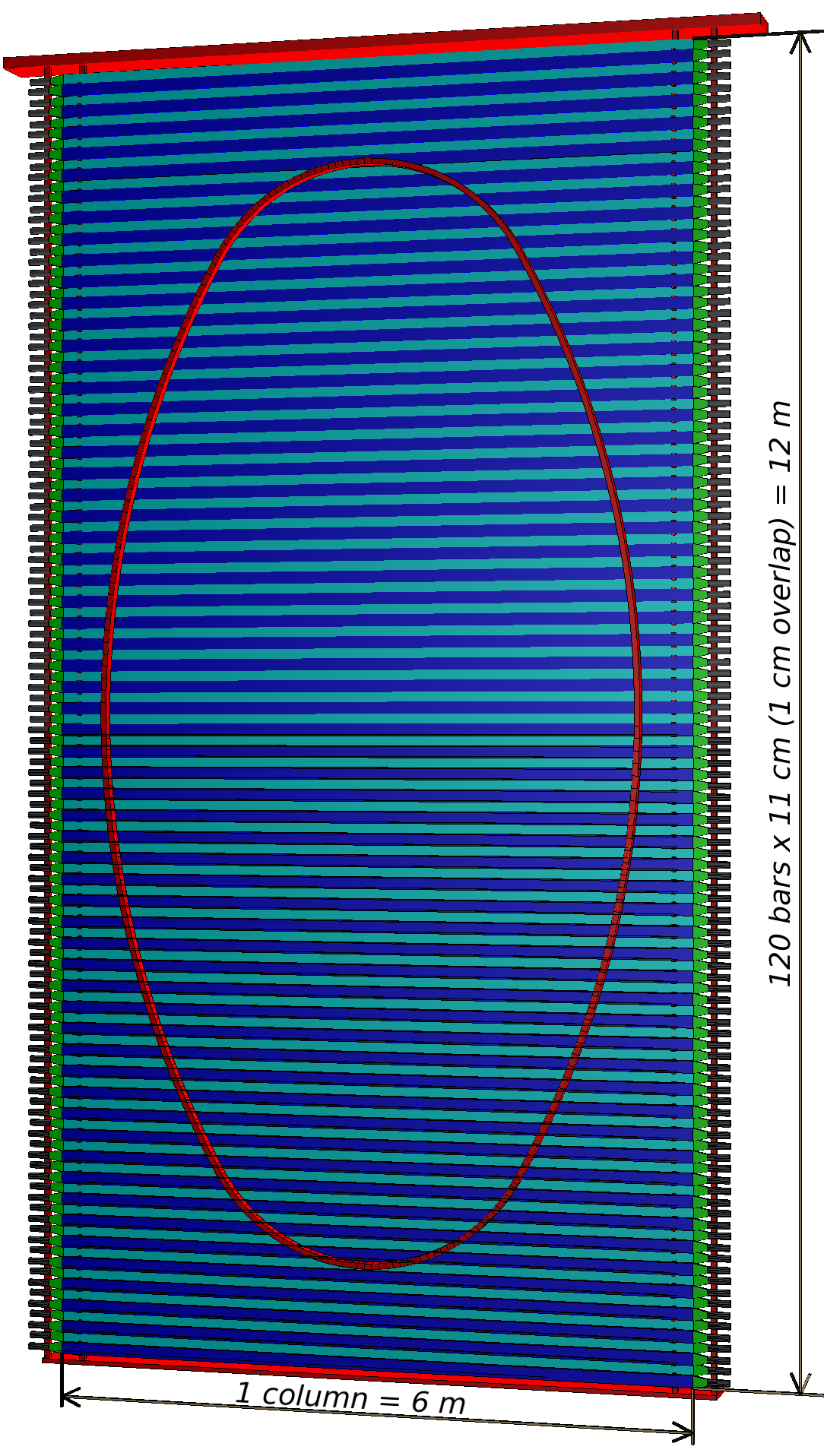}
\caption{
  A schematic front-view of the SHiP timing detector in the plastic-scintillator 
  bar option with PMT light readout. The default two-columns (left) 
  and the single-column (right) setups are presented. 
  The oval at both figures represents the acceptance of the vacuum vessel.
  %Right: rotated view of a section of the detector for the 2-columns (top)
  %and for the single-column (bottom) setups.
}
\label{timing_detector_scint_geometry}
\end{figure}

%\begin{figure}[t]
%\centering
%\includegraphics[width=0.5\linewidth,angle=0]{taggers/timing/fig/PMToption.png}
%\caption{ Baseline design of one column of the SHiP timing detector in the plastic-scintillator bar option with PMT light readout. 
%}
%\label{timing_detector_scint_geometry}
%\end{figure}

The baseline geometry of the proposed detector array comprises two columns of horizontal 
bars 305 cm long, 11 cm wide  and 2 cm thick.
To avoid inactive regions between adjacent bars they are staggered with 1 cm overlap vertically. 
There is also 10 cm overlap between columns.
The schematic front view of the detector is depicted 
in Figure~\ref{timing_detector_scint_geometry} (left). 
%One column is depicted in Figure~\ref{timing_detector_scint_geometry}. 
To cover the full area of the current SHiP muon system, 
the spectrometer timing detector will cover an area of $6\times 12 = 72$ m$^2$, with a total of $2\times 120=240$ bars.
Each bar is read out on both ends, thus requiring 480 PMTs and electronic channels. 
The readout electronics will be located around the edges of the detector module. 
The presence of a residual magnetic field estimated to be \mbox{$\sim$\,0.02 T} 
(see Section~\ref{sec:spectrometermagnet}) requires the use of magnetic shield cases for the PMTs. 
%e.g. cylinders made of Permalloy. 
%The design needs also to allow for the light guides and the PMTs from different columns 
%to overlap in between the columns in a way that minimises dead space. 
These cases (cylinders made of Permalloy) can introduce unwanted dead regions at 
the junction edge of two columns. To avoid that we will use bended light guides 
to settle PMTs behind the active surface of the detector.
As a basic option for PMTs we consider 2'' photomultipliers Fast-Hamamatsu R13089.% \cite{Hamamatsu_R13089}.
%used for the readout in NA61/SHINE \cite{NA612014}.

Taking the support structure into account, the whole detector would weigh about 3 tonnes. 
In Figure~\ref{timing_detector_scint_geometry}\,(right) a single column setup 
is presented. We consider it as a promising alternative to the default 2-columns 
option described above. Use of a single column reduces the number of PMTs by
a factor of two (about 30\% for the overall detector cost). In addition the problem
of a shadow projected by shield cases of PMTs on ECAL does not exist in this case.
%The use of long scintillator bars is not a new. A
It has to be noted that about 20 years ago the OPAL collaboration
had been using 6.8 m bars in the ToF system \cite{Ahmet:1990eg}. The plastic scintillator 
used for the detector had an attenuation length of 2.3 m. Light was collected by 
the XP2262B phototubes. The time resolution, depending on a position along the bar,
ranged from 220 to 280 ps. 
Since then a significant improvement was achieved at all stages of a light collection 
and readout: the attenuation length of EJ200 is a factor of two better; 
the T.T.S. (transit time spread) of modern PMTs is improved by a factor of two;
the use of a waveform digitizer electronics gives an additional boost towards 
smaller values of the time uncertainty.
Thus we have a good chance to obtain the 100 ps resolution with the 6 m long bars.
However, R\&D will be necessary.

An overlapping region between neighboring bars in all the configurations represents approximately 10\%
of their surface. It will be used for the cross-calibration of a time delay
of the light collecting sensors. Since we expect the event rate on a level of
dozens kHz per bar, statistics necessary for the calibration can be collected
within several spills. Another solution for the calibration would be
a light emitting diode (LED) system. A light pulse of LED can be distributed
and delivered to all PMTs by optical fibers. This signal can be used for
the time calibration as well as to monitor the stability during data taking.

%is an optical readout based on a large array of SiPMs connected directly to the scintillator surface on both ends \textit{or inside the plasic itself -- describe here Etam's concept, with figure}, covering as much of the transverse surface as possible. While such a layout has not been used yet in actual experiments due to the current high cost of SiPMs per unit area, the SiPM market is evolving very fast and one might expect much larger detectors to be available at a low price in the near future \textit{quote here the price per mm2 which was proposed to Alexander}. Tests, using counter size up to $120 \times 40 \times 5$ mm$^3$ with three SiPMs ($3 \times 3$ mm$^2$ each) series connected to increase the sensitive area, yield very promising results, with a timing resolution of the order of 60 ps~\cite{Cattaneo2014}. 

%The WLS fibre option is used here as a baseline strategy which is known to provide a timing resolution of 500 ps or slightly better. 
%Since a timing of the order of 100 ps would be needed the large SiPM array configuration is the most interesting option: it is currently the subject of considerable R$\&$D which should yield good results in a 1--3 year timescale.
%If that does not happen, a more traditional design using PMTs is a clear possibility, as it was demonstrated to yield excellent timing resolution in the context of the 86 NA61/SHINE experiment.

\subsubsection*{Expected R\&D}

As described in the previous section, the baseline configuration for the plastic scintillator-based option consists of two columns of relatively short bars, 305 cm long, instrumented with PMTs. All specifications and costs from manufacturers for plastic scintillators and PMTs currently commercialized confirm requirements for timing resolution can be met within budget.The proposed R\&D program aims to further optimize performance and costs by:
%\begin{enumerate}[i]
\begin{itemize}
\item determining the achievable time resolution of $6$ m long bars;
\item studying feasibility of replacing the PMTs at either end of the each bar with SiPM arrays;
\item embedding SiPM arrays throughout the large area plastic scintillators.
\end{itemize}
%\end{enumerate}
Given the previous operation of 6.8 m bars by the OPAL collaboration achieving timing resolutions down to 220 ps, activities would center around validating feasibility of producing 6 m bars with manufacturers and improving the timing resolution with a combination of optimized doping of the plastic scintillator, improved photosensors and the use of waveform digitizers.

%One very promising avenue worth investigating because of the implications for cost reduction is that provided by the replacement of PMTs by SiPMs. SiPMs have far better tolerance to magnetic fields and occupy a much smaller footprint, but until now their cost per unit area has typically been at least on order of magnitude higher than that of traditionally PMTs. However, competition amongst a handful of manufacturers, increase in demand and improvements in processing are all leading to the availability of cheaper SiPMs. It is now possible to purchase SiPMs with a sensitive area of $6 \times 6$ mm$^2$ for 20 Eur/piece. The cost of such devices is now competitive with PMTs, $45$ SiPMs for a cost of 960 CHF would be needed to replace a single PMT, costing 1000 CHF. Moreover the manufacturer SensL also commercializes $8 \times 8$ arrays of such devices, covering $57.4 \times 57.4$ mm$^2$, indicating possibilities for arrays of SiPMs that would couple directly to the bar ends measuring $110 \times 20$ mm$^2$. Whilst those arrays are available now, SensL announces improvements in its process for 2016 by the adoption of through silicon vias (TSV) leading to significant reductions in dead space between adjacent devices on the arrays. Currently the packaging for the device with a sensitive area of  $6 \times 6$ mm$^2$ is $7 \times 7$ mm$^2$, and would be not much larger than the sensitive area with the adoption of TSV.

One very promising prospect is provided by the replacement of PMTs by SiPMs, that can allow an improvement of the performance as well as a reduction of the costs. SiPMs have far better tolerance to magnetic fields and occupy a much smaller footprint, but until now their cost per unit area has typically been at least one order of magnitude higher than that of traditional PMTs. However, competition amongst a handful of manufacturers, increase in demand and improvements in processing are all leading to the availability of cheaper SiPMs. It is now possible to purchase SiPMs with a sensitive area of $6 \times 6$ mm$^2$ for 20 Euro/piece. The cost of such devices is now competitive with PMTs: $45$ SiPMs for a cost of 960 CHF would be needed to replace a single PMT, costing 1000 CHF. Moreover the manufacturer SensL also commercializes $8 \times 8$ arrays of such devices, covering $57.4 \times 57.4$ mm$^2$, indicating possibilities for arrays of SiPMs that would couple directly to the bar ends measuring $110 \times 20$ mm$^2$. Whilst those arrays are available now, SensL announces improvements in its process for 2016 by the adoption of through silicon vias (TSV) leading to significant reductions in dead space between adjacent devices on the arrays. Currently the packaging for the device with a sensitive area of  $6 \times 6$ mm$^2$ is $7 \times 7$ mm$^2$, and would be not much larger than the sensitive area with the adoption of TSV.

%By embedding SiPM arrays throughout the plastic scintillator with regular spacings along the bar length, further improvements in timing resolution are possible and more significantly, better position resolution is achievable. Tests were performed using bars up to $12 \times 4 \times 0.5$ cm$^3$ with three SiPMs ($0.3 \times 0.3$ cm$^2$ each) in series connected directly to the scintillator surface on both bar ends. These yield very promising results, with a timing resolution of the order of $40-60$ ps~\cite{Cattaneo2014}. Such an approach on relatively small surface areas can be extrapolated to much larger areas such as those of the SHiP timing detector.
An other interesting possibility is to embed SiPM arrays throughout the plastic scintillator with regular spacings along the bar length. This would lead in further improvements in timing resolution and more significantly, better position resolution would be achievable. Tests were performed using bars up to $12 \times 4 \times 0.5$ cm$^3$ with three SiPMs ($0.3 \times 0.3$ cm$^2$ each) in series connected directly to the scintillator surface on both bar ends. These yield very promising results, with a timing resolution of the order of $40-60$ ps~\cite{Cattaneo2014}. Such an approach on relatively small surface areas can be extrapolated to much larger areas such as those of the SHiP timing detector.

The fast evolution of the SiPM market opens new possibilities which will be subject of R$\&$D work in the next few years. One can consider the straightforward replacement of the PMTs by SiPMs at each end of the bar as an intermediate step and extrapolate with new concepts, such as the direct inclusion of SiPMs at regular intervals inside the plastic itself. Such tests should yield results within a relatively short time scale, allowing the choice of the technologies well in time for the Technical Design Report.

\subsubsection*{Electronics readout}

%The choice of electronics for the timing detector is in large part informed by previous experience obtained in the NA61/SHINE experiment~\cite{NA612014b}. 
%The development of the new read-out system is based on a DRS4 digitiser developed by PSI~\cite{Ritt2010}. 

The choice of electronics is defined by the requirement to have 
a 100 ps time resolution for the detector.
This precision can be achieved in a measurement of a signal waveform.
As option
for the readout we thus consider the DRS4 chip developed by PSI \cite{Ritt2010}, and the new generation Waveform TDC chip called SAMPIC \cite{SAMPIC}.
%As an option for the readout we consider the SAMPIC digitizer \cite{SAMPIC} 
%and the DRS4 chip developed by PSI \cite{Ritt2010}.
As an alternative we can also apply a traditional approach of splitting 
the photosensor signal and treating separately timing (TDC) and amplitude (ADC) signals.
It has been proven to deliver $50-60$ ps time resolution for 
the   TOF system in MICE\cite{MICE2010}. 
Similar approach with a passive divider which provides an input for 
the charge integration and the time measurements assures precision 
of 110 ps in the ToF system of NA61/SHINE \cite{NA612014}.
%Similar readout systems but with longer 236 cm plastic scintillator bars read out with PMTs at both ends were shown to have time resolutions of order 90 ps ~\cite{Wu2005}. 
However the availability of a digitized waveform can simplify design and yield better time resolutions.

The DRS4 (Domino Ring Sampler, 4-th generation) is a switched capacitor array capable of sampling 8 differential input channels at a sampling rate up to 5 GS/s~\cite{Ritt2010}. The analog waveform is stored in 1024 sampling cells per channel, which are read out by a shift register for external digitization (commercial ADC). 
%%%Excellent timing resolutions of order 50 ps have been reported for 236 cm-long EJ200 plastic scintillator bars, coupled to GDB60 PMTs readout by a DRS4 module operating close to 5 GS/s, and using a variety of techniques such as digital constant fraction discrimination, cross-correlation and amplitude-weighted sliding window~\cite{Wang2015}. These resolutions are comparable to those obtained with LeCroy 10 GS/s oscilloscopes with a similar setup. Other techniques have also been explored for the readout of SiPMs with the DRS4, by shortening the SiPM signal rise time with a clipping capacitance for example ~\cite{Ronzhin2013}. 
%As in case of SAMPIC the internal resolution of the chip is on a level of few picoseconds.
The DRS4 based electronics can be applied in experiments with low trigger rate which require excellent time resolution and pile-up rejection.
%%% conditions satisfied by SHiP.
It was used already in a number of test benches designed to
 evaluate performance of plastic scintillator counters
 \cite{Cattaneo2014,Wang:2015rva,Mineev:2011xp}.

%The SAMPIC  is a R\&D project initially intended to develop a common 
%Waveform and Time to Digital Converter (WTDC) multichannel chip prototype \cite{SAMPIC}
%addressing the need 
SAMPIC is a R\&D project in the frame of which a Waveform and Time to
Digital Converter (WTDC) multichannel chip prototype \cite{SAMPIC} was developed. It was initially intended to address the need
for high precision timing (5 ps RMS) 
required by ATLAS AFP and SuperB FTOF. 
Its natural targets are fast detectors for the time-of-flight measurements 
for particle identification and/or pile-up rejection.
%based on PMTs, diamonds, APDs, SiPMs, and their associated characterization test benches.
%
SAMPIC is a 16 channels ASIC. 
Each channel associates a DLL-based TDC providing a raw time with 
a 64-cell ultra-fast analog memory sampling up to 10 GS/s and allowing fine timing extraction.
% as well as other parameters of the pulse. 
Each channel also integrates a discriminator that can trigger itself independently 
or participate to a more complex trigger. 
Analog data is digitized by an on-chip ADC (8 to 11 bits).  
%and only that corresponding to a region of interest is sent serially to the DAQ. 
The association of the raw and fine timings permits achieving timing resolutions of a few picoseconds. % The short length of the analog memory and the programmable high conversion speed permit reducing the channel deadtime between hits down to ~200 ns, which is almost 3 orders of magnitude lower than DRS4.
At 10 GS/s, the 6.4 ns sampling depth is sufficient to catch the rising edge of a PMT or SiPM signal, and provide high resolution timing information. The relatively small sampling depth allows for low deadtimes, around 2 microseconds when using the ADC in 11-bit mode and well below 1 microsecond in 8-bit mode. 
In order to record the full photosensor signal waveform, it is likely that both an increase in the sampling depth of the SAMPIC and a decrease in typical time constants for photosensors are required. The sampling rate can be lowered to 1.6 GS/s, increasing the sampling depth to 40 ns. In the SiPM option, by combining SiPMs in series, the time constant of the resulting SiPM array can be designed to be compatible with the increased sampling depth. 
Moreover, the next version of SAMPIC will integrate additional functionality such as Time-Over-Threshold measurements of the signal which would complement information on the amplitude of the signal available from its waveform.
The SAMPIC promises to be an excellent candidate option for the timing resolution requirements of this detector with potential for full signal waveform acquisition.
%The DRS4 does not have self-triggering capabilities, so the photosensor signal must be split in order to provide an external trigger to the DRS4 and optimize its readout. It is then possible to use the "Region of Interest (ROI)" functionality to readout a fraction of the 1024 sampling cells. This would minimize the deadtime which is a consequence of the 30 ns/cell readout time. 

%A SAMPIC-based readout scheme is the proposed baseline assuming a particle rate at the level of 25 kHz per scintillator bar. 
%Alternatives such as upgrades to the DRS family, the DRS5, and other types of waveform digitisers can be considered. 

It should be noted that the requirement to read the light output from both ends of each plastic scintillator bar to improve timing resolution also leads to an additional rough position measurement along the bar, by comparing signal amplitude and time at either end of the bar.

%%%%
%The electronics proposed for the Veto-Timing detectors is base on the development of the new read-out system for the NA61/SHINE experiment. It is based on a DRS4 (Domino Ring Sampler, 4-th generation) digitiser originated in PSI~\cite{Ritt2010}. The availability of a digitised waveform provided by DRS4 eliminates the need to separate TDC and ADC channels. The DRS4 based electronics can be applied in experiments with low trigger rate which require excellent time resolution and pile-up rejection, as the case of the SHiP experiment.

%The DRS4 is a switched capacitor array capable to sample 8 differential input channels at a sampling speed up to 5 GHz~\cite{Ritt2010}. The analog waveform is stored in 1024 sampling cells per channel, and can be read out after sampling via a shift register clocked at 33 MHz for external digitisation (commercial ADC). Since this frequency is much lower than the sampling one, it can introduce a dead-time.

%The use of the DRS4 digitiser satisfies the timing requirements of the timing and veto-Veto-Timing detectors.

%% file: taggers/timing/timing_MRPC.tex
\subsection{Detector layout $-$ MRPC option}

An alternative technology for the spectrometer timing detector is the multi gap resistive plate chambers (MRPC).
This technology is already being successfully used in the time-of-flight detector of ALICE \cite{Akindinov:2009zza} and in the Extreme Energy Events project (EEE) \cite{Abbrescia:2008zzc}, dedicated to the study of atmospheric cosmic showers,   in the STAR experiment at RHIC and the FOPI experiment at GSI. All these detector arrays are operating in experiments with a time resolution better than 100 ps. 

The MRPC's are made of stacks of glass plates, separated by  commercial nylon fishing lines used as spacers making the gas gaps. The outer glass plates are coated with resistive paint, and act as high voltage electrodes, while the inner ones are left electrically floating.  Signals induced by charged particles traversing the chamber are induced on segmented readout strips.

\subsubsection{Proposed structure}

\begin{figure}[h]
\centering\includegraphics[width=0.9\linewidth,angle=0]{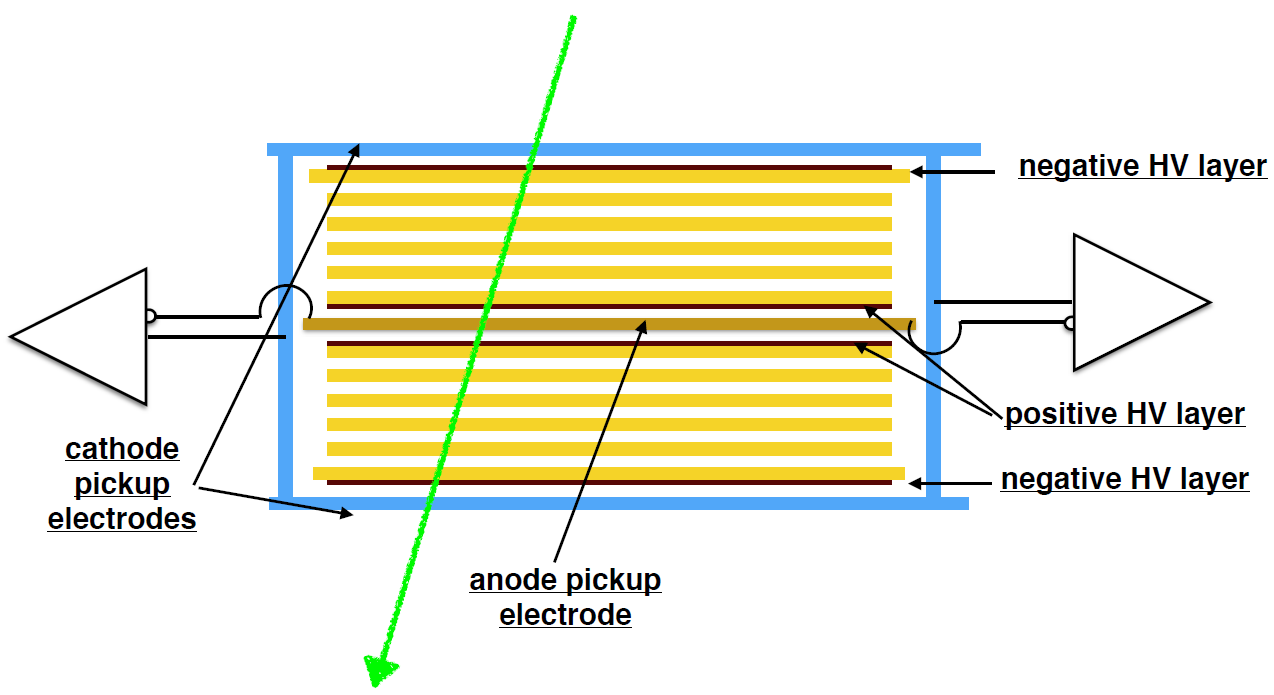}
\caption{Proposed MRPC chamber layout. The glass plates are in yellow.
}
\label{fig:MRPC_proposed}
\end{figure}

The chambers we propose to build  combine the design and the experience of ALICE-TOF and  of the EEE detectors.
An OR-red  double stack configuration is proposed, like in the ALICE-TOF application,  but with 120cm long strips, as in the EEE design.  Strip will be  readout on both sides, again as in the EEE application, with differential readout. A possible structure of the chamber  is shown in Figure~\ref{fig:MRPC_proposed}.
Since readout strips lie longitudinally on the chambers, along the x axis in Figure~\ref{fig:MRPC_proposed}, one coordinate (x) of the incoming particle impact point is given by the difference of the signal arrival times at the two strip edges, while the other (y) is directly obtained from the fired strip position. 
 
The front end electronics will be based on the 8-channel NINO ASICS chips, successfully used in both the ALICE-TOF and EEE projects, which provides a fast digital signal for further processing to a TDC,  with the leading edge used for time pick-off and the  trailing edge  for slewing corrections.

As TDC the 8-channel HPTDC  chip, with  bin width of 25~ps, could be used. Both NINO and HPTDC are available off the shelf. However, the authors of HPTDC are also developing a new TDC chip, again  matched to the NINO needs, with 3ps time bin, which would provide  further improved performance of the whole system. The first prototype will be submitted at the beginning of 2016. %~\cite{CristensenPrivate}. 

 At present  400~$\mu$m glasses in sheets of 120$\times$160cm$^2$ are used for prototypes without problems. %\cite{CrispinPrivate}.  
To be conservative we propose to use chambers of half the size of the active area, i.e.  120$\times$80~cm$^2$, similar to EEE chambers, with 24 readout strips 120cm long (the actual chamber size will be 128$\times$84~cm$^2$). This will allow us to benefit from the tooling for cleaning and handling developed for the EEE chambers.  We will consider during the R\&D phase if it will be possible to make double size chambers. All in all the system should be able to perform with full efficiency and time resolution better than 50~ps. 

\begin{figure}[h]
\centering\includegraphics[width=0.9\linewidth,angle=0]{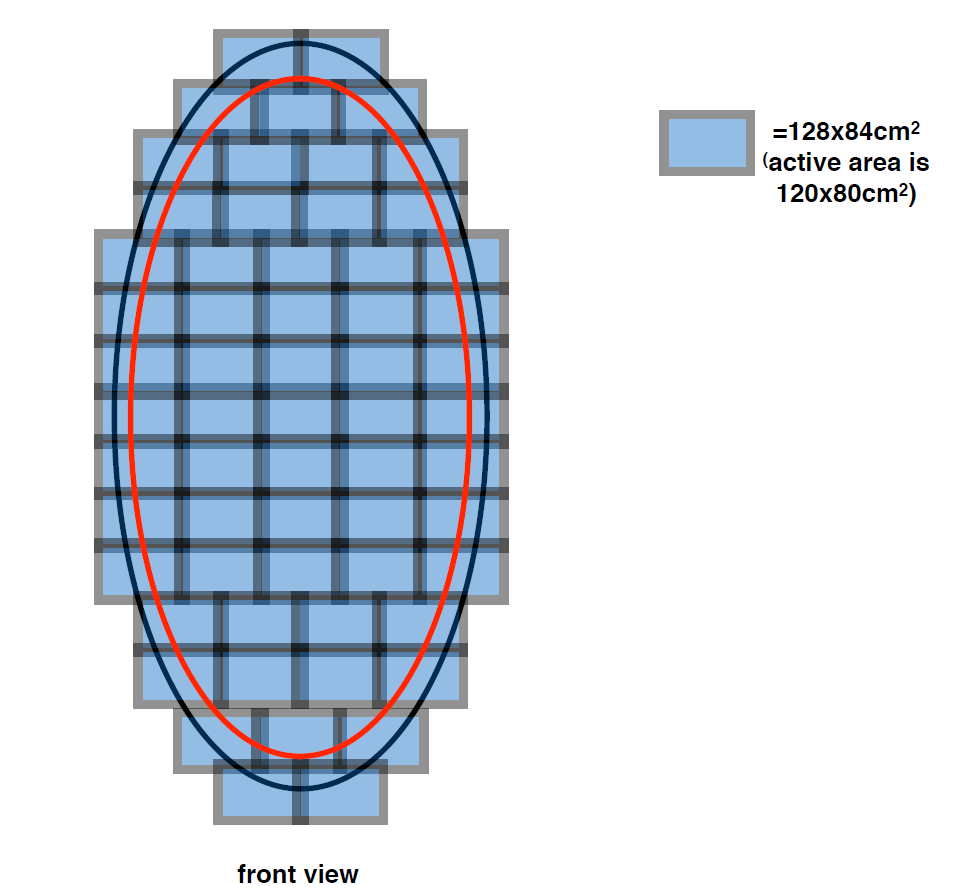}
\caption{Proposed timing  detector  structure with the MRPCs. In red is the outer decay vessel shape and in black is the ECAL shape.
}
\label{fig:MRPC_staggered}
\end{figure}

Due to the high modularity the MRPCs structure can be tailored to cover with good approximation any requested surface. It is conceivable that the surface to be covered is as large as the the ECAL/HCAL surface. Figure~\ref{fig:MRPC_staggered} shows a possible chamber layout, with a staggered structure to avoid dead regions. This layout is achieved with 61 chambers.

%The surface to be equipped with MRPCs in SHIP, is $6 \times 12$ $\rm m^2$ and will be segmented in units as individual chambers, with a staggered layout to avoid dead regions. \\

\subsubsection{Past experience}

The EEE design consisted of single stack structures  characterized by six gas gaps each, around 300 $\mu$m thick, obtained by separating glass plates, 1.1mm and 1.9mm thick, 90$\times$180 cm$^2$ in dimensions. Gas mixture is made of $\rm C_2H_2F_4/SF_6$ mixed in 98/2 proportion, and it is made flowing at a typical rate of 2-3 l/h.
Standard operating voltage ranges around $18\div20$ kV, so that the chambers operate in avalanche saturated mode, and the relative signals are induced on 24 copper strips (per chamber) glued on two
vetronite plates placed on top and bottom of the glass outer electrodes.

The chamber assembly is very straightforward and was performed by supervised high school students.
The EEE-type  chamber  provides good time resolution, below 70~ps, but,  given the  single stack structure, cannot reach full efficiency, and  only reaches about 98.5\% at about 20kV. 

The ALICE TOF detector used instead a double stack of five gas gaps each, with the much smaller chamber size of 120$\times$7.4 cm$^2$ active area. Readout was on one side of the 3.7 cm long and 2.4 cm wide strips. With this detector, test beams showed resolutions of about 50~ps and, thanks to the double stack, with full efficiency\cite{Akindinov:2009zza}.  This detector was placed in the experiment in front of the electromagnetic calorimeter. 

During the first years of operation, the MRPC used in both the ALICE TOF and EEE systems have shown a very stable operation. In the ALICE TOF in particular, currents and rates behaved as expected during the periods of data taking with beam, with no sign of degradation of the detector, allowing also to use the TOF as a trigger detector. 

The actual intrinsic time resolution of the ALICE TOF detector  is 15~ps.  One significant contribution in the ALICE test beams  to the time resolution comes from the HPTDC chip and from the electronic readout chain. In the ALICE experiment other effects contribute to the deterioration of this extraordinary performance such as track extrapolation, which will not be present in SHiP. 

In a beam test a resolution of 20~ps was obtained with a four stack structure\cite{An:2008zzc}.

\subsubsection{Expected R\&D}

The required R$\&$D will be mainly dedicated to the optimization of the mechanical structure, including the number of gas gaps, and single module size and of the readout strips.  Moreover tests will be performed on the optimal choice of the gas mixture. 

\subsubsection{Time alignment}

At present techniques to time align the different channels and chambers are under active study. One possibility would be to time align all the channels of a chamber in by quickly testing all the  chambers in a  test beam before installation and then inter-calibrate the different chambers in the experiment, using background muons. Indeed it is conceivable to run the experiment with the muon filter switched off, so to increase the number of muons going through the detector, and  introduce some partial staggering of the chambers in order to have a part of the readout strips of two different chambers crossed by the same particle. To do this a few more chambers with respect to what described above could be needed. The usage of larger chambers would be of benefit. Another possibility could be to give the different chambers a few degree tilt  with respect to the vertical position and use the cosmic muons crossing different chambers.  It is also possible to design a time calibration system pulsing the readout electronics right at the amplifier input.

%% file: calorimeter/Calorimeters.tex
\section{Calorimetry}
\label{sec:calorimeters}

The SHiP calorimeters handle a number of tasks. They must identify photons, electrons and $\pi^0$-mesons and provide measurements of their energies and positions. Furthermore, the calorimeters must contribute to the identification of charged pions and muons and to provide precise time information for event reconstruction.
Electron/pion identification is necessary for the separation of $\mbox{HNL} \to e^{\pm}\pi^{\mp}$ decays
from the main background channels such as $K^0_S \to \pi^+\pi^-$ and $K^0_L \to \pi^+\pi^-\pi^0$. % and $K^0_L \to e^\pm\pi^\mp\nu$.
Good photon identification and energy measurements are necessary for the correct reconstruction of the $\pi^0$s for several channels. For example, the reconstruction of the $\mbox{HNL} \to l^{\pm} \rho^{\mp} \to l^{\pm} \pi^{\mp}\pi^0$ channels, where $l^\pm$ denotes electron or muon, and for channels involving decays of light mesons with photons in the final state.
For the $\mbox{HNL} \to l^{\pm} \pi^{\mp}\pi^0$ channel, the main background is $K^0_L \to \pi^+\pi^-\pi^0$, which has a branching ratio of  $12.5\%$, that can also be rejected with a precise invariant mass measurement. It is therefore necessary to have an optimal calorimeter energy resolution giving an invariant mass resolution similar to that given by tracking. The calorimeter system also aids in pion/muon discrimination and in the identification of muons, especially in the low momentum region ($p<5\ \gevc$) where a sizeable fraction ($\approx 14\%$) of muons from a 1 \gevctwo\ HNL can be produced but they may not reach all the sensor planes of the muon detector.
%\footnote{\color{red}\it It may be worth noticing that low momenta cover a sizeable fraction of HNL decays, e.g. the range $p<5\gevc$ corresponds to about 14\% of muons from m(HNL)=1\gevctwo\ decays in the $HNL \to \pi^+\mu^-$ channel, as shown in Figure {\bf 4.66} of the MUON section as of TP v0.5 from 19th March.}

The calorimeters are composed of two different elements. An electromagnetic calorimeter (ECAL) is placed at the end of the vacuum vessel right after the timing detector. This is ensued % followed % surrounded
by a hadronic calorimeter (HCAL). The material in front of the calorimeters is kept at the minimum level in order not to spoil the energy and position resolutions.

In the current design, both ECAL and HCAL have an elliptic shape with a 252~cm horizontal semi-axis and 504 cm vertical semi-axis
% corresponding to the baseline SHiP geometry option,
with an acceptance matched to the tracking system.
% They are both placed as close as possible to the vacuum vessel exit window.
The two elements of calorimetry employ similar technologies. Both are based on shower sampling modules with layers of converters and scintillators coupled to wavelength-shifting fibers read out by fast photodetectors.
Geometric matching of acceptance and of lateral detector segmentation (finer granularity for ECAL, larger for HCAL) facilitates the combination of the information from corresponding cells of the two detectors.
The front-end electronics will also largely be common for the whole calorimeter system.

\subsection{Electromagnetic calorimeter}

\subsubsection{Performance requirements}

% The requirements on the electromagnetic calorimeter are quantified with studies of the HNL with a mass range 0.15-2.0 \gevctwo\ with channels involving $e^\pm$, $\pi^0$ and $\gamma$ and also by studies of dark matter candidates for which SHiP has been designed (see Chapter~\ref{sec:requirements}).
The main aims of the electromagnetic calorimeter are: % therefore:
\begin{itemize}
\item to provide electron, photon and pion identification at the offline level;
\item to measure the energy of electrons and photons in the range  0.3~-~70\gev\  with a resolution on $\sigma_E$ better than 10\%.
  The upper limit is defined by the requirement that only a negligible fraction ($<2\%$) of HNL$\to e^\pm\rho^\mp$ events suffers 
from partial energy measurement, while the lower limit is due to the requirement of having a small relative contribution from noise; % to minimize the relative minimal requirement on energy relative uncertainty;
%\footnote{\color{red} Andrey: explain the low threshold}
\item to provide $\pi^0$ reconstruction in the range 0.6~-~100 \gev;
\item to provide timing information on signals at the ns level for signal-event association and background rejection.
\end{itemize}

\noindent
To achieve the above goals the following criteria are imposed on the ECAL design:

\begin{figure}[tbh]
  \setlength{\unitlength}{1mm}
  \begin{center}
  \begin{picture}(180,60)
    \put(0,-5){\includegraphics[width=0.50\linewidth]{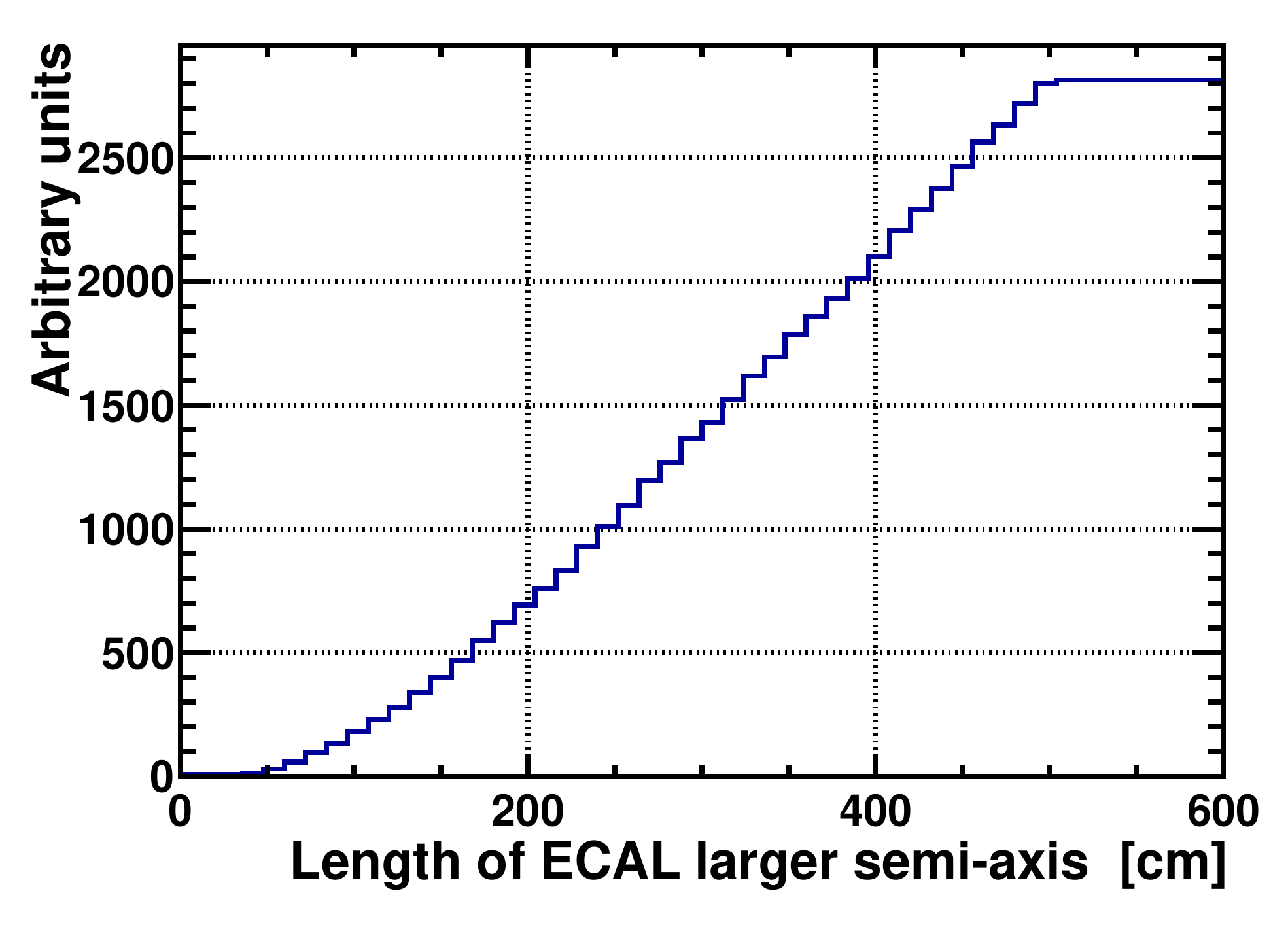}}\put(25,40){(a)}
    \put(87,-5){\includegraphics[width=0.50\linewidth]{./calorimeter/ycalo-ship-2}}\put(112,40){(b)}
  \end{picture}
  \end{center}
  \caption{
    \small Number of reconstructed $\pi^0$ mesons as function of ECAL vertical semi-axis (a).  Number of electron tracks matched to an ECAL cluster vs. the vertical semi-axis (b).
   \label{fig:ecal-size}
  }
\end{figure}

{\bf Acceptance:}
The $z$-position of the upstream face of the ECAL at 37.2~m from the center of the SHiP detector.
To determine the required dimensions of the calorimeter system, the acceptance for the decay $HNL \to e^-(\rho^+ \to \pi^+\pi^0)$ has been analyzed as function of ECAL size. In Figure~\ref{fig:ecal-size} the number of reconstructed $\pi^0$'s (a) and the number of reconstructed electron tracks (b) matched to an ECAL cluster depending on the length of the larger semi-axis of ECAL. As can be seen, both the $\pi^0$ reconstruction efficiency and the electron identification efficiency saturate above the vertical semi-axis size of $\sim500~\mbox{cm}$. Taking into account the acceptance behaviour of Figure~\ref{fig:ecal-size} and the need to minimize the overall costs, the ECAL was chosen to have an elliptical shape of $504\times 1008~\rm{cm}^2$. The outer dimensions of ECAL match those of the decay vessel and cover practically the full acceptance of the tracking system.

\begin{figure}[tb]
  \setlength{\unitlength}{1mm}
  \begin{center}
  \begin{picture}(180,60)
    %\put(0,0){\includegraphics[width=0.40\linewidth]{./calorimeter/ecal_dist_gamma_from_pi0_full.pdf}}\put(15,45){(a)}
    %\put(70,0){\includegraphics[width=0.40\linewidth]{./calorimeter/ecal_dist_gamma_from_pi0.pdf}}\put(78,45){(b)}
    \put(0,-5){\includegraphics[width=0.50\linewidth]{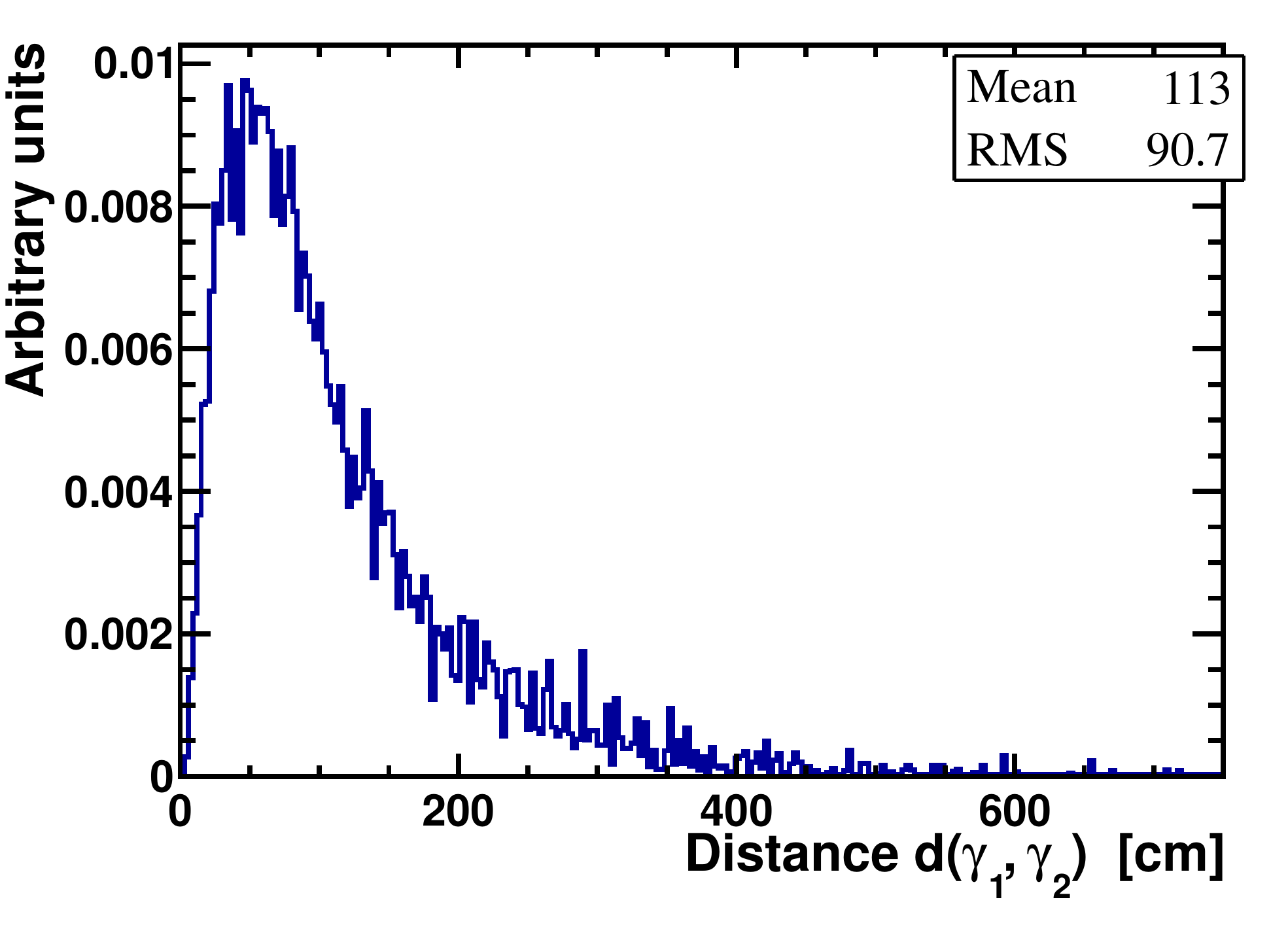}}\put(25,40){(a)}
    \put(87,-5){\includegraphics[width=0.50\linewidth]{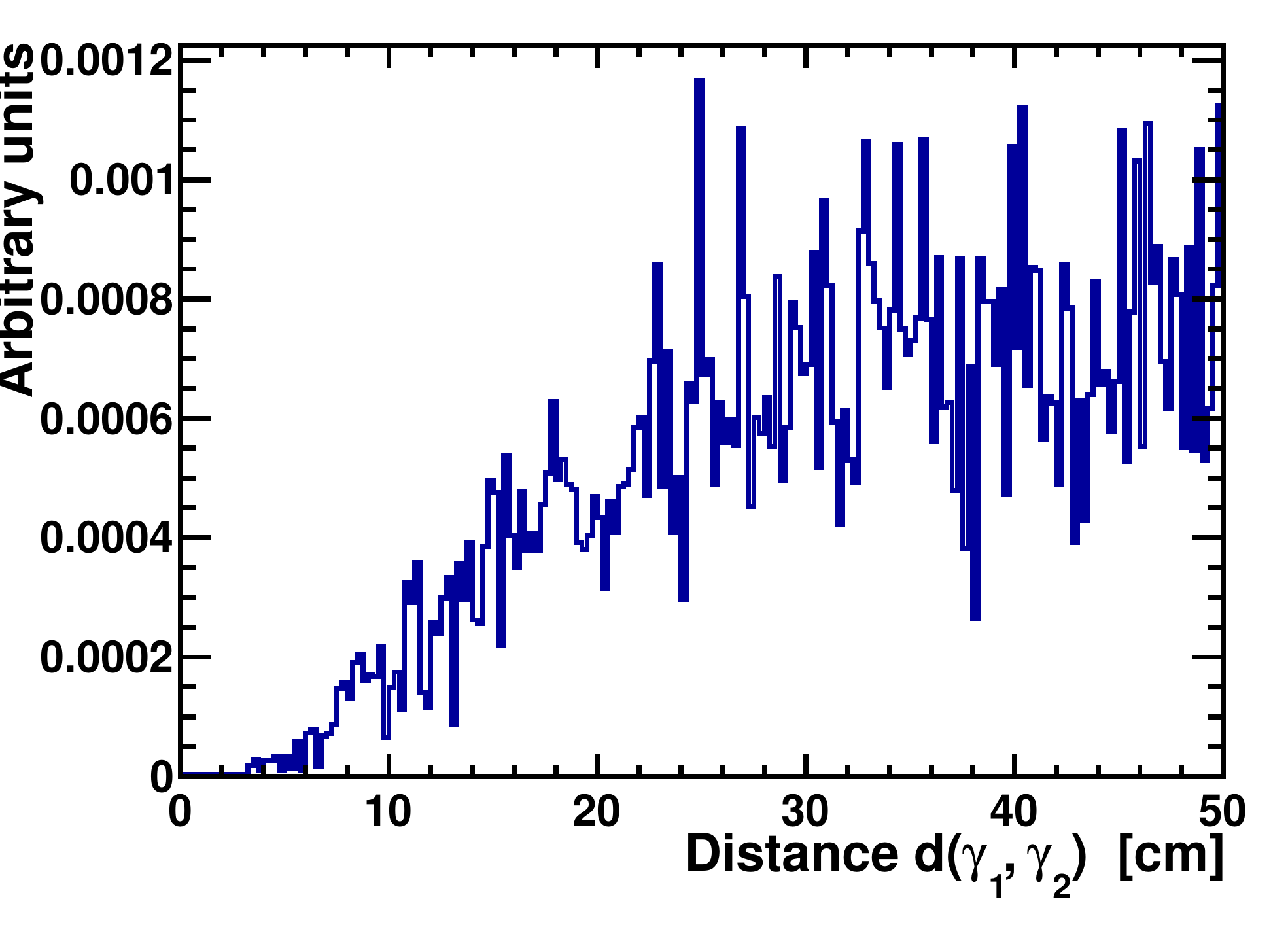}}\put(112,40){(b)}
  \end{picture}
  \end{center}
  \caption{
    \small %make captions a little bit smaller than main text
   Distance between the two photons from a $\pi^0$ from $\mbox{HNL} \to l^{\pm}\rho^{\mp}\to l^{\pm} \pi^{\mp}\pi^0$ decay: (a)~overall distribution and (b) region of small distances $d(\gamma_1,\,\gamma_2)$ up to 50~cm.}
  \label{fig:pi0-distance}
\end{figure}

{\bf Transverse granularity:}
Owing to the expected low occupancy of the experiment, clusters overlapping with background tracks can be neglected. The granularity discussed here has been chosen to provide a good two-cluster position resolution suitable for $\pi^0$ reconstruction for the HNL decays occurring before the tracking system.
Figure~\ref{fig:pi0-distance} shows the distribution of the distance between the two photons from $\pi^0$ for  $m=1$ {\gevctwo} HNL decays before the start of the tracking system, requiring $E_{\gamma} > 0.3~\mbox{\gev}$. % (the energy distribution starts from few \mev).
The minimum separation distance between two photons is as small as a few cm, while the mean distance is as large as 100~cm.
% The mean distance between two photons is about 100~cm and it  starts above a few cm.

%\subsubsection{Energy resolution:}

\subsubsection{Detector technology}
The electromagnetic calorimeter employs the shashlik technique, consisting of a sampling scintillator-lead structure read out by plastic WLS fibres. This technology has been successfully developed and deployed by the HERA-B collaboration\cite{herab:2007} at DESY, the PHENIX collaboration\cite{baz:1996} at BNL
and the LHCb collaboration\cite{lhcb:2010} at CERN. The combination of good energy resolution, fast response time and relatively low cost-to-performance ratio fulfill the requirements of the SHiP electromagnetic calorimeter.

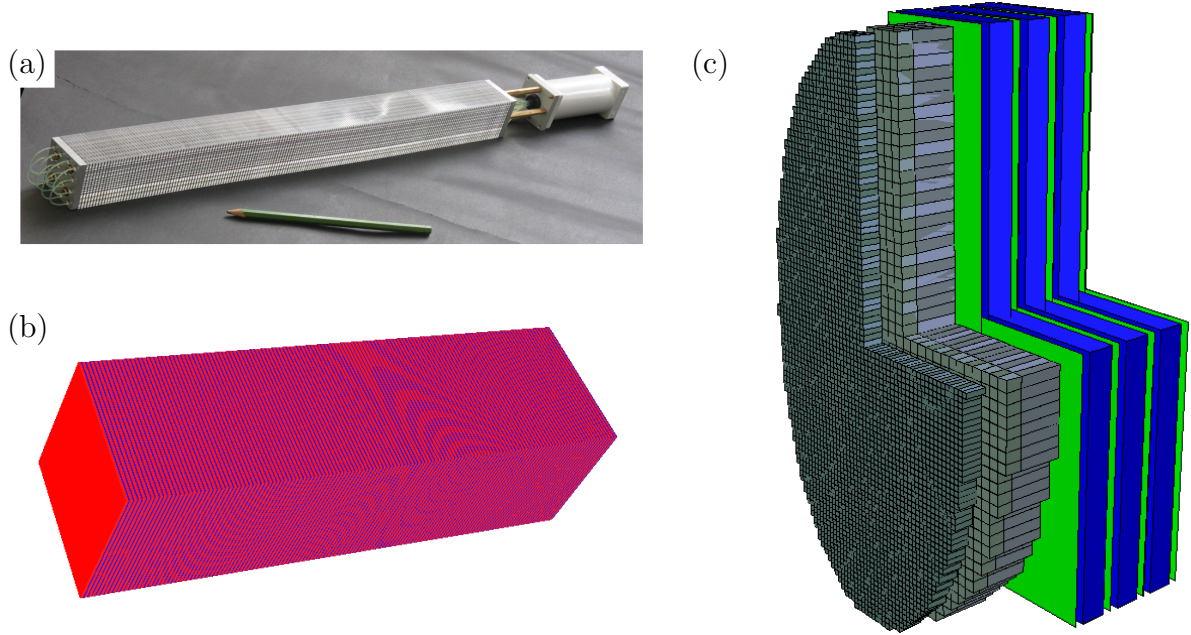
\begin{figure}[tbh]
  \setlength{\unitlength}{1mm}
  \begin{center}
%    \includegraphics[width=0.95\linewidth]{./calorimeter/module_images5.pdf}
%\put(-32,153){(b)}
  \begin{picture}(160,95)
     \put(0,50){\includegraphics[width=0.5\linewidth]{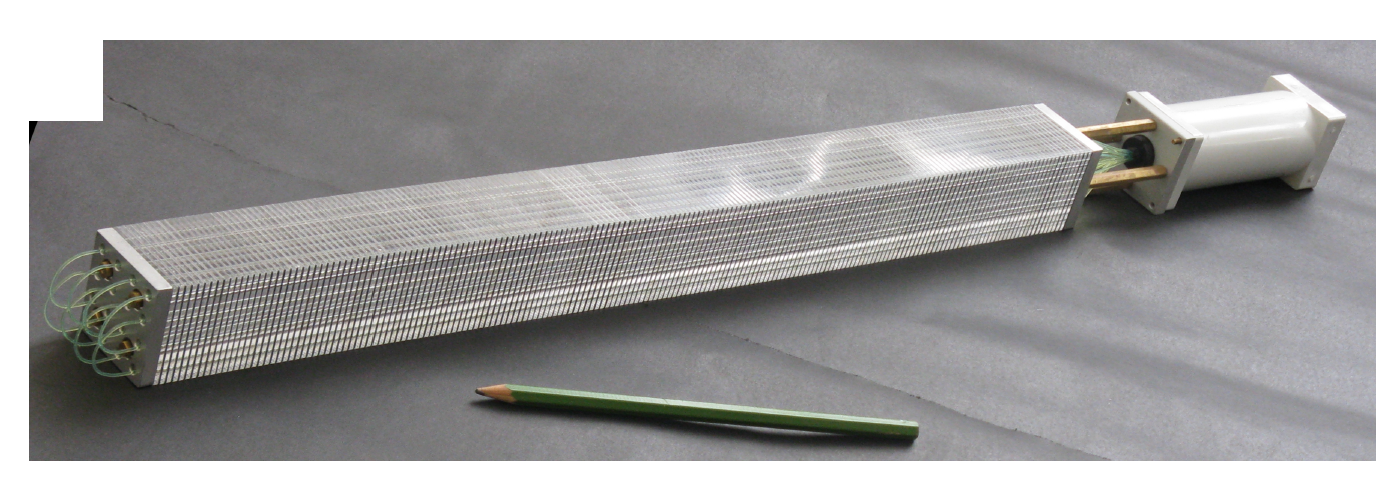}}\put(0,75){(a)}
     \put(90,0){\includegraphics[width=0.45\linewidth]{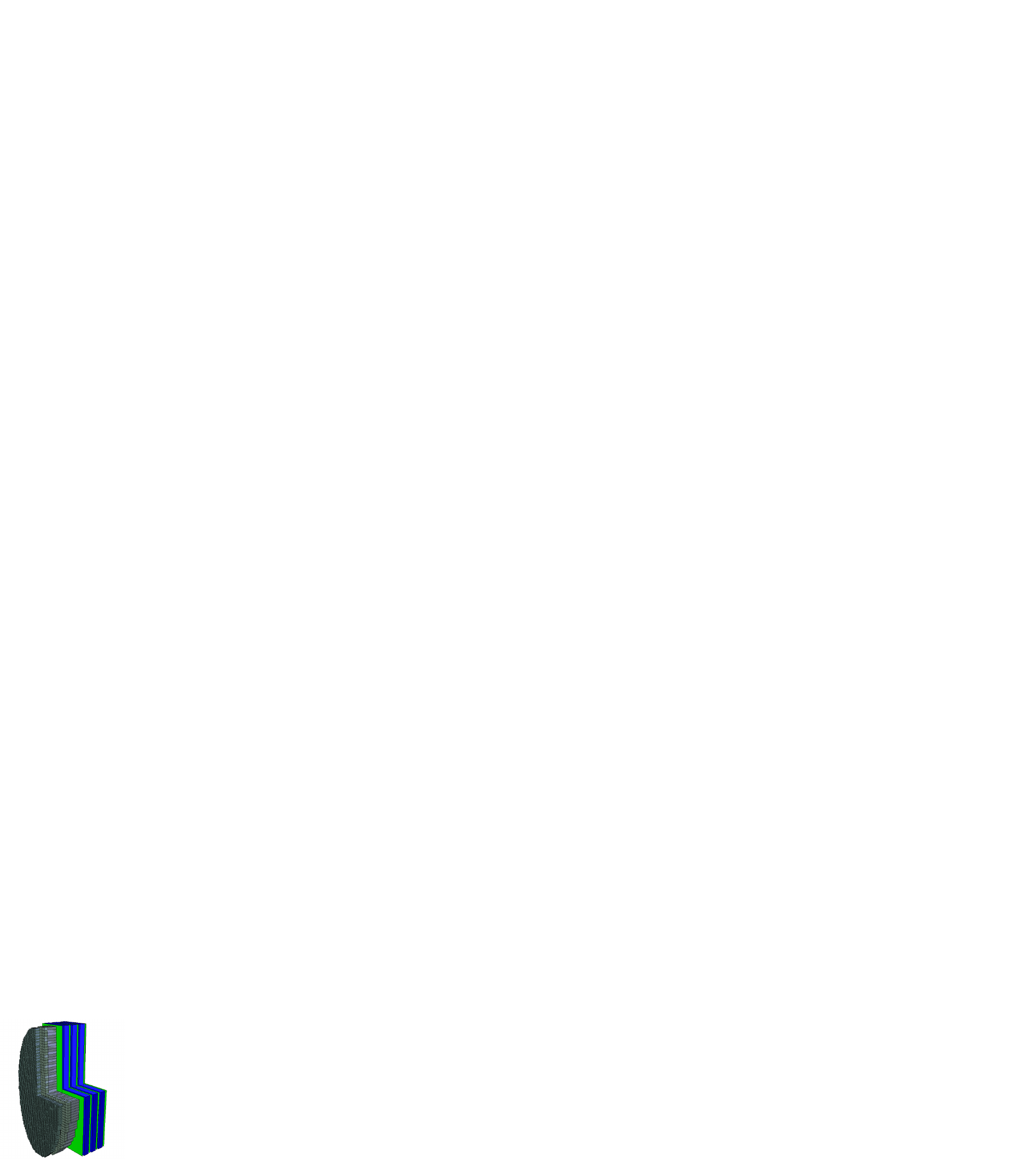}}\put(90,75){(c)} 
     \put(0,0){\includegraphics[width=0.5\linewidth]{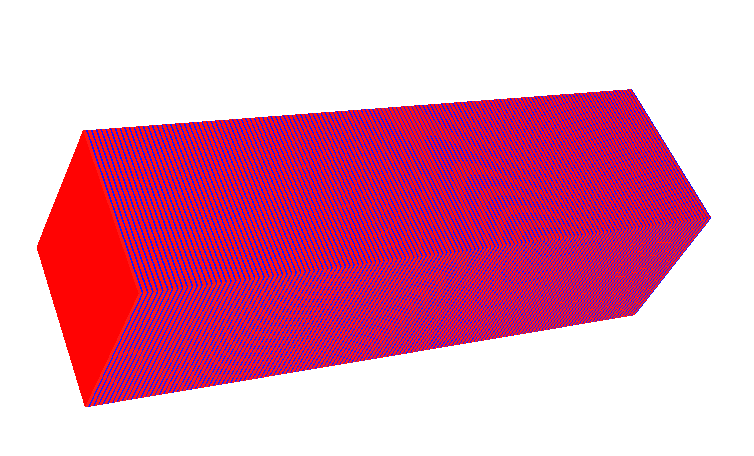}}\put(0,40){(b)} % alternatively ecal-module-lowres.png
     % \put(80,-10){\includegraphics[width=0.5\linewidth]{./calorimeter/ecal-4.png}}\put(100,75){(c)}
     % \includegraphics[width=0.3\linewidth]{./calorimeter/ecal_ellipse6x12m2.png}\put(-32,153){(b)}
     % \includegraphics[width=0.2\linewidth]{hcal.png}\put(-22,153){(c)}
   \end{picture}
    %\vspace*{-1.0cm}
  \end{center}
  \caption{
    \small %make captions a little bit smaller than main text
   Calorimeter system geometry:
   (a) a shashlik calorimeter cell (for display purposes cover foils have been removed),
   (b) Geant4 view of a shashlik ECAL module, and
   (c) % 5x10~m$^2$ elliptic shape ECAL geometry (prospective view with 1/4 removed).
    5x10~m$^2$ elliptic Calorimeter system and its position with respect to the MUON system (3D view with the top-right quarter removed).
  }
  \label{fig:calo-geo}
\end{figure}

Recent developments in shashlik technology have shown that electromagnetic shower energies can be measured with a resolution of $\sigma(E)/E \simeq 6\%/\sqrt{E} $ ($E$ in \gev ) \cite{ecal:mapd}. This performance can be obtained using a sampling structure of 1 mm lead sheets interspersed with 2 mm thick scintillator plates and an accurate design of the light collection by the wavelength shifting (WLS) fibers.
A photo of a shashlik module is shown in Figure~\ref{fig:calo-geo}~(a).
Results of detailed simulation of such a structure using the Geant4 toolkit (shown in Figs.~\ref{fig:calo-geo}~(b, c)) are discussed in the Section~\ref{sec:calorimeters:performance}.

The schematic front view of the ECAL is shown in Figure~\ref{fig:calo-geo}~(c).
The calorimeter is constructed from modules, which are shown in the figure as rectangular blocks with a
transverse dimension of \mbox{12$\times$12~cm$^2$}. Each module is separated into 2$\times$2 light-isolated readout cells.
As shown in Figure~\ref{fig:pi0-distance}~(b) the resulting 6$\times$6~cm$^2$ cells provide nearly 100\% efficiency for two-photon separation. A possible increase of the transversal cell size would introduce some inefficiency but allow substantial reduction in the overall ECAL cost. The modular geometry of the calorimeters facilitates their integration in any geometrical configuration of the SHiP detector. The modules are connected mechanically at the front and backward sides and form a self-supporting structure.

A Geant4 implementation of a calorimeter module is shown in Fig~\ref{fig:calo-geo}~(b). The module is built
from alternating layers of 1~mm thick lead, white reflecting Tyvek paper (2x0.06~mm) and 2~mm thick injection-moulded polystyrene-based scintillator tiles. The complete stack is compressed using stainless steel side covers, pre-tensioned and welded to front and rear steel lids.
In total there are 140 layers resulting in a total depth of 25~$X_0$. The light is read out via 0.6~mm diameter WLS fibers (such as, for example, the BCF-91A from Saint Gobain Cristals) which penetrate the entire length of the module. The fibers belonging to each readout cell are bundled at the end of the module and polished. The density of the fibers is determined via an optimization of light yield, uniformity of response and cost.

The calorimeter has outward dimensions of $\sim 5\times10~\mbox{m}^2$ (or more precisely 504$\times$1008~cm$^2$) and contains  2876 modules arranged in 84 rows  and 42 columns, for a total of 11504 readout channels. The instrumented region is elliptical in shape with an acceptance designed to overlap with the tracking one, see Figure~\ref{fig:calo-geo}~(c).

Fibers from each module cell will be coupled to a photodetector that will be able to provide good linearity in the whole energy range. In addition its signal/noise characteristics allow a precise identification of a minimum ionizing particle signal. This is necessary for the calibration procedure discussed below. Natural photodetector candidates are standard photomultiplier tubes, such as, for example, the FEU-84-3, which is employed in several calorimeters due to its performance and compact size (10.5 cm in length, 3.4 cm in diameter).
Since there no major concerns of radiation damage in the SHiP environment, other candidates, such as
% Silicon PMs or
APDs\cite{ecal:mapd}, will also be considered. Depending on the chosen photodetector technology, some protection against the effects of the magnetic fringe field may be necessary, in particular for the PMT option. In this case, PM tubes will be surrounded by a mu-metal layer.

\subsubsection{Electronic readout chain }

Signals coming out from the photodetectors will be read-out by custom electronic boards equipped with Flash ADCs, FPGAs and gigabit ethernet connection. Signal shaping will be tailored to the characteristics of the chosen photodetector signals searching for a compromise between the total charge range and required measurement accuracy, as well as the timing resolution (at the level of a few ns).
The full scale and calibration requirements drive the ADC range and accuracy, which has to be equivalent to 12 bits for the range with a 0.5 bit accuracy.
The output of the analog chain designed to amplify the voltage signal will be digitized by a FADC with at least an 11 bit resolution and  sampling rate higher than 50 MHz. The exact type and performance of the FADC will depend on the chosen photodetector. Very interesting solutions can be obtained with fast multichannel FADCs, such as, for example, the SAMPIC chip\cite{ecal:sampic} which is able to process 16 readout channels, which allows for an early system integration. The FPGA will measure the charge collected by integrating the samplings in a suitable time window, and the arrival time by interpolating between the ADCs samples\cite{ecal:digi}. Each readout board will process at least 64 individual channels. Since the signal rate will be at the level of few tens of Hz per channel, it will be possible to merge data from all the readout boards in a crate and send the resulting data stream on a single gigabit ethernet line.
The exact number of boards and data lines needed depend on the channel integration in a single board. Using current lower numbers we expect to have in the system about 180 boards on 12 DAQ crates.

\subsubsection{Detector calibration procedure and time alignment}

One of the main concerns of this detector which would be built mostly using well known technology elements is its calibration. This is due to the fact that in the SHiP experiment practically all the standard particles which are produced in the beam interaction are stopped in the target and hadron absorber or are scattered by the muon filter, the only exceptions being neutrinos and a few surviving muons. In the current set-up, during spill time, about 10$^5$ muons per second with an average momentum of 10 \gevc\ are expected to penetrate the vacuum vessel.
This muon flux is enough to perform a continuous ECAL energy calibration using the following two steps:
%\footnote{\em MV: a couple of plots would be welcomed here!}
\begin{enumerate}
\item {\bf Cell equalization:} Muons entering in ECAL cells will give a signal equivalent to an about 400~\mev\ energy deposition with a small dispersion around the central value ($<10\%$). This gives the opportunity to equalize the response of all calorimeter cells. The requirement to be able to measure correctly the MIP signal fixes the minimal ADC position of the MIP peak: 16 counts at 0.5 counts resolution. This also fixes the maximal ADC step to 25 \/ADC count and the bit range to cover the 0-70~\gev\ interval: 12 bits.
\item {\bf Absolute energy scale:} Muons decaying in flight before the magnet are a source of electrons with a broad energy range that can be used to calibrate cells or regions of the calorimeter. The electron momentum measured by the tracking can be compared with the energy deposited on ECAL and they can allow a fine ECAL calibration at a level of 0.5\% statistical uncertainty on a time scale of a week for most of the cells. The overall calibration can be cross-checked on background channels providing $\pi^0$-mesons, such as, for example, $K_L$-mesons produced near the vessel front window.
\end{enumerate}

The time alignment of each readout channel will be performed using electromagnetic showers and muons. The electromagnetic showers involve several cells at the same time (within tenths of ns) and can be used to provide the first inter-calibration of the calorimeter. The matching between the information of a charged particle time (as measured by any timing detector) and the cluster time will provide an absolute and uniform time alignment of all readout cells within few tenths of nanosecond.

\subsubsection{Performance on single particles}
\label{sec:calorimeters:performance}
\begin{figure}[tb]
  \setlength{\unitlength}{1mm}
  \begin{center}
  \begin{picture}(180,60)
    % \put(-10,0){\includegraphics[width=0.55\linewidth]{./calorimeter/eres65-1.png}}\put(20,55){(a)}
    % \put(80,0){\includegraphics[width=0.46\linewidth]{./calorimeter/EoPplot.png}}\put(100,55){(b)}
    \put(0,-5){\includegraphics[width=0.50\linewidth]{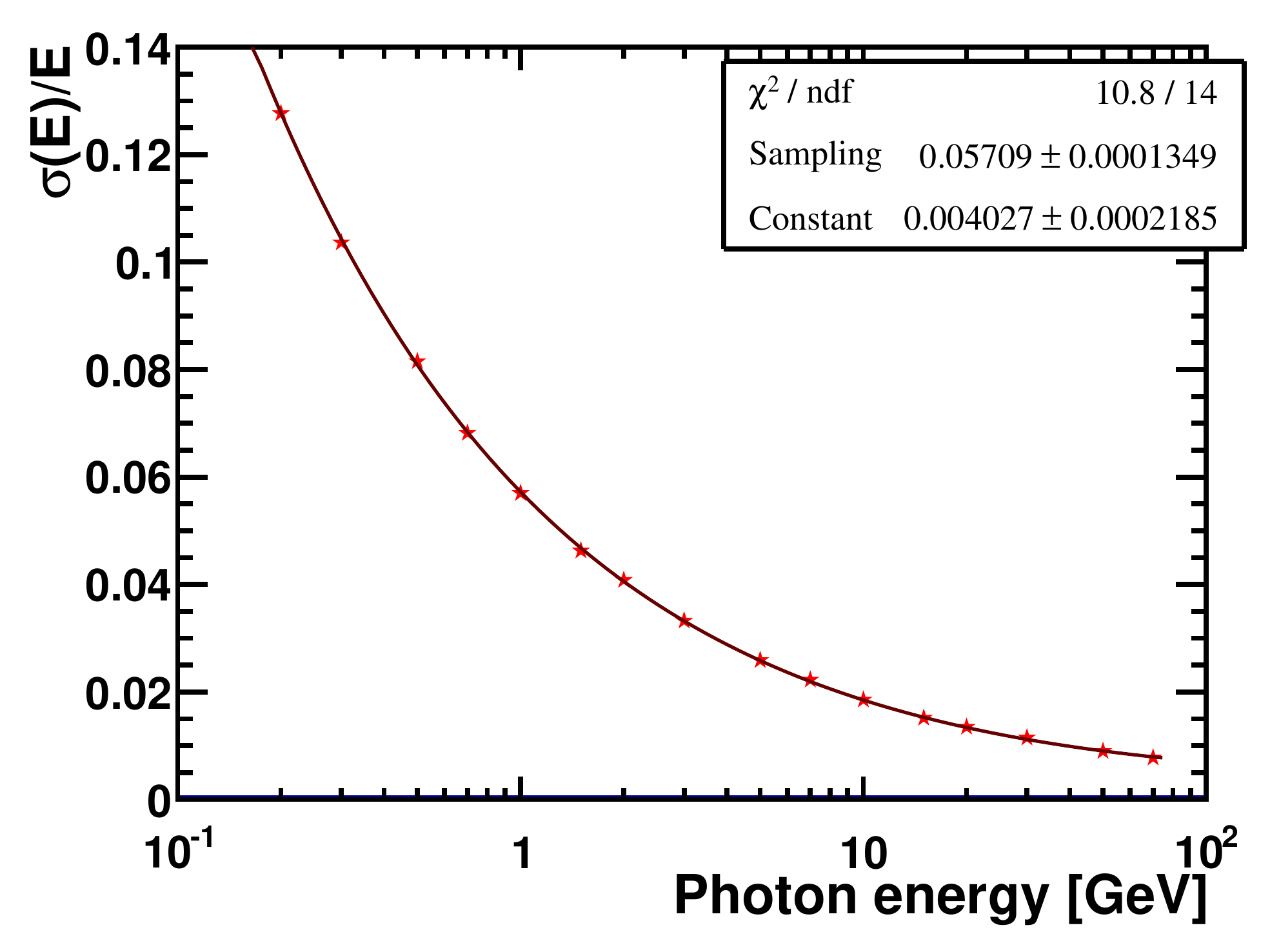}}\put(30,45){(a)}
    \put(85,-5){\includegraphics[width=0.52\linewidth]{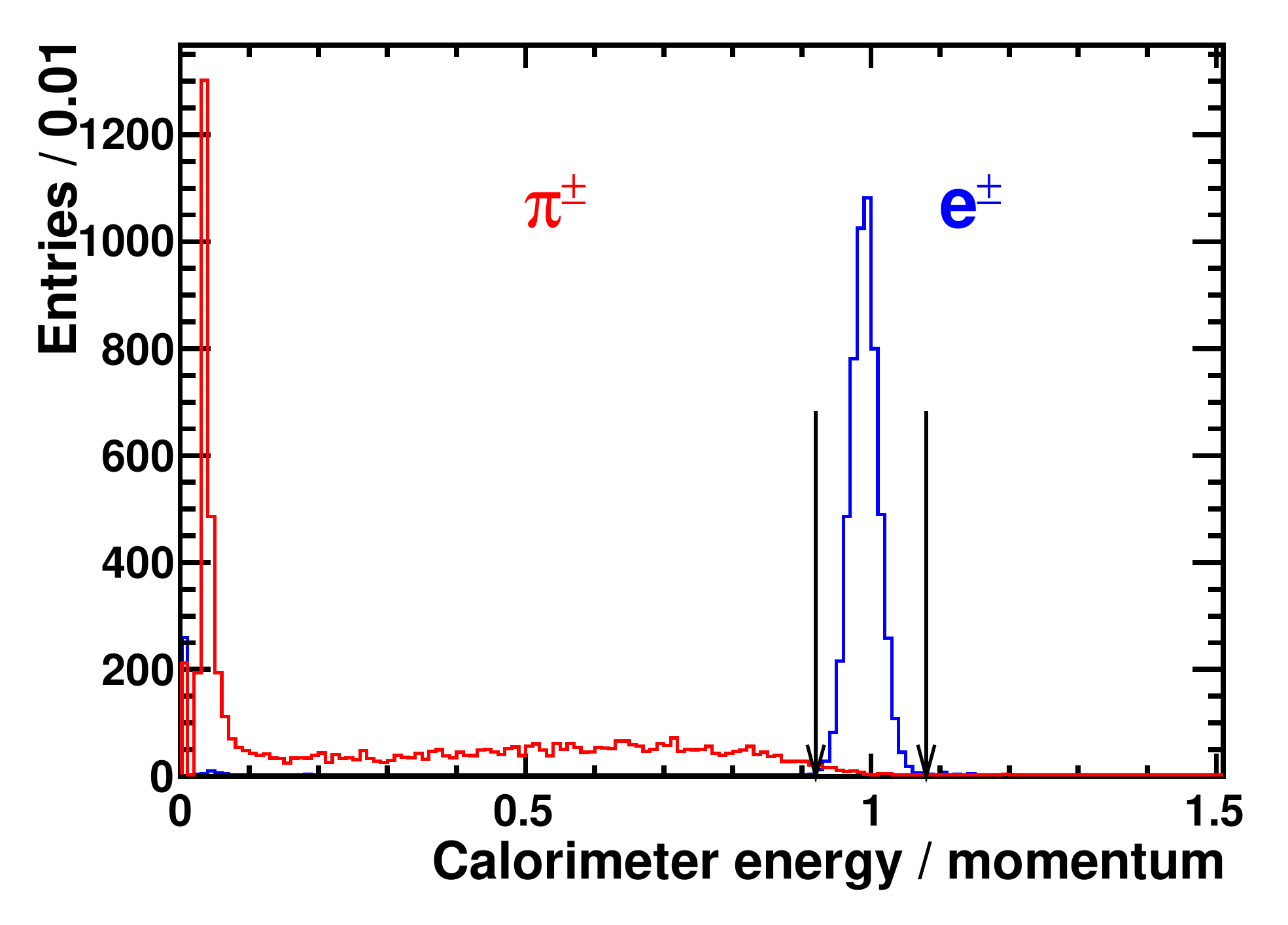}}\put(110,45){(b)}
 %   \vspace*{-1.0cm}
  \end{picture}
  \end{center}
  \caption{
    \small %make captions a little bit smaller than main text
   (a) ECAL relative energy resolution $\sigma(E)/E$ depending on photon energy collected in 5$\times$5 cell clusters and (b) energy over momentum ratio for a 10~\gevc\ momentum sample of electrons (blue) and pions (red) showing the particle identification potential.}
   % a) ECAL relative energy resolution $\sigma(E)/E$ depending on photon energy for 3x3 cell clusters built near local maximum and b) Energy over momentum ratio for a 10 \gevc\ momentum sample of electrons (blue) and pions (red) showing the particle identification potential.}
  \label{fig:ecal-resolution}
\end{figure}

\begin{figure}[tb]
  \setlength{\unitlength}{1mm}
  \begin{center}
  \begin{picture}(85,60)
%    \includegraphics[width=0.47\linewidth]{./calorimeter/SpResol.png}
%    \includegraphics[width=0.47\linewidth]{./calorimeter/spatialres3.png}
%\put(-132,163){(a)}
    % \includegraphics[width=0.47\linewidth]{./calorimeter/pi0.png} % old plot, sigma = 4MeV
    % \includegraphics[width=0.47\linewidth]{./calorimeter/pi0_back-1.png}
    \put(0,-5){\includegraphics[width=0.5\linewidth]{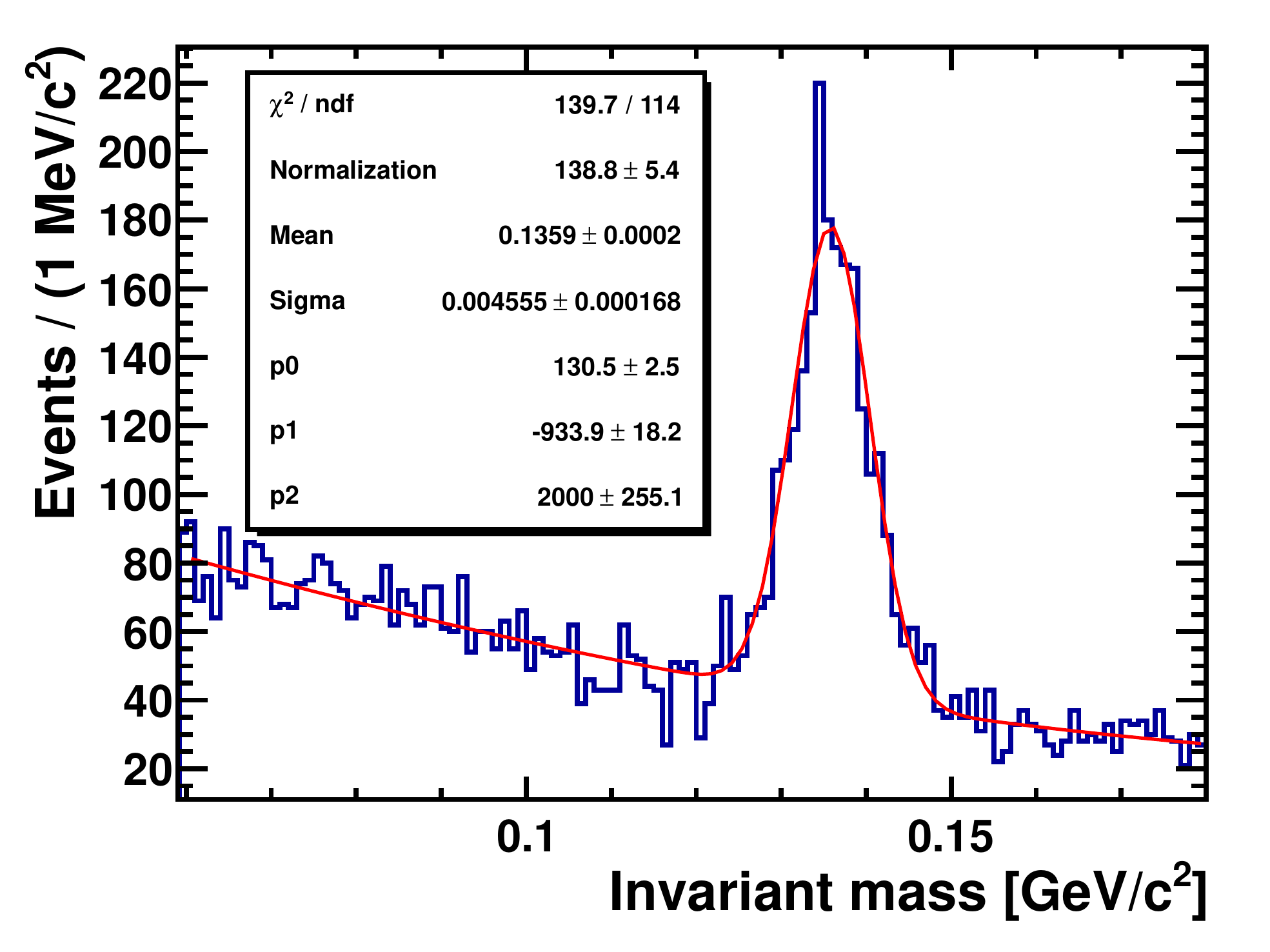}}
%\put(-132,163){(b)}
 %   \vspace*{-1.0cm}
  \end{picture}
  \end{center}
  \caption{
    \small %make captions a little bit smaller than main text
%   a) The spatial resolution on ECAL measured on $\gamma$ samples after S-shape correction and b)
A $\gamma\gamma$ invariant mass spectrum with a clear $\pi^0$ signal with a 5~\mevctwo\ width.}

  \label{fig:ecal-spatial}
\end{figure}

The performance of the calorimeter in SHiP has been studied on single particle samples generated at several fixed energies.  Results are shown in Figure~\ref{fig:ecal-resolution} and in Figure~\ref{fig:ecal-spatial}. The energy resolution on  photons is shown in Figure~\ref{fig:ecal-resolution}~(a), where single photons are generated randomly at the front of the ECAL. The fit to the standard parameterization, $\sigma(E)/E = a/\sqrt{E} \oplus b$,
results in $a = (5.71 \pm 0.01)\%$ and $b = (0.40 \pm 0.02)\%$, which lie well within the requirements described above. The capability of ECAL to distinguish between electron and pions is shown in
Figure~\ref{fig:ecal-resolution}~(b) for samples of electrons and pions with 10\gev\ energy. In both samples particles are generated near the center of SHiP with random directions toward the calorimeter; the momenta of particles ire reconstructed using the tracking system and the calorimeter energies are determined after digitization, clustering and calibration. In these samples, more than 96\% of electrons have been correctly identified in the electron region ($0.92<E/p<1.08$) with the pion contamination below 2\%. The pion rejection can be further improved using the measurements of the cluster shape and position.

ECAL spatial resolutions have been studied on photon samples generated at fixed momentum in the middle of the vacuum vessel. Simulated ECAL signals have been digitized and clustered and a fine tuning procedure of the cluster position evaluated at the shower maximum has been performed. The obtained resolution on the $x$ or $y$ cluster coordinate
%is shown in Figure~\ref{fig:ecal-spatial} (a) and
has been fitted with:
$\sigma_{x,\,y}=0.10\oplus\frac{0.53\%}{\sqrt{E}}\oplus\frac{0.23}{E}$, with $E$ in \gev\ and $\sigma$ in cm for cell sizes of 4 cm width. Results for 6 cm width cells are expected to lie close to the HERA-B results on middle modules\cite{herab:2007}: $\sigma_{x,\,y}=0.28 \oplus\frac{1.35\%}{\sqrt{E}}$,   with $E$ in \gev\ and $\sigma$ in cm.
%\footnote{\color{red}\it The spatial resolution numbers and Fig are probably to be updated for the new cell size!}
A high spatial resolution is an important prerequisite to attain a good invariant mass resolution from $\pi^0$s decays. In Figure~\ref{fig:ecal-spatial} the $\gamma\gamma$ invariant mass of a HNL sample %\footnote{MV to be checked}
is shown. A clear Gaussian peak corresponding to $\pi^0$ is seen. Its width of 5 \mevctwo, obtained in a simulation with 6 cm ECAL cell size, is still dominated by the energy resolution.
	
\subsection{Hadronic calorimeter}

\subsubsection{Performance requirements}
The main aims of the hadron calorimeter are:
\begin{itemize}
\item to provide the pion identification;
\item to provide pion/muon discrimination especially for low momentum particles ($p<5\,$\gevc);
\item to tag neutral particles like $K_L$ and $n$ not seen by other detectors, mainly for background identification and rejection;
\item to provide timing information on signals at the ns level for signal-event association and background rejection.
\end{itemize}

%{\em \small MV: numbers are missing!}
%
\subsubsection{Design considerations}

Several technologies for hadron calorimeters are available, e.g. tile calorimeters. For conceptual simulation studies in this document the shashlik technology has been chosen due its similarity to the ECAL technology and ease of the MC implementation. The performance will be largely independent of the chosen technology and more dependent on the converting and active material types and thicknesses.
In this document, the hadronic calorimeter will therefore be composed of several sampling modules placed behind ECAL with the same acceptance. 
Each module is composed by several layers of converting material alternated with active material. In the simulations, each converting layer is composed by 15 mm thickness of iron and each active layer is made of 5 mm of polystyrene-based scintillator tiles. The module size has been optimized to minimize the total number of independent channels. The longitudinal segmentation can be exploited to maximize the $\pi / \mu$ separation at different momenta keeping the overall amount of material at a minimum in order not to spoil the $\pi / \mu$ separation performed in the muon detector.

To achieve the above aim, a procedure for the optimization of an HCAL longitudinal sampling has been performed using a detailed simulation. This involves the construction of likelihood functions depending on energy deposited in each HCAL station. Various number of layers as well as sampling structures have been studied for one, two or three longitudinal stations and for different thickness in the stations.
A preliminary solution foresees two HCAL stations, the first thin section (H1) with 18 sampling layers followed by a second section (H2) of 48 layers followed by the muon detector. This geometry maximizes the capability to discriminate muons from pions at low momentum with a reduced amount of material in front of the muon detector, such that the high momentum tracks can be efficiently identified by the muon system.  In Table~\ref{tab:hcal} the pion suppression factors ($f$) and the pion misidentification probabilities ($p=1/f$) in the momentum range $1\div 10$~\gevc\ at 95\% muon identification efficiency evaluated on energy deposits on ECAL, H1 and H2 are shown. Further suppression can be obtained using spatial information and shower shape asymmetries.

%The combination of these results with the MUON detector is discussed elsewhere {(\em MV: where exactly?)}.

% table from MIsha Prokudin talk on 2015 02 03 calo meeting
%\begin{table}
%\begin{center}
%\caption{\small Pion suppression factors at 95\% muon identification efficiency, achieved by  ECAL $\oplus$ H1$\oplus$H2 in the current geometry. \label{tab:hcal}}
%\begin{tabular}{|l|c|} \hline\hline
% track &  $\pi$ suppression  \\
%momentum &  ( ECAL $\oplus$ H1$\oplus$H2, 18+48 layers) \\ \hline
% 1  GeV    & 23 \\
% 1.5  GeV  & 32\\
% 2  GeV   &  50\\
% 2.7  GeV  & 120\\
% 3  GeV  &  160\\
% 5  GeV  &  210\\
% 10  GeV  & 250\\
%\hline \hline
%\end{tabular}
%\end{center}
%\end{table}

% table from Victor talk in Neaples  - 2015 02 10
\begin{table}
\begin{center}
\caption{\small Pion suppression factors and pion misidentification probabilities at 95\% muon identification efficiency, achieved by  ECAL $\oplus$ H1$\oplus$H2 in the current geometry.\label{tab:hcal}}
\vspace{0.2mm}
\begin{tabular}{lccccccc} \hline
Track momentum ($\rm GeV/c$)\phantom{\LARGE M}
                                                           &  1 & 1.5 & 2 & 2.7 & 3 & 5 & 10\\ 
\hline
$\pi$ suppression factor                       & 23 & 32 & 50 & 120 & 160 & 210 & 250 \\
$\pi$ misidentification probability (\%) & 4.3 & 3.1 & 0.20 & 0.83 & 0.63 & 0.48 & 0.40 \\
\hline
\end{tabular}
\end{center}
\end{table}

The results for $\pi / \mu$ identification are very sensitive to the Geant tuning.
Test beam measurements are therefore necessary for tuning and validation of the simulation and for the choice of the final HCAL technology and configuration.

%\begin{figure}[htb]
%  \begin{center}
% % old plots
% %    \includegraphics[width=0.22\linewidth]{./calorimeter/hcal.png}\put(-15,190){(a)}
% %    \includegraphics[width=0.4\linewidth]{./calorimeter/ecal-hcal-muon-4.png}
% %\put(-32,190){(b)}
% 
% 
%     \includegraphics[width=0.4\linewidth]{./calorimeter/hcal12-1-geohack.png}\put(-15,190){(a)}
%     \includegraphics[width=0.4\linewidth]{./calorimeter/ecal-hcal12-muon-5-geohack.png}
% \put(-32,190){(b)}
% 
% 
% %\put(-22,153)    %\vspace*{-1.0cm}
%   \end{center}
%   \caption{
%     \small %make captions a little bit smaller than main text
%       (a)  Geant4 drawing of the HCAL wall as implemented in FairShip simulation framework and (b) its position with respect to ECAL and muon system. All detectors are shown with the top-right quarter removed for display purposes.}
%   \label{fig:hcalo-geo}
% \end{figure}
% 

\subsubsection{Detector geometry}
%\footnote{numbers to be checked}

Given the results of the previous studies on the $\pi/\mu$ discrimination and identification, the baseline option for HCAL is a system consisting of  1512 single-cell modules of 24x24 cm$^2$, arranged longitudinally in 2 independently read-out stations. In each station the modules are stacked in  42 rows and 22 columns forming an elliptic shape, see Figure~\ref{fig:calo-geo}~(c). There are two types of HCAL modules, 18-layer modules in the 1$^{\rm st}$ station and 48-layer ones in the 2$^{\rm nd}$ station.
Each layer consists of a 1.5 cm steel absorber and 0.5 cm scintillator tiles wrapped with Tyvek reflecting paper. The total length of the HCAL detector (66 layers) corresponds to about 6.2 hadronic interaction lengths $\lambda_I$.
 As done in the ECAL, the light is collected by \mbox{0.6~mm} diameter WLS fibers penetrating from the front and from the rear of each module. Fibers coming from a module are bundled together and coupled to the same photodetectors used for ECAL.

The longitudinal space taken by the whole Calorimeter system will be about 3.3 m.

\subsubsection{Electronic readout chain }

The HCAL readout chain will need a minimum dynamic range of 13-14 bits to cover the full scale while maintaining good resolution for minimum-ionising particles. The baseline design for the front-end electronics will use analog shapers and commercial FADCs to measure PMT pulses. FPGAs will apply running digital filters to the FADC outputs for both energy extraction and ns-level pulse timing.

In total the HCAL system has 1512 readout channels.
The HCAL readout boards will process approximately 64 channels each, corresponding to a total of about 24 boards hosted in two readout crates. As for the ECAL, it is expected that a single gigabit ethernet link will be sufficient to read out each board to the DAQ system.

%\subsubsection{Detector simulations and/or performances on $\pi/\mu$ separation}

%\subsection{Cost considerations}
%\subsection{R and D program towards the TDR}

%% file: muon/Muon.tex
%=================
\section{Muon detector}
\label{subsec:muondet}
%=================
%\it Describe here physics goal and environment:
%signals: which channels, which kinematics; backgrounds: 1. combinatorial $\mu \mu$  2.$\nu$-induced background ($K_L$ decays, others?);
%and how this translate into requirements for the muon system: efficiency, time resolution, good hadron/muon/electron separation.}

%
%This section describes the performance requirements and the baseline design of the muon system.

%
%The most stringent requirement for the design of the 
%particle identification detectors comes from the possibility to claim a $3\; \sigma$ signal evidence 
%in case of a single event observation in a data sample of $2 \cdot 10^{20}$ protons on target. 
%In fact this goal requires to have $< 0.2 \% $ expected
%background events in the same dataset for a given $U^2$ value.% ({\it refer to the HNL exclusion plot.}).
%Higher levels of background would cause an effective loss of sensitivity in the low $U^2$ range accessible by SHiP 
%and would be equivalent to reduce the number of protons on target. 
%({\it This discussion should be done somewhere else in more quantitative way}).

The muon system, in conjunction with the electromagnetic and hadronic calorimeters,
is designed primarily to identify with high efficiency muons from signal channels as 
$N \to \pi^+ \mu^-, \mu^+ \mu^- \nu_{\mu}$ in the neutrino portal,  $V \to \mu^+ \mu^-$ or $S \to \mu^+ \mu^-$ 
in the vector and scalar portals, respectively, and to separate them
%from charged hadrons, such as pions, kaons and protons.
%
%
%In particular, the muon system is critical in separating signal events such as  
%$N \to \mu^+ \mu^- \nu$, $N\to \pi^+ \mu^-$ or $S/V \to \mu^+ \mu^-$  decays
from $\nu-$ and $\mu-$ induced backgrounds, mostly $K_L  \to \pi^{\pm} \mu^{\mp} \nu_{\mu}$ and $K_S \to \pi^+ \pi^-$
decays originating in the material surrounding the decay volume 
where one or two pions are misidentified as muons. The muon system can also help 
in separating $S/V \to \pi^+ \pi^-$ final states from 
$K_L \to \pi^{\pm} \mu^{\mp} \nu_{\mu}$, where the muon is misidentified as pion.

The momentum and transverse momentum distributions for the decay products of the $N \to \pi^+ \mu^-$ decay
when both tracks are inside the acceptance of the spectrometer are shown in Figure~\ref{fig:HNLpimu}.
About $\sim3\% $ of the pions in the final state decay into muon and neutrino before the first plane of the tracker
and will not be considered as good candidates.
The momentum interval covered by the muon system ranges from $\sim$3~GeV/c up to $\sim$100~GeV/c, 
being the lower threshold defined by the minimum momentum needed for a muon to cross the calorimeter system. 
Muons and pions with momentum lower than $\sim$3 GeV/c
can be identified using the calorimeter system alone (see Section ~{\ref{sec:calorimeters}).

% questi plots vanno rifatti richiedendo l'ellisse e richiedendo che HNL decada prima del primo straw.
\begin{figure}
\includegraphics[width=0.45\textwidth]{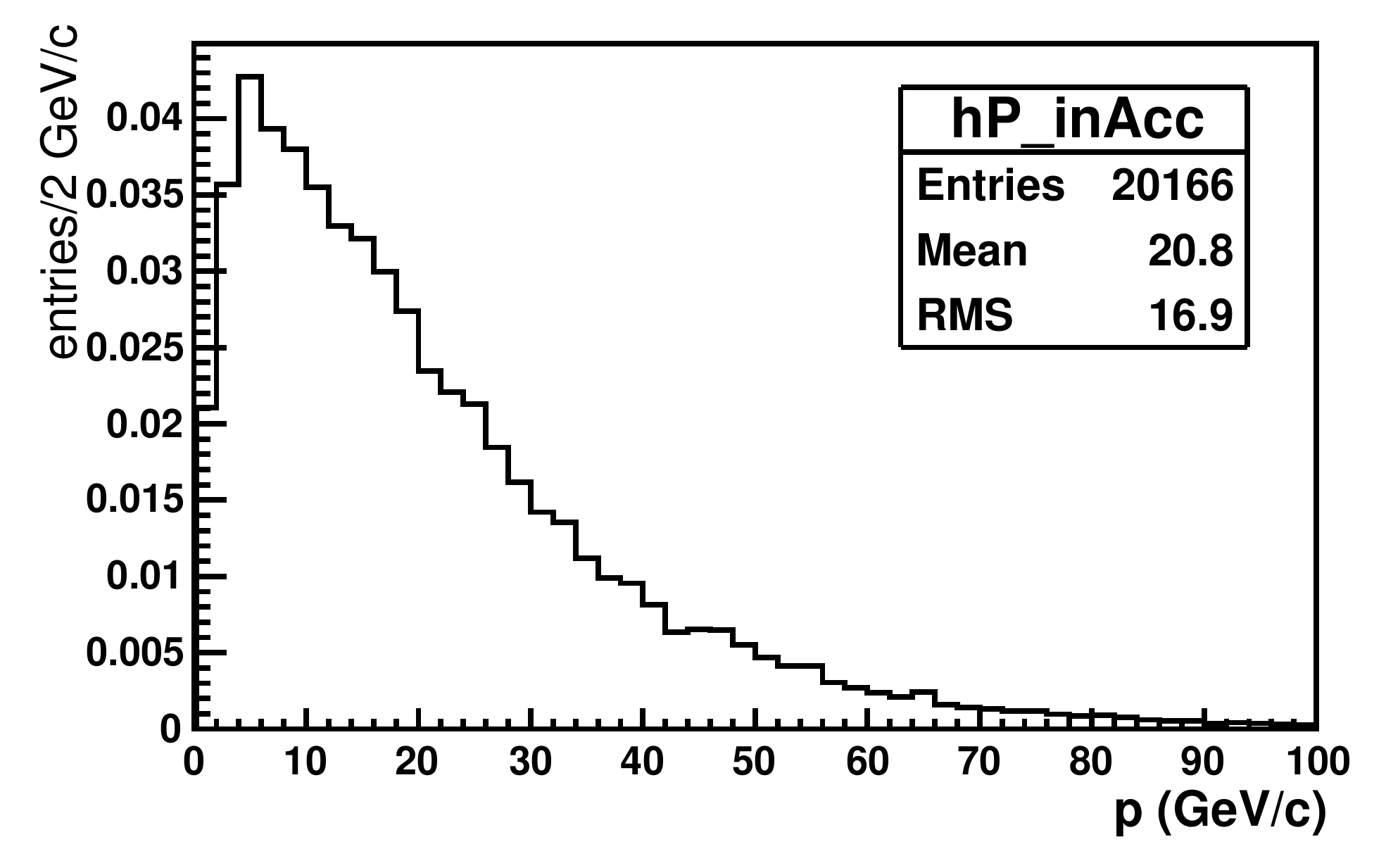}    
\includegraphics[width=0.45\textwidth]{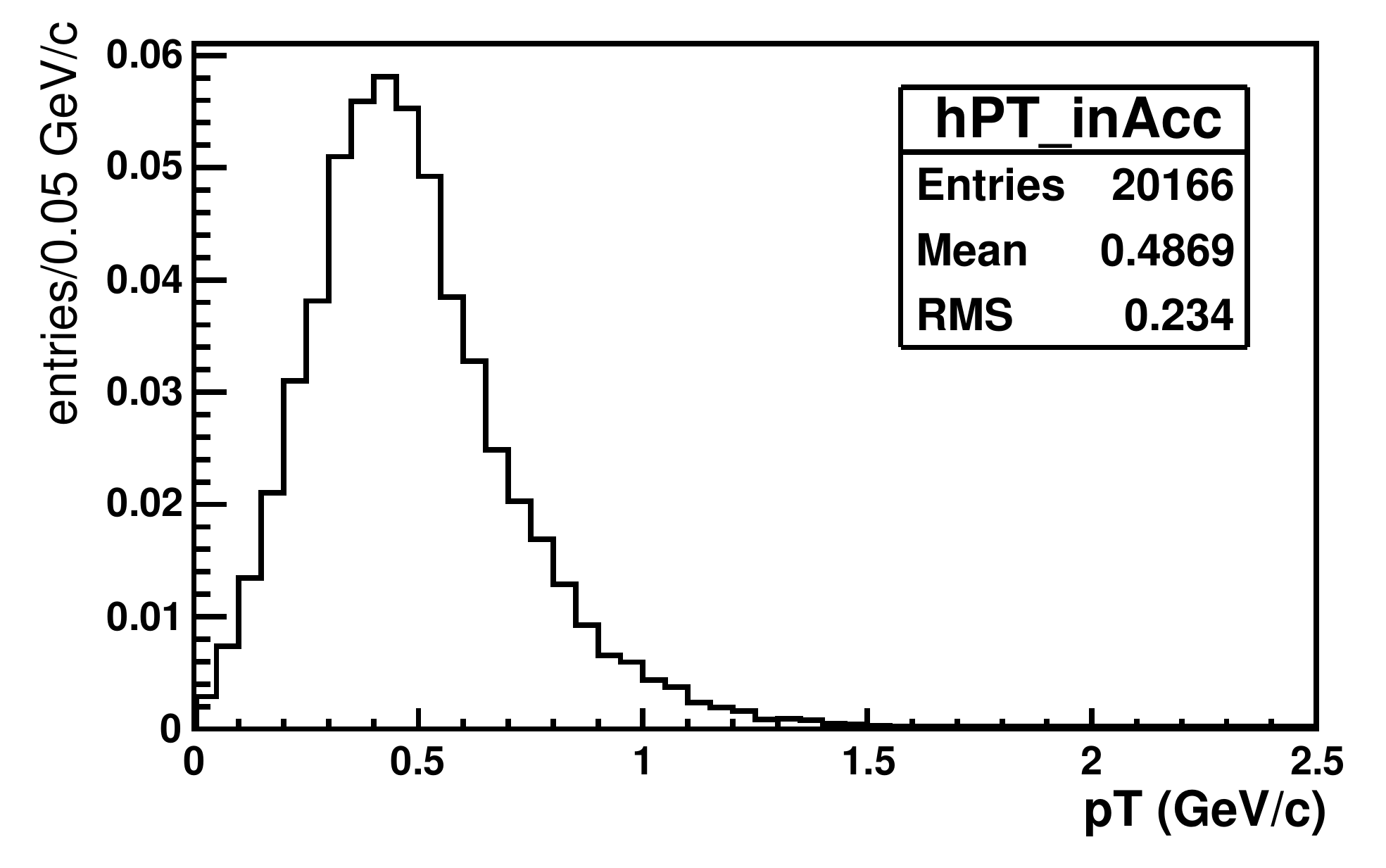}  
\caption{Momentum (left) and transverse momentum (right) of the final states of $N \to \pi^+ \mu^-$ decays for 
$ m(N)  = $ 1 GeV/c$^2$.}
\label{fig:HNLpimu}   
\end{figure}

About 3\% of CC and NC $\nu$-interactions in the material of the muon spectrometer of the $\nu_{\tau}$ detector and in the 
entrance and the walls of the decay vessel  produce a $K_L$ among the fragmentation products.
% among the products of the interactions that enters in the decay volume.
The momentum distribution for $K_L$ mesons that decay in the vacuum tank between the veto station and 
the first straw chamber and have two charged tracks pointing inside the acceptance of the spectrometer is very soft
(Figure~\ref{fig:p_kl} (left)) and even softer is the momentum distribution of the decay products (Figure~\ref{fig:p_kl} (center)); 
when requiring a pion decay  in the chain $K_L \to \pi^+ l^- \nu_l, \pi^+ \to \mu^+ \nu_{\mu}$ , the muons from pion decays 
are all below 1-2~GeV/c  (Figure~\ref{fig:p_kl} (right)) and can be cut away with a small %({\it quantify}) 
loss of efficiency. 

%The particle identification systems have to minimize the $\pi \to \mu$ misidentification probability  when the pion does not decay in flight,
%to reduce the $K_L \to \pi \mu \nu$ ($K_S \to \pi \pi$) events that can mimic signal $N \to \mu \mu \nu$ ($N \to \pi\mu$) events.
%In addition it has to  reduce the $\mu \to \pi$ misidentification probability to reject $K_L \to \pi \mu \nu$ events that can mimic $DP \to \pi \pi$ or $\phi \to \pi \pi$ 
%decays below 10 GeV/c, where most of the $K_L$ decay products are concentrated.

\begin{figure}
\includegraphics[width=0.3\textwidth]{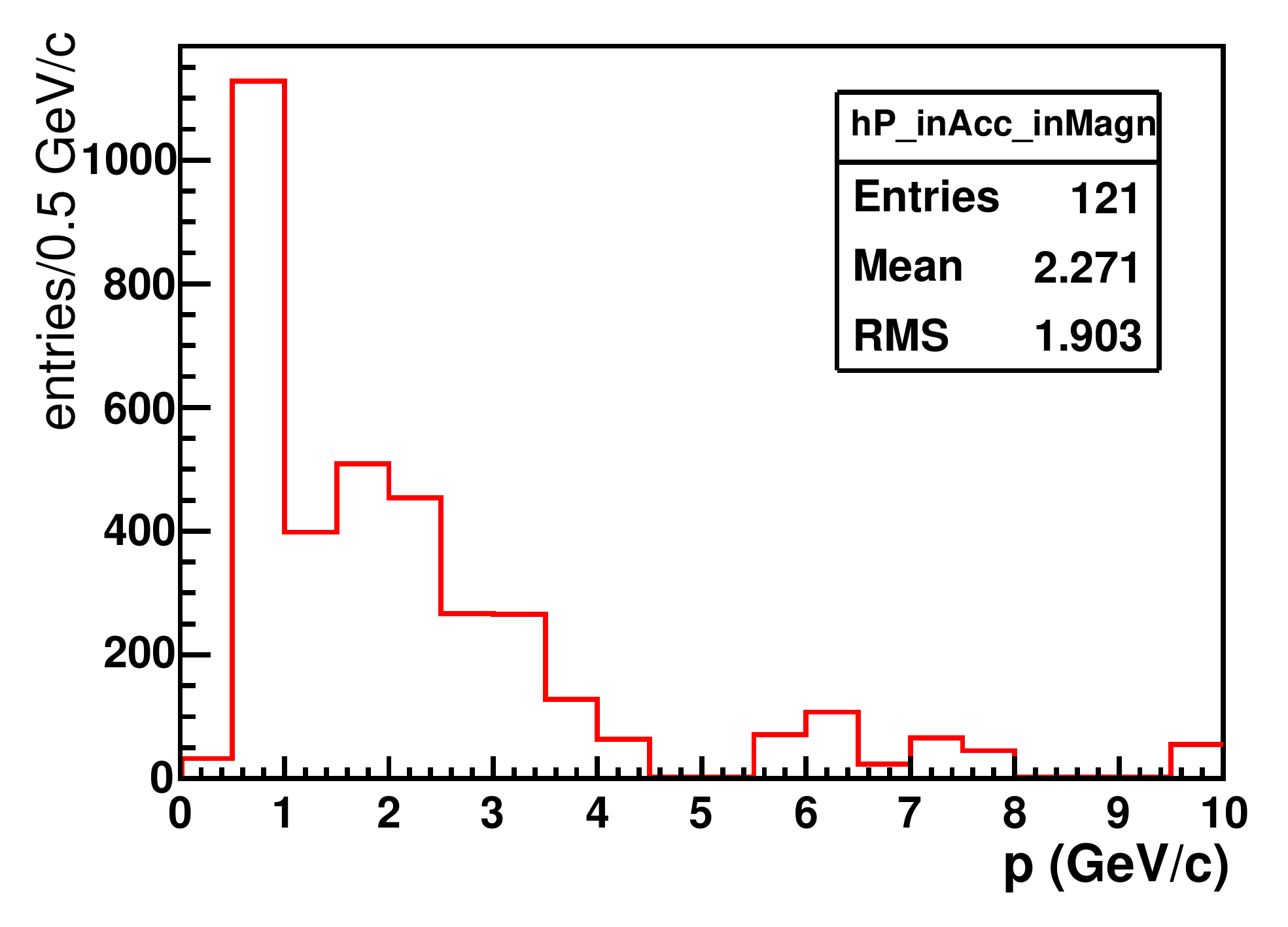}    
\includegraphics[width=0.3\textwidth]{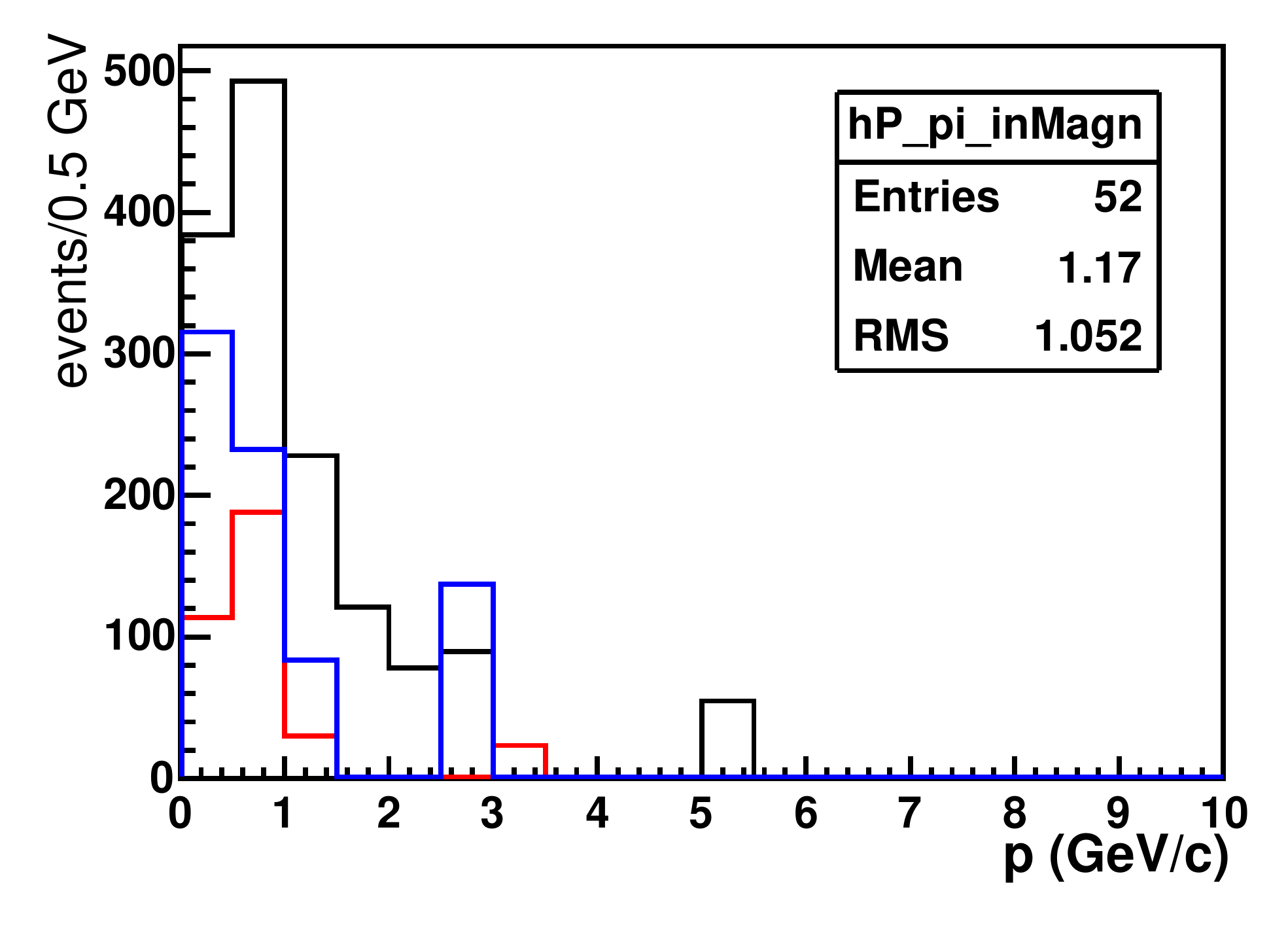}  
\includegraphics[width=0.3\textwidth]{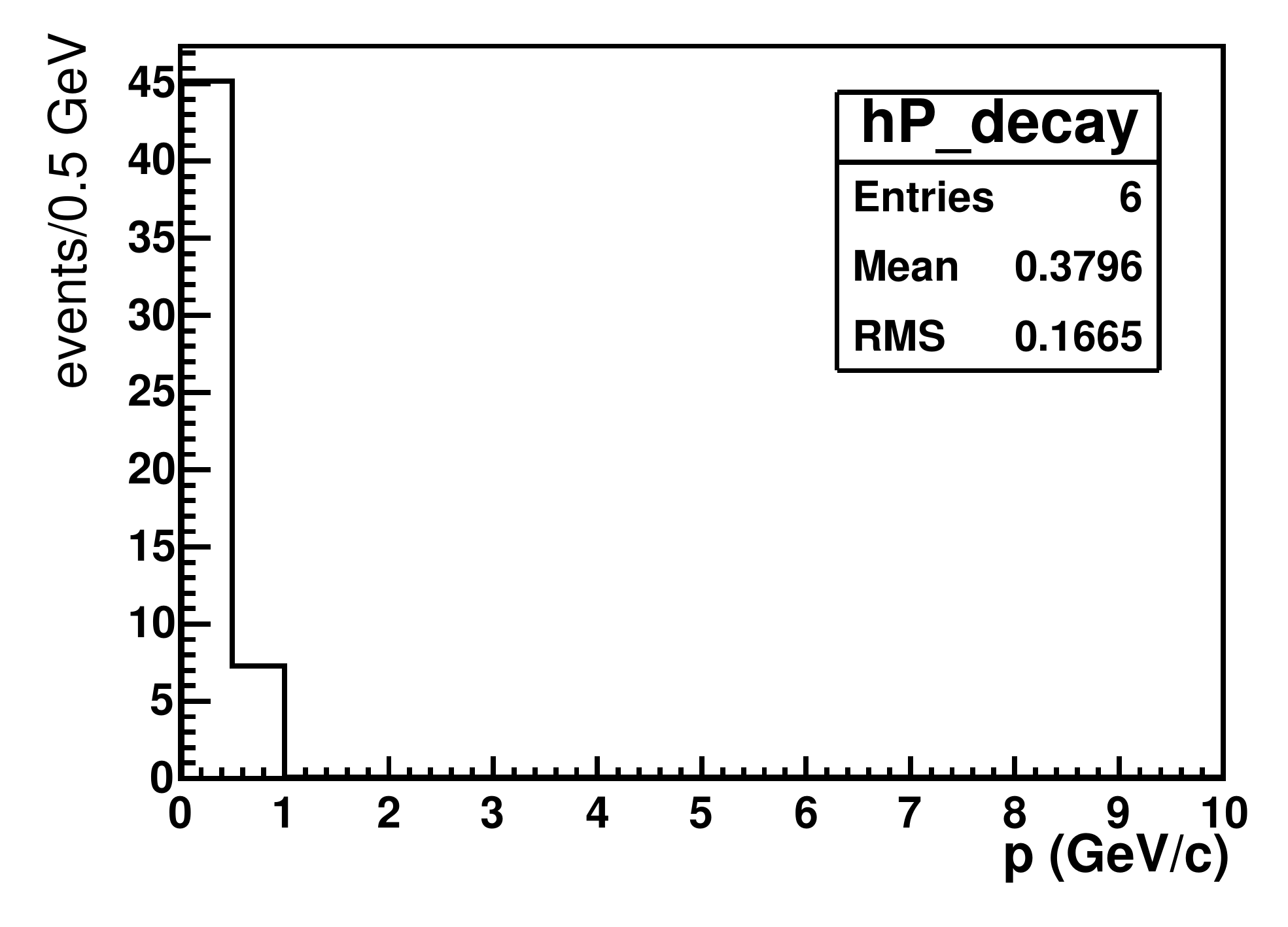}  
\caption{(Left) $K_L$ momentum. (Middle) $\pi$ (black), $\mu$ (red), and $e$ (blue) momentum  from $K_L \to \pi^+ l^- \nu$ with $l^- = \mu^-, e^-$.
(Right) $\mu$ momentum from pion decays from the decay chain $K_L \to \pi^+ l^- \nu, \pi^+ \to \mu^+ \nu_{\mu}$.}
\label{fig:p_kl}   
\end{figure}

%10\% of neutrinos interacting in the  last $\lambda_I$ produce neutrons, $K$ or $\Lambda$ -- is this true?

Finally, the muon system has to separate $N \to \pi^{+} \mu^{-}$ and $N \to  \rho^+ \mu^-$ decays
from random combinations of muon pairs originating from the target when one of the muons is misidentified as pion. 
Random combinations of beam-induced muons that escape the hadron absorber and the active filter
and form a fake vertex inside the decay volume can mimic $N \to \pi^+ \mu^-$ decays if one of the two muons 
is misidentified as pion. 
These are typically high momentum tracks, so the  particle identification system has to
minimize the $\mu \to \pi$ misidentification probability in the entire interesting momentum range.
This background can also be reduced using the time information.
In fact,  beam-induced muons are uniformely distributed over the spill duration ($\sim$ 1 s)
while the decay products of long-lived particles arrive almost simultaneously at the spectrometer.
Simulation studies show that the probability that two combinatorial muons  mimic a decay of a long-lived particle
depends linearly on the time window in which the arrival times of the two muons are recorded: 
for example, a time window of  $(5-10)$ ns corresponds to a probability of
$\sim 10^{-8}$. %({\it from old Hans plot, to be updated}). 
A muon system with a  time resolution $\sigma_t$ of about 1-2 ns will be able to match this level of rejection by selecting 
muons in a time window of $3 \cdot \sigma_t$.
Additional rejection power will come from the timing detector (see Section~\ref{sec:timing_detectors}) that will reach a sub-ns time resolution.

The rate seen by the muon detector is mostly due to the beam-induced muon background. 
Preliminary simulation studies (see Section~\ref{sec:backgroundMuonComb}) show that the flux of muons with $p>3 $ GeV/c is $\sim$ 50 kHz over the entire
muon detector area of (6$\times$12) m$^2$, corresponding to a rate of  $< 0.1 $ Hz/cm$^2$.
Very low radiation tolerance  %({\it how about neutrons?}) 
and a moderate position resolution complete the requirements for the muon detector.

%In the first case, we expect 2660 $\nu_{\mu}$ charged current (CC) 
%interactions per brick in the  opera muon spectrometer per $2 \times 10^{20}$ pot,
%corresponding to $\sim 8 \times 10^6$ $\nu_{\mu}$-CC interactions in the last layer of the opera muon spectrometer.
%About 1.5\% of these interactions produce a $K_L$ that in 30\% of the cases enters into the decay vessel  with an average angle of 0.28 rad
%and an average momentum of $<p> \sim 3.2$ GeV/c corresponding to an average decay length $\lambda \sim 109$ m.
%The average decay path travelled by the $K_L$ into the vacuum vessel is 20 m which corresponds to a probability of 17\% to decay into the vessel.
%Only 33\% of these decays have two tracks reconstructed.
%Hence about $0.5\% \times 17\% \times 33\% \times 8 \cdot 10^6 = 2300 $   $K_L$ decay into the vessel with two reconstructed tracks
%for $2\times 10^{20}$ pot.
% vedere quanti di questi hanno pi/mu e qual e' la distribuzione cinematica del pione con relativi decadimenti in volo.

%=================
\subsection{Detector layout}
\label{subsec:muon_layout}
%=================
%{\it Describe here how many stations and passive filters, 
%which thickness of passive filters, which granularity of the stations (driven by multiple scattering and occupancy);}

The muon detector is placed downstream of the calorimeter systems and comprises four stations of active layers interleaved by three muon 
filters. The detector layout is shown in Figure~\ref{fig:muon_layout}. 

%\vskip -1cm
\begin{figure}[hbt]
\begin{center}
\includegraphics[width=0.5\textwidth]{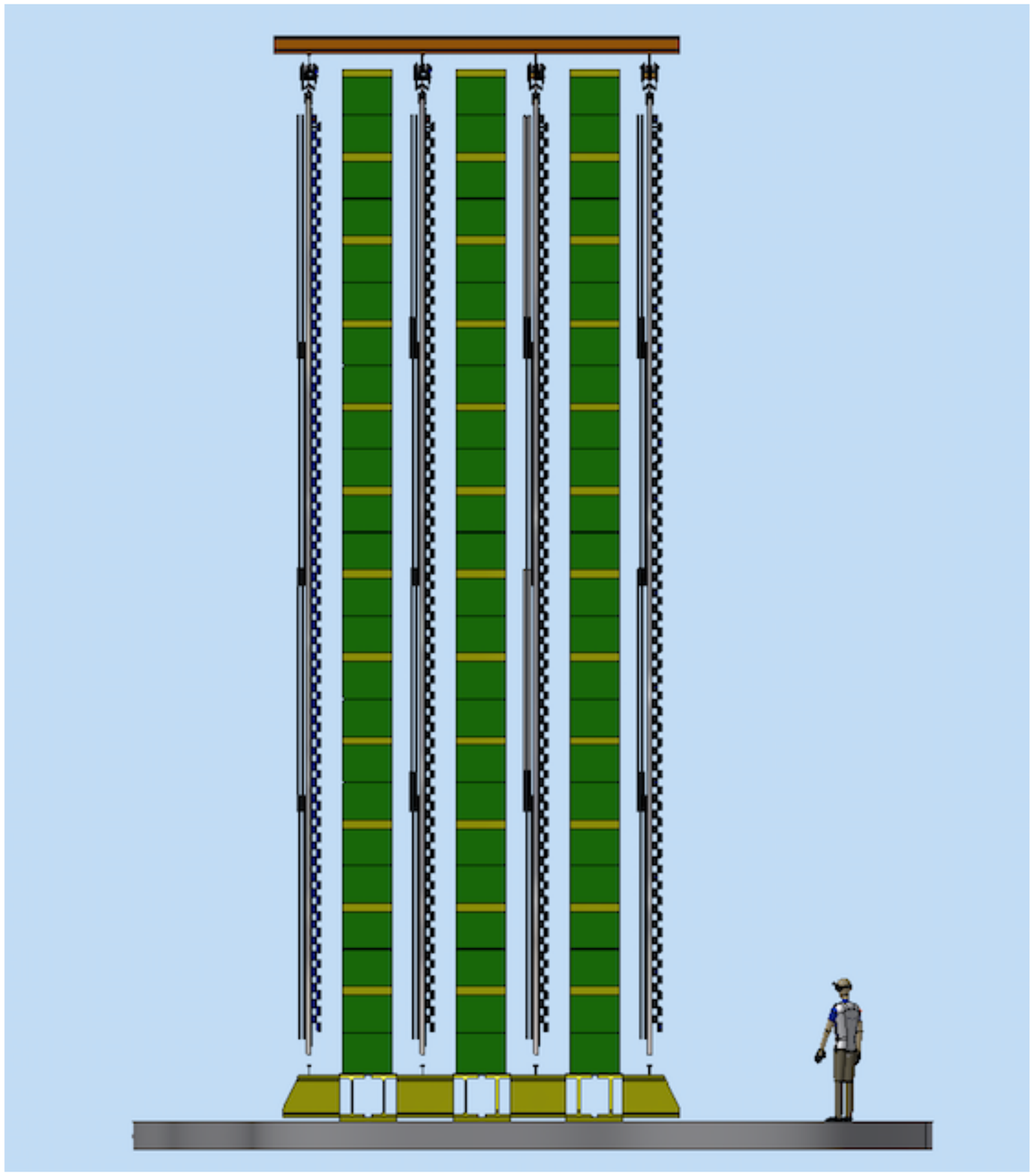}    
\end{center}
\vskip -1cm
\caption{Muon detector layout. Green thick layers are the passive iron filters, grey thin layers are the active modules.
Each active module is defined by two layers (horizontal and vertical) of scintillating bars  and the support aluminum structure.
The beam impinges from the right-hand side.}
\label{fig:muon_layout}   
\end{figure} 

The four stations are 6 m wide, 12 m high  and are placed downstream the calorimeter system.
The amount of material of the calorimeter system corresponds to 6.7 $\lambda_I$. %(in the hcal configuration of 18+48=66 layers) . 
The muon filters are iron walls 60 cm thick corresponding to 3.4 $\lambda_I$ each.
A muon with normal incidence must have an energy at least of 2.6 GeV/c to reach the first muon station 
and at least 5.3 GeV/c to reach the last muon station.
The multiple scattering of  muons in the material of the calorimeter system and the muon filters drives the 
granularity of the system. Preliminary simulation studies show that a readout granularity of 5-10 cm in the transverse directions
is adequate for the interesting momentum range.

%\clearpage
%=======================
\subsection{Active layers technology}
\label{subsec:muon_active_layers}
%=======================
%{\it Extruded scintillating bars with WLS fibres and SiPMs readout:
%description of the basic components (extruded scintillators, WLS fibres, characteristics). Describe other technologies?}

The active detectors of the muon system will cover a surface of $\sim288 $ m$^2$ divided in four stations.
The baseline technology chosen for the active layers is extruded plastic scintillator strips with WLS fibres 
and opto-electronic readout.

These detectors are considered  an established technology  for massive tracking calorimeters in  long-baseline neutrino oscillation experiments. 
The MINOS experiment~\cite{Michael:2008bc} employs extruded bars of $1\times4.1\times800~{\rm cm}^3$ size with 9~m long WLS fibers.    
A fine-grained detector in the Miner$\nu$a experiment~\cite{McFarland:2006pz} is made of triangular-shaped 3.5~m long strips and WLS fibers of 1.2~mm diameter. 
Other experiments that use the same technology are Belle II~\cite{Abe:2010gxa}  and T2K~\cite{Abe:2011ks}.
%All of these detectors use multi-anode PMTs for optical readout. 
%*** aggiungere riferimenti a Belle, T2K, SuperB

This technology is straightforward to operate, as no high voltage or flammable gases
are employed, does not have substantial ageing problems, has high efficiency and good time resolution.
Other reasons for such a choice are: simple transverse segmentation, 
simple and robust construction, potential for distributed production, long-term stability, %ease of time calibration, 
low maintenance, high reliability and  cost effectiveness,
all important aspects for building a large area detector.

The scintillating strips in the SHiP muon detector 
will be 5 (10) cm wide, 3 m long and 2 (1) cm thick. 
Precise definition of the dimensions will be done for the Technical Design Report.
Crossings of horizontal and vertical strips can provide the $x,y$ view 
in each muon station with a readout granularity of  5 (10) cm.
In Figure~\ref{fig:strips_layout} and Figure~\ref{fig:strips_layout2} a possible layout of horizontal and vertical strips for each muon station is shown.

%\vskip -3cm
\begin{figure}[h]
\begin{center}
\includegraphics[width=0.40\textwidth]{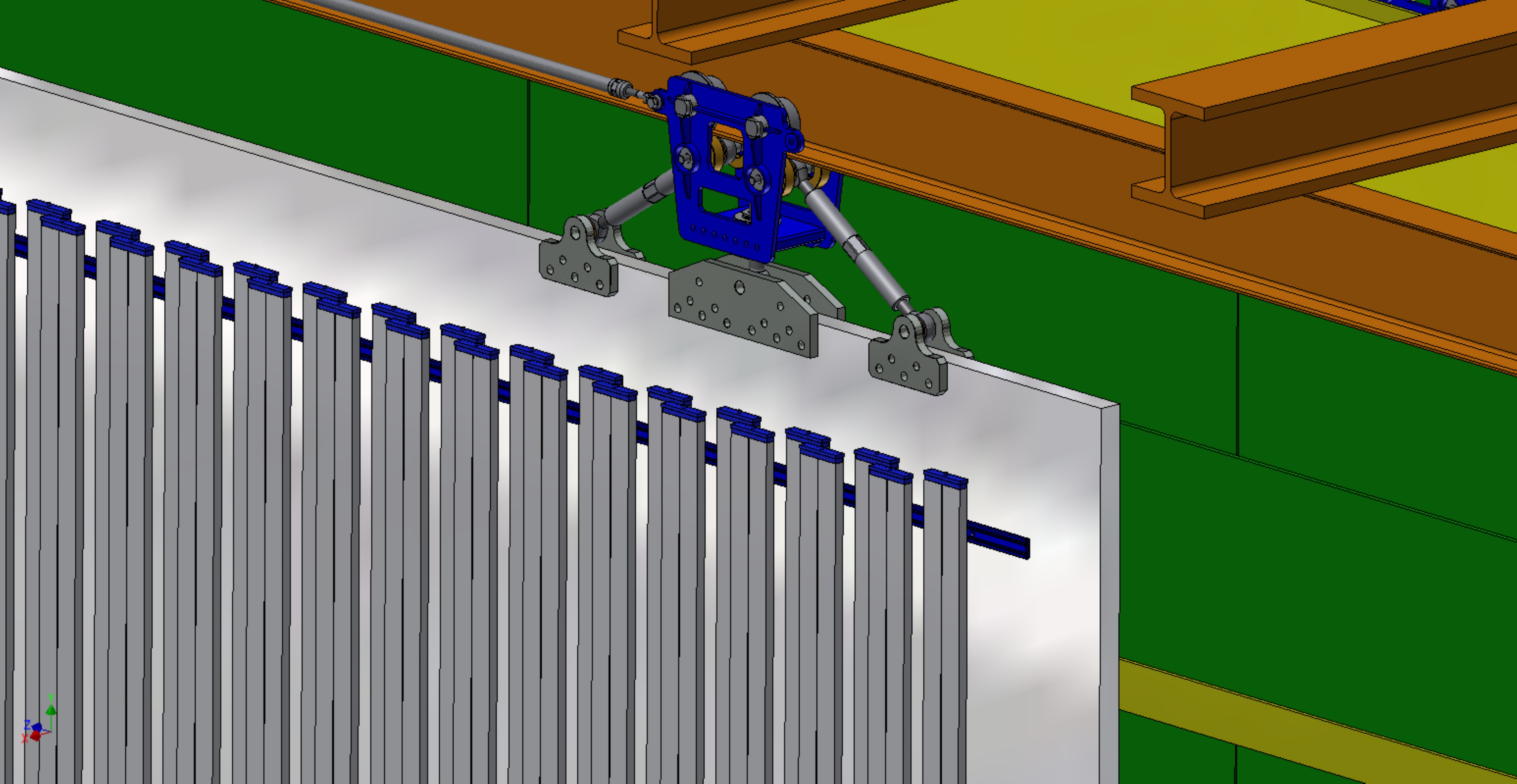}    
\includegraphics[width=0.40\textwidth]{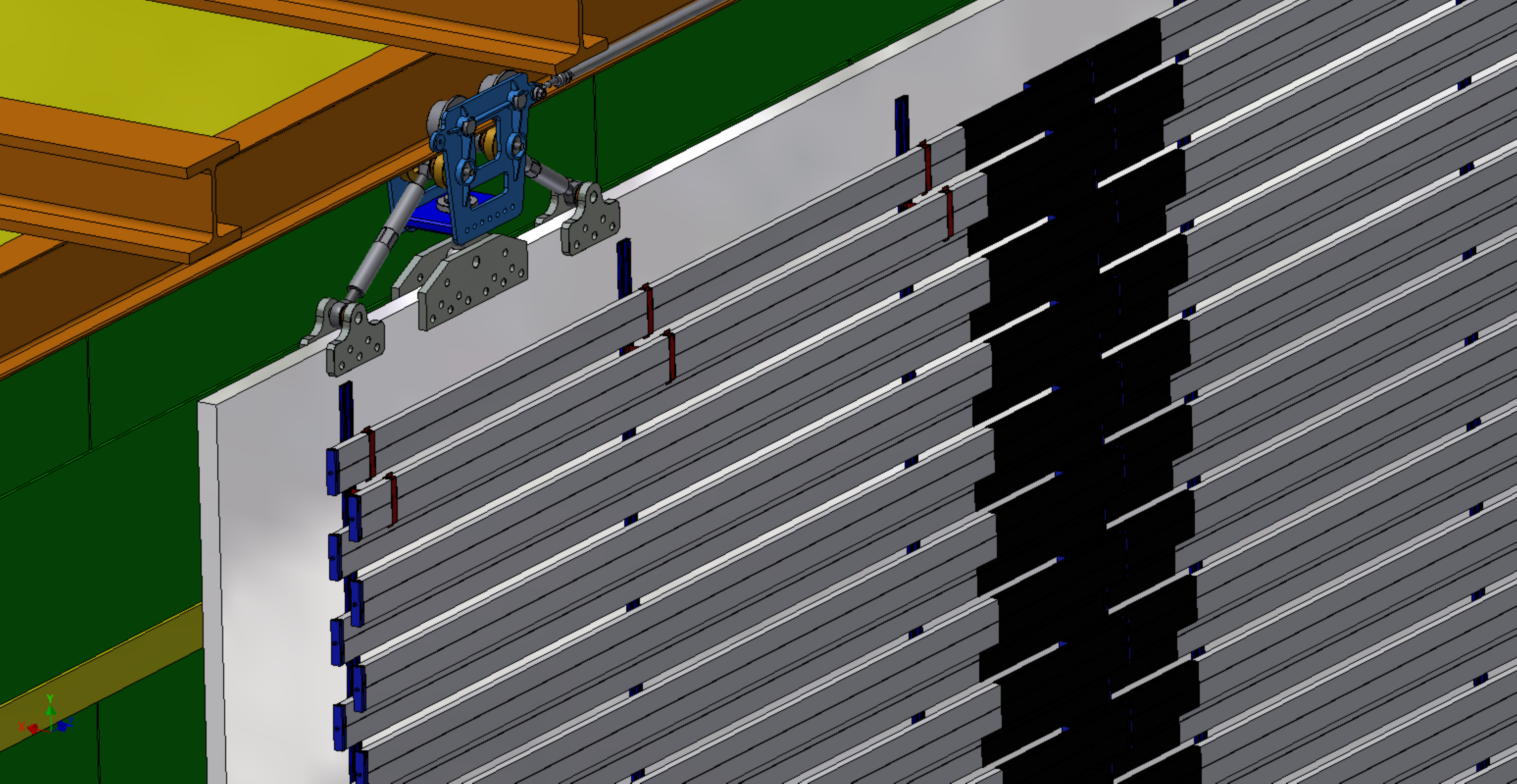}    
%\vskip -3cm
\end{center}
\caption{Scintillating strips layout.}
\label{fig:strips_layout}   
\end{figure}

\begin{figure}[hbt]
\begin{center}
\includegraphics[width=0.6\textwidth]{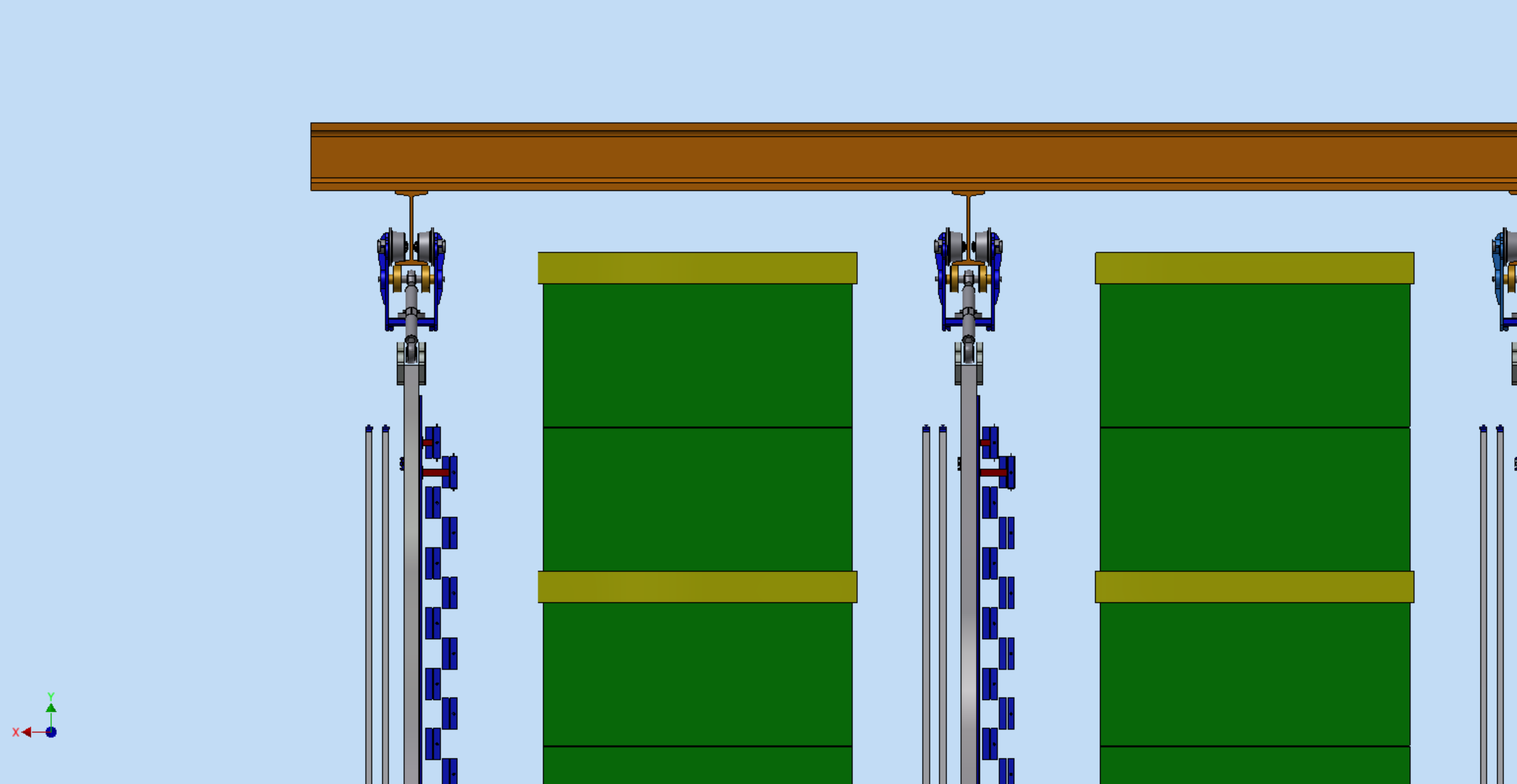}    
\end{center}
%\vskip -2cm
\caption{Muon detector layout: detail.}
\label{fig:strips_layout2}   
\end{figure} 

About 480 vertical strips and 480 horizontal strips 5 cm wide are needed to instrument one muon station in the two views, 3840 strips for the entire system readout at both ends by 7680 photodetectors.
In Table~\ref{tab:muon_layout} the main numbers of the muon detector are summarized.

Given the relatively large area,  a good choice could be the rather inexpensive scintillators produced at the FNAL-NICADD \cite{muon:nicadd, PlaDalmau:2000bf}
facility which are fabricated by extrusion with a thin layer of $Ti O_2$ around the active core. %({\it mention Russian scintillating strips}).
Another possibility could be polystyrene scintillator bars extruded at UNIPLAST plant (Vladimir, Russia). % with the length of 90~cm and with widths of 1, 2, 3 and 4~cm.
In this case, the scintillator composition is  polystyrene doped with 1.5\% of paraterphenyl (PTP) and 0.01\% of POPOP and the 
bars are extruded with thickness of 7 and 10~mm and then covered by a chemical reflector by etching the 
scintillator surface  in a chemical agent that results in the formation of a white micropore deposit over the polystyrene surface~\cite{Kudenko:2001qj}.

Since the attenuation length of the plastic scintillator is rather short ($\sim$35 cm), the light produced
by the particle interaction has to be collected and transported to the photodetectors efficiently
by WLS fibres. These fibers need to have  a good light yield to ensure a high detection efficiency for
fiber lengths of $\sim 3$ m. Possible choices for WLS fibers are those produced by Saint-Gobain
(BCF92)~\cite{muon:bicron} and from Kuraray (Y11-300) ~\cite{muon:kuraray}  factories. 
Both companies produce multiclad fibers with good attenuation length ($\lambda \sim$ 3.5-4 m)
and trapping efficiency ($\sim 10\%$). The fibers from Kuraray have a higher light yield (about
40\% more), while Saint-Gobain fibers have a faster response ($\sim$ 2.7 ns versus $\sim$ 10 ns of the
Kuraray), which ensures a better time resolution. The choice of the manufacturer will be done for the Technical Design Report.

Large volume scintillator detectors %which consist of a large number of readout channels 
require the usage of inexpensive, 
compact, %insensitive to magnetic field 
photosensors with a high efficiency to the green light emitted from  WLS fibers. 
%Multi-pixel Geiger mode avalanche photodiodes  are applied recently for an optical readout in these detectors. 
%Detailed information about such devices and their basic principle of operation can be found in Ref.~\cite{renker}.  
%The first application of such photosensors in a large-scale experiment has been done in the near neutrino detector~\cite{nd280} 
%of the long baseline experiment T2K~\cite{t2k} where approximately 56000 Multipixel Photon Counters (MPPCs)~\cite{mppc} are used. 

Manufacturers have advanced in recent years in developing new generations of multipixel Geiger photodiodes referred as SiPMs.   
Different SiPM types are available on the market from different manufacturers like Hamamatsu, AdvanSiD (FBK), KETEK and SensL
(see Table~\ref{tab:sipm}).
We are planning to test different models in order to find out the solution that fully matches our requirements.
%This section is devoted to SiPMs which are commercially available these days. 
%Photographs of the active area of a few SiPM's are shown in Fig~\ref{fig:photo_sipm}.
%
%\begin{figure}[htb]
%\centering\includegraphics[width=0.6\textwidth]{./muon/fronts_sipm_3.pdf}
%\caption{Photographs of  SiPM's active area.
%KETEK 1x1 mm, 400 pixels  (left);
%MPPC 1.3x1.3 mm, 667 pixels (center)
%AdvanSid 1.2 mm diameter, 660 pixels (right). }
%\label{fig:photo_sipm} 
%\end{figure}
%
%Different SiPM types from different manufacturers have been tested and the results reported in Table~\ref{table:SiPM}.
%{\bf (Ale: update this table with more recent numbers or just remove the table and put a few example in the text)}.
%We have tested a number of SiPM types from different manufacturers. These SiPMs represent
%different generations. We suppose generations for Hamamatsu and SensL were  latest at the date of testing, 
%and probably KETEK and AdvanSiD tested samples are considered to be outdated now. SiPM sensitive surface size 
%is about 1~mm to readout 1~mm WLS fibers. However in the case of SensL we present data for large 3x3~mm type 
%because the tested sample was latest development of the company with improved performance.
%Table~\ref{table:SiPM} summarizes the results. We have to note that photon detection efficiency (PDE) is shown after correction for crosstalk.

\begin{table}[htbp]
\caption{Muon system layout. In the computation we assume scintillator bars 5 cm wide and 2 cm thick
with double sided-readout.}
\label{tab:muon_layout}
\vspace{.1cm}
\begin{center}
\begin{small}
\begin{tabular}{lr}
\hline
number of active stations  & 4 \\
number of views per active station & 2 \\
active stations dimensions  & ($600 \times 1200 \times 2$) cm$^3$ per view \\
                          & (width$\times$length$\times$ thickness)  \\
bar dimensions   & ($5 \times 300 \times 2$) cm$^3$ \\
                          & (width$\times$length$\times$thickness) \\
number of bars   &  3 840 \\
weight of the scintillator & 11.52 t \\
WLS fibres ( 2 fibres/bar) & 23,000 m \\
FEE channels      &  7 680 \\ \hline

number of passive iron filters   & 3 \\
filters dimensions  & $ (600 \times 1200 \times 60 (40)) $ cm$^3$ \\
                          & (width$\times$length$\times$ thickness)  \\   
iron weight   & 1400 (930) t \\
\hline
\end{tabular}
\end{small}
\end{center}
\end{table}

\begin{table}[htbp]
\caption{Parameters of SiPMs from different manufacturers. All types require bias voltage in range from 30 to 70~V.}
\label{tab:sipm}  
\vspace{.1cm}
\begin{center}
\begin{small}
\begin{tabular}{cccccccccc}
\hline
 Parameter        & Hamamatsu & Hamamatsu & KETEK & SensL & AdvanSiD    & AdvanSiD  \\
                 &MPPC       &MPPC       &       &MicroFC&  ASD  &  ASD \\
                 &S12571-025C&S12571-050C&       &-30035 & -RGB1C-P    & -NUV1S-M   \\
\hline
Pixel size, $\mu$m& 25        & 50        & 50    & 35    & 40          &  40  \\
%\hline
Number of pixels  & 1600      & 400       & 576   & 4774  & 625         &  625  \\
%\hline
Sensitive area, mm& 1x1       & 1x1       & 1.2x1.2   & 3x3   & 1x1    &  1x1  \\
%\hline
Gain              & 5x$10^5$ &1.2x$10^6$ &  8x$10^6$   &3x$10^6$ &2.7x$10^6$   & 2.1x$10^6$ \\
%\hline
Dark rate, MHz    & $\sim 0.1 $     & $ \sim 0.1 $     & $\leq 0.4 $   & 0.3-0.8 & $ \leq 0.9 $        &  $\leq 0.2 $  \\
(at T=300 K)  &  & & & & & \\
%\hline
Crosstalk, \%     & 20-25     &  25-30   & 30-50   & $\sim$ 7 &  $ \sim 30 $   &  $ \sim 7 $  \\
%\hline  
 PDE at 520nm, \% & $ \sim 35 $    & $\sim 35 $   & $\sim$36      & $\sim$ 20   &  $\sim$ 32          & $\sim 22$  \\
\hline
\end{tabular}
\end{small}  
\end{center}
\end{table}

%===================================
\subsection{Results from test beam and test stands}
\label{muon_results}
%===================================

%
%{\it Describe here in  a synthetic way the results obtained from test-beam and test-stand at superB 
%(light yield, efficiency, attenuation length, time resolution, etc. etc.}.

Detailed studies were performed on the scintillating bar technology with WLS fibres and SiPM readout 
for the SuperB~\cite{Grauges:2010fi} and T2K~\cite{Abe:2011ks}
experiments. These studies were focused on the detection of Minimum Ionizing Particles (MIP),
where the amount of light produced by the scintillator is typically small. In order to maximize the
efficiency, it is important to optimize the geometry of the detector, the path and the number of
WLS fibers, the optical coupling with the SiPMs and the assembling procedure. Several test stands
were developed to measure the MIP response of prototypes built with various assembling techniques.

%
%Several prototypes were built with different geometries, assembling procedures and optical
%couplings. The light collection efficiency of each prototype was determined by measuring the response
%of the prototype to MIP signals, in this case atmospheric muons \cite{muon:balbi}.

In particular, in the context of the SuperB experiment \cite{Balbi:2014saa},
 the light yield of a 10 mm thick, 45 mm wide and 2 m long scintillator strip from the FNAL-NICADD facility was measured.
A straight groove, 3 mm deep and 1 mm wide, was machined on the top face of the strip. The WLS fiber was inserted down to the bottom
of the groove. The fiber used in this work was the model Y11-300MS with 1 mm diameter produced by
Kuraray \cite{muon:kuraray}. %The end of the fiber connected to the SiPM was polished with a diamond milling machine.
The fiber was glued and adhesive aluminum tape was applied at the end of the fiber and on top of the
groove. In parallel, a complete simulation study based on FLUKA Monte Carlo \cite{muon:fluka} was performed, where the
response of scintillator, photon propagation, collection and detection through WLS fiber and SiPM were nicely
reproduced.

The light yield measured by the SiPM as a function of the position of the impinging particle defined by the trigger position
along the strip is presented in Figure~\ref{fig:attenuation_plot} together with the results from simulation. 
%
%\vskip -2cm
\begin{figure}[htb]
\centering
\includegraphics[width=0.60\columnwidth]{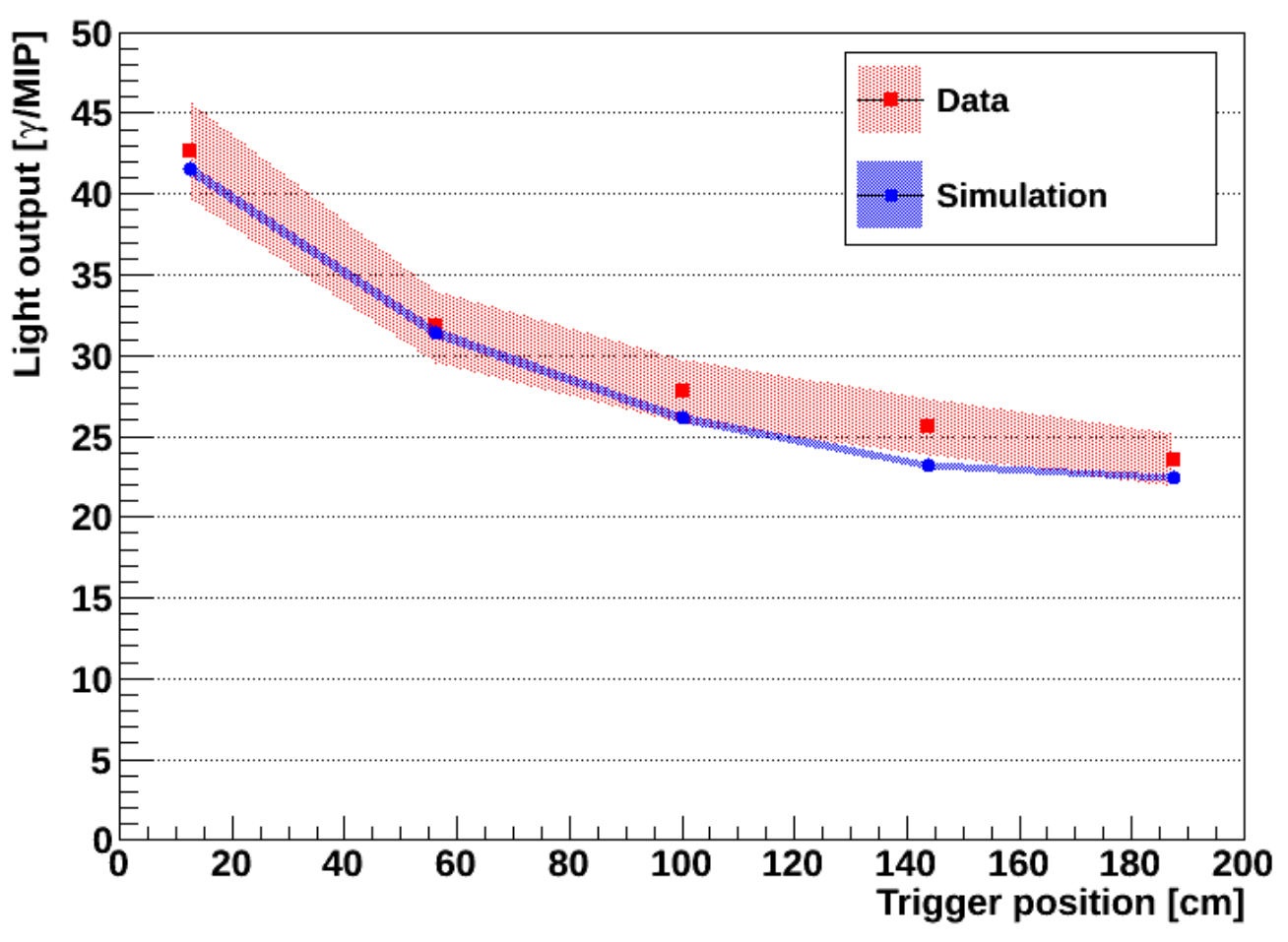}
\caption{Light yield as a function of the position of the incident muon along the scintillator strip. 
The error band for simulated data includes only the statistical uncertainty.}
\label{fig:attenuation_plot}
\end{figure}
%
%As reported by the producer \cite{muon:nicadd1, muon:nicadd2, muon:nicadd3}, the attenuation length in the bulk of the 
%scintillator is very small (about 25 cm). 
The attenuation behaviour of the Y11 fiber shows two components: 
an initial strong attenuation over a distance of about 50 cm, probably dominated by the self
absorption of the fiber, followed by a much lower decrease in light yield ($\lambda \sim 4$ m) in the remaining length.
This has been confirmed also in the literature \cite{Ivashkin:1997sa}.

%This has been measured 
%in literature \cite{muon:kudenko} and is strongly wavelength
%dependent, due mainly to self absorption effects. All these effects, together with the contribution of
%reflection at the far end, play a role in the overall attenuation behaviour of the fibre. As a consequence the measurements show two components, 

In order to compensate for the attenuation of the light signal along the WLS fiber, it is important to
instrument the far end of the fiber with another SiPM. 
%The readout at both ends of the fiber is also
%important for obtaining a time resolution of $o(1)$ ns.

Many tests were performed using scintillating bars extruded at UNIPLAST plant (Vladimir, Russia) 
with a length of 90~cm, widths of 1, 2, 3 and 4~cm and thickness of 7 and 10 mm.
A single 2~mm deep, 1.1~mm wide groove was machined along the center of the bar to  accommodate a 16~m long Y11 fiber.  
For readout the same 1~mm diameter Kuraray Y11(200) fiber (S-type) was used, which was viewed from both sides by Hamamatsu 
MPPC photosensors of the same type that are used in T2K experiment.  The fiber was embedded in a bar using an optical grease. 
The light yield was collected at both fiber ends.
%MPPC bias voltage was adjusted to keep the dark rate below 800~MHz. 
%The overvoltage is about 1.4~V. 
%MPPC photo detection efficiency is expected to be around 28\% for the re-emitted WLS fiber green light, according to published data.

%
%The test bench for detector measurements is shown in Fig~\ref{fig:setup}.
%\begin{figure}[htb]
%\centering\includegraphics[width=0.6\textwidth]{./muon/setup.pdf}
%\caption{Test bench for study of the detector read out using a 16~m long Y11 fiber.}
%\label{fig:setup} 
%\end{figure}

A 16~m long fiber was coiled inside a light tight box with a minimum bending radius of about 30~cm. 
The ionization area within the bars was localized to a $2\times2$ cm$^2$ spot defined by the trigger counter size.  
The position of the  trigger counter was fixed to the center of the 90~cm long bar to optimize scintillation light collection by the fiber.  
%Since a measurement of the 1~cm thick extruded slab yielded an effective attenuation length of the scintillation light in the scintillator of approximately 8.1~cm~\cite{smrd}, we estimate that the full scintillation light collection by a WLS fiber occurs within $\pm$30~cm from the cosmic ray  ionization point.  

The light yield per MIP along the fiber (sum of both ends) for different bars  is shown in Figure~\ref{fig:bar4}.
\begin{figure}[htb]
\centering\includegraphics[width=0.6\textwidth]{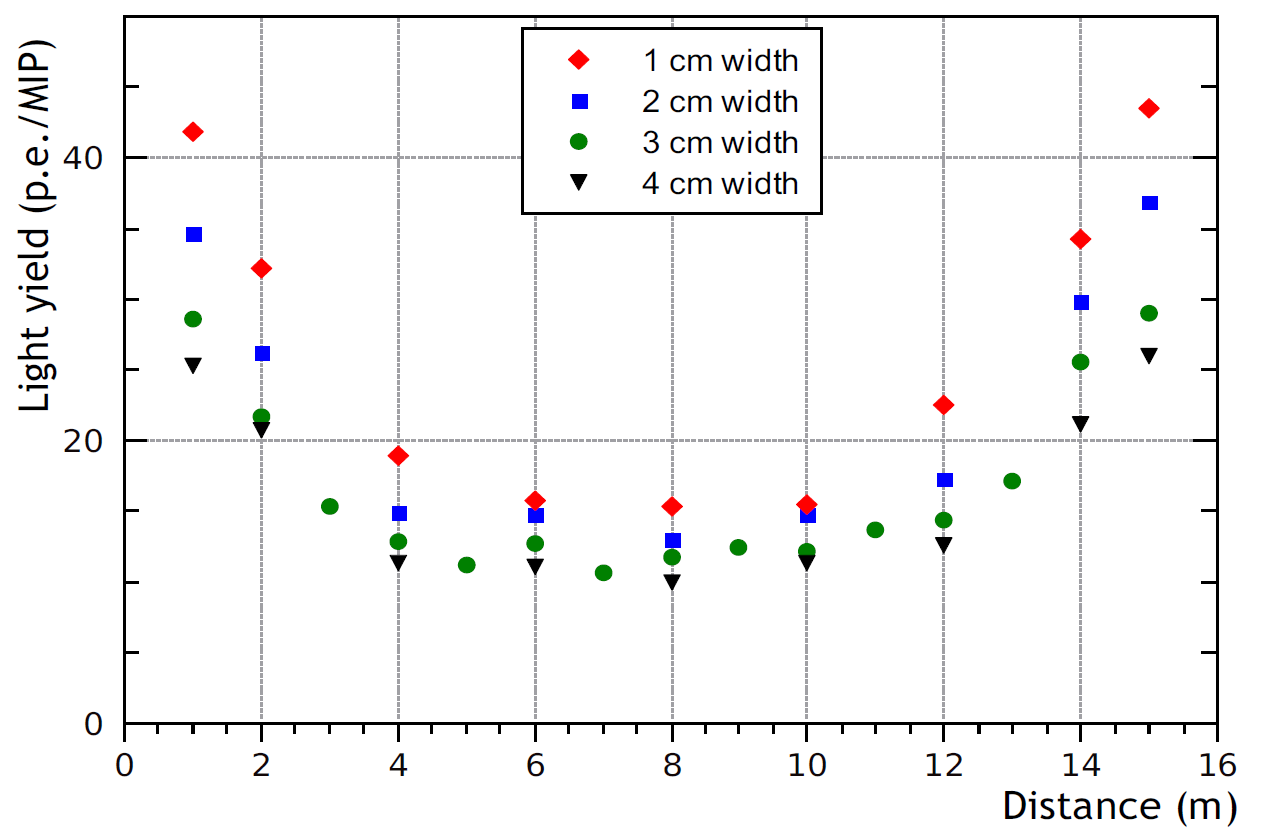}
\caption{Total light yield from both fiber ends  vs the position along the Y11 fiber for 0.7~cm thick bars of different widths of 1, 2, 3 and 4~cm.  
Measurement errors are estimated to be about 7\%.}
\label{fig:bar4} 
\end{figure}
%
%The readout was performed using MPPC and the light yield values have been corrected for dark pulses, optical crosstalk and afterpulsing in the photodiodes. 
%A photoelectron unit was determined as the spacing between neighboring peaks in the measured charge spectrum for MIPs.
The 4~cm wide bar produces in the middle of a 16 m long fibre a light yield of about 10~p.e./MIP when the light is readout at both fiber ends.
This result scaled to a 3 m long bar, as for the SHiP muon detector, gives a light yield of $\sim 40-50$ p.e./MIP.

%We studied how to improve the light
%yield by adding a second fiber on the same scintillator strip following two options. In the first one,
%two fibers are inserted in two parallel grooves in the scintillator, while in the second one both fibers
%are inserted in the same central groove.
%The results are summarized in Tab.~\ref{tab:two_sep}: the light yield of the sum of the two fibers is reported and
%compared with the light yield measured in the configuration with only one fiber.

%\begin{table}[ht]
%\renewcommand{\arraystretch}{1.15}
%\caption{Light collection on short scintillator strip with two fibers}
%\label{tab:two_sep}
%\centering
%\begin{tabular}{cp{1.6cm}p{1.6cm}p{1.6cm}p{2.0cm}}
%\\
%\hline
%& \textbf{Total} [$\gamma$/MIP] & \textbf{Relative to single}\\
%\hline
%Separate grooves & $80 \pm 5$ & + 74\%\\
%\hline
%Same groove  & $82 \pm 5 $ & + 78\%\\
%\hline
%\end{tabular}
%\end{table}

%
%In both cases the light yield is improved by about 75\%, but they differ in manufacturing complexity.
%In the former case two grooves are machined instead of one and the fibers need bending to reach the
%same SiPM. In the latter, the optical coupling to the SiPM is easier, but the glueing procedure is more
%difficult.

The light yield can be further increased by adding more fibres per scintillating bar.
In the contest of the SuperB project, the light yield was studied as a function of the number of fibers and of their position on a scintillator strip
10 mm thick, 45 mm wide and 2 m long.
Preliminary tests using two fibres in the same groove or two separate grooves in the same strip
show similar results, corresponding to an increase of $\sim 75\%$ of the photon yield with respect to a single fibre.

Timing studies were also performed with the same setup and % with the idea of measuring 
%the longitudinal coordinate thanks to the arrival time of the light signal in the WLS+scintillator bar. 
%The above studies were performed on a $1 \times 4.5$ cm$^2$ scintillator bar and 
different configurations (e.g 1,2,3 WLS fibers, 2 different lengths: 30 cm and 200 cm). The fibers were Bicron BCF92, 1 mm in
diameter.
Preliminary results %(the time readout option was soon abandoned (due to a higher cost w.r.t performances??)) 
showed a time resolution of about $1.2- 1.3$ ns obtained with a simple discriminator-TDC
readout.  A similar study will be repeated using fibers Kuraray Y11-300. 
Time resolution can be improved by increasing the light yield (either increasing the thickness of the bar or the number of fibres per bar) and
with a more refined signal processing (see Section~\ref{subsec:muon_fee}).
We are then confident that a time resolution of 1 ns or better can be reached with a suitable R\&D.

%
%Given the much favorable conditions foreseen in the SHIP environment, there is a large margin
%for improvements.
%In fact any increase of light yield due to the increase of the  thickness of the scintillator strip or the number of WLS fibres
%will improve the time resolution. In addition 
%
 
%For example the use of a scintillator strip 2 cm thick will increase the light yield and hence will improve the time resolution.
%Firstly we will use scintillators with a thickness of 2 cm instead of 1 cm, increasing
%significantly the ligh yield; secondly we can use a more refined processing (i.e. a signal digitizer) to
%precisely define the arrival time of the signal. 

%====================
\subsection{Front-end electronics}
\label{subsec:muon_fee}
%====================
%{\it Discuss requirements and  show scheme of FEE.}
The muon readout electronics block diagram is shown in Figure~\ref{FEdiagram}. 

The main tasks required  to the  front-end electronics are:
\begin{itemize}
	\item[-] to allow a fine  biasing of  the SiPM;
	\item[-] to  extract, to amplify and to shape  the detector signal preserving its  time information;
	\item[-] to  measure the arrival time  of each signal with respect  to a master clock;
	\item[-] to format and to transmit the zero-suppressed data to the FARM;
	\item[-] to interface with the TFC and ECS systems (see Section~\ref{sec:online}).
\end{itemize} 

\begin{figure}[h]
\begin{center}
 %  \begin{picture}
  \includegraphics[width=0.95\textwidth]{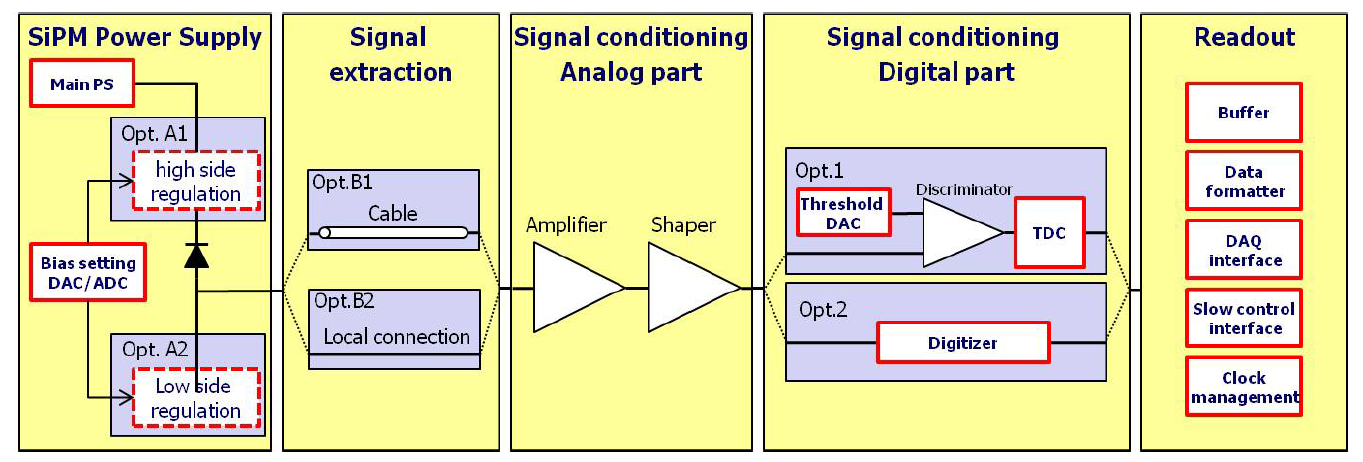}
  % \end{picture}
   \caption{\small Muon readout block diagram}
   \label{FEdiagram}
\end{center}
\end{figure}

The radiation level expected for the detector environment is very low so the radiation hardness of the electronics  is not an issue and off-the-shelf 
components can be used. Also the signal rate of few kHz per channel does not pose any particular requirement while the overall resolution of few ns  
needs a proper signal handling.

The knowledge of the time distribution of photons arriving at the SiPM is, therefore,  
crucial for the design of the read-out electronics. At the present, studies are going on and the final solution is not yet defined, 
but some general statement could anyhow be done.

%As baseline design, the WLS fibers will be read on both side to maximize  the light yield, reducing its dependence on the muon impact point, 
%and to optimize the photon co%llection time along the scintillator strip. Such double-side readout scheme should allow the use of  higher 
%threshold values reducing the SiPM dark noise effect and should increase the  time resolution that strongly depends on the fiber and strip length.

As baseline design, the WLS fibers will be read on both sides to maximize  the light yield. % reducing its dependence on the muon impact point. 
Such double-side readout scheme optimizes  the photon collection time along the scintillator strip and it improves the  time resolution 
that strongly depends on the fiber and strip length. Besides it should allow the use of  higher threshold values reducing the SiPM dark noise effect.

The WLS fibers are directly coupled with the SiPM at the ends of each scintillator bar to maximize the number of converted photoelectrons. 
The SiPM is mounted on a suitable printed circuit board (SiPM-PCB) joined to the scintillator itself and a coaxial cable connects 
the photosensor to the first stage of the electronics chain. The frontend boards is installed in the free accessible location closest to the SiPM-PCB. 
If noise pickup or signal degradation were an issue, the possibility to have the amplification stage on the SiPM-PCB could be considered.

The first electronics stage architecture is, as previously mentioned, strongly influenced by the signal shape, two possible options are considered:

\begin{itemize}
	\item[-] a preamplifier/shaping stage followed by a double threshold discriminator;
	\item[-] a preamplifier/shaping stage followed by a digitizer.
\end{itemize}

In the first case the time measurements is done with a TDC implemented in a FPGA while in the latter case the time must 
be extracted after a proper interpolation of the  acquired samples.   In both cases the acquired data is sent to a second 
electronics stage where they are processed, zero suppressed, formatted, stored in local buffer and sent to the FARM.

The SiPMs need a fine control of their bias voltages to equalize the gains and to compensate temperature variations, 
therefore a channel by channel programmable voltage regulation with remote setting/monitoring will be implemented on the frontend boards. 
A calibration system based on a LED pulsing system will also be developed. 

The electronics system will  be interfaced with the TFC system from which the master clock and the synchronization signal are received and 
with the ECS for all the monitor/settings operations.

In Table~\ref{tab:muon_daq} the quantities relevant for the DAQ system are summarized.
The FEE will consist of 7 680 channels that can be grouped in boards of 16 or 32 channels each (480-240 boards).
The number of channels fired per muon traversing all the active stations is 16  (2 channels/bar $\times $ 2 bars/station (x,y) $\times $ 4 stations).
The data size per channel is given by the channel number (7 680 channels are stored in a word of 13 bits) plus the time information: 
assuming a clock of 10 MHz and a  TDC sensitivity of 100 ps/count, the time information can be stored in a word of 10 bits. 
Hence the data size per channel is 3 bytes.

The time window is defined as the time necessary to collect all the information from a muon crossing the four active stations. This is obtained by adding up the 
time of flight between the first and last active station ($\sim 8 $ ns assuming 40 cm thick passive filters and 30 cm space for the active layers), 
the time of the light emission from a fibre ($\sim 2-3 $ ns) and the time of the propagation of the light into the fibre
when the particle crosses the far end with respect to a FEE channel ( this is about 18 ns for a  300 cm  long fibre, assuming a light propagation into the fibre 
of $\sim$ 17 cm/ns). The time window necessary to collect all the information from the muon system is therefore about 30 ns.

The occupancy is driven by the flux of beam-induced muons seen by the HS muon detector which is about 50 kHz.
%the maximum tolerable rate of beam-induced muons in the emulsions of the $\nu_{\tau}$ detector which is about 100 kHz.
This rate spread-out on a surface of $600 \times 1200$ cm$^2$ gives an occupancy of 0.1 Hz/cm$^2$ corresponding to 150 Hz per scintillating bar
(5 cm wide and 300 cm long). The dark current from the SiPMs is estimated to be about 10 kHz @ 3 p.e. threshold. 
Assuming a zero suppression for channels below threshold,  the expected data size for the muon detector is $\sim$ 3.5 MBytes/s 
(150 Hz/channel $\times$ 7 680 channels $\times$ 3 bytes/channel ). All the channels can be reasonably multiplexed in one single Ethernet line
but probably is convenient to split them in more than one line. Assuming to send data of 24 FEE boards of 16 channels each into one concentrator,
we need five Ethernet links per station, 20 in total.
%This will be defined in the Technical Design Report.

The data volume to configure and monitor each channel can be preliminary estimated to be about 30 bytes/channel. 
All these parameters must be considered very preliminary  and will be defined precisely in the Technical Design Report.

%  La stima del data volume per la configurazione mi sembra impossibile da fare dipende da troppi 
% parametri (architettura, tipo di readout, tipo di indirizzamento,…) possiamo pensare di avere bisogno di settare/monitorare per ciascun canale:
%         setting tensione sipm: 13bit addr +12 bit dato à 3 byte a canale
%         lettura tensione impostata: 13bit addr +12 bit dato à 3 byte a canale
%         setting 2 soglie discriminatore: 13bit addr +12 bit dato à 3 byte per ciascuna soglia (6 byte a canale)
%         lettura 2 soglie discriminatore: 13bit addr +12 bit dato à 3 byte per ciascuna soglia (6 byte a canale)
%         registri di una eventuale FPGA e/o digitizer (per 16 canali) 48 registri da 32 bit à 12 byte canale
%
\begin{table}[htbp]
\caption{Relevant quantities for the DAQ system.}
\label{tab:muon_daq}
\vspace{.1cm}
\begin{center}
\begin{small}
\begin{tabular}{|l|r|}
\hline
FEE channels & 7 680 \\ 
FEE boards (16 or 32 channels/board) & 480 or 240 \\
data size/channel   & 3 Bytes \\
data volume/channel & 30 Bytes \\
time window  & 30 ns \\ 
physical rate/channel    & 300 Hz \\
dark current from SiPMs & 10 kHz @ 3 p.e. threshold \\
total bandwidth in zero suppression mode & 3.5 MBytes/s \\
number of Ethernet links  & 20 (5/station) \\
\hline
\end{tabular}
\end{small}
\end{center}\end{table}

%====================
\subsection{Performance}
\label{muon_performance}
%====================
%{\it Describe here the simulation results: efficiency vs misidentification curves for $\mu/\pi$ with and without hcal. 
%Results shown now are considering only a basic muonID algorithm and no hcal. The curves will improve by the building of a likelihood
%and by including  HCAL. Draw some conclusion about the background levels.}

Charged tracks found with the  tracking system will be matched to the hits in the muon system.
Their identification as muons or hadrons will result from a detailed analysis of the hit patterns in the active muon detector and energy 
deposited in the calorimeters.
A simple algorithm that extrapolates the track direction from the tracking system and  search for hits in the muon system 
within Fields of Interest (FOIs) has been developped and implemented into the FairShip framework.   
The FOIs width depends on the multiple scattering of muons in the material upstream a given station and are, therefore, momentum and station dependent.
The number of stations used in the search for hits in FOI depends on the momentum of the impinging track.
In Figure~\ref{fig:muon_results} the muon identification efficiency (left) and the pion misidentification (right) probability  are shown
as a function of the momentum of the charged tracks. A muon efficiency close to 99\% and a pion misidentification probability of $<0.1\%$
can be  reached for tracks with $p>3$ GeV/c.
These curves are obtained with simulated $N \to \pi^+ \mu^-$ events where the pion
is required to not decay in flight. Only 3\% of pions of the decay chain $N \to \pi^+ \mu^-$ decay in flight and can be simply removed.
For other signal channels, events with pions decaying in flight have to be treated separately using kinematic and pointing requirements
and are not considered in the evaluation of the performance of the particle identification systems.
%The pion misidentification probability in given in the momentum range important for the $K_L \to \pi \mu \nu$ ($K_S \to \pi \pi$) background.

The muon system can be also used to reject muons if used in veto mode (charged tracks not releasing any hit in the muon system are considered as pions). 
In Figure~\ref{fig:pion_results} the pion identification efficiency (left) and the muon misidentification probability (right) are shown as a function of momentum.
A pion identification efficiency of $\sim 100\%$ can be reached for a muon misidentification probability of $\sim 1\%$. 

Further improvements to the muon-hadron separation 
can be obtained by adding more informations like the energy deposited in the calorimeter system, the quality of alignment and the number of the  hits in the muon stations.

The maximum tolerable level of hadron and muon misidentification depends on the signal channel and on the topology of the background.
The overall performance of the hidden particle detector in rejecting different types of backgrounds, using kinematic requirements, 
veto, calorimeter and muon systems is discussed in Section~\ref{sec:backgrounds}.
%in the same momentum range. {\it to be continued}.

\begin{figure}[h]
\includegraphics[width=0.45\textwidth]{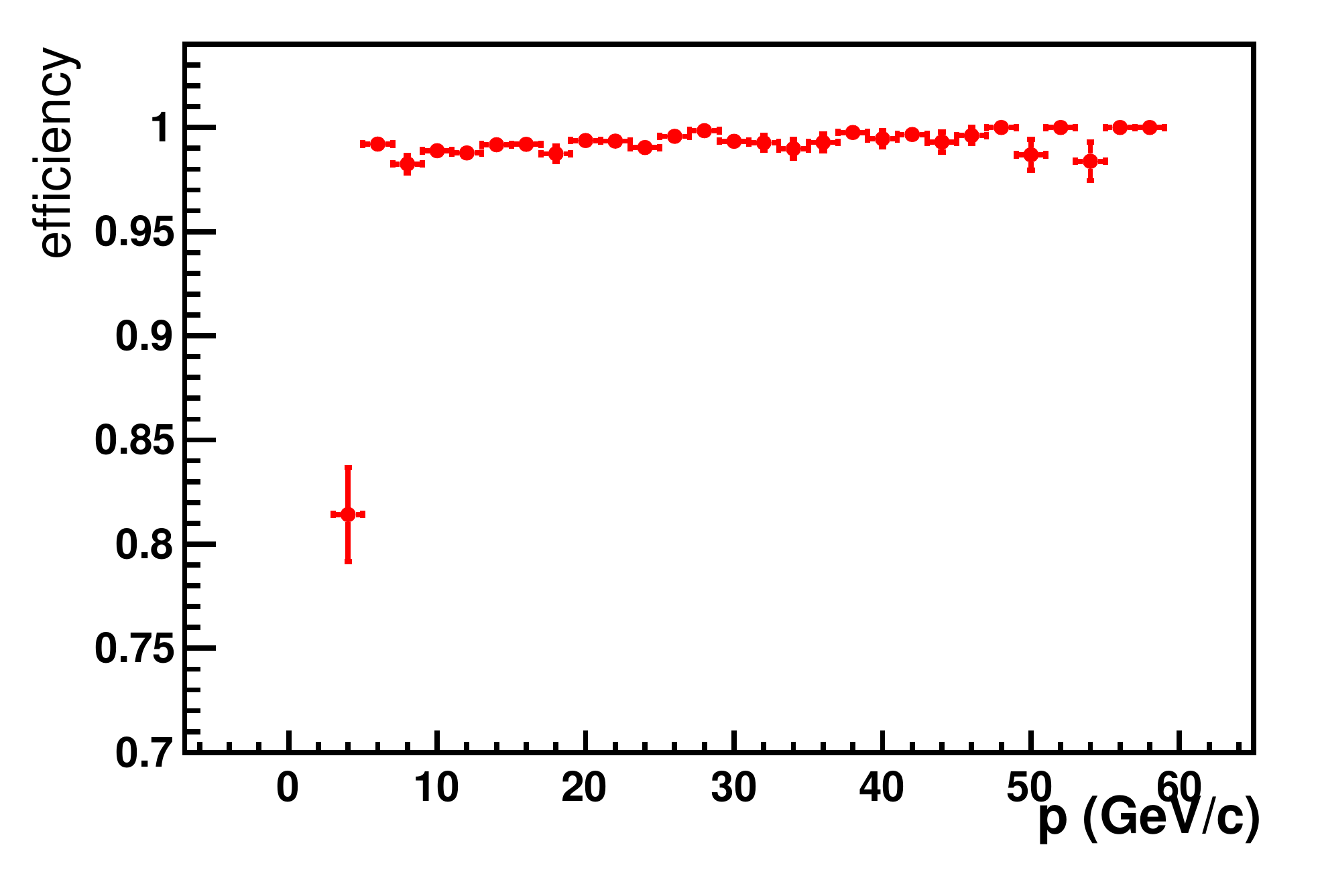}    
\includegraphics[width=0.45\textwidth]{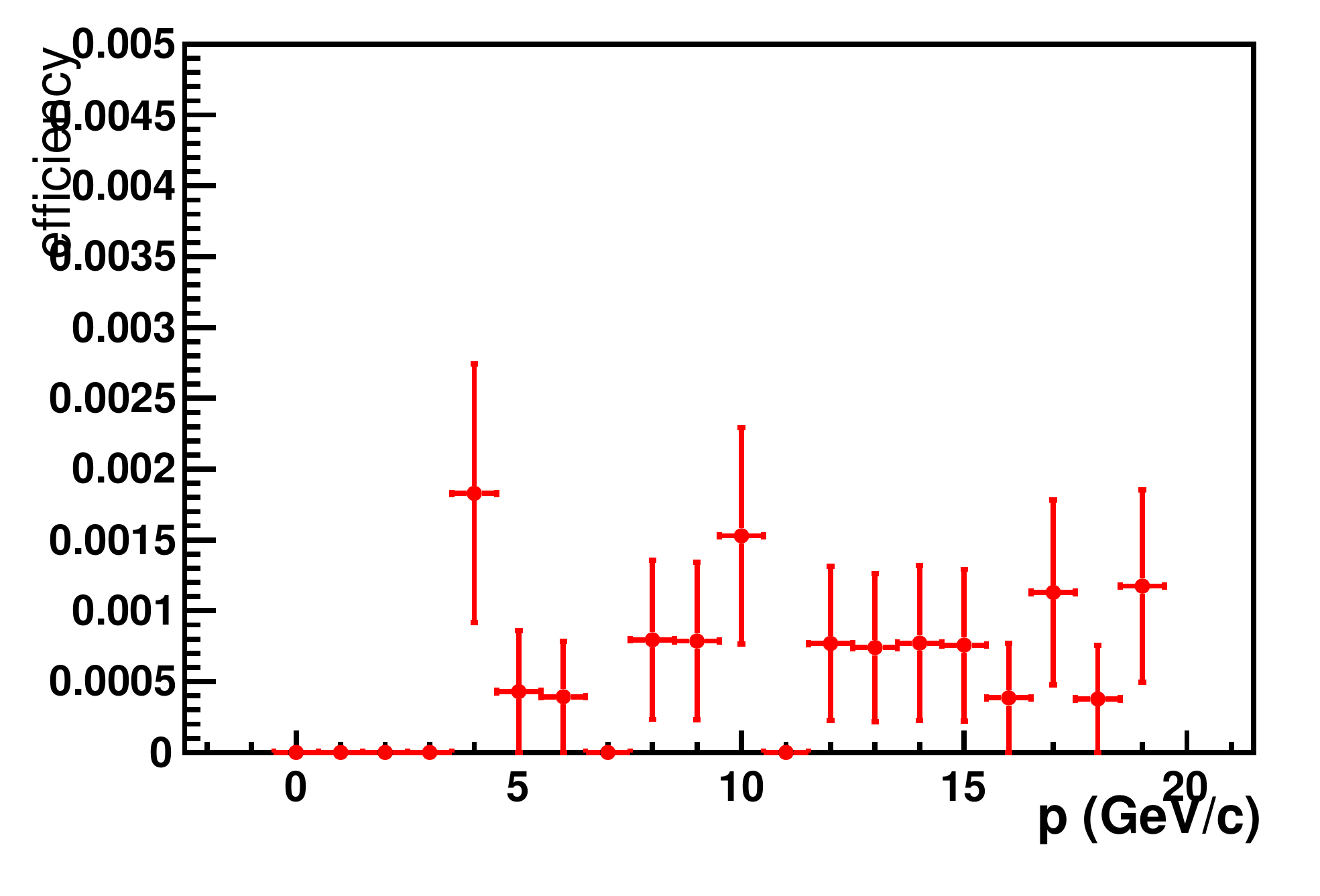}    
\caption{Muon identification efficiency (left) and the pion misidentification probability for pions not decaying in flight (right) as a function of momentum.}
\label{fig:muon_results}   
\end{figure} 

\begin{figure}[h]
\includegraphics[width=0.45\textwidth]{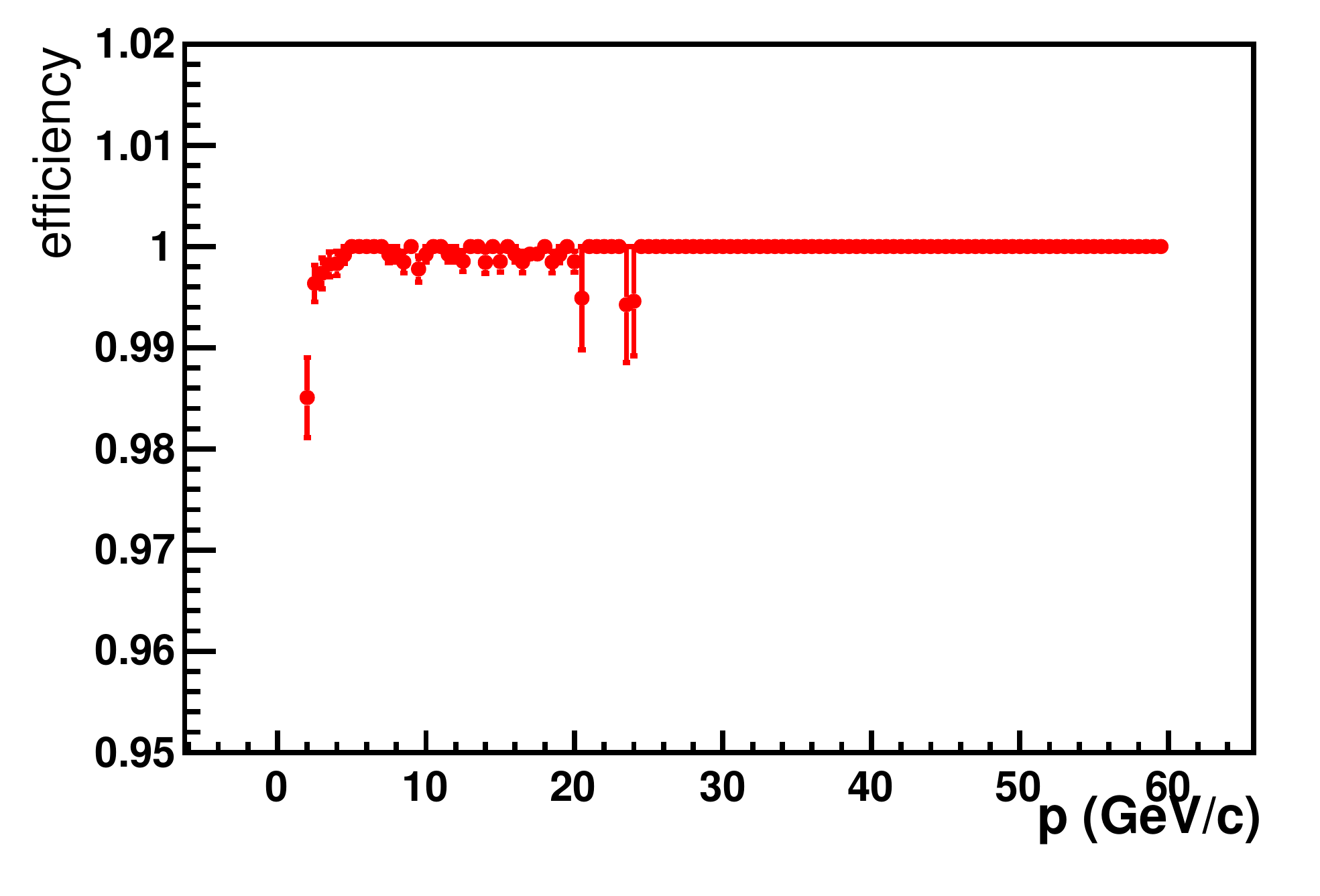}    
\includegraphics[width=0.45\textwidth]{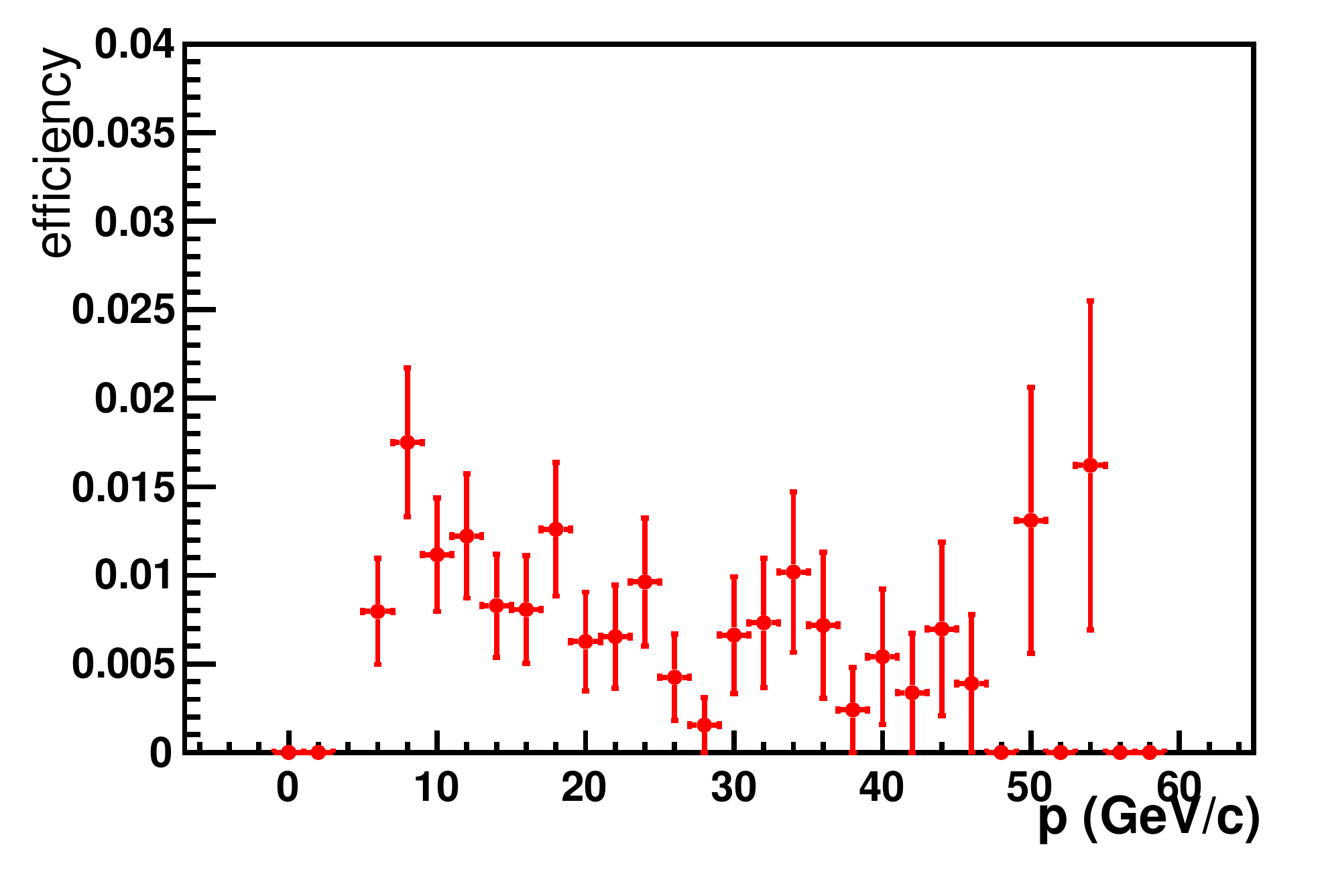}    
\caption{Pion identification efficiency (left) and the muon misidentification probability (right)  as a function of momentum.}
\label{fig:pion_results}   
\end{figure} 

%======================
\subsection{Future R\&D}
\label{muon_RD}
%======================
Further R\&D for the muon system will be needed  with the main goal of  reaching a time resolution of about 1 ns.
This includes:

\begin{itemize}
 \item[-] optimization of the final dimensions of the scintillating bars;
 \item[-] choice of the manufacturer of the scintillating bars;
 \item[-] definition of the number and layout of the WLS fibers;
  \item[-] choice of the manufacturer of the WLS fibers;
 \item[-] test of the SiPMs present on the market;
 \item[-] design and test of prototypes of the front-end electronics.  
\end{itemize}

For this purpose a two weeks long test beam is scheduled at the T9 area of the PS at CERN in October 2015.

%% file: online/Online.tex
\section{Trigger and Online System}
\label{sec:online}

\begin{figure}[h]
\centering
\includegraphics[width=0.55\textwidth,clip=]{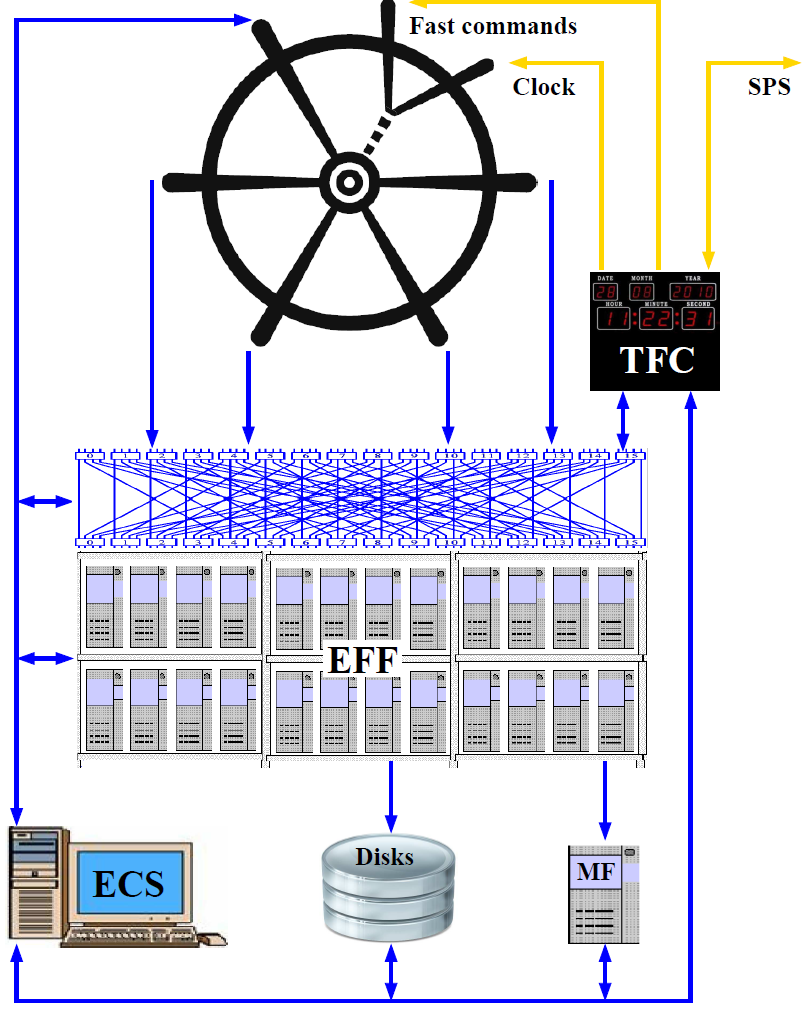}
\caption{The architecture of the online system. The different components are explained in the text.
}
\label{online:Overview}
\end{figure}

The architecture of the Online system is inspired  by the one proposed in the LHCb Trigger and Online Upgrade TDR~\cite{CERN-LHCC-2014-016}.
The philosophy is to implement a trigger-less readout system, which performs
event building of all zero suppressed data and subsequently executes a fully software trigger on an online computer farm. 
Contrary to LHCb the SHiP front-ends (FE) will not be exposed to radiation levels (see Section~\ref{sec:facility}) which require non-commercial solutions for the network. Another simplification compared to LHCb comes from the modest required bandwidth of the system, which suggest the use of a 
commercial switch to link the FE with the computer farm.
Figure~\ref{online:Overview}  shows the overall architecture of the system. 
The Experimental Control System (ECS) connects to all active elements
of SHiP. The Timing and Fast Control system (TFC) generates the clock, which is distributed to the front-ends 
with optical fibres (yellow). All sub-detectors transmit their data over Ethernet (blue) to the Event Filter Farm (EFF) via
a switch. Data selected by the trigger running on the EFF is buffered on disks, and eventually transmitted to the
CERN central storage. The nodes on the EFF declare their availability to the TFC via the switch. The TFC uses fast commands 
over optical fibres to transmit available destinations to the FE.
A fraction of the data is send to the Monitoring Farm (MF) to evaluate the performance of the experiment.

Table~\ref{online:daqsize} summarizes the required data and configuration size for the sub-detectors. 
The total number of Ethernet links shown in table~\ref{online:daqsize} is not reflecting the bandwidth but rather the
partitioning of the FE. 
\begin{table*}[!htp]
\begin{center}
\caption{The amount of data per sub-detector. The time window covers all data
due to one proton on target. The data size is given per particle traversing the 
sub-detector. 
}
\label{online:daqsize}
\begin{tabular}{l r r r r r r r r}\\\hline
Sub-det&Target & RPC &Drift&BckGr&Veto&Spectr&Calo&Muon\\
       &Tracker&     &Tubes& Tagger&Taggers&Tracker&&\\
Section & \ref{sec:target_tracker} & \ref{sec:rpc_tracker} & \ref{sec:drift_tubes} & \ref{sec:taggers} & \ref{sec:upstream_vetotiming}, \ref{sec:timing_detectors} & \ref{sec:tracker} & \ref{sec:calorimeters} & \ref{subsec:muondet}  \\\hline
$\#$-channels &120k&8.2k &4k&1.7k&0.8k&41k&13k&8k\\
Time window (ns) &$350$ &$100$&$1500$&$50$& $50$&$250$&$100$&$30$\\ 
Data size (Bytes) & 28/hit&3/hit&3/hit&4/hit&5/hit&3/hit&160/cl.&3/hit\\
Hits/particle & 10 & 22&40&2&2&40&1&16\\
Ethernet links&few &150&10&few&few&few&16&20\\
\hline
%$\#$ FE-boards &200 & later&1000&60&2300&\\
%Configure size/board (Bytes)& 500& later&320&5k&later&\\
%\hline
\end{tabular}
\end{center}
\end{table*}

The following sections describe the different components of the online system in more detail.

\subsection{Data Acquisition System}
The sub-detectors readout electronics is composed of a front-end which amplifies, shapes and digitises the detector
signal, and a digital back-end which provides processing of 
extracted data, buffering, and communication with data acquisition (DAQ), slow (ECS) and fast (TFC) control systems. 
The digital back-end performs many functions which are common for all sub-detectors, 
and will be implemented in modern FPGAs with high speed transceivers which then sent the data over Ethernet to the switch. 
For low load links aggregation switches will be used to get a reasonable rate/link
into the switch.

The SPS provides a quasi-continuous flux of protons on target (p.o.t.) over its extraction period.
As a result there is no time information available from the SPS around which one could
define a time-window which would contain all data due to one p.o.t.
In addition, the FE will not provide a trigger
which could potentially define such a time window. To enable the software trigger to access
all data corresponding to one p.o.t., the data acquisition system defines
the following time slices: 
\begin{itemize}
\item Frame: this is the smallest time slice which could potentially contain all
data corresponding to one p.o.t. A frame length of 100 ns would suffice for most sub-detectors. 
Using the start of an SPS-spill to synchronize the different sub-detectors, the 
frame number indicates the start of the frame in units of 100 ns since the start of a 
SPS-spill. The data provided by a sub-detector in frame $J$ is:
\begin{center}
\begin{tabular}{|l|l|l|}\hline
Frame-size (2B)& ${J~}(4B)$&Data of frame\\\hline
\end{tabular}
\end{center}
where the fixed length of the different blocks is indicated in Bytes (B).
\item Multi Event Package (MEP): data corresponding to one p.o.t. will inevitably
sometimes be spread over more than one frame. To assure that a computing node receives
all data corresponding to one p.o.t. without having to duplicate all frames, a large number
of frames is combined into a MEP. Consecutive MEPs will have at least one, 
but for some sub-detectors several, frames in common. 
The number of frames in a MEP will be chosen as a compromise between limiting the Ethernet
transport overhead, and the number of frames which needs to be duplicated.
For a reasonable choice of 100 frames/MEP, MEP$_M$ of a sub-detector contains: 
\begin{center}
\begin{tabular}{|l|l|l|l|l|l|l|}\hline
MEP-size(4B)& $M$(4B) &Source-ID(2B) &ECS(4B)&Magic(2B)&Type(4B)&Data\\\hline
\end{tabular}
\end{center}
Data contains:
\begin{center}
\begin{tabular}{|l|l|l|l|l|}\hline
Frame$_{J+0}$& Frame$_{J+1}$&...........&Frame$_{J+100}$&Frame$_{J+101}$\\\hline
\end{tabular}
\end{center}
for the case of one overlapping frame between MEPs. The number of overlapping frames should corresponds to the 
time window and can vary between detectors.
The TFC informs the FE of the destination of each MEP.
A MEP contains some bits to describe the status of this sub-detector (ECS) to provide the trigger with fast information. Each MEP contains a unique bit pattern (Magic) to allow the trigger to check 
the MEP header decoding. Type allows versioning, and can signal 
data which is not zero suppressed or should be used for calibration or monitoring.
\item Event: MEPs with the same number are routed to the same computing node in the EFF.
Inside a MEP the trigger will look for a signal in a seed frame $\pm n$ frames.
For example in MEP$_{M+1}$ the trigger will consider Frame$_{J+101}$ to Frame$_{J+200}$ as 
seed frames. If the trigger selects a group of frames, it will produce an ``Event'', which
will then be written to storage. 
The data format of an event is:
\begin{center}
\begin{tabular}{|l|l|l|l|l|l|l|}\hline
Event-size&Trigger header&Trigger data&MEP$_{\#}$&Frames$_{K-n}$&.......&Frames$_{K+n}$\\\hline
\end{tabular}
\end{center}
Consecutive events are not allowed to have overlapping
frames, but instead will contain more frames than necessary to cover the data of one p.o.t.
Hence, within events selected from a MEP there are no duplicate frames written to storage.
However, frames could still be duplicated if the trigger on a node selects the last frames
in a MEP, while another node selects the first frames of the next MEP. To avoid this,
events which contain the first or last frame of a MEP are written to the same file
during a spill. This file is processed to merge events containing overlapping 
frames, to avoid a possible duplication of events. 
%{\color{blue}\large This post processing is not elegant, but I do not see a more elegant solution for the time being.}
\end{itemize}
Assuming 100 frames/MEP results in an empty event of 632 Bytes/MEP. Hence, for 50 sources a total bandwidth of 
about 3 GBps. From the expected occupancy as given in section \ref{sec:performance}, and the data size as listed
in table~\ref{online:daqsize} it follows that the required bandwidth is an additional GBps. Assuming a similar contribution from noise hits, results in a total estimated bandwidth
of five GBps, and hence the rate with or without p.o.t. will not be very different.
Even by today standards this is a very modest bandwidth, and the necessary Ethernet switch with 100 ports 
could be acquired today.

\subsection{Timing and Fast Control System}
\label{sec:TFC}
The Timing and Fast Control system (TFC) is a central element of the SHiP
DAQ system, as it is responsible for generation and distribution of the
clock, and the distribution of synchronous and asynchronous commands. It is 
the interface between SHiP and the SPS, and it regulates the
load sharing over the various nodes of the Event Filter Farm by transmitting the MEP destinations to the FE. 
The hardware implementation for this system could use PCIe40 boards, the
same backbone as the readout system for the LHCb upgrade. The advantage in the
case of SHiP is that this detector is not in a radioactive environment, therefore
simple Ethernet connectors can be used to distribute the signals from the
system to the Event Filter Farm and to the Front End electronics.
Given the relatively simple nature of the system, it is believed that no
dedicated throttling system is needed to absorb back-pressure from data 
congestion. In the case, expected to be rare, that no CPU is available,
the TFC system would send a ``drop event'' signal to the Front-End,
stopping further processing, and hence this MEP will not be transmitted.
The TFC will measure the accumulated dead time due to dropped MEPs.\par
The 1 s spill from the SPS is followed by 6 s without protons on target, that
will be used by the system to terminate processing the Front-End buffers.
TFC provides FE with the start of SPS spill signal to reset the frame and 
MEP counters. In addition all FE buffers 
should be empty at the start of spill. When buffers still have data, or 
in the case of potential FE buffer overflows, 
the FE are required to truncate the MEP events by setting the truncation 
bit in the ECS-block, and by not transmitting any frame information.
This technique reduces the bandwidth by at least a factor 30, and this way 
the buffers can be flushed before the next SPS spill.

\subsection{Experiment Control System (ECS)}
The ECS is responsible for configuring, monitoring and controlling
all active devices of the experiment. 
The ECS provides the interface between the operator and experiment. It integrates Slow Control for detector
hardware and the Run Control for the DAQ into a single system.
Table~\ref{online:daqsize} gives an estimate
for the necessary configuration size of the whole system.

The ECS is based on the JCOP framework running WinCC Open Architecture Service. The ECS is built with standard
components using protocols supported by the JCOP framework.
Standardized interfaces supporting
protocols such as I2C, JTAG, SPI or CANOpen will be made available for the sub-detectors which will ease the
maintainability over the life-time of the experiment. Since radiation hardness is not required standard
components used for industrial automation can be used. If there is a need for monitoring high granularity data in the FE
this will be made possible via the switch where the ECS data can be mixed with MEPs and forwarded to the MF.

SHiP will use a configuration database which is local at the experimental facility and integrated with
the ECS. A relational database for detector conditions will be made available for online
and offline use. The choice of technology for configuration and conditions database will follow the
recommendations for CERN experiments.

\subsection{Event Filter Farm}
The Event Filter Farm (EFF) is the brain of the SHiP trigger system. It 
provides the necessary computing resources to select 
interesting events for off-line storage with the Trigger, running algorithms
as similar as possible to the offline ones.
It consists of a farm of multi-core Personal Computers (PCs), and nodes signal 
their availability to the TFC via the switch to assure proper load balancing.
The size of this farm can be estimated from the average computing time for each
MEP. We assume 10 ms/MEP, which would contain roughly 10 tracks.
As a comparison, LHCb assumes 13 ms/event with ~100 tracks in 2019. 
The MEP rate is 100 kHz, hence we need 1000 processes running
simultaneously at the EFF.
Servers in the LHCb upgrade are expected to run 400 processes simultaneously, 
hence SHiP would only require three servers.

\subsection{Trigger}
The SHiP Trigger is an application in C++, of which $\sim 1000$ copies run on the EFF. 
The code is based on the same software as used throughout SHiP. 
The trigger uses, where time allows, the same reconstruction algorithms as used off-line. 

It consists of trigger lines, each line being optimized to cover a certain
class of events of interest. Each trigger line is configured by a python script,
which defines the basic reconstruction steps for objects on which event selections can be based, 
for instance straw chamber tracks, or tracks identified as a muon. 
Each line contains the selection parameters and where applicable down scaling fractions. 
All lines operate independently. Typically one line will be configured to select
events with two tracks reconstructed in the straw chambers which form a vertex, while neither of the
two is tagged as originating outside the decay vessel. Another line will select muons
which point back to the emulsion.

A combination of lines form a unique trigger with its associated Trigger Configuration Key (TCK), which is
encoded for every event in the raw data. The architecture of the trigger assures that while
lines are independent, all reconstruction algorithms are executed only once per event.
A typical TCK will contain at least one line per physics channel. 
Additional lines are for luminosity measurements, counting truncated MEPs, pre-scaled physics lines
with looser cuts and lines selecting calibration and monitoring data for fast feedback
on the quality of the data.

With an event size of $\sim 10$ kB, and a very open trigger essentially selecting all
events with at least one well reconstructed track, the estimated amount of storage space
will be of the order of a PB/year.

\subsection{Monitoring}
An independent computing node for monitoring purposes will run in parallel with the EFF. The trigger will
contain algorithms to select events suitable for monitoring, which will then be routed to the MF
via the switch.
The monitoring nodes will be less strict than the EFF with respect to
which code is allowed to run. The MF allows
online monitoring of sub-detector performance and
general online monitoring of data quality.
The online monitoring data is generally stored in histograms hence the monitoring node requires little
storage. No raw data will be stored in the node.

%\subsection{Online System cost breakdown}
%For internal use for the time being, to be merged with all others in the final cost+schedule.
%
%The following institutes will share the responsibilities for the Online System: CERN, NBI, UCL and Uppsala.
%\begin{table}[h!]
%\begin{center}
%\caption{Cost/Responsibilities of the Online System}
%\label{online:cost}
%\begin{tabular}{l r r}\\\hline
%Item&Units&Cost (kCHF)\\\hline
%Switch&2&40\\
%Ethernet links&100&20\\
%TFC and optical links&&50 \\
%ECS&&50 \\
%EFF $\&$ Monitoring&4&20\\
%Storage & 1 PB &?\\
%Infrastructure&-&50\\
%\hline
%Total & &230\\\hline
%\end{tabular}
%\end{center}
%\end{table}

\newpage

%% file: computing/Computing.tex
\section{Offline computing}
\label{sec:computing}
As offline computing framework SHiP selected the \textsc{FairRoot}~\cite{AlTurany:2012gc} framework
that will be used for simulation, reconstruction, analysis and event display. 
\textsc{FairRoot} is an ideal turn-key solution for small (and large) experiments that
do not have enough person power for a full fledged computing group or experiments that do not
see the need to develop their own offline computing framework. \textsc{FairRoot} is being
used by all FAIR~\cite{computing:FAIR} experiments and it is the basis of the
O2~\cite{Ananya:2014kya} combined online and offline framework for the ALICE experiment.

FairRoot's main idea is to provide a single package with generic mechanisms to deal
with most commonly used computing tasks in HEP. FairRoot allows the physicist to:
\begin{itemize}
   \item Focus on physics deliverables while reusing well-tested software components.
   \item Not drown in low-level details but use pre-built and well-tested code for common tasks.
   \item Concentrate on detector performance details, while avoiding pure software
         engineering issues like storage, retrieval, code organization, etc.
\end{itemize}

\subsection{Basic functionalities}
\begin{figure}[h]
\centering
\includegraphics[width=1.0\textwidth,clip=]{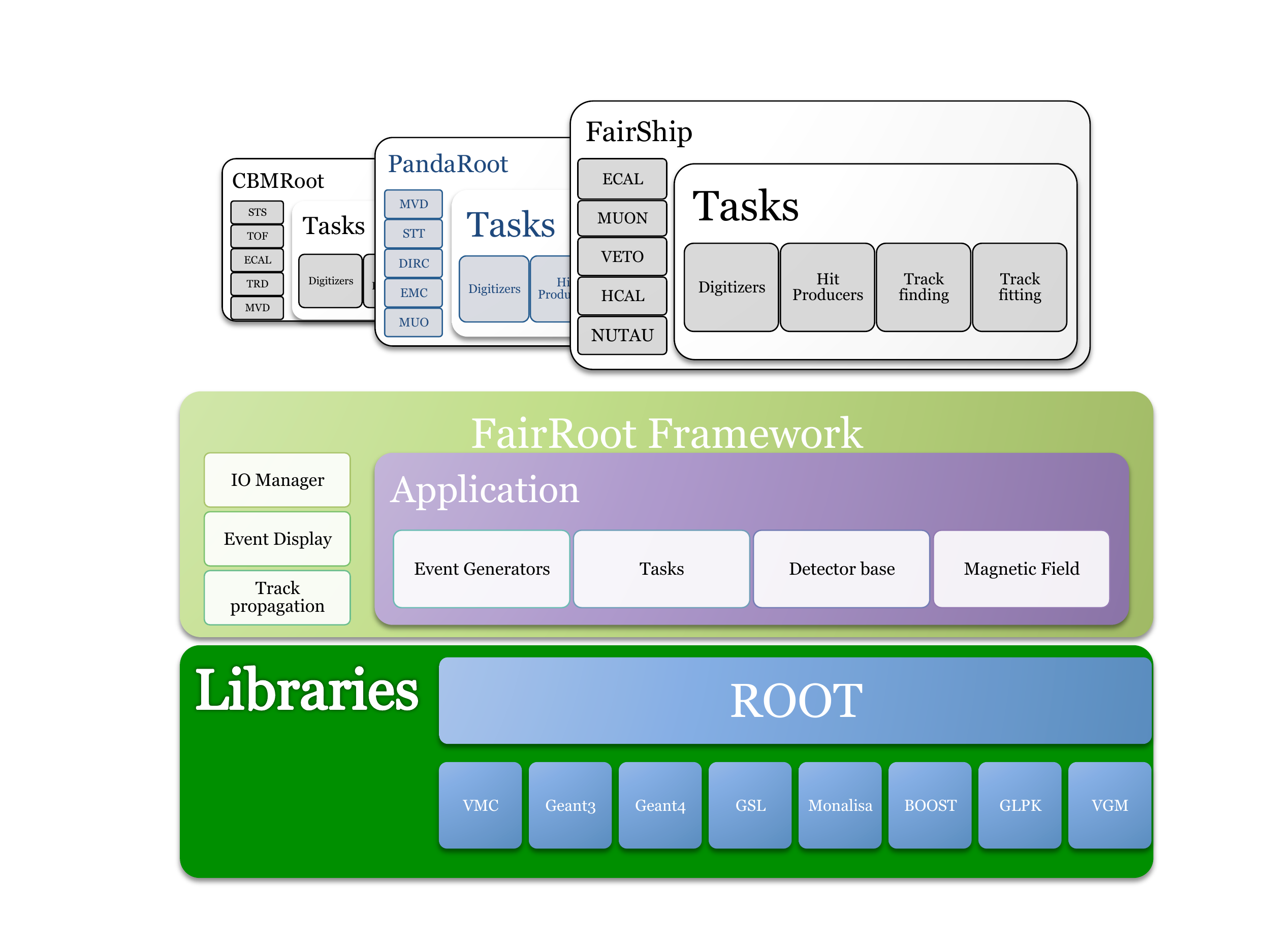}
\caption{The architecture of the FairRoot system.}
\label{computing:FairRoot-design}
\end{figure}

The \textsc{FairRoot} framework is fully based on the ROOT~\cite{root} system. The user
can generate simulated data and/or perform analysis with the same framework. Using the ROOT
Virtual Monte Carlo (VMC) interface different simulation transport engines are supported
while keeping the user code independent of a specific transport code. Currently Geant3
and Geant4 are supported. The framework delivers base classes which enable the users
to construct their detectors and/or analysis tasks in a simple way. Detectors are
defined using the ROOT {\tt TGeo} classes. This single geometrical description is used by
FairRoot for all supported tasks, providing a single coherent view of the detector.
It also provides some special purpose functionality like track visualization.
An interface for reading magnetic field maps is also implemented.
Figure~\ref{computing:FairRoot-design} shows the overall architecture of \textsc{FairRoot}. 

\subsection{Input/Output procedure}
The storage of all information collected by the different sensitive detectors is done on
an event by event basis (an event means in this context one interaction between one beam
particle and the target). All persistent objects are serialized and stored into binary
ROOT files. An interface class ({\tt CbmMCPoint}) is provided to define the structure of
a registered hit in a detector. Each detector can then provide a more specific implementation
following the {\tt CbmMCPoint} API. All registered hits will be collected into dedicated lists,
one list corresponding to one detector entity. The ROOT class {\tt TTree} is used to organize
the output data into an ``ntuple like'' data structure. In the analysis case, the
{\tt CbmRootManager} provides methods to read this information. A partial input/output mechanism
is also supported.

\subsection{Parameter definition}
In order to analyze the simulated data, several numerical parameters are needed, like, calibration/digitization parameters or geometrical positions of detectors.
One common characteristic of most of these parameters is that they will go through
several different versions corresponding, for example, to changes in the detector
definitions or any other conditions. This makes it necessary to have a parameter
repository with a well-defined versioning system. The runtime database (realized
through the {\tt CbmRuntimeDb} class) is such a repository. Different inputs are
supported: ASCII, ROOT files and Oracle DB.

\subsection{Implementation of the algorithms}
The analysis (reconstruction) is organized in tasks. For each event we need to accomplish
various tasks or reconstruction algorithms. The {\tt CbmTask} is an abstract class defining
a generic API allowing to execute one task and to navigate through a list of tasks.
Users can create their own algorithms inheriting from {\tt CbmTask}. Each task defines the
relevant input data and parameters and creates its particular output data during the
initialization phase. During the execution phase, the relevant input data and
parameters are retrieved from the input file and the output data objects are stored
in the output file.

\subsection{FairShip code}
All SHiP specific code is collected in a package called FairShip. The FairShip code is
basically only a specialization of the generic base classes provided by FairRoot, as described
above. With in addition some specific auxiliary libraries needed by SHiP like genfit.

\subsection{Packaging}
The complete SHiP offline computing environment consists of three packages:
\begin{itemize}
   \item FairSoft: all packages needed by FairRoot, like ROOT, Geant3, Geant4, etc, etc.
   \item FairRoot: the FairRoot framework proper.
   \item FairShip: the SHiP specific implementation of the FairRoot classes.
\end{itemize}
These three packages are maintained on GitHub~\cite{computing:github} and available
under the LGPL Open Source license. The FairRoot code can be installed and run on all major Linux distributions and on OSX. The more platforms are supported the better for the code quality as different compilers always find different issues in the code.

For running simulation and analysis jobs at larger scale a special framework called SkyGrid was designed and developed. It allows for the optimal usage of hardware computing resources at the same time it gives user flexibility to choose arbitrary set of libraries and modules his/her job depends on. Those possibilities come from the design of SkyGrid that is based on lightweight virtualization technology called Docker~\cite{computing:docker} and on job scheduling system. Job scheduling is based on auction principles and the actual job can be sent to a variety of execution environments that owner of a computational cluster chooses on his own. At the moment YARN~\cite{computing:yarn} and vanilla execution containers are supported. 

SkyGrid has been deployed on Yandex site and it had united tens (from 20 to 60) of heterogeneous machines to run SHiP simulation and analysis jobs. 
Support for containerization of user jobs allows also for running FairShip software (that depends on plenty of other packages) on any environment that has support for virtualization (so it is possible to run the analysis scripts not only on Mac OSX or Linux, but on Windows or on BSD-like systems). 

SkyGrid had been used for various simulations for the SHiP experiment. In the end it allowed for better understanding of the physics processes happening inside the detector as well as for finding better detector geometry that improves performance/cost ratio for it.

\newpage

%% file: performance/Performance.tex
\chapter{Physics Performance}
\label{sec:performance}

This section reviews the SHiP physics performance for representative physics processes
taken as examples to demonstrate the SHiP physics reach. The full physics programme
is addressed in the Physics Proposal~\cite{PP} and includes an extended list of signal final states
demanding the detection of neutral particles and better particle identification which can
be optimized al later stage, given to the open geometry of the SHiP detector.

The SHiP physics sensitivities are evaluated on the basis of the full simulation and
reconstruction package, \textsc{FairShip} (Section~\ref{sec:simulation}), which is being gradually 
elaborated to achieve even better performance. The evaluation of certain types of backgrounds requires
larger data samples, the simulation of which is currently ongoing.

The current state of understanding of the SHiP sensitivities is presented separately for
Hidden Sector particles (Section~\ref{sec:sensitivity}) and for physics channels with tau neutrinos (Section~\ref{sec:yield})

\section{MC simulation}
\input{simulation/fairship.tex}

\input{simulation/muon_simulation.tex}
\input{sensitivity/ToyMC.tex}

\input{sensitivity/background.tex}

\input{sensitivity/Sensitivity.tex}

\input{nusensitivity/nutauphysics}

%% file: simulation/fairship.tex
%%%%\section{MC simulation}

\label{sec:simulation}

The overall simulation and reconstruction software called \textsc{FairShip} is based on the \textsc{FairRoot}~\cite{AlTurany:2012gc} package, a lightweight software framework based on ROOT~\cite{root}.

For the physics performance studies events are produced using the Monte\,Carlo generators:
PYTHIA v8~\cite{pythia8} for the initial proton fixed target interaction,
GENIE~\cite{genie} for inelastic neutrino interactions, PYTHIA6~\cite{pythia6} for inelastic muon interactions
and GEANT4~\cite{Agostinelli:2002hh} for following the particles through the detector setup.

Default values are used for the physics parameters of PYTHIA, with the exception of the signal generation of HNLs,
where in all leptonic and semi-leptonic decay modes of charm particles the neutrinos have been replaced by a HNL particle, with its
mass, lifetime and decay channels. Based on the mass of the HNL particle and its couplings, all possible decay modes with their
corresponding amplitudes are calculated and used to determine the lifetime of the HNL.

The SHiP detector response is simulated in the GEANT4 framework. The detector geometry is implemented
using the TGEO package within ROOT. 	

A crucial issue for the overall setup is the performance of the muon shield. It needs to reduce the flux of muons
in the decay vessel and detector region by six orders of magnitude. 
    % don't know why people don't like to add this comment: and function up to $350$\,GeV of muon momentum.
To verify the realism of the simulation, a setup has been simulated reflecting the CHARM beam-dump experiment~\cite{Bergsma:1990yc}, see Section~\ref{sec:muon_simulation}. Good agreement between the experimental numbers and the simulation
has been achieved giving confidence in the simulation results for the active and passive shielding.

\begin{figure}[ht]
\centering
\includegraphics[width=0.6\textwidth,clip=]{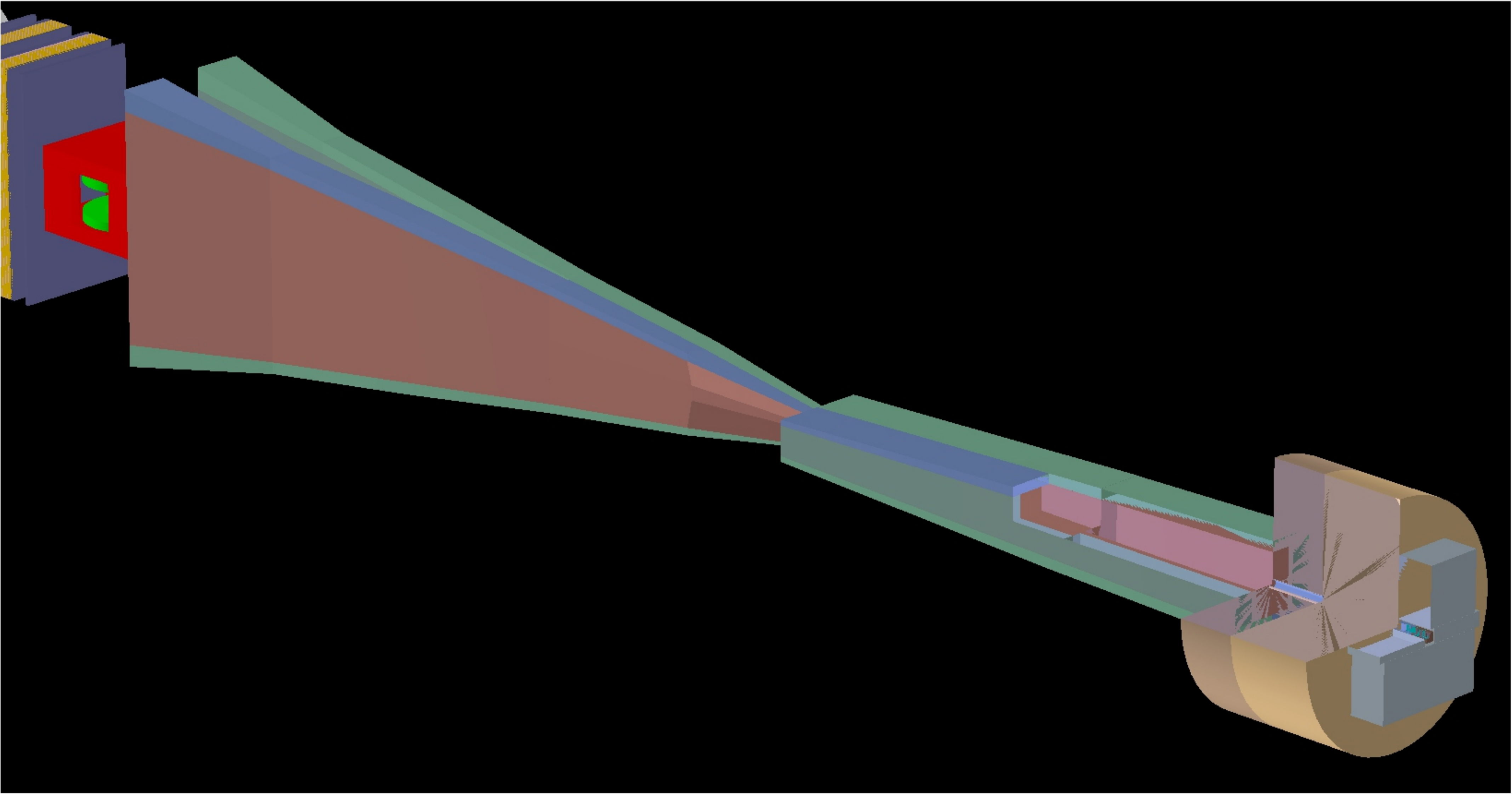}
\caption{The first part of the setup, target area followed by a 5~m iron hadron absorber and the active muon shield.
The different colors of the magnets represent the different magnetic field directions. } \label{fig:fairship-det1}
\end{figure}

\begin{figure}[ht]
\centering
\includegraphics[width=0.6\textwidth,clip=]{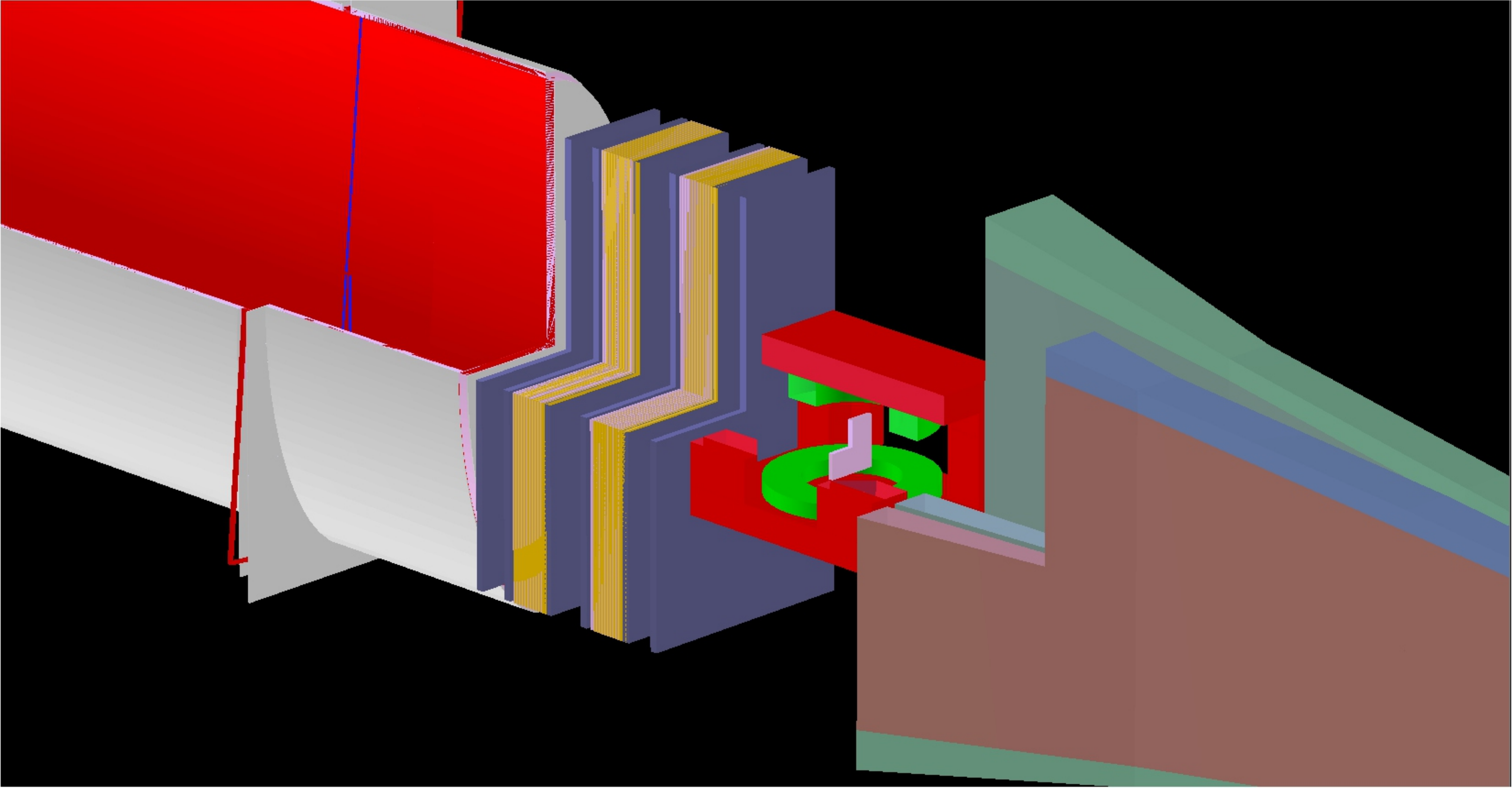}
\caption{The second part of the setup, including instrumentation for the $\nu_{\tau}$ measurements. Goliath magnet containing the emulsion and target tracker,
followed by the muon magnetic spectrometer of the tau neutrino detector.} \label{fig:fairship-det2}
\end{figure}

\begin{figure}[ht]
\centering
\includegraphics[width=0.6\textwidth,clip=]{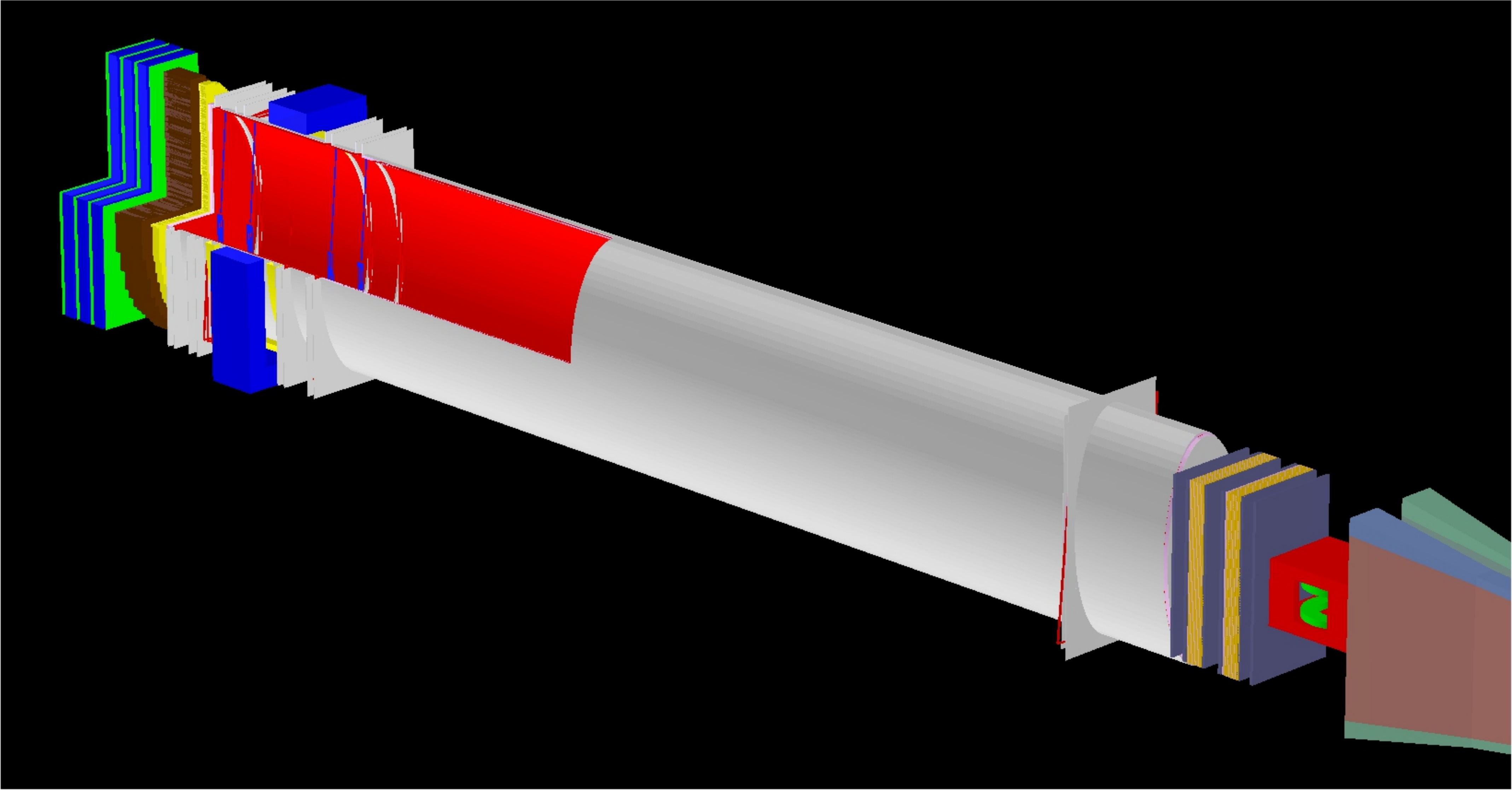}
\caption{The third part of the setup, showing the instrumentation for the hidden particle detection, consists of the vacuum vessel, with the straw veto tagger, 
followed by the HS spectrometer inside, the electromagnetic and hadronic calorimeters outside, timing station in front, and muon system at the end.
} \label{fig:fairship-det3}
\end{figure}

\begin{figure}[ht]
\centering
\includegraphics[width=0.6\textwidth,clip=]{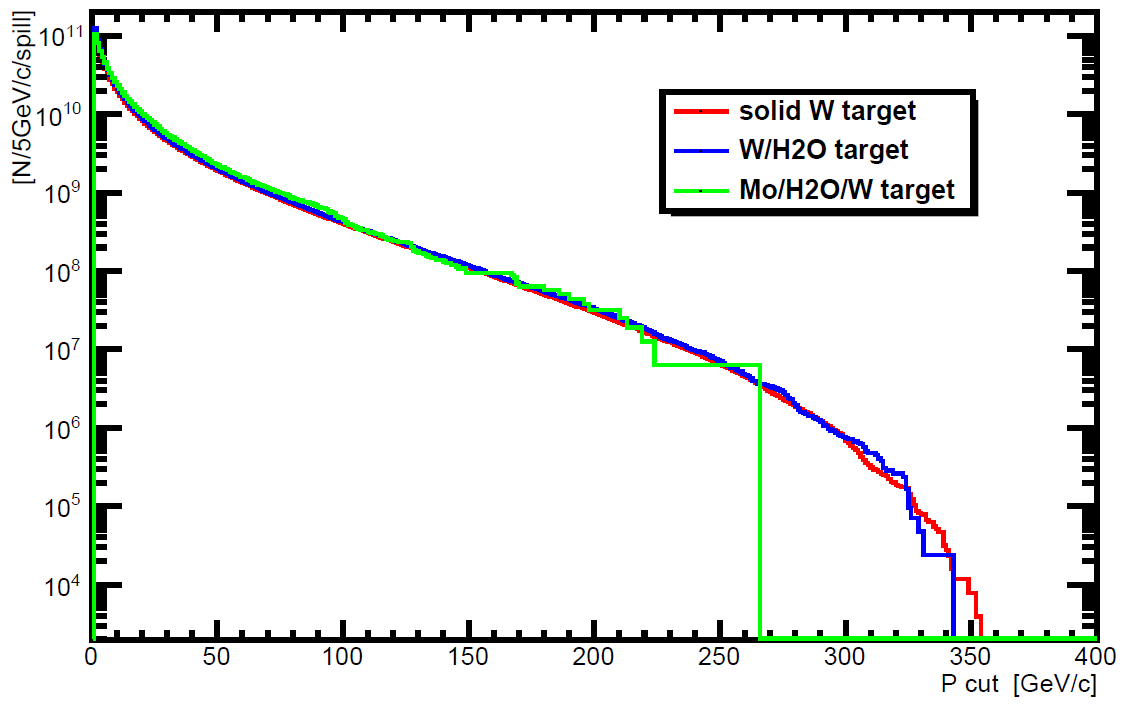}
\caption{The muon rate after the target and absorber is shown for the different target configurations studied during the optimization. Most of statistics for muon background studies come from simulations with a solid tungsten target. The difference to the baseline design is small ($<20$\%), so that the full statistics can be used to optimize the active muon shield.} 
\label{fig:fairship-targetComparison}
\end{figure}

Figures~\ref{fig:fairship-det1} - \ref{fig:fairship-det3} show the details of the simulation of the SHiP facility implemented in \textsc{FairShip}: Figure~\ref{fig:fairship-det1} shows the target system and the active filter as described in Chapter~\ref{sec:facility}, Figure~\ref{fig:fairship-det2} shows the tau neutrino detector (section~\ref{sec:nutaudet}, and Figure~\ref{fig:fairship-det3} shows the Hidden Particle detector as described in Chapter~\ref{sec:detector}. The whole setup is placed in a cavern surrounded by concrete and rock. Following the developments of the CERN task force (Section~\ref{sec:target}), different versions of the proton target have been implemented. For most of the muon background studies, a target with a total length of $50$\,cm of tungsten interleaved with $4$ slits of $1$\,cm filled with water followed by a 5m hadron absorber made of iron has been simulated. A more recent version of the target based on a mixture of 
molybdenum and tungsten, also with water cooling, has been used to study the impact on the increase of the muon and neutrino fluxes due to the reduction of the target density. Figure~\ref{fig:fairship-targetComparison} shows a comparison of the muon fluxes for the different target configurations. Differences are well within $20\%$ which allows to combine the high statistics data simulated with different target configurations for background studies. Two options are implemented for the muon shielding, a passive and an active shielding. The passive shielding is based on
a $40$\,m long conical shape made of tungsten ($\sim 100$\,t), surrounded by lead and continued with lead up to a total length of
$70$\,m. The active shielding (Section~\ref{sec:muon_shield}) is based on conventional warm magnets with a field of $1.8$\,Tesla. 

The simulation of the Tracking Detector (Section~\ref{sec:tracker}) includes the details of individual straw tubes, their envelope together with  the gas and tungsten wire inside.
The drift time is determined from the position of the particle trajectory in the straw tube and smeared according to the
expected spatial resolution. The energy deposition in the ECAL and HCAL is calculated in the GEANT framework with energy thresholds of $0.1\,$MeV for electrons and photons and $0.3\,$MeV for hadrons. With these threshold values the shower development in the media of the calorimeters including energy leakage at the boundaries can be accurately simulated.

Track parameters are obtained by an iterative procedure based on the Kalman filter algorithm where continuous energy losses and
multiple scattering are taken into account~\cite{Höppner2010518}. The same tool is also used to retrieve track parameters upstream and downstream of the magnetic field. The reconstruction in the ECAL starts from finding local maxima of the energy deposition. A $5\times 5$ cells peak area is build near each maximum. An ECAL cluster is defined as a set of peak areas with common cells. The reconstruction produces a reconstructed object from each maximum of the cluster. If two neighbouring maxima in the cluster have no common neighbor cells then a simple energy splitting algorithm is used. Otherwise a dedicated unfolding procedure is required. Finally energy deposition after splitting is converted to reconstructed energy using cluster calibration. The kinematic parameters of a photon can be calculated from the reconstructed objects and an assumption about the origin point of the photon. Charged tracks reconstructed with the tracking system are matched to the hits in the muon system. Their identification as muons or hadrons results from a detailed analysis of the hit patterns in the active muon detector and energy deposited in the calorimeters. A simple algorithm that extrapolates the track direction from the tracking system and search for hits in the muon system within Fields of Interest (FOIs) is used.

The signal simulation uses Pythia8 as the primary event generator simulating proton (400~GeV) on fixed target collisions with the option "HardQCD::hardccbar  = on". Only the HNL decay products are given to GEANT4 for further processing. The simulations for the muon and neutrino induced backgrounds are very CPU intensive processes and splitted in several steps. In the first step, Pythia8 is used as the 
primary event generator simulating proton on fixed target collisions running in minimum bias mode. Long-lived particles, kaons, 
pions and strange baryons, are set stable. Particles with energies exceeding a given threshold are given to GEANT4 for further processing of interactions in the target or close-by hadron absorber, or their decay. During a several months campaign, more than 3 billions of such events were produced. 
In a second step, these events are used either as input in a two dimensional fast Monte Carlo simulation used for optimization of the muon shield,  or in a detailed GEANT4 simulation to determine the remaining muon background in the decay volume and SHiP detectors. For simulating muon interactions in the material with high statistics, the Pythia6 MC generator is used. The products of these interactions are placed along the muon trajectory taking into account the material distribution and further processed by GEANT4 to simulate the response of the SHiP detectors.        

As a by product of the production runs for the muon background, also the neutrino fluxes are recorded in the first step.
The GENIE MC event generator is used to provide a template of neutrino interactions. These events are distributed along a neutrino flight direction using the same procedure as for muons described before. 

%% file: simulation/muon_simulation.tex
\subsection{Validation of the simulated muon spectrum}
\label{sec:muon_simulation}

The validity of the simulation has been verified by comparing simulation predictions to data from the CHARM beam-dump experiment at CERN~\cite{Bergsma:1990yc}. This experiment used 400\,GeV protons impinging on a copper target, the length of which limited the escape of particles from hadronic cascades. 
The particles leaving the target then consisted mostly of neutrinos and muons. 
The muons were stopped in a 200\,m long iron shield which had several gaps where the muon flux was monitored by a set of solid state detectors. 
For the present study, the beam dump target, muon shield and this muon flux monitoring set-up were simulated in the \textsc{FairShip} simulation framework.
Four combinations of different physics models were simulated to cover the muon energy range which spans eV-TeV~\cite{Ivanchenko:2010zz}.
Two different electromagnetic generators were used: The EMV generator is known to have limited accuracy in  simulating processes involving the multiple scattering of charged particles. The EMX generator improves the accuracy by applying production thresholds on secondary particles produced in the simulation of the 
photoelectric effect and Compton scattering. Two different Quantum Chromo Dynamic (QCD) string models were used to describe interactions of nucleons, pions and kaons at energies above 15\,GeV: the standard Quark Gluon String (QGS) model and the FTF model where the FRITIOF description of string excitation and fragmentation is used~\cite{Agostinelli:2002hh}.
The number of muons recorded by each solid state detector per $10^7$ protons on target is shown in Table~\ref{tab:muon_valid}, along with the results of the simulation in the various configurations~\cite{Bergsma:1990yc}.
The observed number of hits is only known extrapolated to the entire area transverse to the beam line. 
The systematic error associated with this extrapolation is $1.1$\%~\cite{Bergsma:1990yc}.
The data and simulation are in very good agreement, giving confidence in the use of the simulated muon rates.
The observed results were found to be insensitive to the simulation of the surrounding cavern. 
The muon energy threshold considered was $E_{\mu}>0.5$\,GeV where the CHARM data contained no muons with $E_{\mu}>5$\,GeV.

\begin{table}[htb]
\caption{The number of muons recorded by each solid state detector per $10^{7}$ protons on target. Data from the CHARM experiment are shown together with four different types of \textsc{FairShip} simulations.}
\label{tab:muon_valid}
\centering 
\begin{tabular}{@{}crrrrr@{}} \\ \hline
\multirow{2}{*}{Type of simulation:} & \multicolumn{2}{c}{EMV} & \multicolumn{2}{c}{EMX} & \multicolumn{1}{c}{CHARM} \\  
                                    &  QGSP      & FTFP       & QGSP       & FTFP       &  \multicolumn{1}{c}{data}    \\ \midrule
%(lr){2-5}                           &            &            &            &            &                             \\   
Pit 1                               & 8419 & 9225 & 8583 & 9226 & 8200 \\
Pit 2                               & 624  & 630  & 697  & 645  & 655  \\
Pit 3                               & 147  & 168  & 208  & 165  & 137  \\
Pit 4                               & 36   & 55    & 37   & 45    & 33.1  \\
Pit 5                               & 14    & 8     & 4     & 9     & 6.1   \\ \hline
\end{tabular}
\label{tab:compared}
\end{table}

%% file: sensitivity/ToyMC.tex
\subsection{Toy Monte Carlo}

In order to extrapolate the results of the full simulation and scan a large number of points in the sensitivity plane of the various signals,  a fast simulation (toy MC) is used.  The \textsc{FairShip} software is used to generate HNLs of given masses and couplings, and to compute the geometrical acceptance, ${\cal A\times P_{\rm vtx}}$ (more details can be found in Section~\ref{sec:sensitivity}), which is then compared to the Toy Monte Carlo. 
A comparison between the toy MC and the full simulation is shown in Table~\ref{tab:sigcomp} for the $N\rightarrow \mu \pi $ decay channel and for HNL masses ranging from 0.3 GeV/c$^2$ to 1.1 GeV/c$^2$ and various couplings. The results show good agreement between the fast and full simulation. 
To assess the impact of the reconstruction and the selection on the signal, a correction to the fast simulation is applied using the full simulation as a function of the mass of the HNL, for two body and three body decays.

\begin{table}[h]
\caption{\label{tab:sigcomp} Comparison of signal acceptances for the $H\!N\!L\rightarrow \mu \pi $ decay channel using the toy Monte Carlo and \textsc{FairShip}. Most ratios are compatible with unity within the uncertainty. }
\footnotesize
\begin{tabular}{rrrrccccc}
\hline
$m_{H\!N\!L}$ & $U^2_e$ & $U^2_{\mu}$ & $U^2_{\tau}$ & Model & toy MC  & \textsc{FairShip} & \textsc{FairShip} & Ratio \\ 

[GeV/c$^2$] & & & & &   ${\cal A\times P_{\rm vtx}}$ & raw &  ${\cal A\times P_{\rm vtx}}$ & toy/\textsc{FairShip} \\ \hline
 1.1 & 6.25e-10 & 1e-08 & 2.375e-09 & 2 & 2.96e-05 & 226 & 2.64e-05 & 1.12  \\ 
 1.1 & 3.125e-09 & 5e-08 & 1.1875e-08 & 2 & 1.44e-04& 232 &  1.45e-04 & 0.99  \\  
 1.0 & 4.47e-10 & 7.15e-09 & 1.70e-09 & 2 & 5.76e-06&  214 & 5.55e-06 &  1.04 \\ 
 1.0 & 6.25e-09 & 1e-07 & 2.375e-08 & 2 &  8.04e-05&  217 & 7.85e-05  & 1.02 \\ 
 1.0 & 1e-09 & 1.923e-11 & 1.923e-11 & 1 &7.23e-07  &  215 & 7.07e-07 & 1.02  \\ 
 1.0 & 9.09e-10 & 1e-08 & 1e-08 & 5 & 1.04e-05& 206  & 8.98e-06 & 1.16 \\ 
 0.6 & 6.25e-10 & 1e-08 & 2.375e-09 &2  & 4.48e-07&  200 &  5.63e-07 & 0.80  \\ 
 0.6 & 3.64e-09 & 4e-08  & 4e-08 & 5 &  2.33e-06&  212 & 3.11e-06 & 0.75 \\ 
 0.5 & 6.25e-10 & 1e-08 & 2.375e-09 & 2 & 1.89e-07&  189 &  2.26e-07 & 0.84 \\ 
 0.5 & 6.25e-09 & 1e-07 & 2.375e-08 & 2 & 1.85e-06&  188 & 2.62e-06 & 0.71 \\ 
 0.5 & 6.25e-08 & 1e-06 & 2.375e-07 & 2 & 1.79e-05&  209 & 2.58e-05 & 0.69  \\ 
 0.5 & 1e-08 & 1.923e-10 & 1.923e-10 & 1 & 1.86e-07& 194  & 2.05e-07  & 0.91  \\ 
 0.3 & 5e-09 & 8e-08 & 1.9e-08 & 2 & 1.09e-07&  185 & 1.25e-07 & 0.87  \\ 
\hline
\end{tabular}
\end{table}

%% file: sensitivity/background.tex
\def\klpp{$K^0_L\to \pi^{-}\pi^+$}
\def\kspp{$K^0_S\to \pi^{-}\pi^+$}
\def\klpopopo{$K^0_L\to 3\pi^{0}$}
\def\kspopo{$K^0_S\to 2\pi^{0}$}
\def\LpPi{$\Lambda \to p \pi$}
\def\klpmn{$K^0_L\to \pi^{-}\mu^+\nu_{\mu}(+\pi^0$) }
\def\klpm{$K^0_L\to \pi^{-}\mu^+\nu_{\mu}$ }
\def\klpppio{$K^0_L\to \pi^{-}\pi^+\pi^{0}$}
\def\klpppion{$K^0_L\to \pi^{-}\pi^+\pi^{0}(+\pi^0$)}
\def\klpe{$K^0_L\to \pi^{-}e^+\nu_{e}$}

\section{Sensitivity to Hidden Sector particles}
\label{sec:sensitivity_intro}

The sections below discuss the background evaluation and the sensitivity to a few 
selected physics channels which have experimental signatures representative 
for the whole SHiP physics programme.  One of the main challenges for experiments 
looking for very weakly interacting particles is the suppression of the background. 
The goal of the SHiP experiment is to control the background to a level of 0.1 
expected background events for the assumed integrated exposure of $2\cdot 10^{20}$ 
protons on target. This will allow probing a variety of BSM models. An extensive list 
of these models and the impact of SHiP are discussed in detail in the SHiP Physics 
Proposal~\cite{PP}. 

The procedure described in Section~\ref{sec:sensitivity} for determining the signal 
yield in the case of HNLs has been applied to all of the representative physics channels.
The sensitivities can be interpreted as 90\% C.L. if no events are observed in the 
experiment. They can also be interpreted as 3$\sigma$ evidence for the signal if two 
events are observed with a background level corresponding to 0.1~events.

The background evaluation and the sensitivities are normalized to 5$\cdot$10$^{13}$ 
protons on target per spill, and an integrated exposure 
of 2$\cdot$10$^{20}$ protons on target. At nominal conditions, this
may be accumulated in a period of five years of operation at the SPS
with the assumption of $4\cdot 10^6$ SPS data taking cycles per year.
The fiducial decay volume used in the evaluation of background 
and signal sensitivity is defined by the interior of the vacuum vessel starting 
from 5~m downstream of the entrance window and up to the first straw tracker station. 

The calculation of backgrounds and sensitivities is based on the \textsc{FairShip} 
full simulation software. The toy Monte Carlo has been used to boost
statistics of the signal and for
rapidly interpolating between the results of the full simulation.

%--------------------------------------
\subsection{Background evaluation}
\label{sec:backgrounds}
%--------------------------------------

This section presents an evaluation of the three main sources of background as introduced in 
Section~\ref{sec:hs_reqs}. These are

\begin{itemize}
\item[-] {\bf Neutrino- and muon-induced backgrounds:} neutrinos and
  muons coming from hadron decays produced in the proton interactions can interact inelastically
  with the material surrounding the HS decay volume. These interactions can
  generate particles that enter the decay volume and mimic signal events.

\item[-] {\bf Random combinations of tracks} 
in the fiducial volume from muons, or other charged particles 
from interactions in the proximity of the detector, which enter the decay volume and together fake
decay vertices which resemble signal events.

\item[-] {\bf Cosmic muons} entering the decay volume
 can create background in the same way as the previous two classes.

\end{itemize}

Table~\ref{tab:backgrounds} shows an extensive list of the physics signals and their main backgrounds. 

% Long-lived particles can be produced in the interactions of neutrinos and muons coming 
% from the proton target and from cosmic rays.

\begin{table}[htb]
\scriptsize
\begin{center}
\caption{\footnotesize Signal and background channels for the Hidden Sector
  detector. The last column  lists the cuts which can be used to
  suppress the backgrounds. 
The abbreviations for the physics channels correspond to: HNL=Heavy Neutral Lepton,
NEU=neutralino, 
DP=Dark Photon, 
PNGB= Pseudo-Nambu Goldston Boson,
HP= Higgs Portal,
CS=Chern-Simons,
HSU= Hidden SUSY,
RDM=random di-muons from the target,
The abbreviations used for techniques to reject the backgrounds correspond to: 
IP=impact parameter at the target,
CPV= charged particle veto,
NT=neutrino interaction tagger,
VP= photon veto (i.e. if there is a photon around),
TI=timing cuts with timing detector,
P=total momentum cuts of the daughters, 
POA=1 particle outside acceptance,
PID($\mu\pi$)=probability that a $\mu$ is misidentified as $\pi$ or kaon.}

\label{tab:backgrounds}
\begin{tabular*}{1\textwidth}{llll}
\hline
Signature  & Physics &Backgrounds   &  Cuts  \\ 
\hline
$\pi^-\mu^+$,$K^-\mu^+$ & HNL,NEU & \pbox{20cm}{ RDM, \\ \klpm} & \pbox{20cm}{ IP,TI,PID($\mu\pi$)\\P,IP,NT }\\ %\hline
 & & & \\
$\pi^-\pi^0\mu^+$ & HNL($\to \rho^-\mu^+$) & \pbox{20cm}{\klpmn, \\\klpppio}      &\pbox{20cm}{ P,IP,NT,TI,\\ P,IP,NT,PID($\pi\mu$) }\\ %\hline
 & & & \\
$\pi^-e^+$,$K^-e^+$ & HNL, NEU & \klpe  &P,IP,NT \\ %\hline
 & & & \\
$\pi^-\pi^0e^+$ & HNL($\to \rho^-e^+$) & \pbox{20cm}{\klpe, \\\klpppio}      &P,IP,NT,TI,PID($\pi e$)  \\ %\hline
 & & & \\
$\mu^-e^+$+$p^{miss}$ & HNL,HP($\to \tau\tau$) &\pbox{20cm}{ \klpm, \\ \klpe}    &P,NT, PID($\pi\mu,\pi e$)\\ %\hline
 & & & \\
%\rowcolor{yellow}
$\mu^-\mu^+$+$p^{miss}$ & HNL,HP($\to \tau\tau$) & \pbox{20cm}{ RDM, \\ \klpm}  & \pbox{20cm}{ TI\\P,NT, PID($\pi\mu$) } \\ %\hline
 & & & \\
%\rowcolor{yellow}
$\mu^-\mu^+$ & DP,PNGB,HP & \pbox{20cm}{ RDM, \\ \klpm} & \pbox{20cm}{ TI,IP\\P,NT, IP, PID($\pi\mu$) } \\ %\hline
 & & & \\
$\mu^-\mu^+\gamma$ & CS & \pbox{20cm}{\klpppio, \\ \klpmn }   &P,IP,NT, PID($\pi\mu$),TI,VP\\ %\hline
 & & & \\
$e^-e^+$+$p^{miss}$ & HNL,HP & \klpe &    P,NT, PID($\pi e$)\\ %\hline
 & & & \\
$e^-e^+$ & DP,PNGB,HP &\klpe  &  P,IP,NT, PID($\pi e$)\\ %\hline
 & & & \\
$\pi^-\pi^+$ & DP,PNGB,HP & \pbox{20cm}{\klpm, \\\klpe\\
  \klpppio,\\\klpp}  &P,NT,\pbox{20cm}{PID($\mu\pi$),IP\\PID($ e\pi$),IP\\ POA,IP \\}\\ %\hline
 & & & \\
$\pi^-\pi^+$+$p^{miss}$ & \pbox{20cm}{DP,PNGB,\\ HP($\to \tau\tau$),\\ HSU,HNL($\to\rho^0\nu$)} & \pbox{20cm}{\klpm, \\ \klpe, \\ \klpppio,\\ \klpp,\kspp,\LpPi}  &P,NT,\pbox{20cm}{PID($\mu\pi$),\\PID($ e\pi$),\\ POA \\}\\ %\hline
 & & & \\
$K^+K^-$ & DP,PNGB, HP &\pbox{20cm}{\klpm, \\\klpe\\ \klpppio,\\\klpp,\kspp,\LpPi} &P,NT, PID($\pi\mu,\pi e$),IP \\ %\hline
 & & & \\
$\pi^+\pi^-\pi^0$ & \pbox{20cm}{ DP,PNGB,HP,\\HNL($\eta\nu$)}&\klpppio   &P,IP,NT \\ %\hline
 & & & \\
$\pi^+\pi^-\pi^0 \pi^0$ & DP,PNGB,HP& \klpppion     &P,IP,NT,TI\\ %\hline
 & & & \\
$\pi^+\pi^-\pi^0 \pi^0 \pi^0$ & PNGB($\to\pi\pi\eta$)  & $-$   & $-$   \\ %\hline
 & & & \\
$\pi^+\pi^-\gamma \gamma$ & PNGB($\to\pi\pi\eta$) &\klpppio     &P, IP, NT,M($\gamma\gamma$)\\ %\hline
 & & & \\
$\pi^+\pi^-\pi^+ \pi^-$ & DP,PNGB,HP& $-$  &$-$ \\ %\hline
 & & & \\
$\pi^+\pi^-\mu^+ \mu^-$ & HSU & $-$   &$-$ \\ %\hline
 & & & \\
$\pi^+\pi^-e^+ e^-$ &HSU & $-$  &$-$ \\ %\hline
 & & & \\
$\mu^+\mu^-\mu^+ \mu^-$ & HSU&  $-$  &$-$ \\ %\hline
 & & & \\
$\mu^+\mu^-e^+ e^-$ & HSU&  $-$  &$-$ \\ 
\hline
%\hline\hline
%$K^0_S \nu$ & NEU && & & & & &  &\\ \hline
%$\gamma \gamma$ & NEU && & & & & &  &\\ \hline
%$\pi^0\pi^0$ & DP,PNGB, HP&\pbox{20cm}{\klpopopo,\\ \kspopo }& & &  && &  &P,NT\\ \hline
\end{tabular*}
\end{center}
\end{table}

\clearpage

\input{sensitivity/neutrinoBkg.tex}

\input{sensitivity/muonDIS.tex}

\input{sensitivity/muonComb.tex}

\input{sensitivity/CosmicBkg.tex}

\input{sensitivity/backgroundSummary.tex}

%% file: sensitivity/neutrinoBkg.tex
%----------------------------------------------------
\subsubsection{Neutrino induced background}
\label{sec:backgroundNu}
%----------------------------------------------------
Neutrinos coming from the interactions of the primary protons and surviving the hadron absorber and the muon shield
can interact inelastically with the material surrounding the decay volume and generate long-lived $V^0$  particles among the interaction products.
The $V^0$ decays can mimic the topology and modes of the hidden particles decays in the fiducial volume.

The flux of neutrinos is estimated to be $1.0 \cdot 10^{11}$
neutrinos per spill with momentum between 2 GeV and 100~GeV and within
100mrad, while we expect $ 7.3\cdot 10^{10}$ antineutrinos in the
same momentum and angular range. This corresponds to  $N_{\nu} \sim
4\cdot 10^{17}$ neutrinos and  $N_{\bar{\nu}} \sim
3\cdot 10^{17}$  anti-neutrinos per $N_{pot} \sim 2\cdot 10^{20}$ 
protons on target. These numbers are dominated by the muon neutrinos coming from the decays of pions and kaons produced in the proton interactions.
The decays of charm and beauty hadrons contribute about $\sim 10\%$ of the total neutrino sample.

%
%The number of neutrino interactions is given by:
%\begin{equation}
%N=\delta_N \times \sigma_{\nu}(E)\times \phi_{\nu}\times V \times T,  
%\end{equation}
%where $\delta_N$ is the density of nucleons, $\sigma_{\nu}(E)$ is the
%neutrino cross section~\cite{}, $\phi_{\nu}$ is the neutrino flux, $V$
%is the volume and $T$ is the running period.  

A large statistics neutrino flux has been generated with the method
described in details in Section~\ref{sec:simulation}. 
Pythia8~\cite{pythia8} is used to generate proton-nucleon interaction
and then the event is passed to GEANT4. The software
GENIE~\cite{genie} is used to simulate
the neutrino charge current (CC) and neutral current (NC) interactions
with the detector material. 
These interactions have been distributed
in the material surrounding the decay volume.
The interaction products have been further processed by GEANT4 to 
simulate the response of the detectors, as described in Section~\ref{sec:simulation}. Each event is then 
reweighted by the probability that the interaction happens in a given material by taking 
into account the density of the material and the energy dependence of
the inelastic cross section.

The neutrino interactions mostly take place in the muon magnetic spectrometer of the tau neutrino detector,
the entrance window of the vacuum vessel and the surrounding walls of the vacuum vessel.
The probability that neutrinos interact with the residual gas inside
the decay volume is negligible.
Given an average neutrino energy of $5\,$GeV, at atmospheric pressure $1.2\cdot 10^{5}$ neutrino scattering
interactions are expected in $50\,$m of air in five years. In order to
have an expectation of less than one neutrino interaction in data
sample corresponding to $2\cdot 10^{20}$ protons on target, the vacuum
pressure is set at $10^{-6}\,$bar. A more detailed simulation of
$2\cdot 10^{5}$ neutrino interaction inside the decay volume shows
that many of these background events can be rejected by making use of
the surrounding veto tagger and the geometric requirement
of pointing back to the primary proton target, which would allow relaxing 
the requirements on the vacuum pressure\footnote{Assuming the muon
  scattering in the air is under control.}. This will be studied further.

Overall we expect $\sim 10^7$ neutrino interactions for $N_{pot} = 2 \cdot 10^{20}$.
Among them about $10^4$ events have two tracks of opposite charge
reconstructed in the HS spectrometer as potential signal candidates.

%The positions of the interactions producing these events are displayed in Figure~\ref{fig:nu_interactions}.
The number of interactions occurring in a given subdetector is shown in
Table~\ref{tab:interaction1} for neutrino induced background. For the
anti-neutrinos we expect a factor three less events, due to the
smaller flux and cross section. 
More than 80\% of the neutrino interactions occur in walls of the
vacuum vessel (including the liquid scintillator and the ribs). The locations of the interactions 
of the neutrinos which are not rejected by the veto systems, as well as the ones surviving the offline selection, 
are also shown. 
%Figure~\ref{fig:nu_products} shows the the charged (left) and the neutral (right) particles 
%directly produced in the neutrino interactions.
%Figure~\ref{fig:nu_charge_neu} (left) and Figure~\ref{fig:nu_charge_neu} (right) show the types of charged and neutral particles produced
%in these interactions. 
%{\it On average each interaction produces {\it XXX} charged tracks and {\it YYY} neutral particles.}
%

%\begin{figure} [htp]
%  \centering
%  \includegraphics[width = 0.5\textwidth]{./sensitivity/IEall.pdf}\\%}
%  \includegraphics[width = 0.5\textwidth]{./sensitivity/IEveto.pdf}%}
%  \includegraphics[width = 0.5\textwidth]{./sensitivity/IEoff.pdf}%}\\
%  \caption{Positions of the neutrino interactions corresponding to $N_{pot} = 2 \cdot 10^{20}$ for events with two tracks of 
%    opposite charge reconstructed in the HS spectrometer (top). Bottom plots show the same sample after the veto (left) and selection (right) requirements.}
%\label{fig:nu_interactions}   
%\end{figure}

\begin{table}
   \centering
   \caption{Number of neutrino interactions produced in the material
     of the experiment for $N_{pot} = 2 \cdot 10^{20}$. 
The first column shows the origin of the neutrino interaction,
labelled as ``Detector''. The second column shows the number of
interactions with at least two charged tracks reconstructed in the HS
spectrometer. Number of remaining events  
 after applying the veto 
            requirements  are shown in the third column, and after
            applying the selection described in the text in the fourth column.}
            \label{tab:interaction1}
\vspace{2mm}
   \begin{tabular}{lrrr}
       \hline
       Detector & Reconstructed (\%) & Not Vetoed (\%) & Selected (\%) \\
       \hline
       $\nu$ detector & 940 (9.4) & 0 (0.0)& 54 (97.2)\\
       Vessel lids & 42 (0.4)& 0 (0.0)& 1 (1.5)\\
       Vessel walls & 8859 (88.2) & 37 (71.6)& 1 (1.2)\\
       Tracking system & 202 (2.0) & 14.7 (28.4)& 0 (0.0)\\
       %cave & 0 (0.0) & 0 (0.0)& 0 (0.0)\\
       \hline
       \textbf{Total} & \textbf{10044}  & \textbf{52} & \textbf{56} \\
       \hline
   \end{tabular}
\end{table}

% \begin{table}
%    \centering
%    \caption{Number of anti-neutrino interactions produced in the material of the experiment for $N_{pot} = 2 \cdot
%      10^{20}$ and with at least two charged tracks reconstructed in the HS spectrometer (first column), after applying 
%             the veto requirements (second column), and after applying the selection described in the text (third column).}
%             \label{tab:interaction2}
%   \begin{tabular}{lrrr}
%       \hline
%       Detector & Reconstructed (\%) & Not vetoed (\%) & Selected (\%) \\
%       \hline
%       $\nu$ detector & 39 (16.2) & 0 (0.0)& 2 (98.4)\\
%       vessel lids & 1 (0.3)& 0 (0.0)& 0 (1.6)\\
%       vessel walls & 197 (82.3) & 1 (83.0)& 0 (0.0)\\
%       tracking system & 3 (1.3) & 0.3 (17.0)& 0 (1.6)\\
%       cave & 0 (0.0) & 0 (0.0)& 0 (0.0)\\
%       \hline
%       \textbf{Total} & \textbf{239}  & \textbf{2} & \textbf{2} \\
%        \hline
%   \end{tabular}
% \end{table}

%\begin{figure}[htb]
%\includegraphics[width=0.49\textwidth]{./sensitivity/c_product_idChargedPart_fromNu_recoed.pdf}    
%\includegraphics[width=0.49\textwidth]{./sensitivity/c_product_idNeutrPart_fromNu_recoed.pdf}    
%\caption{Charged  (left) and neutral (right)  particles directly produced in the neutrino interactions corresponding to $N_{pot} = 2\cdot 10^{20}$.}
%\label{fig:nu_products}   
%\end{figure} 

The topology of the products of the neutrino interactions is such that relatively loose selection cuts
allow efficient rejection. In general the interaction products do not point to the
target, do not have a reconstructed vertex inside the decay volume, and have very poor track quality.
The requirement of having two high quality ($\chi^2$/nDoF$<2.5$ and nDoF$>10$) reconstructed tracks of 
opposite charge in the spectrometer, forming a vertex with a distance-of-closest approach DOCA$<$30~cm 
inside the decay volume and pointing to the target with an impact parameter $IP<2.5$~m, 
allows rejecting about 99.4\% of the reconstructed neutrino
background, 
as can be deduced from the total numbers 
in Table~\ref{tab:interaction1} (fourth column). The comparison of the distributions of the observables used in the offline 
selection for the HNL signal and the neutrino background is described in Section~\ref{sec:effectreco}.

Several background taggers and veto systems in the SHiP detector are able to detect the interaction products of the neutrino 
inelastic interactions and further reduce this background. The first veto detector is the active planes 
in the muon magnetic spectrometer of the tau neutrino detector, followed by the upstream veto tagger 
located in front of the vacuum vessel. The upstream veto tagger provides redundancy against  background 
initiated by interactions of neutrinos  in the material of the muon magnetic spectrometer.
Background induced by interactions in the entrance window of the vacuum vessel is instead tagged by 
the straw veto tagger located in vacuum $\sim$5~m downstream of the entrance window. 
%The SVT  has very  low material budget, at the expense of a slightly lower tagging efficiency.
Finally the liquid scintillator surrounding the vacuum vessel vetoes interactions taking place around 
the entire decay volume. These background taggers are conservatively assumed to have a detection efficiency of 90\%.

The background taggers and the veto systems are very effective in rejecting neutrino interactions. The requirement of having at least 
one background tagger or veto system with a positive response\footnote{This requirement means having at least one hit in one of 
the active layer of the tau neutrino detector or in the upstream veto tagger or straw veto tagger or a signal 
of at least 45 MeV in one of the cells of the surrounding liquid scintillator.} together with a loose requirement on the pointing of the interaction products to the target\footnote{The impact parameter of the reconstructed track pair with respect to the target is required to be less than 10~m.}, 
rejects about 99.5\% of this background. The third column (``Not Vetoed'') in Table~\ref{tab:interaction1} 
shows the number of the neutrino interactions in the material of the different parts of the detector which 
survives the veto requirements.  
%The charged and the neutral particles produced in the neutrino interactions 
%that remain after the veto and after the selection requirements are shown in Figure~\ref{fig:nu_charge_neu_after_veto_sel}.
%Figure~\ref{fig:nu_charge_neu_after_veto_sel} shows the type of charged and neutral particles remaining after the veto and the selection requirements.

% \begin{figure}[hbt]
%   \centering
%   \includegraphics[width=.45\textwidth]{./sensitivity/c_product_idChargedPart_fromNu_not-vetoed.pdf}
%   \includegraphics[width=.45\textwidth]{./sensitivity/c_product_idChargedPart_fromNu_selected.pdf}\\
%   \includegraphics[width=.45\textwidth]{./sensitivity/c_product_idNeutrPart_fromNu_not-vetoed.pdf}
%   \includegraphics[width=.45\textwidth]{./sensitivity/c_product_idNeutrPart_fromNu_selected.pdf}
%   \caption{Charged (top) and neutral (bottom) particles produced by the neutrino inelastic scattering 
%            corresponding to $N_{pot} = 2 \cdot 10^{20}$. Left plot shows the distribution of particle  
%            species after the application of the veto requirements and right plot after the selection 
%            requirements.}
% \label{fig:nu_charge_neu_after_veto_sel}
% \end{figure}

\begin{figure}[htb]
\includegraphics[width=0.49\textwidth]{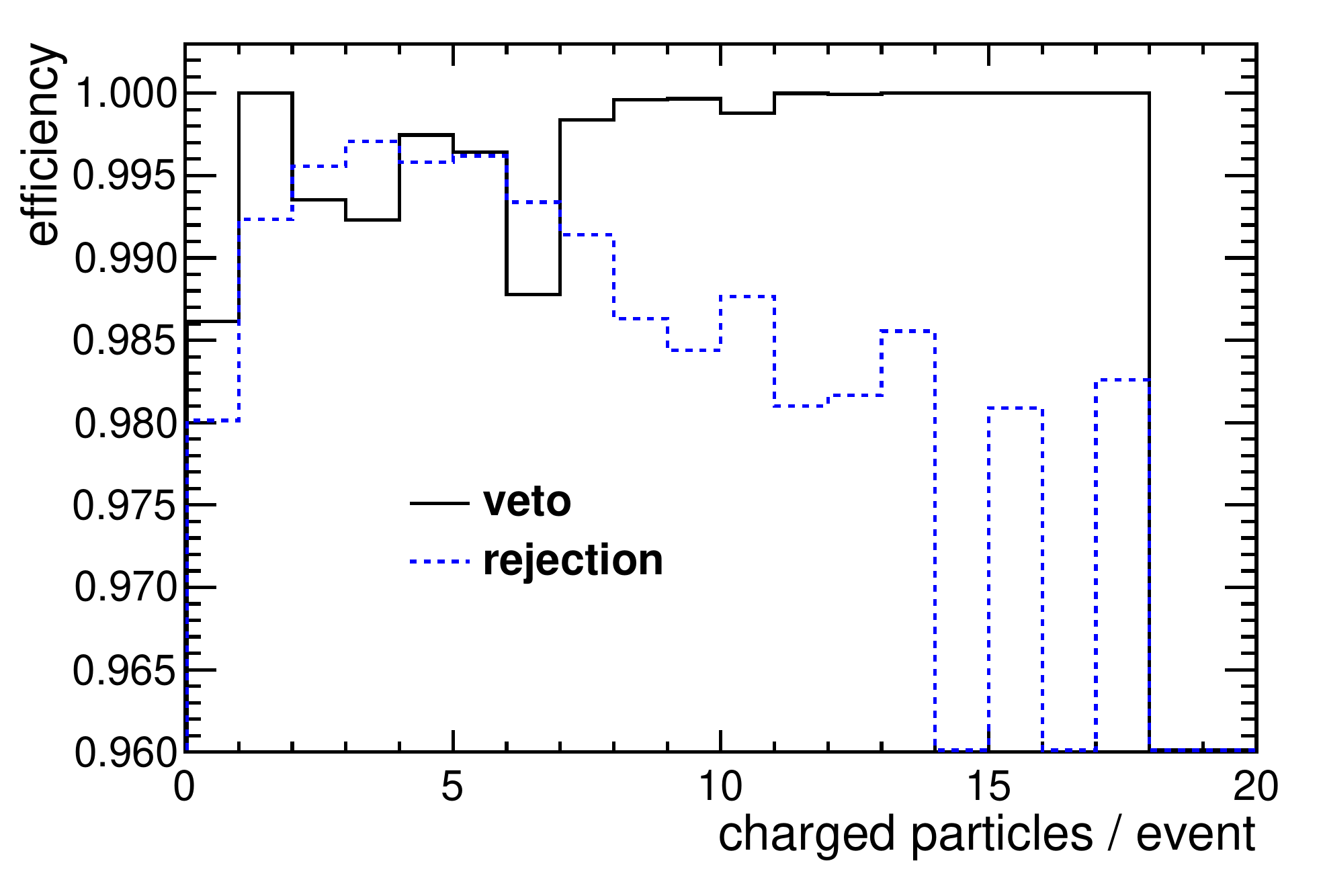}    
\includegraphics[width=0.49\textwidth]{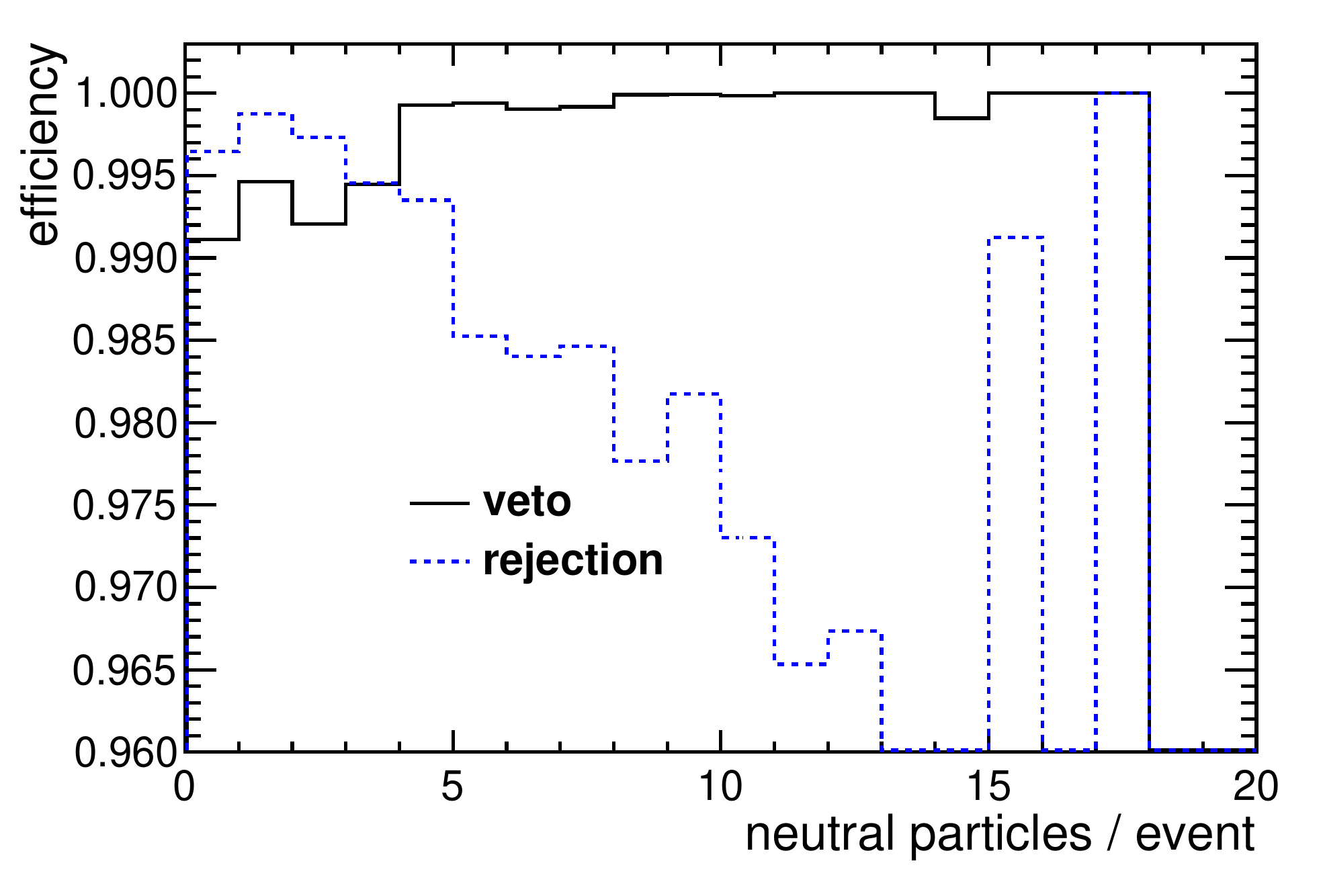}    
\caption{Rejection efficiency of the selection requirements and the efficiency of the veto 
requirements as a function of the charged and neutral multiplicities of the neutrino 
interactions.}
\label{fig:eff_topo_veto}   
\end{figure}

The rejection efficiency of the selection requirements and the efficiency of the veto requirements as 
a function of the charged and the neutral particle multiplicities in the neutrino interactions are shown 
in Figure~\ref{fig:eff_topo_veto}. 
As expected, the efficiency of the veto requirement increases with the number of particles in the event.
The rejection of the selection requirements instead is higher for a lower particle multiplicity. Hence, 
it complements the veto requirement. The combination of the selection and the veto requirements allows 
reducing the neutrino induced background to zero. The overall set of requirements is redundant and can 
be used for various cross checks.

% increasing the reliability of the entire procedure.

%% file: sensitivity/muonDIS.tex
\subsubsection{Background from muon inelastic scattering}
\label{sec:backgroundMuonDIS}

\begin{figure}[ht]
\centering
\includegraphics[width=0.6\textwidth,clip=]{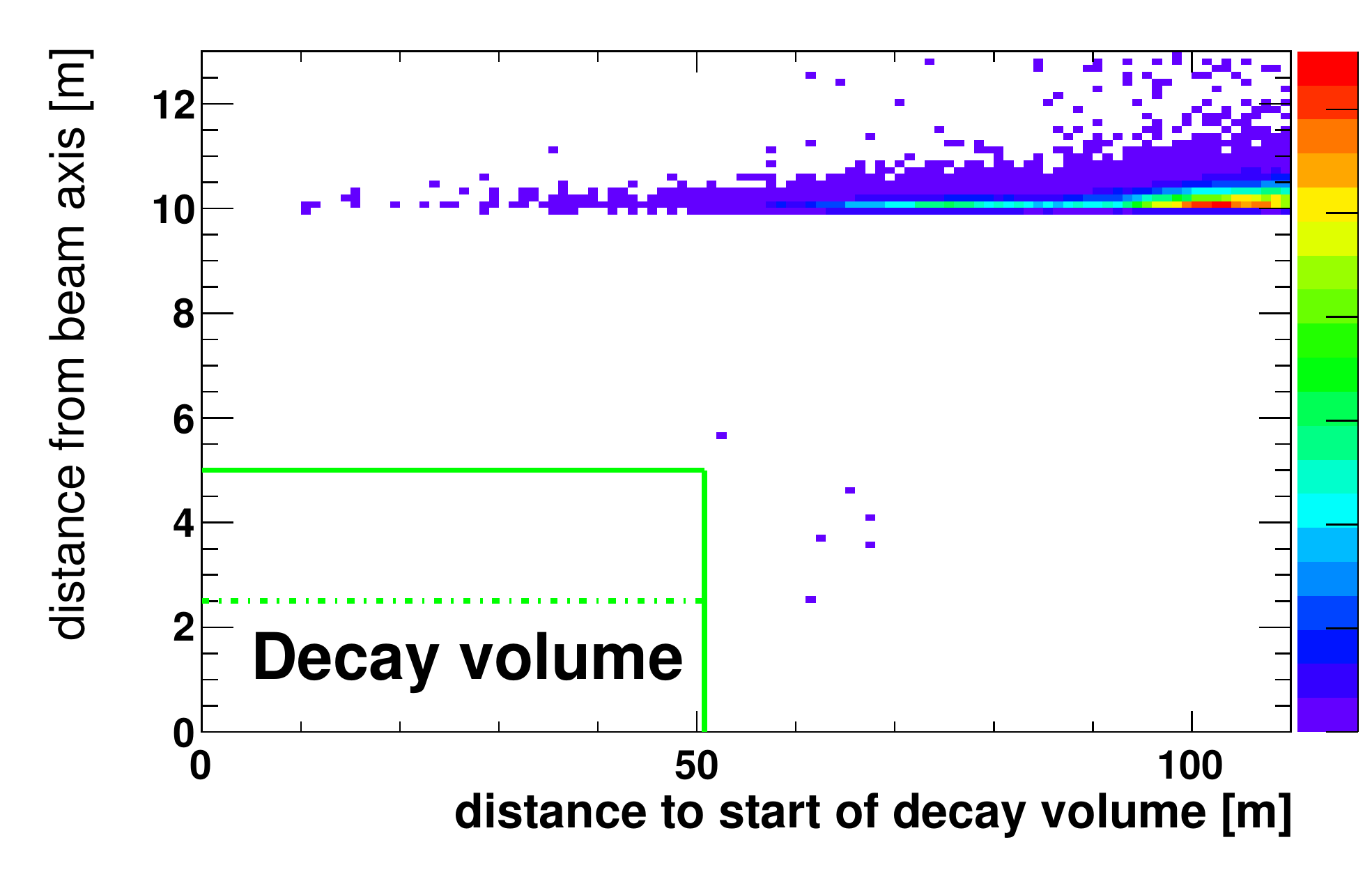}
\caption{Distribution of the muon interaction point in the concrete walls of the experimental hall as function of $\Delta z$, distance to the start of the decay volume and transverse distance to the beam axis.}
 \label{fig:muonDIS-muonConcreterhoZ}
\end{figure}

\begin{figure}[ht]
\centering
\includegraphics[width=0.95\textwidth,clip=]{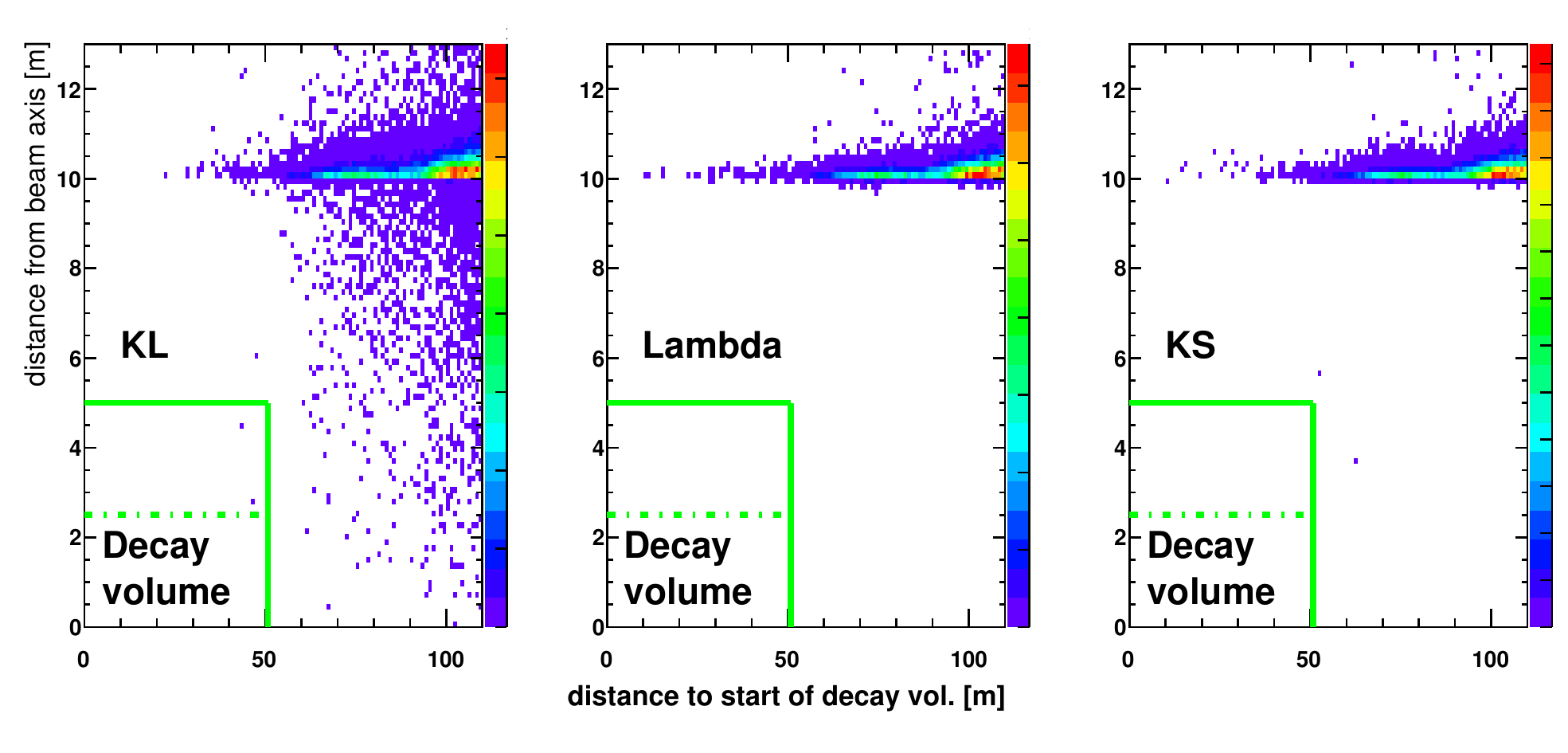}
\caption{Distribution of the V$^0$ end vertex as function of $\Delta z$ (left), distance to the start of the decay volume (middle), and transverse distance to the beam axis (right).}
 \label{fig:muonDIS-V0Decay_point}
\end{figure}

\begin{figure}[ht]
\centering
\includegraphics[width=0.7\textwidth,clip=]{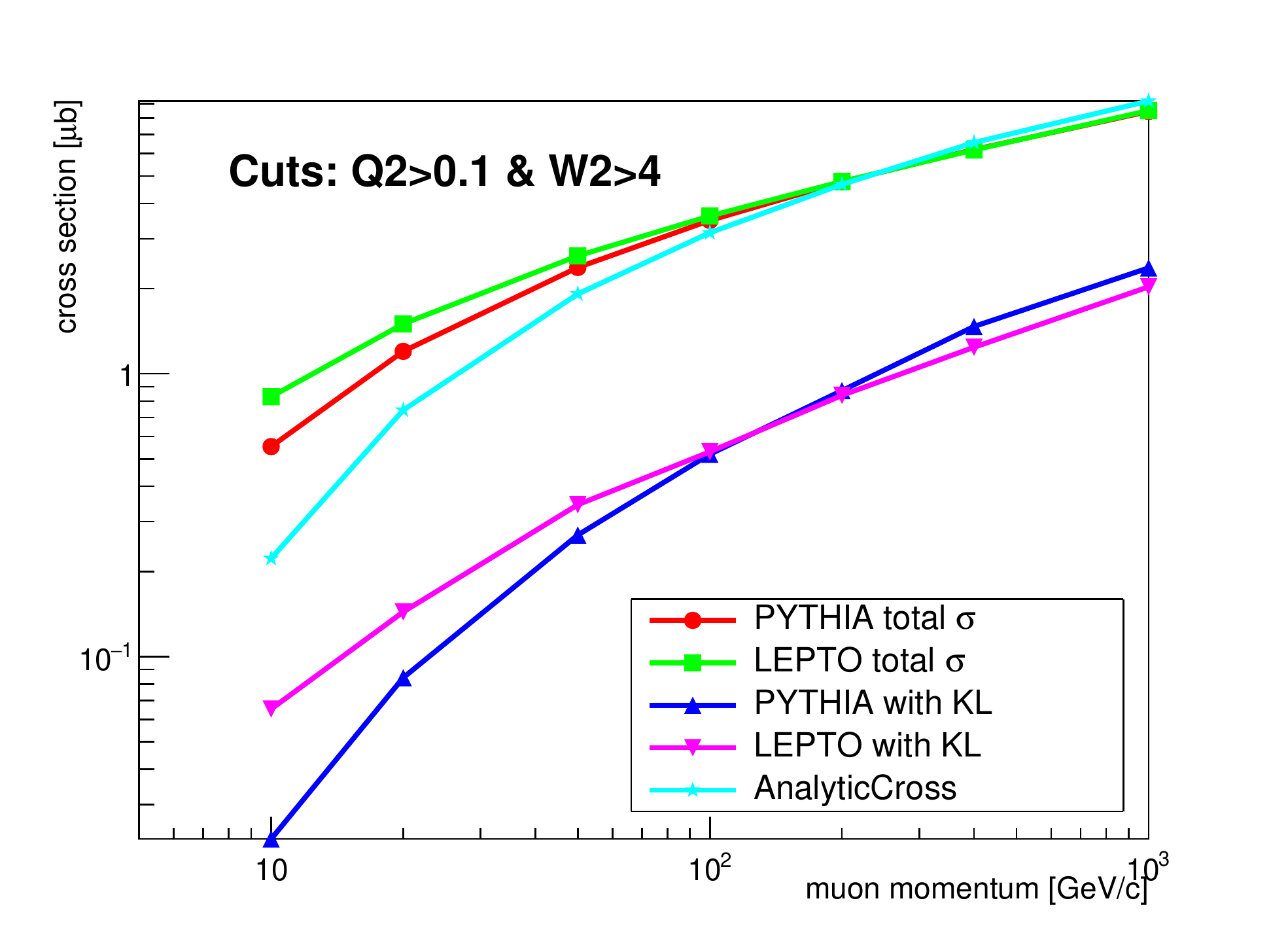}
\caption{Inelastic scattering cross section for muons on nucleons as funtion of muon momentum using an analytic calculation~\cite{muCrossAnalytic}, Pythia6~\cite{pythia6} and LEPTO~\cite{Ingelman:1996mq} generator with special settings. The fraction of the cross section with $\mathrm{K_L}$ in the final state is also shown.} \label{fig:muonDIS-cross}
\end{figure}

Essentially all muons will eventually reach the cavern wall. Due to the design of the active shielding, which matches the momentum of a muon with the necessary $\int$Bdl to miss the decay volume and the SHiP spectrometer, most of the muons hit the cavern wall with a shallow angle downstream of the decay volume (Figure~\ref{fig:muonDIS-muonConcreterhoZ}). V$^0$ particles ($K_L$, $K_S$, $\Lambda$) produced in muon inelastic scattering with nucleons of the concrete walls preferentially travel even further downstream or stop in the concrete, see Figure~\ref{fig:muonDIS-V0Decay_point}. Simulating such events by placing the muon interaction events simulated with Pythia 6~\cite{pythia6} at the place where the muons hit the concrete walls shows no induced background activity in the SHiP spectrometer. Folding the flux of muons with the cross section for inelastic collisions (Figure~\ref{fig:muonDIS-cross}) as function of the muon momentum, the simulated data set corresponds to about $2.5\cdot 10^{17}$ protons on target. Although this is still a factor $1000$ below the total statistics of the experiment, there are no signs that this is causing a serious background. The study will be continued when more details about the material distribution in the experimental hall is known.       

A second source for such background events are muons which are not sufficiently deflected and which hit material close to the entry of the decay volume. This background is similar to the one caused by neutrino inelastic interactions. The requirement for the design of the muon shield is to reduce the muon rate to a level that this background becomes similar to the irreducible background from neutrino interactions. Making the simple model, that only interactions in the last interaction lengths close to the decay volume produce V$^0$ particles which eventually decay inside, a rate of $5\cdot 10^3\,$muons with $E=100\,$GeV per spill would produce about as many muon interactions as neutrino interactions. Assuming a veto efficiency of $90\%$, a rate of $\sim 50\cdot 10^3\,$muons per spill can be safely tolerated.
%From EOI, neutrino CC(NC) rate of $600(200)10^3$ per interaction length per $2\cdot 10^{20}$ protons on target. 

\begin{figure}[ht]
\centering
\includegraphics[width=0.7\textwidth,clip=]{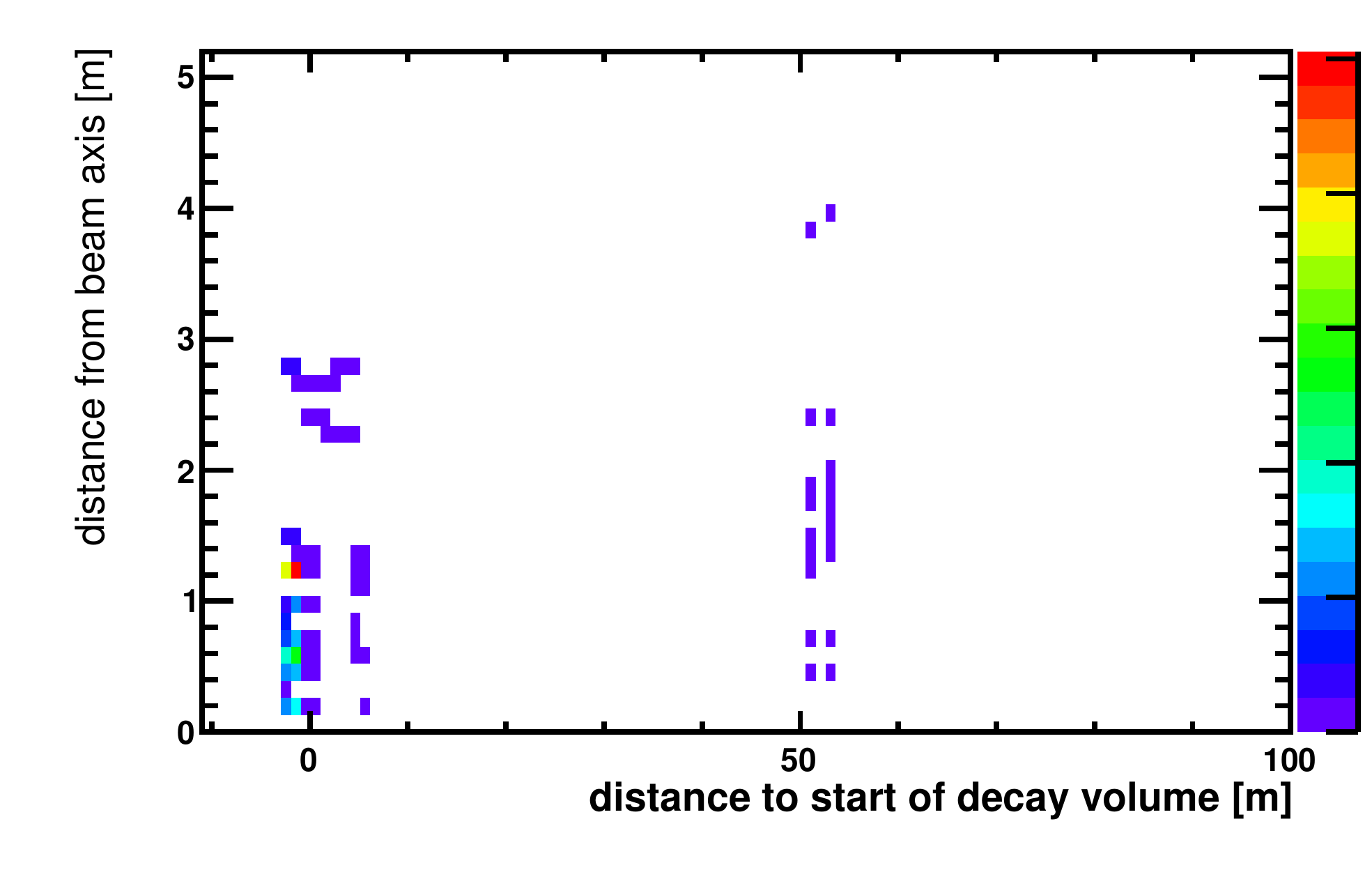}
\caption{Distribution of the muon interaction point as function of $\Delta z$, distance to the start of the decay volume and transverse distance to the beam axis.}
 \label{fig:muDIS_murhoz}
\end{figure}

\begin{figure}[ht]
\centering
\includegraphics[width=0.95\textwidth,clip=]{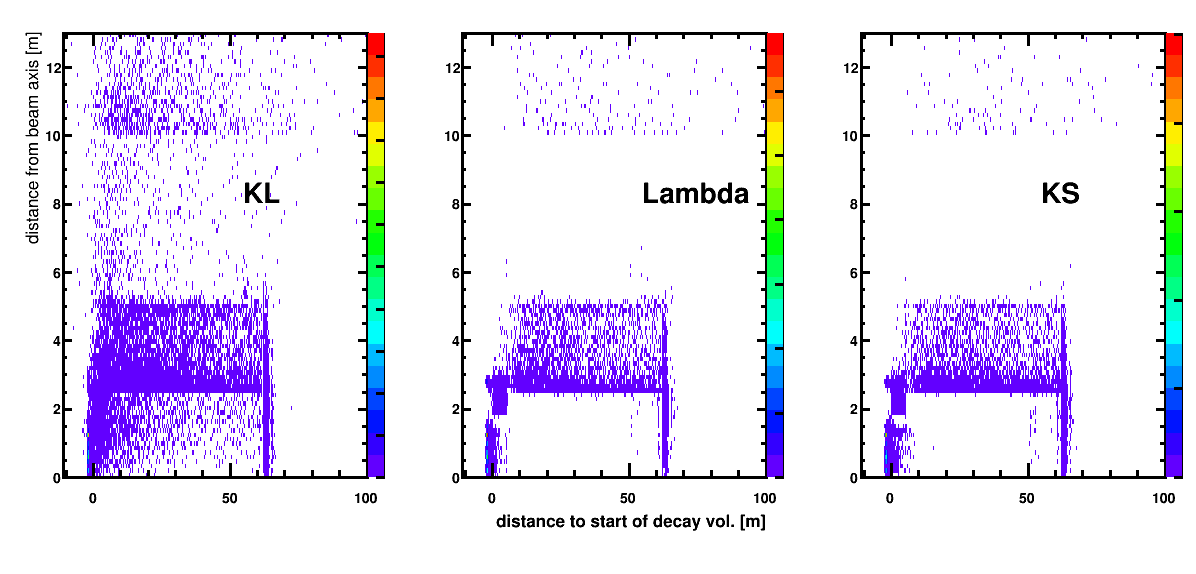}
\caption{Distribution of the V$^0$ end vertex as function of $\Delta z$, distance to the start of the decay volume and transverse distance to the beam axis.}
 \label{fig:V0DISDecay_point}
\end{figure}

\begin{figure}[ht]
\centering
\includegraphics[width=0.7\textwidth,clip=]{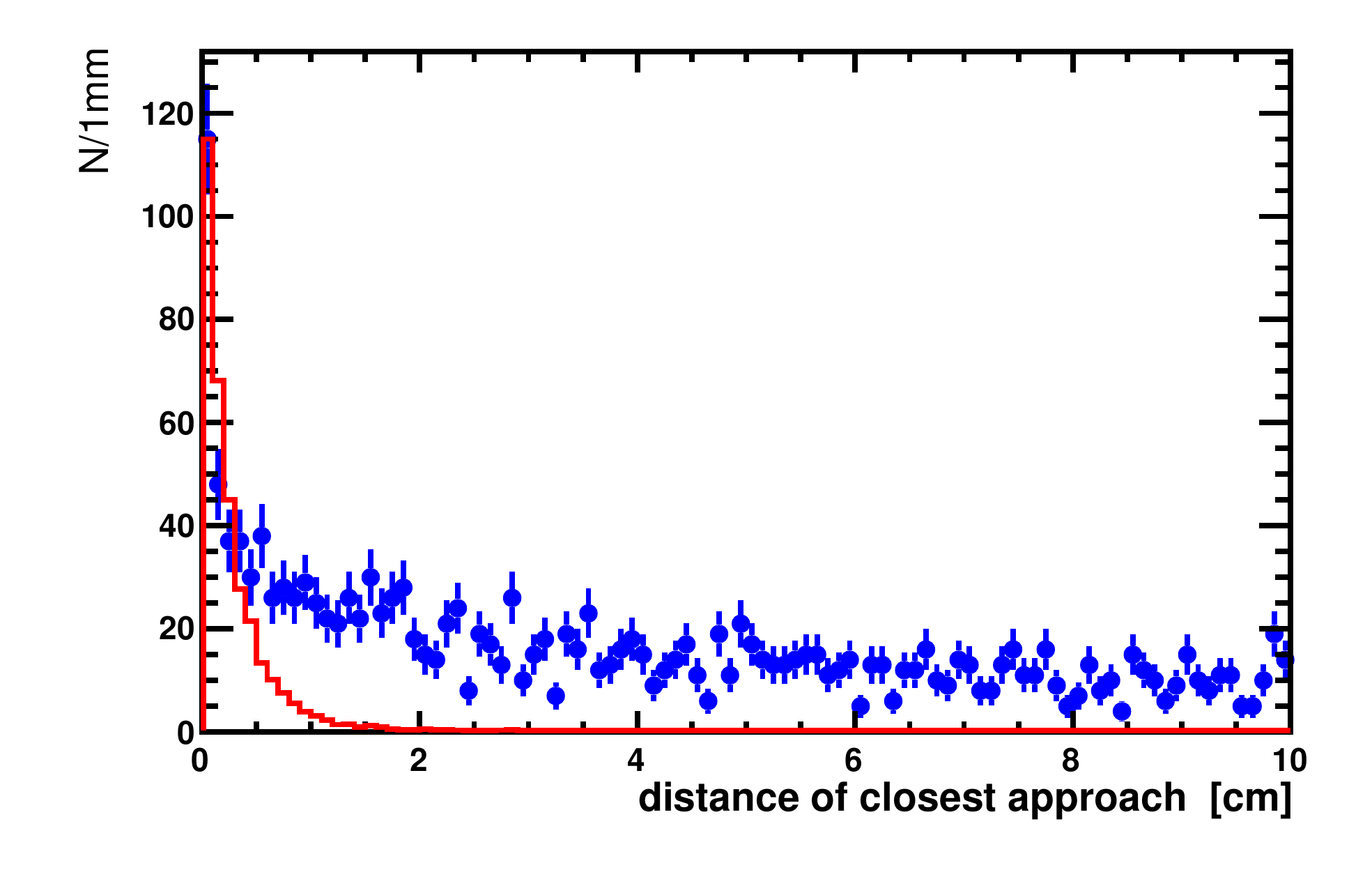}
\caption{Distance of closest approach of two tracks for signal (red) and background (blue).}
 \label{fig:muDIS_DOCAwithSignal}
\end{figure}

\begin{figure}[ht]
\centering
\includegraphics[width=0.7\textwidth,clip=]{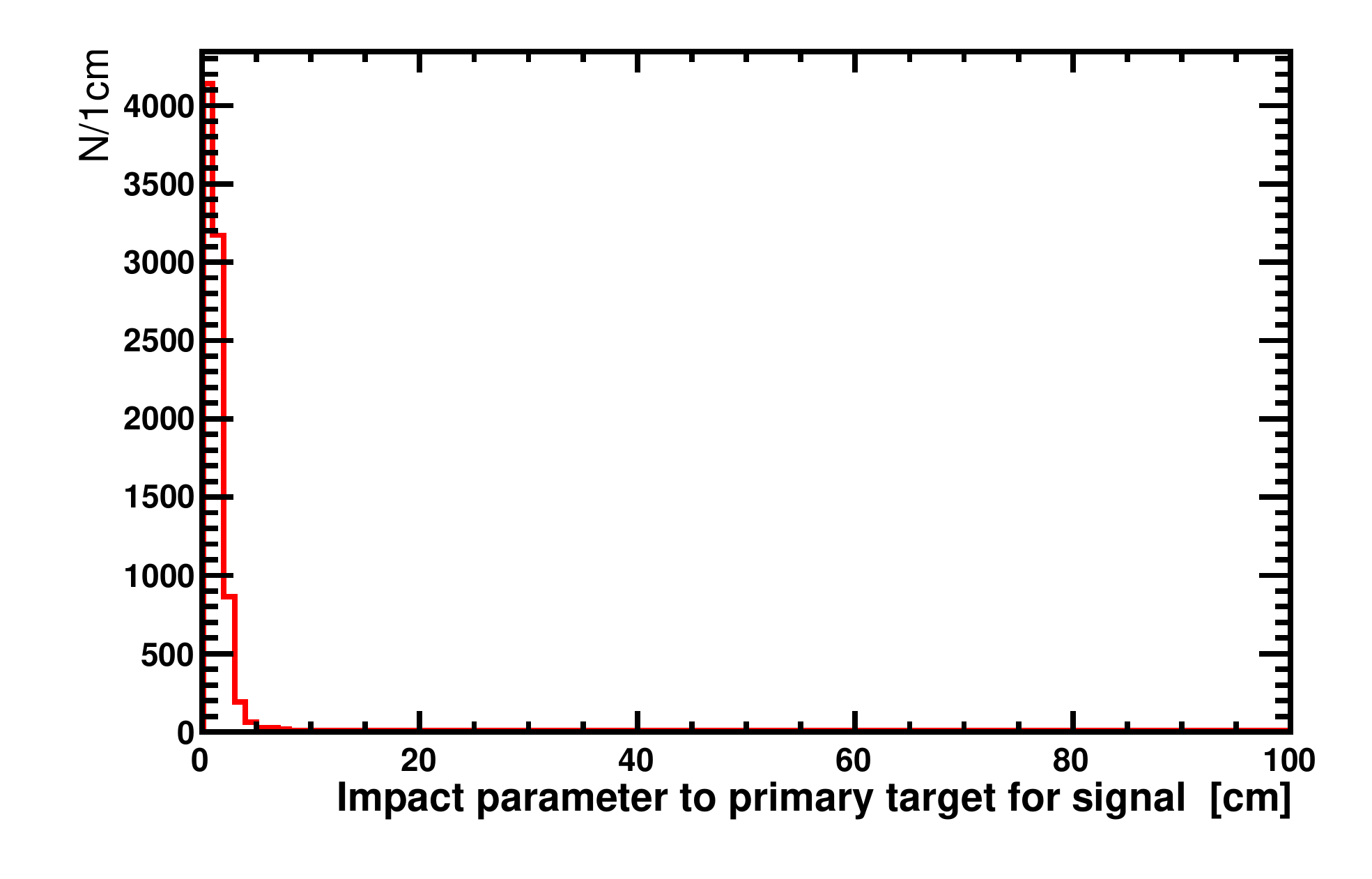}
\caption{Impact parameter with respect to the primary target for fully reconstructed signal (HNL) events.}
 \label{fig:muDIS_signalIP0}
\end{figure}

\begin{figure}[ht]
\centering
\includegraphics[width=0.7\textwidth,clip=]{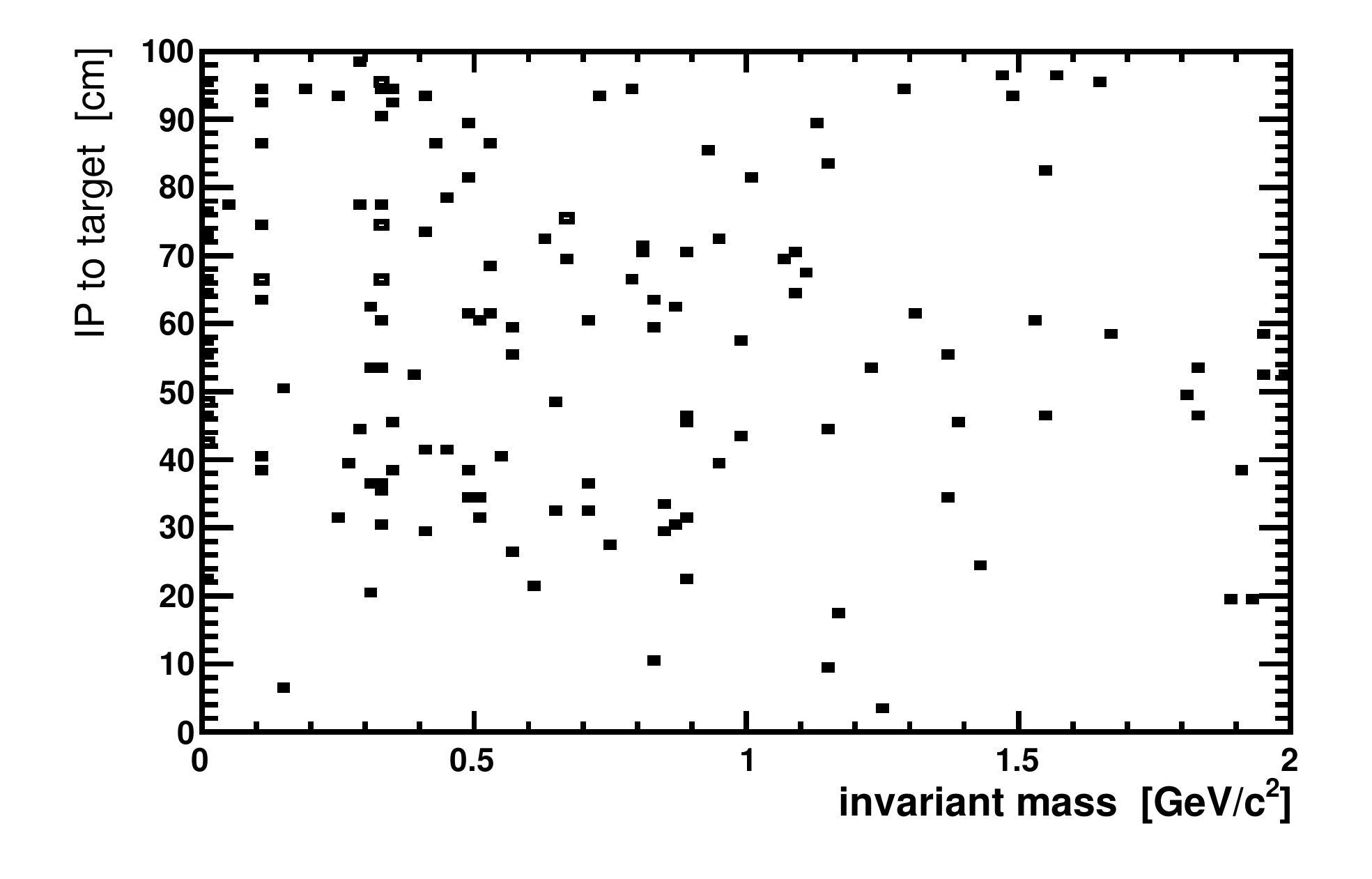}
\caption{Impact parameter of 2-body combinations of opposite charge with respect to the primary target as function of their invariant mass.}
 \label{fig:muDIS_invMassIP}
\end{figure}

\begin{figure}[ht]
\centering
\includegraphics[width=0.75\textwidth,clip=]{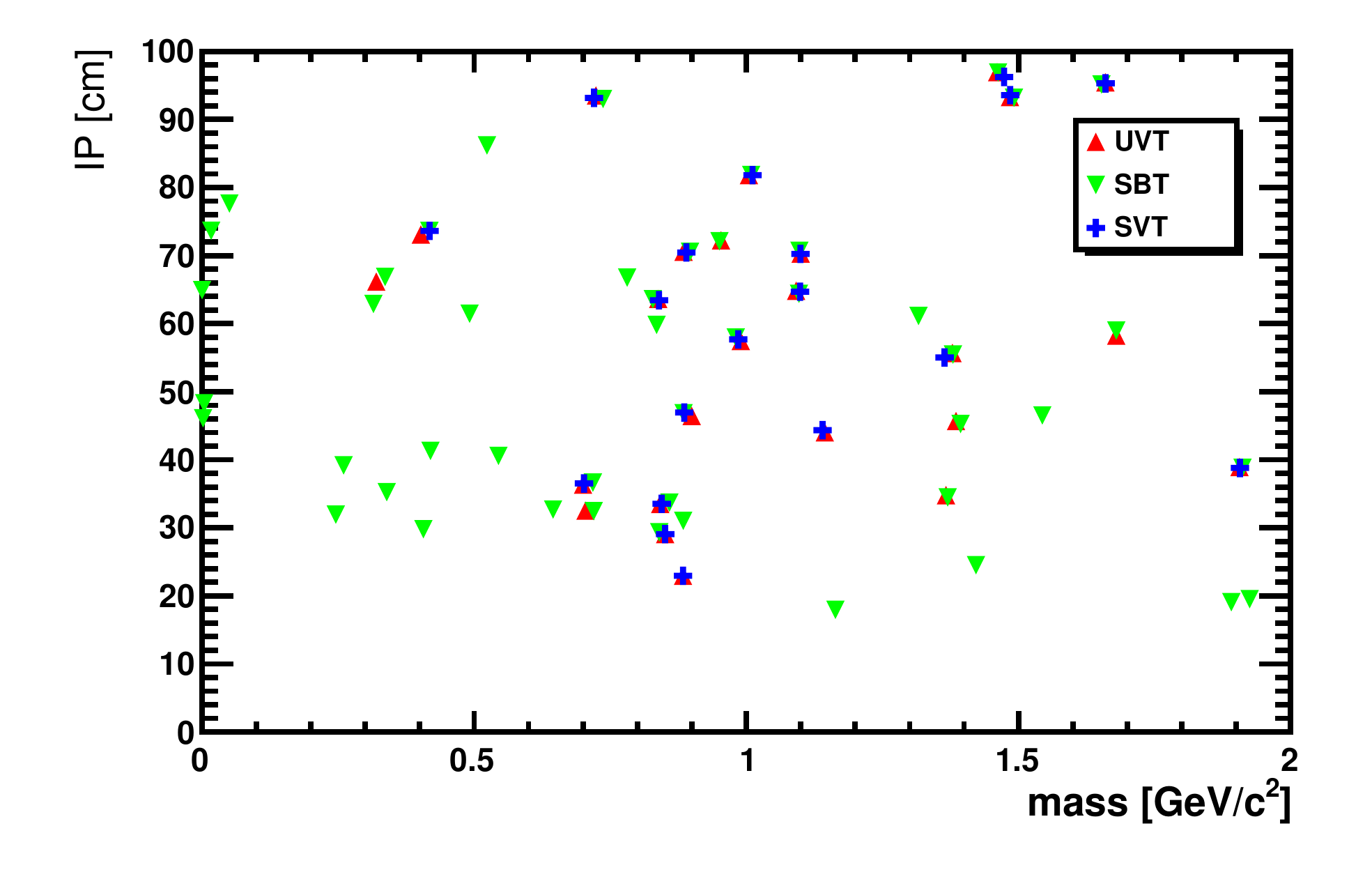}
\caption{Impact parameter of 2-body combinations of opposite charge with respect to the primary target as function of their invariant mass applying different veto requirements,
no signal in the Upstream Veto Tagger, no signal in the Surrounding Background Tagger or no signal in the Straw Veto Tagger.}
 \label{fig:muDIS_invMassIPVeto}
\end{figure}

From the full simulation of the residual muon background, we observe a rate of about $7000$ of fully reconstructed muons per spill inside the SHiP spectrometer with energies up to $200\,$ GeV. For each of these muons, we generate $10000$ muon interaction events with Pythia6, which we distribute along the muon flight path proportional to the material density seen by the muon. The products of these interactions are then further processed with the FairShip simulation respectively Geant4, followed by a track and vertex reconstruction. The distribution of the muon interaction as function of the distance to the entrance of the decay volume and the transerve distance to a virtual beam line is shown in Figure~\ref{fig:muDIS_murhoz}. Most of the interactions occur in the region of the tau neutrino detector. Figure~\ref{fig:V0DISDecay_point} shows the endpoint of the V$^0$ particles. As expected, mainly the $\mathrm{K_L}$ has decays inside the decay volume. Only 2-track combinations are retained which have a distance of closest approach of less than 1~cm. The mean for signal events is $0.4$~cm (Figure~\ref{fig:muDIS_DOCAwithSignal}). The invariant mass distribution of these candidates together with their impact parameter with respect to the primary target is shown in Figure~\ref{fig:muDIS_invMassIP}. While signal events have a mean impact parameter of $<3$~cm (Figure~\ref{fig:muDIS_signalIP0}), these events do not point back to the primary target. Requiring  $IP<10$~cm, we are left with 4 events. Additional background rejection is available by using the information of the Veto detectors, the Upstream Veto Tagger, the Straw Veto Tagger and the  
Surrounding Background Tagger. Requiring no count in any of these Veto detectors removes all of the remaining events (Figure~\ref{fig:muDIS_invMassIPVeto}). Given the limited amount of statistics we have simulated so far ($\mathcal{O}(3\cdot 10^{20}$)) protons on target, we can give an upper limit of $<2$ background events at 90\%~C.L. for $2\cdot 10^{20}\,$ protons on target.             


\clearpage

%% file: sensitivity/muonComb.tex
\subsubsection{Muon combinatorial background}
\label{sec:backgroundMuonComb}

Random combinations of muons which enter the vacuum vessel, either by back-scattering
in the surrounding cavern walls or due to the imperfection of the muon shield, may 
mimic signal events. The rate of fully reconstructed muons inside the SHiP spectrometer 
has been estimated with the full \textsc{FairShip} simulation to 7~kHz during the spill.
In order to study the expected combinatorial background the fully simulated residual muons
(${\cal O}(10)$), which enter the decay volume, are used to seed a toy MC to simulate the 
expected number of muons in the decay volume ($3.5\cdot 10^{10}$ muons, that is 
${\cal O}(10^{12})$ muon pairs) in the five years of data taking. The inherent systematic 
uncertainty of the toy MC generation requires further studies.

A basic set of kinematic and topological criteria, which are in line with the selection 
employed in the signal sensitivity studies in Section~\ref{sec:sensitivity}, allows reducing 
the number of fake dimuons. The cuts are summarized in Table~\ref{tab:bkgMuCombCuts}.

\begin{table}[htb]
\begin{center}
\label{tab:bkgMuCombCuts}
\caption{Effect of the loose offline pre-selection on the muon combinatorial background.}
\vspace{2mm}
\begin{tabular}{ll}\hline
Cut                                          & Value \\
\hline
Track $P$                                    & $>1.5~\gevc$   \\
Track $\chi^{2}/\mathrm{ndof}$               & $<25$           \\
dimuon DOCA                                  & $<1$~cm          \\
dimuon vertex                                & fiducial              \\
dimuon mass                                  &  $>0.2~\gevctwo$ \\
IP w.r.t target                              &  $<2.5$~m          \\
\hline
Efficiency                                    & $10^{-4}$ \\
\hline
\end{tabular}
\end{center}
\end{table}

The first requirement specifically designed to suppress combinatorial background is the 
use of the timing veto detectors described in Section~\ref{sec:timing_detectors}. 
Using a timing window of 340~ps, corresponding to three times the resolution 
of the timing detector, the number of dimuon vertices with the muons in the window 
integrated over the assumed running time of the experiment is ${\cal O}(10)$. 
Hence, the timing requirement contributes a factor $\sim 10^{-7}$ to the combinatorial background suppression.

The residual combinatorial background is further suppressed by
employing information from the upstream veto detectors, as discussed in 
Section~\ref{sec:upstream_vetotiming}, as well as the surrounding background 
tagger, as described in Section~\ref{sec:taggers}. This allows another reduction 
factor of $10^{-4}$.  Consequently, the muon combinatorial background is reduced 
to $<0.1$ events. Further studies are required in order to fully understand the sources of muons 
penetrating into the decay volume. 
%If 
%needed an additional suppression can also be obtained by requiring that both
%tracks leave signals in the first and the fourth straw tracker stations, as well 
%as in the calorimeters. 

%  crossings with more than two muons, over the lifetime of the experiment.

% As demonstrated the combination of the basic set of kinematic and topological criteria 
% in conjunction with the timing and spatial vetoes can bring this background under control.
% However, 

%% file: sensitivity/CosmicBkg.tex
\subsubsection{Cosmic muon background}
\label{sec:backgroundCosmics}

Muon background from cosmic rays is generated to reproduce the measured flux at sea level  
of $174~\unit{m^{-2}s^{-1}}$ ~\cite{Kempa:2006mr}.

%The momentum ($p$) distribution being given 
% by the following parametrizations and shown in Figure~\ref{fig:CosmicsFlux} was used to 
% simulate the cosmic muon background:
% \begin{eqnarray*}
% \frac{d^2\phi(p)}{dp\ d\Omega} [\unit{m^{-2}s^{-1}(GeV/c)^{-1}sr^{-1}}]&=& 22p^{-0.6-0.285\ln(p)} \text{ for $0.1\unit{GeV}<p<100\unit{GeV}$} \\
% \frac{d^2\phi(p)}{dp\ d\Omega} [\unit{m^{-2}s^{-1}(GeV/c)^{-1}sr^{-1}}]&=& 1400p^{-2.7}\left( \frac{1}{1+\frac{p}{115}}+\frac{0.054}{1+\frac{p}{850}}\right) \text{for $100 \unit{GeV} < p < 1 \unit{TeV}$}%~\cite{CMBG-formula} 
% \end{eqnarray*}

\begin{figure}[tb]
\begin{center}
\includegraphics[width=0.6\linewidth]{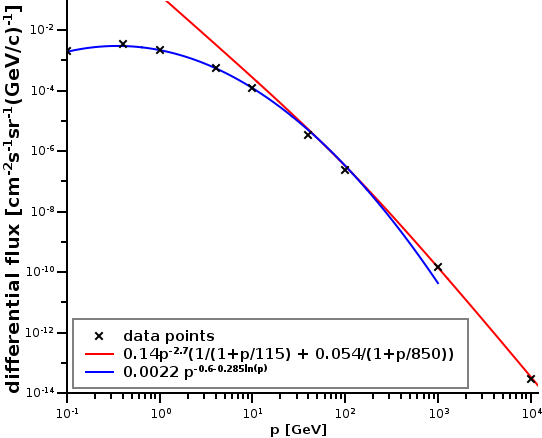}
\caption{Differential flux of cosmic muons as a function of the muon momentum
         as used in the simulation. The data points are taken from~\cite{Kempa:2006mr}.
         Muon momenta below 100~GeV/c are modelled by the parametrisation shown in blue.
         The parametrisation (in red) for momenta above 100~GeV/c is taken from~\cite{Agashe:2014kda}.}
\label{fig:CosmicsFlux}
\end{center}
\end{figure}

The ratio of antimuons over muons is assumed to be
\begin{eqnarray*}
\frac{N_{\mu^{+}}}{N_{\mu^{-}}} = 1.278 %~\cite{CMBG-formula}.
\end{eqnarray*}

% The angular distribution is taken as $\cos^2(\theta_{zenith})$%~\cite{CMBG-formula}. \\
% \noindent
% Muons from cosmic rays are being produced in an area much larger than the experimental hall. \\[1ex]

A total of $90 \cdot 10^{6}$ events, corresponding to 180 SPS spills of 1~s length, 
have been generated over a large area of $30 \times 90~{\rm m}^{2}$ above the 
experimental cavern following a flat momentum distribution. Each event is then weighted 
according to the momentum distribution of cosmic muons (see Fig.~\ref{fig:CosmicsFlux}) 
in such a way that the sea-level flux of $174~\unit{m^{-2}s^{-1}}$ is preserved.
From these simulated events we find that the rate in the surround background tagger is about 32 kHz.
The surround background tagger is considered as triggered if the energy deposition in the liquid 
scintillator volume in at least one segment is larger than 45 MeV, which is about the energy loss 
of a minimum ionizing particle (MIP) in a 30 cm thick layer of the liquid scintillator.
We consider a signal-like candidate coming from a cosmic muon event if a pair of opposite 
sign charged tracks is found. These tracks have to have hits in at least three tracking stations
with at least 25 hits and a $\chi^{2}/dof < 25$ per track. 
From the generated samples, $9300$ events with opposite sign charged tracks 
are expected in the five years of SHiP running time. The pairs of these reconstructed tracks are 
typically a muon and an electron or a positron or an electron and positron pair. In all of these events
the incoming muons have momenta above 100~GeV/c.
Requiring a distance of closest approach (DOCA) of the two tracks less than $10~\unit{cm}$ and requiring 
the two-track vertex to be inside the fiducial decay volume, the number of expected events is reduced to 
$612$ events. 
In all these events, it is found that sufficient energy is deposited in the liquid scintillator to be 
triggered (assuming a tagger efficiency of $100~\%$). In the majority of these events, two and more 
(up to 15) segments have triggered.
Considering a tagger-segment efficiency $\epsilon_{\rm seg}$ smaller than $100~\%$, the probability 
to veto such an event is given by $1 - (1-\epsilon_{\rm seg})^{n_{\rm seg}}$, where $n_{\rm seg}$ is 
the number of triggered background tagger segments in case of $100~\%$ efficiency. Assuming a probability 
of $\epsilon_{\rm seg}=99~\%$ the expected number of two-track events in five years of SHiP data taking, 
that is $5\cdot 10^6$ spills, is estimated to be $0.15$.  
All events initiated by muons from cosmic rays with a two-track signature show an impact parameter to the SHiP 
target $>33~\unit{m}$. Thus, it is straight-forward to exclude these events with the standard selection.

(Deep) Inelastic Scattering (DIS) events of cosmic muons are of particular concern.
In order to overcome the large gap between the number of simulated spills and the total number of spills 
during the five year runtime period we simulated a dedicated sample of DIS 
events with muons from the cosmic ray spectrum. The emphasis of these DIS studies was put on three 
specific parts of the experimental setup: 
\begin{itemize}
\item the upper experimental hall concrete walls
\item the material of the vacuum vessel, the liquid scintillator tank and the supporting ribs
\item the lower experimental hall concrete walls
\end{itemize}
The number of expected DIS interactions induced by muons from cosmic rays is estimated using
\begin{eqnarray*}
N_{DIS} = \frac{\int N_{\mu} \cdot \sigma_{DIS}(p) \cdot N_{spills} \cdot  \rho_{material} \cdot l\cdot \phi(p) dp d\Omega}{\int \phi(p) dpd\Omega} \text{,}
\end{eqnarray*}
where $N_{\mu}$ is the number of incident muons per spill in a place where subproducts of the 
DIS interaction are capable to reach the vacuum volume, $N_{spills}$ is the number of spills 
during the runtime of the experiment, $\rho_{material}$ is the mass density of the material in 
which the DIS interaction point takes place, $l$ is the length of material passed in the respective 
part where for the concrete walls we choose about a nuclear interaction length since for much
larger lengths V$^0$ particles would likely be absorbed before decaying into charged particles, 
$\sigma_{DIS}(p)$ gives the cross section for DIS of muons on protons taken from~\cite{Agashe:2014kda}, 
and $\phi(p)$ is the spectrum of the cosmic muons given above. 
Incident muons are simulated with momenta between 10~GeV/c and 1~TeV/c. 
With the aforementioned rate of background tagger activity initiated by muons from cosmic rays 
and five years of SHiP runtime the expected number of DIS interactions results in:

% and material densities and lengths (concrete: $\rho = 2.4~\unit{g\ cm^{-3}}$, $l = 30~\unit{cm}$; 
% iron: $\rho = 7.7~\unit{g\ cm^{-3}}$, $l = 3~\unit{cm}$; aluminium: $\rho = 2.7~\unit{g\ cm^{-3}}$, 
% $l = 0.8~\unit{cm}$) 

% To simplify the computation, 
% $\sigma_{DIS}(p)$ is parametrized as $[\log_{10}(p/\unit{GeV})]^2$, which is a conservative approach
% since it overestimates the cross section at smaller energies. 
% \noindent

\begin{itemize}
\item lower concrete wall: $N_{DIS}^{lower} \approx  10.1~\unit{M}$ events 
\item upper concrete wall: $N_{DIS}^{upper} \approx 10.1~\unit{M}$ events
\item vacuum vessel: $N_{DIS}^{VV} = N_{DIS}^{inner} + N_{DIS}^{outer} \approx 3.23~\unit{M} + 0.31~\unit{M} = 3.54~\unit{M}$ events
\end{itemize}
The DIS interactions are simulated with Pythia6 and the produced particles are given 
to GEANT4 and tracked through the detector simulation within FairShip. The results are:
% With the statistics simulated we estimate the potential dangerousness of DIS interactions 
% by cosmic muons. \\[2ex]
% So far, we have simulated 2.3~M simulated DIS interactions in the lower concrete wall,
% 400k DIS interactions inside the upper concrete wall of the cavern, and 830k DIS interactions 
% inside the vessel material. 

\begin{itemize}
\item For the 2.3~M simulated DIS interactions in the lower concrete wall of the cavern 
      we do not see any reconstructed two-track event. 
\item With 400k simulated DIS interactions inside the upper concrete wall of the cavern we 
      find $2.7$ reconstructed two-track events for the five years runtime. Applying the 
      cuts on DOCA and the vertex location as described above reduces this background 
      expectation to $0.3$ events.
      These remaining background events are characterized by a large number
      of background tagger segments hit (about 15) and is therefore expected to be vetoed with a 
      very high probability. In addition, most of these events show an impact parameter 
      to the target $>10$~m. 
\item From 830k simulated DIS interactions inside the vessel material we estimate the 
      number of reconstructed two-track events $0.07$. Applying the cuts on DOCA and 
      the vertex location as described above reduces this background expectation to $0.028$.
      Out of those there are events (expected $0.001$) that do not trigger the background 
      tagger. Here again, all reconstructed vertices show an impact parameter to the target 
      $>10$~m providing an additional handle to exclude them as potential signal events.
\end{itemize}

%\begin{thebibliography}{99}
%\bibitem{CMBG-spectrum} J.~Kempa, {\it The muon energy spectra at various geomagnetic latitudes}, 29th International Cosmic Ray Conference,  V6, P57-60, Pune, India, 2005
%\bibitem{CMBG-formula}  K.~A.~Olive et al. (Particle Data Group), Chin. Phys. C, 38,  090001 (2014). , Chapter 28 on Cosmic Rays
%\end{thebibliography}

%% file: sensitivity/backgroundSummary.tex
\subsubsection{Summary of background evaluation}

The background sources have been studied with the full simulation. Given the large number of protons 
on target it is challenging to produce sufficient statistics for all background sources. As described in 
the previous sections, different strategies are adopted to boost the statistics for the different 
background sources. 

\begin{itemize}
\item Neutrino background: force all neutrinos to interact in the detector
\item Muon DIS: force all muons to interact in the material 
\item Muon combinatorial: we assume factorization of the timing veto, the spatial veto and the rejection efficiency
\item Cosmics muon background: we assume factorization between veto efficiency and rejection efficiency
\end{itemize}

\begin{table}[htb]
   \centering
   \caption{Expected number of background events extrapolated from the
     Monte Carlo simulation and re-weighted to correspond to 5 years
     of running of the experiment. For all background sources with
     sufficient large statistics we expect $<0.1$ background events,
     while for some region of the phase space we are limited by the
     available statistics ( "0 MC" indicates zero background events
     observed in the MC). The statistical factor is the ratio between
     the equivalent Monte Carlo statistics and the total expected
     number of events. When the statistical factor is $\leq 1$, a
     90\%C.L. upper limit on the background can be obtained dividing
     2.3 by the statistical factor. With our statistical power no
     evidence of any irreducible background is found.}
            \label{tab:SummaryBkg}
\vspace{2mm}
   \begin{tabular}{lrr}
       \hline
       Background source & Statistical factor  & Expected background \\
       \hline
       $\nu$ ($p>10.0$~GeV/c) & $35.$ &  $<0.07$ \\
       $\nu$ ($4.0$~GeV/c$<p<10.0$~GeV/c) & $\sim 1$& $0$ (MC)\\
       $\nu$ ($2.0$~GeV/c$<p<4.0$~GeV/c) & $0.07$ & $0$ (MC)\\
       $\mu$ DIS HS & $\sim 1$ & $0$ (MC)\\
       $\mu$ DIS wall&  $0.001$ & $0$ (MC) \\
       $\mu$ Combinatorial & $10^{4}$ & $<0.1$ \\
       $\mu$ Cosmics ($p<100$~GeV/c) & $0.2$ & $0$ (MC)\\
       $\mu$ Cosmics ($p>100$~GeV/c) & $800.$ & $<0.1$ \\
       $\mu$ Cosmics DIS ($p>100$~GeV/c) & $10^3$ & $<0.1$ \\
       $\mu$ Cosmics DIS ($10$~GeV/c$<p<100$~GeV/c) & $\sim 1$ & $0$ (MC)\\
       \hline
   \end{tabular}
\end{table}

Table~\ref{tab:SummaryBkg} shows the expected number of background
events for each source estimated with the Monte Carlo simulation and
scaled to five years of running of the experiment.
The statistical factor is the ratio between the statistics generated for that particular background and the number 
of expected background events in five years of running. When the
statistical factor is $\leq 1$ the 90\% ~C.L. upper limit on the
number of background events can be calculated by dividing 2.3 by the
statistical factor.
No systematic 
uncertainty is assigned due to the assumptions made in the procedure to boost the statistics. This will be studied
further.

The neutrino background is divided in three regions of neutrino momentum which have different statistical 
significances. In the region at low momentum, between $2.0$~GeV/c~$<p<4.0$~GeV/c, 14 times less statistics was generated 
than what is expected in the experiment. This is due to the exponential drop of the neutrino energy spectrum,
which is only partially compensated by the fact that the cross section for the neutrino DIS rises linearly with the 
energy. Assuming the same rejection power for the events at low momentum as has been determined for events at 
high momentum, this background is reduced to a negligible
level. 

Similarly, the muon cosmics background is generated with different
statistical significances in two different momentum regions: below $100$~GeV/c and
above $100$~GeV/c. The lack of statistics below for cosmic muons with a
momentum $100$~GeV/c is reflected into an upper limit of about $12$
events. 
%It should be noticed that, given that all surviving cosmic
%muon events have large momentum and can be easily rejected using
%simple selection criteria, we do not expect a significant background
%from this source. 
However, the rate of cosmic muons is comparable to the rate of muons, from the proton target, penetrating the decay 
volume. Following the analysis of the muon combinatorial background and the background from muon deep inelastic
scattering, it is evident that the background from cosmic is also negligible.

The background from muon inelastic interactions in the cavern walls is also lacking statistics. This background has 
been generated with a statistics of about 1000 times less than what is expected in the five years of running. 
No background candidates are found with the \textsc{FairShip} full simulation\footnote{The background for DIS HS and DIS wall 
have currently only been estimated for fully reconstructed final states. 
Background evaluation for partially reconstructed final states is ongoing.}. 
In addition, several additional
requirements can be applied to this background, which are likely to reduce it to a negligible level. Dedicated 
studies to boost the statistics are ongoing at the moment. For the muon
inelastic scattering coming from the front and interacting with the HS
spectrometer, we generated a statistics about equivalent to the
expected in five years of running. Assuming the same veto capabilities
as for neutrino background, we expect for this background $<0.1$ events. 

In summary, we have identified several sources of backgrounds and studied with the full simulation. We have no evidence that any of these background will have a significant
impact on the experiment, and we do not have any evidence for any
source of irreducible background. 
For some of the background sources we have currently limited statistics available. 
Studies with larger statistics are ongoing, but extrapolating from the high statistics samples we do not expect any major background.
 In the following we assume a level of background of
$0.1$ events for the entire run of the experiment.

%% file: sensitivity/Sensitivity.tex
\subsection{Estimation of signal yields}
\label{sec:sensitivity}

The same procedure for estimating the signal yield, as describe below for the HNLs, has been applied to all 
the representative physics signals. In the case of the HNLs, the yield depends on the hierarchy of the 
active neutrino masses and on the relative strength of the HNL couplings to the three SM flavours 
$U^2_e, U^2_\mu, U^2_\tau$. Five scenarios which conform 
to existing theoretical studies are considered\cite{Gorbunov:2007ak,Canetti:2010aw}:

\begin{enumerate}[I.]
\item $U_e^2:U_{\mu}^2:U_{\tau}^2\sim 52:1:1$, inverted hierarchy of
  active neutrino masses\cite{Gorbunov:2007ak} % Gorbunov, Shap 2007
\item $U_e^2:U_{\mu}^2:U_{\tau}^2\sim 1:16:3.8$, normal hierarchy of
  active neutrino masses\cite{Gorbunov:2007ak} % Gorbunov, Shap 2007
\item $U_e^2:U_{\mu}^2:U_{\tau}^2\sim 0.061:1:4.3$, normal hierarchy of
  active neutrino masses\cite{Gorbunov:2007ak} % Gorbunov, Shap 2007
\item $U_e^2:U_{\mu}^2:U_{\tau}^2\sim 48:1:1$, inverted hierarchy of
  active neutrino masses\cite{Canetti:2010aw} % Canetti, Shap 2010
\item $U_e^2:U_{\mu}^2:U_{\tau}^2\sim 1:11:11$, normal hierarchy of
  active neutrino masses\cite{Canetti:2010aw} % Canetti, Shap 2010
\end{enumerate}

%The sensitivity limits obtained for the scenarios above are shown and discussed in %Section~\ref{sec:HNLsplots}.

Scenario II with a total coupling to the SM of $U^2=9.3\cdot 10^{-9}$ and a 
HNL mass of 1~GeV/c$^2$ was chosen as a \emph{benchmark} scenario to investigate SHiP's acceptance in detail. 

The number of HNLs that are detectable at SHiP depends on the HNL production rate 
and the detector acceptance. It is given by:
\begin{equation}\label{equ:nHNL}
n(H\!N\!L) = N(\text{p.o.t})\times \chi(pp\rightarrow H\!N\!L) \times {\cal P}_{\text{vtx}}  \times {\cal A_{\rm tot}}(H\!N\!L\rightarrow \text{visible}) 
\end{equation} 
where
\begin{itemize}
\item $N(\text{p.o.t})=2\cdot 10^{20}$ is the number of protons on target expected in five years of SHiP operation at nominal conditions. 
\item $\chi(pp\rightarrow H\!N\!L)$ is the total production rate of HNLs per proton interaction. It is equal to:
\begin{align}
\chi(pp\rightarrow H\!N\!L) = \;& 2 \times \left[ \chi(pp\rightarrow c\bar{c}) \times {\cal BR}(c\rightarrow H\!N\!L)\right.\notag\\
 &\left. +\, \chi(pp\rightarrow b\bar{b}) \times {\cal BR}(b\rightarrow H\!N\!L) \right] \times U^2
\end{align}
where $\chi(pp\rightarrow c\bar{c})=1.7\cdot 10^{-3}$ and $\chi(pp\rightarrow b\bar{b})=1.6\cdot 10^{-7}$ are the production fractions of charm and beauty mesons for a 400~GeV proton beam colliding on a molybdenum target. HNLs are mainly produced in $D_{(s)}$ meson decays, but $B_{(s)}$ meson decays also contribute and they are the only source of HNLs for masses above 2 GeV/c$^2$. The fractions of heavy meson decays into HNLs ${\cal BR}(c\to H\!N\!L)$ and $ {\cal BR}(b\to H\!N\!L)$, parametrized according to \cite{Gorbunov:2007ak}, take into account all of the dominant decay channels of $D_{(s)}$ and $B_{(s)}$ mesons into HNLs which are kinematically allowed:
		\begin{align}
			&D\to K\ell +H\!N\!L\notag\\
			&D_s\to \ell +H\!N\!L\notag\\
			&D_s\to\tau \nu_\tau \text{ followed by } \tau\to\ell\nu +H\!N\!L\text{ or }\tau\to\pi +H\!N\!L\notag\\
			&B\to\ell +H\!N\!L\notag\\
			&B\to D\ell +H\!N\!L\notag\\
			&B_s\to D_s \ell +H\!N\!L\label{eqn:prodchans}
		\end{align}

The widths of these channels are parametrized according to Ref.~\cite{Gorbunov:2007ak} 
as a function of the HNL mass and couplings\footnote{Other decays with
smaller branching ratios are not included. Further studies to include
all other $b\to u$ transitions is ongoing.}. The factor two is added to take into account 
the fact that each of the quarks in the pair can hadronize individually and can result in 
the production of a HNL. $U^2 = U_{e}^2 + U_{\mu}^2 + U_{\tau}^2$ represents the total coupling between 
the HNLs and the SM flavours. It is a free parameter, but the relative ratios between the 
different lepton flavours are fixed for every scenario I - V. 
		
\item ${\cal P}_{\rm vtx}$ is the probability that the decay vertex of a HNL of a given mass and couplings is 
located within the fiducial volume of SHiP. 
The HNL lifetime is estimated as the sum of the widths of its main decay channels such as
	$H\!N\!L \to 3\nu,\, \pi^0\nu,\, \pi^\pm\ell,\, \rho^0\nu,\, \rho^\pm\ell,\, \ell^+ \ell^-\nu$.
The branching ratios for these channels are parametrized according to Ref.~\cite{Gorbunov:2007ak} and shown 
in Figure~\ref{fig:BRHNL} as a function of the HNL mass. The formulas in Ref.~\cite{Gorbunov:2007ak} 
are valid up to $m_{H\!N\!L}\sim 1$~GeV/c$^2$. If the HNL mass is much larger than the QCD scale, 
$m_{H\!N\!L} \gg \lambda_{QCD}$, the two quarks from $H\!N\!L \to q\bar{q}\nu$ decays tend to hadronize individually. 
For masses in the region of 1 - 5 GeV/c$^2$, the inclusive $HNL \to q\bar{q}\nu$ decay width is 
extrapolated from the parametrisation of $H\!N\!L \to \ell^+ \ell^-\nu$ with appropriate corrections.

 \begin{figure}
 \centering
\includegraphics[width=13cm]{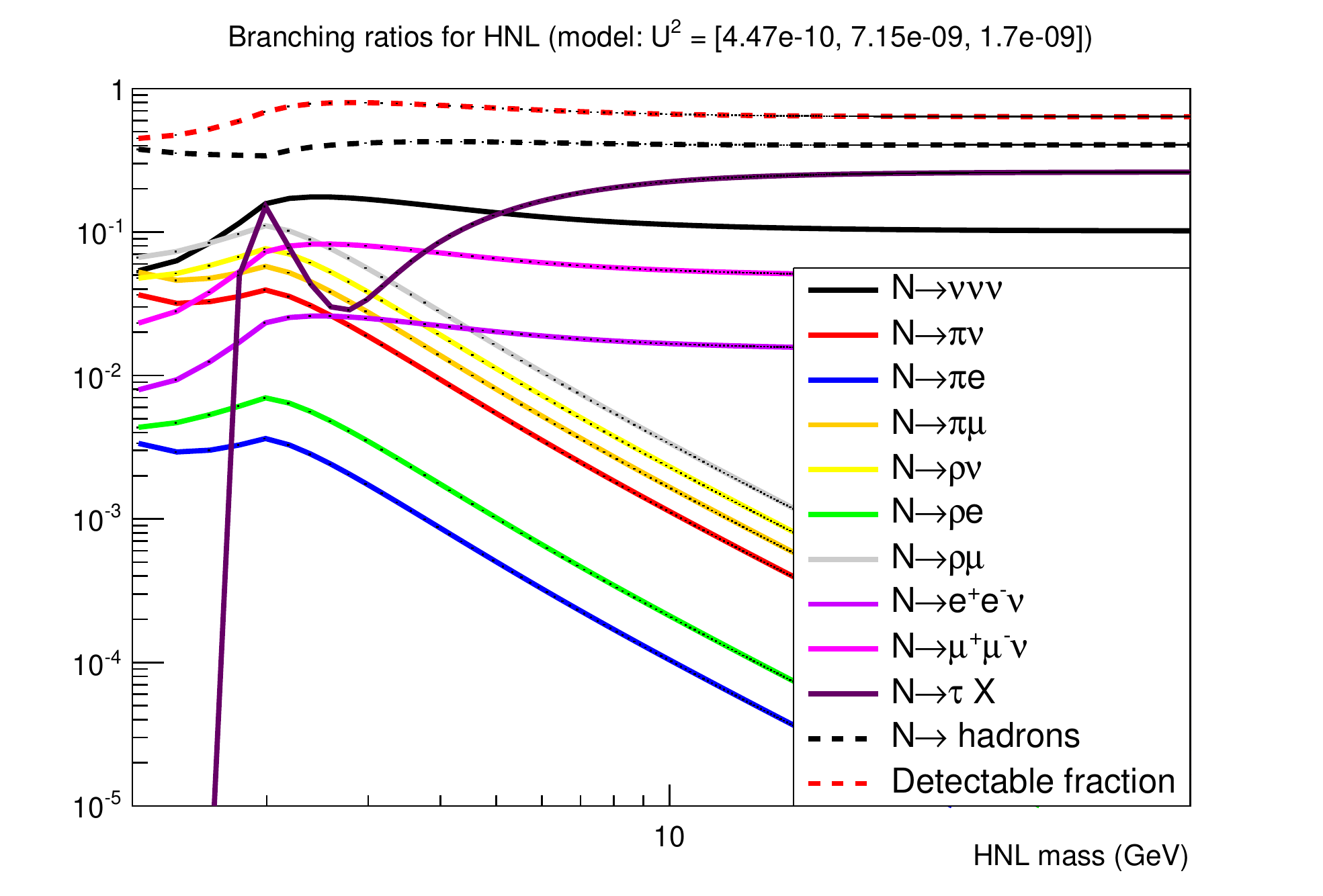}
 \caption{\label{fig:BRHNL} HNL branching ratios as a function of its mass for the benchmark scenario.}
 \end{figure}
\item  ${\cal A_{\rm tot}}(H\!N\!L\rightarrow \text{visible})$ is the detector acceptance of all the visible final states, 
i.e. the fraction of HNLs decaying in the fiducial volume of SHiP that results in a detectable final state. It is equal to: 
\begin{equation}\label{eqn:hnlacc}
{\cal A_{\rm tot}}(H\!N\!L \rightarrow \text{visible})=\sum_{i=\text{visible channel}} {\cal BR}(H\!N\!L \rightarrow i) \times {\cal A}(i)
\end{equation}
where the index $i$ runs over the final states with two charged particles. The estimation of the geometrical 
acceptance ${\cal A}(i)$ is explained in detail in Section~\ref{sec:detailedA}.

\end{itemize}

%%%%%%%%%%%%%%%%%%%%%%%%%%%%%%%%%%%%%%%%%%%%%%%%%%%%%%%%%%%%%%%%%%%%%%%%%%%%%
%%%%%%%%%%%%%%%%%%%%%%%%%%%%%%%%%%%%%%%%%%%%%%%%%%%%%%%%%%%%%%%%%%%%%%%%%%%%%
%%%%%%%%%%%%%%%%%%%%%%%%%%%%%%%%%%%%%%%%%%%%%%%%%%%%%%%%%%%%%%%%%%%%%%%%%%%%%

%%%%%%%%%%%%%%%%%%%%%%%%%%%%%%%%%%%%%%%%%%%%%%%%%%%%%%%%%%%%%%%%%%%%%%%%%%%%%

\subsubsection{Detector acceptance}\label{sec:detailedA}

The visible fraction of HNL $\mathcal{A}_{tot}(H\!N\!L \to \text{visible})$ 
is a function of the branching ratio and of the final state acceptance for the visible HNL decay channels 
(see Equation~\ref{eqn:hnlacc}). All the decay channels producing two charged particles in the final state 
are considered detectable. The decay chain $H\!N\!L\to\rho^0\nu$ followed by 
$\rho^0\to \pi^+\pi^-$ is also included. The final states with $\pi^0$'s are conservatively marked 
as not reconstructable currently and will be addressed later.

The final state acceptance is computed as
\begin{equation}
\mathcal{A}(i) = \frac{\text{\# reconstructable}}{\text{\# simulated}}
\end{equation}

Simulated HNLs are marked as reconstructable if they satisfy the following requirements: 
\begin{itemize}
\item ``Normalization'': in order to match the toy Monte Carlo, only HNLs with decay vertex between the entrance and the exit lid of the vacuum vessel.
\item ``Fiducial'' : position of the HNL decay vertex is within the fiducial volume as defined in Section~\ref{sec:sensitivity_intro}
% \item ``$z_{\text{straw veto}} < z_{\rm vtx}<z_{\rm out}$'':  the decay vertex is located downstream of the straw veto tagger.
% \item ``vertex in ellipse'': the $x,y$ position of the HNL decay vertex is inside the elliptical fiducial volume ($r_x = 250$~cm, $r_y = 500$~cm).
\item ``Straw 1, 2, 4'': both HNL daughters leave signals in both straw stations before the magnet (1 and 2) and in station 4 after the magnet. 
% These hits are within the elliptical fiducial volume ($r_x = 250$~cm, $r_y = 500$~cm).
\item ``ECAL" : some energy is deposited in the ECAL (only for $H\!N\!L \rightarrow \mu\pi$ and $H\!N\!L \rightarrow ee\nu$).
\item ``Muon station 1, 2" : muons from HNL decays leave signals in the first two muon stations (only for $H\!N\!L \rightarrow \mu\pi$ and $H\!N\!L \rightarrow \mu\mu\nu$). 
\end{itemize}

The number of HNL candidates and the total experimental acceptance at different stages of the selection 
are given in Tables~\ref{tab:cutflowtruth} to~\ref{tab:cutflowtruth3} 
for $H\!N\!L \rightarrow \mu\pi$ and $H\!N\!L \rightarrow \mu\mu\nu$ in the benchmark scenario, and for 
$H\!N\!L \rightarrow ee\nu$ in a scenario with couplings of the order of $10^{-6}$ and with an HNL mass of 100~MeV/$c^2$. 
The error on the acceptance is estimated to about 10~\% by generating various independent 
HNL samples with the toy MC.

\begin{table}[!!!!h]
\caption{\label{tab:cutflowtruth} Number of HNL candidates at the different selection stages for the $H\!N\!L \rightarrow \mu\pi$  decay channel in the benchmark scenario
for $m_{HNL} = 1$~GeV/c$^2$.}
\centering
\vspace{2mm}
\begin{tabular}{lccc}
\hline
\textbf{Selection} & \textbf{Entries}  & \textbf{Acceptance} & \textbf{Rel. loss [\%]} \\  \hline
Normalization & 1065 & 9.07e-05 &0 \\ 
%\tiny{$z_{\text{straw veto}} < z_{\rm vtx}<z_{\rm out}$}  & 874 & 7.51e-05 &  17    \\  
Fiducial & 430 & 1.93e-05 &  74 \\ 
Straw 1 and 2  & 259 & 7.47e-06 & 61 \\ 
Straw 4 & 221 & 5.78e-06 & 23   \\ 
ECAL  & 221 & 5.78e-06 & 0  \\ 
Muon station 1  & 217 & 5.65e-06 & 2  \\
Muon station 2  & 214 & 5.55e-06 & 2  \\ 
\hline
\end{tabular}
\end{table}

\begin{table}[!!!!h]
\caption{\label{tab:cutflowtruth2} Number of HNL candidates at the different selection stages for the $H\!N\!L\rightarrow \mu\mu\nu$ decay channel in the benchmark scenario
for $m_{HNL} = 1$~GeV/c$^2$.}
\centering
\vspace{2mm}
\begin{tabular}{lccc}
\hline
\textbf{Selection} & \textbf{Entries}  & \textbf{Acceptance} & \textbf{Rel. loss [\%]} \\  \hline
Normalization & 1143 &  9.30e-05& 0  \\ 
%\tiny{$z_{\text{straw veto}} < z_{\rm vtx}<z_{\rm out}$}  & 927 & 7.66e-05 & 18  \\ 
Fiducial &  450 & 1.88e-05 & 75 \\ 
Straw 1 and 2  &  301 & 9.21e-06 & 51  \\ 
Straw 4 & 221 & 7.13e-06 &  23  \\ 
%ECAL  & 221 & 7.13e-06 & 0 \\  \hline
Muon station 1  &  230 & 6.33e-06 & 11  \\ 
Muon station 2  &  220 & 5.88e-06 &  7  \\  
\hline
\end{tabular}
\end{table}

\begin{table}[!!!!h]
\caption{\label{tab:cutflowtruth3} Number of HNL candidates at different stages of the selection for the $H\!N\!L\rightarrow ee\nu$ decay channel in a scenario with couplings $\mathcal{O}(10^{-6})$ and $m_{H\!N\!L} = 100$~MeV/$c^2$.} 
\centering
\vspace{2mm}
\begin{tabular}{lccc}
\hline
\textbf{Selection} & \textbf{Entries}  & \textbf{Acceptance} & \textbf{Rel. loss [\%]} \\  \hline
Normalization & 1112 & 9.72e-09 & 0 \\ 
%\tiny{$z_{\text{straw veto}} < z_{\rm vtx}<z_{\rm out}$}  &  941 & 8.74e-09 & 10 \\  
Fiducial &  302 & 8.00e-10 & 91 \\ 
Straw 1 and 2  &  213 & 4.58e-10 & 43 \\  
Straw 4 &  161 & 2.42e-10 & 47 \\  
ECAL  &  161  & 2.42e-10 & 0\\  
%muon station 1  &  161 & 2.42e-10  & 0 \\  \hline
%muon station 2  &161 & 2.42e-10 & 0 \\  \hline
\hline
\end{tabular}
\end{table}

\subsubsection{Reconstruction and selection efficiencies}\label{sec:effectreco}

To assess the impact of the reconstruction procedure on the signal acceptance, the same track reconstruction algorithm, as described in 
Section~\ref{sec:Tracker performance studies from simulation}, is applied to the generated signal samples using \textsc{FairShip}. The 
impact of the loose offline pre-selections applied to reject background in Section~\ref{sec:backgrounds} is estimated by using the selection 
devised against the background from neutrino interactions, since they closely mimic the topology of HNL decays.

\begin{figure}[!!!!h]
 \centering
 \includegraphics[width=6cm]{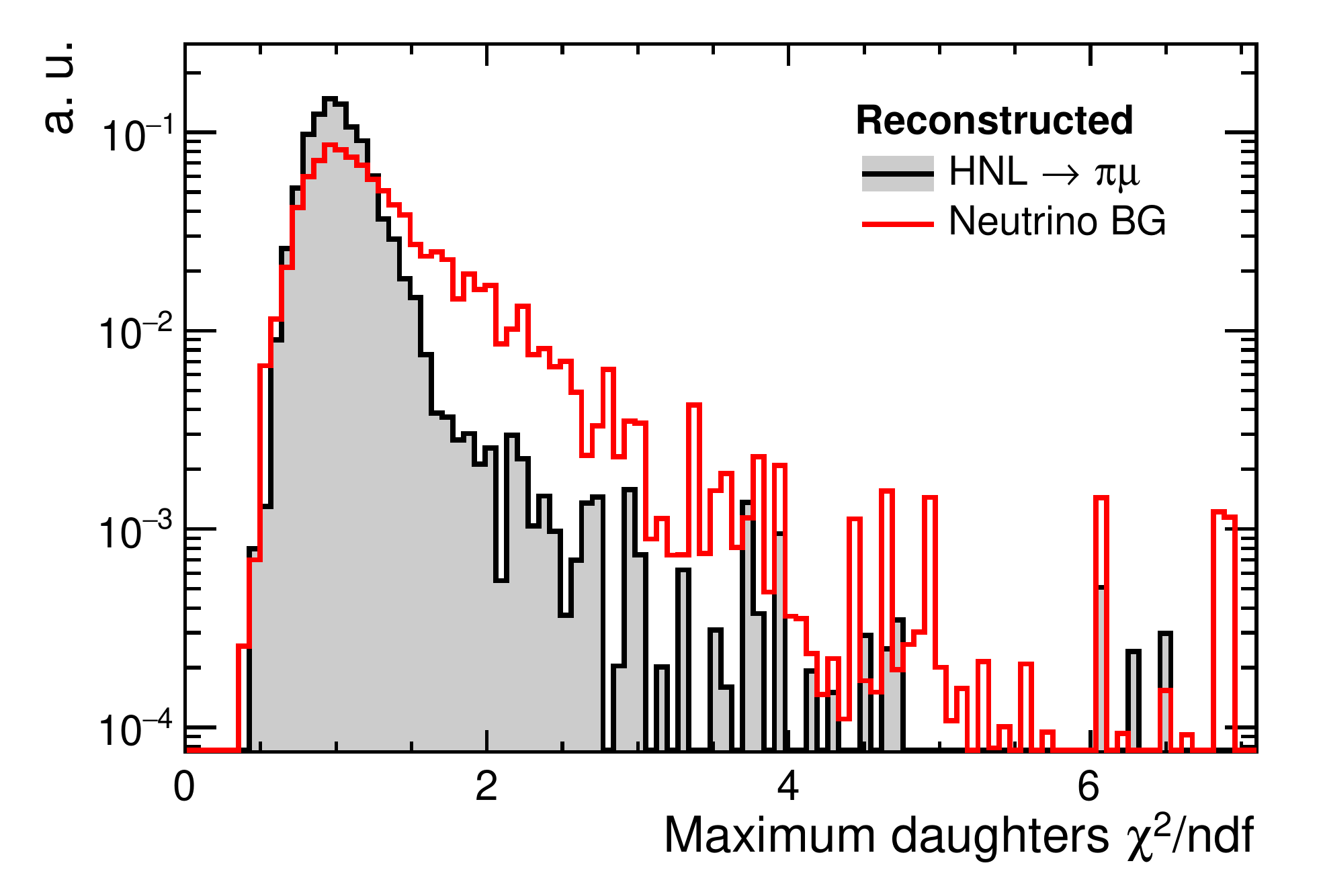}\includegraphics[width=6cm]{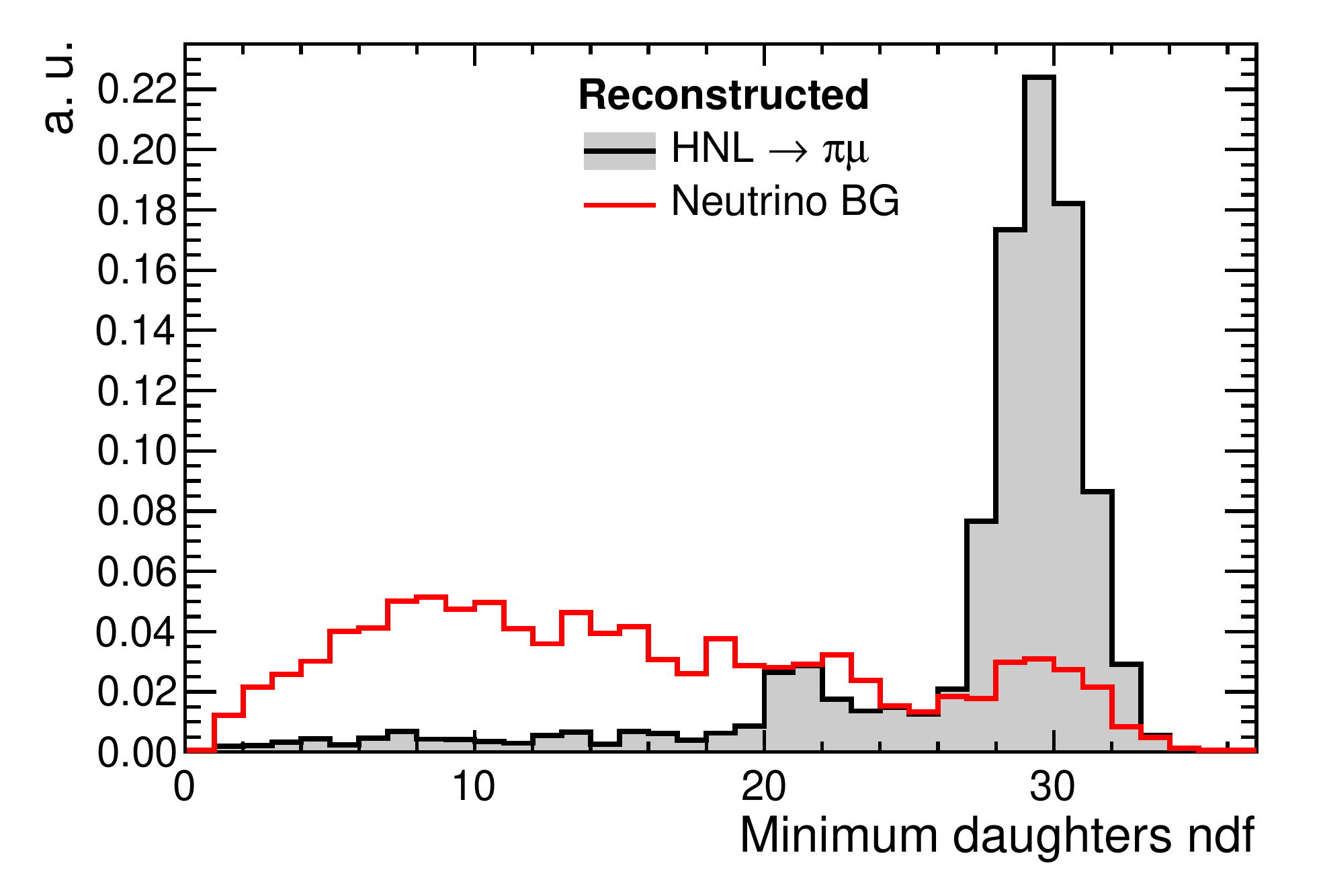}
 \includegraphics[width=6cm]{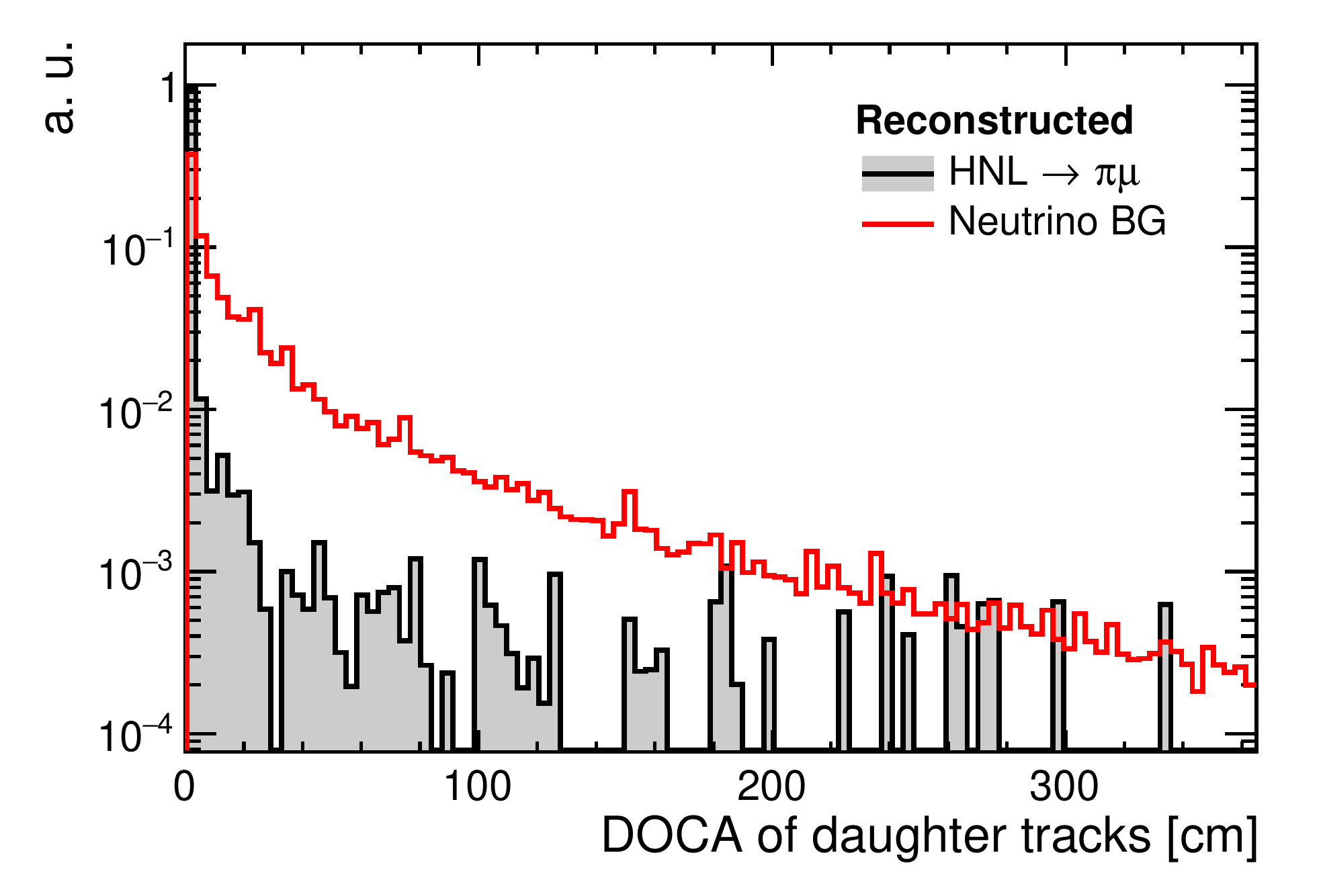}\includegraphics[width=6cm]{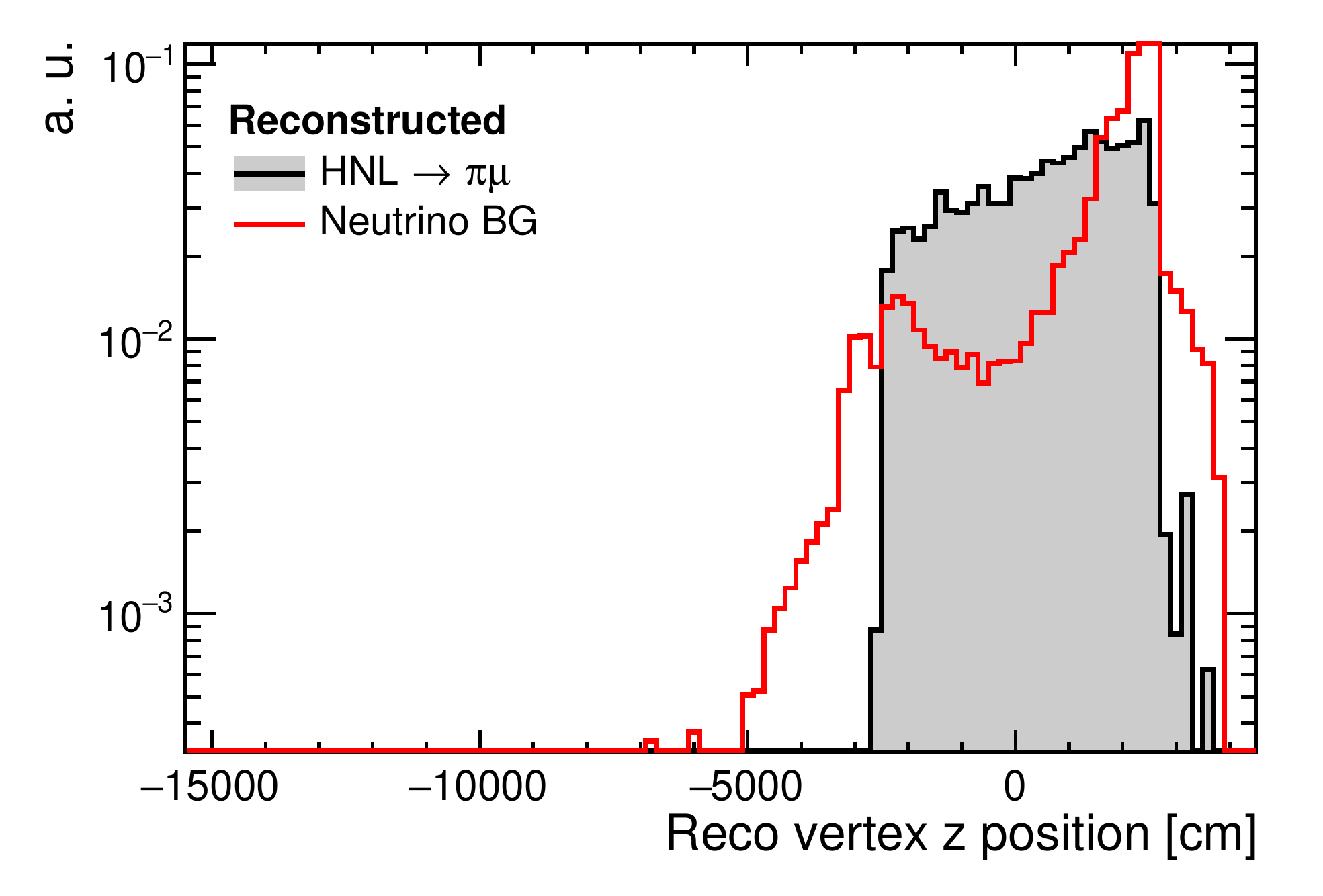}
 \includegraphics[width=6cm]{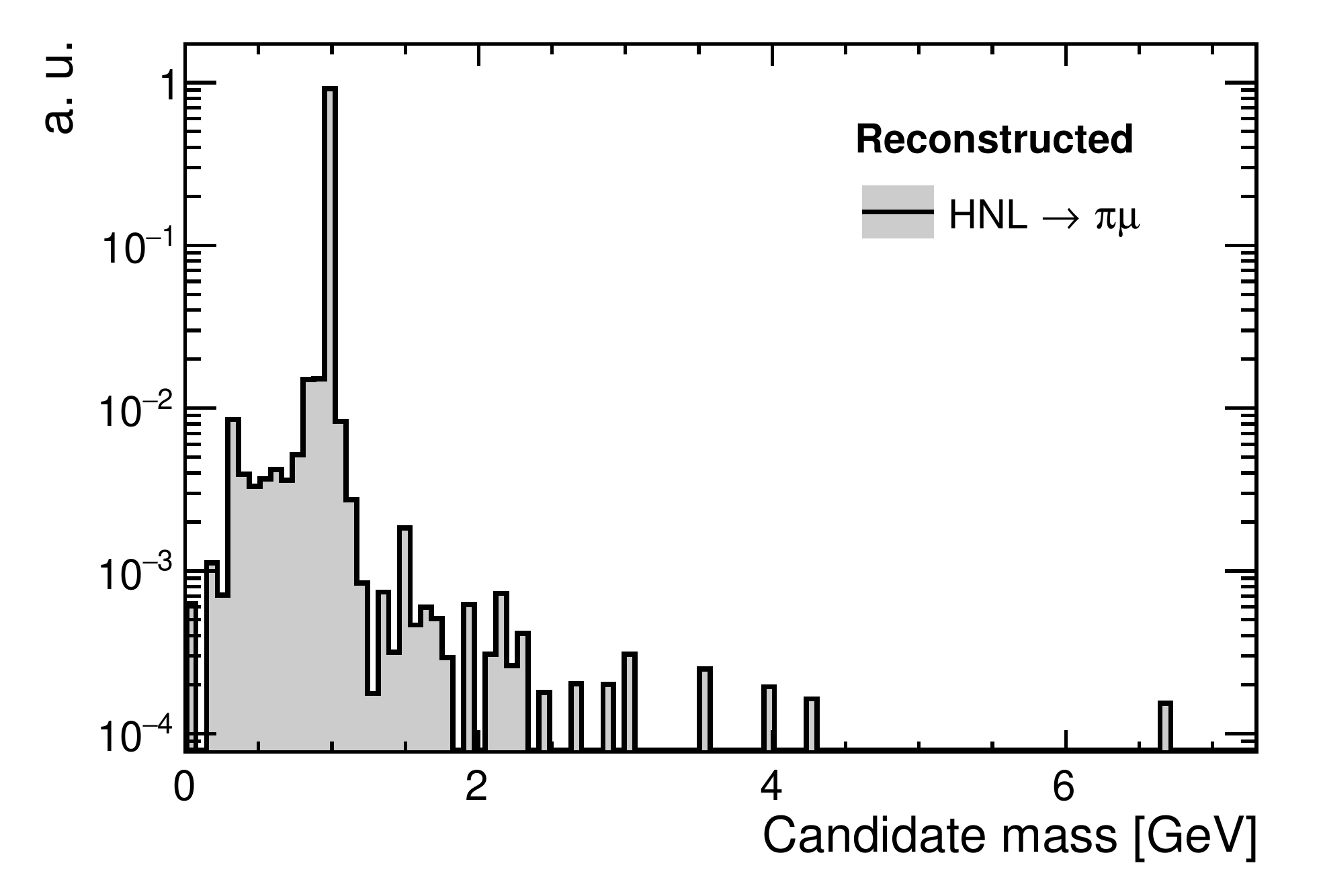}\includegraphics[width=6cm]{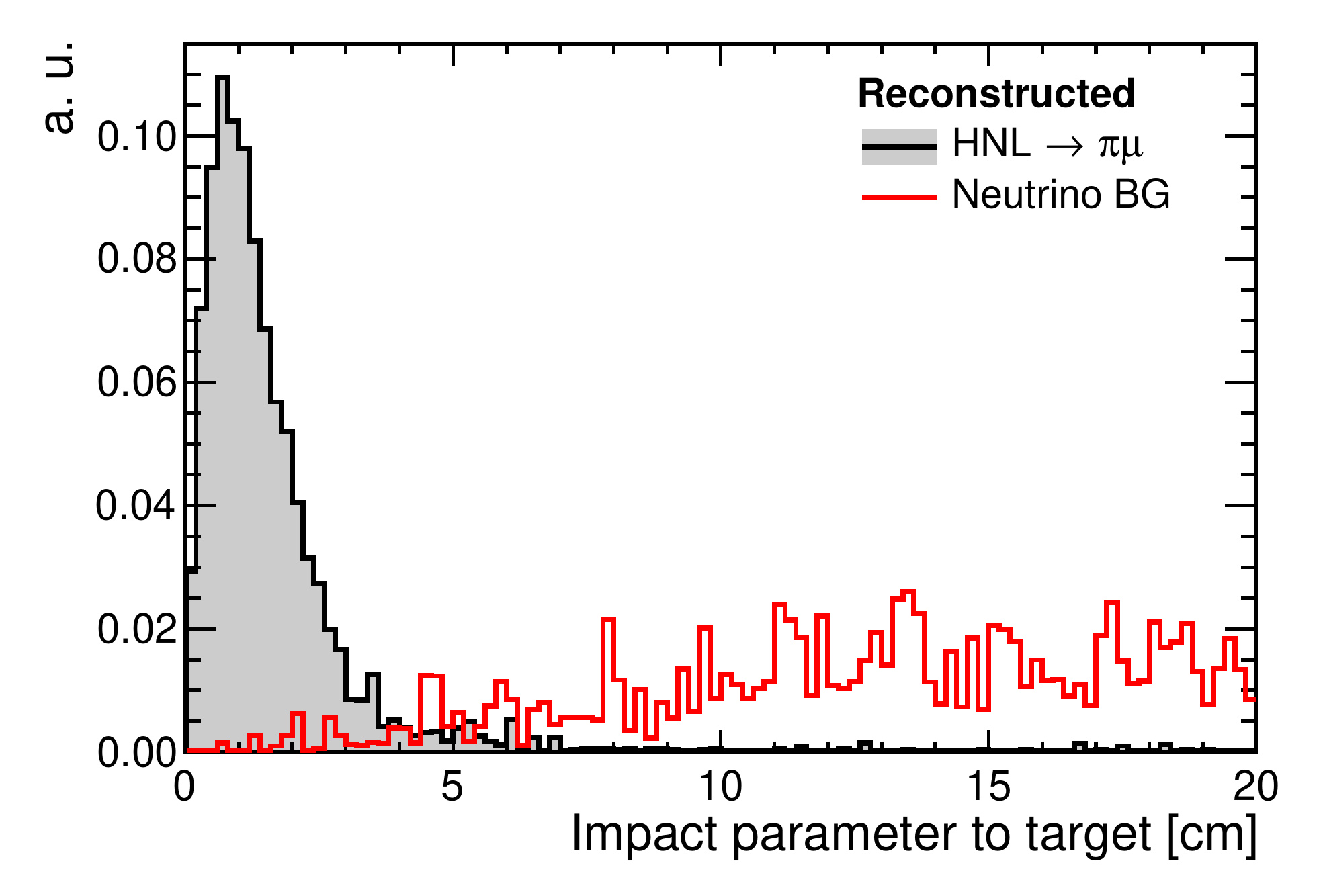}
 \includegraphics[width=6cm]{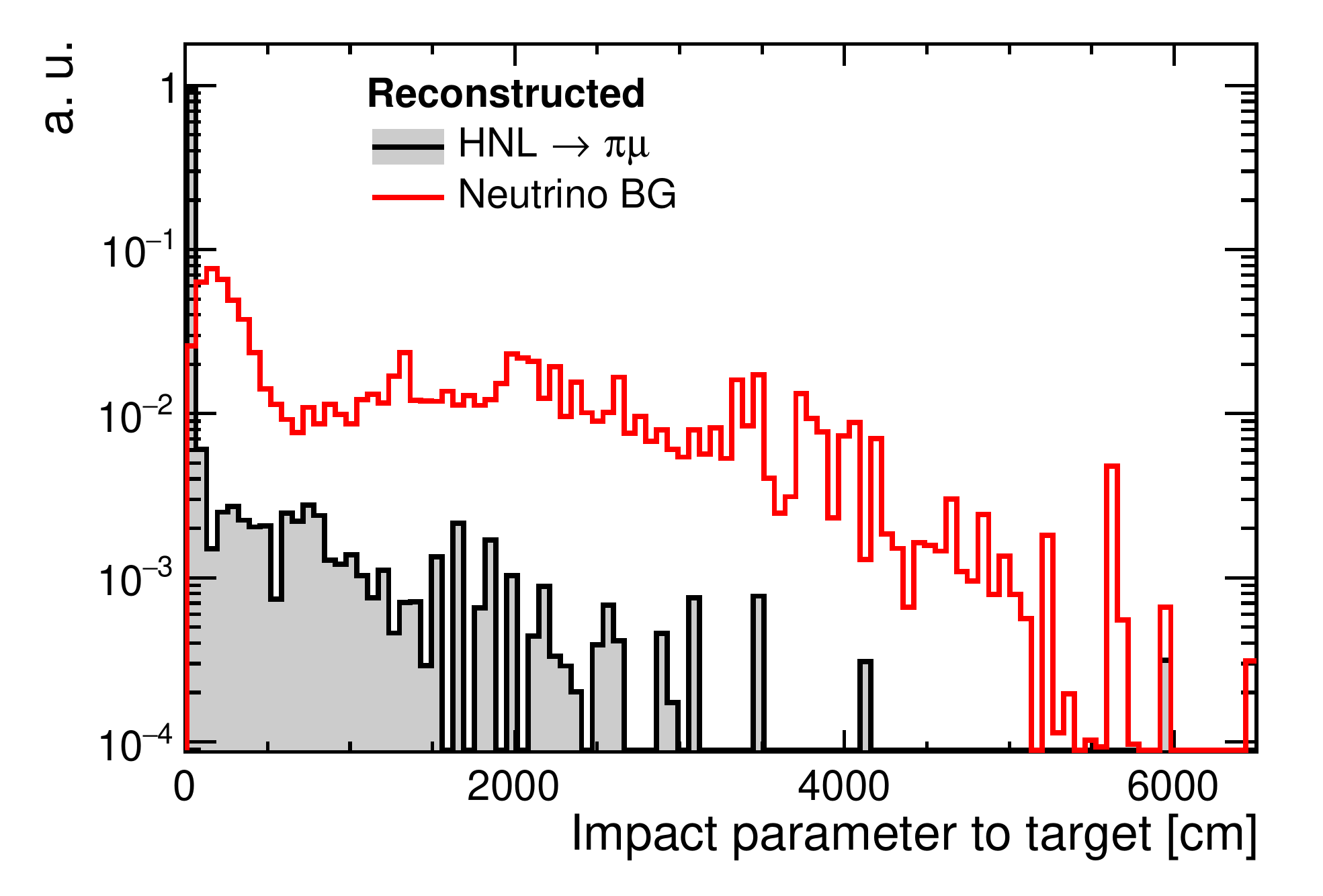}
 \caption{\label{fig:recodist1pimu} $\chi^2/ndf$, $ndf$, distance of closest approach of the daughter tracks, $z$ position of the decay vertex, reconstructed candidate mass, and impact parameter to the target distributions for 2-track signal candidates in the $H\!N\!L\rightarrow \mu \pi$ channel (shaded area). The solid red line represents neutrino induced reconstructed background events.}
 \end{figure}

\begin{figure}[!!!!h]
 \centering
 \includegraphics[width=6cm]{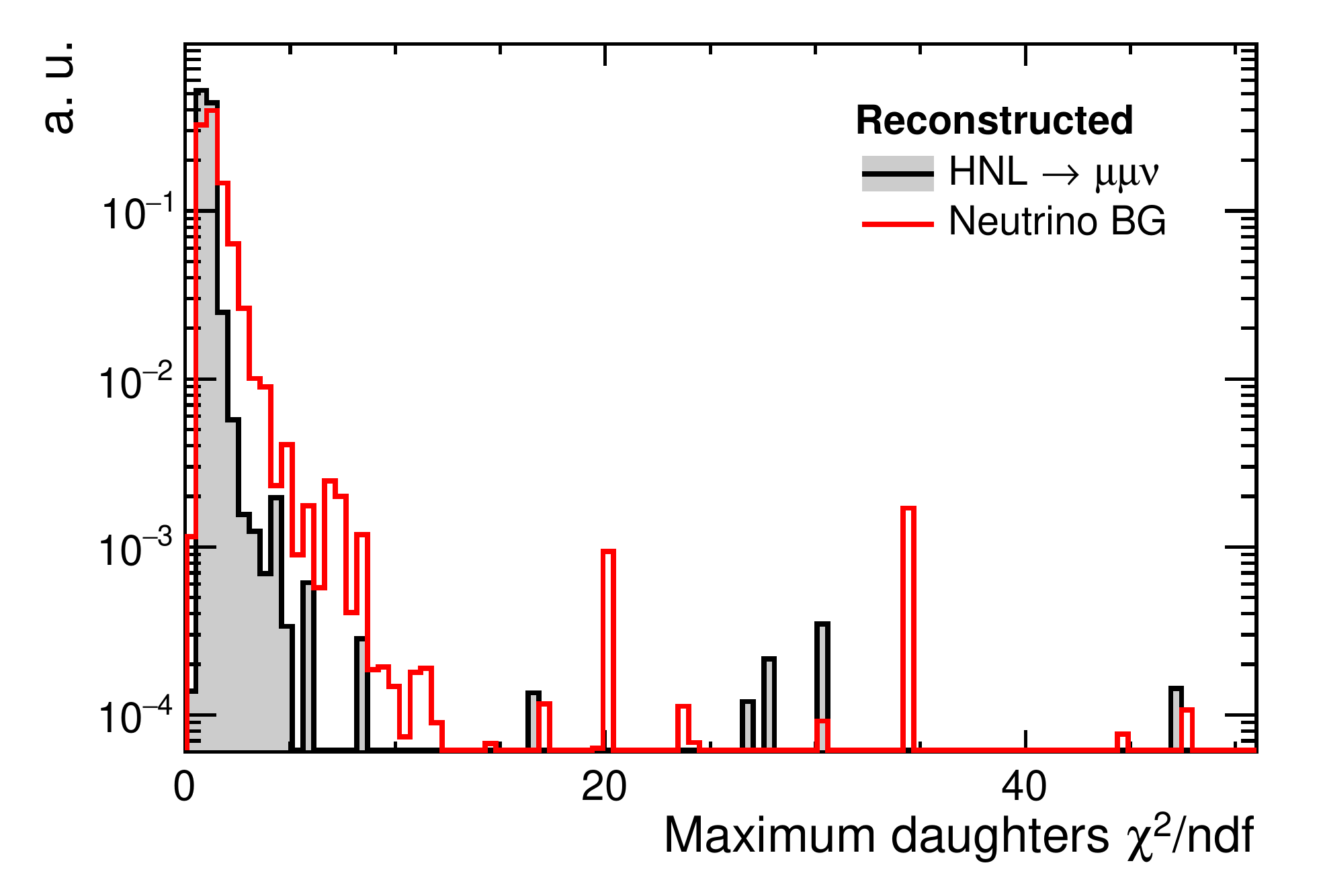}\includegraphics[width=6cm]{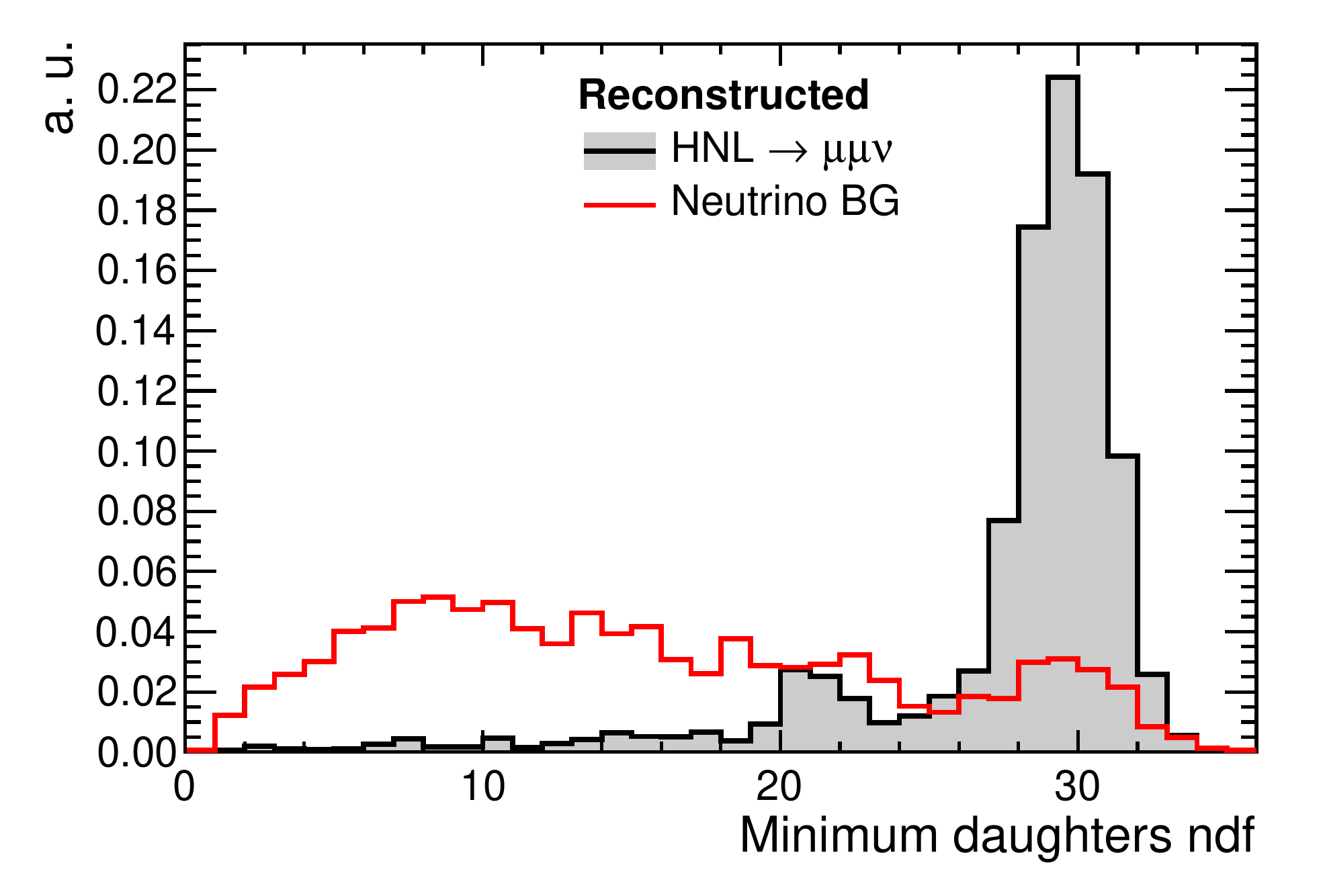}
 \includegraphics[width=6cm]{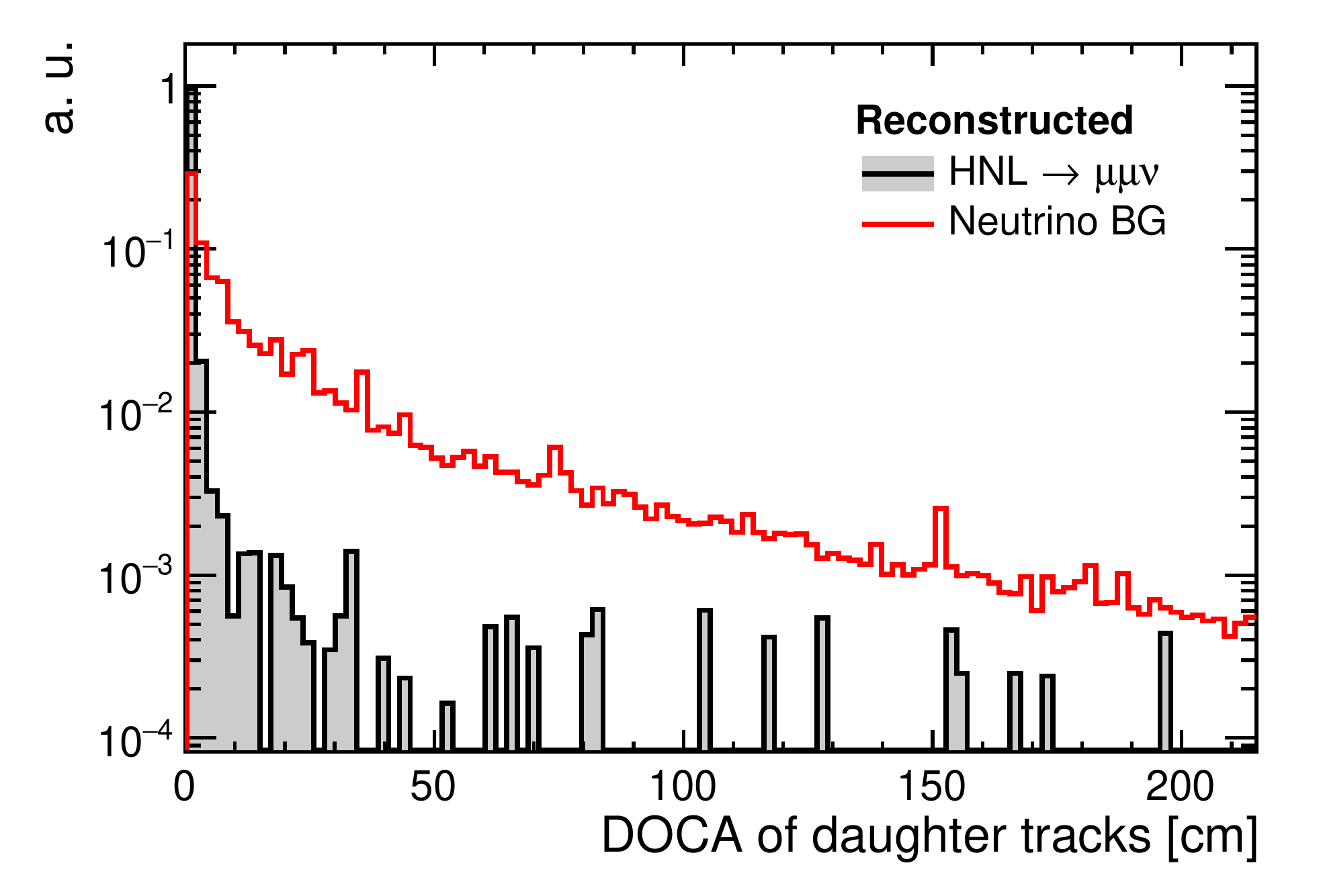}\includegraphics[width=6cm]{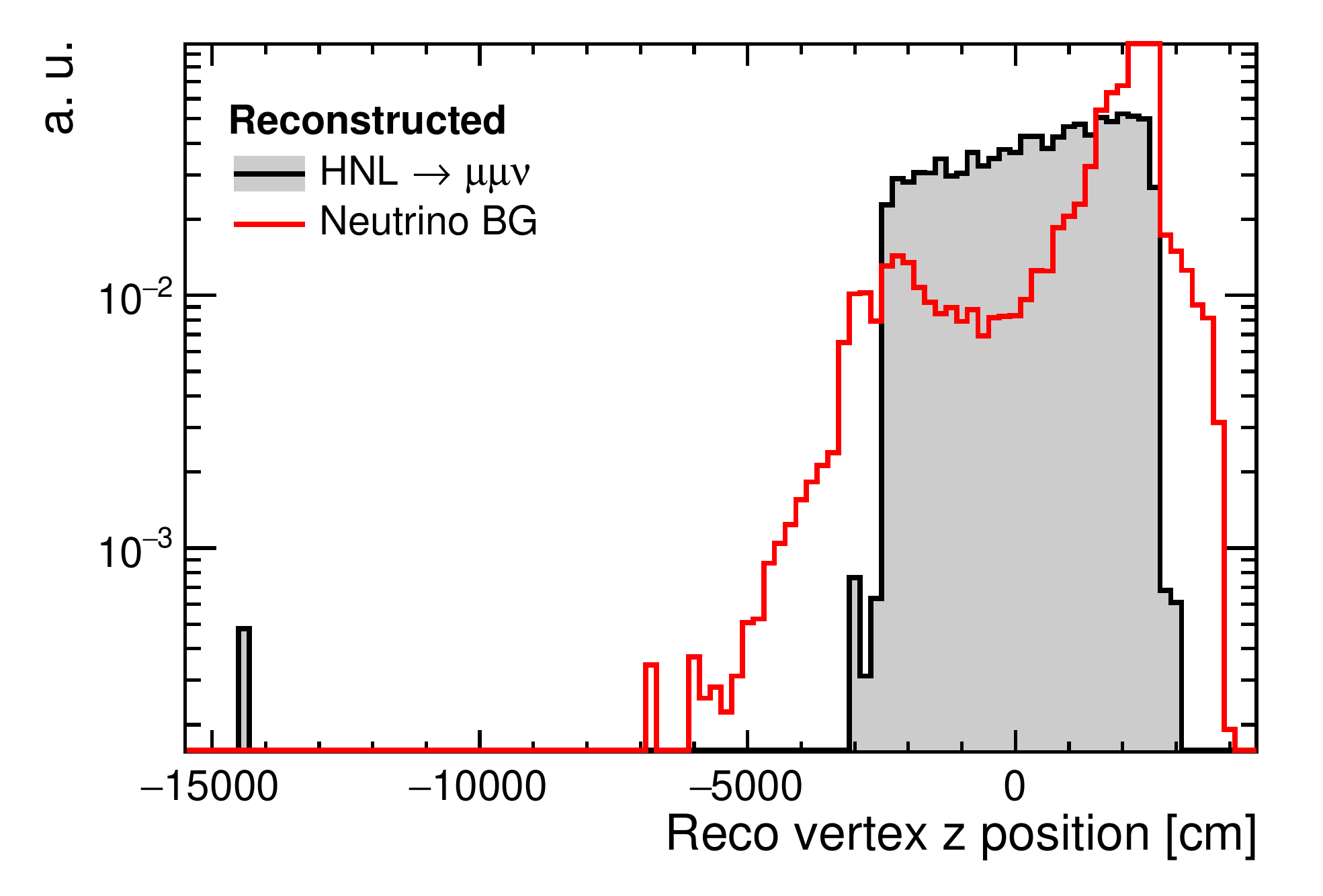}
 \includegraphics[width=6cm]{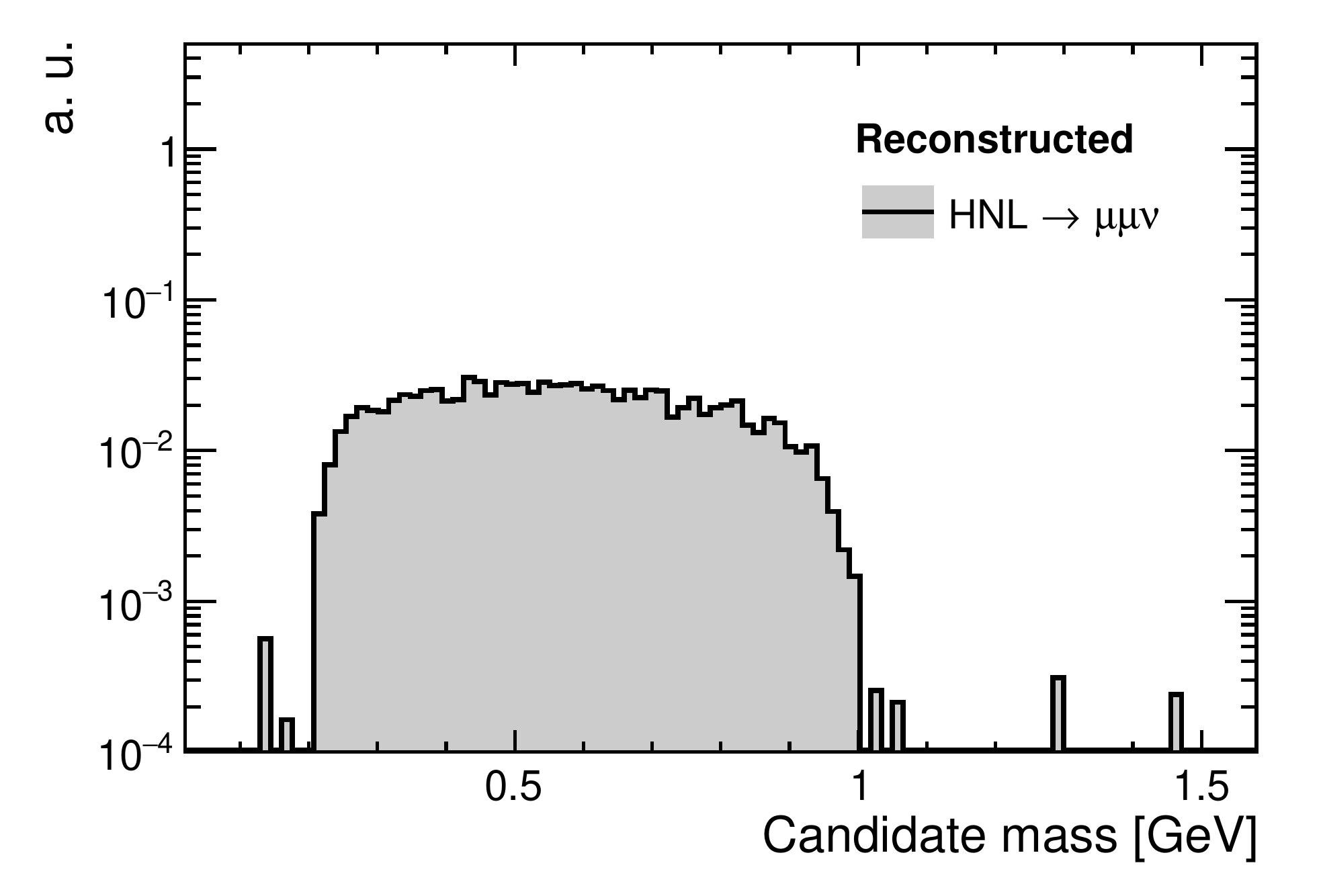}\includegraphics[width=6cm]{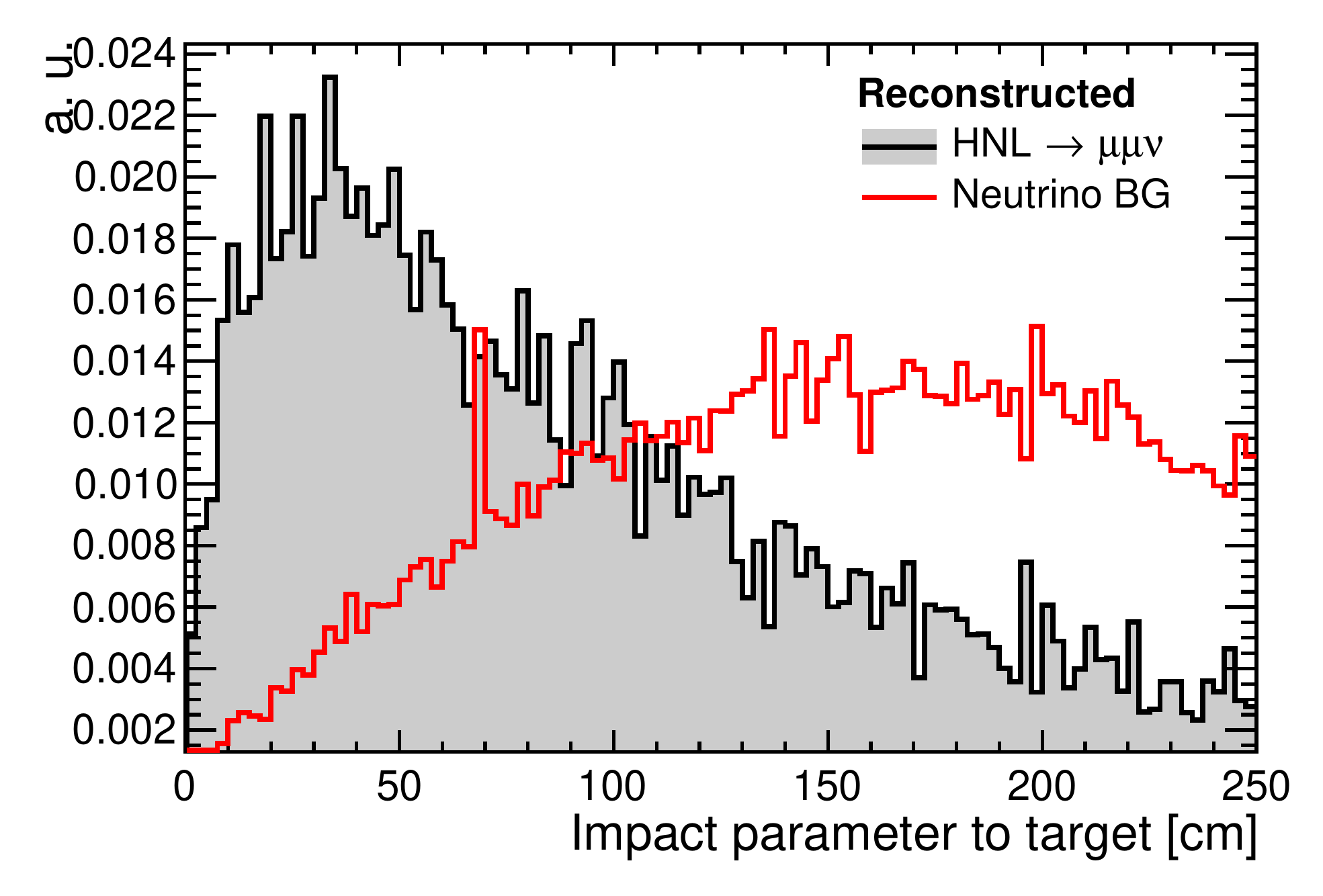}
 \includegraphics[width=6cm]{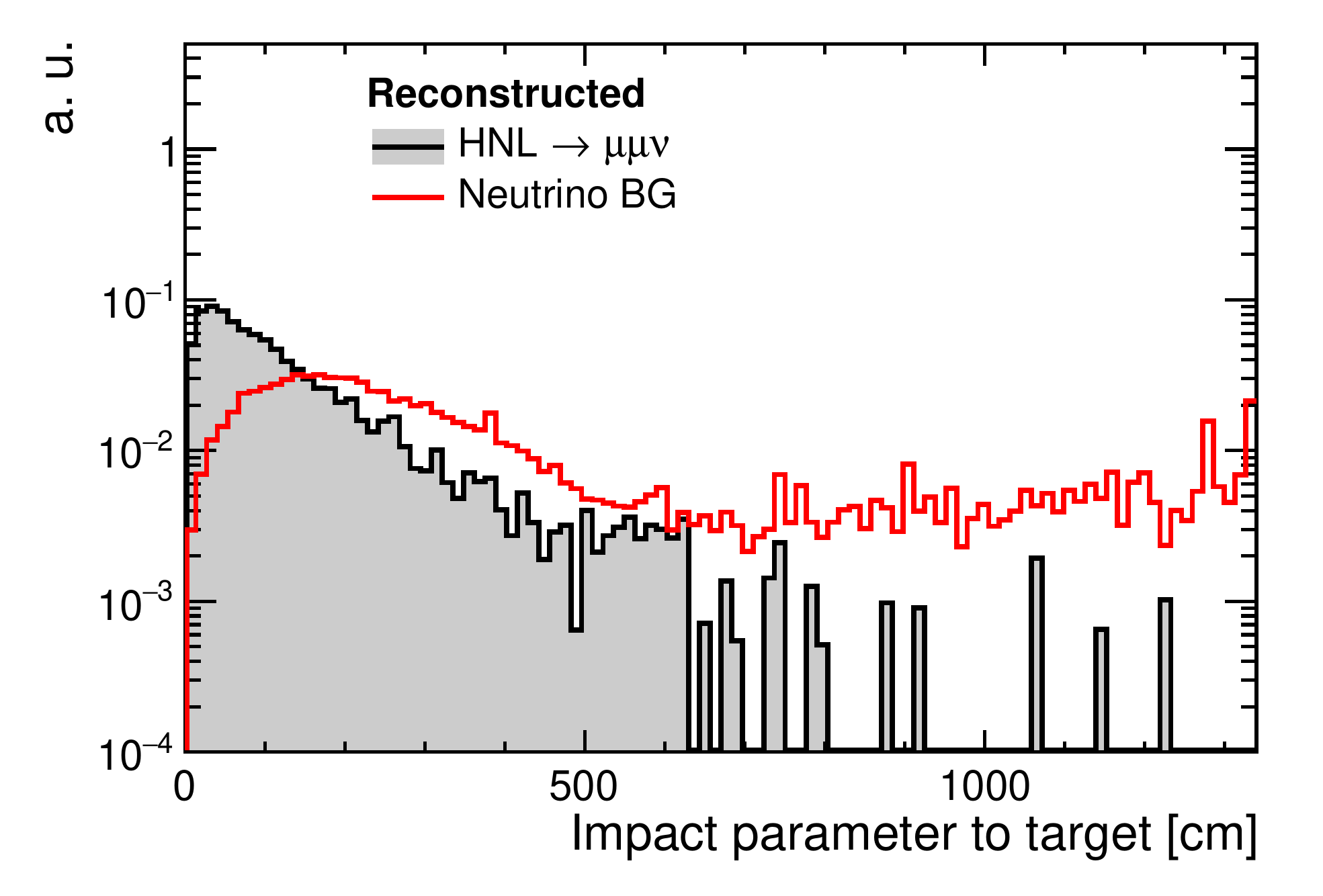}
 \caption{\label{fig:recodist1mumunu} $\chi^2/ndf$, $ndf$, distance of closest approach of the daughter tracks, $z$ position of the decay vertex, reconstructed candidate mass, and impact parameter to the target distributions for 2-track signal candidates in the $H\!N\!L\rightarrow \mu \mu \nu$ channel (shaded area). The solid red line represents neutrino induced reconstructed background events.}
 \end{figure}

Figures~\ref{fig:recodist1pimu} and \ref{fig:recodist1mumunu} show distributions of $\chi^2/ndf$, $ndf$, 
distance of closest approach of the daughter tracks, $z$ position of the decay vertex, reconstructed 
candidate mass and reconstructed mass for events with $\chi^2/ndf<5$ and $ndf>25$, and impact parameter to the target. 
These observables are shown for $H\!N\!L\rightarrow \mu \pi$ and  $H\!N\!L\rightarrow\mu\mu\nu$ superimposed on the 
same observables for neutrino induced background. No further selection is applied.

% \begin{figure}[!!!!h]
%  \centering
%  \includegraphics[width=6cm]{sensitivity/acceptance/tiny-chi2-all}\includegraphics[width=6cm]{sensitivity/acceptance/tiny-ndf-all}
%  \includegraphics[width=6cm]{sensitivity/acceptance/tiny-doca-all}\includegraphics[width=6cm]{sensitivity/acceptance/tiny-z-all}
%  \includegraphics[width=6cm]{sensitivity/acceptance/tiny-mass-all}\includegraphics[width=6cm]{sensitivity/acceptance/tiny-mass2-all}
%  \includegraphics[width=6cm]{sensitivity/acceptance/tiny-IP0-all}
%  \caption{\label{fig:recodist1eenu} $\chi^2/ndf$, $ndf$, distance of closest approach of the daughter tracks, $z$ position of the decay vertex, reconstructed candidate mass and reconstructed m% ass for events with $\chi^2/ndf<5$ and $ndf>15$, and impact parameter to the target distributions for 2-track signal candidates in the $H\!N\!L\rightarrow e e \nu$ channel (solid black line). % The red line represents neutrino induced reconstructed background events.}
%  \end{figure}
 
The following criteria have been devised for the neutrino background in order to select good quality tracks and to help reduce the level of background:
%The HNL candidates are selected as follows: 
\begin{itemize}
\item ``Event reconstructed'': the track fit has converged for both tracks
\item ``Event not vetoed'': the event is not tagged by any veto detector and the reconstructed candidate has an impact parameter with respect to the beam dump target smaller than 10~m
\item ``$\chi^2/\text{ndf}<5$'': both daughter tracks have a reduced $\chi^2$ smaller than 5. 
\item ``ndf $>$ 25'': both tracks have a number of degrees of freedom greater than 25. This ensures that we have enough measurements to perform a solid track fit.
\item ``Fiducial'' : position of the HNL decay vertex is within the fiducial volume as defined in Section~\ref{sec:sensitivity_intro}
% \item ``z$_A{\text{straw veto}} < z_{\rm vtx} < z_{trk1}$'': the decay vertex is located downstream of the straw veto tagger.
% \item ``Vertex $\in$ ellipse'': the $x,y$ position of the HNL decay vertex is inside the elliptical fiducial volume ($r_x = 250$~cm, $r_y = 500$~cm). 
\item ``Tracks'': both HNL daughters leave a signal in one of the straw stations before the magnet (1 or 2) and one of the straw stations after the magnet (3 or 4). % These hits are within the elliptical fiducial volume ($r_x = 250$~cm, $r_y = 500$~cm).
\item ``150 MeV in Ecal'': at least 150 MeV are deposited in the ECAL (only for the $H\!N\!L\rightarrow \pi \mu$ and $H\!N\!L\rightarrow e e \nu$ channels).
\item ``$n$ muon(s) in muon station $m$'': $n$ muons from HNL decays are recorded in the $m^{th}$ muon station. 
\item ``DOCA $<$ 30 cm'': the distance of closest approach (DOCA) between the two daughter tracks is smaller than 30~cm. This selection is very loose: the reconstruction algorithm currently included in \textsc{FairShip} allows achieving an average DOCA of few~cm for reconstructed candidates corresponding to simulated HNLs (see Section~\ref{sec:Tracker performance studies from simulation}).
\item ``IP $<$ 2.5 m'': the impact parameter to the target (IP) is smaller than 2.5~m. This requirement was found to be the most effective in reducing the background contamination. 
\end{itemize}

The number of HNL candidates selected after the different requirements is given in Tables~\ref{tab:recpimu} and \ref{tab:recmmnu} for signal samples, and in Table~\ref{tab:recnubkg} for a neutrino background sample roughly corresponding to the real number of events from neutrino interactions expected in five years of SHiP operation (see Section~\ref{sec:backgroundNu}).

\begin{table}[!!!!h]
	\begin{center}
		\caption{ Effect of the offline selection on $HNL \to \pi \mu$ \label{tab:recpimu}}
\vspace{2mm}
		\begin{tabular}{lccc}
			\hline
			\textbf{Selection} & \textbf{Entries} & \textbf{Acceptance} & \textbf{Selection efficiency} \\
			\hline
			Event reconstructed  &  4471  &  $6.43 \cdot 10^{-6}$  &  - \\
%			\hline
			Event not vetoed  &  3540  &  $4.87 \cdot 10^{-6}$  &  75.8 \% \\
%			\hline
			$\chi^2/\text{N.d.f.} < 5 $  &  3540  &  $4.87 \cdot 10^{-6}$  &  100.0 \% \\
%			\hline
			$\text{N.d.f.} > 25$  &  3249  &  $4.37 \cdot 10^{-6}$  &  89.7 \% \\
%			\hline
			Vtx in fiducial vol.  &  3224  &  $4.34 \cdot 10^{-6}$  &  99.3 \% \\
%			\hline
			Tracks $\in$ fiducial vol.  &  3223  &  $4.34 \cdot 10^{-6}$  &  100.0 \% \\
%			\hline
			150~MeV in Ecal  &  3223  &  $4.34 \cdot 10^{-6}$  &  100.0 \% \\
%			\hline
			1 muon in 1$^{\text{st}}$ muon station  &  3201  &  $4.3 \cdot 10^{-6}$  &  99.1 \% \\
%			\hline
			1 muon in 2$^{\text{nd}}$ muon station  &  3156  &  $4.22 \cdot 10^{-6}$  &  98.2 \% \\
%			\hline
			DOCA $<$ 30~cm  &  3155  &  $4.22 \cdot 10^{-6}$  &  100.0 \% \\
%			\hline
			IP $<$ 2.5 m  &  3155  &  $4.22 \cdot 10^{-6}$  &  100.0 \% \\
			\hline
		\end{tabular}
	\end{center}
	\end{table}

	\begin{table}[!!!!h]
	\begin{center}
		\caption{ Effect of the offline selection on $HNL \to \mu \mu \nu$ \label{tab:recmmnu}}
\vspace{2mm}
		\begin{tabular}{lccc}
			\hline
			\textbf{Selection} & \textbf{Entries} & \textbf{Acceptance} & \textbf{Selection efficiency} \\
			\hline
			Event reconstructed  &  4856  &  $7.64 \cdot 10^{-6}$  &  - \\
%			\hline
			Event not vetoed  &  4072  &  $6.13 \cdot 10^{-6}$  &  80.3 \% \\
%			\hline
			$\chi^2/\text{N.d.f.} < 5 $  &  4067  &  $6.12 \cdot 10^{-6}$  &  99.8 \% \\
%			\hline
			$\text{N.d.f.} > 25$  &  3718  &  $5.39 \cdot 10^{-6}$  &  88.1 \% \\
%			\hline
			Vtx in fiducial vol.  &  3685  &  $5.34 \cdot 10^{-6}$  &  99.1 \% \\
%			\hline
			Tracks $\in$ fiducial vol.  &  3685  &  $5.34 \cdot 10^{-6}$  &  100.0 \% \\
%			\hline
			2 muons in 1$^{\text{st}}$ muon station  &  3559  &  $5.03 \cdot 10^{-6}$  &  94.1 \% \\
%			\hline
			2 muons in 2$^{\text{nd}}$ muon station  &  3373  &  $4.6 \cdot 10^{-6}$  &  91.6 \% \\
%			\hline
			DOCA $<$ 30~cm  &  3371  &  $4.6 \cdot 10^{-6}$  &  99.9 \% \\
%			\hline
			IP $<$ 2.5 m  &  3207  &  $4.16 \cdot 10^{-6}$  &  90.4 \% \\
			\hline
		\end{tabular}
	\end{center}
	\end{table}

% 	\begin{table}[!!!!h]
%		\centering 
%		\begin{tabular}{|l|c|c|c|}
%			\hline
%			\textbf{Selection} & \textbf{Entries} & \textbf{Acceptance} & \textbf{Selection efficiency} \\
%      			\hline
%			Event reconstructed  &  2820  &  $1.83 \cdot 10^{-10}$  &  - \\
%			\hline
%			Event not vetoed  &  2433  &  $1.56 \cdot 10^{-10}$  &  85.5 \% \\
%			\hline
%			$\chi^2/\text{N.d.f.} < 5 $  &  2426  &  $1.55 \cdot 10^{-10}$  &  99.6 \% \\
%			\hline
%			$\text{N.d.f.} > 15$  &  2388  &  $1.52 \cdot 10^{-10}$  &  97.7 \% \\
%			\hline
%			$z_{straw\, veto} < z_{vtx} < z_{trk1}$  &  2095  &  $1.34 \cdot 10^{-10}$  &  88.3 \% \\
%			\hline
%			Vertex $\in $ ellipse  &  2095  &  $1.34 \cdot 10^{-10}$  &  100.0 \% \\
%			\hline
%			Tracks $\in$ ellipse  &  2095  &  $1.34 \cdot 10^{-10}$  &  100.0 \% \\
%			\hline
%			150~MeV in Ecal  &  2092  &  $1.34 \cdot 10^{-10}$  &  99.6 \% \\
%			\hline
%			DOCA $<$ 30~cm  &  2090  &  $1.33 \cdot 10^{-10}$  &  99.9 \% \\
%			\hline
%			IP $<$ 2.5 m  &  2089  &  $1.33 \cdot 10^{-10}$  &  99.7 \% \\
%			\hline
%		\end{tabular}
%		\caption{Effect of the offline selection on  $HNL \to e e \nu$ using the signal model of Table \ref{tab:sigcomp3}. \label{tab:receenu}}
%	\end{table}

\begin{table}[!!!!h]
	\begin{center}
		\caption{ Effect of the offline selection on neutrino-induced background \label{tab:recnubkg}}
\vspace{2mm}
		\begin{tabular}{lccc}
			\hline
			\textbf{Selection} & \textbf{Entries} & \textbf{Events / 5 years} & \textbf{Selection efficiency} \\
			\hline
			Event reconstructed  &  107828  &  $1 \cdot 10^{4}$  &  - \\
%			\hline
			Event not vetoed  &  254  &  51.8  &  0.5 \% \\
%			\hline
			$\chi^2/\text{N.d.f.} < 5 $  &  253  &  51.7  &  99.9 \% \\
%			\hline
			$\text{N.d.f.} > 25$  &  74  &  8.98  &  17.4 \% \\
%			\hline
			Vtx in fiducial vol.  &  10  &  3.94  &  43.8 \% \\
%			\hline
			Tracks $\in$ fiducial vol.  &  10  &  3.94  &  100.0 \% \\
%			\hline
			150~MeV in Ecal  &  9  &  2.01  &  51.1 \% \\
%			\hline
			1 muon in 1$^{\text{st}}$ muon station  &  8  &  2.01  &  99.7 \% \\
%			\hline
			1 muon in 2$^{\text{nd}}$ muon station  &  8  &  2.01  &  100.0 \% \\
%			\hline
			DOCA $<$ 30~cm  &  7  &  1.72  &  85.5 \% \\
%			\hline
			IP $<$ 2.5 m  &  0  &  0  &  0.0 \% \\
			\hline
		\end{tabular}
	\end{center}
	\end{table}

%The resulting acceptance after all selection criteria are applied is compared with the raw acceptance %computed with the toy Monte Carlo. The ratio between these two values is applied as efficiency factor to the %toy, in order to take into account the reconstruction and selection efficiencies when providing estimates for %SHiP's sensitivity to HNLs. 
% It is found to be $\sim$70$\%$ for HNL masses above twice that of the muons, and 40$\%$ below.\\

%%%%%%%%%%%%%%%%%%%%%%%%%%%%%%%%%%%%%%%%%%%%%%%%%%%%%%%%%%%%%%%%%%%%%%%%%%%%%
%%%%%%%%%%%%%%%%%%%%%%%%%%%%%%%%%%%%%%%%%%%%%%%%%%%%%%%%%%%%%%%%%%%%%%%%%%%%%
%%%%%%%%%%%%%%%%%%%%%%%%%%%%%%%%%%%%%%%%%%%%%%%%%%%%%%%%%%%%%%%%%%%%%%%%%%%%%

\input{sensitivity/HNLs}

\input{sensitivity/LLP}

%% file: sensitivity/HNLs.tex
\subsection{Sensitivity to HNLs}
\label{sec:HNLsplots}

The signal acceptance results obtained with \textsc{FairShip}
and the toy Monte Carlo, without applying further offline selection criteria, are in agreement within errors. The
signal acceptance for the benchmark scenario in the $\pi\mu$ channel is found to be ${\cal
  A}=(5.8 \pm 1.8) \cdot 10^{-6}$ for the toy Monte Carlo, and ${\cal
  A}=(5.6 \pm 0.6) \cdot 10^{-6}$ for \textsc{Fairship}. 
The toy Monte Carlo is then corrected for the reconstruction and
selection efficiencies and used to determine the sensitivity.
%The signal acceptance
%decreases to ${\cal  A}=(4.4 \pm 1.8) \cdot 10^{-6}$ at the reconstruction level, following the offline selections.
%Following the validation of the toy Monte Carlo against 
%\textsc{FairShip}, it is used to assess the sensitivity contours in
%the HNL mass-couplings parameter space. 
%The toy
%Monte Carlo technique has the advantage of being faster and easier to
%configure compared to the full \textsc{FairShip} simulation. It
%can also be used to completely determine the expected number of
%events in five years of SHiP operation, because it includes
%classes to estimate both the rate of HNLs produced at the target and
%the acceptance to the HNL decay products. 
%Therefore the toy Monte Carlo is used in order to interpolate between the results of the full simulation.
% Correction factors obtained from the full simulation are applied
% to take into account the reconstruction and selection efficiencies.

\begin{figure}[tb]
\begin{center}
\includegraphics[width=0.42\linewidth]{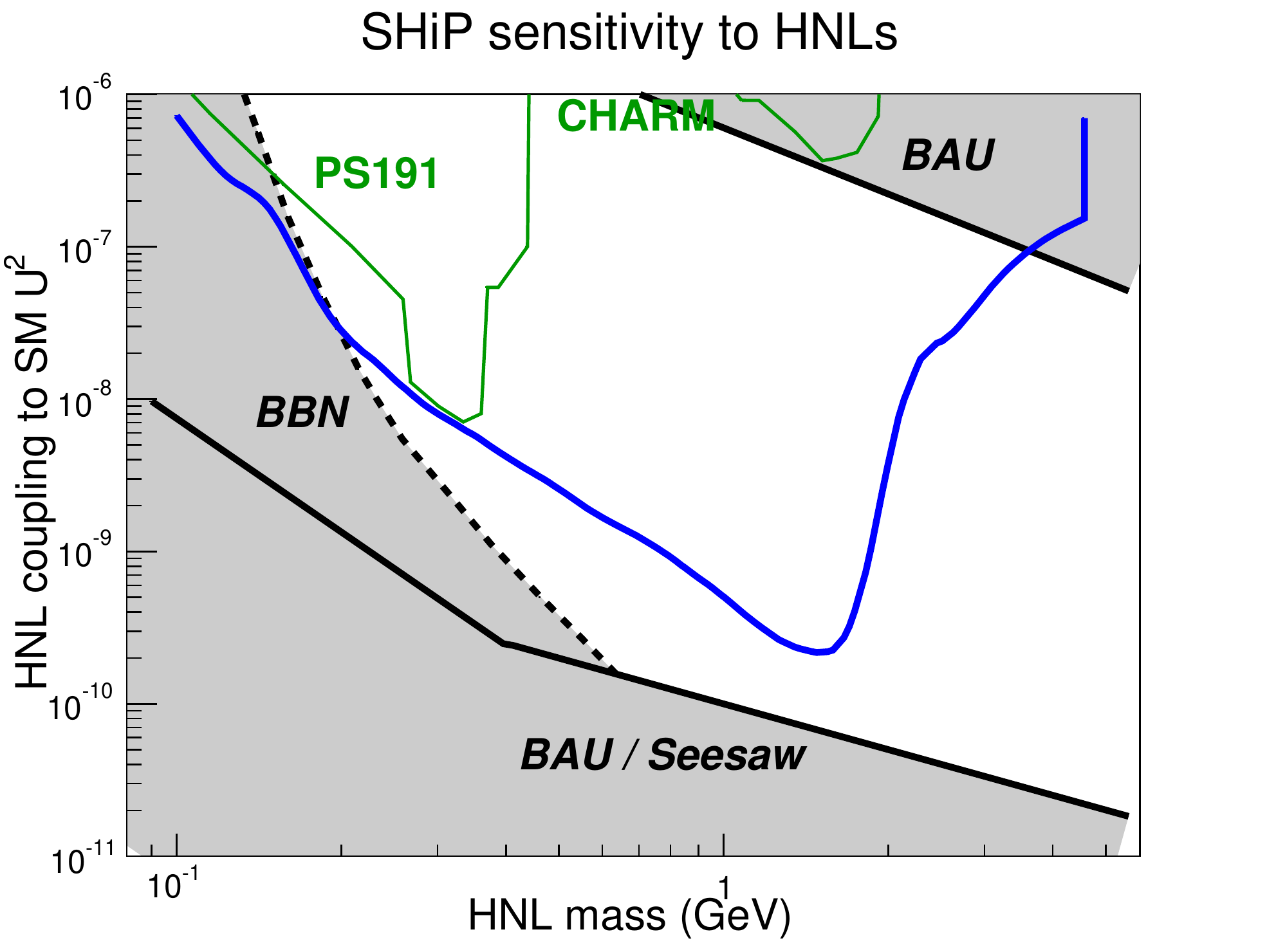}
\includegraphics[width=0.42\linewidth]{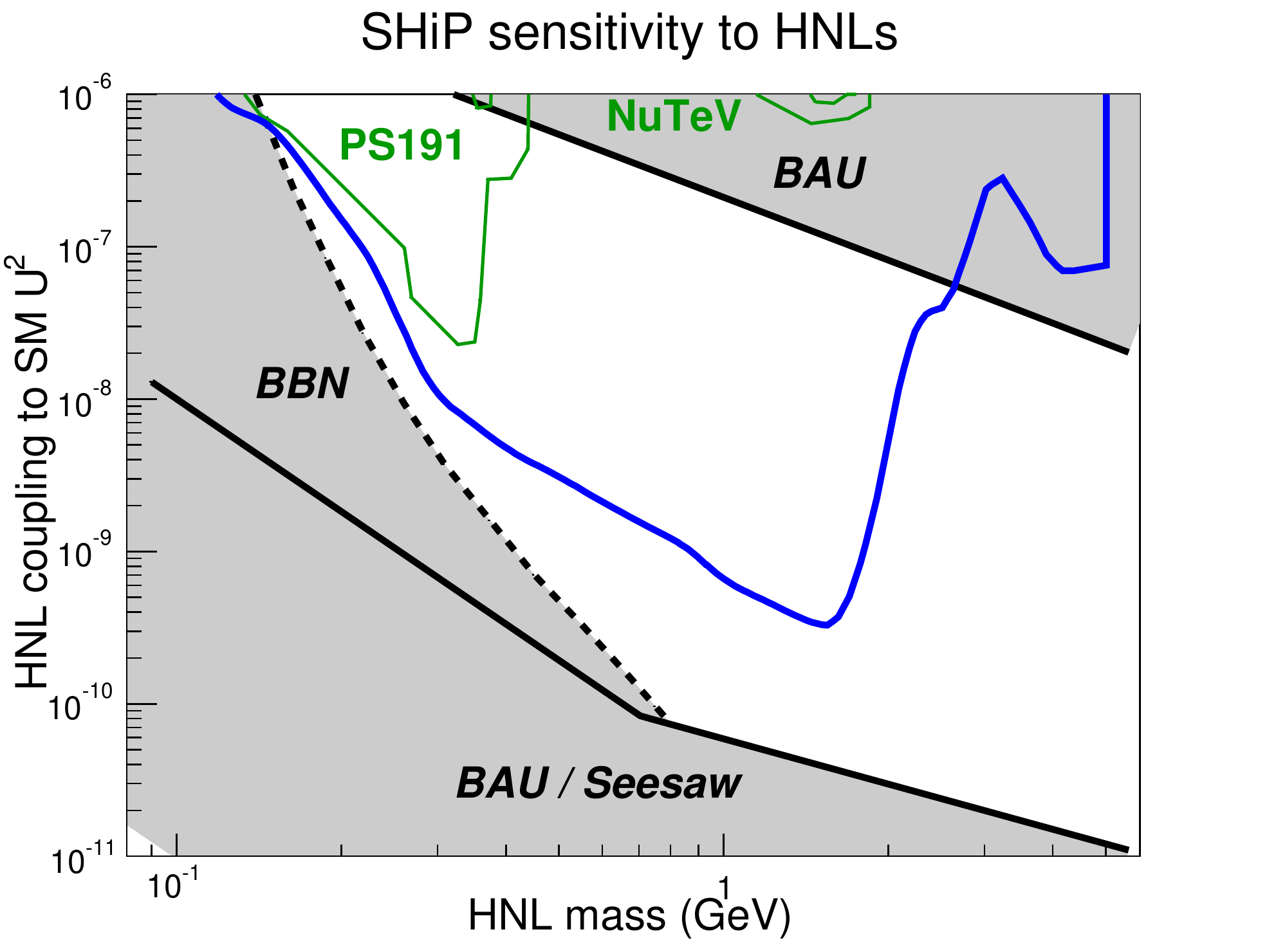}
\includegraphics[width=0.42\linewidth]{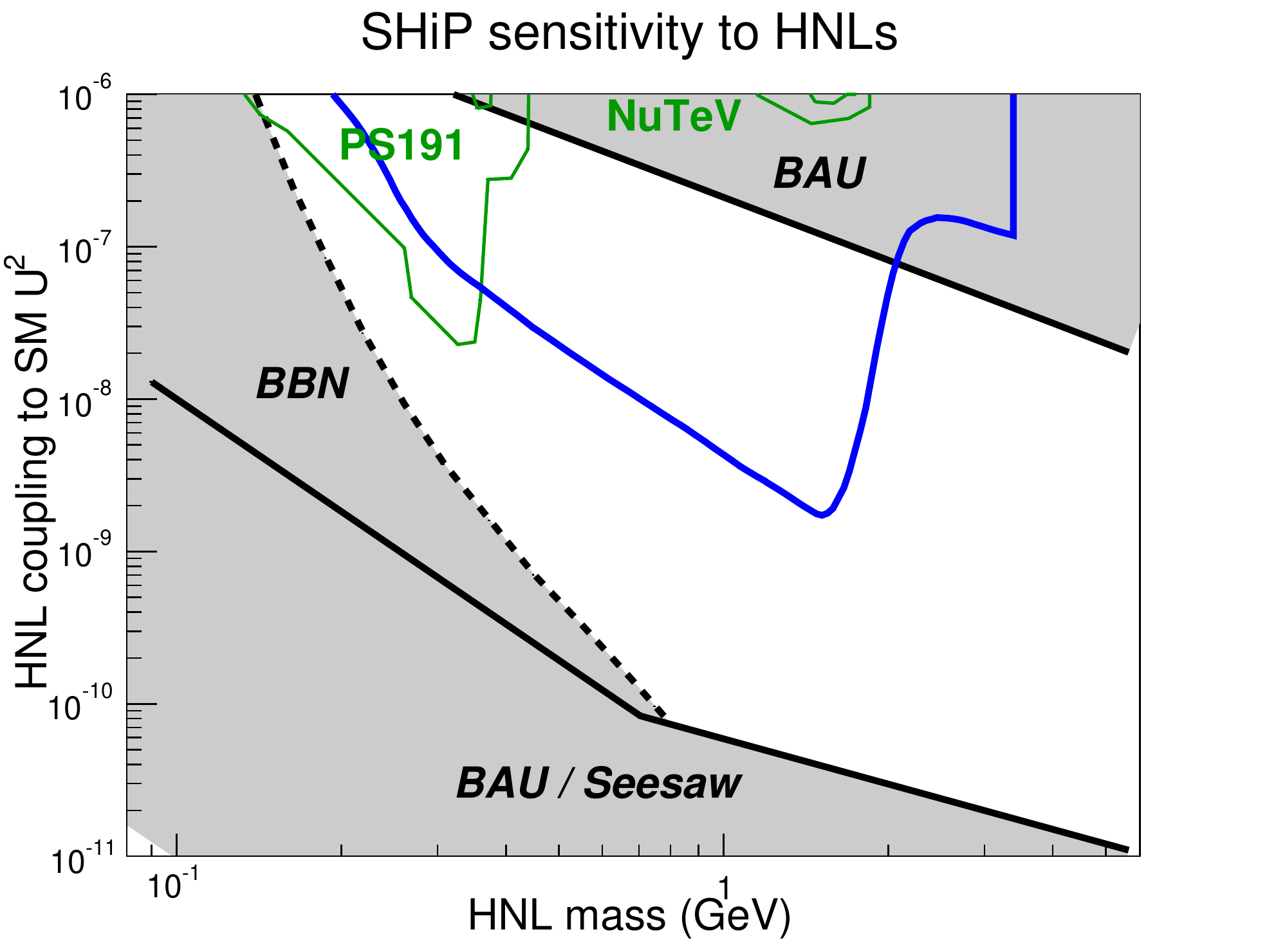}
\caption{Sensitivity regions in the parameter space of the $\nu$MSM, for three scenarios where $U_e^2$, $U_{\mu}^2$ and
$U_{\tau}^2$ dominate respectively (models I, II and III of Ref.~\cite{Gorbunov:2007ak}).}
\label{fig:Uf2limits}
\end{center}
\end{figure}

\begin{figure}[tb]
\begin{center}
\includegraphics[width=0.42\linewidth]{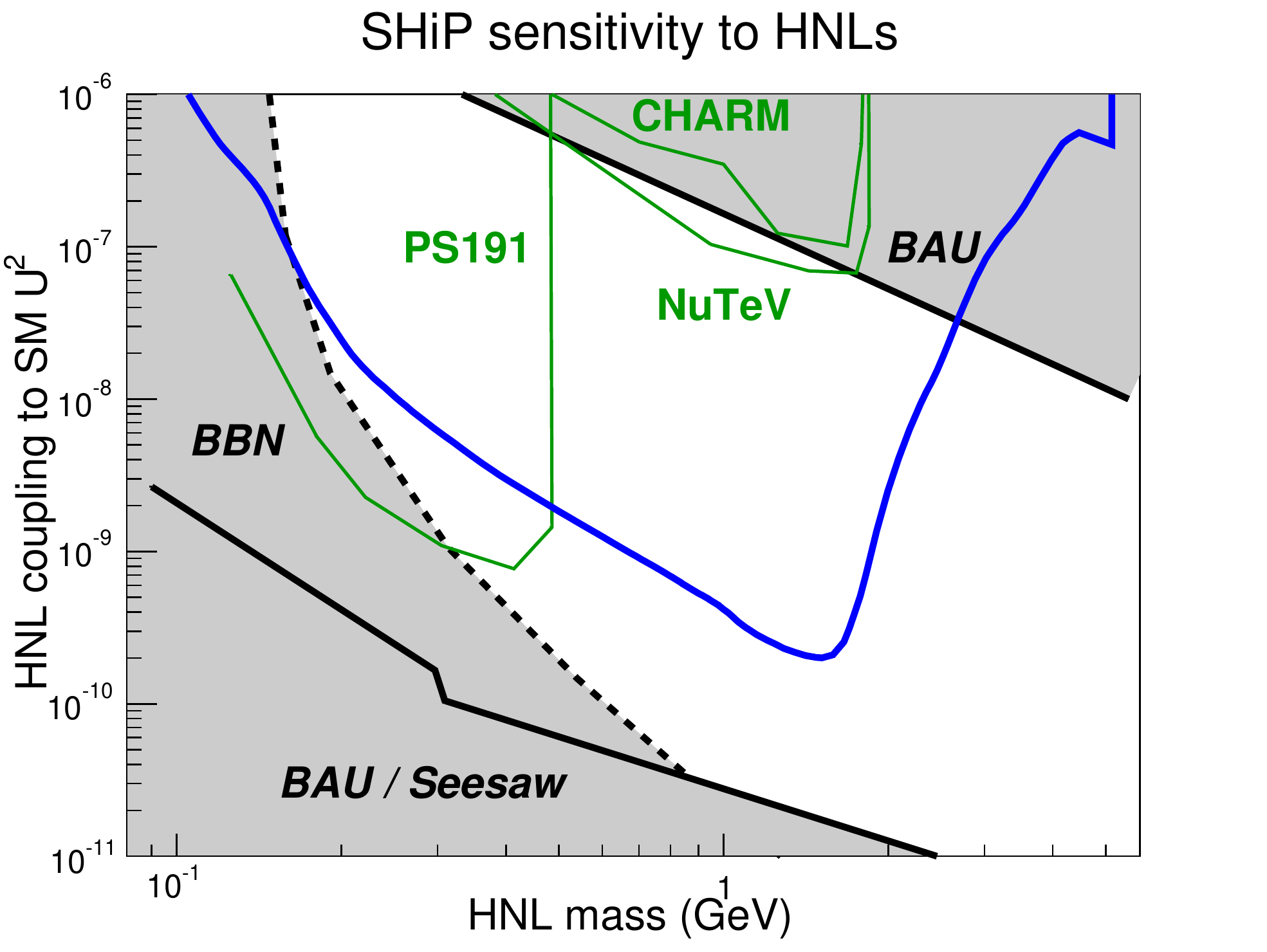}
\includegraphics[width=0.42\linewidth]{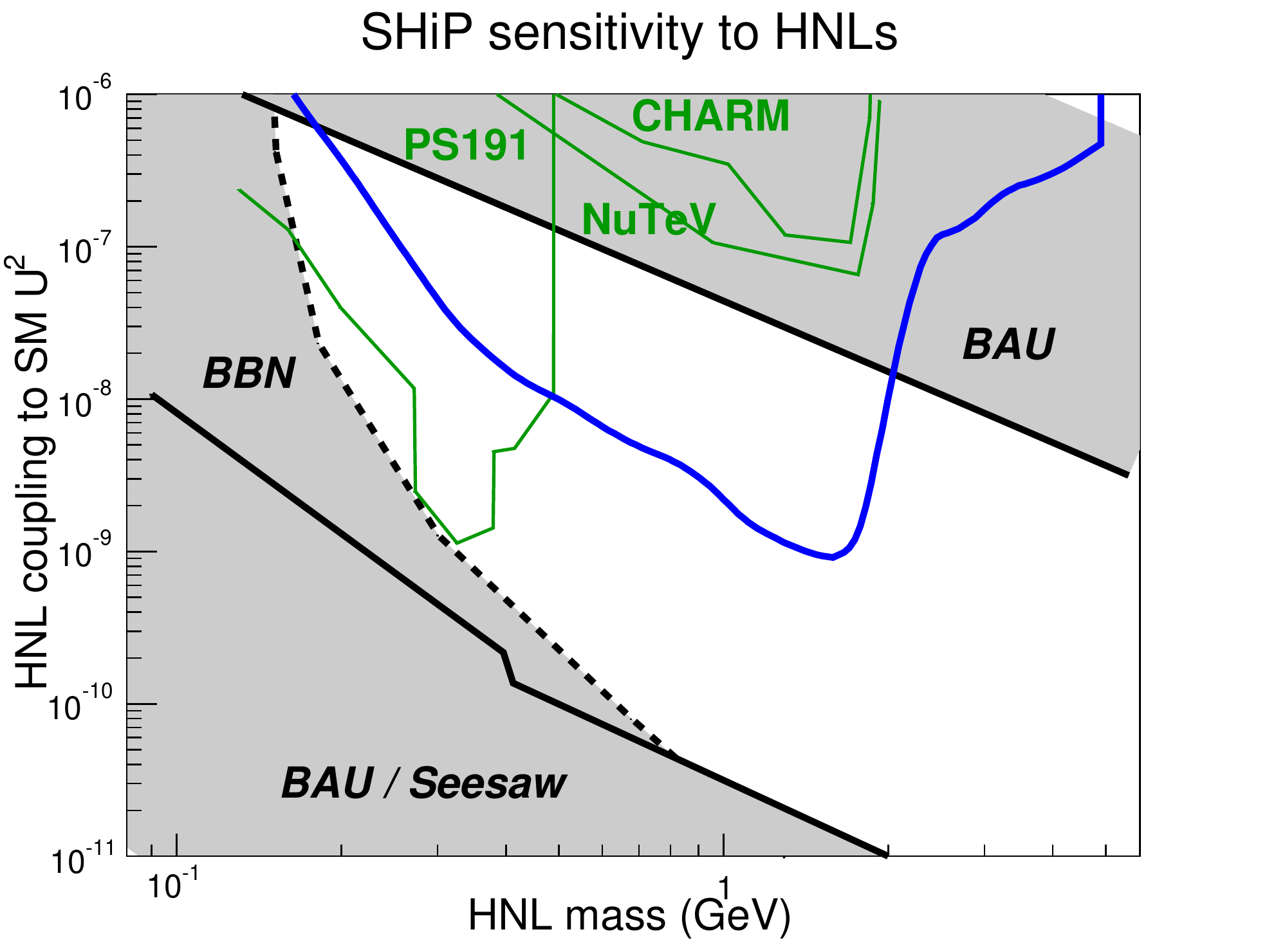}
\caption{Sensitivity regions in the parameter space of the $\nu$MSM, for scenarios IV and V of Ref.~\cite{Canetti:2010aw}, for which baryogenesis was numerically proven.}
\label{fig:U2limitsBAU}
\end{center}
\end{figure}

The sensitivities to scenarios where $U_e^2$, $U_{\mu}^2$ and
$U_{\tau}^2$ dominate (Scenarios I, II and III of Ref.~\cite{Gorbunov:2007ak}) are
shown in Figure~\ref{fig:Uf2limits}.
In addition, the sensitivity to two scenarios IV and V of
Ref.~\cite{Canetti:2010aw}, which are particularly interesting for 
baryogenesis, is shown
in Figure~\ref{fig:U2limitsBAU}.

 Assuming a level of background of $0.1$ events, the curves of
Figure~\ref{fig:Uf2limits} and Figure~\ref{fig:U2limitsBAU} can be interpreted as $3\sigma$
  evidences if two events are observed.

% \begin{thebibliography}{1} 
%  \bibitem{bib:GorbShap2007}D.~Gorbunov and M.~Shaposhnikov, JHEP 0701 (2007) 015, arXiv:0705.1729 [hep-ph]
% \bibitem{bib:CaneShap2010}L.~Canetti and M.~Shaposhnikov, JCAP 1009 (2010) 001 
% \end{thebibliography}
 
\clearpage

%% file: sensitivity/LLP.tex
\subsection{Sensitivity to Dark Photon}
\label{sec:sensitivityDP}

Several models with Hidden Sector predict the existence of massive dark photons ($\gamma^{\prime}$) that mix
  with the SM photon, via a kinetic mixing~\cite{Essig:2013lka}. 
The dark photon is the mediator between the
  visible SM sector and the Hidden Sector. The Lagrangian is given by:

\begin{equation}
{\cal L}_{eff} = {\cal L}_{SM} + {\cal L}_{hidden} +
\frac{\epsilon}{2} A^{\prime}_{\mu\nu}F^{\mu\nu},
\end{equation}

where $\epsilon$ is the mixing between the dark photon and the SM photon. 
Experimentally, dark photons can be produced in decays of mesons
  that decay into photons. The following decays give the main
  contributions at SHiP: $\pi^0 \to \gamma^{\prime} \gamma$, $\eta\to
  \gamma^{\prime}\gamma$, $\omega \to \pi^0 \gamma^{\prime}$ and
  $\eta^{\prime}\to \gamma^{\prime}\gamma$. The second production mechanism is analogous to proton
  bremsstrahlung. The differential production rate of
  $\gamma^{\prime}$ can be calculated in the Weizs\"acker-Williams
  approximation. Indicating the proton momentum carried away in the
  direction of the incoming proton by the
  $\gamma^{\prime}$ with $p_{\parallel}=z P$, and with $p_{\perp}$ the transverse momentum of the dark photon,
  we have:
\begin{equation}
\frac{dN}{dz
  dp_{\perp}^2}=\frac{\sigma_{pA}(s^{\prime})}{\sigma_{pA}(s)} w_{ba
  (z, p_{\perp}^2)}, 
\end{equation}
where $s^{\prime}= 2m_p (E_p-E_{\gamma^{\prime}})$ and
$s=2m_pE_p$. 
%The hadronic cross-section is parametrized as~\cite{} 
%\begin{equation}
%\frac{\sigma_{pp}(s)}{\unit{mb}}=35.45+0.308 \log^2\left(
 % \frac{s}{28.94 \unit{GeV^2}} \right)^{0.545}-33.34\left( \frac{1\unit{GeV^2}}{s^{\prime}}\right)^{0.458}.
%\end{equation} 
The factor $\omega_{ba}$ is given by
\begin{equation}
\begin{split}
\omega_{ba}(z, p^2_{\perp})=&\frac{\epsilon^2 \alpha_{QED}}{2\pi
  H}\left[ \frac{1+(1-z)^2}{z}-2z(1-z)\left(
    \frac{2m_p^2+m_{\gamma^{\prime}}^2}{H}-z^2\frac{2m_p^4}{H^2}\right)
\right. \\ &+
  \left. 2z(1-z)(z+(1-z)^2)\frac{m_p^2m_{\gamma^{\prime}}^2}{H^2}+
  2z(1-z)^2\frac{m_{\gamma^{\prime}}^2}{H^2}\right], 
\end{split}
\end{equation}

\noindent
where
$H(p_{\perp}^2,z)=p_{\perp}^2+(1-z)m_{\gamma^{\prime}}^2+z^2m^2_p$. 
The hadronic cross section is related to the proton-proton cross
section by $\sigma_{pA}(s)=f(A)\sigma_{pp}(s)$. The production rate of
the dark photon is then multiplied by the form factor
$F_1^2(m_{\gamma^{\prime}})=(1+\frac{m_{\gamma^{\prime}}^2}{m_D^2})^{-2}$,
where the Dirac mass is defined as a function of the Dirac radius as
$m_D^2=12/r_D^2$ where $r_D=0.8$~fm~\cite{Gorbunov:2014wqa}. The form factor $F_1$ takes into account
that above $\Lambda_{QCD}$ our approximation breaks and we should
consider parton bremsstrahlung rather than proton bremsstrahlung. 
%\begin{equation}
%\frac{\sigma_{p\gamma^{\prime}}(s^{\prime})}{\sigma_{p\gamma^{\prime}}(s)}=\frac{\sigma_{pp}(s^{\prime})}{\sigma_{pp}(s)}\times
%\end{equation}

Dark photons decay into pair of SM particles by mixing again with the
  SM photon. The dark photon decay width into leptons is given by 
\begin{equation}
\Gamma(\gamma^{\prime}\to \ell^+ \ell^-) = \frac{1}{3} \alpha_{QED}
m_{\gamma^{\prime}}\epsilon^2\sqrt{1-\frac{4m^2_{\ell}}{m^2_{\gamma^{\prime}}}}\left(
  1+\frac{2m^2_{\ell}}{m^2_{\gamma^{\prime}}} \right), 
\end{equation}
 
\noindent
where $m_{\ell}$ is the lepton mass. The partial decay width into
hadrons is given by 

\begin{equation}
\Gamma (\gamma^{\prime}\to h \overline{h}) =
\frac{1}{3}\alpha_{QED}m_{\gamma^{\prime}}\epsilon^2
R(m_{\gamma^{\prime}}), 
\end{equation}

\noindent
where 
\begin{equation}
R(\sqrt{s})=\frac{\sigma (e^+e^-\to \text{hadrons})}{\sigma(e^+e^- \to
  \mu^+ \mu^-)}.
\end{equation}

\begin{figure}[tb]
\begin{center}
\includegraphics[width=0.75\linewidth]{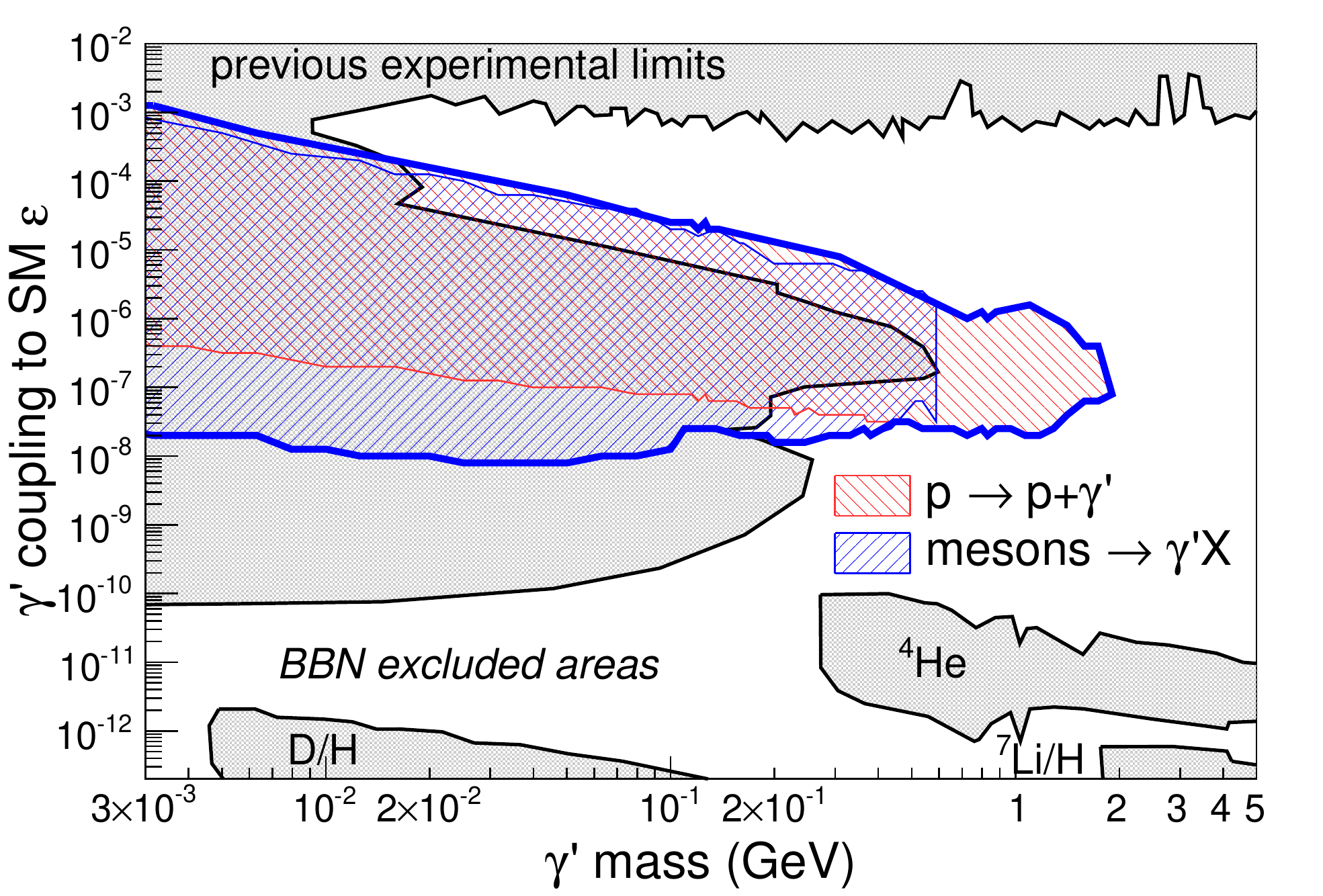}
\caption{SHiP sensitivity to the dark photon as a function of the
  mixing with the SM photon $\epsilon$ and the dark photon mass. The
  production via mesons and the proton bremsstrahlung are shown separately.}
\label{fig:DarkPhoton}
\end{center}
\end{figure}

The SHiP sensitivity to dark photons (Fig.~\ref{fig:DarkPhoton}) is obtained with the same fast
simulation that was used for the HNLs, and the same procedure
described in the previous section. 
%On the left figure the sensitivity of meson decays and proton
%  bremsstrahlung are shown separately, while in the right figure they
%  are shown combined. 

\input{sensitivity/Scalars}

%% file: sensitivity/Scalars.tex
\clearpage
%====================================
\subsection{Sensitivity to light hidden scalars mixing with the Higgs}
\label{sec:sensitivityLS}
%====================================

\vskip 2mm
Additional scalars are required to exist in various extensions of the SM. %~\cite{ref1, ref2, ref3...2,3,4,5,7,8 di 1310.8042v1}
Speculations regarding a possible Higgs portal coupling to new, weakly-coupled 'hidden' scalar particles have intensified after the 
discovery of the Higgs boson. The SHiP experiment has the potential to probe this portal provided that the new states connected to it are 
light with masses of $\cal O$(GeV/c$^2$). The purpose of this chapter is to survey the SHiP sensitivity to a light hidden scalar particle $S$, 
which mixes with the SM Higgs with a parameter $\sin^2 \theta$ and which decays only to SM particles with a coupling constant $\sin^2 \theta$ 
compared to the SM Higgs.

\vskip 2mm
The production of the light hidden scalars is expected to occur mostly via meson decays, namely $B$ and $K$ mesons as $D$ decays are highly 
suppressed by the CKM mechanism.
%\begin{center}
%\includegraphics[width=0.3\textwidth]{Diagram.pdf}
%\end{center}
At a fixed target experiment the production of light hidden scalars is given by \cite{Clarke:2013aya}:
\[
 N_{S} \simeq N_{pot} \times ( 2 \cdot \chi_{s} \cdot f_{S1} \cdot 0.5 \cdot BR(K^{\pm} \to \pi^+ S) + 
                                               2 \cdot  \chi_{s} \cdot f_{S2} \cdot 0.25 \cdot BR(K_L \to \pi^0 S) +
                                              2 \cdot \chi_b \cdot BR(B \to X_s S))  
\]
\noindent
where:
\begin{itemize}
\item[-]  $N_{pot}$ is the total number of protons on target;
\item[-]  $\chi_s = {\sigma_{pp \to s \overline{s} X}/ \sigma_{pp \to Y}}$ and 
$\chi_b = \sigma_{pp \to b \overline{b} X}/\sigma_{pp \to Y}$ are the ratios of the production cross sections of $s$ and $b$ quarks with respect to the total $pp$ cross section;
\item[-] $f_{S1}$ and $f_{S2}$ are the fractions of $K^+$  and $K_L$  that decay before being absorbed in the target;
\item[-] $BR(K^{\pm} \to \pi^{\pm} S)$, $BR(K_L \to \pi^0 S)$, $BR(B \to X_s S)$ are the branching fractions of $K^{\pm}$, $K_L$ and $B$ into the scalar particle
(see Refs.~\cite{Batell:2009jf,Schmidt-Hoberg:2013hba} for explicit formulae).
\end{itemize}

% \begin{itemize}
%\item[-] 
% \item[-] $\chi_s = 1/7, \chi_b = 1.7 \times 10^{-7}$ are the production fractions of $s$ and $b$ quarks respectively for protons on target of 400 GeV/c ;
% \item[-] $BR(K^+ \to \pi^+ S) \simeq sin^2(\theta) \times 0.002 \times { 2 |\vec{p_{S}}| \over m_K}$;
% \item[-] $ BR(K_L \to \pi^+ S) = BR(K^+ \to \pi^+ S) \times {\Gamma(K^+) \over \Gamma(K_L)}$;
% \item[-] $ BR(B^+ \to K^+ S) \simeq \sin^2(\theta)   \times 0.5 \times { 2 |\vec{p_{S}}| \over m_B} \times F(q^2)$;
%\item[-] $F(q^2) = 0.33 / (1-q^2/38 GeV^2)$, $q^2 = m_{S}$.
%\end{itemize}

\noindent
For $m_S < 2 m_{\mu}$, the light scalar decays almost exclusively to $e^+ e^-$. 
Above $2 m_{\mu}$, the decay  $S \to \mu^+ \mu^-$ takes over until the $2 m_{\pi} \simeq 280$ MeV/c$^2$ threshold.
For heavier scalars, additional decay channels are open ($\pi \pi$, $KK$, $\eta \eta$, $\tau \tau$, $DD$, ..). 
Currently there are significant theoretical uncertainties in the computation of their rates~\cite{Voloshin:1985tc, Raby:1988qf, Truong:1989my, Donoghue:1990xh}.   
%In Fig.~\ref{fig:BRmm} the branching fraction of the scalar into muon pairs is shown.

\vskip 2mm
The partial decay width of the new light scalar into muon pairs is given by:
\begin{equation}
\Gamma (S \to \mu \mu) =   {\sin^2\theta \, m^2_{\mu} m_{S} \over 8 \pi v^2}  \beta_{\mu}^3
\end{equation}
\noindent
where $m_{S}$ is the mass of the scalar particle, $v \sim 246$ GeV is the Higgs condensate, and $\beta_{\mu} = \sqrt{1-4{m^2_{\mu}/m^2_{S}}}$ is a kinematic factor.
%For $m_{S} = m_{B^+} - m_{K^+}$ the following channels are open:
%$ee$, $\mu \mu$, $\pi \pi$, $\eta \eta$, $DD$, $\tau \tau$}  and the total width of the scalar particle will be given by
%$ \Gamma_{\rm tot} = (\Gamma_{ee} + \Gamma_{\mu\mu} + \Gamma_{\pi \pi} + \Gamma_{\eta \eta} + \Gamma_{DD} + \Gamma_{\tau\tau} $.

%\begin{figure}[tb]
%\begin{center}
%\includegraphics[width=0.8\linewidth]{sensitivity/BRmm.pdf}
%\caption{}
%\label{fig:BRmm}
%\end{center}
%\end{figure}

\vskip 2mm
In the computation of the total decay width and following the prescription in Ref.~\cite{Schmidt-Hoberg:2013hba}, the following relations for the partial decay widths
are assumed:

\begin{equation}
 \Gamma_{\mu\mu} : \Gamma_{\pi \pi} : \Gamma_{KK} : \Gamma_{\eta \eta} : \Gamma_{DD} : \Gamma_{gg} = 
m_{\mu}^2 \beta_{\mu}^3 : 3  (m_{u}^2 + m_{d}^2) \beta^3_{\pi} : 3 {9 \over 13} m^2_s \beta^3_K : 3 {4 \over13} m^2_s \beta^3_{\eta} :
m^2_c \beta^3_D : m^2_{\tau} \beta^3_{\tau}.
\end{equation}

\begin{figure}[tb]
\begin{center}
\includegraphics[width=0.6\linewidth]{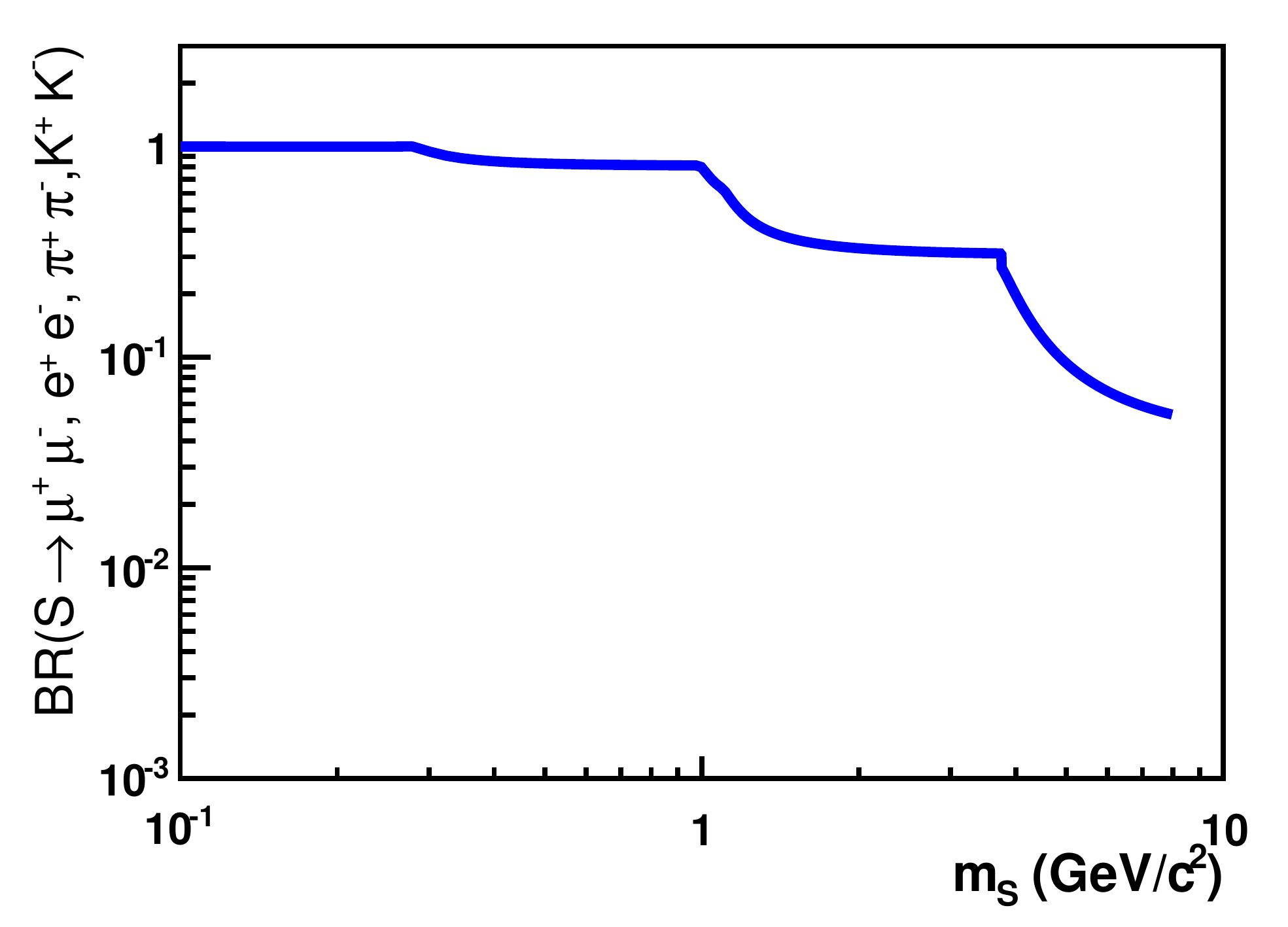}
\caption{ Total branching fraction for decays into $e^+e^-,\mu^+\mu^-, \pi^+\pi^-, K^+K^-$ final states as a function of the mass for a light scalar particle mixing with the Higgs.}
\label{fig:BR}
\end{center}
\end{figure}

\vskip 2mm
SHiP will be able to reconstruct all the decay final states. As an
example, the SHiP sensitivity to light scalar particles  
decaying only to two-body final states,  $S \to X^+ X^-$ where $X = e, \mu, \pi, K$, is computed.
Figure~\ref{fig:BR} shows the total branching fraction in two-body final state as a function of the mass of the scalar particle in the mass range interesting for SHiP.

%---------------------------
%---------------------------

\vskip 2mm
The number of events of $S \to X^+ X^-$ reconstructed in the HS detector is given by:
\begin{equation}
 N_{obs} = N_{S} \times \sum_{X=ee,\mu\mu,\pi\pi,KK} BR(S \to X^+ X^-)  \times {\cal A}(X) \times \epsilon (X)
\label{eq:scalar}
\end{equation}
\noindent 
where the acceptance $\cal A$ for a light scalar particle %of a given mass $m_S$, coupling parameter $\sin^2 \theta$ and 
decaying into a final state $X$
%\[
 %{\cal A}(\sin^2 \theta, X) = \left( e^{-Dmin/\beta \gamma c \tau} - e^{-Dmax/\beta\gamma c \tau} \right ) \times P(X^+, X^-) 
%{\cal A}(\sin^2 \theta, m_S, X) = \int_{\text{SHiP}} e^{-l/\gamma c\tau}\, dl
%.P_{HNL} \times P(X^+, X^-) 
%\]
%\noindent
is given by the product of the probability that the light scalar particle decays inside the fiducial volume
and  the probability  ${\cal P}_{vtx}(X^+, X^-)$ that the two charged 
tracks in the final state are reconstructed in the HS magnetic spectrometer. %, as explained in Section~\ref{sec:detailedP}.
The efficiency $\epsilon (X)$ is the product of the reconstruction and selection efficiencies for two-body decays.
This has been estimated to $40\%$ below the $2 \cdot m_{\mu}$ threshold and 70\% above this threshold (see Section~\ref{sec:effectreco}).

% \footnote{The fiducial volume is defined in the $x,y$ direction by the ellipse of the vacuum tank and in the $z$ direction by the position of the 
% straw veto placed at 5 m downstream of the entrance window of the decay vessel and by the position 
% of the first straw tracker station.} 

The toy Monte Carlo technique is used to assess the sensitivity contours in the mass-coupling parameter space.
This is done by generating light scalar particles from $B$ and $K$ decays of different masses and different values of the coupling parameter, letting them decay in the chosen final states,
and computing the corresponding acceptance as described in Section~\ref{sec:sensitivity}.
%Assuming no events observed, the upper limit on the mixing parameter $\sin^2\theta$ at 90\% C.L. as a function of 
%the mass of the light scalar particle can be computed by using toys experiments.

\begin{figure}[tb]
\begin{center}
\includegraphics[width=0.7\linewidth]{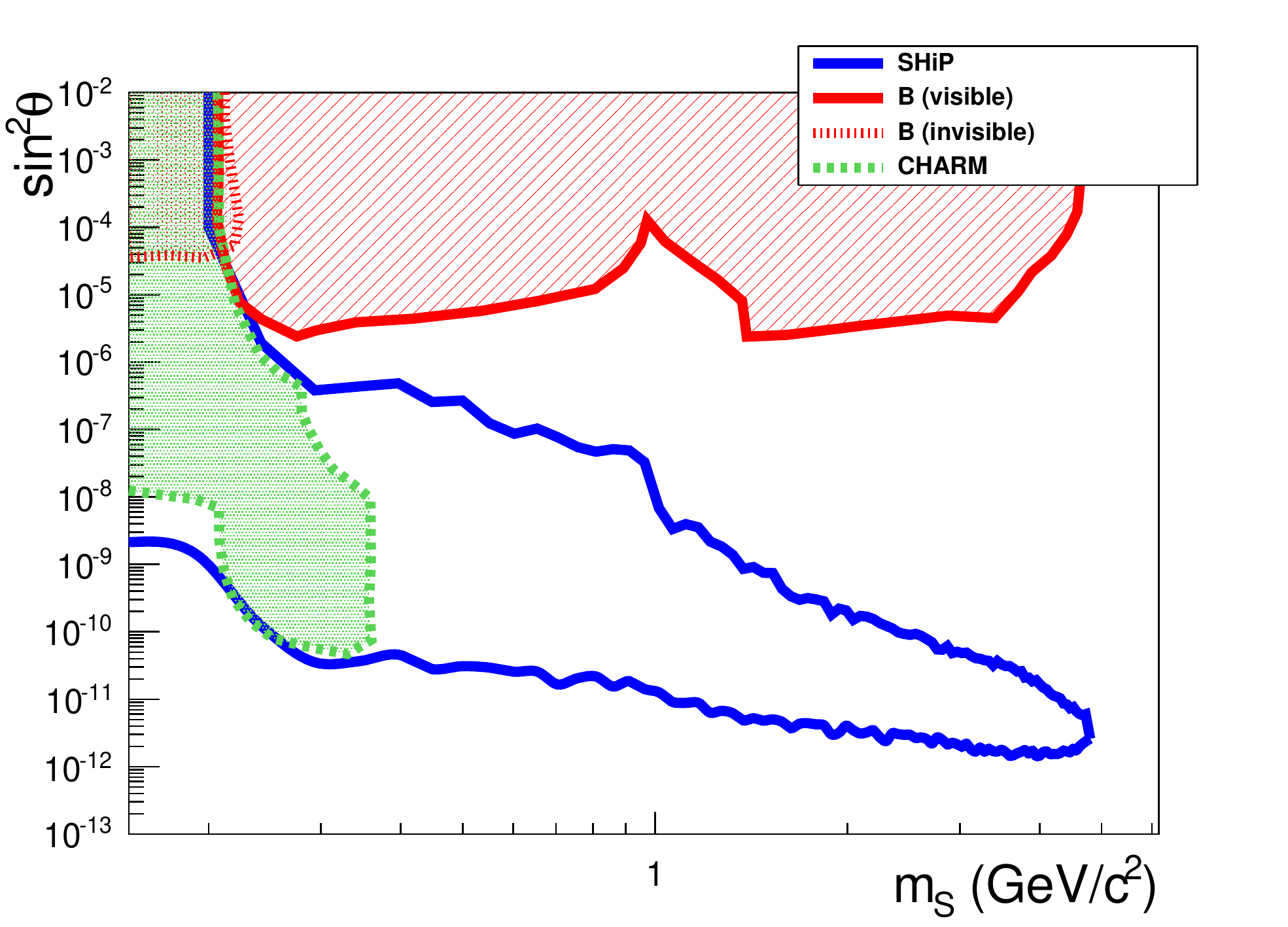}
\caption{SHiP exclusion limit at 90 \% C.L.  for a light hidden scalar particle of mass $m_S$ coupling to the Higgs with $sin^2 \theta$ mixing parameter and decaying in $e^+e^-,\mu^+\mu^-, \pi^+\pi^-, K^+K^-$ final states (solid blue line). Red dashed area is the excluded region from $B$-factories in the visible modes, red dotted area is the excluded region from $B$-factories in the invisible modes and green shaded area in the exclusion region from the CHARM experiment.}
\label{fig:LightInflaton}
\end{center}
\end{figure}

\vskip 2mm
The sensitivity contour is shown in Figure~\ref{fig:LightInflaton}. This plot shows the exclusion limit at 90\% CL in case no event is observed
 or the $3 \cdot \sigma$ evidence in case two events are observed with 0.1 expected background events. The computation is
based on the following input values: $N_{\rm pot} = 2 \cdot 10^{20}$, 
%$\sigma_{b \overline{b}} \sim 3.0$ nb ~\cite{Lourenco},   
%$\sigma_{pp} = 10.7$ mb,   
$\chi_b \sim 1.6 \cdot 10^{-7}$,  $\chi_S \sim 1/7$, $f_{S1} \sim 0.2 \%$ and $f_{S2} \sim 0.02\%$.
The kaon decays dominate below the $m_K - m_{\pi}$ threshold while the $B$ decays dominate above this threshold. Despite the huge
kaon production cross section, the impact of the kaons in the SHiP sensitivity is largely reduced by the fact that most of them 
are stopped in the target before decaying.
For example, for $m_S$ = 4 GeV/c$^2$ and $BR(B \to S K) \sim sin^2 \theta \sim 2\cdot 10^{-12}$ , following Eq.~\ref{eq:scalar} and 
Figure~\ref{fig:BR}, we have $N_{\rm obs} = N_{\rm pot} \times 2 \chi_b \times BR(B \to S K) \times BR(S\to X^+X^-, X=e,\mu,\pi,K) \times {\cal A} = 0.3 -3.2 $ for ${\cal A}$ ranging from 1\% to 10\%.

\vskip 2mm
Notice that the SHiP experiment, with a sample of $N_{pot} \sim 2 \cdot 10^{20}$ protons on target collected in five years 
of data taking will probe a unique range of couplings and masses, thus complementing existing limits from rare $B$
meson decays $B \to  K S$ from $B$-factories, with $S$ decaying in the visible dilepton channel~\cite{Wei:2009zv,Aubert:2008ps} and the invisible
channel~\cite{Chen:2007zk,delAmoSanchez:2010bk} and limits from the CHARM experiment~\cite{Bergsma:1985qz}.

%% file: nusensitivity/nutauphysics.tex
\section{Physics with $\nu_{\tau}$}
\label{sec:yield}

\begin{figure}
\centering
\includegraphics[scale=0.5]{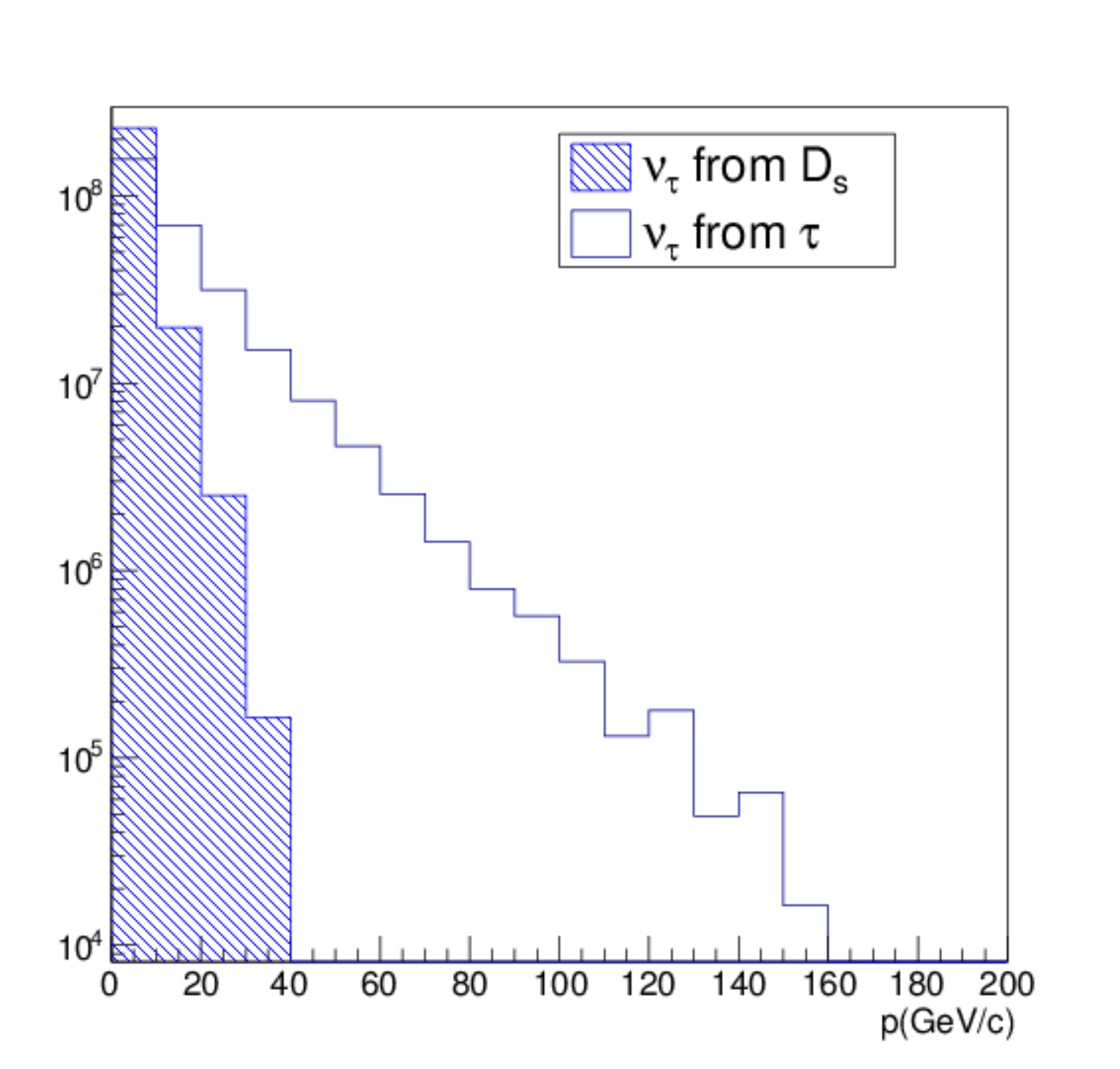}
\caption{Momentum distribution of $\nu_\tau$ produced in $\tau$  and $D_s$ decays.}
\label{fig:nutau_prod}
\end{figure}

Tau neutrinos are copiously produced in the leptonic decay of a $D_s^-$ meson into $\tau^-$ and $\bar{\nu}_\tau$, and the subsequent decay of the $\tau^-$ into a $\nu_\tau$. The production of $D_s^\pm$ through proton interactions in the target is symmetric and therefore an equal contribution is given by $D_s^+$. Due to the different kinematics of the decays, the energy distribution of $\nu_\tau$ from $D_s^+$ is much softer when compared to that from $\tau^-$, as shown in Figure~\ref{fig:nutau_prod}.

The number of $\nu_\tau$ and $\bar{\nu}_\tau$  emerging from the molybdenum target
can be estimated as follows:
\begin{equation}
N_{\nu_{\tau} + \bar{\nu}_{\tau}} = 4 N_p \dfrac{\sigma_{c\bar{c}}}{\sigma_{pN}} f_{D_s} Br(D_s \rightarrow \tau)  = 2.85 \cdot 10^{-5} N_p
\label{eq:flux}
\end{equation}
 where 
\begin{itemize}
\item $N_p$ is the number of interacting protons (all incoming ones);
\item$\sigma_{c\bar{c}} = 18.1 \pm1.7$ $\mu$barn~\cite{doublecharm} is the associated charm production  per nucleon.
\item $\sigma_{pN} = 10.7$ mbarn is the hadronic cross section per nucleon in a Mo target. The inelastic hadronic cross section per nucleon on a target with $A$ nucleons can be expressed as $\sigma_{pN} = \sigma_{pA}/A = 1/\lambda_{int} \rho N_A$ where $\lambda_{int}$ is the nuclear interaction length, $\rho$ is the target density and $N_A$ is the Avogadro number. The inelastic cross section $pA$ shows the dependence $A^{0.71}$ on the mass number~\cite{massnumber}; 
\item $f_{D_s} = (7.7 \pm 0.6 ^{+0.5}_{-0.4})\%$~\cite{f_ds} is the fraction of $D_s$ mesons produced;
\item $Br(D_s \rightarrow \tau) = (5.54 \pm 0.24)$\%~\cite{pdg} is the $D_s$ branching ratio into $\tau$;
\item the factor 4 accounts for the charm pair production and for the two $\nu_\tau$ produced per $D_s$ decay.
\end{itemize}
For 2$\cdot$10$^{20}$ protons on target, Equation~\ref{eq:flux} gives $N_{\nu_{\tau} + \bar{\nu}_{\tau}} = 5.7 \cdot 10^{15}$. 

The neutrino flux obtained with Equation~\ref{eq:flux}, together with values based on experimental measurements, was used for the normalization. The full kinematics of the $D_s$ decay was included in the simulation performed with Pythia8.

\begin{figure}
\centering
\includegraphics[scale=0.35]{requirements/spectrum_had_stop.pdf}
\includegraphics[scale=0.35]{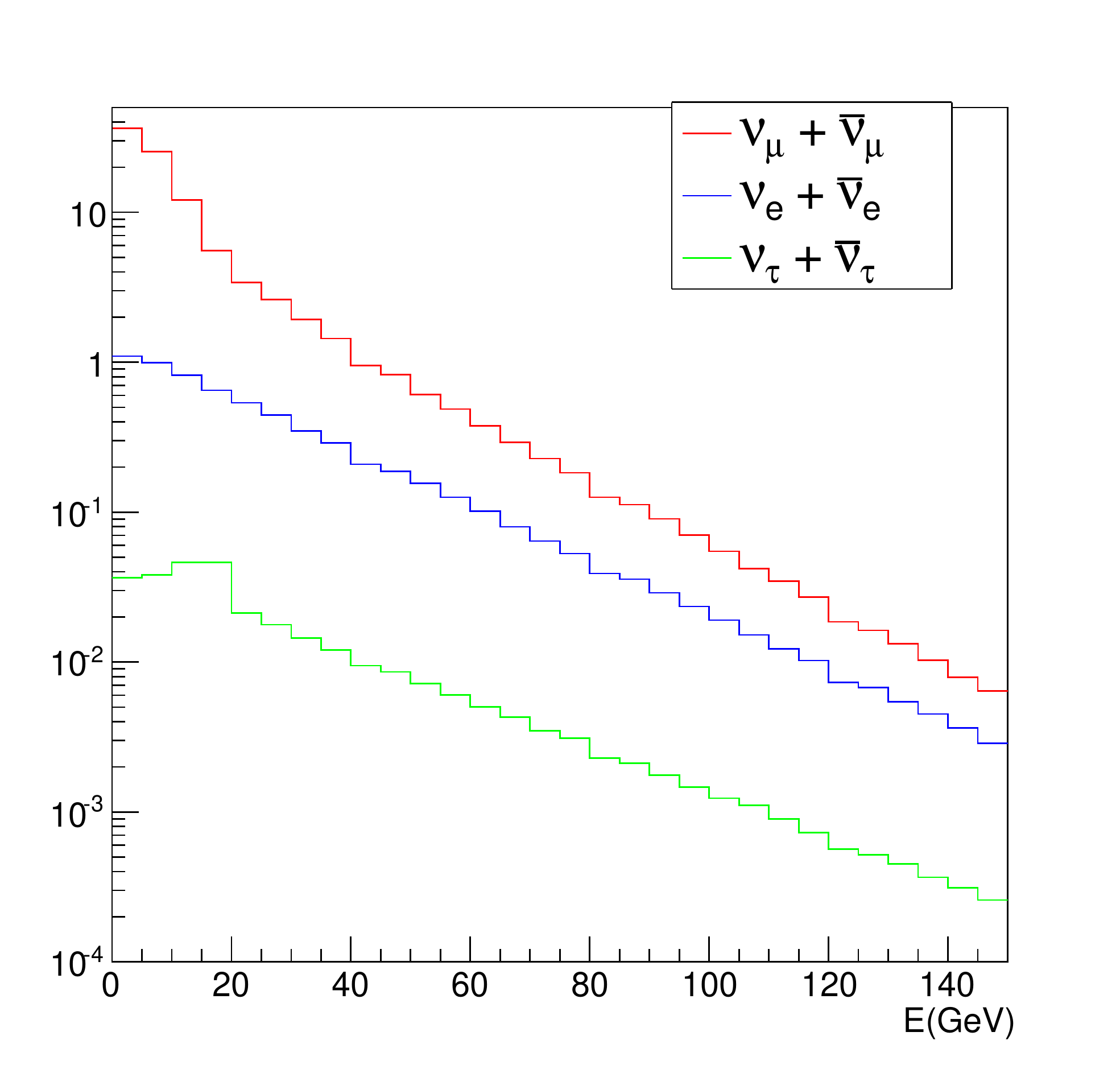}
\caption{Energy spectra of the different neutrino flavors produced in the proton target (left) and reaching the neutrino target (right). A 0.5 GeV cut is applied for $\nu_\mu$ and $\nu_e$. The total number of neutrinos is normalized to 100.}
\label{fig:spectrum2}
\end{figure}

Unlike the tau neutrinos produced by charmed hadrons decays only, electron and muon neutrinos are also induced by the decay of soft $\pi$ and $K$ produced as secondary particles in proton interactions. A Geant4 based simulation of the proton target and of the hadron stopper provides the spectra shown in Figure~\ref{fig:spectrum2} and the yields reported in Tab.~\ref{tab:neutrino_all} once an energy threshold of 0.5 GeV is applied. It is worth noting that neutrinos with an energy lower than 0.5 GeV have a very small cross section and a wider angular distribution, thus leading to a negligible contribution.
\begin{figure}
\centering
\includegraphics[scale=0.6]{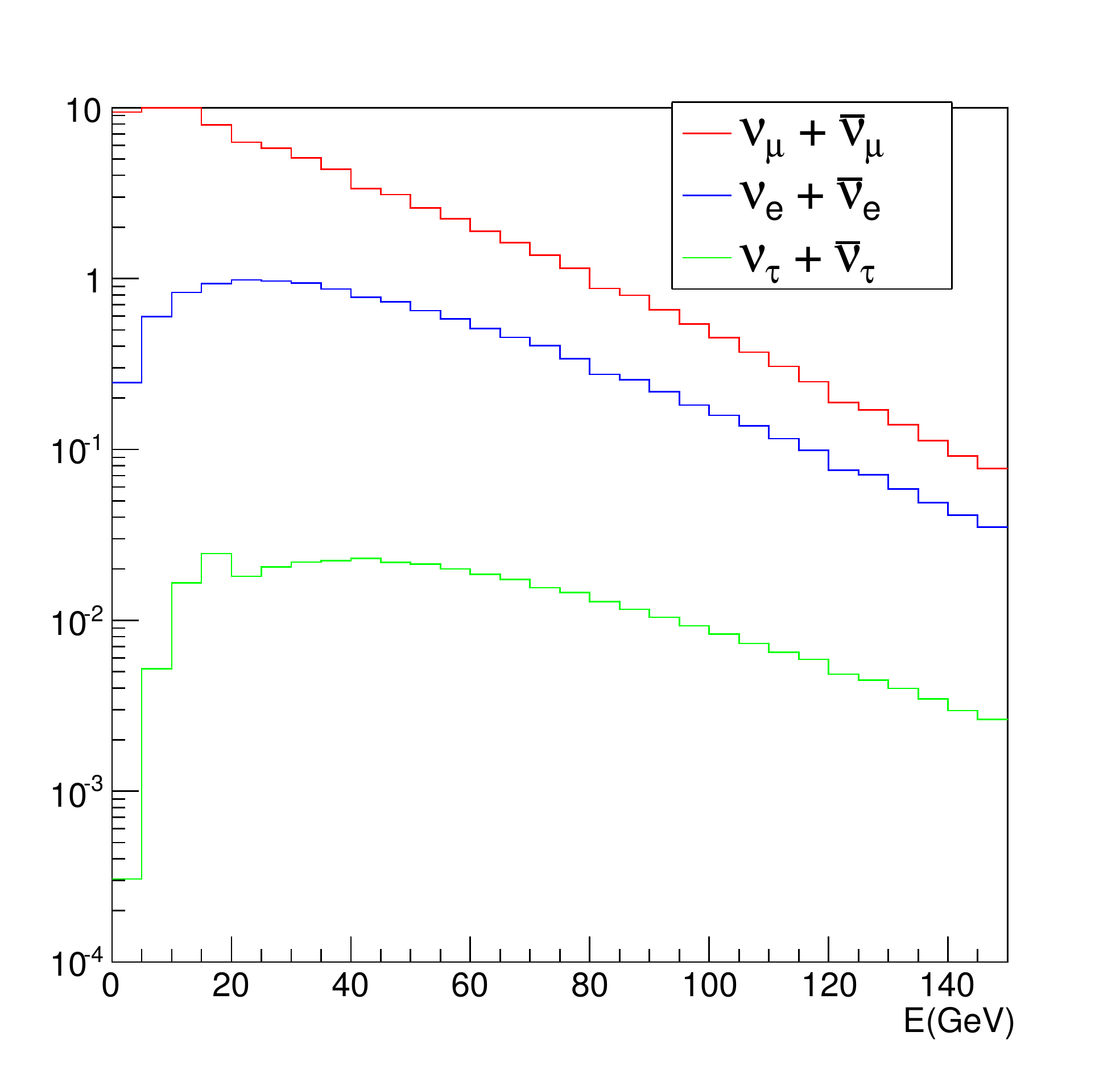}
\caption{Energy spectrum of the different neutrino flavors interacting in the neutrino detector. A 0.5 GeV cut is applied for $\nu_\mu$ and $\nu_e$. The total number of neutrinos is normalized to 100.}
\label{fig:spectrum_int}
\end{figure}

The distance of the neutrino detector from the downstream end of the proton target ($\sim56.5$~m) as well as its compact transversal area ($\sim1.3$~m$^2$) reduce the neutrino flux to about 1.3~\% of the produced neutrino flux. The acceptance for the $\nu_\tau$ component amounts to 5~\%.  The energy spectrum at the neutrino detector shows higher mean values because the low momentum component goes out of the geometrical acceptance. The average energies become about 8 GeV, 28 GeV and 28 GeV for muon, electron and tau neutrinos, respectively, as reported in the central column of Tab.~\ref{tab:neutrino_all} and in Figure~\ref{fig:spectrum2} (right).

\begin{figure}
\centering
\includegraphics[scale=0.35]{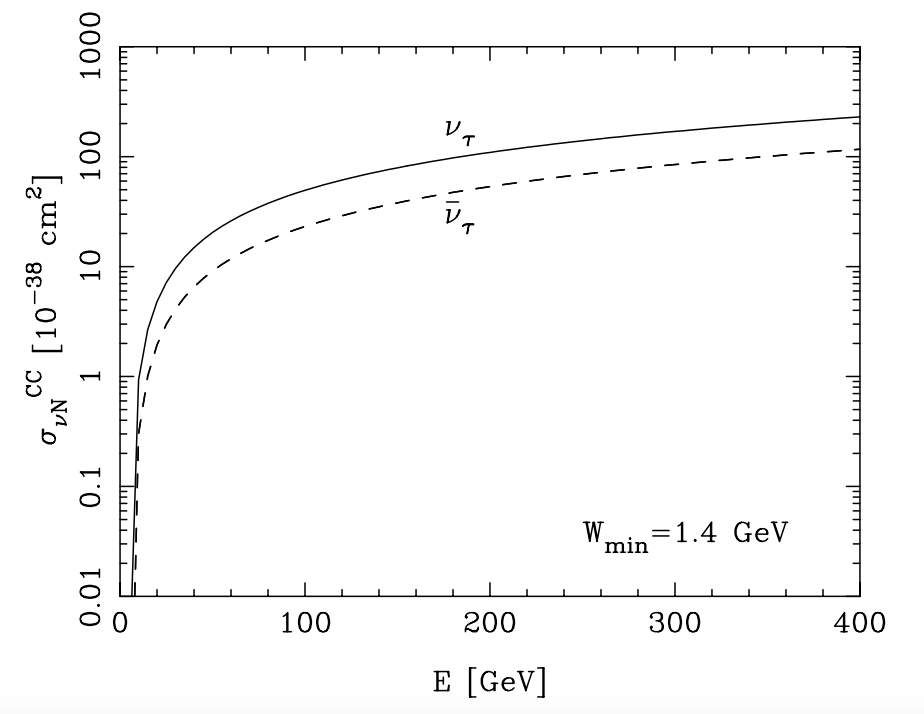}
\caption{Charged-current deep inelastic $\nu_\tau$ and $\bar{\nu}_\tau$ cross sections~\cite{reno}.}
\label{fig:cross_sec}
\end{figure}

The number of charged-current (CC) deep-inelastic (DIS) interactions in the neutrino target is evaluated by the convolution of  the neutrino spectrum at the detector site with the cross section.
The CC deep-inelastic neutrino cross section per nucleon, expressed in cm$^2$, can be written as $\sigma_{\nu_l}=  (6.75 \pm 0.09)\cdot10^{-39} E$ (GeV) $(l=\mu,e)$~\cite{minos}. Anti-neutrino deep inelastic cross section is one half of the neutrino cross section. The $\nu_\tau$ and $\bar{\nu}_\tau$ cross sections used are shown in Figure~\ref{fig:cross_sec}~\cite{reno}. The corrections for a non-isoscalar target were not taken into account in the calculation. 

The expected  number of charged-current deep-inelastic neutrino interactions in a $9.6$ ton detector in five years run is reported in the right column of Tab.~\ref{tab:neutrino_all}; the energy spectra are shown in Figure~\ref{fig:spectrum_int}. The contribution of quasi-elastic (QE) interactions amounts to 13\% of the DIS component for muon neutrinos,  7\% for electronic and about 5\% for tau neutrinos as it is reported in Tab.~\ref{tab:qe}. Neglecting the non-isoscalarity of the target leads to an overestimation of the number of anti-neutrino interactions and to an underestimation of the number of neutrinos. 
\begin{table}[htdp]
\begin{center}
\caption{Integrated neutrino yield for $2\cdot10^{20}$ p.o.t. for the  different neutrino flavors: at the proton target (left), at the neutrino target (middle) and charged-current deep-inelastic interactions (right). An energy  cut ($E>0.5$ GeV) is applied for muon and electron neutrinos; no energy cut is applied for tau neutrinos.}
\label{tab:neutrino_all}
\begin{tabular}{c | c c | c c | c c  }
\hline
& $<$E$>$  & Beam & $<$E$>$ & Neutrino  & $<$E$>$ &  CC DIS\\
&       (GeV) & dump & (GeV) & target & (GeV) & interactions \\
 \hline
 $N_{\nu_e}$     & 3 & $2.1 \cdot 10^{17}$  & 28 & $3.6 \cdot 10^{15}$  & 46 & $2.5 \cdot 10^{5}$ \\
 $N_{\nu_\mu}$ & 1.4 & $4.4 \cdot 10^{18}$  & 8 & $ 5.2 \cdot 10^{16}$   & 29 & $1.7 \cdot 10^{6}$ \\
 $N_{\nu_\tau}$ & 9 & $2.8 \cdot 10^{15}$  & 28 & $1.4 \cdot 10^{14}$  & 59 & $6.7 \cdot 10^{3}$ \\
 $N_{\overline{\nu}_e}$       & 4 &$1.6  \cdot 10^{17}$  & 27 &$2.7 \cdot 10^{15}$  & 46 &$9.0 \cdot 10^{4}$ \\ 
 $N_{\overline{\nu}_\mu}$  & 1.5 & $2.8 \cdot 10^{18}$  & 8 & $4.0 \cdot 10^{16}$   & 28 & $6.7 \cdot 10^{5}$ \\
 $N_{\overline{\nu}_\tau}$   & 8 & $2.8 \cdot 10^{15}$  & 26 & $1.4 \cdot 10^{14}$ & 58 & $3.4 \cdot 10^{3}$ \\
 \hline
\end{tabular}
\end{center}
\end{table}
\begin{table}[htdp]
\begin{center}
\caption{Charged-current quasi-elastic neutrino interactions expected in five years run for electron, muon and tau neutrinos.}
\label{tab:qe}
\begin{tabular}{c c }
\hline
 & CC QE interactions \\
 \hline
$\nu_\mu$ & $194\cdot10^3$ \\
$\nu_e$ & $14 \cdot10^3$  \\
$\nu_{\tau}$ & 300 \\
$\overline{\nu}_\mu$ &  $119 \cdot10^3$\\
$\overline{\nu}_e$ & $9.4 \cdot10^3$\\
$\bar{\nu}_{\tau}$ & 180 \\
 \hline
\end{tabular}
\end{center}
\end{table}

\subsection{Neutrino detection}
The neutrino detector has the unique capability of detecting all three neutrino flavors and of distinguishing neutrinos from anti-neutrinos. The detection of neutrino interaction vertices is described together with the flavour identification. The neutrino flavour is determined by identifying  the flavour of the primary charged lepton produced in neutrino charged-current interactions. 
The lepton identification is also used to classify the tracks produced in the $\tau$ decay and, therefore, to identify the decay channel.

The identification of $\nu_\tau$ and $\overline{\nu}_\tau$ interactions requires, as a first step, the detection of both the neutrino interaction and the $\tau$ decay vertices. Two procedures, the event location and decay search, are applied. 
The event location consists of the reconstruction of the neutrino interaction vertex and of the definition of its three-dimensional position with micrometric accuracy. The decay search procedure aims at the detection of the $\tau$ decay vertex.

The detection efficiencies and the backgrounds (Section~\ref{sec:background_nu}) have been determined using the full simulation.

The geometrical efficiency $\epsilon_{geom}$ includes losses due to edge effects in the brick. 
We conservatively used 1 mm from the transverse edges and 5 mm from the downstream edge of the brick~\cite{Agafonova:2014Khd}. 
Edge effects are independent on the $\tau$ decay channel and cause a fiducial volume inefficiency of 4\%, thus giving $\epsilon_{geom} = 96$\%.

We  assume that the location of a neutrino interaction requires the presence of least one track with a momentum lager than 1 GeV/c attached to the primary vertex, with a slope $\tan\theta<1$.
The resulting location efficiency, $\epsilon_{loc}$, is 87\%.

The criteria used to select $\tau$ candidates are purely topological. They are based on the measurement of the kink angle and of the impact parameter of the daughter track with respect to the primary vertex. Candidates are selected if the kink angle is larger that 20 mrad and the impact parameter is larger than 10 $\mu$m. The decay search efficiency $\epsilon_{ds}$ ranges from 67\% to 76\% for the different channels as it can be seen in Tab.~\ref{tab:eff_all}. The higher values correspond to the $\tau \to 3h$ channel, where the presence of three decay products enhances the probability to select the decay topology.

The overall topological $\tau$ detection efficiency is given by $
\epsilon_{tot} = \epsilon_{geom} \cdot \epsilon_{loc} \cdot \epsilon_{ds}$
and it is also reported in Tab~\ref{tab:eff_all} for the different $\tau$ decay channels.

\begin{table}[htdp]
\begin{center}
 \caption{Decay search and overall efficiencies for the different $\tau$ decay channels.}
 \label{tab:eff_all}
\vspace{2mm}
\begin{tabular}{c c c}
\hline
  decay channel & $\epsilon_{ds}$ (\%) & $\epsilon_{tot}$ (\%) \\
 \hline
 $\tau \to \mu$  &  72  & 60 \\
 $\tau \to h$      &  74  & 62 \\
 $\tau \to 3h$    &  76  & 63 \\
  $\tau \to e$     &  67  & 56 \\
 \hline
 \end{tabular}
 \end{center}
  \end{table}

The purity of the $\nu_{\tau}$ sample can be further improved using the neutrino flavour identification and the discrimination power between neutrinos from anti-neutrinos. The neutrino flavor is determined by identifying  the primary lepton produced in neutrino charged-current interactions. 
The lepton identification is also used to classify the tracks produced in the $\tau$ decay and, therefore, to identify the decay channel.
 
\subsubsection{Electron identification}

The high granularity of the emulsions allows the electromagnetic shower identification, hence the separation of electrons and pions, by exploiting their different behavior in the ECC.
Electrons quickly develop an electromagnetic shower in lead. Moreover, at the relevant energies of several GeV, when electrons pass through
a high density material, their energy strongly decreases whilst it remains almost constant for pions.
Therefore, angular deviations along the track show different longitudinal
profiles for electrons and pions. A method based on
this approach is possible with an ECC, as described in~\cite{ELE-PI,Kodama:2003ze}. Moreover, electrons are clearly separated by $\gamma$ rays: indeed, $\gamma$'s are identified by the detection of their conversion point where an electron pair is produced, thanks to the micrometric accuracy of the emulsions and to the high sampling of the brick. 

The electromagnetic shower is identified by counting the number of track segments in a cone around the track. The dimensions of the cone are optimized in order to include the largest number of cascade electron pairs while minimizing the background tracks.
This method, largely exploited in the OPERA experiment~\cite{nue}, provides an electron identification efficiency of 90\% both for electrons produced in $\nu_e$ CC interactions and for electrons produced in the $\tau$ decay.
 
The information retrieved from the Target Tracker planes might improve the identification efficiency for electrons produced in the most downstream part of the brick.    

\subsubsection{Muon identification}

A high muon identification efficiency is needed both for the selection of the $\tau \to \mu$ decay channel and for the background rejection. Charmed hadrons produced in $\nu_\mu$ CC interactions  constitute a background in the $\nu_\tau$ search if the primary muon is not identified.

The muon identification is performed by the Target Trackers (TT), the Drift Tube Tracker (DTT) and the muon spectrometer of the tau neutrino detector.

\begin{figure}
\centering
\includegraphics[scale=0.45]{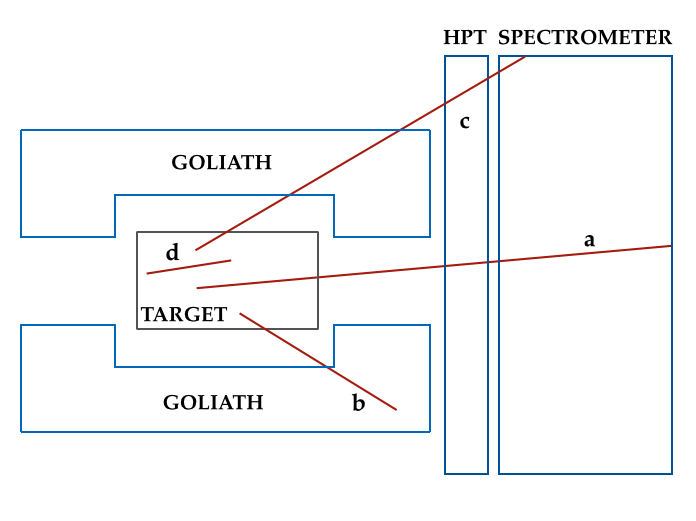}
\caption{Schematic drawing of the different muon topologies.}
\label{fig:muonid}
\end{figure}

The different muon topologies that can occur are shown schematically in Figure~\ref{fig:muonid}. About 71\% of primary muons from neutrino interactions with charm production reach the spectrometer without crossing the passive material of the Goliath magnet (a) while 24\% enter the Goliath magnet iron.  The remaining muons (5\%) either stop within the target (d) or exit aside of the target without crossing the Goliath.
The 24\% of muons  entering the Goliath magnet is made of 7\% stopping within it (b) and 16\% passing through and reaching the DTT planes immediately downstream of the target (c). 
About 1\% does not fall within the DTT acceptance.  
 
A muon candidate is selected according to the following criteria:
\begin{itemize}
\item the track enters  the muon spectrometer without crossing the passive material of the Goliath magnet and traverses at least three iron slabs,
\item the track enters  the DTT plane immediately downstream of the target after having crossed the passive material of the Goliath magnet,
\item the track has a momentum smaller than 2 GeV/c and the product of its length in the neutrino target and the density along its path is larger than 300 g/cm$^2$.
\end{itemize}
The above selection results in a muon identification efficiency of about 90\% for both $\nu_\mu$ charm events  $(\epsilon_{\mu}^{charm})$ and for the muonic decay channel of the $\tau$ lepton $(\epsilon_{\mu}^{tau})$.

The muon identification efficiency for low momentum muons (i.e.~those stopping within the target) may increase by exploiting the correlation between its range and momentum. This potential improvement is not considered in the present evaluation of the performances. 
The probability for an hadron at the primary vertex to be identified as a muon is $\eta_{\mu}=1.5\%$.
%, that corresponds to a loss of efficiency. 
%$\epsilon_{fake-\mu}=1.5\%$
\subsubsection{Charge measurement}
The charge measurement of $\tau$ decay products is used to separate $\nu_\tau$ and $\overline{\nu}_\tau$.
The charge of the hadrons is measured  by the Compact Emulsion Spectrometer (CES) and that of muons by the magnetic spectrometer and the CES. The electronic decay channel of the $\tau$ lepton is currently not considered for the discrimination of $\nu_\tau$ and $\overline{\nu}_\tau$.

Hadrons produced in the $\tau$ lepton decay can undergo hadronic re-interactions in the brick. Due also to the geometrical acceptance, about 75\% of them reach the CES, where their charge can be determined by the track curvature in the magnetic field. On the other hand, the muon charge is mainly measured by the muon spectrometer. When the muon does not reach the magnetic spectrometer, the information of the CES is used. 

The efficiencies for correct charge assignment of a single hadron $\epsilon_{charge}^{h}$, of a muon $\epsilon_{charge}^{\mu}$ and of a particle decaying in three hadrons $\epsilon_{charge}^{3h}$ are as follows
\begin{equation}
 \epsilon_{charge}^{h}= 70 \% \qquad  \epsilon_{charge}^{\mu}=94\% \qquad  \epsilon_{charge}^{3h}=49\%.
 \end{equation}
The  charge misidentification probabilities for a single hadron $\omega_{charge}^{h}$,  for a muon $\omega_{charge}^{\mu}$ and for a particle decaying in three hadrons $\omega_{charge}^{3h}$ are as follows
 \begin{equation}
\omega_{charge}^{h}= 0.5 \% \qquad \omega_{charge}^{\mu}=1.5\% \qquad \omega_{charge}^{3h}= 1.0 \% .
\end{equation}

\subsection{Background evaluation}
\label{sec:background_nu}
The $\nu_\tau$ search is based on the observation of a topology with two vertices. The physics processes showing the same topology are hadron re-interactions and charmed particles decays.

The hadronic interaction background, affecting the $\tau \to h$ and $\tau \to 3h$ channels, is strongly reduced by requiring 
a single-prong or a three-prong topology and the absence of nuclear fragments up to $\tan\theta=3$. 
The application of loose cuts on the transverse momentum at the secondary vertex ($p_T>0.1$ GeV/c) and on the momentum of the product ($p_{dau}>1$ GeV/c) in the single-hadron channel makes the hadronic re-interaction background negligible with respect to the background induced by charmed hadrons. 

The charmed hadron production in $\nu_\mu (\overline{\nu}_\mu)$ and $\nu_e (\overline{\nu}_e)$ interactions when the primary lepton is not identified is the main background source.
Indeed, the mass and the flight length of charmed hadrons are
similar to those of the $\tau$ and therefore selection criteria have similar efficiencies to detect
the decay topology. Having the possibility to measure the charge of the decay daughter, charmed hadrons produced in neutrino interactions contribute mainly to the $\overline{\nu}_\tau$ background while those produced in anti-neutrino interactions to the $\nu_\tau$ background.

\begin{figure}
\centering
\includegraphics[scale=0.6]{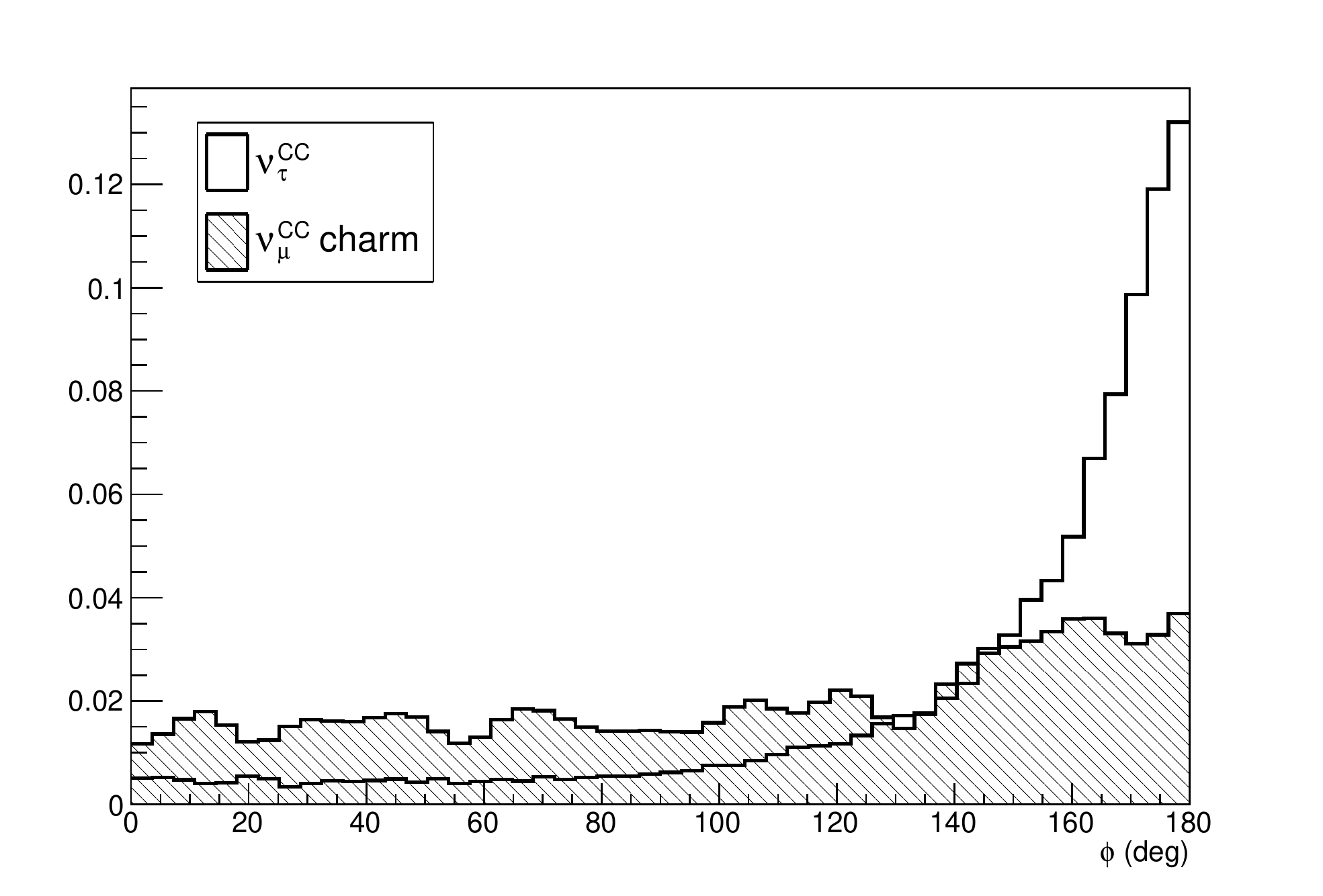}
\caption{Distribution of the $\phi$ variable for the $\nu_\tau$ signal (white) and the charmed hadron background (shaded).}
\label{fig:phi}
\end{figure}

In $\nu_\tau$ interactions, the $\tau$ lepton tends to be back to back to the hadronic jet in the plane perpendicular to the neutrino incoming direction. On the contrary, the charmed hadron tends to be in the middle of the hadronic jet with the not identified muon flying in the opposite direction. Exploiting this feature, one can define the $\phi$ variable as the angle in the transverse plane between the decaying particle and the hadron jet after having removed the particle with the largest $\phi$ value. The $\phi$ distribution is shown in Figure~\ref{fig:phi} for the $\nu_\tau$ signal and the charmed hadron background:
a cut on $\phi>\pi/2$ suppresses this background source with minimal signal loss.
\\
The kinematical selection efficiency for signal, $\epsilon_{kin}^{\tau \to i}$, and background, $\epsilon_{kin}^{C \to i}$, amounts to
\begin{displaymath}
\epsilon_{kin}^{\tau \to h} = 70\% \qquad  \epsilon_{kin}^{\tau \to 3h} = 69\%\\
\end{displaymath}  
\begin{displaymath}
\epsilon_{kin}^{C \to h} = 35\% \qquad  \epsilon_{kin}^{C \to 3h} = 32\%.\\
\end{displaymath}  
The expected charm yield w.r.t.~the neutrino charged current interactions ($\sigma_{charm}/\sigma_{\nu_{k CC}}$) is estimated by a convolution of the cross section measured by
the CHORUS experiment~\cite{chorus_res} with the SHiP energy spectrum. The same energy dependence for neutrinos and anti-neutrinos is assumed. The fractions are reported in Tab.~\ref{tab:sigma_charm} for the different channels.

\begin{table}[htdp]
\begin{center}
\caption{Relative charm production yield per neutrino CC interactions, as expected at the energies of the SHiP experiment, for electron and muon neutrinos.}
\label{tab:sigma_charm}
\vspace{2mm}
\begin{tabular}{c  c}
\hline
 charm fractions & ($\%$) \\
 \hline\
 $\sigma_{charm} / \sigma_{\nu_{\mu CC}}$  & 4.1 \\
 $\sigma_{charm} / \sigma_{\overline{\nu}_{\mu CC}}$  & 4.1 \\
 $\sigma_{charm} / \sigma_{\nu_{e CC}}$  & 6.0 \\
 $\sigma_{charm} / \sigma_{\overline{\nu}_{e CC}}$  & 6.0 \\
 \hline
\end{tabular}
\end{center}
\end{table}

We report the estimate of charm background separately for the different decay channels. The background in the $\tau\to h$ and $\tau \to 3h$ decay channels is due to the decay of charged charmed particles $C^\pm (D^{\pm}$, $D_s^{\pm}$, $\Lambda_c^{+}$ and  $\overline{\Lambda}_c^{-}$) in the hadronic channels. The corresponding branching ratios are $Br(C^{\pm} \to h^{\pm}) = 53.5\%$ and $Br(C^{\pm} \to 3h) = 30.8\%$~\cite{chorus_res,pdg}.

The expected background yield in the $\tau \to i$ $(i=h, 3h)$ decay channel can be written as 
\begin{equation}
\begin{split}
N^{bg}_{\nu_\tau} (\tau \to i) = & \sum_{k=\mu,e} N_{\nu_k}   \frac{\sigma_{charm}}{\sigma_{\nu_{k CC}}}  \epsilon_{tot}  (1-\epsilon^{charm}_{k})   f_{C^\pm}  Br(C\to i)   \omega^{i}_{charge}  \epsilon_{kin}^{C \to i} (1-\eta_{\mu})\\
+ &\sum_{k=\mu,e} N_{\overline{\nu}_k}   \frac{\sigma_{charm}}{\sigma_{\overline{\nu}_{k CC}}}  \epsilon_{tot}  (1-\epsilon^{charm}_{k})  f_{C^\pm}  Br(C\to i)  \epsilon^{i}_{charge}  \epsilon_{kin}^{C \to i}  (1-\eta_{\mu})
 \\
\end{split}
\label{eq:nutau}
\end{equation}
where $N_{ \nu_k (\overline{\nu}_k)}$ is the number of neutrino (anti-neutrino) interactions with flavor~$k$ (see Tab.~\ref{tab:neutrino_all}), $f_{C^+} = 56$\% is the fraction of charged charmed hadrons~\cite{chorus}, and $f_{C^-} = 28$\% is the corresponding fraction in anti-neutrino interactions~\cite{Onengut:2004dy}.  The term $(1-\epsilon^{charm}_{k})$ gives the probability of misidentifying the primary lepton while $(1-\eta_{\mu})$ gives the probability that none of the hadrons at the primary vertex is wrongly identified as a muon. The background to  $\overline{\nu}_\tau$ in these channels is obtained from the formula~\ref{eq:nutau}, replacing $N_{\nu_k}$ with $N_{\overline{\nu}_k}$, and vice-versa.

The background in the $\tau \to \mu$ decay channel consists of  charmed hadrons decaying into a muon. The branching ratio of this process amounts to $Br(C^{\pm} \to \mu^{\pm})=5.3\%$~\cite{chorus_res,pdg}. The charm background in this channel is largely suppressed by requiring that only one muon is reconstructed in the electronic detectors and that it is in agreement with the slopes of the $\tau$ decay daughter. 

For the muonic decay channel of $\nu_\tau$ interactions, the expected charm background can be written as 
\begin{equation}
\begin{split}
N^{bg}_{\nu_\tau} (\tau \to \mu) = & \sum_{k=\mu,e} N_{\nu_k}   \frac{\sigma_{charm}}{\sigma_{\nu_{k CC}}}  \epsilon_{tot}  (1-\epsilon^{charm}_{k})   f_{C^\pm}  Br(C\to \mu)   \omega^{\mu}_{charge}  (1-\eta_{\mu})\\
+ &\sum_{k=\mu,e} N_{\overline{\nu}_k}   \frac{\sigma_{charm}}{\sigma_{\overline{\nu}_{k CC}}}  \epsilon_{tot}  (1-\epsilon^{charm}_{k})  f_{C^\pm}  Br(C\to \mu)  \epsilon^{\mu}_{charge} (1-\eta_{\mu}) \\
\end{split}
\end{equation}
Also in this case, the background to the $\overline{\nu}_\tau$ is obtained replacing $N_{\nu_k}$ with 
$N_{\overline{\nu}_k}$, and vice-versa.

Similarly, it is possible to estimate the background yield in the electronic decay channel. In this case, 
the branching ratio is $Br(C^{\pm} \to e^{\pm})=5.1\%$~\cite{chorus_res,pdg} and, since the charge of the decay product is not measurable, 
only the background contribution to the inclusive $\nu_\tau$ and $\overline{\nu}_\tau$ signal can be evaluated. 

The overall charm background for $\nu_\tau$ and $\bar{\nu}_\tau$ in the different decay channels is reported in Tab.~\ref{tab:nutau_all}, together with the corresponding signal to background ratio $R$.

\subsection{Estimation of the signal yield}
The expected number of $\nu_\tau$ and $\overline{\nu}_\tau$ interactions in the muonic and hadronic ($i=h,3h$)  $\tau$ decay channels can be computed as
\begin{equation} 
\begin{split}
N^{exp}_{\nu_\tau (\overline{\nu}_\tau) } ( \tau\rightarrow \mu) & =  N_{\nu_\tau (\overline{\nu}_\tau) } Br(\tau\to \mu)  \epsilon^{\tau \to \mu}_{tot} \epsilon^{tau}_{\mu} \epsilon^{\mu}_{charge} (1-\eta_{\mu}) \\\\
N^{exp}_{\nu_\tau (\overline{\nu}_\tau) } ( \tau\rightarrow i)   & =  N_{\nu_\tau (\overline{\nu}_\tau)} Br(\tau\to i)   \epsilon^{\tau \to i}_{tot}   \epsilon^{i}_{charge} \epsilon_{kin}^{\tau \to i} (1-\eta_{\mu})\\\\ 
\end{split}
\label{eq:tausignal}
\end{equation}
where $N_{\nu_\tau (\overline{\nu}_\tau)}$ is the number of $\nu_\tau (\overline{\nu}_\tau)$ interactions in the neutrino target.
Since the charge of the electron is not measurable, only an inclusive measurement of $\nu_\tau$ and $\overline{\nu}_\tau$
is possible in the $\tau \to e$ decay channel. The expected number of tau neutrinos and anti-neutrinos in the $\tau \rightarrow e$ decay channel are computed with a similar expression as in Eq.~\ref{eq:tausignal} and about 850 interactions are expected with an overall background of 160 events.

About 1800 $\nu_\tau$ and 900 $\overline{\nu}_\tau$ detected interactions are expected in five years run, as reported in Tab.~\ref{tab:nutau_all}.

\begin{table}[htdp]
\begin{center}
 \caption{Expected number of $\nu_\tau$ and $\overline{\nu}_\tau$, $N^{exp}$, and charm background events, $N^{bg}$, observed in
the different $\tau$ decay channels, except for the $\tau \to e$, where
the lepton number cannot be determined. The signal to background ratio
$R$ is also reported.}
 \label{tab:nutau_all}
\vspace{2mm}
\begin{tabular}{c |c c c| c c c }
\hline
  decay channel &  \multicolumn{3}{c}{$\nu_\tau$}  &
\multicolumn{3}{c}{$\overline{\nu}_\tau$}  \\
                         &  $N^{exp}$  &  $N^{bg}$  & $R$  & $N^{exp}$
&  $N^{bg}$  & $R$\\
 \hline
 $\tau \to \mu$  & 570 &  30   & 19 & 290 & 140 & 2\\
 $\tau \to h$      & 990 &  80   & 12 & 500 & 380 & 1.3\\
 $\tau \to 3h$    & 210 &  30   & 7   & 110 & 140  & 0.8\\
 \hline
 Total & 1770 & 140 & 13 & 900 & 660  & 1.4 \\
 \hline
 \end{tabular}
 \end{center}
 \end{table}

\subsection{Structure functions $F_4$ and $F_5$}

The charged-current $\nu_\tau$ $(\overline{\nu}_\tau)$ differential cross section is represented by a standard set of five structure functions:  
\begin{displaymath}
\begin{split}
\frac{d^2\sigma^{\nu(\overline{\nu})}}{dxdy} &= \frac{G_F^2 M E_\nu}{\pi (1+Q^2/M^2_W)^2}
 \biggl(
  (y^2x + \frac{m_\tau^2y}{2E_\nu M})F_1 + 
 \left[ (1-\frac{m_\tau^2}{4E_\nu^2}) - (1+\frac{Mx}{2E_\nu}) \right] F_2 \\
 & \pm  \left[ xy(1-\frac{y}{2}) - \frac{m_\tau^2y}{4E_\nu M}\right] F_3 +
 \frac{m_\tau^2(m_\tau^2+Q^2)}{4E_\nu^2M^2x} F_4 -
 \frac{m_\tau^2}{E_\nu M} F_5  
  \biggr), \\
\end{split}
\end{displaymath}
 where $\{x, y, Q^2\}$ are the standard DIS kinematic variables related through $Q^2 = 2M_NE_\nu x y$.

The structure functions $F_4$ and $F_5$, pointed out by Albright and Jarlskog in Ref.~\cite{aj}, are neglected in muon neutrino interactions because of a suppression factor depending on the
square of the charged lepton mass divided by the nucleon mass times neutrino energy. Given the higher mass value of the $\tau$ lepton, $F_4$ and $F_5$ structure functions contribute, instead, to the tau neutrino cross section.
At leading order, in the limit of massless quarks and target hadrons, $F_4=0$ and $2xF_5=F_2$, 
where $x$ is the Bjorken-$x$ variable (Albright-Jarlskog relations). Calculations at NLO show that $F_4$ is about 1\% of $F_5$ \cite{reno}. 

With the statistics of tau neutrino interactions collected in five years run, the SHiP experiment will have the unique capability of being sensitive to $F_4$ and $F_5$. 
The hypothesis of $F_4 = F_5 = 0$ would result in an increase of the $\nu_\tau$ and $\overline{\nu}_\tau$ charged-current deep-inelastic cross sections and consequently, of the number of expected $\nu_\tau$ and $\overline{\nu}_\tau$ interactions.
\begin{figure}
\begin{center}
\includegraphics[width=0.95\linewidth]{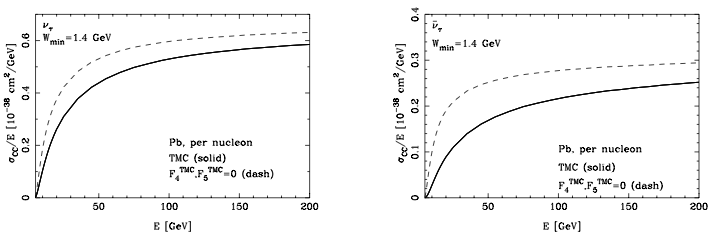}
\end{center}
\caption{$\nu_\tau$ (right) and $\overline{\nu}_\tau$ (left) CC DIS cross section  predicted by the SM (solid) and in the $F_4=F_5=0$ hypothesis (dashed)~\cite{reno}. }
\label{fig:xsec_f5}
\end{figure}

Figure~\ref{fig:xsec_f5} shows that the difference between the cross sections in the $F_4=F_5=0$ hypothesis  and the SM prediction is larger for lower neutrino energies.  This behavior reflects in the energy dependence of the variable $r$, defined as the ratio between the cross section in the two hypotheses: it is higher for lower neutrino energies, where the discrepancy of the two curves is larger, and decreases, tending to one, for higher energies, where the contribution of $F_4$ and $F_5$ becomes negligible. 

\begin{figure}
\centering
\includegraphics[scale=0.8]{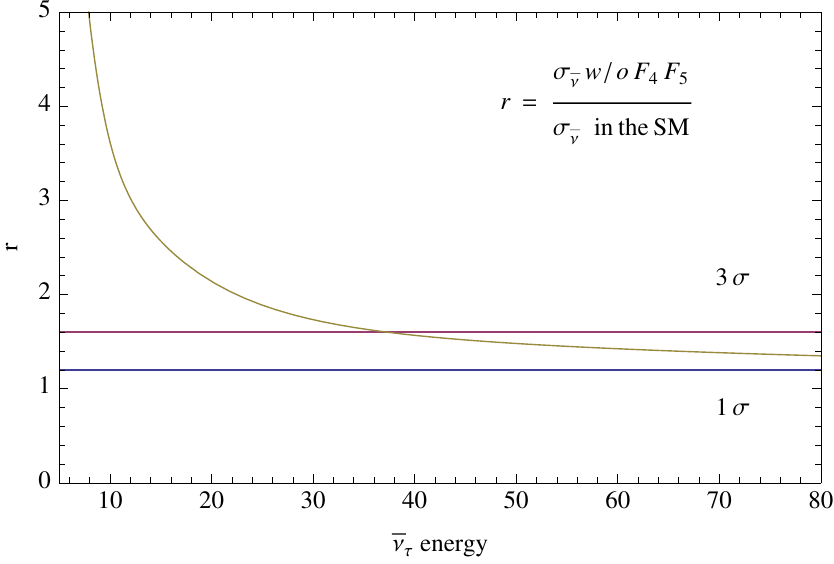}
\includegraphics[scale=0.8]{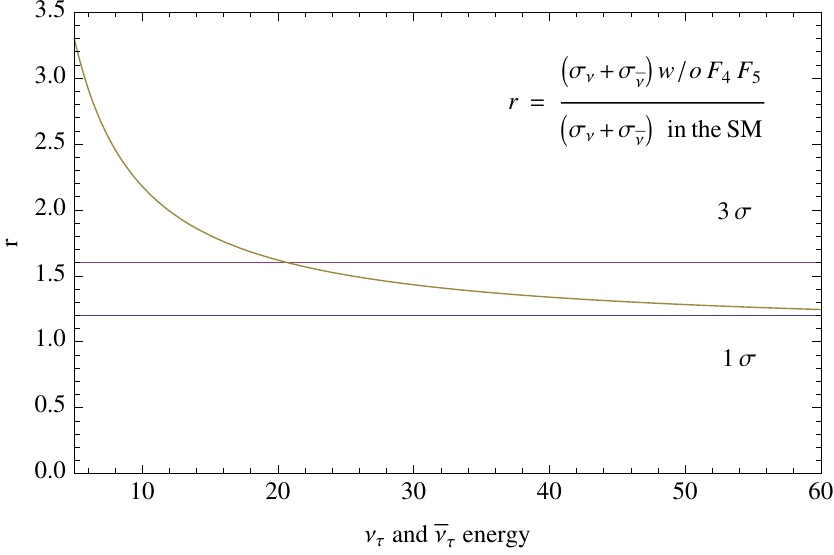}
\caption{Energy dependence of the ratio $r$ between the DIS cross section in the $F_4=F_5=0$ hypothesis and the SM prediction, for $\overline{\nu}_\tau$ (left) and for the sum of $\nu_\tau$ and $\overline{\nu}_\tau$ (right).}
\label{fig:f4f5}
\end{figure}

The ratio $r$ is reported for $\overline{\nu}_\tau$ in the left plot of Figure~\ref{fig:f4f5}. The evidence for a non-zero value of $F_4$ and $F_5$ with a significance of 3~$\sigma$ requires the ratio $r$ to be larger than 1.6, given the uncertainty of 20\% on the incoming neutrino flux. 
This condition is satisfied for $E_{\overline{\nu}_\tau}<38$ GeV, where we expect to observe about 300 $\overline{\nu}_\tau$ interactions.

The ratio $r$ was estimated also for the sum of $\nu_\tau$ and $\overline{\nu}_\tau$. The right plot of Figure~\ref{fig:f4f5} shows that a significance of 3~$\sigma$ is achieved in this case for neutrino energies below 20 GeV. The total number of neutrino interactions is expected to be 420 in this case, accounting also for the $\tau \rightarrow e$ channel. 

\subsection{Neutrino induced charm production}
\label{sec:nucharm}
The first evidence of charmed hadron production in neutrino interactions was produced in 1974 through the observation of opposite sign dimuons~\cite{benvenuti}. Since then, it has been assessed that high energy neutrino interactions make charmed hadron production at the level of a few percent and therefore they constitute a powerful tool to study charm physics. Unlike colliding beams, neutrino scattering produces charmed hadrons also via peculiar processes like quasi-elastic and diffractive productions which makes them a unique tool for exclusive charm studies~\cite{charmrep}. 

Since its discovery, charmed hadron production in neutrino interactions has been studied in two ways: dilepton studies with both calorimeter and bubble chamber technique on one side and nuclear emulsion experiments with the visual observation of charmed hadron decays on the other side. The main advantage of nuclear emulsions is that very loose kinematical cuts may be applied since the charmed particle is identified through the observation of its decay. 
This translates into a  good sensitivity to the slow-rescaling threshold behaviour and consequently to the charm quark mass.

\begin{figure}[h]
\centering
\includegraphics[width=0.4\textwidth]{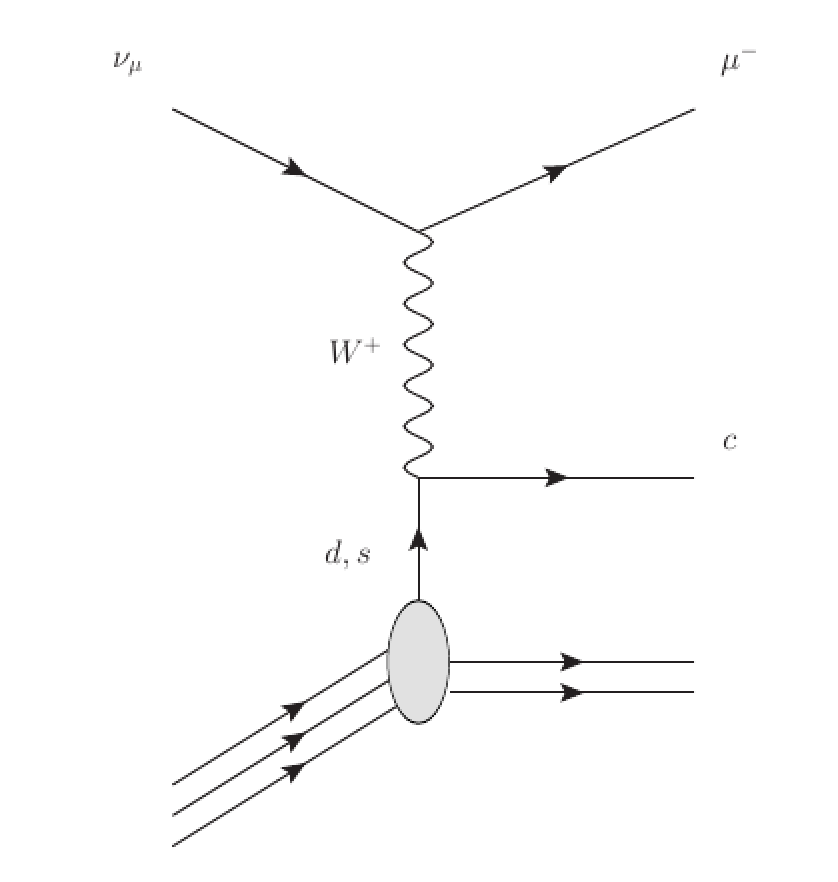}\quad
\includegraphics[width = 0.5\textwidth]{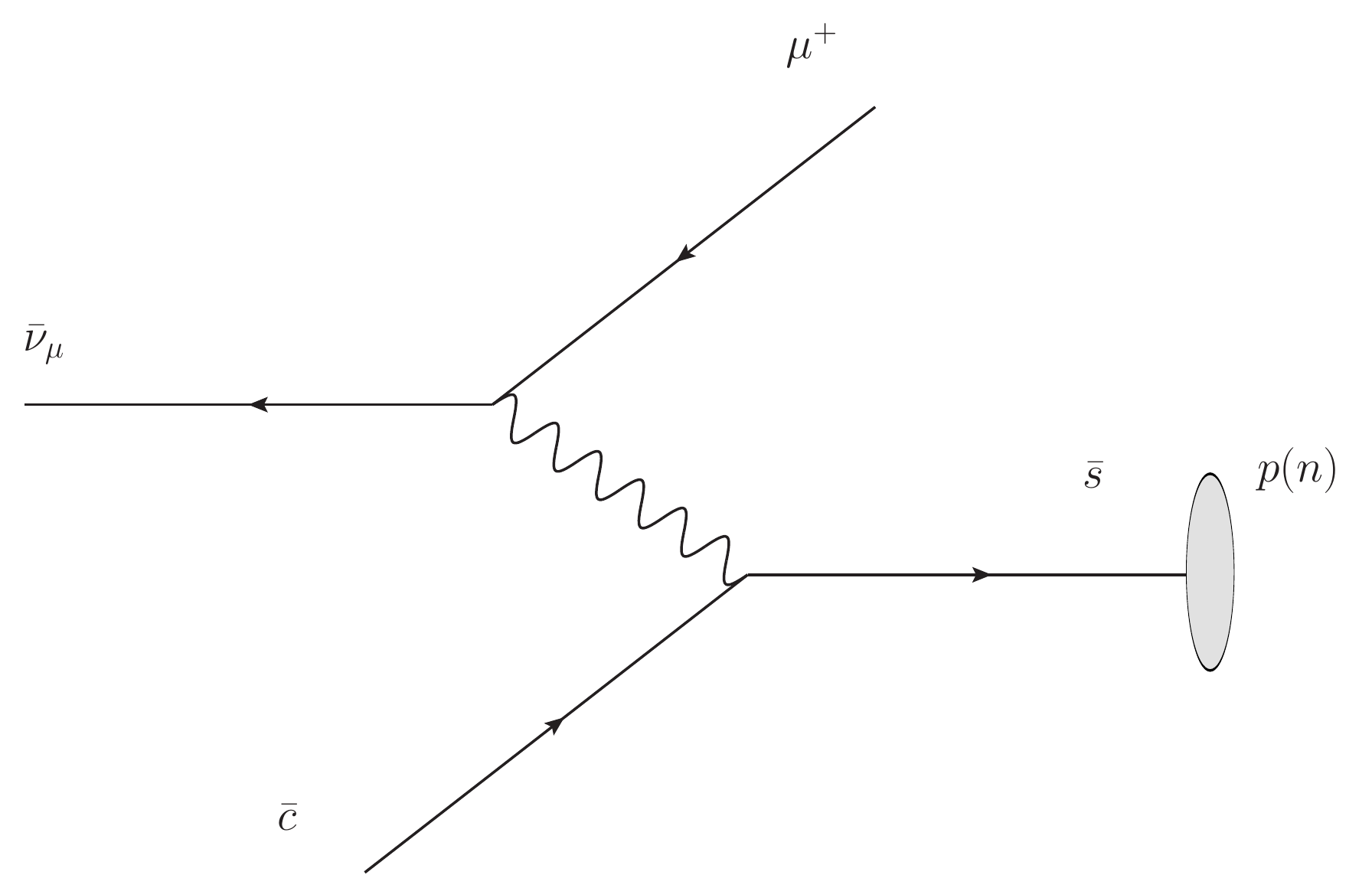}
\caption{Diagram for charm production in neutrino (left) and anti-neutrino (right) charged-current interactions. }
\label{fig:charmFeynmann}
\end{figure}

Figure~\ref{fig:charmFeynmann} shows the Feynman diagrams for the production of charmed hadrons in neutrino and anti-neutrino interactions. 
Data on charmed hadron production in neutrino and anti-neutrino charged-current interactions were reported by many experiments:
CDHS~\cite{cdhs}, CCFR~\cite{ccfr}, CHARM~\cite{charm}, CHARMII~\cite{charm2}, NuTeV~\cite{nutev}, BEBC~\cite{bebc}, NOMAD~\cite{nomad}, E531~\cite{e531} and CHORUS~\cite{chorus_res}.
Experiments based on calorimetric technology identify charmed hadrons produced in $\nu$ interactions only in the muonic decay channel, where two opposite charged muons are expected in the final state, by applying a 5 GeV cut on the minimum muon momentum to reduce the background from punch-through pions. The nuclear emulsion technique, instead, identifies charmed hadrons through the observation of a two vertex topology: the first one produced by neutrino charged-current interactions, the second one produced by the charmed hadron decay, occurring within a few millimetres.  The decays of charmed hadrons are therefore accessible, without any kinematical cut.

Despite the fact that 1.280.000 $\nu_{\mu}$ and 270.000 $\bar{\nu}_{\mu}$ charged-current events were collected  by the NuTeV/CCFR collaboration, the detection of charm production was limited by the selection based on the muon decay channel (on average less than 10\% branching ratio) and by the cut of 5~GeV applied for the muon identification. For these reasons, only 5.102 $\nu_{\mu}$ and 1.458 $\bar{\nu}_{\mu}$ events were identified as charm production in charged current interactions. The emulsion experiment with the largest neutrino flux was CHORUS~\cite{chorus_res}. Out of a sample of 143742 neutrino-induced charged-current interaction detected, 2013 charm candidates coming from $\nu_{\mu}$ and 32 coming from $\bar{\nu}_{\mu}$ interactions were reported. No charm candidate from electron neutrino interactions was ever reported. 

The relative charm production yield in muon and electron neutrino and anti-neutrino interactions expected at SHiP energies is reported in Table~\ref{tab:sigma_charm}. In 5 years run, more than $10^{5}$ neutrino induced charmed hadrons are expected, as reported in Table~\ref{tab:charmevts}. It is worth noting that the total charm yield exceeds the statistics available in previous experiments by more than one order of magnitude. Therefore all the studies on charm physics performed with neutrino interactions will be revised with improved accuracy and some channels inaccessible in the past will be explored.

\begin{table}[h]
\centering
\caption{Summary of the contributions of the different neutrino species to neutrino induced charm events.}
\label{tab:charmevts}
\vspace{2mm}
\begin{tabular}{cc}
\hline 
& Expected events \\ 
\hline 
$\nu_{\mu}$ &  6.8 $\cdot 10^4$  \\ 
$\nu_{e}$ & 1.5 $\cdot 10^4$\\ 
$\bar{\nu_{\mu}}$ & 2.7 $\cdot 10^4$\\ 
$\bar{\nu_{e}}$ &  5.4 $\cdot 10^3$\\ 
\hline 
Total     & 1.1 $\cdot 10^5$\\
\hline
\end{tabular} 
\end{table}

Unlike neutrino scattering where the presence of valence quarks favours the $d$-quark as neutrino target thus compensating the large suppression provided by the Cabibbo angle, 
charmed hadron production in anti-neutrino interactions selects anti-strange quark in the nucleon, as sketched in Figure~\ref{fig:charmFeynmann}. Precise knowledge of the strangeness is an important information for many precision tests of the SM as well as for BSM searches at the LHC. For instance, the precise measurement of the $W$ mass depends on the knowledge of the strangeness contents in the proton~\cite{bozzi}. The strange see quark determination by all the PDF groups relies mainly on the dimuon data collected by the NuTeV/CCFR collaboration.

A simulated SHiP data sample normalized to the expected statistics was used to estimate the distribution of muon neutrino interactions with charm production. 
Simulated data were divided in five bins for the $x$ variable, three for the $y$ variable and three for the neutrino energies. A systematic uncertainty for all the bins was conservatively assumed to be as large as 5\% and uncorrelated. Moreover, only muon neutrinos were considered in this analysis. The statistical uncertainty in each bin will reduce from $10\div 20$\%, as measured by previous experiments, to $2-3$\% for charm production and $4-5$\% for anti-charm production. 

\begin{figure}[h]
\includegraphics[width =0.95\textwidth]{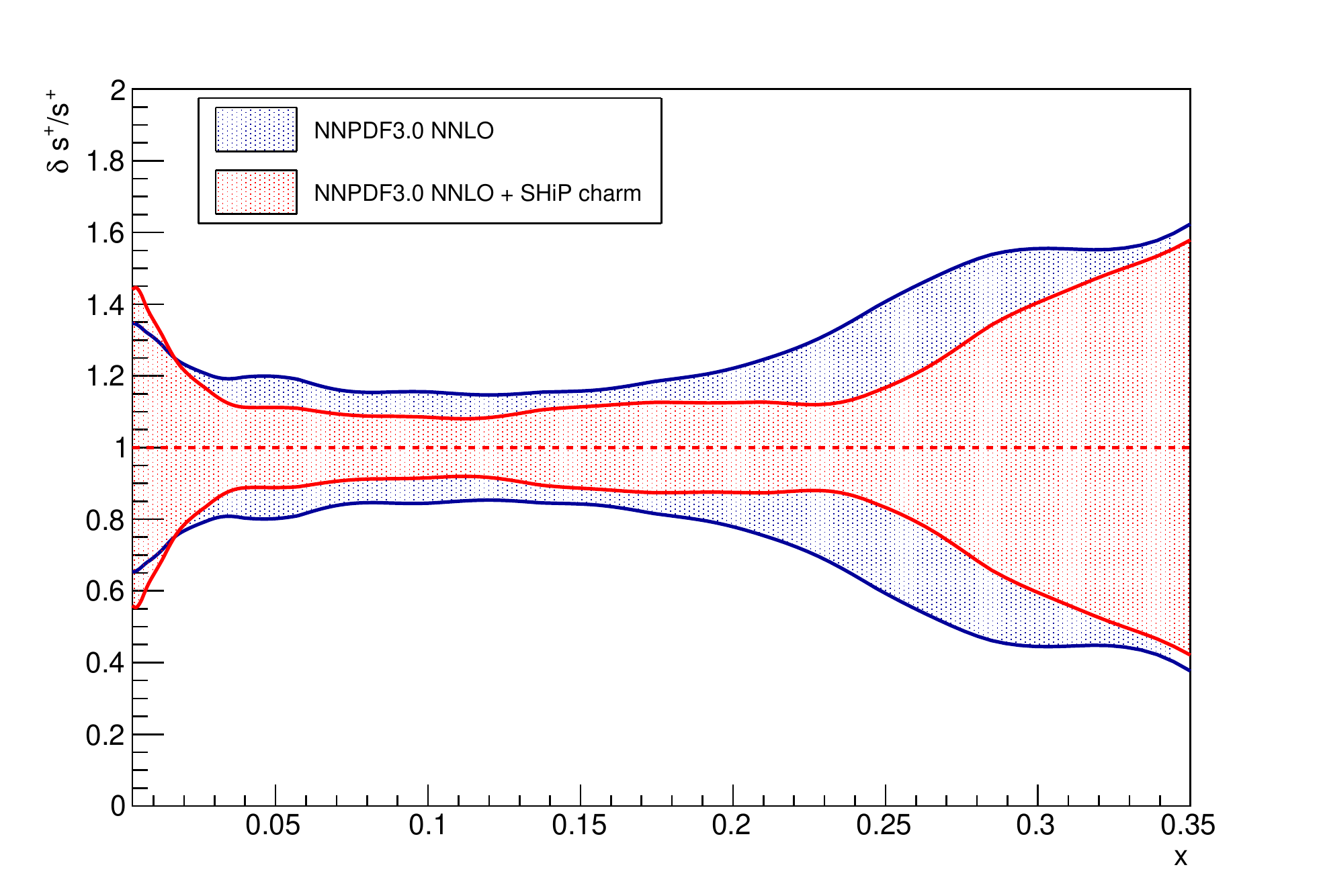} 
\caption{The blue line shows the present accuracy in the distribution of $s^+ = s + \bar{s}$ for $x$ ranging from 0 to 0.35. The red contour shows the improvement obtained with SHiP. }
\label{fig:strangeness_plus}
\end{figure}

\begin{figure}
\includegraphics[width =0.95\textwidth]{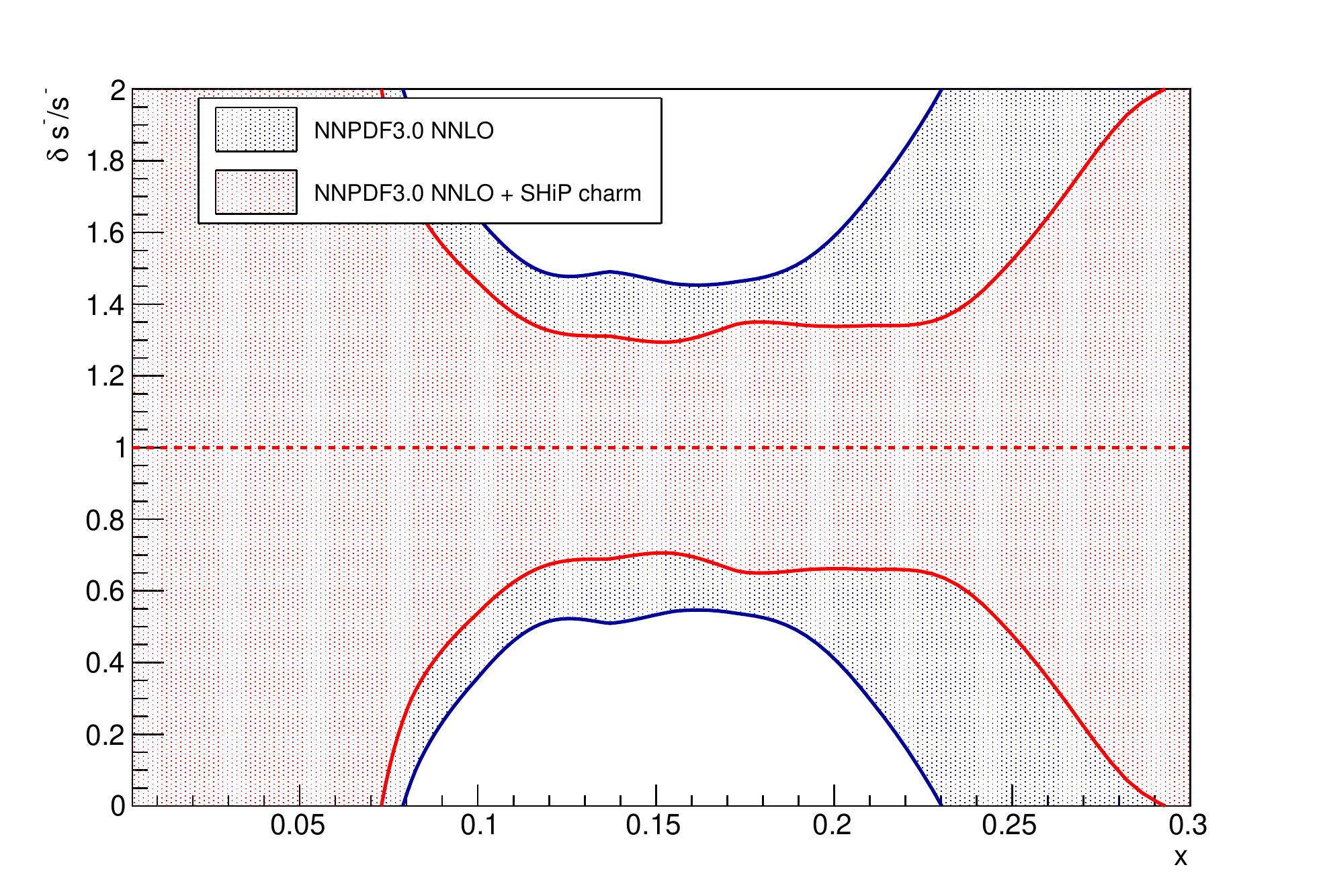}
\caption{The blue line shows the present accuracy in the distribution of $s^- = s - \bar{s}$ for $x$ ranging from 0 to 0.3. The red contour shows the improvement obtained with SHiP. }
\label{fig:strangeness_mins}
\end{figure}

The potential impact of  simulated charm data was assessed by adding them to the NNPDF3.0 NNLO fit~\cite{Ball:2014uwa}, following the approach described in Refs.~\cite{Ball:2010gb,Ball:2011gg}. The constraining power of the SHiP pseudodata is shown in Figure~\ref{fig:strangeness_plus} for the $s^+$ variable defined as $s^+ = s(x) + \bar{s}(x)$. The vertical axis reports $1+\Delta s^+/s^+$ where $\Delta s^+$ is the accuracy on $s^+$, in such a way that the difference to unity indicates the accuracy. The horizontal axis is the Bjorken variable $x$. A significant improvement on this variable is achieved in the $x$ range between $0.03$ and $0.35$. Figure~\ref{fig:strangeness_mins} shows the improvement achieved on the variable $s^- = s -\bar{s}$ versus the $x$ variable. The vertical axis reports $1+\Delta s^-/s^-$. A significant gain with SHIP data is obtained in the $x$ range between $0.08$ and $0.30$.

\subsection{$\nu_{\tau}$ magnetic moment}
\label{sec:MagMom}
While for $\nu_{\mu}$ and $\nu_{e}$ a non-zero magnetic moment has been excluded down to $\mu_{\nu} < 6.9 \cdot 10^{-10} \mu_{B}$ \cite{pdg} and to $\mu_{\nu} < 2.9 \cdot 10^{-11} \mu_{B}$ \cite{pdg} respectively. % , the current upper limit for $\nu_{\tau}$ is set by the DONUT experiment to be $3.9 \cdot 10^{-7} \mu_{B}$~\cite{nutauMM} with only 9 observed events. 
With a few thousands $\nu_{\tau}$ charged-current interactions in the neutrino target, the SHiP experiment can considerably constrain the $\nu_{\tau}$ magnetic moment. 

 The presence of a non-zero magnetic moment adds an extra-component to the elastic cross section for the process $\nu + e^{-} \rightarrow \nu + e^{-}$ due to photon exchange, leading to an anomalous increase of the measured value of this cross section. In this process, the scattering angle  in the laboratory frame of the outgoing electron with respect to the direction of the incoming neutrino is limited by kinematic constraints \cite{KinConstraintMM}:
\begin{equation}
\theta^{2}_{\nu - e} < \frac{2m_e}{E_e}.
\label{eq:ThetaEkink}
\end{equation}
Therefore, if the energy of the electron is above 1~GeV, $\theta_{\nu - e}$ must be below 30 mrad.
\noindent Processes showing the same topology and thus constituting background to signal observation are:
\begin{itemize}
\item neutrino elastic scattering (ES) with electrons of the target
\begin{itemize}
	\item $\nu_{\mu}$, $\bar{\nu}_{\mu}$, $\nu_{e}$ and $\bar{\nu}_e$ neutral current interactions : 
				\[\nu_x + e^{-} \rightarrow \nu_x + e^{-} \]
	\item $\nu_e$ and $\bar{\nu}_e$ charged current interactions:
				\[\nu_e (\bar{\nu}_e) + e^{-} \rightarrow e^{-} + \nu_e (\bar{\nu}_e)\]
\end{itemize}
\item electron neutrino and anti-neutrino quasi elastic scattering (QE) with nucleons of the target:
	\[\nu_e + n \rightarrow e^{-} +p \]
		\[\bar{\nu}_e + p \rightarrow e^{+} +n \]
 	when protons in the final state are not revealed in the neutrino detector.
\item charged current deep-inelastic interactions (DIS) of electron neutrinos and anti-neutrinos with nucleons in the neutrino target with no revealed hadrons in the final state:
	\[\nu_e (\bar{\nu}_e) + N \rightarrow e^{-} (e^{+}) + X\]
\item electron neutrino and anti-neutrino resonant processes.
\end{itemize}
In nuclear emulsion films all charged particles with momenta larger than 100~MeV$/c$ (170 MeV$/c$ for protons) are detectable if the projections of their slope $\theta$ satifies the condition $\tan \theta_x < 3$ and $\tan \theta_y < 3$.  $\gamma$ rays are detectable if their momentum is above 100~MeV$/c$ within the same angular range as for charged particles.  
  
\begin{figure}[h]
\centering
\includegraphics[width = 0.45\textwidth]{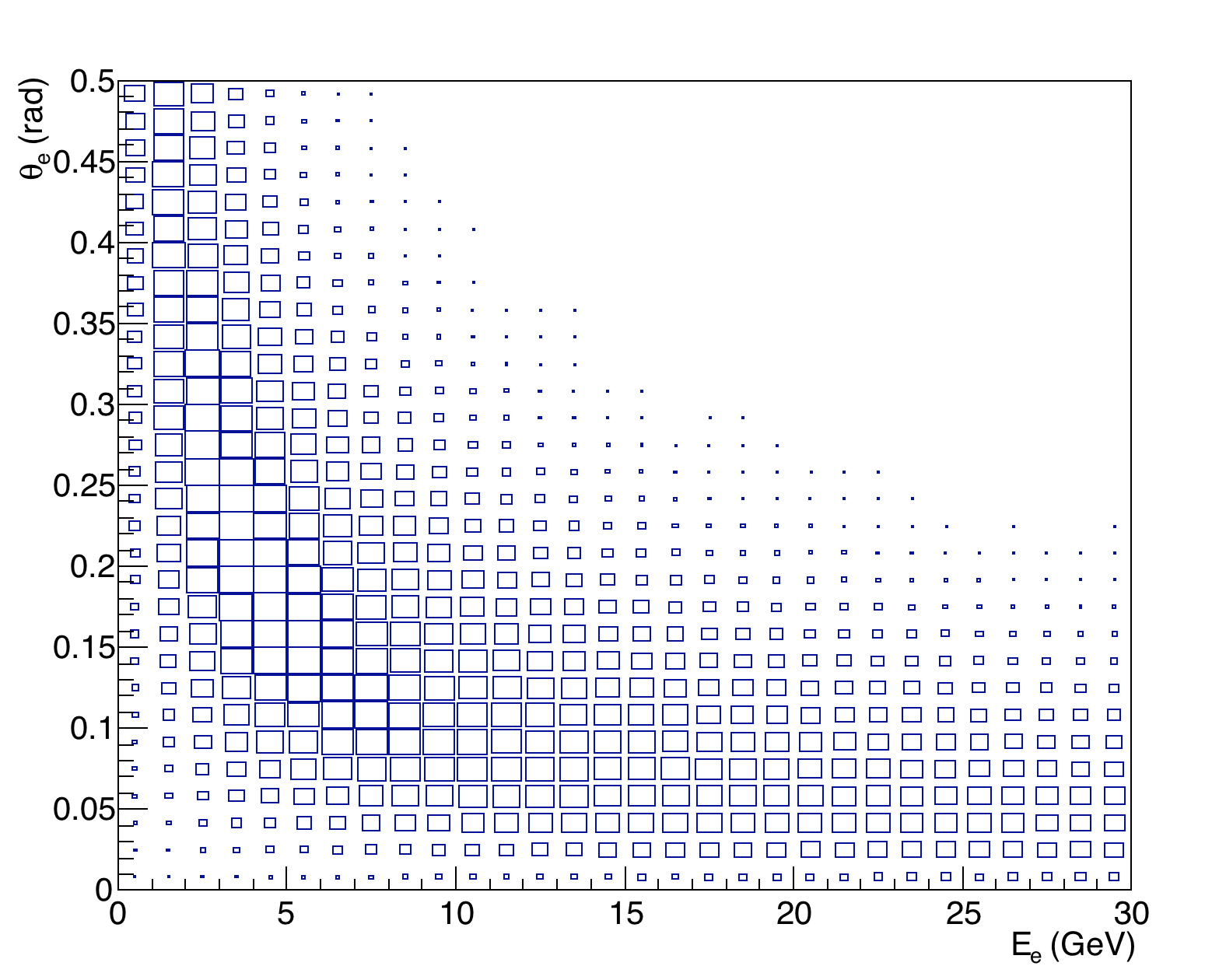} 
\caption{Correlation between the electron scattering angle and the electron energy for $\nu_{e}$ deep-inelastic scattering processes. }
\label{fig:ThetaE}
\end{figure}

Taking into account the proper energy spectrum expected in SHiP, electron neutrino interactions in lead have been generated with the GENIE event generator to estimate the background yield due to the above-described sources. Only those events satisfying the following criteria are selected as background:
 \begin{itemize}
\item the electron is the only visible particle in the final state;
\item electron energy is above 1 GeV;
\item electron scattering angle $\theta_{\nu - e}$ is below 30 mrad.
\end{itemize}
For deep-inelastic processes, the correlation between the electron scattering angle and its energy is reported in Figure~\ref{fig:ThetaE}. Requiring an electron energy above 1 GeV and a scattering angle below 30 mrad, only a small fraction (8.4$\cdot 10^{-4}$) of the overall $\nu_e$ deep-inelastic scattering events is selected. Applying the same cuts to electron anti-neutrinos, the fraction of selected events amounts to 5.8$\cdot 10^{-3}$. The difference of the rejection power for neutrinos and anti-neutrinos comes from the different preferred neutrino   target, being neutrons for neutrinos and protons for anti-neutrinos. Therefore, a proton is often present in the final state of neutrino interactions, thus making the deep-inelastic interaction easier to be detected.  
All contributions from the different background processes are summarised in Tab.~\ref{tab:bkg}.

\begin{table}
\centering
\caption{Different sources of  background for the observation of the anomalous magnetic moment of the tau neutrino.}
\label{tab:bkg}
\vspace{2mm}
\begin{tabular}{cc}
\hline 
& $E_e >$ 1 GeV  \\ 
\hline 
DIS & 732 \\ 
ES & 386 \\ 
QE & 1795 \\ 
RES & 646 \\ 
\hline 
Total & 3559\\ 
 \hline 
\end{tabular} 
\end{table}

The statistical uncertainty on the background yield is 60 events. The  systematic uncertainty comes mostly from the knowledge of the neutrino flux. An optimisation of this analysis as a function of the systematic uncertainty is ongoing. The systematic uncertainty can be considerably improved by studying the deep-inelastic interactions and their energy spectrum. As an example, we report in the following the sensitivity to the magnetic moment if the systematic uncertainty could be reduced down to 5\%.  In this case, the evidence for the tau neutrino anomalous magnetic moment with a significance of 3~$\sigma$ requires the observation of an excess of about 540 events over the background. 
The number of expected events for a magnetic moment $\mu_{\nu}$ can be computed as
\begin{equation}
n_{evt} = \frac{\mu_{\nu}^2}{\mu_{B}^2}  \int \Phi_{\nu_{\tau}} \sigma^{\mu} N_{nucl} dE 
\label{eq:expevts}
\end{equation}
where $\Phi_{\nu_{\tau}}$ is the tau neutrino flux at the target, $N_{nucl}$ is the number of nucleons and $\sigma^{\mu}$ is the contribution to the cross section due to the anomalous magnetic moment~\cite{Domogatsky:1971tu}. Hence, a region down to a magnetic moment $\mu_{\nu} = 1.5\cdot 10^{-7}$ could be explored.

\subsection{Direct Dark Matter detection}

A systematic study of the detection of light dark matter produced via the dark photon, $A$, portal has been reported in several papers~\cite{Batell:2009di,Izaguirre:2013uxa,Batell:2014mga,Kahn:2014sra}.
The most stringent constraints follow from the LSND experiment, which collected $1.8 \cdot 10^{23}$ protons on target. In a proton dump, the dark matter particle $\chi$ can be produced in the decay of the dark photon.
As discussed in the SHiP Physics Proposal~\cite{PP}, there are various production mechanisms  of the dark photon in proton beam dump experiments: pseudo-scalar meson decays, proton bremsstrahlung and QCD production. 
% For example, a typical production scheme of the dark matter particle $\chi$ when the dark photon $A$ is produced by $\pi^0$ decays ($m_A < m_{\pi^0}$) is the following:
% \begin{equation}
% pp \to \pi^0 + X, ~~ \pi^0  \to A \gamma, ~~ A \to \chi\bar\chi. 
% \end{equation}

 The $\chi$ particles might be detected through the scattering off electrons and/or nuclei.
The experimental signature of dark matter interactions is  neutral current interaction on both electrons and nuclei in a massive detector to cope with the low cross section. 
The studies described in this section for the SHiP neutrino detector are based on the identification of the scattering of dark matter particles off electrons. Scattering on nuclei will be the subject for a future study.
 
The model of kinetic mixing mediation in Ref.~\cite{deNiverville:2012ij} was assumed  to derive the sensitivity  and  signal distributions were calculated accordingly. 
 
Background sources for this search are similar to the search for the $\nu_\tau$ anomalous magnetic moment (see Section~\ref{sec:MagMom}): neutral current $\nu_\mu$ and $\nu_e$ scattering on electrons, and charged current elastic, resonant and deep inelastic  $\nu_e$ scattering off nuclei. 

The GENIE MonteCarlo was used to generate  background events. The main variables to separate signal from background are the electron energy and the angle with respect to the neutrino direction and the number of detectable particles at the neutrino interaction vertex. Assuming that all the interactions occur in the lead of the neutrino detector, charged particles would be reconstructed if their momentum is above 100~MeV/c and 170~MeV/c, respectively for pion and kaons and for protons. Photons are detected if their momentum is above 100~MeV/c. The uncertainty due to the unknown origin of the neutrino in the beam dump is about 1~mrad and  the electron angle is reconstructed in the detector with a resolution of 3~mrad, thereby dominating the total angular resolution. 

\begin{figure}[h]
\begin{center}
\includegraphics[width=0.48\linewidth]{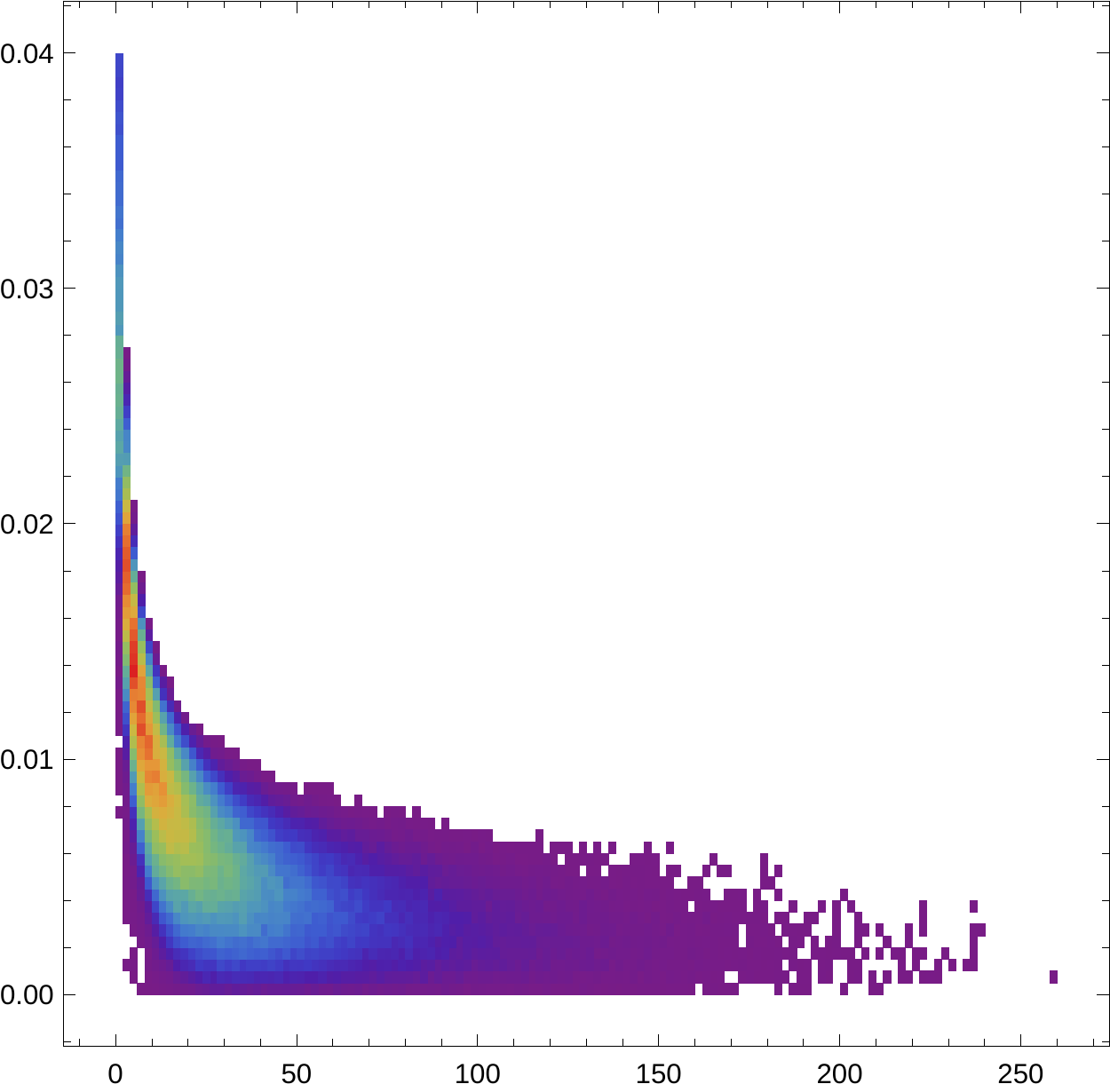}
\includegraphics[width=0.48\linewidth]{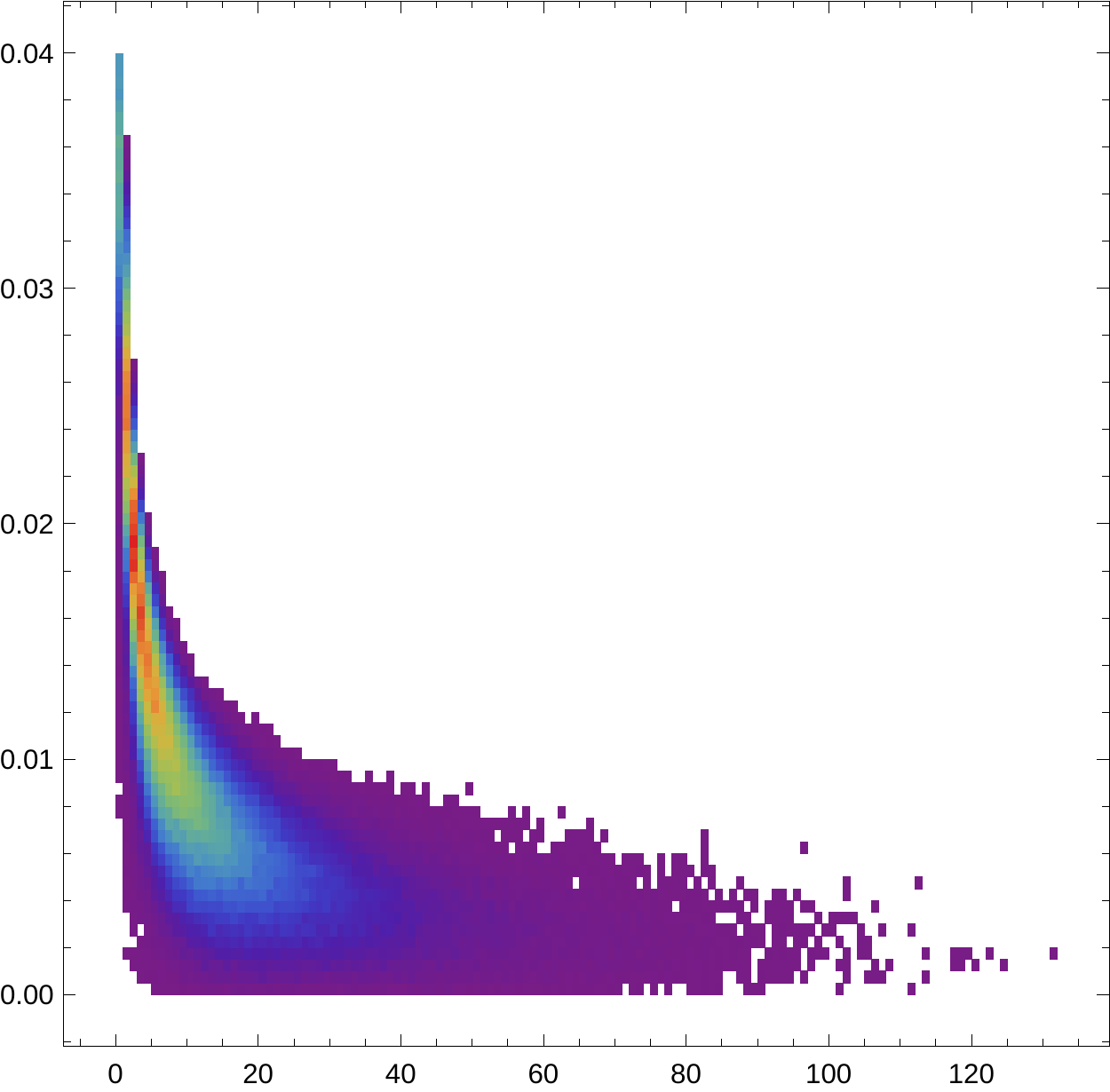}
\caption{Electron scattering angle (mrad) versus the  electron energy (GeV) for signal candidates, (a) in a model with $m_A = 800$~MeV/c$^2$ and $m_\chi = 200$~MeV/c$^2$  and (b) with $m_A = 405$~MeV/c$^2$ and $m_\chi = 200$~MeV/c$^2$.}
\label{fig:sig_DM}
\end{center}
\end{figure}

Figure~\ref{fig:sig_DM} shows the correlation between the electron energy and the angle for signal candidates, in a model (a) with the dark photon mass $m_A = 800$~MeV/c$^2$ and the dark matter mass $m_\chi = 200$~MeV/c$^2$ and (b) with $m_A = 405$~MeV/c$^2$ and $m_\chi = 200$~MeV/c$^2$. Most of the signal is concentrated at low energies, below 20~GeV and with angles between 10~mrad and 20~mrad. This information is used in the selection.

\begin{figure}[h]
\begin{center}
\includegraphics[width=0.45\linewidth]{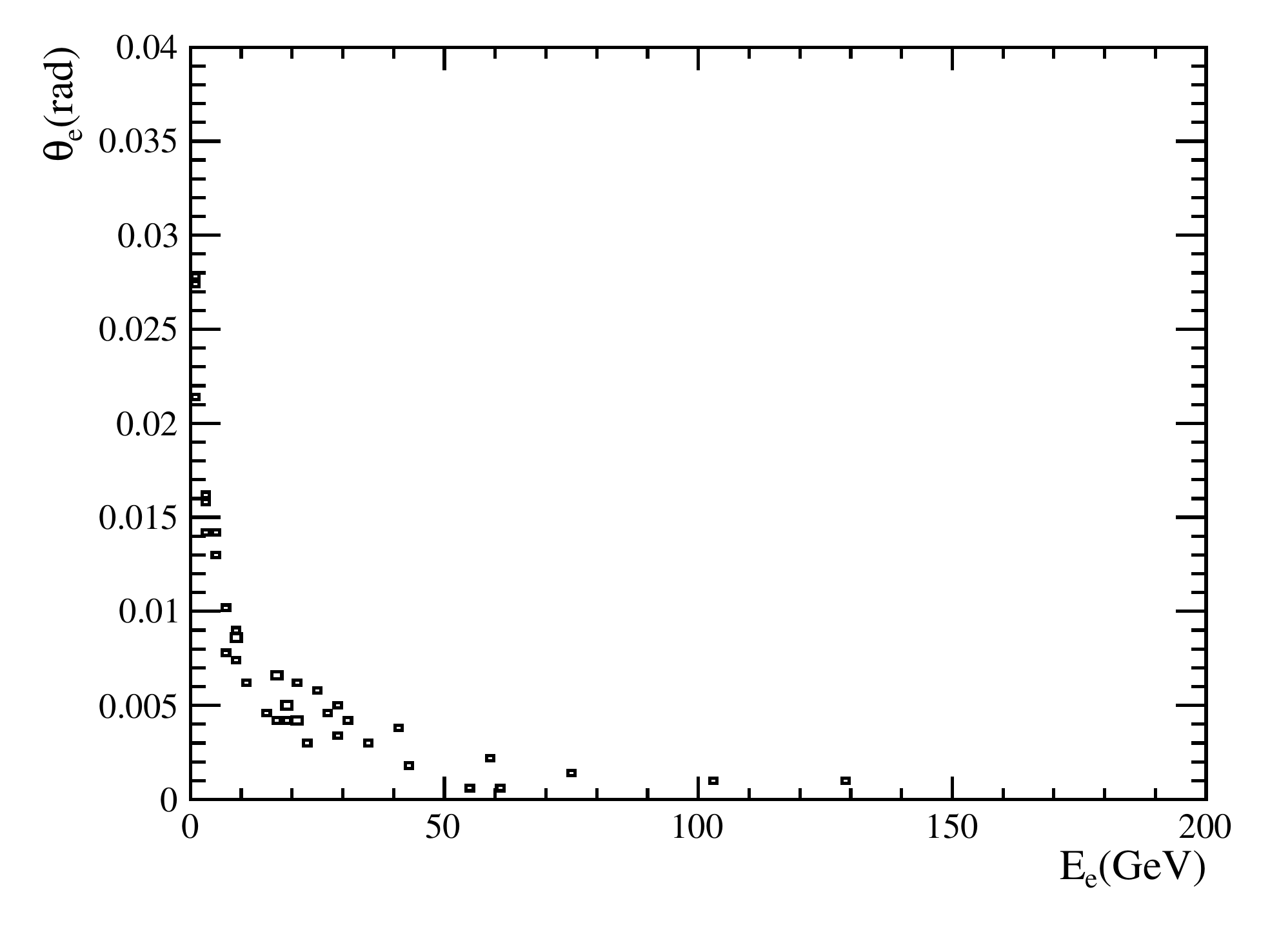}
\includegraphics[width=0.45\linewidth]{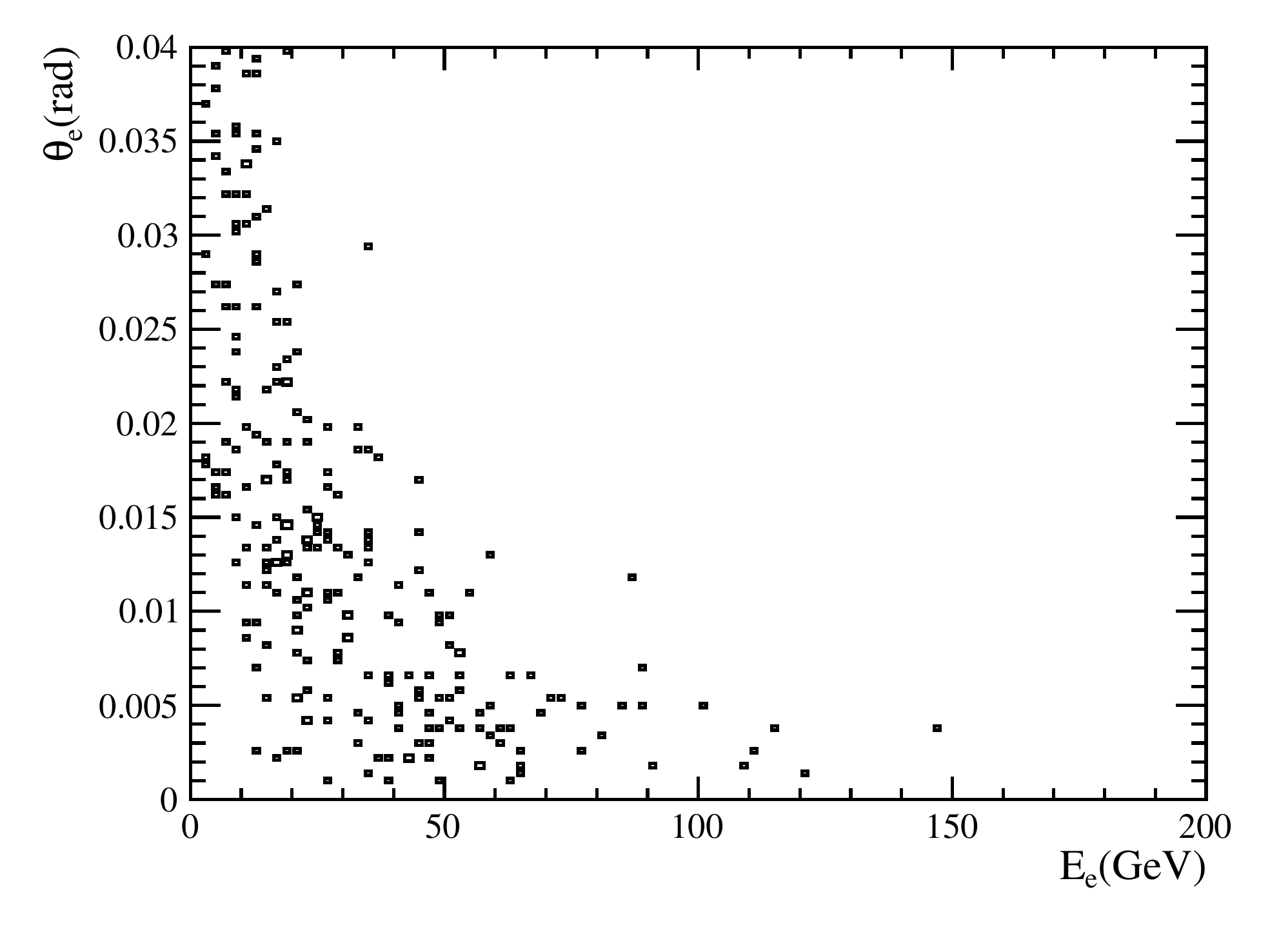}\\
\includegraphics[width=0.45\linewidth]{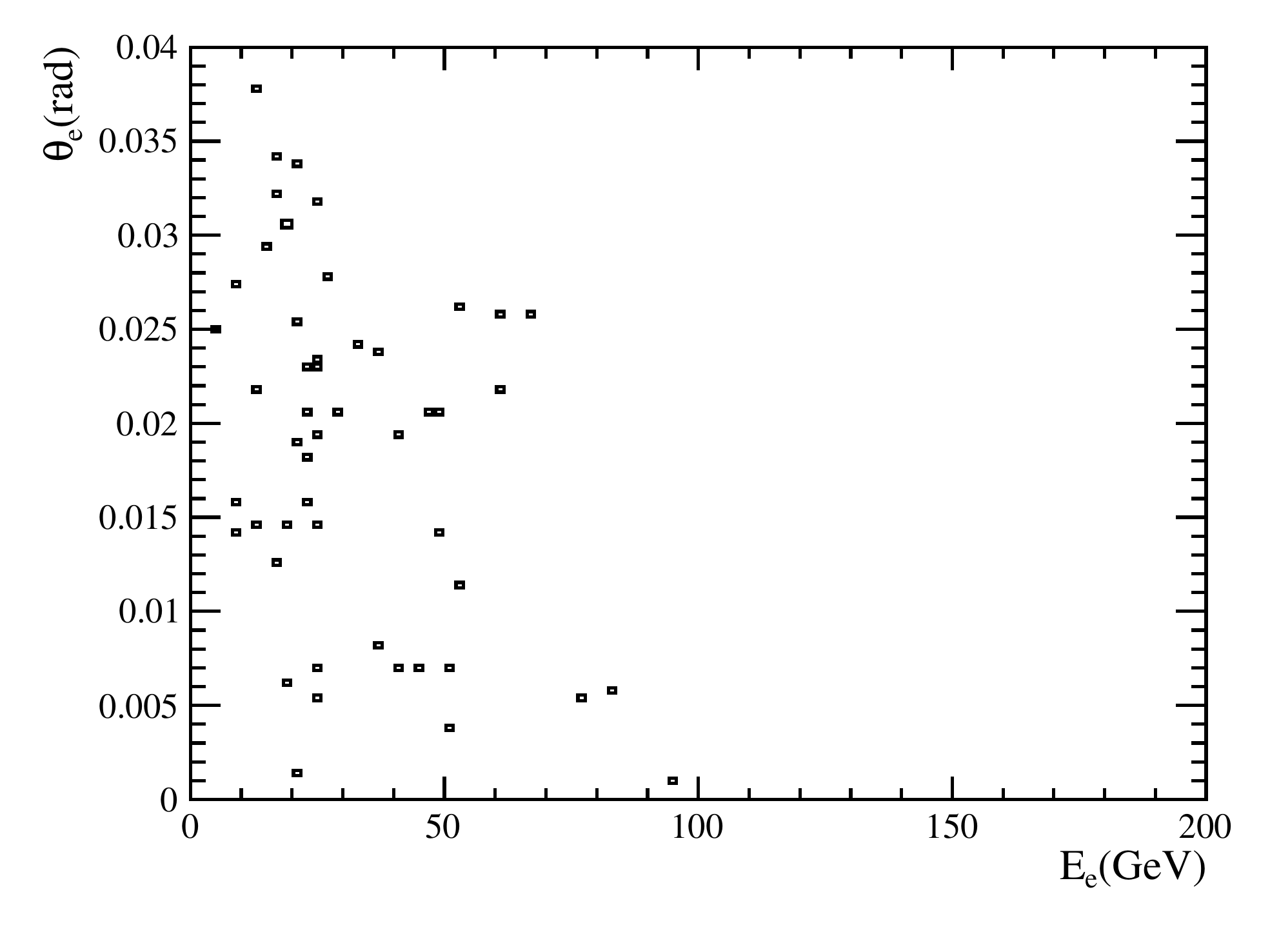}
\includegraphics[width=0.45\linewidth]{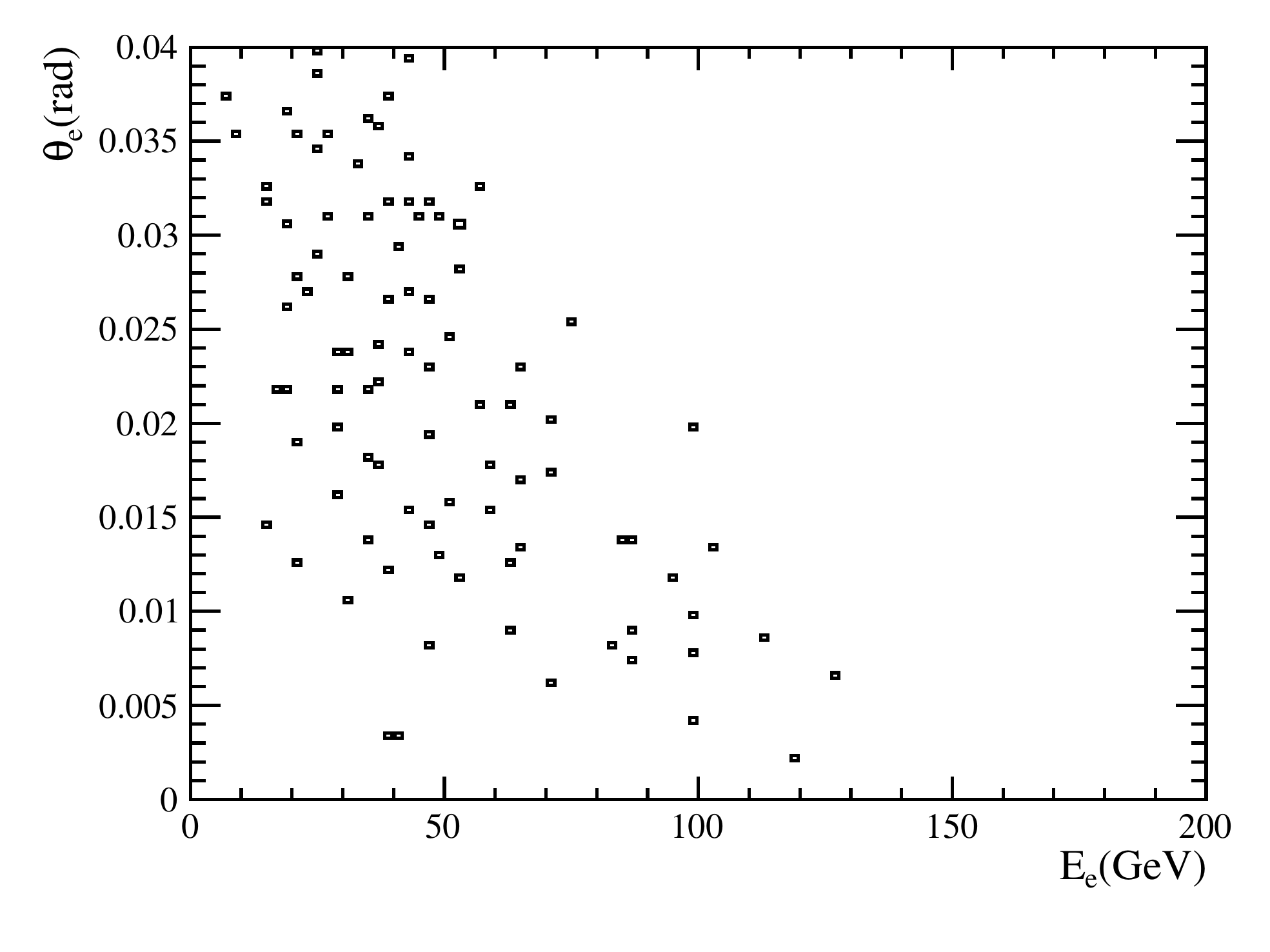}
\caption{Electron scattering angle versus the  electron energy  for elastic $\nu_e$ scattering on electrons (top-left), quasi-elastic (top-right), deep inelastic (bottom-left) and  resonant scattering (bottom-right) for candidates without visible tracks at the neutrino interaction vertex.}
\label{fig:bidi_DM}
\end{center}
\end{figure}

Figure~\ref{fig:bidi_DM} shows, for the  four neutrino scattering processes studied, for $\nu_e$ scattering, the correlation plots between the electron energy and the electron angle, with respect to the neutrino direction. 
The above-mentioned angular and energy cuts were applied to suppress the background. Table~\ref{tab:DM} shows, for the various background contributions, the number of expected events after cuts on a sample corresponding to the whole SHiP data-taking.The $\nu_\mu$ and $\nu_e$ fluxes  will be constrained from data with the study of $\nu_\mu$ and $\nu_e$ charged current deep inelastic scattering events, assuming Standard Model cross sections. The signal efficiency after cuts was estimated to be about~50\%.  Since only several possible combinations of dark photon and dark matter masses have been considered here, this background estimate has to be taken as an order of magnitude calculation rather than a precise computation of the sensitivity. A more detailed analysis is left for future studies.

\begin{table}[H]
\begin{center}
\label{tab:DM}
\caption{Number of  background events  to the dark matter search after cuts, from neutrino interactions with 2$\cdot 10^{20}$ p.o.t.}
\vspace{2mm}
\begin{tabular}{llllll}
\hline
  & $\nu_e$ & $\bar{\nu}_e$ & $\nu_\mu$ & $\bar{\nu}_{\mu}$   & all  \\
\hline
Elastic scattering on $e^-$   & 16 & 2 & 20 & 18 & 56\\
Quasi - elastic scattering  & 105 & 73&   & & 178 \\
Resonant scattering  & 13& 27 & & & 40\\
Deep inelastic scattering  &3 & 7& & & 10\\
\hline
Total               & 137 & 109 & 20  &18  & 284\\
\hline
\end{tabular}
\end{center}
\end{table}

\begin{figure}[H]
\centerline{\includegraphics[width=0.55\textwidth]{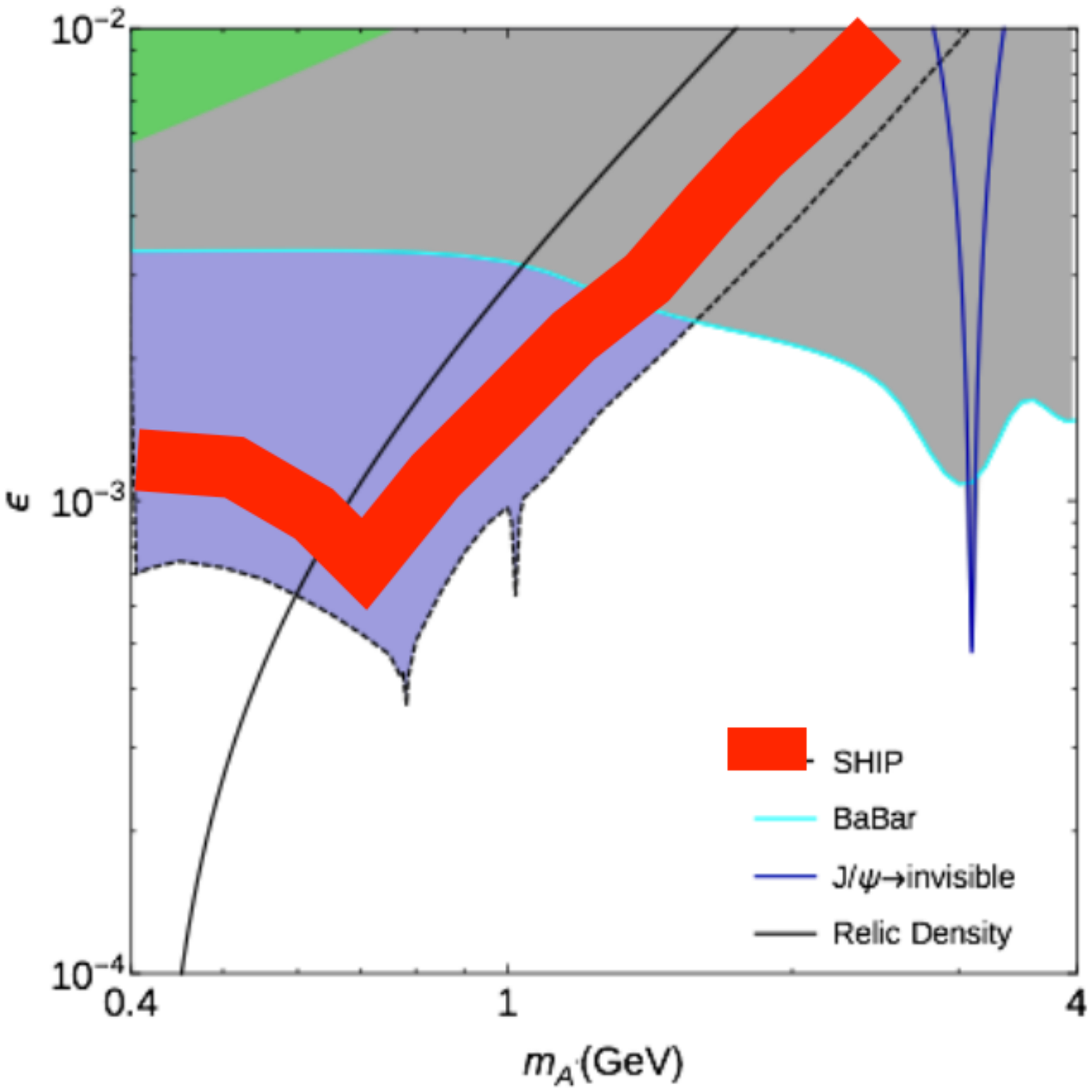}}
\caption{SHiP sensitivity to dark matter signal versus $m_A$. The red line gives the upper limit at 90\% C.L. The blue contour corresponds to 10 events sensitivity. The other lines  give constraints on the scattering of the 
light dark matter produced in $e$ and $p$ collisions with fixed targets. $\alpha_D= 0.1$, $m_\chi = 200$ MeV/c$^2$. $\varepsilon$ is the coupling constant of the dark photon to the electromagnetic current. }
\label{fig:VPfig6}
\end{figure}

With this number of background events and the above mentioned signal efficiency, we can estimate the upper limit at 90\%~CL~on dark matter signal, taking into account that the coupling $\varepsilon$ depends on the number of events with the fourth power. The 90\% CL upper limit is shown by the red line in Figure~\ref{fig:VPfig6}. 
The sensitivity is quite considerable and compares well with
previous experiments. 
An additional handle to suppress the background could come from studying the space distribution of the neutrino interaction vertex and will be a subject of future studies. 

In addition, in $U(1)_B$ models~\cite{Batell:2014yra} where a leptophobic mediator replaces the dark photon,
 the constraints from lepton-driven experiments (BaBar and E137) will be significantly relaxed and
SHiP will produce a strong additional reach to LSND and MiniBooNE.

%% file: cost_schedule/Cost_Schedule.tex
\chapter{Project Planning and Cost}
\label{sec:project}

\section{Project schedule}
\label{sec:schedule}

The SHiP project time line is motivated by the timeliness and the wide interest in the SHiP physics case, 
which uniquely extends the exploration at the intensity frontier. At the same time, the schedule has 
been defined to respect the operating schedule of the CERN accelerator complex, as well as the 
technical and financial constraints. In order to cause minimal interference to the 
operation of the existing facilities, the strategy for the SHiP project schedule is based on taking 
maximum advantage of the upcoming Long Shutdowns (LS). The start of data taking at the SHiP facility is 
planned for 2025 which is in time with the start of the LHC high luminosity phase in Run 4.

The relatively extensive civil engineering required for the SHiP facility is largely the main 
driver of the overall project schedule. Figure~\ref{fig:facility_wp} in Section~\ref{sec:civil_engineering} 
shows the four Work Packages (WP) prepared for the project schedule. The only point of interference with the 
existing facilities on the North Area concerns WP1, which consists of the construction of the junction 
cavern and the first 60~m of the SHiP beam extraction line. It includes the installation of the new 
splitter/switch in TDC2 and the SHiP bending magnets as well as services and sufficient shielding to 
work downstream when the North Area operation resumes. Considering that LS2 with 
its 21-months stop of the SPS beam to the North Area in 2018-2020 is the longest in the CERN accelerator 
schedule up to 2035, the SHiP project schedule targets this for WP1. The rest of the construction 
works may be performed during LHC Run 3 with the aim of commissioning and starting operation of the SHiP 
facility in 2025. This results in a realistic schedule with 10 years from submission 
of the Technical Proposal to full nominal physics data taking. 

\clearpage

\begin{sidewaysfigure}[htb]
\begin{center}
\includegraphics[width=0.99\linewidth]{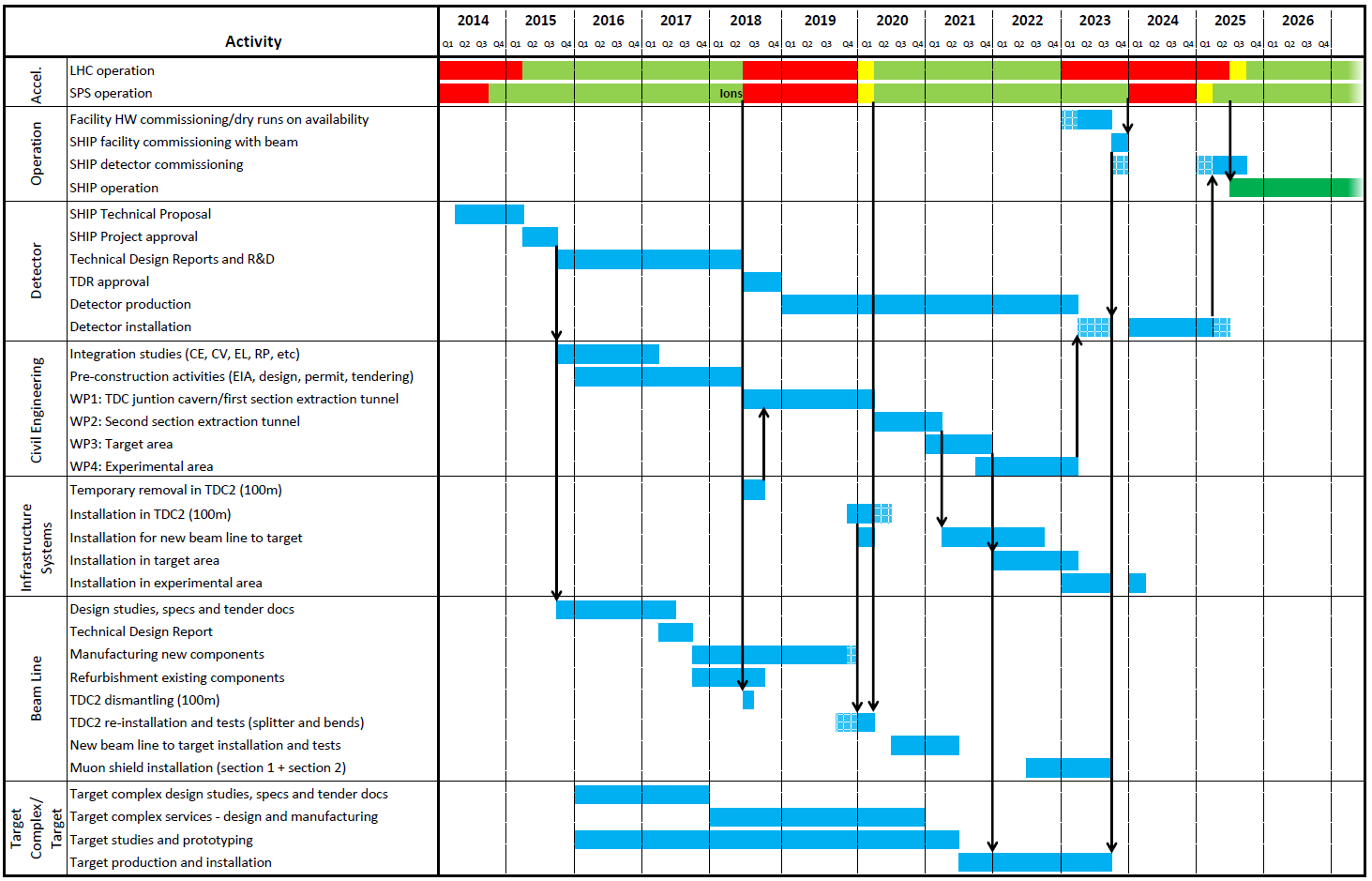}
\caption{Project schedule for the SHiP facility and detector.}
\label{fig:project_schedule}
\end{center}
\end{sidewaysfigure}

\clearpage

In order to commission efficiently the SHiP detector and rapidly reach nominal operation in 2025, it 
is crucial to validate the facility before the end of Run 3 and before the completion of the detector 
installation. In particular this includes:

\begin{enumerate}
\item{Setup and optimization of the SPS extraction for SHiP.}
\item{Validation and performance checks of the new TDC2 splitter/switch.}
\item{Validation and performance measurements of the target and validation of the operational aspects of the target and the target complex.}
\item{Validation of the performance of the muon shield.}
\item{Commissioning of selected detector components useful for the measurements of the residual particle fluxes.}
\item{Measurements of the physics background for the experiment with large area background counters.}
\end{enumerate}

Thus, the project schedule targets the delivery of the facility, including the muon shield and a 
set of background counters by beginning of 2023. This will then be followed by the last
phase of hardware commissioning of the facility and a few months of operation with beam at the 
end of 2023.

The detailed SHiP project time line is shown in Figure~\ref{fig:project_schedule}. It has been elaborated 
in collaboration with CERN's engineering, accelerator, and 
safety groups. This work is documented in the annexes listed in Appendix~\ref{sec:support_documents}.
The key milestones are:

\begin{itemize}
\item{Start of the global integration and design phase in 2016}
\item{Submission of TDRs in 2018}
\item{Start of civil engineering in 2018}
\item{Preparation of the detector production, including modules '0' in 2018}
\item{Start of detector production 2019}
\item{Start of detector installation in 2023}
\item{Facility commissioning run at the end of 2023}
\item{Detector commissioning and start of physics data taking in 2025}
\end{itemize}

Thus the schedule leaves three years for the detector R\&D and the preparation of the TDRs, and four 
years for the detector production phase.

The schedule relies critically on finalizing all preparatory works for the civil engineering before 
mid-2018, including the integration and the detailed design. It is assumed that an Environmental Impact 
Assessment is prepared and approved by the local authorities in parallel, prior to the submission 
of the building permit for the project and the tendering for the civil engineering contracts, to allow 
construction works to commence. The integration study with all concerned parties needs to be completed 
before the start of the detailed civil engineering design. The proposed work package structure 
allows the design and the integration studies to run in parallel work package by work package so that 
the civil engineering may start efficiently by no later than the start of LS2.

%This involves setting up a dedicated integration team looking at all aspects of the project 
%(beam line, detectors, HVAC, electricity, transport, radiation protection, safety, access etc.). 

Without optimization and without introducing significant parallel activities, the duration of the 
civil engineering for WP1 has been estimated at 21 months.  In this respect the current schedule 
does not allow sufficient time for the dismantling of the current beam line, infrastructure and 
services, and the reinstallation of the infrastructure and services, and the new splitter/switch 
and the SHiP bending magnets. The dismantling and the reinstallation have been estimated at four 
months and six months, respectively. 
However, WP1 involves large works outside of TDC2 which may be done in parallel with the 
dismantling, that is, the excavation of the first section of the SHiP extraction tunnel and the 
preparatory works for the construction of the junction cavern. It is also assumed that cooldown 
can at least partly be ensured by a North Area ion run at the end of Run 2. In the last phase, WP1 engages in 
the construction of the surface structures which may partly be done in parallel with reinstallation. 
Only a more comprehensive planning of the LS2 works will show if WP1 remits the start of the 
North Area physics after LS2 by three to six months. However, this should not have impact on the rest of 
the SHiP project schedule.

\begin{table}[htb]
\begin{center}
\caption{Resources required in late 2015 and in 2016 for starting the critical design activities in order 
to respect the project schedule.The manpower amounts to 10\% of the total manpower required to prepare 
the facility (Table~\ref{tab:SHiP_manpower}).\vspace{0.3mm}}
\label{tab:SHiP_earlyRandD}
\begin{tabular}{lccrr}
\hline
                         	        & \multicolumn{2}{c}{\bf Manpower (FTE)} & \multicolumn{2}{r}{\bf Cost (kCHF)} \\
{\bf Activity}                                & {\bf 2015 (0.5~year)} & {\bf 2016}              & {\bf 2015}    & {\bf 2016} \\ 
\hline
Integration studies (CV, EL, RP, HSE)   &    1.0           &   3.0            &         &   0    \\
Pre-construction activities (CE)        &  0.25 engineer   &   1.5 engineer   &         & 1000   \\
                                        &  0.5 fellow      &   1.0 fellow     &         &        \\
                                        &  0.5 draughtman  &   0.5 draughtman &         &        \\
Beam splitter/switch MSSB-S             &  0.5 fellow      &   1.0 fellow     &         &  100   \\
Target assembly                         &                  &   0.2            &         &   90   \\
Target water cooled shielding           &                  &   0.2            &         &   60   \\
\hline
{\bf Total manpower / Cost}                   &    {\bf 2.8}      &   {\bf 7.4}      &   {\bf 0}     &  {\bf 1250}  \\
{\bf Manpower by category}                        & \multicolumn{2}{c}{\bf 4.5(fellow)/5.7(staff)} &     &        \\
\hline
\end{tabular}
\end{center}
\end{table}

\subsection{Initial preparatory and R\&D activities on critical path}

Clearly the civil engineering works in LS2 are critical and therefore require a prompt start of the preparatory studies 
already at the end of 2015. In addition, the new splitter/switch in TDC2 must be ready for LS2 since 
they are needed for normal North Area operation, and can only be installed after a cooldown period 
to allow removal of the existing highly activated devices. The R\&D, the design, the production and 
the testing of these magnets are estimated at a total of 48 months (Section~\ref{sec:splitter_RandD}), 
implying a start of the design studies by the end of 2015.

The extensive R\&D and prototyping of the target and the target complex, as outlined in 
Section~\ref{sec:target_RandD}, require the design phase to start in 2016. 

Table~\ref{tab:SHiP_earlyRandD} summarizes the minimal set of resources which are required for the
 critical initial integration and design studies in order to respect the SHiP project schedule. The 
project schedule therefore requires an approval of these in 2015.

The reuse of the components of the OPERA detector for the SHiP facility must be clarified during 2016. 
 
\section{Cost and resources}
\label{sec:cost}

The cost estimate shown in Table~\ref{tab:SHiP_overall_cost} for the SHiP facility and the detector 
is based on a detailed breakdown of each item in the conceptual design, as described in this document. 
The breakdown of the detector per sub-system is shown in Table~\ref{tab:SHiP_detector_cost}. The detailed 
breakdown and resource requirements for the different items related to the facility can be found in the 
corresponding annexes listed in Appendix~\ref{sec:support_documents}.

\begin{table}[htb]
\begin{center}
\caption{Overall cost of the SHiP facility and the detectors.}
\label{tab:SHiP_overall_cost}
\vspace{2mm}
\begin{tabular}{lrr}
\hline
{\bf Item}                               & \multicolumn{2}{r}{\bf Cost (MCHF)} \\
\hline
{\bf Facility}                           &        &  {\bf   135.8} \\
\,\,\,\,\,\, Civil engineering           &  57.4  &               \\
\,\,\,\,\,\, Infrastructure and services &  22.0  &               \\
\,\,\,\,\,\, Extraction and beamline     &  21.0  &               \\
\,\,\,\,\,\, Target and target complex   &  24.0  &               \\
\,\,\,\,\,\, Muon shield                 &  11.4  &               \\
{\bf Detector}                           &        &  {\bf 58.7}   \\
\,\,\,\,\,\, Tau neutrino detector       &  11.6  &               \\
\,\,\,\,\,\, Hidden Sector detector      &  46.8  &               \\
\,\,\,\,\,\, Computing and online system &  0.2   &               \\
\hline
{\bf Grand total}                        &        &  {\bf 194.5} \\
\hline
\end{tabular}
\end{center}
\end{table}

\begin{table}[htb]
\begin{center}
\caption{Breakdown of the cost of the SHiP detectors.\vspace{0.3mm}}
\label{tab:SHiP_detector_cost}
\vspace{2mm}
\begin{tabular}{lrr}
\hline
{\bf Item}                                     & \multicolumn{2}{r}{\bf Cost (MCHF)} \\
\hline
{\bf Tau neutrino detector}              &        &  {\bf   11.6} \\
\,\,\,\,\,\, Active neutrino target      &   6.8  &               \\
\,\,\,\,\,\, Fibre tracker               &   2.5  &               \\
\,\,\,\,\,\, Muon magnetic spectrometer  &   2.3  &               \\
{\bf Hidden Sector detector}             &        &  {\bf   46.8} \\
\,\,\,\,\,\, HS vacuum vessel            &  11.7  &               \\
\,\,\,\,\,\, Surround background tagger  &   2.1  &               \\
\,\,\,\,\,\, Upstream veto tagger        &   0.1  &               \\ 
\,\,\,\,\,\, Straw veto tagger           &   0.8  &               \\
\,\,\,\,\,\, Spectrometer straw tracker  &   6.4  &               \\
\,\,\,\,\,\, Spectrometer magnet         &   5.3  &               \\
\,\,\,\,\,\, Spectrometer timing detector&   0.5  &               \\
\,\,\,\,\,\, Electromagnetic calorimeter &  10.2  &               \\ 
\,\,\,\,\,\, Hadronic calorimeter        &   4.8  &               \\
\,\,\,\,\,\, Muon detector               &   2.5  &               \\
\,\,\,\,\,\, Muon iron filter            &   2.3  &               \\
{\bf Computing and online system}          &        &  {\bf  0.2}  \\
\hline
{\bf Total detectors}                      &        &  {\bf 58.7}   \\
\hline
\end{tabular}
\end{center}
\end{table}

The cost estimate has been based on the situation of the Swiss Francs in mid-March 2015 (1 CHF = 0.93 Euro = 1 USD). 
Wherever applicable, the material cost estimate includes the industrial support labour. With the exception 
of civil engineering, the costs include no contingencies. All infrastructure systems have been 
included. The costs have all been cross-checked with similar installations built or studied at 
CERN. The estimate of the CERN staff manpower is shown in Table~\ref{tab:SHiP_manpower}. The cost of
fellows has been included in the cost of the facility.}

\begin{table}[htb]
\begin{center}
\caption{Required man power for constructing and setting up the SHiP facility. The cost of
fellows has been included in the cost of the facility in Table~\ref{tab:SHiP_overall_cost}.\vspace{0.3mm}}
\label{tab:SHiP_manpower}
\vspace{2mm}
\begin{tabular}{lrr}
\hline
{\bf Item}                                     & {\bf Staff} &  {\bf Fellows}   \\
                                               & {\bf FTE}   &  {\bf (MCHF)}    \\
\hline
{\bf Civil engineering}                        & {\bf 10}    &  {\bf 0.7}      \\
{\,\,\,\,\,\,Site investigation}               &             &                 \\
\,\,\,\,\,\,Junction cavern                  &             &                 \\
\,\,\,\,\,\,Extraction tunnel                &             &                 \\
\,\,\,\,\,\,Target complex                   &      10     &     0.7         \\
\,\,\,\,\,\,Muon shield section 1            &             &                 \\
\,\,\,\,\,\,Experimental area                &             &                 \\
\,\,\,\,\,\,New access road                  &             &                 \\
{\bf Infrastructure and services}              & {\bf 17.4}  &  {\bf 1.5}      \\
\,\,\,\,\,\,Cooling plants                   &       3     &                 \\
\,\,\,\,\,\,Ventilation plants               &       2     &                 \\
\,\,\,\,\,\,Electrical infrastructure        &       1.3   &     0.8         \\
\,\,\,\,\,\,Access \& safety                 &       1.5   &                 \\
\,\,\,\,\,\,Safety systems                   &       1.3   &                 \\
\,\,\,\,\,\,Radiation protection             &       4     &     0.7         \\
\,\,\,\,\,\,Transport: cranes,lifts, tooling &       1.3   &                 \\
\,\,\,\,\,\,Temporary removal/installation TDC2 &    3     &                 \\
{\bf Extraction and beamline}                  & {\bf 31.5}  &  {\bf 1.5}      \\
\,\,\,\,\,\,Extraction upgrades              &       1     &     0.1         \\
\,\,\,\,\,\,MSSB-S splitter/switch           &       3     &     0.2         \\
\,\,\,\,\,\,Other magnets                    &       2     &                 \\
\,\,\,\,\,\,Powering, incl. cables           &      12     &                 \\
\,\,\,\,\,\,Beam vacuum                      &       2     &     0.2         \\
\,\,\,\,\,\,Beam instrumentation             &       4     &     0.4         \\
\,\,\,\,\,\,Interlocks                       &       1     &                 \\
\,\,\,\,\,\,Other beam line costs            &       6.5   &     0.6         \\
{\bf Target and target complex}                & {\bf 35}    &  {\bf 2.6}      \\
\,\,\,\,\,\,Target \& exchange system        &             &     0.6         \\
\,\,\,\,\,\,Hadron absorber                  &             &     0.6         \\
\,\,\,\,\,\,Removable shielding              &             &                 \\
\,\,\,\,\,\,Helium enclosures                &      35     &     0.4         \\
\,\,\,\,\,\,Controls                         &             &     0.4         \\
\,\,\,\,\,\,Prototypes and testing           &             &     0.6         \\
{\bf Muon shield}                            & {\bf  1.8}  &  {\bf }         \\
\,\,\,\,\,\,Muon shield magnets              &       1.8   &                 \\
{\bf Detector}                               & {\bf  4}    &                 \\
\,\,\,\,\,\,Detector installation            &       4     &                 \\
{\bf Integration}                            & {\bf  3}    &                 \\
\,\,\,\,\,\,Integration and coordination     &       3     &                 \\
\hline
{\bf Total}                                  & {\bf 102.7} & {\bf 6.3}      \\
\hline
\end{tabular}
\end{center}
\end{table}

The production and the construction of the detector is based on the concept of deliverables, which 
includes the detector components and assembly, the associated electronics and infrastructure systems, 
as well as the transport to CERN and the specific operations related to the installation. 
The process of preparing a dedicated MoU has started in the Collaboration which will define the 
boundaries between responsibilities for the deliverables. It will also outline the financial 
strategy for the operation of the facility and the detectors.

\begin{figure}[htb]
\begin{center}
\includegraphics[width=0.9\linewidth]{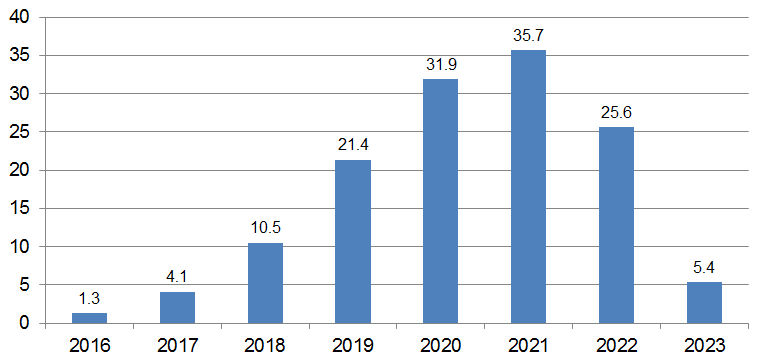}
\caption{Overall cost profile for the construction of the SHiP facility in MCHF (Total 135.8 MCHF). 
The detectors costs are not included.}
\label{fig:cost_profile}
\end{center}
\end{figure}

The schedule shown in Figure~\ref{fig:project_schedule} allows defining the cost profile for the 
SHiP project shown in Figure~\ref{fig:cost_profile}.

% CERN, as host laboratory, will provide the necessary services
% in terms of ventilation, power distribution, cooling water, gas distribution, survey network, 
% internal transport and cranes operation, as well as the Technical Coordination for the installation. 

\clearpage

\section{Organization of the Collaboration}
\label{sec:organization}

The members of the Collaboration have equal status in conducting the Experiment, including full voting 
rights and the right to be considered for appointment to official functions related to the Experiment. 
Currently, the Country Representatives Board (CRB) is the forum for decision and policy-making, and represents
the ultimate authority of SHiP. 

The CRB strives to reach consensus whenever possible. It elects its own Chairperson, as well as the Spokesperson, 
the CERN Local Contact person (or Technical Coordinator), and the Secretary. The current representatives per country
are listed in Table~\ref{tab:CRB}. The present (interim) appointments are:

\begin{tabular}{ll}
\\
Spokesperson: & A. Golutvin (Imperial College London)\\
CERN Local Contact person: & R. Jacobsson (CERN)\\
Chairperson of the CRB: & E. van Herwijnen (CERN)\\
\\
\end{tabular}

The election and voting procedure are described in the constitution.
The Spokesperson represents the Collaboration to the outside and lead the Collaboration in all day-to-day matters. 
The Spokesperson is elected for a term of 3 years and he/she can be re-elected several times.
The CERN Local Contact person is appointed by the CRB. The CERN Local Contact person takes care in particular 
of coordinating and overseeing the preparation of the different parts of the apparatus until completion of the 
construction of the experiment. He/She will also be responsible for the supervision of the maintenance of the 
apparatus and future upgrades.

The executive responsibility rests with the Spokesperson and the CERN Local Contact person. 
They are advised by the Working Group (WG) Conveners, composed of the coordinators of the major subsystems.
The body of WG Conveners is actively involved in the preparation and running of the Experiment. It also acts as 
steering, conference and publication committee.

% ~\footnote{ 
% M. Shaposhnikov (Theory), D. Gorbunov (Theory), M. Ferro-Luzzi (Tracking), G. de Lellis ($\nu_\tau$ detector),
% W. Bonivento (PID), G. Lanfranchi (Muon), M. Villa (Calorimeter), V. Egorychev (Calorimeter), B. Storaci (Timing and front tagger),
% H. Lacker (Backgroud tagger), A. Malinin (Vacuum vessel), H. Dijkstra (Online \& trigger), N. Serra (Physics), F. Rademakers
% (Computing), T. Ruf (Simulation), A. Golutvin (Spokesperson),
% R. Jacobsson (Facility, CERN Local Contact person), E. van Herwijnen (secretary).}. 

The Plenary meeting, chaired by the Spokesperson, is open to everyone in the Collaboration.
Physics and technical issues are discussed in the Plenary meeting before any major decision is taken.
Decisions taken by the CRB are reported at the Plenary meeting.

\section{Responsibilities}
\label{sec:responsibilities}

Since the Collaboration is still experiencing substantial growth the detailed assignment of the responsibilities 
may still change
until after the approval of the experiment. The capabilities and resources of the institutes are the major criteria, 
in line with
their
interests and R\&D activities in preparation for the Technical Proposal.
The current assignment of responsibilities and interests are summarized in Table~\ref{tab:interests}.
\begin{table*}[p]
\caption{Present SHiP Country Representative Board members}
\label{tab:CRB}
\vspace{2mm}
\centering
\begin{tabular}{llll}
\hline
{\bf Country}& {\bf Representative}& {\bf Country}& {\bf Representative} \\
\hline
Bulgaria& R. Tsenov& Sweden& D. Milstead/R. Brenner\\ 
Chile & H. Hakobyan& CERN & H. Dijkstra\\ 
Denmark& S. Xella&   Switzerland& N. Serra\\ 
France& J. Chauveau& Turkey & A.M. G\"uler\\ 
Germany& H. Lacker& United Kingdom&M. Campanelli \\ 
Italy& W. Bonivento& Ukraine& I. Kadenko \\ 
Japan& M. Komatsu& United States of America& G. Mitselmakher \\
Russia& N. Polukhina/V. Shevchenko&  &   \\ 
\hline  
\end{tabular}
\end{table*}

%\begin{table*}[t]

\begin{table*}[t]
%\scriptsize
\caption{Interests expressed by the institutes in the construction of SHiP components.}
\label{tab:interests}
\vspace{2mm}
\centering
\begin{tabular}{ll}
\hline
{\bf Component}&{\bf Institutes}\\ \hline
Beamline and target & CERN\\ 
Infrastructure & CERN \\ 
Muon shield & RAL, Imperial College, Warwick \\ 
HS vacuum vessel & NRC KI \\  
Straw tracker & CERN, JINR, MEPhI, PNPI\\ 
ECAL & ITEP, Orsay, IHEP, INFN-Bologna \\ 
HCAL& ITEP, IHEP, INFN-Bologna, Stockholm\\ 
Muon& INFN-Bologna, INFN-Cagliari, INFN-Lab. Naz. Frascati,\\ 
&INFN-Ferrara, INR RAS, MEPhi\\ 
Surrounding background tagger& Berlin, LPNHE, MEPhI\\ 
Timing detector and upstream veto& Z\"urich, Geneva, INFN-Cagliari, Orsay, LPNHE\\ 
$\nu_\tau$ detector emulsion target, &INFN-Naples, INFN-Bari, INFN-Lab. Naz. Gran Sasso, \\
&Nagoya, Nihon, Aichi, Kobe, Moscow SU, \\
&Lebedev, Toho, Middle East Technical University, Ankara\\ 
$\nu_\tau$ detector tracker & NRC KI, INFN-Lab. Naz. Frascati \\ 
$\nu_\tau$ detector magnet & INFN-Lab. Naz. Frascati, INFN-Bari, INFN-Naples, \\
&INFN-Roma \\ 
$\nu_\tau$ tracking system (RPC) & INFN-Lab. Naz. Frascati, INFN-Bari, \\
&INFN-Lab. Naz. Gran Sasso, INFN-Naples, INFN-Roma \\ 
$\nu_\tau$ tracking system (drift tubes) & Hamburg \\ 
Online computing & CERN, Niels Bohr, Uppsala, UCL, YSDA, LPHNE \\ 
Offline computing & CERN, YSDA \\ 
MC simulation & CERN, Sofia, INFN-Cagliari, INFN-Lab. Naz. Frascati,\\ 
&INFN-Napoli, Z\"urich, Geneva and EPFL Lausanne, \\
&Valparaiso, Berlin, PNPI, NRC KI, SINP MSU, MEPhI, \\
&Middle East Technical University, Ankara, Bristol, YSDA,\\
&Imperial College, Florida, Kyiv\\ 
\hline
\end{tabular}
\end{table*}

%% file: annexes/Annexes.tex
\begin{appendices}

%\addcontentsline{toc}{chapter}{List of support documents}

\chapter{List of support documents}

\label{sec:support_documents}

\subsection*{Annex 1: An experiment to Search for Hidden Particles (SHiP) at the SPS North Area}
\label{sec:annex_task_force}
EDMS: https://edms.cern.ch/document/1369559 (SHiP-TP-2015-A1) \\

This document reports on a preliminary evaluation of the feasibility and the implications of the SHiP facility at CERN 
by a joint study group between the Accelerator Sector, the HSE unit and members of the SHiP experiment, that was convened by the 
Enlarged CERN Directorate on February 4, 2014. It was aimed at producing a layout, a timeline, and an estimate of the resources
required to setup the facility, as well as identifying issues requiring further investigations.

In view of the encouragement to produce a Technical Proposal and a Physics Proposal, it was agreed to allow the joint
study group to contribute to the preparation of the Technical Proposal by investigating further the main issues
identified in the preliminary report. These corresponded to finalizing

\begin{itemize}
\item{a working point for the SHiP facility in terms of the beam parameters and the SPS configuration,}
\item{a strategy for the SPS extraction and the beam transfer up to the target including the required splitter/switch at the North Area,} 
\item{a first complete design of the target and the target complex,}
\item{the conclusions on the situation with respect to the radiological aspects and the associated actions required to respect the CERN regulations,}
\item{and updating the civil engineering for the final layout of the facility, the work plan, and a revised cost estimate.}
\end{itemize}

These studies are reported on in detail in the documents listed in Annex 2 - 6. As such, those documents supersede the 
information in the preliminary report in Annex 1 on those particular subjects.

\subsection*{Annex 2: The SPS beam parameters, the operational cycle, and proton sharing with the SHiP facility}
\label{sec:annex_beam_parameters}
EDMS: https://edms.cern.ch/document/1498984 (SHiP-TP-2015-A2) \\

\subsection*{Annex 3: Extraction and beam transfer for the SHiP facility}
\label{sec:annex_beam_transfer}
EDMS: https://edms.cern.ch/document/1495859 (SHiP-TP-2015-A3) \\

\subsection*{Annex 4: Design of the SHiP target and target complex}
\label{sec:annex_target}
EDMS: https://edms.cern.ch/document/1465053 (SHiP-TP-2015-A4) \\

\subsection*{Annex 5: Radiation protection studies for the SHiP facility}
\label{sec:annex_rp}
EDMS: https://edms.cern.ch/document/CERN-RP-2015-020-REPORTS-TN (SHiP-TP-2015-A5) \\

\subsection*{Annex 6: Civil engineering for the SHiP facility}
\label{sec:annex_ce}
EDMS: https://edms.cern.ch/document/1499253 (SHiP-TP-2015-A6) \\

\chapter{Other facilities}
\label{app:others}

Hidden particles could in principle be searched for below the charm mass with already existing 
or planned fixed target experiments. These experiments 
would have to cope with the same background issues as SHiP, and would therefore need to be 
extensively modified involving significant costs. Some modifications required are also incompatible 
with the planned physics programmes for these experiments. Nevertheless, for comparison the sensitivity
reach of these experiments have been investigated assuming that the background is fully dealt with.
 
Among the existing experiments, the most suited for searching for HNLs could be the NA62 experiment 
at the CERN SPS. It has a similar structure to SHiP with a long vacuum tank followed by a spectrometer 
with particle identification detectors. It is conceivable that, after the current run which 
is dedicated to the search for the $K^+\to\pi^+\nu\bar{\nu}$ decay, this experiment could be adapted
to a search for HNLs. At present, its maximum proton beam intensity on target is limited to 
$4\cdot 10^{12}$ protons per spill with a 30~s cycle. This leads to an annual integrated intensity that is a 
factor of 40 smaller than SHiP. Its acceptance to HNLs was calculated assuming that the 
detector, starting from the front window of the vacuum tank, is moved to a distance of 80~m
from the target. This is required by the need of a muon shielding. As a result the geometrical
acceptance is about 1/8th of SHiP for HNLs with mass of 1~GeV/c$^2$. Consequently, the statistical 
sensitivity reach in $U^2$ would be more than a factor 15 worse than SHIP, assuming no background.

%\begin{figure}[htb]
%\begin{center}
%\includegraphics[width=0.6\linewidth]{requirements/req_cross_section.png}
%\caption{Total $c\overline{c}$ production cross-sections at fixed-target energies~\cite{doublecharm}.}
%\label{fig:req_cross_section}
%\end{center}
%\end{figure}

One of the possible future beamlines is the 120~GeV beams at FNAL, with the expected PIP-II 
upgrade to 1.2MW by 2023 and possibly 2.3MW by 2030. This is a conventional neutrino beam aimed 
at studying long baseline neutrino oscillations. 
The double charm production cross-section as a function of the centre-of-mass energy at fixed 
target experiments is shown in Figure~\ref{fig:req_cross_section}; at 120~GeV ($E_{cms}\sim$15.0~GeV)
the charm cross section is about 1/8th of that at %400~GeV 
($E_{cms}\sim$27.4~GeV). The planned LBNE experiment on the FNAL beam line includes a near detector 
located at 500~m downstream of the target, which could also be the location of an experiment 
searching for HNLs. A case study for an HNL search has been performed~\cite{Adams:2013qkq}, ignoring
all backgrounds. However, the conventional neutrino beam produces huge backgrounds 
from active muon neutrinos, estimated to be several orders of magnitude larger than the background at SHiP 
in the detector acceptance. Therefore, in particular below the kaon mass, reaching zero background 
from $V^0$ production from muon neutrinos would be impossible. Above the kaon mass the statistical 
sensitivity reach in $U^2$ even with zero background would be more than a factor ten worse than at SHiP. 

% Hidden particles can also be searched for at the CERN LHC. However, due the need of going to the 
% very high luminosities to reach values of $U^2$ below $10^{-7}$, the background suppression will
% be insufficient for the simple topologies of the HNL decays.

% Fixed target experiments of the type described above could in principle be performed
% with beams of different energies. We use for this discussion the beams proposed in the past at FNAL 
% and KEK such as the 800 and 120~GeV beams at FNAL with 1$\cdot$10$^{19}$ and 
% 4$\cdot$10$^{19}$ protons on target, respectively, or the 30~GeV beam at KEK with 1$\cdot$10$^{21}$ 
% protons on target. At FNAL, the 800~GeV beam would give a similar production of charm mesons to that 
% of the SHiP facility, i.e. the lower proton intensity would be approximately compensated by the 
% increase in the charm cross-section at the higher energy~\cite{adams2009}. However, a significantly 
% more complicated and longer muon shield would be required due to the higher beam energy, leading 
% to a significant loss of acceptance. The FNAL 120~GeV beam would have a factor ten lower event yield 
% than in the proposed SHiP experiment at the SPS, while the KEK beam would have a 
% factor $1.5 - 2$ lower yield. The latter estimate has a large uncertainty due to the poor knowledge of the 
% charm cross section at low energies.
 
Hidden particles can also be searched for at the CERN LHC. The sensitivity of the colliding beam 
experiments is estimated assuming a luminosity of 1000~fb$^{-1}$ and an energy of 14~TeV, as is 
foreseen in three to four years of operation in the high luminosity phase. For the sake of comparison, 
a 100~m long and a 5~m diameter cylindrical decay volume is hypothetically located 60~m away from an 
equivalent interaction region and 50~mrad off-axis. The overall 
yield for HNL would approximately be a factor of 200 smaller in the collider detectors than 
in the SHiP detector.

A $e^+e^-$ flavour factory with a mean luminosity of $10^{36}\rm ~cm^{-2}s^{-1}$ would produce $\sim 2\cdot10^{9}$ $B$-mesons
which decay semi-leptonically per year. A HNL of 1 GeV with $U_{\mu}^2=10^{-8}$ has a $c\tau$ of over 50 km, hence direct 
detection of HNL decays is excluded even for B-mesons produced at rest. Detection via missing mass would require much larger statistics due to 
large background levels.

\end{appendices}